\documentclass[10pt,a4paper]{book}
\usepackage[a4paper, total={6in, 8in}]{geometry}
\usepackage{wrapfig}

\usepackage[utf8]{inputenc}
\usepackage[vietnamese]{babel}
\usepackage[titletoc]{appendix}

\addto\captionsvietnamese{%

}
\usepackage{mathrsfs}
\usepackage{rotating}

\usepackage{graphicx}
\usepackage{hyperref}

\usepackage{multicol}

\usepackage{etoolbox}
\patchcmd{\abstract}{\titlepage}{\titlepage 
\addcontentsline{toc}{chapter}{Abstract}}{}{}
\usepackage{imakeidx}
\makeindex

\usepackage{amstext,latexsym,amsbsy,amssymb,amsmath,amsthm}
\usepackage[absolute,overlay]{textpos}
\usepackage{amsmath,bm}
\usepackage{amsmath,amssymb,mathrsfs}
\usepackage{bbm}
\usepackage{bbold}
\usepackage{mathbbol}
\usepackage{dsfont}
\usepackage{listings}
\usepackage[framed,numbered,autolinebreaks,useliterate]{mcode}

\newcommand{\m}[1]{\mathbf{#1}}
\newcommand{\mc}[1]{\mathcal{#1}}
\newcommand{\mb}[1]{\mathbb{#1}}
\newcommand{\mcl}[1]{\boldsymbol{\mathcal{#1}}}

\newcommand{\bmm}[1]{\boldsymbol{#1}}

\usepackage{enumitem}

\usepackage[caption=0]{subfig}
\usepackage{multirow}
\usepackage{longtable}
\usepackage[leftcaption]{sidecap}
\usepackage{optidef}
\usepackage{booktabs}
\usepackage{tabularx}

\theoremstyle{plain}
\newtheorem{thm}{Theorem}[chapter] 
\theoremstyle{definition}
\newtheorem{Definition}{Định nghĩa}[chapter] 
\newtheorem{example}{Ví dụ}[chapter]
\newtheorem{exercise}{Bài tập}[chapter]
\newtheorem{remark}{Ghi chú}[chapter]
\newtheorem{theorem}[thm]{Định lý}
\newtheorem{lemma}[thm]{Bổ đề}
\newtheorem{corollary}[thm]{Hệ quả}

\usepackage[most]{tcolorbox}
\usepackage{xcolor}

\newtcolorbox{story}[1]{%
  enhanced,
  breakable,
  title={#1},
  colback=blue!5,
  colframe=blue!50!black,
  fonttitle=\bfseries\itshape,
  borderline west={3pt}{0pt}{blue!50!black},
  boxrule=0pt,
  left=8pt,
  right=6pt,
  top=6pt,
  bottom=6pt
}

\usepackage{algorithmic}

\usepackage{savesym}
\savesymbol{AND}
\savesymbol{OR}
\savesymbol{NOT}
\savesymbol{TO}
\savesymbol{COMMENT}
\savesymbol{BODY}
\savesymbol{IF}
\savesymbol{ELSE}
\savesymbol{ELSIF}
\savesymbol{FOR}
\savesymbol{WHILE}

\usepackage{tikz}
\usepackage{tikz,tkz-euclide}
\usetikzlibrary{calc}
\usetikzlibrary{decorations.markings}
\tikzset{>=latex} 
\usetikzlibrary{backgrounds}
\usetikzlibrary{positioning}
\usetikzlibrary{shapes.geometric, calc}
\tikzset{
    position/.style args={#1:#2 from #3}{
        at=(#3.#1), anchor=#1+180, shift=(#1:#2)
    }
}
\usetikzlibrary{patterns}
\usepackage{float}
\usepackage{pgfplots}
\pgfplotsset{compat=1.18}

\usepackage[ruled,linesnumbered]{algorithm2e}

\usetikzlibrary{shapes,arrows}
\tikzstyle{decision} = [diamond, draw, fill=blue!20, text width=4.5em, text badly centered, node distance=3cm, inner sep=0pt]
\tikzstyle{block1} = [rectangle, draw, text width=8em, text centered, minimum height=4em]
\tikzstyle{block2} = [rectangle, draw, text width=3em, text centered, minimum height=4em]
\tikzstyle{block3} = [rectangle, draw, text width=11em, text centered, minimum height=12em, dashed,black]
\tikzstyle{block4} = [rectangle, draw, text width=11em, text centered, minimum height=18em, dashed,black]
\tikzstyle{block5} = [rectangle, draw, text width=11em, text centered, minimum height=32em, dashed,black]
\tikzstyle{block6} = [rectangle, draw, text width=11em, text centered, minimum height=18.5em, dashed,black]
\tikzstyle{block7} = [rectangle, draw, text width=11em, text centered, minimum height=11.8em, dashed,black]
\tikzstyle{line01} = [draw, -latex']
\tikzstyle{line02} = [draw, latex'-latex']

\usepackage{nomencl}
\makenomenclature

\author{Trịnh Hoàng Minh, Nguyễn Minh Hiệu}
\title{Điều khiển hệ đa tác tử}

\begin{document}
%
\begingroup
\thispagestyle{empty} 
\begin{tikzpicture}[remember picture,overlay]
\node[inner sep=0pt] (background) at (current page.center) {\includegraphics[width=\paperwidth]{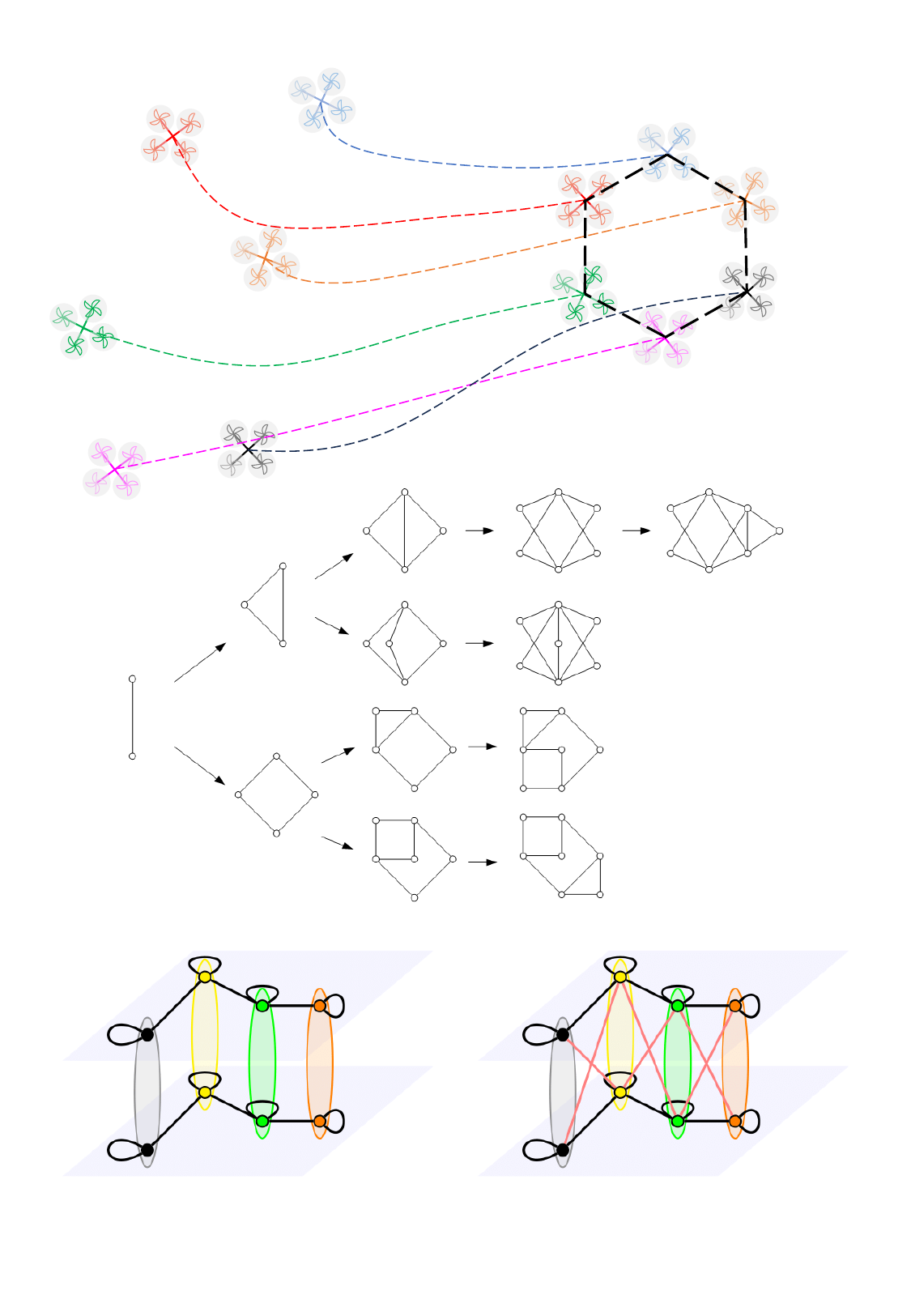}};
\node[inner sep=0pt] (background) at (current page.center) {};
\draw (current page.center) node [fill=blue!10,fill opacity=0.8,text opacity=1,inner sep=1cm]{\Huge\centering\bfseries\sffamily\parbox[c][][t]{\paperwidth}{\centering \textcolor{black}{ĐIỀU KHIỂN HỆ ĐA TÁC TỬ}\\[16pt]
{\Large \textcolor{black}{Trịnh Hoàng Minh, Nguyễn Minh Hiệu}}}};
\end{tikzpicture}
\vfill
\endgroup
\maketitle


\begin{center}\large

  \emph{Điều khiển hệ đa tác tử} \\
  Trịnh Hoàng Minh, Nguyễn Minh Hiệu\\
  Phiên bản: 02/2026
    
\end{center}

{\normalsize
  
  \noindent\emph{Trang web của sách:} các phiên bản điện tử, code mô phỏng, và cập nhật của tài liệu này được phân phối tại website:
  \begin{center}
  \url{https://sites.google.com/view/minhhoangtrinh/dieu-khien-he-da-tac-tu}
  \end{center}
  \bigskip

  \noindent\emph{Bản quyền:} Bản quyền tài liệu này thuộc về các tác giả. Độc giả không được sao chép lại nội dung trong tài liệu này với mục đích thương mại, nhưng có thể sử dụng với mục đích học tập, nghiên cứu.
  
  \bigskip

\chapter*{Preface}
\addcontentsline{toc}{chapter}{Preface}
Since the early 2000s, control of multiagent systems has attracted significant research interest, with applications ranging from natural collective behaviors and social dynamics to engineered systems such as autonomous vehicles, sensor networks, and smart grids.

Although research on multi-agent systems has diversified into numerous specialized directions, textbooks-including those in English-that provide a systematic treatment of the fundamental principles of multi-agent system control remain scarce. The material presented in this book has been developed and used in teaching since 2021, initially as a concise Vietnamese-language reference for the courses Networked Control Systems and Control of Multi-Agent Systems at Hanoi University of Science and Technology. The book focuses on a selection of fundamental  topics of broad and continuing interest in the field. The complexity of several topics is asymptotic to that encountered in research-level studies, however, the analysis is presented in a step-by-step manner to facilitate access to commonly used methods and tools.

The material is divided into three main parts. Part I introduces multiagent systems and basic graph-theoretic concepts. Part II addresses the design and analysis of linear consensus algorithms. Part III covers selected applications and research directions, including formation control, network localization, distributed optimization, opinion dynamics, and matrix-weighted networks. Each chapter concludes with notes on notable researchers in this field, further reading, and exercises.

This book cannot be completed without the encouragement, support and suggestions from families, colleagues and friends. The authors appreciate feedback from readers to further improve the content of the book.\\
Trịnh Hoàng Minh (\texttt{trinhhoangminhbk@gmail.com}), Tayan Polytechnic, Hanoi, Vietnam  \\ 
Nguyễn Minh Hiệu (\texttt{nmhieulc1@gmail.com}), HCLTech Vietnam - ANZx, Hochiminh City , Vietnam

\chapter*{Lời nói đầu}

\addcontentsline{toc}{chapter}{Lời nói đầu}
Điều khiển hệ đa tác tử là một hướng nghiên cứu đã và đang được quan tâm trên thế giới từ khoảng đầu những năm 2000. Nội dung nghiên cứu bao gồm các hệ đa tác tử trong tự nhiên (hiện tượng tụ bầy ở chim, cá, đồng bộ hóa ở côn trùng,...), các hiện tượng xã hội (mạng xã hội, mạng học thuật, mô hình lan truyền dịch bệnh,...), và trong kĩ thuật (hệ các robot tự hành, mạng cảm biến, lưới điện và hệ thống giao thông thông minh, hệ tính toán phân tán,...).

Mặc dù nghiên cứu về các hệ đa tác tử hiện nay đã phát triển theo nhiều hướng chuyên sâu khác nhau, số lượng giáo trình—kể cả bằng tiếng Anh—bao quát một cách hệ thống các kiến thức cơ bản về điều khiển hệ đa tác tử vẫn còn hạn chế. Tài liệu này được biên soạn và đưa vào giảng dạy từ năm 2021, với mục tiêu ban đầu là cung cấp một nguồn tham khảo ngắn gọn bằng tiếng Việt cho học viên hai học phần \emph{Điều khiển nối mạng} và \emph{Điều khiển hệ đa tác tử} tại Đại học Bách Khoa Hà Nội. Nhóm tác giả lựa chọn và trình bày một số vấn đề đang nhận được sự quan tâm rộng rãi trong lĩnh vực hệ đa tác tử và trình bày dưới dạng một sách giáo trình. Nội dung một số chủ đề có mức độ phức tạp tiệm cận các công bố khoa học, nhưng được chia nhỏ và trình bày thành từng bước để giúp độc giả dễ tiếp cận các phương pháp và công cụ phân tích thường dùng.

Nội dung tài liệu này được chia thành ba phần chính. Phần I giới thiệu về hệ đa tác tử và cung cấp một số kiến thức cơ bản về lý thuyết đồ thị. Phần II tập trung vào thiết kế, phân tích các luật đồng thuận và đồng bộ hóa. Phần III giới thiệu về một số ứng dụng của hệ đa tác tử bao gồm điều khiển đội hình, định vị mạng cảm biến, tính toán và tối ưu phân tán với ứng dụng trong bài toán sản xuất-phân phối, một số mô hình động học quan điểm, và mô hình đồ thị trọng số ma trận. Các ghi chú mở rộng về tài liệu tham khảo, bài tập, và giới thiệu một số chuyên gia trong lĩnh vực cũng được trình bày tại mỗi chương. Phần phụ lục cung cấp một số kiến thức liên quan để độc giả tiện tra cứu.

Cuốn sách này sẽ không thể hoàn thành nếu thiếu sự động viên, hỗ trợ và góp ý của gia đình, bạn bè và đồng nghiệp. Nhóm tác giả hi vọng sẽ nhận được phản hồi của độc giả để tiếp tục hoàn thiện nội dung cuốn sách.\\
Trịnh Hoàng Minh (\texttt{trinhhoangminhbk@gmail.com}), Tayan Polytechnic, Hà Nội \\ 
Nguyễn Minh Hiệu (\texttt{nmhieulc1@gmail.com}), HCLTech Vietnam - ANZx, Thành phố Hồ Chí Minh

\clearpage

\tableofcontents

\listoffigures

\listoftables


\chapter*{Danh mục kí hiệu}
\addcontentsline{toc}{chapter}{Danh mục kí hiệu}
\vspace{.8cm}
\renewcommand{\nomname}{Danh mục kí hiệu} 
\renewcommand{\nompreamble}{Các kí hiệu sau đây sẽ được sử dụng trong tài liệu.}
\begin{tabular}{lp{.749\textwidth}}
  $\mb{N}, \mb{Z}, \mb{R}, \mb{C}$ & Tập hợp các số tự nhiên, số nguyên, số thực, số phức  \\
  $\mb{N}_+, \mb{Z}_+, \mb{R}_+$ & Tập hợp các số tự nhiên dương, số nguyên dương, số thực dương  \\
  $\mb{R}^d, \mb{C}^d$ & Không gian các vector $d$ chiều với các phần tử nhận giá trị thực, phức \\
  $\mb{R}^{m \times n}, \mb{C}^{m \times n}$ & Không gian các ma trận kích thước $m \times n$ với các phần tử nhận giá trị thực, phức \\
  $\alpha, \beta, \gamma, \ldots$, $a, b, c, \ldots$ & Các đại lượng vô hướng hoặc các hàm nhận giá trị vô hướng \\
  $\m{a}, \m{b}, \m{c}, \ldots$ & Các vector \\
  $\m{A}, \m{B}, \m{C}, \ldots$ & Các ma trận \\
  $\mc{A}, \mc{B}, \mc{C}, \ldots$ & Các không gian vector con hay tập con của $\mb{R}^d$\\
  ${A}, {B}, {C}, \ldots$ & Các đồ thị hoặc các tập hợp liên quan đến đồ thị\\
  $\imath, \jmath $ & Đơn vị ảo $\imath^2=\jmath^2=-1$\\
  ${\rm Re}(s), {\rm Im}(s)$ & Phần thực, phần ảo của số phức $s$\\
  $\mathtt{e}$ & Hằng số Euler ($\mathtt{e} \approx 2,71828$) \\
  $\m{A}^\top$ & Chuyển vị của ma trận $\m{A}$\\
  $\m{A}^{-1}$ & Nghịch đảo của ma trận $\m{A}$\\
  $\text{det}(\m{A})$ & Định thức của ma trận $\m{A}$ \\
  $\text{trace}(\m{A})$ & Vết của ma trận $\m{A}$ \\
  $\text{ker}(\m{A})$ & Không gian nhân (không gian rỗng) của ma trận $\m{A}$\\
  $\text{span}(\m{a}_1,\ldots, \m{a}_n)$ & Không gian tuyến tính sinh bởi các vector $\m{a}_i,~i=1,\ldots,n$\\
  $\text{im}(\m{A})$ & Không gian ảnh của ma trận $\m{A}$\\
  $\text{dim}(\mc{A})$ & Số chiều của không gian $\mc{A}$\\
  $\text{diag}(\m{a})$ & Ma trận đường chéo có các phần tử trên đường chéo là các phần tử của vector $\m{a}$\\
  $\text{blkdiag}(\m{A}_k)$ & Ma trận đường chéo khối với các ma trận $\m{A}_k$ trên đường chéo chính\\
  $\text{rank}(\m{A})$ & Hạng của ma trận $\m{A}$\\
  $\|\m{a}\|_l, ~\|\m{A}\|_l$ & Chuẩn-$l$ của vector $\m{a}$ và ma trận $\m{A}$\\
  $\|\m{a}\|, ~\|\m{A}\|$ & Chuẩn-2 (hay chuẩn Euclid) của vector $a$ và ma trận $\m{A}$\\
  $|\cdot|$ & Modun của một đại lượng vô hướng, hoặc lực lượng của một tập hợp\\
  $\m{1}_n, \m{1}_{n \times m}$ & Vector kích thước $n\times 1$ và ma trận kích thước $n\times m$ với tất cả phần tử $1$\\
  $\m{0}_n, \m{0}_{n \times m}$ & Vector kích thước $n\times 1$,  ma trận kích thước $n\times m$ với tất cả phần tử $0$\\
  $\m{0}$ & Ma trận với tất cả phần tử $0$ có kích thước phù hợp với ngữ cảnh\\
  $\m{I}_n$ & Ma trận đơn vị kích thước $n\times n$\\
  $\otimes$ & Tích Kronecker\\
  $^g\Sigma$ & Hệ qui chiếu toàn cục\\
  $^i\Sigma$ & Hệ qui chiếu riêng (địa phương) của tác tử $i$\\
  $\m{a}_i, \m{b}_i, \m{c}_i, \ldots$ & Các vector liên quan tới tác tử $i$ viết trong hệ qui chiếu toàn cục $^g\Sigma$\\
  $\m{a}_i^i, \m{b}_i^i, \m{c}_i^i, \ldots$ & Các vector liên quan tới tác tử $i$ viết trong hệ qui chiếu riêng $^i\Sigma$\\
  $\m{a}_{ij},\m{b}_{ij},\m{c}_{ij}, \ldots$ & Các vector biến tương đối giữa hai tác tử $i$ và $j$  viết trong hệ qui chiếu $^g\Sigma$\\
  $\m{a}_{ij}^i,\m{b}_{ij}^i,\m{c}_{ij}^i, \ldots$ & Các vector biến tương đối giữa hai tác tử $i$ và $j$  viết trong hệ qui chiếu  $^i\Sigma$\\
  $\m{a}^*,\m{b}^*,\m{c}^*,\ldots$ & Các vector đặt
\end{tabular}
%
%
\part{Cơ sở}
\chapter{Giới thiệu về hệ đa tác tử}
\label{chap:introduction}
\section{Ví dụ và định nghĩa về hệ đa tác tử}
\label{c1:sec1_Intro}
Sự phát triển của các ngành khoa học và kĩ thuật đã mở rộng hiểu biết và khả năng can thiệp, biến đổi tự nhiên để phục vụ đời sống con người, đồng thời liên tục đặt ra những vấn đề và thách thức mới. Hệ đa tác tử là khái niệm chung để chỉ một loạt các đối tượng nghiên cứu  xuất hiện trong nhiều lĩnh vực khác nhau, từ thuần túy lý thuyết tới công nghệ ứng dụng. Chúng ra sẽ xem xét một số ví dụ về hệ đa tác tử trước khi đưa ra một định nghĩa cụ thể về hệ đa tác tử.
\begin{itemize}
\item \textbf{Tác tử phần mềm}: các thành phần trong một phần mềm được thiết kế như những modun riêng biệt, có trao đổi thông tin với nhau để cùng giải quyết một tác vụ (Hình \ref{fig:c1_examples_a}).

\item \textbf{Mạng học thuật}: Khác với các nhà toán học thường thực hiện nghiên cứu độc lập và ít hợp tác nghiên cứu, nhà toán học Paul Erdos (1913 - 1996)) có số lượng công bố khoa học và số lượng đồng tác giả rất lớn. Người ta gọi một người có chỉ số Erdos là 1 nếu là đồng tác giả một bài báo với Erdos. Người có chỉ số Erdos bằng 2 khi người đó là đồng tác giả với một người có chỉ số Erdos bằng 1 nhưng không là đồng tác giả với Erdos. Với định nghĩa trên, và do toán học là nền tảng của nhiều ngành khoa học kĩ thuật khác nhau, mỗi nhà khoa học có thể tra cứu chỉ số Erdos của bản thân\footnote{Thông tin về số Erdos có thể tra cứu tại: \url{https://mathscinet.ams.org/mathscinet/freetools/collab-dist}}. Số Erdos phần nào cho biết về mức độ gần xa của nhà nghiên cứu với toán học và với các nhà toán học cùng thời của Erdos. Mạng lưới học thuật là một hệ đa tác tử bao gồm các học giả (tác tử) liên kết thông qua các công bố khoa học chung. (Hình \ref{fig:c1_examples_b})

\item \textbf{Mạng lưới truyền tải điện}: Lưới điện ở mỗi địa phương là một tác tử, có công suất phát điện, yêu cầu tải riêng và tách biệt nhau về mặt địa lý. Lưới điện toàn quốc kết nối các lưới điện địa phương. Cơ quan điều độ ở mỗi địa phương có thể điều phối hoạt động sản xuất, mua bán điện với các địa phương khác, và phân phối điện tại địa phương để duy trì cân bằng công suất theo thời gian thực (Hình \ref{fig:c1_examples_c}).

\item \textbf{Chim di cư theo đội hình}: Mỗi cá thể trong bầy là một tác tử. Bầy chim di cư bay theo đội hình chữ V, trong đó một tác tử trong bầy (được luân phiên thay đổi trong hành trình) làm nhiệm vụ dẫn đường. Một số giả thuyết được chứng minh qua dữ liệu thực tế chứng tỏ việc giữ đội hình bay dạng chữ V giúp các tác tử bay sau tiết kiệm năng lượng thông qua các cơ chế tương hỗ khí động học. Tuy nhiên, dữ liệu chỉ ra rằng phần lớn năng lượng tiết kiệm là nhờ tác tử bay sau có thể ở trạng thái nghỉ một phần khi không mất năng lượng định hướng và tìm đường \cite{Weimerskirch2001energy}. Những tác tử bị lạc bầy thường không thể tự hoàn thành việc di cư (Hình \ref{fig:c1_examples_d}).

\end{itemize}

\begin{figure}[th!]
\centering
\subfloat[]{
\includegraphics[height = 7.2cm]{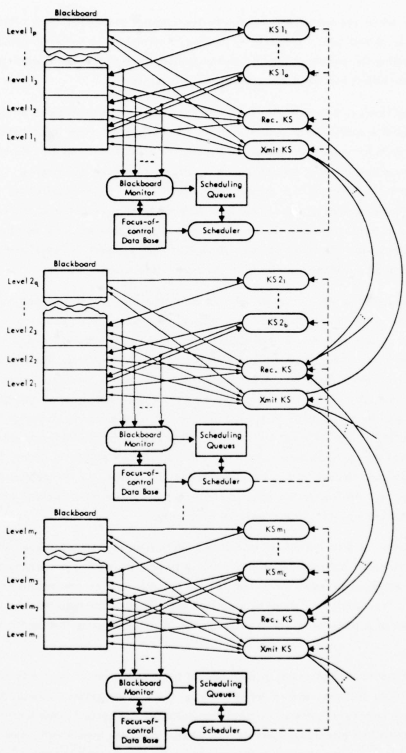}
\label{fig:c1_examples_a}
}
\hfill
\subfloat[]{
\includegraphics[height = 7.2cm]{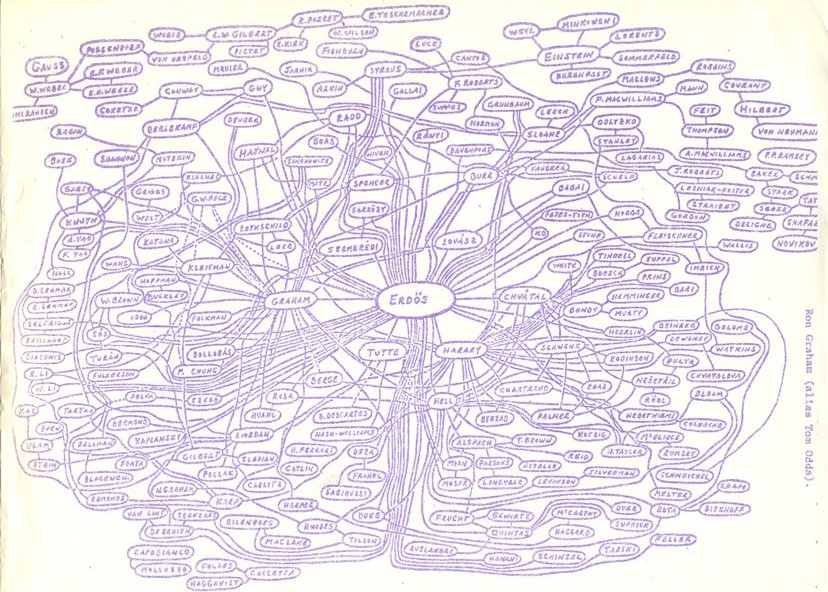}
\label{fig:c1_examples_b}
}
\\
\subfloat[]{
\includegraphics[height = 5cm]{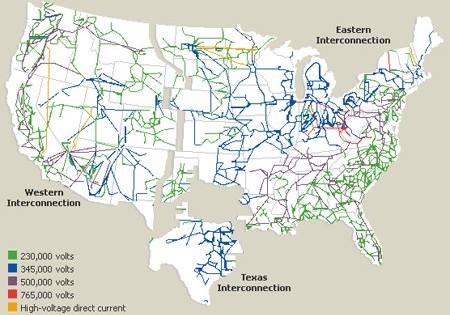}
\label{fig:c1_examples_c}
}
\hfill
\subfloat[]{
\includegraphics[height = 5cm]{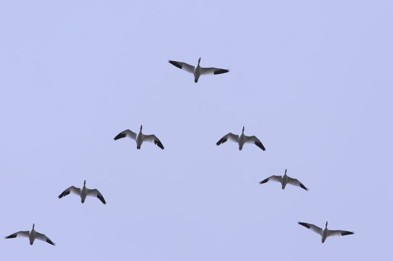}
\label{fig:c1_examples_d}
}
\caption{(a) Sơ đồ mô tả Hearsay II - tác tử phần mềm phân tán trong ứng dụng nhận hiểu lời nói (1980) \cite{Lesser1980distributed}; (b) Đồ thị mô tả mạng hợp tác học thuật của P. Erd\"{o}s; (c) Lưới điện Hợp chúng quốc Hoa Kỳ (nguồn: Bách khoa toàn thư Encarta của Microsoft); (d) Chim di cư bay theo đội hình chữ V (Nguồn: Google image). \label{fig:c1_examples}}
\end{figure}

Trong tài liệu này, \emph{hệ đa tác tử} (multiagent system) được định nghĩa là một hệ thống bao gồm các hệ con nhỏ hơn (gọi là các \emph{tác tử}), và các hệ con này có tương tác với nhau. Những tác tử thường tách biệt với nhau về vật lý hoặc/và về thông tin, có cấu trúc và hành vi tương tự nhau. Trong hệ đa tác tử, người ta thường quan tâm tới hai yếu tố: tương tác giữa các tác tử (tương tác địa phương) và hành vi chung của toàn hệ, chỉ xuất hiện khi số lượng tác tử trong hệ đủ lớn.

Trước khi đi vào nội dung chính của tài liệu, chúng ta sẽ điểm qua một số thuật ngữ thường được sử dụng có ý nghĩa gần với khái niệm hệ đa tác tử. Mặc dù các thuật ngữ được dùng khá rộng rãi trong các tài liệu học thuật, hiện tại nhiều thuật ngữ còn trong quá trình nghiên cứu, vẫn đang phát triển và do đó chưa có định nghĩa đúng tuyệt đối, chỉ có những định nghĩa chấp nhận được. Những định nghĩa trong tài liệu này là theo quan điểm của tác giả dựa trên một số tài liệu học thuật, từ điển, và wikipedia.

\subsection{Một số khái niệm liên quan tới hệ đa tác tử}
Theo wikipedia\footnote{Tham khảo: \url{https://en.wikipedia.org/wiki/System}}, \emph{hệ thống} là tập hợp các phần tử có liên kết và tương tác với nhau theo một số qui tắc và cấu trúc tạo thành một thể thống nhất và thường thực hiện một chức năng hay nhiệm vụ chung. Một hệ thống thường có tương tác với môi trường bên ngoài, có thể qua tín hiệu vào, tín hiệu ra, có chức năng và có các bất định. \index{hệ thống} \index{tác tử} \index{tác tử thông minh}

Thuật ngữ \emph{tác tử thông minh} (intelligent agent) có nguồn gốc từ nghiên cứu về trí tuệ nhân tạo (artificial intelligence), là một khái niệm trừu tượng hóa về phần mềm độc lập có khả năng thực hiện một mục tiêu thông qua các tác vụ cơ bản như biểu diễn thông tin, thực hiện suy luận logic và tính toán. Thuật ngữ \emph{hệ nhiều tác tử} (multiple agents, multiagent system) xuất hiện trong \cite{Konolige1980} vào năm 1980, trong đó các tác tử lập kế hoạch có xét tới hành động (action) của các tác tử khác. Thuật ngữ này tiếp nối khái niệm về trí thông minh nhân tạo phân tán (distributed AI) được sử dụng từ những năm 1970, là một mở rộng tự nhiên khi xem xét tới quan hệ và ảnh hưởng lẫn nhau giữa các tác tử khi cùng giải quyết một vấn đề. \index{tác tử thông minh}

Trong hệ đa tác tử, các tác tử có khả năng tự tổ chức (self-organization) để thực hiện nhiệm vụ chung mà không cần tới tác động từ bên ngoài. Khả năng tự tổ chức là một tính chất đặc trưng của hệ đa tác tử. Khái niệm \emph{hệ tự tổ chức} (self-organized system), do đó, đôi khi được dùng như phép hoán dụ để chỉ hệ đa tác tử. \index{hệ!tự tổ chức}

Một thuật ngữ được sử dụng khá phổ biến hiện nay trong cộng đồng nghiên cứu về điều khiển và khoa học mạng lưới (network science) là \emph{hệ thống nối mạng} (network system). Mặc dù thuật ngữ này tương đối mới, bản chất của khái niệm - việc kết nối các hệ thống để xây dựng các hệ thống lớn hơn (các hệ thống của hệ thống (system of systems)) - đã được thực hiện bởi từ rất lâu trước khi hệ nối mạng trở thành một hướng nghiên cứu nghiêm túc. Để ví dụ, nước sạch được dẫn từ nguồn ở núi cao tới thành phố qua kênh dẫn, phân phối tới dân cư trong thành phố, và thu gom để xả thải sau sử dụng đã được thực hiện bởi các nhà quản lý từ khi các đô thị lớn ra đời. Các thành phố đã được kết nối bởi mạng lưới giao thông đồng thời với sự xuất hiện các quốc gia cổ đại. Mạng điện thành phố, hay hệ thống điện tín giữa các thành phố đã được xây dựng sau phát minh về đèn điện và máy phát điện của Edison, khi lý thuyết về mạch điện còn sơ khai. Mạng internet đã được phát triển nhanh chóng từ một số mạng riêng cục bộ \cite{Barabasi2003linked}. \index{hệ thống!nối mạng} \index{khoa học mạng lưới} \index{điều khiển!hệ nối mạng}

Sự xuất hiện và bùng nổ của mạng internet làm nảy sinh các khái niệm \emph{hệ quy mô lớn} (large-scale systems) và \emph{mạng phức hợp} (complex network). Khái niệm hệ quy mô lớn không chỉ đề cập đến số lượng các hệ con trong hệ là rất lớn mà còn để ám chỉ hệ thể hiện khác biệt khi được xem xét và phân tích ở các qui mô khác nhau. Mạng cục bộ (mạng LAN) có tổ chức đơn giản và thường có thể quản lý tập trung bởi một máy chủ. Trong khi đó, mạng internet có quy mô lớn, liên kết nhiều mạng độc lập, vận hành ngay cả khi một phần mạng gặp sự cố. Khái niệm hệ phức hợp nhấn mạnh vào sự phức tạp trong tìm kiếm cách thức mô hình hóa và phương pháp phân tích hệ. Bộ não là một hệ phức hợp tạo bởi hàng triệu nơ-ron thần kinh. Tuy nhiên, việc mô phỏng hoạt động từng nơ-ron không giúp ích nhiều trong giải thích hoạt động của não người, do kết nối và cơ chế phối hợp hoạt động của các nơ ron là rất phức tạp. Một cách khác được sử dụng là phân chia não thành các vùng chức năng, phân tích hoạt động của từng vùng và cách các vùng não phối hợp trong khi đối tượng thực hiện các nhiệm vụ. Các liên kết trong mạng phức hợp thường đa chiều, phi tuyến, tuân theo một số qui luật xác suất và tồn tại trong thời gian nhất định. \index{hệ!quy mô lớn} \index{hệ!phức hợp}

\begin{figure}[t!]
\centering
\includegraphics[width=.8\textwidth]{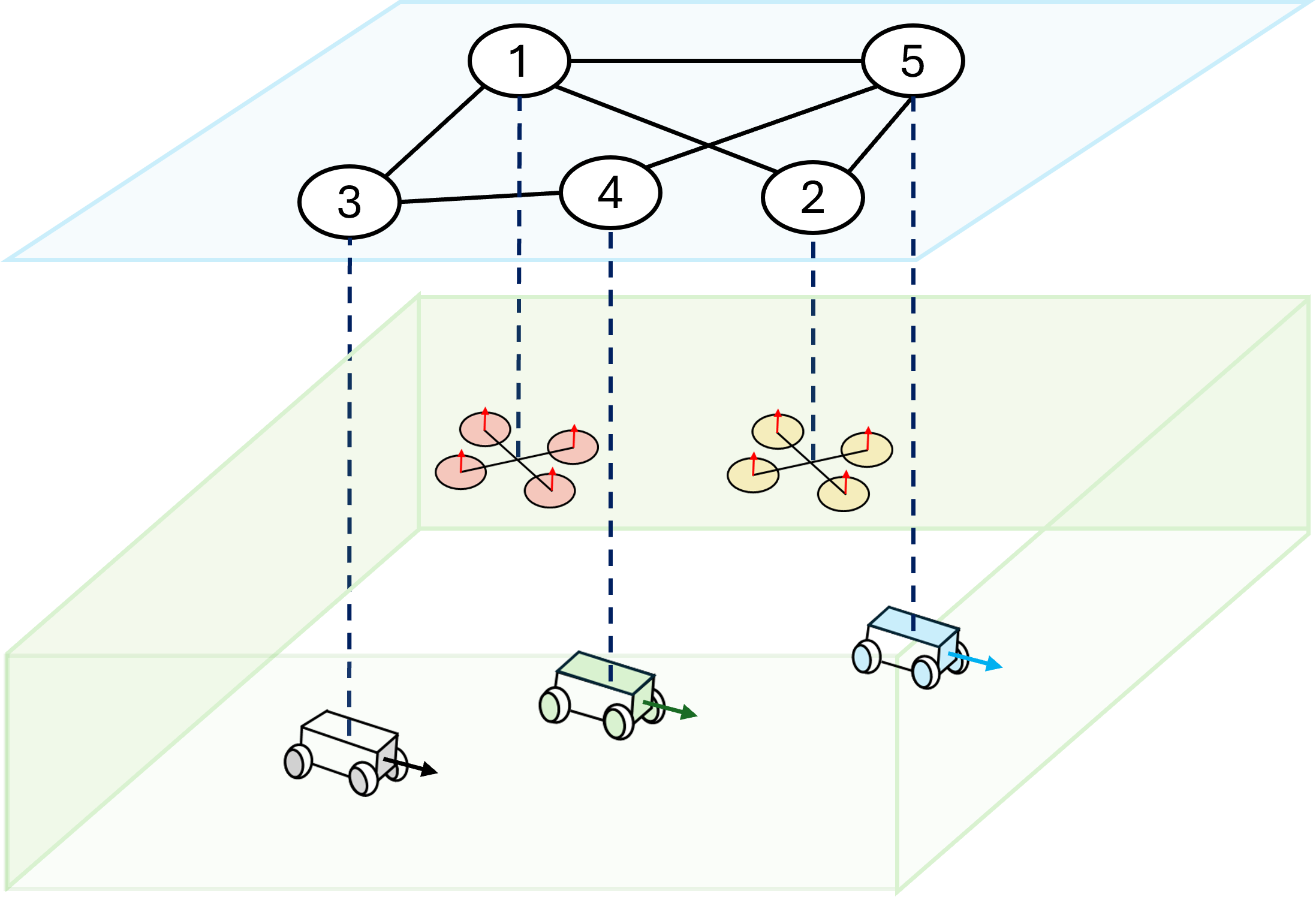}
\caption{Ví dụ về hệ thống không gian mạng - vật lý. Ở tầng thông tin, mỗi tác tử thu thập hoặc chia sẻ thông tin với tác tử láng giềng về vị trí hoặc/và vận tốc để tính toán quỹ đạo và/hoặc vận tốc đặt. Ở tầng vật lý, mỗi tác tử có thể có mô hình động học hoặc động lực học khác nhau. Mỗi bộ điều khiển riêng xác định luật điều khiển cho mỗi tác tử để bám theo quỹ đạo đặt.}
\label{fig:cyberPhysicalSystem}
\end{figure}

Cuối cùng, khái niệm về \emph{hệ thống!không gian mạng - vật lý} (cyber-physical system)\index{hệ thống!không gian mạng - vật lý}, là một thuật ngữ kĩ thuật chỉ các hệ đa tác tử có liên kết vật lý và được vận hành dựa trên sự tích hợp chặt chẽ của mạng cảm biến (quan sát quá trình vật lý), truyền thông (chia sẻ thông tin), và phần mềm chức năng (có thể là giám sát, điều khiển). Hệ thống không gian mạng - vật lý chỉ khả thi khi nền tảng kĩ thuật về truyền thông, cảm biến, lưu trữ dữ liệu, và tính toán điều khiển đủ để hiện thực hóa các nhiệm vụ của hệ đa tác tử với độ trễ thấp. Lưu ý rằng các tác tử thường cách biệt về địa lý và không đồng nhất về vật lý.

\subsection{Một số khái niệm liên quan tới điều khiển hệ đa tác tử}
Trong điều khiển các hệ đa tác tử, chúng ta dành nhiều quan tâm tới các thuật toán điều khiển phân tán. Theo Đại Từ điển về Hệ thống và Điều khiển \cite{Baillieul2015encyclopedia}, dựa trên cấu trúc thông tin và truyền thông trong hệ đa tác tử, chúng ta có các khái niệm:
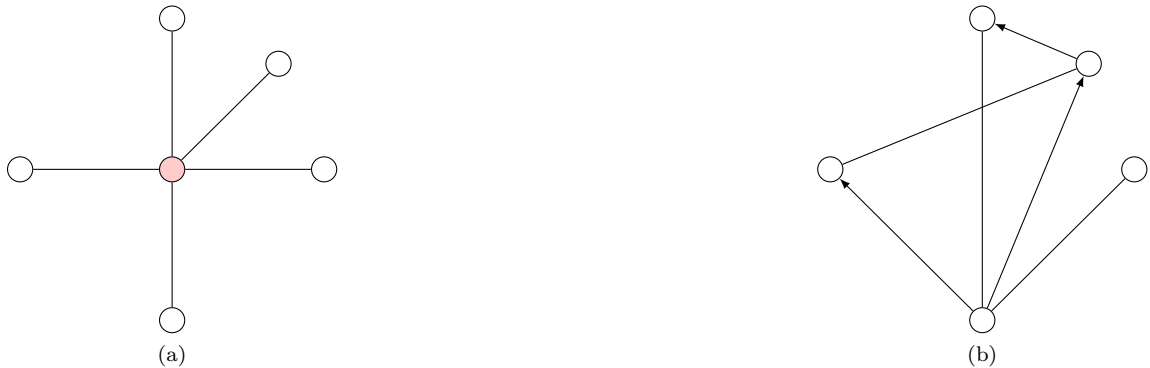
\begin{figure}
\centering
\subfloat[]{
\begin{tikzpicture}[
  vertex/.style={circle, draw, minimum size=2mm},
  edge/.style={-}
]
\node[vertex,fill=red!20] (v0) at (0,0) { };

\node[vertex] (v1) at (2,0)   { };
\node[vertex] (v2) at (0,2)   { };
\node[vertex] (v3) at (-2,0)  { };
\node[vertex] (v4) at (0,-2)  { };
\node[vertex] (v5) at (1.4,1.4) { };

\foreach \i in {1,...,5}
  \draw[edge] (v0) -- (v\i);

\end{tikzpicture}
}
\hfill
\subfloat[]{
\begin{tikzpicture}[
  vertex/.style={circle, draw, minimum size=2mm},
  edge/.style={-}
]

\node[vertex] (v1) at (2,0)   { };
\node[vertex] (v2) at (0,2)   { };
\node[vertex] (v3) at (-2,0)  { };
\node[vertex] (v4) at (0,-2)  { };
\node[vertex] (v5) at (1.4,1.4) { };

\draw[edge,->] (v1) -- (v4) -- (v3);
\draw[edge,->] (v2) -- (v4) -- (v5);
\draw[edge,->] (v3) -- (v5) -- (v2);
\end{tikzpicture}
}
\caption{(a) Cấu trúc thông tin trong điều khiển phi tập trung: bộ điều khiển trung tâm ở vị trí chính giữa (nền đỏ), các bộ điều khiển riêng của các tác tử là các hình tròn nhỏ nền trắng, mỗi đoạn thẳng thể hiện một đường trao đổi thông tin (tác tử gửi thông tin tới bộ điều khiển trung tâm, bộ điều khiển trung tâm gửi tín hiệu điều khiển cấp cao tới tác tử); (b) Cấu trúc thông tin trong điều khiển phân tán: mỗi bộ điều khiển riêng thể hiện bởi một hình tròn nhỏ nền trắng, mỗi mũi tên giữa hai bộ điều khiển thể hiện thông tin được trao đổi theo một chiều, và mỗi đoạn thẳng thể hiện một đường trao đổi thông tin hai chiều. \label{fig:c1_controlStructures}}
\end{figure}
\begin{itemize}
\item \textbf{Điều khiển tập trung}: hệ có một bộ điều khiển (trung tâm) thu thập, xử lý thông tin từ tất cả các tác tử từ đó tính toán và gửi đi tín hiệu điều khiển tương ứng cho từng tác tử.\index{điều khiển!tập trung}

\item \textbf{Điều khiển phi tập trung}: mỗi tác tử có bộ điều khiển địa phương (điều khiển cấp thấp), tồn tại một bộ điều khiển trung tâm thu thập một số thông tin về hoạt động của các tác tử đồng thời tính toán các tín hiệu điều khiển (cấp cao) cho mỗi bộ điều khiển riêng. Ví dụ như trong hệ thống điều khiển phối hợp chuyển động các robot công nghiệp, mỗi robot có bộ điều khiển vị trí, lực riêng. Bộ điều khiển trung tâm lập lịch để hệ robot vận hành theo từng công đoạn. \index{điều khiển!phi tập trung}

\item \textbf{Điều khiển phân tán}: mỗi tác tử có khả năng đo đạc, truyền thông với một vài tác tử khác trong hệ từ đó tự quyết định, tính toán tín hiệu điều khiển riêng. \index{điều khiển!phân tán}
\end{itemize}

Hình \ref{fig:c1_controlStructures} mô tả cấu trúc thông tin trong điều khiển phi tập trung và điều khiển phân tán. Cấu trúc thông tin ở điều khiển phi tập trung có dạng hình sao với bộ điều khiển trung tâm ở vị trí chính giữa, thể hiện rằng thông tin trao đổi giữa các tác tử nếu có đều phải thông qua bộ điều khiển trung tâm. Ở điều khiển phân tán, các tác tử có thể trực tiếp trao đổi thông tin với một số tác tử khác. Cấu trúc thông tin ở điều khiển phân tán không có ràng buộc về cấu hình cụ thể, đồng thời sự trao đổi thông tin giữa hai tác tử có thể là một chiều (một tác tử chỉ nhận, một tác tử chỉ gửi) hoặc hai chiều (hai tác tử nhận và gửi thông tin từ nhau).

Chú ý rằng trong nghiên cứu về hệ đa tác tử, nếu các tác tử chia sẻ chung tài nguyên (cơ sở dữ liệu lưu trữ ở cùng một máy chủ) nhưng tự tính toán luật điều khiển thì hệ vẫn được phân loại là điều khiển phi tập trung chứ chưa được phân loại là điều khiển phân tán. Trong bài toán điều khiển đội hình trong Chương \ref{chap:formation}, một thuật toán được phân loại là hoàn toàn phân tán còn yêu cầu các điều kiện bổ sung về lượng thông tin địa phương mà mỗi tác tử có thể sử dụng.

Một số thuật ngữ thường xuyên được sử dụng với nghĩa gần với điều khiển hệ đa tác tử bao gồm:
\begin{itemize}
\item \textbf{Điều phối hệ đa tác tử} (coordination): \index{điều phối hệ đa tác tử} có ý nghĩa rằng vận động trong hệ nhằm đạt mục tiêu chung được thực hiện bởi ít nhất hai tác tử, nhấn mạnh thiết lập tương tác giữa các tác tử.

\item \textbf{Điều khiển hợp tác/bất hợp tác} (cooperative/non-cooperative control): Tương tác giữa các tác tử có thể phân chia thành tương tác hợp tác và tương tác cạnh tranh. Do mỗi tác tử có một mục tiêu riêng, tùy theo phương pháp thiết kế thuật toán điều khiển mà vận động của từng tác tử có thể làm lợi hoặc làm thiệt hại cho mục tiêu của tác tử khác. Tuy nhiên, từ phương diện của người thiết kế điều khiển bên ngoài, việc chọn sử dụng tương tác hợp tác hay bất hợp tác là để đảm bảo mục tiêu chung của cả hệ được hoàn thành. \index{điều khiển!hợp tác} \index{điều khiển!bất hợp tác}

\item \textbf{Thuật toán bầy đàn} (swarm algorithm):\index{thuật toán!bầy đàn} Các hệ đa tác tử thường phức tạp và khó mô tả chính xác, do đó, một nhóm các thuật toán cho hệ đa tác tử được thiết kế dựa trên trên học hỏi và bắt chước từ các hiện tượng bầy đàn trong tự nhiên. 
\end{itemize}
Như vậy, điều phối bao hàm cả điều khiển hợp tác và bất hợp tác. Hành vi và luật điều khiển riêng của mỗi tác tử ngoài các biến trạng thái của tác tử đó còn chứa các biến trạng thái của một số tác tử khác trong hệ. Những thuật toán bầy đàn như ant-colony, bee-colony, partical swarm là các thuật toán tối ưu hiệu quả được sử dụng rộng rãi trong khoa học máy tính, tuy nhiên hiện chưa nằm trong trọng tâm nghiên cứu trong điều khiển hệ đa tác tử. Mặc dù tự nhiên là cảm hứng để mô hình hóa và thiết kế thuật toán điều khiển hệ đa tác tử \cite{Olfati2006flocking}, tài liệu này dành phần lớn quan tâm tới các thuật toán chính xác, có thể chứng minh được bằng toán học và thường là thiết kế được dựa trên mô hình.

\section{Điều khiển phân tán hệ đa tác tử}
\label{c1:s2_controlMAS}

Với mục tiêu chung là thiết kế các hệ thống gồm nhiều phần tử tự động thực hiện các tác vụ thay con người, hai nhánh nghiên cứu đã được phát triển tương đối độc lập: nhánh điều khiển phi tập trung/phân tán dựa trên nền tảng lý thuyết điều khiển cho các hệ động học từ những năm 1970 và nhánh công nghệ thông tin dựa trên thiết lập thuật toán và hệ thống các qui tắc xử lý thông tin song song, phân tán từ những năm 1980. 

Các hệ thống điều khiển phân tán (Distributed Control System) được phát triển từ những năm 1970 và dần được sử dụng rộng rãi trong công nghiệp (điều khiển quá trình) \cite{Hoang2001BUS}. Các ứng dụng của cơ sở dữ liệu phân tán hiện nay có được sử dụng trong cuộc sống hàng ngày có thể kể đến hệ thống giao dịch (transaction) tài chính, điện toán đám mây và các dịch vụ web, ứng dụng mạng xã hội \cite{Nguyen2004CSDL}.

Một số ứng dụng điều khiển đội hình (xe tự hành, máy bay không người lái, vệ tinh nhân tạo) nằm tại phần giao của hai nhánh nghiên cứu đã được công bố từ những năm 1990 bởi cộng đồng nghiên cứu robotics và hàng không vũ trụ \cite{Sugihara1990,Brock1992,Robertson1999,Young2001control}. Ứng dụng điều khiển đội hình được nghiên cứu, hiện thực hóa phục vụ dân sự nhờ sự phát triển và phổ biến của các công nghệ như định vị GPS, truyền thông không dây, cảm biến, vi điều khiển và hệ nhúng.

Nghiên cứu về điều khiển hệ đa tác tử hiện tại bắt đầu từ khoảng đầu những năm 2000 \cite{Fax2004,Olfati2004consensus}, khi vai trò của lý thuyết đồ thị, thuật toán đồng thuận\footnote{Giao thức đồng thuận (consensus/agreement protocol) là thuật ngữ được sử dụng rộng rãi trong các thuật toán phân tán ở lĩnh vực Khoa học máy tính. Khái niệm ``đồng thuận'' cũng xuất hiện trong nghiên cứu xã hội học \cite{degroot1974reaching} và sẽ được đề cập cụ thể hơn tại Chương \ref{chap:social_network}.}, và lý thuyết điều khiển được xác định tương đối rõ ràng.

Ở mục này, chúng ta sẽ đề cập cụ thể hơn về mục tiêu, nhiệm vụ, phạm vi, và một số vấn đề được quan tâm trong điều khiển hệ đa tác tử.

\subsection{Mục tiêu và nhiệm vụ}
Đầu tiên, chúng ta sẽ tìm cách phát biểu một bài toán điều khiển một hệ đa tác tử. Xét hệ gồm $n\geq 2$ tác tử cần giải quyết một Bài toán $A$ - gọi là bài toán toàn cục (hoặc mục tiêu toàn cục) của cả hệ. Các giả thiết bao gồm:  
\begin{itemize}
\item \textbf{Bài toán/mục tiêu cục bộ}: Thông tin về Bài toán $A$ được chia nhỏ thành các bài toán cục bộ $A_1,A_2,\ldots,A_n$ và phân chia cho từng tác tử trong hệ. Các bài toán cục bộ có thể không tách biệt nhau, tuy nhiên việc tất cả các bài toán cục bộ được giải quyết cần dẫn tới một lời giải cho Bài toán toàn cục $A$.

\item \textbf{Cấu trúc thông tin}: Các tác tử có khả năng tiếp nhận và/hoặc trao đổi thông tin với các tác tử khác trong hệ (và có thể cả với môi trường bên ngoài). 

\item \textbf{Tác tử}: Mỗi tác tử $i=1,\ldots,n$ trong hệ được giả thiết có khả năng lưu trữ thông tin về bài toán riêng $A_i$ và thông tin liên quan tới lời giải riêng dưới dạng một tập hợp các thông tin cục bộ $X_i$. Thông tin được tiếp nhận cũng là một phần trong thông tin cục bộ mà tác tử $i$ sở hữu. Hơn nữa, mỗi tác tử có khả năng tính toán, quyết định và thay đổi thông tin cục bộ dựa trên các thuật toán điều khiển.
\end{itemize}
Điều khiển hệ đa tác tử bao gồm các nhiệm vụ: mô hình hóa hệ đa tác tử, xác định bài toán toàn cục và thiết kế bài toán cục bộ, thiết lập cấu trúc thông tin, và thiết kế thuật toán điều khiển dựa trên thông tin cục bộ nhằm thực hiện các bài toán cục bộ $A_i$, từ đó thực hiện bài toán toàn cục $A$.

Thông tin trong hệ đa tác tử bao gồm thông tin về các đại lượng vật lý (mang thông tin động học/động lực học của tác tử) hoặc biến tính toán (phục vụ cho quá trình ra quyết định). Do đó, tác tử có thể thu thập  thông tin liên quan tới tác tử khác nhờ đo đạc các biến tương đối hay thông qua trao đổi qua mạng truyền thông. Tương tự, việc thay đổi các biến trạng thái có thể dựa trên cơ cấu chấp hành đối với biến vật lý và/ hoặc cập nhật biến trong bộ nhớ với biến tính toán.

Các phương án điều khiển hệ đa tác tử cần được kiểm chứng bằng phân tích toán học, mô phỏng số, hiện thực hóa trên các hệ thống thử nghiệm và sản phẩm thực tế. Các chương tiếp theo trong tài liệu này trình bày việc mô hình hóa, thiết kế điều khiển kết hợp phân tích ổn định và hành vi tiệm cận cho một số hệ đa tác tử. Chú ý rằng từ góc độ của người làm điều khiển lý thuyết thì khâu phân tích hệ có vai trò quan trọng nhất. Không có bài báo nào về điều khiển hệ đa tác tử thiếu các phân tích toán học. 

\begin{example}[Tụ bầy ở chim và cá] \label{VD:1.1}
Sự tương đồng về hiện tượng tụ bầy ở chim (flocking) và chuyển động nhiệt hỗn loạn của các phân tử khí đã được quan tâm bởi các nhà vật từ cuối những năm 1980. Một mô hình đơn giản lấy cảm hứng từ tự nhiên đã được đề xuất bởi Reynolds \cite{Reynolds1987flocks}. Trong mô hình này, mỗi tác tử (được gọi là một boid) di chuyển trong không gian ba chiều tuân theo ba qui tắc đơn giản là chia tách, căn chỉnh, và gắn kết. Với ba luật đơn giản trên, Reynolds đã mô phỏng lại được một loạt các hiện tượng quan sát được trong tự nhiên.

Một mô hình khác được đề xuất bởi nhóm nghiên cứu của nhà vật lý học Vics\'{e}k \cite{Vicsek1995novel} mô tả hệ các tác tử chuyển động trên mặt phẳng với cùng tốc độ nhưng với góc hướng khác nhau. Mỗi tác tử cập nhật hướng di chuyển tại mỗi thời điểm gián đoạn cách đều nhau dựa trên trung bình cộng về góc hướng của tác tử đó và các tác tử lân cận và một thành phần nhiễu từ môi trường. Mô phỏng cho thấy nếu như nhiễu là không đáng kể thì các tác tử sẽ di chuyển theo cùng một hướng sau thời gian đủ lâu. Tuy nhiên, với nhiễu là tương đối lớn, chuyển động của hệ có tính ngẫu nhiên, với hướng chuyển động không đồng nhất. Mô hình Vics\'{e}k là một trong những ví dụ đầu tiên về hệ động học nhiều tác tử được xem xét dựa trên lý thuyết ổn định \cite{Jadbabaie2003coordination}.
\end{example}

\begin{story}{Brian D. O. Anderson và Điều khiển phân tán hệ đa tác tử}
Brian D. O. Anderson nhận bằng Cử nhân ngành Toán và Kĩ thuật Điện tại Đại học Sydney, Úc (1962, 1964), và bằng TS ngành Điện tại Đại học Stanford, Hoa Kỳ (1966). Sau thòi gian ngắn làm việc tại ĐH Stanford và trong công nghiệp ở USA, Brian Anderson lần lượt giữ vị trí giáo sư tại Đại học Newcastle (1967 - 1981), và Đại học Quốc gia Úc (1981-nay), nơi ông thành lập Bộ môn Kĩ thuật Hệ thống và điều hành Trường Khoa học và Kĩ thuật Thông tin - một đơn vị chuyên về nghiên cứu thuộc Đại học Quốc gia Úc. Giáo sư Anderson giữ vị trí Chủ tịch Hiệp hội Điều khiển Quốc tế (IFAC) từ 1990 - 1993, Chủ tịch Viện Hàn lâm Khoa học Úc (1998 - 2002), và là thành viên của các tổ chức học thuật và quỹ nghiên cứu uy tín của Úc, Vương quốc Anh, và Hoa Kỳ. 

Nghiên cứu của Brian Anderson trải dài ở các lĩnh vực lý thuyết mạch, xử lý tín hiệu, lý thuyết điều khiển. Từ khoảng đầu những năm 2000, ông là một trong những người đặt nền móng cho các nghiên cứu về điều khiển đội hình và định vị mạng cảm biến, cũng như mô hình động học ý kiến hiện nay. Các nghiên cứu về điều khiển đội hình và định vị mạng cảm biến của ông có hợp tác với Bộ quốc phòng Úc, cũng như hợp tác quốc tế với các Cơ sở học thuật uy tín khắp thế giới.

Các công trình nghiên cứu của Anderson và cộng sự chỉ ra vai trò của lý thuyết cứng và các biến hình học như khoảng cách, góc hướng, góc sai lệch trong thiết kế cấu trúc thông tin của đội hình cũng như mạng cảm biến. Ông coi trọng việc trao đổi học thuật và tăng cường kết nối trong cộng đồng nghiên cứu điều khiển – tự động hóa, giữa các đồng nghiệp cũng như với sinh viên có cùng quan tâm nghiên cứu.
\end{story}

\subsection{Phạm vi}
Theo tài liệu \cite{MesbahiEgerstedt},\footnote{\cite{MesbahiEgerstedt} là một trong những sách tham khảo xuất hiện sớm nhất về điều khiển hệ đa tác tử (2008). Những kết quả và ý tưởng trong tài liệu này vẫn mang tính thời sự trong các nghiên cứu về điều khiển hệ đa tác tử ở hiện tại.} điều khiển hệ đa tác tử được phát triển dựa trên của ba lĩnh vực chính: lý thuyết đồ thị, tối ưu hóa, và lý thuyết điều khiển, trong đó:
\begin{itemize}
\item \textbf{Lý thuyết đồ thị}: là công cụ để mô hình hóa các tương tác trong hệ đa tác tử. Lưu ý rằng đồ thị mô tả cấu trúc thông tin trong hệ, trực tiếp quyết định liệu mục tiêu của hệ có thể đạt được hay không.

\item \textbf{Tối ưu hóa}: Bài toán toàn cục và bài toán cục bộ thường được lượng hóa bằng các hàm mục tiêu. Tối ưu hóa là công cụ để thiết lập bài toán, từ đó đề xuất luật điều khiển (cấp cao) cho hệ đa tác tử. Hơn nữa, với các tham số ảnh hưởng tới chất lượng hệ thống,\footnote{Ví dụ, ở chương \ref{chap:consensus}, chúng ta sẽ thấy rằng tốc độ tiệm cận tới giá trị đồng thuận phụ thuộc vào giá trị riêng dương nhỏ nhất của ma trận Laplace của đồ thị.} tối ưu hóa là cần thiết trong tổng hợp (synthesis) hệ đa tác tử.

\item \textbf{Lý thuyết điều khiển}: Các tác tử và tương tác giữa các tác tử được mô tả bởi các quá trình động học và động lực học (phương trình vi phân). Lý thuyết điều khiển cung cấp công cụ để phân tích ổn định, các phương pháp thiết kế để hệ thực hiện được mục tiêu dưới ảnh hưởng của bất định (ở bên trong hệ cũng như từ môi trường bên ngoài), và đánh giá chất lượng các hệ đa tác tử.
\end{itemize}

Hiện tại, chưa có nhiều kết quả từ lý thuyết đồ thị tìm được ứng dụng phù hợp trong hệ đa tác tử \cite{Zelazo2018graph}. Ảnh hưởng của tối ưu hóa trong điều khiển hệ đa tác tử ngày một nhiều, nhưng chưa quá sâu. Trong khi đó, những kết quả mới từ lý thuyết điều khiển được ứng dụng trong hệ đa tác tử tương đối nhanh. Điều này một phần là do cộng đồng nghiên cứu về hệ đa tác tử phần lớn tiếp cận hướng nghiên cứu này với nền tảng kĩ thuật, và do đó phần lớn sự quan tâm được dành cho các ứng dụng thực tế. Trong khi đó, nghiên cứu về điều khiển đa tác tử chưa được quan tâm nhiều từ cộng đồng nghiên cứu về tổ hợp và lý thuyết đồ thị.

\subsection{Một số chủ đề nghiên cứu}
Bảng \ref{tab:c1_MAS_issue} cung cấp mô tả ngắn gọn về một số vấn đề trong nghiên cứu hệ đa tác tử cùng trích dẫn tới một bài báo trong chủ đề tương ứng. Tài liệu này sẽ dành lượng lớn cho thuật toán đồng thuận do rất nhiều bài toán trong hệ đa tác tử được phát triển từ thuật toán đơn giản này. 

\begin{table}[th]
\caption{Một số ứng dụng của hệ đa tác tử}
\label{tab:c1_MAS_issue}
\begin{tabular}{@{}p{0.08\textwidth}p{0.22\textwidth}p{0.6\textwidth}@{}}
\toprule
Thứ tự & Vấn đề                & Mô tả                           \\ \midrule
1  & Đồng thuận  \cite{Olfati2007consensuspieee}          & Biến trạng thái của các tác tử bằng nhau     \\
2  & Đồng bộ hóa \cite{Li2009consensus}   & Các biến trạng thái hoặc biến đầu ra của các tác tử có cùng một quĩ đạo (nghiệm của phương trình vi phân) \\
3  & Điều khiển đội hình \cite{Anderson2008}  & Các tác tử phân bố tại các điểm trong không gian, vị trí tương đối của các tác tử thỏa mãn một số ràng buộc hình học. Các vấn đề liên quan bao gồm: tránh va chạm, giữ liên kết, căn chính hệ qui chiếu, thiết kế và thay đổi đội hình đặt \\ 
4  & Định vị mạng cảm biến \cite{Aspnes2006} & Ước lượng vị trí các nút mạng dựa trên một số nút tham chiếu và biến đo hình học \\
5 & Kết hợp dữ liệu cảm biến (Sensor fusion) \cite{Xiao2005scheme} & Ứng dụng của đồng thuận trong mạng cảm biến: xác định ước lượng cực đại hợp lý từ một số quan sát (giá trị đo) của các cảm biến khác nhau, hoặc ước lượng các tham số toàn cục của hệ \\
6 & Tối ưu phân tán \cite{Nedic2018distributed} & Tối thiểu hóa hàm mục tiêu toàn cục kết hợp với các ràng buộc địa phương. \\
7 & Sản xuất - phân phối tài nguyên phân tán \cite{Molzahn2017} & Ứng dụng của tối ưu phân tán trong xác định lượng sản xuất phù hợp phân phối có tính đến các ràng buộc về tài nguyên trong các hệ compartmental như hệ thống điện, nước, khí đốt, hệ thống nhiệt tòa nhà, giao thông thông minh \\
8 & Ổn định tần số và công suất lưới điện \cite{Simpson2013synchronization} & Ứng dụng của thuật toán đồng thuận trong điều khiển hệ gồm nhiều máy phát với công suất khác nhau \\
9 & Thuật toán dựa trên lý thuyết trò chơi \cite{Marden2018game} & Hệ đa tác tử tìm điểm cân bằng Nash trong lý thuyết trò chơi thông qua tối ưu các mục tiêu riêng đối lập nhau \\ 
10 & Mạng xã hội \cite{Proskurnikov2016} & Nghiên cứu các mô hình đơn giản hóa của mạng xã hội, tập trung vào phân tích sự phụ thuộc của hành vi tiệm cận của mô hình đối với các tham số \\
11 & Mạng đa lớp (Multilayer network) \cite{Kivela2014} & Mô hình mạng với nhiều lớp và các tương tác cùng lớp và khác lớp \\ \hline
\end{tabular}
\end{table}

Ngoài các thuật toán cơ bản đã được áp dụng vào hệ đa tác tử, một số  phương pháp và xu hướng thiết kế mở rộng cho hệ đa tác tử (ảnh hưởng từ  các lý thuyết điều khiển và tính toán phân tán) bao gồm:
\begin{itemize}
\item Thiết kế bám (tracking): Thuật toán đảm bảo chỉ tiêu chất lượng của hệ đa tác tử khi bài toán toàn cục có tham số thay đổi theo thời gian.
\item Thiết kế bền vững (robustness): Thuật toán đảm bảo chỉ tiêu chất lượng của hệ đa tác tử duy trì trong miền cho phép khi tính tới ảnh hưởng của nhiễu và bất định.
\item Thiết kế thích nghi (adaptive): Thuật toán có khả năng thay đổi tham số hoặc cấu trúc khi điều kiện vận hành thay đổi.
\item Thiết kế tối ưu (optimal): Thuật toán dựa trên tối ưu hóa một số chỉ tiêu chất lượng (hàm chi phí) gắn với quá trình động học của hệ.
\item Thiết kế chịu lỗi (fault-tolerance): Thuật toán đảm bảo hoạt động của hệ đa tác tử khi có sự cố (sai lệch cảm biến, hỏng một phần cơ cấu chấp hành, sai lệch thông tin trao đổi qua mạng), có thể thông qua cơ chế phát hiện - cô lập lỗi - thay thế hoặc qua điều khiển bền vững.
\item Thiết kế duy trì và phục hồi (resilience): Thuật toán đảm bảo hoạt động của hệ đa tác tử khi có sự cố, đồng thời có cơ chế thích nghi và tái phục hồi sau sự cố.
\item Thiết kế riêng tư (privacy): Thuật toán giới hạn thông tin trao đổi giữa các tác tử.
\item Thiết kế an toàn (security): Thuật toán giảm khả năng thông tin trao đổi giữa các tác tử trong mạng bị đánh cắp (có thể dựa trên các thuật toán mã hóa - giải mã (encryption/decryption),...
\end{itemize}

Độc giả quan tâm tới các chủ đề không được đề cập trong tài liệu này có thể dễ dàng tìm kiếm với các từ khóa liên quan. 

\section{Cấu trúc của tài liệu}
Tài liệu này được chia thành ba phần: cơ sở, hệ đồng thuận, và các ứng dụng của hệ đa tác tử. Trong phần I, sau khi định nghĩa về hệ đa tác tử, cơ sở về lý thuyết đồ thị sẽ được trình bày ở chương ~\ref{chap:graphTheory}. Tiếp theo, ở phần II, một số kết quả quan trọng trong phân tích hệ đồng thuận sẽ được trình bày. Những kết quả về hệ đồng thuận ở chương \ref{chap:consensus}--\ref{chap:lyapunov} là cơ sở lý thuyết cho các nghiên cứu về điều khiển hệ đa tác tử. Những ứng dụng của hệ đồng thuận trong các bài toán như điều khiển đội hình, định vị mạng cảm biến, tối ưu phân tán, và một số mô hình động học quan điểm trong nghiên cứu mạng xã hội sẽ được lần lượt giới thiệu. Cuối phần III, ở Chương \ref{chap:MWC}, tác giả giới thiệu về hướng nghiên cứu của bản thân về đồ thị trọng số ma trận và thuật toán đồng thuận trọng số ma trận.

Một số kết quả về đại số tuyến tính, tối ưu lồi, lý thuyết điều khiển  được trình bày trong phần phụ lục để độc giả tiện tham khảo. Mã nguồn các ví dụ mô phỏng sử dụng phần mềm MATLAB\textsuperscript{\texttrademark} (phiên bản R2022b) được cung cấp để hỗ trợ độc giả tiếp cận các kết quả lý thuyết trong tài liệu.

\section{Ghi chú và tài liệu tham khảo}
Nội dung của tài liệu này yêu cầu người đọc đã có kiến thức về đại số tuyến tính \cite{Strang1988}, cơ bản về giải tích hàm nhiều biến và phương trình vi phân. Người đọc có thể tự bổ sung những kiến thức liên quan về điều khiển tuyến tính, điều khiển phi tuyến, và tối ưu hóa trong quá trình đọc tài liệu.

Ngoại trừ nội dung về tối ưu phân tán và hệ đồng thuận trọng số ma trận, các nội dung trong tài liệu đã được giảng dạy trong thời lượng một học kỳ (15 tuần) ở bậc đại học và cao học các ngành Cơ-điện tử và ngành Điện (Điều khiển - Tự động hóa). Độc giả tự tìm hiểu có thể tập trung vào phần I, II để nắm được cốt lõi chủ đề. Lưu ý rằng mỗi ứng dụng ở phần III hoàn toàn có thể trình bày và giảng dạy như một môn học riêng. Tài liệu này giới hạn ở việc trình bày bài toán, một số thuật toán và phân tích toán học trong mỗi chủ đề.

Một số tài liệu tham khảo tiếng Anh về điều khiển hệ đa tác tử, điều khiển nối mạng, hệ đồng thuận bao gồm \cite{Shamma2008cooperative,MesbahiEgerstedt,Ren2008springer,Qu2009cooperative,Bullo2019lectures}.
Trong các tài liệu này, tác giả khuyến nghị người đọc với nền tảng toán kĩ thuật tham khảo \cite{Bullo2019lectures}. Bản dịch tiếng Việt \cite{Trinh2024} của giáo trình \cite{Bullo2019lectures} cũng được phân phối trên website của tác giả (F. Bullo).

\chapter{Lý thuyết đồ thị}
\label{chap:graphTheory}
Đồ thị là một cấu trúc rời rạc gồm các đỉnh và các cạnh nối các đỉnh đó. Lý thuyết đồ thị được phát triển bởi L. Euler, với bài toán Bảy cây cầu ở K\"{o}nigsberg. Cho đến đầu thế kỷ 20, khi giáo trình đầu tiên về lý thuyết đồ thị được xuất bản bởi D. K\"{o}nig \cite{Konig2001theorie} năm 1936, lý thuyết đồ thị vẫn chưa được xem là một nhánh nghiên cứu nghiêm túc của toán học. Lý thuyết đồ thị dần được quan tâm rộng rãi nhờ các ứng dụng trong khoa học máy tính từ giữa thế kỷ 20.\index{lý thuyết!đồ thị}

Đồ thị được sử dụng để mô tả nhiều bài toán trong các lĩnh vực khác nhau. 
Mục tiêu của chương này là trình bày một số kết quả thường dùng của lý thuyết đồ thị trong nghiên cứu về điều khiển hệ đa tác tử. Đầu tiên, các định nghĩa và kết quả cơ bản về đồ thị được giới thiệu trong mục~\ref{sec2:do_thi}. Tiếp theo, mục~\ref{sec2:dai_so_do_thi} trình bày các cấu trúc đại số để mô tả đồ thị, ví dụ như ma trận kề, ma trận liên thuộc, và ma trận Laplace. Các kết quả trình bày ở chương \ref{chap:graphTheory} sẽ được sử dụng xuyên suốt trong tài liệu để mô tả và phân tích các hệ đa tác tử.

\section{Đồ thị}
\label{sec2:do_thi}
\subsection{Đồ thị vô hướng}
Một đồ thị đơn, hữu hạn, vô hướng (hay gọi ngắn gọn là một \index{đồ thị}\emph{đồ thị}) $G = ({V}, {E})$, gồm một \emph{tập đỉnh} \index{tập!đỉnh} ${V} = \{v_1, v_2, \ldots, v_n\}$ với $|{V}| = n>0$ phần tử, và một \emph{tập cạnh} \index{tập!cạnh} ${E} = \{(v_i,v_j)|~ i, j = 1, \ldots, n, i \neq j\} \in {V} \times {V}$ với $|{E}| = m$ phần tử. Ta gọi $v_i \in {V}$ và $(v_i,v_j) \in {E}$ tương ứng là môt \emph{đỉnh} và một \emph{cạnh} của đồ thị ${G}$. Do đồ thị là vô hướng nên nếu có $(v_i,v_j) \in {E}$ thì cũng có $(v_j,v_i) \in {E}$.\footnote{Bởi vậy, với đồ thị vô hướng ${G}$, ta chỉ cần viết $(v_i,v_j)$ và không phân biệt giữa $(v_i,v_j)$ và $(v_j,v_i)$.} Khi có nhiều đồ thị khác nhau, ta kí hiệu tập đỉnh và tập cạnh tương ứng của đồ thị ${G}$ bởi ${V}({G})$ và ${E}({G})$. Trong một số ngữ cảnh, để đơn giản, ta có thể kí hiệu đỉnh $v_i$ bởi $i$, cạnh $(v_i, v_j)$ bởi $(i,j)$ hoặc $e_{ij}$. \index{đồ thị!vô hướng}

Mỗi đồ thị ${G}$ có một biểu diễn hình học tương ứng, gồm các vòng tròn nhỏ biểu diễn các đỉnh $v_i \in {V}$, và các đoạn thẳng (hay các cung) nối $v_i$ với $v_j$ nếu $(v_i,v_j) \in {E}$. 
\begin{example} \label{eg:2.1}
Trên hình~\ref{fig:c2_f1_dothi}: 
\begin{itemize}
    \item Đồ thị ${G}_1 = ({V}_1,{E}_1)$ có tập đỉnh ${V}_1 = \{v_1, v_2, v_3, v_4\}$ và  tập cạnh \[{E}_1 = \{(v_1,v_2), (v_2,v_3), (v_3,v_4), (v_4,v_1), (v_1,v_3)\}.\]
    \item Đồ thị ${G}_2=({V}_2,{E}_2)$ có tập đỉnh ${V}_2 = \{u_1, u_2, u_3, u_4, u_5\}$ và tập cạnh \[{E}_2 = \{(u_1,u_2), (u_2,u_3), (u_3,u_4), (u_4,u_5), (u_5,u_1)\}.\]
    \item Đồ thị ${G}_3=({V}_3,{E}_3)$ có tập đỉnh ${V}_3 = \{v_1, v_2, v_3, v_4, v_5\}$ và tập cạnh \[{E}_3 = \{(v_1,v_3), (v_3,v_5), (v_5,v_2), (v_2,v_4), (v_4,v_1)\}.\]
\end{itemize}
\end{example}
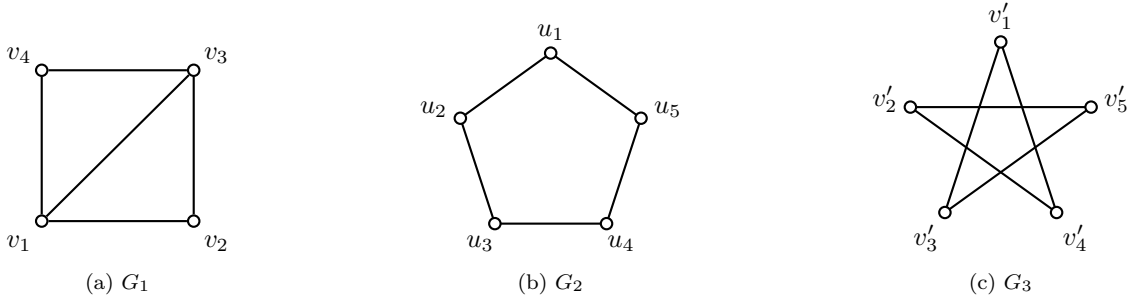
\begin{figure}
\centering
\subfloat[${G}_1$]{
\begin{tikzpicture}[
roundnode/.style={circle, draw=black, thick, minimum size=1.5mm,inner sep= 0.25mm},
squarednode/.style={rectangle, draw=red!60, fill=red!5, very thick, minimum size=5mm},
]
\node[roundnode]   (v2)   at   (0,0) {};
\node[roundnode]   (v3)   at   (0,2) {};
\node[roundnode]   (v4)   at   (-2,2) {};
\node[roundnode]   (v1)   at   (-2,0) {};
\draw[-, thick] (v1)--(v2)--(v3)--(v4)--(v1)--(v3);
\node (a1) at (0.3,-0.3){$v_2$};
\node (a2) at (0.3,2.2){$v_3$};
\node (a3) at (-2.3,2.2){$v_4$};
\node (a4) at (-2.3,-0.3){$v_1$};
\end{tikzpicture}
} \hfill
\subfloat[${G}_2$]{
\begin{tikzpicture}
\node (A) [draw=none,regular polygon, regular polygon sides=5, minimum size=2.6cm,outer sep=0pt] {};
\foreach \n in {1,...,5} {
    \node at (A.corner \n) [anchor=360/5*(\n-1)+270] {$u_{\n}$};}
\node (B) [draw,thick,regular polygon, regular polygon sides=5, minimum size=2.5cm,outer sep=0pt] {};
\foreach \n in {1,...,5} {
    \draw[fill=white,thick] (B.corner \n) circle[radius=.75mm];}
\end{tikzpicture}
} \hfill
\subfloat[${G}_3$]{
\begin{tikzpicture}
\node (A) [draw=none,regular polygon, regular polygon sides=5, minimum size=2.6cm,outer sep=0pt] {};
\foreach \n in {1,...,5} {
    \node at (A.corner \n) [anchor=360/5*(\n-1)+270] {$v_{\n}'$};}
\node (B) [draw=none,thick,regular polygon, regular polygon sides=5, minimum size=2.5cm,outer sep=0pt] {};
\draw[-, thick] (B.corner 1)--(B.corner 3);
\draw[-, thick] (B.corner 2)--(B.corner 4);
\draw[-, thick] (B.corner 3)--(B.corner 5);
\draw[-, thick] (B.corner 5)--(B.corner 2);
\draw[-, thick] (B.corner 4)--(B.corner 1);
\foreach \n in {1,...,5} {
    \draw[fill=white,thick] (B.corner \n) circle[radius=.75mm];}
\end{tikzpicture}
}
\caption{Một số ví dụ về đồ thị vô hướng.}
\label{fig:c2_f1_dothi}
\end{figure}

Giả sử $(v_i,v_j) \in E(G)$ thì ta nói $v_i$, $v_j$ là hai đỉnh kề nhau \index{đỉnh kề} (kí hiệu $v_i \sim v_j$), và đỉnh $v_i$ gọi là \emph{liên thuộc} với cạnh $(v_i,v_j)$. Hai cạnh phân biệt có chung một đỉnh gọi là hai cạnh kề\index{cạnh kề}. Hai đồ thị là đẳng cấu nếu tồn tại một song ánh giữa hai tập đỉnh mà bảo toàn quan hệ liền kề. Từ nay, ta không phân biệt giữa hai đồ thị đẳng cấu \index{đồ thị!đẳng cấu}. Nếu ${G}$ và ${H}$ là hai đồ thị đẳng cấu, ta viết ${G} \cong {H}$ hoặc đơn giản là ${G} = {H}$.

\begin{example} \label{eg:2.2}
Xét hai đồ thị ${G}_2$ và ${G}_3$ trên Hình~\ref{fig:c2_f1_dothi}(b) và (c). Định nghĩa ánh xạ $f:{V}_2 \to {V}_3$ như sau: $f(v_1) = v_1'$, $f(v_2) = v_3'$, $f(v_3) = v_5'$, $f(v_4) = v_2'$, và $f(v_5) = v_4'$. Dễ thấy $f$ là một song ánh giữa ${V}_1$ và ${V}_2$. Hơn nữa, có thể kiểm tra:
\begin{itemize}
\item $(v_1,v_2) \in {E}_1$ thì $(f(v_1),f(v_2)) = (v_1',v_3') \in {E}_2$,
\item $(v_2,v_3) \in {E}_1$ thì $(f(v_3),f(v_3)) = (v_3',v_5') \in {E}_2$, 
\item $(v_3,v_4) \in {E}_1$ thì $(f(v_3),f(v_4)) = (v_5',v_2') \in {E}_2$, 
\item $(v_4,v_5) \in {E}_1$ thì $(f(v_4),f(v_5)) = (v_2',v_4') \in {E}_2$, 
\item $(v_5,v_1) \in {E}_1$ thì $(f(v_5),f(v_1)) = (v_4',v_1') \in {E}_2$.
\end{itemize}
Như vậy $f$ bảo toàn quan hệ liền kề và ta đi đến kết luận rằng ${G}_2 \cong {G}_3$.
\end{example}

Một đồ thị ${G}' = ({V}',{E}')$ là một \emph{đồ thị con} \index{đồ thị!con} của ${G} = ({V},{E})$ nếu ${V}' \subseteq {V}$ và ${E}' \subset {E}$. Trong trường hợp này ta viết ${G}' \subseteq {G}$. Nếu ${V}' \subset {V}$ thì đồ thị $({V}',{E} \cap {V}'\times {V}')$ là một \emph{đồ thị con dẫn xuất} từ ${V}'$, và kí hiệu bởi ${G}[{V}']$. Đồ thị ${H}$ là một đồ thị con dẫn xuất \index{đồ thị!con dẫn xuất} của ${G}$ nếu ${H} \subset {G}$ và ${H} = {G}[{V}({H})]$. Một phân hoạch của đồ thị $G$ là tập $V_1,\ldots,V_k \subset V$, $k\ge 2$, sao cho $V_1\cup V_2 \cup \ldots \cup V_k = V$ và $V_i \cap V_j = \emptyset, \forall i,j \in \{1,\ldots,k\},~i\neq j$.\index{phân hoạch} 

Chúng ta định nghĩa \emph{tập láng giềng} \index{tập!láng giềng} của đỉnh $v_i$ của đồ thị ${G}$ là tập chứa các đỉnh kề với $v_i$, nghĩa là ${N}(v_i) = \{v_j|~ (v_i,v_j) \in {E}\}$. \emph{Bậc} của đỉnh $v_i$ là số phần tử của \index{bậc} tập láng giềng ${N}(v_i)$, tức là ${\rm deg}(v_i) = |{N}(v_i)|$. Ta cũng có thể định nghĩa bậc của một đỉnh trong đồ thị là số cạnh liên thuộc với nó. Khi chỉ có một đồ thị ${G}$, ta có thể dùng kí hiệu rút gọn ${N}(v_i) = {N}_i$ và ${\rm deg}(v_i) = {\rm deg}_i$. Khi có nhiều đồ thị khác nhau, ta thêm kí hiệu đồ thị như một chỉ số dưới. Ví dụ nếu ${H}$ là một đồ thị con được dẫn xuất của ${G}$ và ${v} \in {H}$ thì 
\begin{equation*}
{N}_{{H}}({v}) = N_{{G}}({v}) \cap {V}({H}),~\text{và } \text{deg}_{{H}}(v) = |{N}_{{H}}(v)|.
\end{equation*}

Với tập ${V}' \subset {V}$, ta định nghĩa tập ${N}({V}') = \cup \{{N}(v)|~ v \in {V}'\}$. \emph{Bậc tối thiểu}\index{bậc!tối thiểu} của các đỉnh của ${G}$ được kí hiệu bởi $\delta({G})$ và \emph{bậc tối đa}\index{bậc!tối đa} được kí hiệu bởi $\Delta({G})$. Nếu $\delta({G}) = \Delta({G}) = k$, tức là mọi đỉnh của ${G}$ đều có bậc $k$, thì ${G}$ gọi là một đồ thị đều bậc $k$ \index{đồ thị!đều} hoặc đồ thị chính quy bậc $k$.\index{đồ thị!chính quy} Đồ thị chính quy mạnh là đồ thị đều\index{đồ thị!chính quy mạnh} mà mọi cặp đỉnh kề nhau có cùng số láng giềng chung và mọi cặp đỉnh không kề nhau có cùng số láng giềng chung. Đồ thị đều bậc $k = |{V}|-1 = n-1$ gọi là \emph{đồ thị đầy đủ} bậc $n$, kí hiệu bởi ${K}_n$.\index{đồ thị!đầy đủ} Hình \ref{fig:c1_regularGraphs} minh họa các khái niệm đồ thị chính quy (a), đồ thị chính quy mạnh (b), và đồ thị đều (là một đồ thị chính quy mạnh đặc biệt).

\begin{figure}[th!]
\centering
\subfloat[$C_6$]{
\begin{tikzpicture}[scale=.8]
  \foreach \i/\angle in {1/120,2/20,3/-20,4/-120,5/-150,6/150} {
    \node[circle,draw] (v\i) at (\angle:2) {};
  }
  \foreach \i/\j in {1/2,2/3,3/4,4/5,5/6,6/1} {
    \draw (v\i) -- (v\j);
  }
\end{tikzpicture}
}
\hfill
\subfloat[]{
\begin{tikzpicture}[scale=.8]
  \foreach \i/\angle in {1/120,2/20,3/-20,4/-120,5/-150,6/150} {
    \node[circle,draw] (v\i) at (\angle:2) {};
  }
  \foreach \i/\j in {1/2,2/3,3/4,4/5,5/6,6/1} {
    \draw (v\i) -- (v\j);
  }
  \foreach \i/\j in {1/4,2/5,3/6} {
    \draw (v\i) -- (v\j);
  }
\end{tikzpicture}
}
\hfill
\subfloat[$K_6$]{
\begin{tikzpicture}[scale=0.8]
  \foreach \i/\angle in {1/90,2/20,3/-20,4/-90,5/-150,6/150} {
    \node[circle, draw, inner sep=2pt] (v\i) at (\angle:2) { };
  }
  \foreach \i in {1,...,6} {
    \foreach \j in {\i,...,6} {
      \ifnum\i<\j
        \draw (v\i) -- (v\j);
      \fi
    }
  }
\end{tikzpicture}
}
\caption{(a) Đồ thị chính quy nhưng không chính quy mạnh; (b) và (c): Đồ thị chính quy mạnh.}
\label{fig:c1_regularGraphs}
\end{figure}
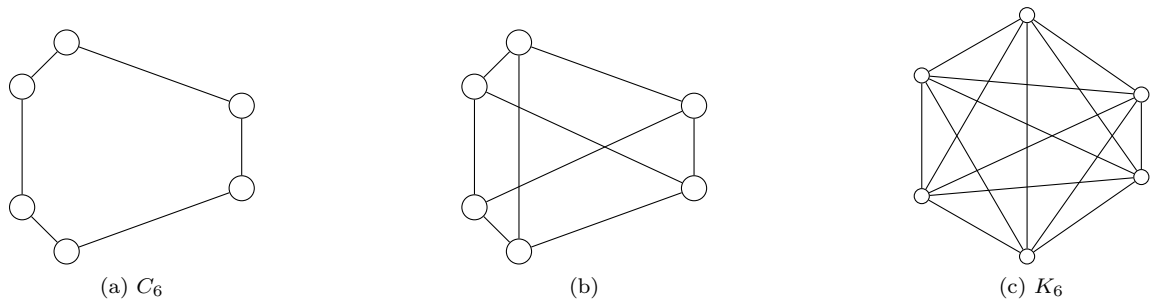

Nếu ${E}' \subset {E}({G})$ thì ${G} - {E}' \triangleq ({V},{E}({G})\setminus {E}')$ gọi là đồ thị thu hẹp \index{đồ thị!thu hẹp} từ ${G}$ sau khi ta xóa đi các cạnh trong ${E}'$. Tương tự, nếu ${V}' \subset {V}({G})$ thì ${G} - {V}'$ là đồ thị thu hẹp từ ${G}$ sau khi xóa đi các đỉnh thuộc ${V}'$, với qui ước rằng khi một đỉnh $v \in {V}'$ bị xóa đi thì các cạnh kề với ${v}$ cũng bị xóa đi. Nói cách khác, nếu ${G} = ({V}, {E})$ thì ${G} - {V}' = ({V}\setminus {V}', {E} \cap ({V}\setminus {V}')\times ({V}\setminus {V}'))$. Nếu ${V}' = \{v\}$, ta thường viết ${G} - v$ thay cho ${G} - \{v\}$. Tương tự, ta viết ${G} - e$ thay cho ${G} - \{e\}$. Với đồ thị ${H} \subset {G}$, ta có thể viết ${G} - {H}$ thay cho ${G} - {V}({H})$. Nếu cạnh $e \in {V}\times {V} \setminus {E}$ thì đồ thị mở rộng \index{đồ thị!mở rộng} từ ${G}$ bằng cách thêm vào cạnh $e$ được kí hiệu là ${G} + {e} = ({V},{E} \cup \{e\})$. Ta cũng có những định nghĩa tương tự cho đồ thị mở rộng nhờ thêm đỉnh.

Với ${V} = \{v_1, \ldots, v_n\}$, $\{{\rm deg}(v_i)\}_1^n$ gọi là \emph{chuỗi bậc} của ${G}$. Thông thường, ta sắp xếp các đỉnh sau cho chuỗi bậc là đơn điệu tăng hoặc đơn điệu giảm. Dễ thấy rằng \index{chuỗi bậc}
\begin{equation*}
\sum_{i=1}^n \text{deg}(v_i) = 2 |{E}(G)| = 2m,
\end{equation*}
do đó tổng chuỗi bậc $\sum_{i=1}^n \text{deg}(v_i)$ luôn luôn là một số chẵn.

Với $u$ và $v$ là hai đỉnh không nhất thiết trùng nhau của ${G}$, một \emph{đường đi} \index{đường đi} $W$ nối $u-v$ là một chuỗi luân phiên giữa đỉnh và cạnh, $u_1, e_1, u_2, e_2, \ldots, u_l, e_l, u_{l+1}$, với $u_1 = u$ là đỉnh đầu, $u_{l+1} = v$ là đỉnh cuối, và các cạnh $e_i = (u_i,u_{i+1}) \in {E}({G}), 1 \leq i \leq l$. Thông thường, ta kí hiệu $W = v_1v_2\ldots v_{l+1}$ bởi dưới dạng này có thể xác định rõ các cạnh của $W$. Độ dài của lối mòn $W$ này là $l$. Tập đỉnh và tập cạnh của $W$ được kí hiệu lần lượt là $V(W) = \{v_i| ~ i = 1, \ldots, l+1\}$ và $E(W) = \{e_i| ~i = 1, \ldots, l+1\}$. Một đường đi có tất cả các cạnh khác nhau gọi là một \emph{lối mòn}\index{lối mòn}. Một đường đi có tất cả các đỉnh đôi một khác nhau gọi là một \emph{đường đi đơn}\index{đường đi!đơn}. \index{độ dài đường đi}

Một đường đi có đỉnh đầu và đỉnh cuối trùng nhau được gọi là một \emph{chu trình}\index{chu trình}, một lối mòn có đỉnh đầu và đỉnh cuối trùng nhau gọi là một \emph{mạch} (circuit)\index{mạch}, và một mạch có mọi đỉnh phân biệt ngoại trừ đỉnh đầu và đỉnh cuối gọi là một \emph{chu trình đơn}. Chu trình đơn độ dài $l \geq 3$ được kí hiệu bởi $C_n$. \index{chu trình!đơn} Một chu trình thường được cho dưới dạng $v_1v_2\ldots v_l$ (thay cho $v_1v_2\ldots v_lv_1$). Ta thường đồng nhất đường đi đơn $P$ và chu trình đơn $C$ tương ứng với các đồ thị $(V(P),E(P))$ và $(V(C),E(C))$. Như vậy, $v_1v_2\ldots v_{l+1}$ và $v_{l+1}v_l \ldots v_1$ kí hiệu cùng một đường đi. Tương tự, $v_1v_2\ldots v_l$ và $v_2v_3\ldots v_lv_1$ kí hiệu cùng một chu trình. 

Một đồ thị là \index{liên thông} \emph{liên thông} nếu tồn tại ít nhất một đường đi giữa hai đỉnh bất kì của đồ thị. Ngược lại, ta gọi đồ thị là không liên thông. Một \emph{thành phân liên thông} \index{thành phần!liên thông} của $G$ là một đồ thị con $H = (V(H), E(G) \cap V(H) \times V(H))$ sao cho $H$ là liên thông và không có đường đi nào trong $G$ nối $H$ với các đỉnh $v \in V(G)\setminus V(H)$.

\begin{figure}[t!]
    \centering
    \input{Tikz/c2_vertex_cutset.tex}
    \caption{Ví dụ về các lát cắt cạnh của đồ thị $G$ gồm 5 đỉnh và 8 cạnh, $\kappa_0(G)=1$.}
    \label{fig:chap1_vertex_cutset}
\end{figure}

Một đồ thị liên thông không chứa chu trình nào gọi là một \emph{cây}\index{cây}. Một đồ thị không chứa chu trình nào gọi là một \index{rừng}\emph{rừng}. Như vậy, một rừng chứa một hay nhiều thành phần liên thông, mỗi thành phần là một đồ thị con không chứa chu trình.

Một cây chứa tất cả các đỉnh của đồ thị gọi là một \index{cây!bao trùm}cây bao trùm. Một cây có $n$ đỉnh thì luôn có $n-1$ cạnh. Đường đi giữa hai đỉnh bất kì trong một cây là duy nhất. Một rừng gồm $n$ đỉnh và $c$ thành phần có $n-c$ cạnh. Rõ ràng, đồ thị chứa cây bao trùm thì liên thông.

Khoảng cách trên đồ thị giữa hai đỉnh $u$, $v$, kí hiệu bởi $d(u,v)$ là độ dài nhỏ nhất của một đường đi nối $u$ với $v$. Nếu như không tồn tại một đường đi nào nối $u-v$, hay nói cách khác, $u$ và $v$ thuộc về hai thành phần khác nhau, ta qui ước $d(u,v) = \infty$.\index{khoảng cách đồ thị} \emph{Đường kính của đồ thị} là khoảng cách lớn nhất giữa hai đỉnh trong đồ thị.\index{đường kính của đồ thị}

\begin{remark}
Trong các định nghĩa ở trên, ta không xét đồ thị có chứa \emph{khuyên} (một cạnh nối một đỉnh với chính nó) và giả thiết rằng giữa hai đỉnh bất kì chỉ được nối bởi một cạnh (không có các cạnh \emph{cạnh bội}). Một đồ thị có cạnh bội gọi là một đa đồ thị. Một đa đồ thị có chứa khuyên gọi là một giả đồ thị.\index{khuyên}
\end{remark}

Giả sử $G=(V,E)$ là một đồ thị liên thông. Mỗi \emph{lát cắt đỉnh} là một tập con các đỉnh của $V$ mà khi loại bỏ các đỉnh này thì đồ thị mất tính liên thông. \emph{chỉ số!liên kết đỉnh} của $G$, kí hiệu $\kappa_0(G)$ là số đỉnh nhỏ nhất trong mọi lát cắt đỉnh của $G$\index{chỉ số!liên kết đỉnh}. Tương tự, ta định nghĩa \emph{lát cắt cạnh}\index{lát cắt!cạnh} là một tập con của $E$ mà nếu loại các cạnh này sẽ làm đồ thị $G$ mất liên thông. Số cạnh tối thiểu trong mọi lát cắt cạnh của $G$ gọi là \emph{chỉ số!liên kết cạnh} và kí hiệu bởi $\kappa_1(G)$. Ví dụ về các lát cắt đỉnh và lát cắt cạnh được cho như trong Hình~\ref{fig:chap1_vertex_cutset} và~\ref{fig:chap1_edge_cutset}. \index{lát cắt!đỉnh}

\begin{figure}[t!]
    \centering
    \subfloat[]{
\resizebox{2.5cm}{!}
{
\begin{tikzpicture}[
roundnode/.style={circle, draw=black, thick, minimum size=2mm,inner sep= 0.25mm},
squarednode/.style={rectangle, draw=black, thick, minimum size=3.5mm,inner sep= 0.25mm},
]
    
    \node (v1) at (0,1.5) {};
    \node (v2) at (0,-1.5) {};
    \node (v3) at (4,1.5) {};
    \node (v4) at (4,-1.5) {};
    
    \node[roundnode] (u1) at (0,0) { }; %
    \node[roundnode] (u2) at (1,1) { };%
    \node[roundnode] (u3) at (1,-1) { };%
    \node[roundnode] (u4) at (2,0) { };%
    \node[roundnode] (u5) at (3,1) { };%
    \node[roundnode] (u6) at (3,-1) { };%
    
    \draw[-, very thick] (u1)--(u2)--(u3)--(u4)--(u5)--(u6)--(u4)--(u2);
    \draw[-, very thick] (u1)--(u3);
\end{tikzpicture}}}
\hfill
\subfloat[]{
\resizebox{2.5cm}{!}
{
\begin{tikzpicture}[
roundnode/.style={circle, draw=black, thick, minimum size=2mm,inner sep= 0.25mm},
squarednode/.style={rectangle, draw=black, thick, minimum size=3.5mm,inner sep= 0.25mm},
]
    \draw[fill=red!10, draw=red!50!white, ultra thick, dashed] plot [smooth cycle, tension=.6] coordinates { 
    (.1,0.5) (1,.75) (1.9,.5) (1,0.25) }; 
    
    \node (v1) at (0,1.5) {};
    \node (v2) at (0,-1.5) {};
    \node (v3) at (4,1.5) {};
    \node (v4) at (4,-1.5) {};
    
    \node[roundnode] (u1) at (0,0) { }; %
    \node[roundnode] (u2) at (1,1) { };%
    \node[roundnode] (u3) at (1,-1) { };%
    \node[roundnode] (u4) at (2,0) { };%
    \node[roundnode] (u5) at (3,1) { };%
    \node[roundnode] (u6) at (3,-1) { };%
    
    \draw[-, very thick] (u1)--(u3)--(u4)--(u5)--(u6)--(u4);
    \draw[-, very thick, color = red] (u1)--(u2);
    \draw[-, very thick, color = red] (u3)--(u2);
    \draw[-, very thick, color = red] (u4)--(u2);
\end{tikzpicture}}}
\hfill
\subfloat[]{
\resizebox{2.5cm}{!}
{
\begin{tikzpicture}[
roundnode/.style={circle, draw=black, thick, minimum size=2mm,inner sep= 0.25mm},
squarednode/.style={rectangle, draw=black, thick, minimum size=3.5mm,inner sep= 0.25mm},
]
    \draw[fill=red!10, draw=red!50!white, ultra thick, dashed] plot [smooth cycle, tension=.6] coordinates { 
    (2.25,0.25) (2.75,0.25) (2,.75) (1.25,0) (2,-.75)  (2.75,-0.25) (2.25,-0.25) (2,-0.5) (1.5,0) (2,0.5)}; 
    
    \node (v1) at (0,1.5) {};
    \node (v2) at (0,-1.5) {};
    \node (v3) at (4,1.5) {};
    \node (v4) at (4,-1.5) {};
    
    \node[roundnode] (u1) at (0,0) { }; %
    \node[roundnode] (u2) at (1,1) { };%
    \node[roundnode] (u3) at (1,-1) { };%
    \node[roundnode] (u4) at (2,0) { };%
    \node[roundnode] (u5) at (3,1) { };%
    \node[roundnode] (u6) at (3,-1) { };%
    
    \draw[-, very thick] (u1)--(u2)--(u3)--(u1);
    \draw[-, very thick, color = red] (u4)--(u2);
    \draw[-, very thick, color = red] (u3)--(u4);
    \draw[-, very thick, color = red] (u4)--(u5);
    \draw[-, very thick, color = red] (u4)--(u6);
    \draw[-, very thick] (u6)--(u5);
\end{tikzpicture}}}
\hfill
\subfloat[]{
\resizebox{2.5cm}{!}
{
\begin{tikzpicture}[
roundnode/.style={circle, draw=black, thick, minimum size=2mm,inner sep= 0.25mm},
squarednode/.style={rectangle, draw=black, thick, minimum size=3.5mm,inner sep= 0.25mm},
]
    \draw[fill=red!10, draw=red!50!white, ultra thick, dashed] plot [smooth cycle, tension=.6] coordinates {(2.1,.75) (3,0.25) (3.5,-.5) (2.5,0)}; 
    
    \node (v1) at (0,1.5) {};
    \node (v2) at (0,-1.5) {};
    \node (v3) at (4,1.5) {};
    \node (v4) at (4,-1.5) {};
    
    \node[roundnode] (u1) at (0,0) { }; %
    \node[roundnode] (u2) at (1,1) { };%
    \node[roundnode] (u3) at (1,-1) { };%
    \node[roundnode] (u4) at (2,0) { };%
    \node[roundnode] (u5) at (3,1) { };%
    \node[roundnode] (u6) at (3,-1) { };%
    
    \draw[-, very thick] (u1)--(u2)--(u3)--(u1);
    \draw[-, very thick] (u4)--(u2);
    \draw[-, very thick] (u3)--(u4);
    \draw[-, very thick, color = red] (u4)--(u5);
    \draw[-, very thick] (u4)--(u6);
    \draw[-, very thick, color = red] (u6)--(u5);
\end{tikzpicture}}}
\hfill
\subfloat[]{
\resizebox{2.5cm}{!}
{
\begin{tikzpicture}[
roundnode/.style={circle, draw=black, thick, minimum size=2mm,inner sep= 0.25mm},
squarednode/.style={rectangle, draw=black, thick, minimum size=3.5mm,inner sep= 0.25mm},
]
    \draw[fill=red!10, draw=red!50!white, ultra thick, dashed] plot [smooth cycle, tension=.6] coordinates { 
    (.3,0) (.5,0.9) (0.7,0) (0.5,-0.9)}; 
    
    \node (v1) at (0,1.5) {};
    \node (v2) at (0,-1.5) {};
    \node (v3) at (4,1.5) {};
    \node (v4) at (4,-1.5) {};
    
    \node[roundnode] (u1) at (0,0) { }; %
    \node[roundnode] (u2) at (1,1) { };%
    \node[roundnode] (u3) at (1,-1) { };%
    \node[roundnode] (u4) at (2,0) { };%
    \node[roundnode] (u5) at (3,1) { };%
    \node[roundnode] (u6) at (3,-1) { };%
    
    \draw[-, very thick] (u4)--(u2)--(u3)--(u4);
    \draw[-, very thick, color = red] (u1)--(u2);
    \draw[-, very thick, color = red] (u3)--(u1);
    \draw[-, very thick] (u4)--(u5);
    \draw[-, very thick] (u4)--(u6);
    \draw[-, very thick] (u6)--(u5);
\end{tikzpicture}}}
    \caption{Ví dụ về các lát cắt đỉnh của đồ thị $G$ gồm 5 đỉnh và 8 cạnh, $\kappa_1(G)=2$.}
    \label{fig:chap1_edge_cutset}
\end{figure}
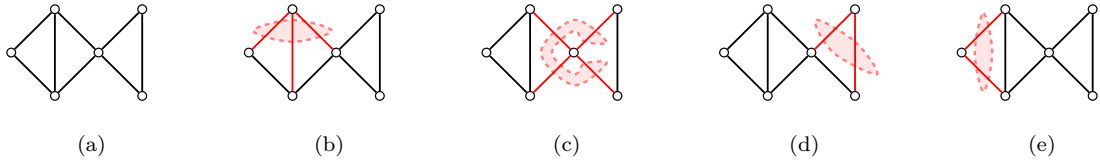

Với $S,T\subset V$ là một phân hoạch nhị nguyên bất kỳ của đồ thị liên thông $G$ ($S\cup T = V$ và $S \cap T = \emptyset$), kí hiệu $\mathcal{C}_G$ là tập gồm mọi các lát cắt cạnh $C(S,T) \subseteq E(G)$. Độ lớn của lát cắt $C(S,T)$ là số phần tử của $C(S,T)$. Giả sử $|S|\ge |T|$, ta nói $T$ bị cắt khỏi $G$. Hiệu quả của phép cắt $C(S,T)$ là tỉ số $\frac{|C(S,T)|}{|T|}=\frac{|C(S,T)|}{\min(|S|,|T|)}$. Hằng số Cheeger của đồ thị được định nghĩa bởi
\begin{align}\label{eq:c1_cheeger_const}
h(G)=\min_{C(S,T) \in \mc{C}_G} \frac{|C(S,T)|}{|T|}>0
\end{align}
là một chỉ số về mức độ bền vững của đồ thị liên thông. Nếu $h(G)$ nhỏ, tồn tại phân hoạch nhị nguyên với kích thước hai thành phần không quá khác biệt, liên kết với nhau bởi một số lượng nhỏ các cạnh. Ngược lại, nếu $h(G)$ lớn, số cạnh trong đồ thị $G$ sẽ phân bố đối xứng hơn trong đồ thị, lát cắt cạnh của hai phân hoạch bất kỳ của tập đỉnh $V(G)$ có kích thước tương đối lớn. Chú ý rằng khái niệm ``lớn'' và ``nhỏ'' tùy thuộc vào đồ thị được xét (so sánh $h(G)$ với giá trị cực đại của hiệu quả của phép cắt).

\subsection{Đồ thị hữu hướng}
Một \index{đồ thị!hữu hướng} đồ thị hữu hướng $G$ định nghĩa bởi một tập đỉnh $V = V(G)$ cùng với một tập các cạnh hữu hướng $E = E(G)\in V \times V$. Một cạnh có hướng $(u,v) \in E$ (với $u \neq v$) được biểu diễn hình học bởi một cung có hướng từ đỉnh $u$ tới đỉnh $v$. Khi làm việc với đồ thị hữu hướng, $(u,v)$ và $(v,u)$ là hai cạnh khác nhau và có thể cùng tồn tại. Hơn nữa, nếu $(u,v) \in E$ thì không suy ra được rằng $(v,u) \in E$. Hầu hết các định nghĩa cho đồ thị vô hướng có thể được mở rộng ngay cho đồ thị có hướng. Với mỗi đỉnh $u \in V$, ta định nghĩa bậc-ra \index{bậc-ra} của đỉnh $u$ là số cạnh có hướng xuất phát từ $u$, kí hiệu $\text{deg}^+(u)$. Định nghĩa tập láng giềng-ra \index{tập!láng giềng-ra} của đỉnh $u$ bởi $N^+(u) = \{v|~ (u,v) \in E\}$ thì ta có $\text{deg}^+(u)=|N^+(u)|$.\footnote{Chú ý rằng một số tài liệu sử dụng kí hiệu $\text{deg}_{out}(u)$ và $N_{out}(u)$ với cùng ý nghĩa.} Hoàn toàn tương tự, ta có thể định nghĩa bậc-vào \index{bậc-vào} $\text{deg}^-(u)$ và tập láng giềng-vào \index{tập!láng giềng-vào} $N^-(u)$ của đỉnh $u$. Nếu đồ thị $G$ có $\text{deg}^+(u) = \text{deg}^-(u)$ với mọi $u \in V$ thì $G$ gọi là một đồ thị cân bằng.

Một đường đi hữu hướng \index{đường đi!hữu hướng} $u_0u_1\ldots u_k$ là một đường đi chứa các cạnh hữu hướng $(u_i,u_{i+1}) \in E$. Với một đồ thị đơn, vô hướng $H$, ta có thể xây dựng một đồ thị hữu hướng $G$ bằng cách gán cho mỗi cạnh của $H$ một hướng nhất định. Đồ thị $G$ gọi là một đồ thị được định hướng \index{đồ thị!được định hướng} của $H$. Ngược lại, với một đồ thị hữu hướng $G$, ta có thể loại bỏ hướng của tất cả các cạnh của $G$ và thu được đồ thị vô hướng $H$.\footnote{Nếu tồn tại hai cạnh $(u,v)$ và $(v,u)$ trong $G$ thì ta coi chúng là một cạnh trong $H$.} Nếu đồ thị vô hướng $H$ là liên thông, thì $G$ là một đồ thị \index{liên thông yếu} liên thông yếu. Nếu với mỗi cặp đỉnh $u, v \in V(G), u \neq v$ đều tồn tại một đường đi hữu hướng $u-v$, thì $G$ gọi là một đồ thị \index{liên thông mạnh} liên thông mạnh. Có thể thấy rằng khi đồ thị là vô hướng thì việc xét hai khái niệm liên thông yếu và liên thông mạnh là tương đương.

Nếu trong đồ thị hữu hướng $G = (V,E)$, tồn tại một đỉnh $u \in V$ sao cho với mỗi đỉnh $v \in E, v \neq u$, ta đều tìm được một đường đi hữu hướng từ $u$ tới $v$ trong $G$, thì đỉnh $u$ gọi là một \emph{gốc-ra}\index{gốc-ra} của $G$, và $G$ gọi là một đồ thị có gốc-ra tại $u$. Khi đó, $G$ chứa một cây bao trùm hữu hướng $T$ sao cho từ gốc-ra $u$ có thể đi đến tất cả các đỉnh khác trong $T$. Tương tự, nếu trong $G$ tồn tại một đỉnh $u \in V$ sao cho với mọi $v \in E, v \neq u$, ta đều tìm được một đường đi hữu hướng trong $G$ từ $v$ tới $u$, thì đỉnh $u$ gọi là một gốc-vào của $G$, còn $G$ được gọi là một đồ thị có gốc-vào tại $u$. Khi đó, $G$ chứa một cây bao trùm \index{cây!bao trùm hữu hướng} hữu hướng $T$ sao cho gốc-vào $u$ có thể đi tới được từ tất cả các đỉnh khác trong $T$. Với một đồ thị liên thông mạnh thì mỗi đỉnh của đồ thị vừa là một gốc-vào và cũng là một gốc-ra.

Với mỗi đỉnh $u$ bất kì của $G$, ta luôn có thể tìm được một đồ thị con, liên thông mạnh, tối đa $H$ của $G$ sao cho $u \in V(H)$. Ở đây, thuật ngữ ``tối đa'' mang ý nghĩa là không tồn tại một đồ thị con liên thông mạnh $K$ nào của $G$ sao cho $H$ là một đồ thị con dẫn xuất của $K$. Chú ý rằng đồ thị $H$ có thể chỉ chứa một đỉnh $u$. Với mọi đồ thị $G$, ta luôn có thể phân hoạch $G$ thành các đồ thị con liên thông mạnh tối đa\index{đồ thị!con!liên thông mạnh tối đa}. Mỗi đồ thị con này gọi là một thành phần của $G$. Do tính tối đa của mỗi thành phần, phân hoạch này của $G$ là duy nhất. \index{phân hoạch}

\subsection{Đồ thị có trọng số}
Ta có thể định nghĩa một đồ thị có trọng số $G$ bởi một bộ ba $({V},{E},{A})$, trong đó ngoài tập đỉnh ${V}$ và tập cạnh ${E}$, ta có thêm tập trọng số ${A} = \{\omega_{ij} \in \mb{R}_{+}|~ i \neq j, i, j \in V\}$. Ứng với mỗi cạnh $(i,j)\in E$ có một trọng số $\omega_{ij} > 0$ tương ứng. Trong khi đó, nếu $(i,j)\notin E$ thì trọng số $\omega_{ij}$ được cho bằng 0. Nhờ có trọng số, ngoài tính liên kết trong đồ thị, ta có thể đánh giá tương đối về mức độ  quan trọng (hay mạnh yếu) của các cạnh (liên kết) trong đồ thị.

\section{Đại số đồ thị}
\label{sec2:dai_so_do_thi}
\subsection{Một số ma trận của đồ thị}
\subsubsection{Ma trận bậc và ma trận kề}
Xét đồ thị $G=(V,E,A)$, ta định nghĩa \index{ma trận!kề} \emph{ma trận kề} $\m{A}(G) =[a_{ij}]_{n \times n} \in \mb{R}^{n \times n}$ của $G$ bởi:
\begin{equation}
a_{ij} = \left\lbrace \begin{array}{ll}
\omega_{ji}, & \text{nếu } (v_j,v_i) \in {E},\\
0, & \text{các trường hợp khác.} \\
\end{array} \right.
\end{equation}
\index{ma trận!bậc} \emph{Ma trận bậc} $\m{D}(G)$ được định nghĩa bởi 
\begin{align}
\m{D}(G) &= \text{diag}(\text{deg}(v_1),\ldots,\text{deg}(v_n)) \nonumber\\
&= \begin{bmatrix}
\text{deg}(v_1)& 0 & \cdots & 0\\
0 & \text{deg}(v_2) & \cdots & 0 \\
\vdots & \vdots & \ddots & \vdots\\
0 & 0 & \cdots & \text{deg}(v_n)
\end{bmatrix},
\end{align}
với $\text{deg}(v_k)=\text{deg}^{-}(v_k)=\sum_{j\in N_i} a_{ij}$. Khi chỉ có đồ thị $G$, ta sẽ viết gọn $\m{A}(G) = \m{A}$ và $\m{D}(G) = \m{D}$. 

\begin{example} \label{eg:2.3}
Ma trận kề và ma trận bậc của đồ thị trên hình~\ref{fig:c2_f1_dothi}(a) được cho bởi:
\begin{align*}
\begin{array}{*{20}{c}}
  {}&{\begin{array}{*{20}{c}}
  {{v_1}}&{{v_2}}&{{v_3}}&{{v_4}}
\end{array}} \\ 
  {\m{A}(G_1) = }&{\left[ {\begin{array}{*{20}{c}}
  0&1&1&1 \\ 
  1&0&1&0 \\ 
  1&1&0&1 \\ 
  1&0&1&0 
\end{array}} \right]} 
\end{array}\begin{array}{*{20}{c}}
  {} \\ 
  {\begin{array}{*{20}{c}}
  {{v_1}} \\ 
  {{v_2}} \\ 
  {{v_3}} \\ 
  {{v_4}} 
\end{array}} 
\end{array},\qquad \m{D}(G_1) = \begin{bmatrix}
3& 0 & 0 & 0\\
0 & 2 & 0 & 0 \\
0 & 0 & 3 & 0\\
0 & 0 & 0 & 2
\end{bmatrix}.
\end{align*}
\end{example}
Dễ thấy ma trận kề $\m{A} = \m{A}^\top$ luôn có các phần tử trên đường chéo chính bằng 0, và các phần tử khác đều không âm. Hơn nữa, nếu đồ thị $G$ là vô hướng thì ma trận kề là đối xứng $\m{A} = \m{A}^\top$.

\subsubsection{Ma trận liên thuộc}
Giả sử đồ thị $G=(V,E,A)$ có $|V| = n$ đỉnh và $|E| = m$ cạnh. Xét một đồ thị định hướng bất kì của $G$ và đánh số các cạnh của đồ thị này bởi $e_1, \ldots, e_m$, \index{ma trận!liên thuộc} \emph{ma trận liên thuộc} $\m{H} = [h_{ki}]_{m \times n} \in \mb{R}^{m \times n}$ biểu diễn mối liên hệ giữa các đỉnh và các cạnh của đồ thị $G$. Mỗi hàng của ma trận liên thuộc ứng với một cạnh của $E$ trong khi mỗi cột tương ứng với một đỉnh trong $V$. Cụ thể, các phần tử của ma trận liên thuộc được xác định bởi
\begin{equation}
h_{ki} = \left\lbrace \begin{array}{rl}
-1, & \text{nếu } e_k = (v_i,v_j),\\
1, & \text{nếu } e_k = (v_j,v_i),\\
0, & \text{trường hợp khác.}
\end{array}\right.
\end{equation}
\begin{example} \label{eg:2.4}
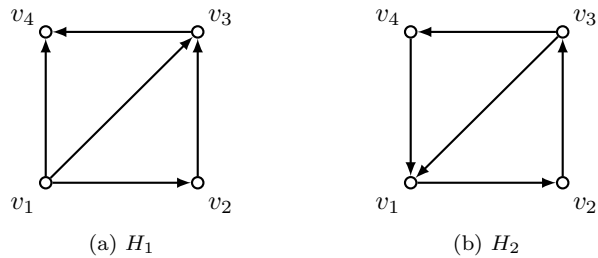
\begin{figure}[h!]
\centering
\subfloat[${H}_1$]{
\centering
\begin{tikzpicture}[
roundnode/.style={circle, draw=black, thick, minimum size=1.5mm,inner sep= 0.25mm},
squarednode/.style={rectangle, draw=red!60, fill=red!5, very thick, minimum size=5mm},
]
\node[roundnode]   (v2)   at   (0,0) {};
\node[roundnode]   (v3)   at   (0,2) {};
\node[roundnode]   (v4)   at   (-2,2) {};
\node[roundnode]   (v1)   at   (-2,0) {};
\draw[->, thick] (v1)--(v2);
\draw[->, thick] (v1)--(v3);
\draw[->, thick] (v1)--(v4);
\draw[->, thick] (v2)--(v3);
\draw[->, thick] (v3)--(v4);
\node (a1) at (0.3,-0.3){$v_2$};
\node (a2) at (0.3,2.2){$v_3$};
\node (a3) at (-2.3,2.2){$v_4$};
\node (a4) at (-2.3,-0.3){$v_1$};
\end{tikzpicture}
}\qquad \qquad
\subfloat[${{H}_2}$]{
\begin{tikzpicture}[
roundnode/.style={circle, draw=black, thick, minimum size=1.5mm,inner sep= 0.25mm},
squarednode/.style={rectangle, draw=red!60, fill=red!5, very thick, minimum size=5mm},
]
\node[roundnode]   (v2)   at   (0,0) {};
\node[roundnode]   (v3)   at   (0,2) {};
\node[roundnode]   (v4)   at   (-2,2) {};
\node[roundnode]   (v1)   at   (-2,0) {};
\draw[->, thick] (v1)--(v2);
\draw[->, thick] (v3)--(v1);
\draw[->, thick] (v4)--(v1);
\draw[->, thick] (v2)--(v3);
\draw[->, thick] (v3)--(v4);
\node (a1) at (0.3,-0.3){$v_2$};
\node (a2) at (0.3,2.2){$v_3$};
\node (a3) at (-2.3,2.2){$v_4$};
\node (a4) at (-2.3,-0.3){$v_1$};
\end{tikzpicture}
}
\caption{Một số đồ thị định hướng khác nhau của đồ thị $G$ trên Hình~\ref{fig:c2_f1_dothi}(a).}
\label{fig:c2_eg2.5}
\end{figure}
Ma trận liên thuộc của đồ thị biểu diễn trên Hình~\ref{fig:c2_eg2.5}(a) được cho bởi:
\[\begin{array}{*{20}{c}}
  {}&{\begin{array}{*{20}{c}}
  {{\text{  }}{v_1}}&{{\text{ }}{v_2}}&{{\text{ }}{v_3}}&{~{v_4}{\text{ }}} 
\end{array}} \\ 
  {\m{H}(H_1) = }&{\left[ {\begin{array}{*{20}{c}}
  { - 1}&1&0&0 \\ 
  { - 1}&0&1&0 \\ 
  { - 1}&0&0&1 \\ 
  0&{ - 1}&1&0 \\ 
  0&0&{ - 1}&1 
\end{array}} \right]} 
\end{array}{\text{  }}\begin{array}{*{20}{c}}
  {} \\ 
  {{e_1} = ({v_1},{v_2})} \\ 
  {{e_2} = ({v_1},{v_3})} \\ 
  {{e_3} = ({v_1},{v_4})} \\ 
  {{e_4} = ({v_2},{v_3})} \\ 
  {{e_5} = ({v_3},{v_4})} 
\end{array}\]
\end{example}
Mặc dù với mỗi cách định hướng khác nhau của $G$, dấu của các hàng của ma trận $\m{H}$ sẽ thay đổi tương ứng. Tuy nhiên, hạng của ma trận $\m{H}$ là không phụ thuộc vào cách ta định hướng đồ thị $G$.

Ma trận liên thuộc cho ta thông tin về cấu trúc của đồ thị $G$. Cụ thể, nếu đồ thị $G$ là liên thông thì không gian rỗng của ma trận $\m{H}$ được sinh bởi vector $\m{1}_n = [1, \ldots, 1]^\top$, nói cách khác $\text{ker}(\m{H}) = \text{span}\{\m{1}_n\}$. Trong khi đó, không gian nhân của ma trận $\m{H}^\top$ liên hệ với \index{không gian!chu trình} \emph{không gian chu trình} \footnote{cycle space} của đồ thị $G$. 

Xét một đồ thị định hướng $H$ của $G$ và giả sử rằng $G$ là liên thông. Trong đồ thị ${H}$, một \index{vector đường đi đánh dấu} vector đường đi đánh dấu\footnote{signed path vector} $\m{z}$ tương ứng với một đường đi hữu hướng $P$ sao cho phần tử thứ $i$ của $\m{z} = [z_1, \ldots, z_m]^\top$ nhận giá trị:
\begin{equation*}
z_{i} = \left\lbrace \begin{array}{cl}
+1, & \text{nếu cạnh thứ $i$ được đi thuận hướng trong $P$},\\
-1, & \text{nếu cạnh thứ $i$ được đi ngược hướng trong $P$},\\
0, & \text{nếu cạnh thứ $i$ không thuộc $P$.}
\end{array}\right.
\end{equation*}
Với một đường đi $P$ với đỉnh đầu và cuối khác nhau trong đồ thị ${H}$ mô tả bởi vector $\m{z}$, vector $\m{y} = \m{H}^\top \m{z}$ nhận giá trị $-1$ nếu đỉnh $i$ là đỉnh xuất phát của đường đi, $1$ nếu đỉnh $i$ là đỉnh kết thúc, và $0$ trong các trường hợp khác. Không gian rỗng của $\m{H}^\top$ được sinh bởi các vector đường đi đánh dấu độc lập tuyến tính tương ứng với các \emph{chu trình độc lập} của $H$. Gọi $\mu$ là số chu trình độc lập trong ${H}$ (cũng như của $G$) thì $\mu = {\rm dim}({\rm ker}(\m{H}^\top))$. Do đó, ${\rm rank}(\m{H}^\top) = \text{dim}(\text{im}(\m{H}^\top)) - \text{dim}({\rm ker}(\m{H}^\top)) = m - \mu$. Mặt khác, với $G$ liên thông thì ${\rm rank}(\m{H}) = {\rm dim}({\rm im}(\m{H})) - {\rm dim}({\rm ker}(\m{H})) = n - 1$. Từ ${\rm rank}(\m{H}) = {\rm rank}(\m{H}^\top)$, ta suy ra công thức \index{chu trình!độc lập}
\begin{equation}
    \mu = m - n + 1. \label{eq:c2_cycleSpace}
\end{equation}
Trong trường hợp $G$ không liên thông và có $c$ thành phần, ${\rm dim}({\rm ker}(\m{H})) = c$. Khi đó, công thức tính số chu trình độc lập trở thành $\mu = m - n + c$. Chú ý rằng công thức \eqref{eq:c2_cycleSpace} cũng có thể suy ra như sau: 

Xét một cây bao trùm bất kỳ ${T}$ của $G$ gồm $n-1$ cạnh. Rõ ràng, ${T}$ không chứa bất kỳ chu trình nào. Với mỗi cạnh $e \in E(G)\setminus E({T})$ được thêm vào đồ thị thì ta có một chu trình độc lập tương ứng. Vì vậy, số chu trình độc lập trong $G$ bằng $|E(G)\setminus E({T})|=m-n+1$.

\begin{example} \label{eg:2.5}
Xét ma trận liên thuộc trong ví dụ~\ref{eg:2.4} (đồ thị trên hình~\ref{fig:c2_eg2.5}(a)). Không gian nhân của $\m{H}$ được sinh bởi vector $\m{1}_4$ do đồ thị là liên thông yếu. Dễ thấy không gian nhân của $\m{H}^\top$ chứa các vector
\begin{itemize}
\item $\m{z}_1 = [1, -1, 0, 1, 0]^\top$ tương ứng với chu trình đi qua các đỉnh $v_1$, $v_2$, $v_3$, $v_1$;
\item $\m{z}_2 = [0, 1, -1, 0, 1]^\top$ tương ứng với chu trình đi qua các đỉnh $v_1$, $v_2$, $v_3$, $v_1$;
\item $\m{z}_3 = [1, 0, -1, 1, 1]^\top$ tương ứng với chu trình đi qua các đỉnh $v_1$, $v_2$, $v_3$, $v_4$, $v_1$.
\end{itemize}
Do $\m{z}_3 = \m{z}_1 + \m{z}_2$, $\m{z}_3$ phụ thuộc tuyến tính với $\m{z}_1$ và $\m{z}_2$. Như vậy $\mu(H) = 2$ và ${\rm ker}(\m{H}^\top) = {\rm span}\{\m{z}_1, \m{z}_2\}$. Có thể kiểm tra lại rằng $\mu = m - n +1 = 5 - 4 + 1 = 2$.
\end{example}

\subsubsection{Ma trận Laplace của đồ thị vô hướng}
\index{ma trận!Laplace} Ma trận Laplace $\mcl{L} = [l_{ij}]_{n \times n} \in \mb{R}^{n\times n}$ của đồ thị vô hướng $G$ được định nghĩa bởi:
\begin{equation} \label{eq:c2_Laplacian_def}
\mcl{L} \triangleq \m{D} - \m{A}.
\end{equation}
Nói cách khác, các phần tử của ma trận Laplace có thể được định nghĩa từ các phần tử của ma trận $\m{A}$ như sau:
\begin{align*}
l_{ij} = \left\lbrace \begin{array}{cl}
-a_{ij}, & \text{nếu } i, j \in {V},~i\neq j,\\
\sum_{j=1}^n a_{ij}, & \text{nếu } i = j.
\end{array}\right.
\end{align*} 
Chúng ta cũng có thể định nghĩa ma trận $\mcl{L}$ dựa trên ma trận liên thuộc. Với một định hướng bất kỳ của $G$, ta có một ma trận liên thuộc $\m{H}$ tương ứng. Ma trận Laplace có thể viết dưới dạng:
\begin{align} \label{eq:c2_Laplacian_incidence}
\mcl{L} = \m{H}^\top \m{H}.
\end{align}
Chứng minh công thức \eqref{eq:c2_Laplacian_incidence} là nội dung Bài tập \ref{ex:2_15}. Như vậy, với $G$ là đồ thị vô hướng thì ma trận Laplace $\mcl{L}$ là đối xứng và bán xác định dương. Do đó, các trị riêng của $\mcl{L}$ là các số thực và ta có thể sắp xếp chúng theo thứ tự tăng dần như sau:
\begin{equation}
\lambda_1(G) \leq \lambda_2(G) \leq \ldots \leq \lambda_n(G).
\end{equation}
Mặc dù có thể định nghĩa từ ma trận $\m{H}$ (xem phương trình \eqref{eq:c2_Laplacian_incidence}), ma trận $\mcl{L}$ không phụ thuộc vào cách ta chọn định hướng các cạnh khi biểu diễn ma trận $\m{H}$ (phương trình \eqref{eq:c2_Laplacian_def}).

\begin{theorem}[Các tính chất của ma trận Laplace của đồ thị vô hướng]\label{thm:ch2_th1_Laplacian}
Xét đồ thị vô hướng $G=(V,E)$ với ma trận Laplace $\mcl{L}$. Ta có:
\begin{enumerate}
\item $\mcl{L}$ là đối xứng: $\mcl{L} = \mcl{L}^\top$.
\item $\mcl{L}$ là bán xác định dương. Các giá trị riêng của $\mcl{L}$, sắp xếp theo thứ tự tăng dần, thỏa mãn: $$0 = \lambda_1 \leq \lambda_2 \leq \ldots \leq \lambda_n.$$
\item $\mcl{L}$ thuộc tập các ma trận thuộc lớp M\index{M-ma trận}. Mỗi ma trận thuộc lớp M thỏa mãn $\m{M}=[\m{M}]_{ij} \in \mb{R}^{n \times n}$ có các phần tử ngoài đường chéo không dương ($[\m{M}]_{ij} \leq 0, \forall i \neq j$), nhưng phần thực của các trị riêng của nó đều không âm ($\lambda_i(\m{M}) \geq 0, \forall i = 1, \ldots, n$).
\item Tổng các hàng và các cột của $\mcl{L}$ đều bằng $0$. Nói cách khác, ma trận $\mcl{L}$ và $\mcl{L}^\top$ đều nhận vector $\m{1}_n$ là một vector riêng ứng với trị riêng $\lambda_1(G) = 0$.
\item Đồ thị $G$ là liên thông khi và chỉ khi $\lambda_2(G) > 0$. Giá trị $\lambda_2(G)$ còn gọi là là giá trị riêng Fiedler hay trị số liên thông của đồ thị $G$.
\item Nếu $G$ gồm $c$ thành phần liên thông, ta có thể đánh số các đỉnh của $G$ sao cho ma trận $\mcl{L}$ có thể viết dưới dạng một ma trận khối đường chéo $\mcl{L} = \text{blkdiag}(\mcl{L}_1, \ldots, \mcl{L}_c)$, trong đó $\mcl{L}_i, i = 1, \ldots, c$, là những ma trận Laplace tương ứng của mỗi thành phần liên thông trong $G$. Số chiều của không gian rỗng của $\mcl{L}$ bằng với số thành phần liên thông của đồ thị ($\text{dim}(\text{ker}(\mcl{L})) = c$).
\item Vết của ma trận Laplace bằng hai lần số cạnh của $G$, nói cách khác $\text{trace}(\mcl{L}) = 2m$.
\item (Định lý ma trận - cây) Kí hiệu $\mcl{L}_{v_i}$ là ma trận thu được từ $\mcl{L}$ sau khi xóa đi hàng và cột ứng với đỉnh $v_i$ bất kì của đồ thị. Số cây bao trùm đồ thị $G$ được tính bởi $\tau(G) = \text{det}(\mcl{L}_{v_i})$.
\item Phương trình $\mcl{L}\m{x}=\m{b}$ có nghiệm khi và chỉ khi $\m{b}^\top \m{1}_n={0}$. Nghiệm của phương trình có dạng $\m{x}=\alpha \m{1}_n+ \mcl{L}^\dagger \m{b}$, trong đó $\alpha \in \mb{R}$ và $\mcl{L}^\dagger=\m{P}\text{diag}(0,\lambda_2^{-1},\ldots,\lambda_n^{-1})\m{P}^\top$ là ma trận giả nghịch đảo Moore-Penrose của $\mcl{L}$, $\m{P} = [\m{v}_1, \ldots, \m{v}_n]$ với $\m{v}_k$ là các vector riêng đã được chuẩn hóa tương ứng với các giá trị riêng $\lambda_1,\ldots, \lambda_n$ của ma trận $\mcl{L}$. Ma trận $\mcl{L}^{\dagger}$ là đối xứng, bán xác định dương, có tổng hàng và tổng cột bằng 0, và thỏa mãn $\mcl{L}^{\dagger}\mcl{L}=\mcl{L}\mcl{L}^{\dagger}=\m{I}_n - \frac{1}{n}\m{1}_n \m{1}_n^\top$.
\item Giả sử ${G}$ là đồ thị liên thông và $G'$ là đồ thị thu được từ $G$ sau khi xóa đi các đỉnh $V_1={1,\ldots, l}$, $l\ge 1$, thì ma trận Laplace có thể viết dưới dạng:
\begin{equation}
    \mcl{L} = \begin{bmatrix}
    \mcl{L}_{11} & \mcl{L}_{12}\\
    \mcl{L}_{21} & \mcl{L}_{22}
    \end{bmatrix},
\end{equation}
trong đó $\mcl{L}_{11} \in \mb{R}^{l\times l}$, $\mcl{L}_{12}=\mcl{L}_{12}^\top=\mcl{L}_{21} \in \mb{R}^{l\times(n-l)}$, và $\mcl{L}_{22}=\mcl{L}(G')-\text{diag}(\mcl{L}_{21}\m{1}_l) \in \mb{R}^{(n-l)\times(n-l)}$ là một ma trận đối xứng, xác định dương. Khi $l=1$ thì $\mcl{L}_{22}$ còn có tên gọi là ma trận Laplace nối đất,\index{ma trận!Laplace nối đất} với ma trận nghịch đảo gồm các phần tử không âm thỏa mãn $(\mcl{L}_{22}^{-1})_{ij}=(\m{e}_i - \m{e}_1)^\top \mcl{L}^\dagger (\m{e}_i - \m{e}_1)$, trong đó $\m{e}_i=[0,\ldots,0,1,0,\ldots,0]^\top \in \mb{R}^n$ là vector đơn vị với các phần tử bằng 0 ngoại trừ phần tử thứ  $i$ có giá trị bằng $1$.
\item \cite{Klein1993} Điện trở hiệu dụng $r_{ij}^{\text{eff}}$ giữa hai đỉnh $i,j$ trong đồ thị liên thông $G$ được cho bởi $r_{ij}^{\text{eff}}=(\m{e}_i-\m{e}_j)^\top \mcl{L}^{\dagger}(\m{e}_i-\m{e}_j) = \mcl{L}^{\dagger}_{ii} + \mcl{L}^{\dagger}_{jj} - 2 \mcl{L}^{\dagger}_{ij}$. Điện trở hiệu dụng là một metric trong đồ thị, thỏa mãn các tính chất: (a) Không âm: $r_{ij}^{\text{eff}}\ge 0, \forall i, j \in V$ và $r_{ij}^{\text{eff}}=0$ khi và chỉ khi $i=j$; (b) Đối xứng: $r_{ij}^{\text{eff}}=r_{ji}^{\text{eff}}, \forall i, j \in V$; (c) Bất đẳng thức tam giác: $r_{ij}^{\text{eff}} \le r_{ik}^{\text{eff}} + r_{kj}^{\text{eff}}, \forall i, j, k \in V$.
\end{enumerate}
\end{theorem}

\begin{proof}
\begin{enumerate}
\item[1-4.] Các tính chất này được suy từ cách định nghĩa ma trận Laplace theo các  phương trình~\eqref{eq:c2_Laplacian_def}--\eqref{eq:c2_Laplacian_incidence}.

\item[5.] Từ phương trình~\eqref{eq:c2_Laplacian_incidence}, ta có $\text{ker}(\mcl{L}) = \text{ker}(\m{H}^\top\m{H}) = \text{ker}(\m{H})$. Do không gian nhân của $\m{H}$ được sinh bởi vector $\m{1}_n$ khi và chỉ khi $G$ là liên thông, ta có điều phải chứng minh.

\item[6.] Tương tự tính chất 5, ta có thể đánh số các cạnh của từng thành phần liên thông $G_k = (V_k,E_k)$, $k=1,\ldots, c$ trong $G$ sao cho $e_1,\ldots, e_{m_1}$ thuộc $G_1$, $e_{m_1+1}, \ldots, e_{m_1+m_2}$ thuộc $G_2$, \ldots, và $e_{\sum_{k=1}^{c-1}m_k+1},\ldots, e_{\sum_{k=1}^cm_{k}}$, với $\sum_{k=1}^cm_{k}=m$. Do các thành phần liên thông không có đỉnh chung, ma trận liên thuộc của đồ thị có dạng:
\begin{align}
    \m{H} = \begin{bmatrix}
    \m{H}_1 & \m{0}   & \cdots & \m{0}\\
    \m{0}   & \m{H}_2 & \ddots & \vdots \\
    \vdots  & \ddots  & \ddots & \m{0}\\
    \m{0}   & \cdots  & \m{0}  & \m{H}_c
    \end{bmatrix}
\end{align}
Do mỗi thành phần là liên thông, định nghĩa $$\m{v}_k = \text{vec}(\m{0}_{|V_1|}, \ldots, \m{0}_{|V_{k-1}|}, \m{1}_{|V_k|}, \m{0}_{|V_{k+1}|}, \ldots, \m{0}_{|V_{c}|}),~k=1,\ldots, c,$$ thì 
$$\m{H}\m{v}_k = \begin{bmatrix} \m{0}_{|V_{1}|}\\
\vdots \\
\m{H}_k \m{1}_{|V_k|} \\
\vdots\\
\m{0}_{|V_{c}|}
\end{bmatrix} = \m{0}_{n}, ~\text{do } \m{H}_k \m{1}_{|V_k|}= \m{0}_{|V_k|}.$$
Như vậy, dim$(\text{ker}(\m{H})) = c$. Do 
\begin{align}
    \mcl{L} &= \m{H}^\top\m{H} = \begin{bmatrix}
    \m{H}_1^\top \m{H}_1 & \m{0}   & \cdots & \m{0}\\
    \m{0}   & \m{H}_2^\top \m{H}_2 & \ddots & \vdots\\
    \vdots  & \ddots  & \ddots & \m{0} \\
    \m{0}   & \cdots  & \m{0}  & \m{H}_c^\top \m{H}_c
    \end{bmatrix} \nonumber \\
    &= {\rm blkdiag}(\mcl{L}_1,\ldots,\mcl{L}_c).
\end{align}
nên cũng có ${\rm dim}({\rm ker}(\m{H})) = {\rm dim}({\rm ker}(\m{H}^\top \m{H}))={\rm ker}(\mcl{L}) = c$. 
\item[7.] Từ phương trình~\eqref{eq:c2_Laplacian_def}, ta có ${\rm trace}(\mcl{L}) = \sum_{i=1}^n {\rm deg}(v_i) = 2m$.

\item[8.] Không mất tính tổng quát, ta có thể chọn $i=1$. Đầu tiên, ta chứng minh kết quả sau: \textit{Xét tập $S$ gồm $n-1$ cạnh bất kì của đồ thị $G$. Nếu các cạnh của $S$ không tạo thành một cây bao trùm của $G$ thì  ${\rm det}(\m{H}^\top_{v_1}[S]) = 0$ ($\m{H}^\top_{v_1}[S]$ là ma trận con của $\m{H}^\top$ sau khi xóa đi hàng 1 và các cột tương ứng với các cạnh không thuộc $S$). Ngược lại, nếu các cạnh của $S$ tạo thành một cây bao trùm của $G$ thì ${\rm det}(\m{H}^\top_{v_1}[S]) = \pm 1$.}

Thật vậy, nếu các cạnh của $S$ không tạo thành một cây bao trùm của $G$ thì một tập các cạnh (giả thiết không kề với $v_1$) trong $S$ sẽ tạo thành của một chu trình trong $G$. Chọn  vector đường đi đánh dấu $\m{z}$ tương ứng với chu trình này thì $\m{H}^\top_{v_1}[S] \m{z}= \m{0}_m$. Điều này chứng tỏ các cột của $\m{H}^\top_{v_1}[S]$ phụ thuộc tuyến tính, từ đó ta suy ra ${\rm det}(\m{H}^\top_{v_1}[S]) = 0$.

Giả sử các cạnh của $S$ lập thành một cây bao trùm $T$ của $G$. Với $e$ là một cạnh của $T$ liên thuộc với $v_1$ thì cột tương ứng của $e$ trong  ma trận $\m{H}^\top_{v_1}[S]$ chỉ chứa duy nhất một phần tử khác 0 (nhận giá trị $\pm 1$). Trong ma trận $\m{H}^\top_{v_1}[S]$, sau khi xóa đi hàng và cột tương ứng với phần tử khác $0$ của $e$, ta thu được một ma trận $\m{H}^{'\top}_{v_1} \in \mb{R}^{(n-2)\times (n-2)}$. Chú ý rằng ${\rm det}(\m{H}^\top_{v_1}[S])= \pm {\rm det}(\m{H}^{'\top}_{v_1}[S])$. Gọi $T'=T \setminus e$ là cây thu được từ $T$ sau khi thu hẹp cạnh $e$ vào đỉnh $v$. Khi đó $\m{H}^{'\top}_{v_1}[S]$ chính là ma trận $\m{H}^{'\top}_{v_1}(T')$ sau khi xóa đi hàng tương ứng với đỉnh $u$. Do đó, theo nguyên tắc quy nạp theo số đỉnh $n$  (trường hợp $n=2$ dễ thấy ${\rm det}(\m{H}^\top_{v_1}[S]) = \pm 1$), ta có ${\rm det}(\m{H}^{'\top}_{v_1}) = \pm 1$.

Tiếp theo, ta chứng minh định lý ma trận - cây. Do $\mcl{L} = \m{H}^\top \m{H}$, ta suy ra $\mcl{L}_{v_1} = \m{H}^\top_{v_1}\m{H}_{v_1}$. Theo định lý Cauchy-Binet: 
\begin{equation*}
{\rm det}(\mcl{L}_{v_1}) = \sum_{S} {\rm det}(\m{H}^\top_{v_1}[S]) {\rm det}(\m{H}_{v_1}[S]),
\end{equation*}
với $S$ là một tập bất kì gồm $n-1$ phần tử trong tập $\{1, \ldots, m\}$. Do ${\rm det}(\m{H}^\top_{v_1}[S]) = {\rm det}(\m{H}_{v_1}[S])$, ta suy ra
\[{\rm det}(\mcl{L}_{v_1}) = \sum_{S} ({\rm det}(\m{H}^\top_{v_1}[S]))^2.\]
Áp dụng kết quả vừa chứng minh ở trên, với mỗi tập $S$ tạo thành một cây bao trùm của $G$ thì ${\rm det}(\m{H}^\top_{v_1}[S]) = \pm 1$, trong khi ngược lại thì ${\rm det}(\m{H}^\top_{v_1}[S]) = 0$. Do đó, tổng ở vế phải của đẳng thức trên đúng bằng số cây bao trùm của đồ thị $G$.
\item[9.] Do ma trận $\mcl{L}$ là suy biến với ker$(\mcl{L})= {\rm im}(\m{1}_n)$ nên  im$(\mcl{L})\perp {\rm im}(\m{1}_n)$. Do đó, phương trình $\mcl{L}\m{x}=\m{b}$ có nghiệm khi và chỉ khi $\m{b} \in {\rm im}(\mcl{L})$, tức là $\m{1}_n^\top \m{b} = 0$. Nghiệm của phương trình được cho bởi:
\begin{align}
    \m{x} = \alpha \m{1}_n + \mcl{L}^{\dagger} \m{b},
\end{align}
với $\alpha \in \mb{R}$ và $\mcl{L}^{\dagger} \m{b} \perp \text{im}(\m{1}_n)$. 
Nhân trực tiếp hai ma trận $\mcl{L} = \m{P}\text{diag}(0,\lambda_2,\ldots,\lambda_n)\m{P}^\top$ và $\mcl{L}^\dagger = \m{P}\text{diag}(0,\lambda_2^{-1},\ldots,\lambda_n^{-1})\m{P}^\top$, ta thu được điều phải chứng minh. 
\item[10.] Từ phương trình $\mcl{L}_{22}=\mcl{L}(G')-\text{diag}(\mcl{L}_{21}\m{1}_l)$ ta có ngay $\mcl{L}_{22}$ là đối xứng. Do đồ thị $G$ là liên thông, $\m{B}=- {\rm diag}(\mcl{L}_{21}\m{1}_l) = {\rm diag}(b_1,\ldots,b_{n-1})$ là một ma trận đường chéo với $b_{i} = \sum_{j=1}^la_{ij}\ge 0$ và tồn tại ít nhất một phần tử $b_{i}>0$.

Với mọi vector riêng $\m{v}_k \in \mb{R}^{n-1}$ của $\mcl{L}(G')$ ứng với giá trị riêng $\lambda_k \ge 0$ thì $\m{v}_k^\top \mcl{L}_{22} \m{v}_k = \lambda_k + \sum_{j=1}^{n-1} b_j v_{kj}^2>0$. Do span$(\m{v}_1,\ldots, \m{v}_{n-1}) = \mb{R}^{(n-1)\times(n-1)}$ nên ta suy ra $\mcl{L}_{22}$ là ma trận xác định dương.

\item[11.] (a) Do ma trận giả nghịch đảo $\mcl{L}^{\dagger}$ là đối xứng và bán xác định dương nên $r_{ij}^{\rm eff}$ là không âm. Hơn nữa, khi $i\ne j$ thì do $(\m{e}_i-\m{e}_j) \perp {\rm ker}(\mcl{L}^{\rm dagger}) = {\rm im}(\m{1}_n)$ nên $r_{ij}^{\rm eff}>0$. Vậy $r_{ij}^{\rm eff}=0$ khi và chỉ khi $i=j$. 

(b) Do $(\m{e}_i-\m{e}_j)^\top\mcl{L}^{\dagger}(\m{e}_i-\m{e}_j) = (\m{e}_j-\m{e}_i)^\top\mcl{L}^{\dagger}(\m{e}_j-\m{e}_i)$, ta có $r_{ij}^{\rm eff}=r_{ji}^{\rm eff}$.

(c) Chúng ta sẽ sử dụng kết quả sau (nội dung của Bài tập \ref{ex:2_18}) trong chứng minh \cite{Ellens2011}: ``Xét đồ thị vô hướng liên thông $G$ với ma trận Laplace $\mcl{L}$. Giả sử phương trình $\mcl{L}\m{V}=\m{e}_i - \m{e}_j$ có nghiệm $\m{V}=[v_1,\ldots,v_n]^\top \in \mb{R}^n$ thì $v_i \ge v_k \ge v_j, \forall k \in V \setminus \{i, j\}$.''

Ta có các phương trình điện áp khi đưa vào và lấy ra các dòng điện cường độ 1 và -1
\begin{align}
\m{L}\m{V}_1 &= \m{e}_i-\m{e}_j,~\m{V}_1 = [v_{11},\ldots,v_{1n}]^\top \qquad (\text{vào } i \text{ ra } j)\\
\m{L}\m{V}_2 &= \m{e}_j-\m{e}_k,~\m{V}_2 = [v_{21},\ldots,v_{2n}]^\top \qquad (\text{vào } j \text{ ra } k).
\end{align}
Từ hai phương trình trên suy ra 
\begin{align} \label{eq:c2_dong_dien_gop}
\m{L}(\m{V}_1+\m{V}_2) = \m{e}_i - \m{e}_k,
\end{align}
hay $\m{V}_3 = \m{V}_1 + \m{V}_2$ là vector điện áp ở các đỉnh khi đưa vào và lấy ra tại đỉnh $i$ và đỉnh $k$ dòng điện cường độ 1 và -1. 
Từ phương trình \eqref{eq:c2_dong_dien_gop} suy ra
\begin{align} \label{eq:c2_dien_ap_gop}
v_{1i} + v_{2i} - (v_{1k} + v_{2k}) = (\m{e}_i - \m{e}_k)^\top\m{L}^{\dagger}\m{L}(\m{V}_1+\m{V}_2) = r_{ik}^{\rm eff}.
\end{align}
Áp dụng kết quả ở đầu chứng minh, ta có $v_{1k}\ge v_{1j}$ và $v_{2i} \le v_{2j}$. Thay hai bất đẳng thức trên vào \eqref{eq:c2_dien_ap_gop} thu được
\begin{align*}
r_{ik}^{\rm eff} &\leq (v_{1i} - v_{1j}) + (v_{2j}  - v_{2k}) \\
& = (\m{e}_i - \m{e}_k)^\top\m{L}^{\dagger}\m{L}\m{V}_1 + (\m{e}_j - \m{e}_k)^\top\m{L}^{\dagger}\m{L}\m{V}_2 \\
& = r_{ij}^{\rm eff} + r_{jk}^{\rm eff}.
\end{align*}
\end{enumerate}
\end{proof}
\subsection{Ma trận Laplace của đồ thị hữu hướng}
Với đồ thị hữu hướng, có trọng số $G = (V,E,A)$, ma trận kề (với trọng số) được định nghĩa bởi \index{ma trận!kề}
\begin{equation}
[\m{A}]_{ij} = \left\lbrace \begin{array}{ll}
a_{ij}, & \text{nếu } (v_j,v_i) \in {E},\\
0, & \text{các trường hợp khác.} \\
\end{array} \right.
\end{equation}
Ma trận bậc (vào) được định nghĩa bởi \[\m{D} = \text{diag}(\text{deg}^{-}(v_i)),\] với  $\text{deg}^{-}(v_i) = \sum_{j \in N^{-}(v_i)} a_{ij}$. Với định nghĩa này, ta có  \index{ma trận!bậc vào}
\begin{equation}
\m{D} = \text{diag}(\m{A} \m{1}_n).
\end{equation}
Ma trận Laplace được định nghĩa bởi \index{ma trận!Laplace}
\begin{equation}\label{eq:c2_Laplace_weight}
\mcl{L} \triangleq \m{D} - \m{A} = \text{diag}(\m{A} \m{1}_n) - \m{A}.
\end{equation}
Với định nghĩa này, ta vẫn có $\mcl{L} \m{1}_n = \text{diag}(\m{A} \m{1}_n)\m{1}_n - \m{A}\m{1}_n = \m{0}_n$, hay $\m{1}_n \in \text{ker}(\mcl{L})$ và $\mcl{L}$ luôn có một giá trị riêng bằng $0$. Tuy nhiên, $\mcl{L}$ lúc này không đối xứng nên tổng các cột của $\mcl{L}$ có thể khác 0. 

\begin{story}{Mehran Mesbahi, Magnus Ergestedt và cuốn sách ``Graph Theoretic Methods in Multiagent Networks''}
Mehran Mesbahi là giáo sư tại Khoa Hàng không Vũ trụ, Đại học Washington, Seattle. Nghiên cứu của Mesbahi tập trung vào điều khiển nối mạng, tối ưu hóa, và hàng không vũ trụ. Magnus Egerstedt là giáo sư tại Trường Kĩ thuật Henry Samueli tại Đại học California, Irvine. Các đóng góp khoa học của Egerstedt liên quan tới lý thuyết hệ lai liên tục - rời rạc, điều khiển hệ đa tác tử và robotics. 

``Graph Theoretic Methods in Multiagent Networks''\cite{MesbahiEgerstedt} là một trong những sách chuyên khảo về điều khiển hệ đa tác tử xuất hiện sớm nhất (2010). Sách được viết để phục vụ giảng dạy sau đại học, tuy nhiên phần nội dung cơ sở có thể dùng làm tài liệu tự học cho người chưa có kiến thức về điều khiển. Từ khi xuất hiện đến nay, các nội dung trong mười ba chương của sách là đề tài của rất nhiều nghiên cứu, công bố khoa học trong lĩnh vực điều khiển hệ đa tác tử. Cuốn sách cũng đề cập đến rất nhiều ứng dụng thực tế, có phần đi trước thời đại của hệ đa tác tử trong hàng không vũ trụ và robotics.

Người đọc giáo trình mong muốn thử nghiệm thuật toán trên hệ nhiều robot có thể tìm hiểu về Robotarium - testbed được phát triển bởi Egerstedt khi ông là giáo sư tại Georgia Tech: \url{https://www.robotarium.gatech.edu/}.
\end{story}

Định lý ma trận - cây\footnote{matrix-tree theorem} cho đồ thị hữu hướng được phát biểu như sau:
\begin{theorem}[Định lý ma trận - cây \index{định lý!ma trận - cây} \cite{Tutte1984Graph,West1996introduction}] \label{thm:c2_matrix_tree} Với $v$ là một đỉnh bất kì của đồ thị hữu hướng có trọng số $G$, ta có
\begin{align*}
{\rm det}(\mcl{L}_v(G)) = \sum_{T \in \mc{T}_v} \prod_{(v_j,v_i) \in T} a_{ij},
\end{align*}
trong đó $\mc{T}_v$ là tập hợp các cây bao trùm của $G$ có gốc ra tại $v$, $\prod_{(v_j,v_i) \in T} a_{ij}$ là tích trọng số của các cạnh thuộc cây bao trùm $T$, và $\mcl{L}_v(G)$ là ma trận thu được từ $\mcl{L}(G)$ sau khi xóa đi hàng và cột tương ứng với đỉnh $v$.
\end{theorem}

Đối với một đồ thị hữu hướng, có trọng số $G$, điều kiện cần và đủ để $G$ chứa một cây bao trùm có gốc ra (hay $G$ là một đồ thị có gốc ra) được cho trong định lý sau đây:
\begin{theorem} \label{thm:2.3}
Một đồ thị hữu hướng $G$ với $n$ đỉnh chứa một cây bao trùm có gốc ra khi và chỉ khi $\text{rank}(\mcl{L}) = n-1$. Khi đó, $\text{ker}(\mcl{L}) = \text{im}(\m{1}_n)$.
\end{theorem}
\begin{proof}
Điều cần chứng minh tương đương với việc đa thức đặc tính của $\mcl{L}$ nhận $0$ là một nghiệm đơn. Viết đa thức đặc tính dưới dạng
\begin{equation}
p_{G} (\lambda) = \lambda^n + \alpha_{n-1} \lambda^{n-1} + \ldots + \alpha_1 \lambda + \alpha_0,
\end{equation}
thì $\alpha_0 = 0$ do ma trận Laplace luôn có một giá trị riêng bằng $0$. Bởi vậy, $\text{rank}(\mcl{L}) = n - 1$ khi và chỉ khi $\alpha_1 \neq 0$. Mặt khác,
\begin{align*}
\alpha_1 = \sum_{v} \text{det}(\mcl{L}_v),
\end{align*}
với $\mcl{L}_v$ là ma trận thu được từ $\mcl{L}$ sau khi xóa đi hàng và cột thứ $v$. Từ định lý ma trận - cây, ta có $\text{det}(\mcl{L}_v)\neq 0$ khi và chỉ khi tồn tại một cây bao trùm có gốc ra tại $v \in G$. Cuối cùng, do $\mcl{L}\m{1}_n = \m{0}_n$ luôn đúng, nếu $\text{rank}(\mcl{L}(G)) = n-1$, ta suy ra $\text{ker}(\mcl{L}(G)) = \text{im}(\m{1}_n)$.
\end{proof}

Một số tính chất của ma trận Laplace của đồ thị hữu hướng, có trọng số $G$ được tóm tắt trong định lý sau:
\begin{theorem}[Ma trận Laplace của đồ thị hữu hướng, có trọng số]\label{thm:c2_Laplace_directed} Giả thiết rằng $G = (V, E, A)$ là một đồ thị hữu hướng, có trọng số và có gốc-ra. Khi đó, 
\begin{enumerate}
\item $\mcl{L}$ nhận $\m{1}_n$ là một vector riêng bên phải ứng với giá trị riêng $\lambda_1 = 0$. Các giá trị riêng khác của $\mcl{L}$ thỏa mãn $\text{Re}(\lambda_i) > 0, i = 2, \ldots, n$. Các giá trị riêng của $\mcl{L}$ đều nằm trong đĩa tròn tâm $\Delta+j0$, bán kính $\Delta = \max_{i=1,\ldots,n} \text{deg}^-(v_i)$ trên mặt phẳng phức (Hình~\ref{fig:ch2_gerchgorin}). 

\item  Nếu $G$ là liên thông mạnh thì ma trận $\mcl{L}$ là tối giản, tức là không tồn tại ma trận hoán vị $\m{P}$ để $\m{P}\mcl{L}\m{P}^\top$ có dạng ma trận khối đường chéo trên. Nếu như $G$ có gốc ra, thì tồn tại ma trận hoán vị đưa ma trận $\mcl{L}$ về dạng:
\begin{align} \label{eq:c2_laplace_permutation1}
    \mcl{L}' = \begin{bmatrix}
        \mcl{L}_{11} & \mcl{L}_{12} & \ldots & \mcl{L}_{1k}\\
        \m{0} & \mcl{L}_{22} & \ldots & \mcl{L}_{2k} \\
        \vdots & \vdots & \ddots & \vdots \\
        \m{0} & \m{0} &\ldots & \mcl{L}_{kk}
    \end{bmatrix}
\end{align}
trong đó $\mcl{L}_{ii}, i = 1, \ldots, k-1$, là tối giản, có ít nhất một hàng với tổng hàng dương, và $\mcl{L}_{kk}$ là tối giản hoặc bằng 0.
\item Đồ thị $G$ là liên thông mạnh khi và chỉ khi tồn tại vector $\bm{\gamma} = [\gamma_1, \ldots, \gamma_n]^\top$ thỏa mãn $\bm{\gamma}^\top \mcl{L} = \m{0}^\top_n$, $\bm{\gamma}^\top\m{1}_n = 1$, và $\gamma_i > 0, ~i=1, \ldots, n$.
\end{enumerate}
\end{theorem}

\begin{proof}
\begin{enumerate}
    \item Áp dụng Định lý Gerschgorin\index{định lý!Gerschgorin} (xem Phụ lục \ref{append:matrix_theory}) cho ma trận $\mcl{L}$, các giá trị riêng của $\mcl{L}$ nằm trong miền hợp của các đĩa tròn trên mặt phẳng phức cho bởi
\begin{align}
    B_i = \{ s \in \mb{C}|~ |s - l_{ii}| \leq R_i \},
\end{align}
trong đó $R_i = \sum_{j\in {N}_i} |l_{ij}| = \sum_{j\in {N}_i} |a_{ij}| = l_{ii} = {\rm deg}^-(v_i)$. Các đĩa tròn này đều giao với trục ảo tại gốc tọa độ và đều nằm trong đĩa tròn $B$ tâm $\Delta + j0$, bán kính $R = \Delta = \max_i {\rm deg}^-(v_i)$. Theo Định lý~\ref{thm:2.3}, ma trận $\mcl{L}$ chỉ có thể có duy nhất một giá trị riêng bằng 0 nên ta suy ra các giá trị riêng còn lại của $\mcl{L}$ đều phải nằm bên phải trục ảo, tức là ${\rm Re}(\lambda_i) > 0$, $\forall i = 2, \ldots, n$.

\begin{SCfigure}[.7][t!]
\caption{Các giá trị riêng của $\mcl{L}$ nằm trong đĩa tròn $B$ tâm $\Delta+j0$, bán kính $\Delta = \max_i \text{deg}^-(v_i)$ (miền màu đỏ). Các trị riêng của $-\mcl{L}$ nằm trong đĩa tròn $B'$ đối xứng với $B$ qua trục ảo (miền phía trái trục ảo). 
\label{fig:ch2_gerchgorin}}
\hspace{1cm}
\begin{tikzpicture}[>=latex, scale=1]
\coordinate (Ot) at (2,0) {};
\node (O) at (2,0) {$+$};
\node (O1) at (-2,0) {$+$};
\node[anchor=north east] at (0, 0) (a1) {$0$};
\draw[color=red, fill=red!20, opacity = .50, thick] (O) circle (2 cm);
\draw[color=blue, fill=blue!20, opacity = .50, thick] (O1) circle (2 cm);
\coordinate[position=30:{2 cm} from Ot] (oc);
\draw[thick,->] (-5,0) -- (5,0) node[anchor=north] {$\sigma$};
\draw[thick,->] (0,-2.5) -- (0,2.5) node[anchor=east] {$j\omega$};
\draw[thick,<->] (Ot) -- (oc) node[midway,above=1 mm] {$\Delta$};
\node[anchor=north] at (2,0)    (a) {$\Delta + 0j$};
\node[anchor=north] at (-2,0)   (a) {$-\Delta + 0j$};
\node[anchor=north] at (2,-1)   (a) {$B$};
\node[anchor=north] at (-2,-1)  (a) {$B'$};
\end{tikzpicture}
\end{SCfigure}

\item Giả sử $G$ là đồ thị liên thông mạnh và tồn tại ma trận hoán vị $\m{P}$ (tương ứng với một cách dán nhãn lại các đỉnh trong $V$) sao cho $\mcl{L}$ có thể biểu diễn dưới dạng
\begin{align} \label{eq:c2_laplace_permutation}
    \mcl{L}' = \begin{bmatrix}
        \mcl{L}_{11} & \mcl{L}_{12} \\
        \m{0} & \mcl{L}_{22}
    \end{bmatrix},
\end{align}
trong đó $\mcl{L}_{11} \in \mb{R}^{l\times l}$ và $\mcl{L}_{22} \in \mb{R}^{(n-l) \times (n-l)}$ với $l \ge 1$. 
Dễ thấy từ cấu trúc của ma trận $\mcl{L}'$, các đỉnh $v_{1}, \ldots, v_{l}$ không có đường đi có hướng nào tới các đỉnh $v_{l+1}, \ldots, v_{n}$. Từ đây suy ra $G$ không thể là đồ thị liên thông mạnh, trái với giả thiết ban đầu. 

Nếu $G$ là một đồ thị có gốc ra và không liên thông mạnh, ta có thể đánh số các đỉnh trong thành phần liên thông mạnh tối đa chứa các gốc ra của $G$ bởi $v_{l+1}, \ldots, v_{n}$. Khi đó, ta có dạng ma trận Laplace tương ứng ở dạng \eqref{eq:c2_laplace_permutation}. Tiếp theo, ta coi thành phần liên thông mạnh tối đa chứa gốc ra của $G$ như một đỉnh $v_{l+1}$  và đánh số các đỉnh trong thành phần liên thông mạnh tối đa kề với $v_{l+1}$ bởi $v_{l_1},\ldots,v_l$ thì ma trận Laplace tương ứng lúc này có dạng:
\begin{align} 
    \mcl{L}' = \begin{bmatrix}
        \mcl{L}_{11}' & \mcl{L}_{12}' & \mcl{L}_{13}'\\
        \m{0} & \mcl{L}_{22}' & \mcl{L}_{23}'\\
        \m{0} & \m{0} & \mcl{L}_{33}'
    \end{bmatrix}.
\end{align}
Tiếp tục quá trình đánh số như vậy sẽ thu được ma trận Laplace như ở phương trình~\eqref{eq:c2_laplace_permutation1}.

\item Giả sử $G$ là đồ thị liên thông mạnh. Định nghĩa ma trận $\m{B} = \rho\m{I}_n - \mcl{L}$ với $\rho > \max_{i=1,\ldots,n} l_{ii}$ thì $\m{B}$ là một ma trận không âm\index{ma trận!không âm}. Các phần tử nằm ngoài đường chéo chính của $\m{B}$ giống hệt với các phần tử nằm ngoài đường chéo chính của $\m{A}$, do đó đồ thị nhận $\m{B}$ là ma trận kề cũng là một đồ thị liên thông mạnh. Từ đây suy ra $\m{B}$ là không rút gọn được. Thêm nữa, $\m{B}\m{1}_n = ( \rho\m{I}_n - \mcl{L})\m{1}_n = \rho \m{1}_n$. Theo lý thuyết Perron-Frobenius (xem phụ lục \ref{append:matrix_theory}), $\rho$,  giá trị riêng trội của $\m{B}$, là một giá trị riêng đơn giản\index{giá trị riêng đơn giản} và vector riêng bên trái $\bmm{\gamma}$ của $\m{B}$ có thể được chọn là vector duy nhất thỏa mãn $\bmm{\gamma}^\top \m{B} = \rho \bmm{\gamma}^\top$, $\gamma_i >0,\forall i=1,\ldots,n$, và  $\bmm{\gamma}^\top \m{1}_n = 1$. Từ $\bmm{\gamma}^\top \m{B} = \rho \bmm{\gamma}^\top$ suy ra $\bmm{\gamma}^\top \mcl{L} = \m{0}_n^\top$, hay $\bmm{\gamma}$ là vector riêng bên trái của $\mcl{L}$ ứng với giá trị riêng 0.

Ngược lại, giả sử $\mcl{L}$ có duy nhất vector riêng bên trái $\bmm{\gamma}$ thỏa mãn $\bmm{\gamma}^\top \mcl{L} = \m{0}_n^\top$, $\gamma_i >0,\forall i=1,\ldots,n$, và  $\bmm{\gamma}^\top \m{1}_n = 1$ nhưng $G$ không liên thông mạnh. Do $G$ không liên thông mạnh, tồn tại một cách đánh số các đỉnh của $G$ sao cho ma trận $\mcl{L}$ có thể biểu diễn ở dạng \eqref{eq:c2_laplace_permutation}, trong đó $\mcl{L}_{22}$ là ma trận Laplace tương ứng với thành phần liên thông mạnh tối đa chứa các gốc ra.\index{thành phần!liên thông mạnh tối đa chứa các gốc ra}

Từ chứng minh phần thuận, tồn tại $\bmm{\zeta}$ là vector riêng bên trái duy nhất ứng với giá trị riêng $0$ của ma trận $\mcl{L}_{22}$ với $\zeta_i >0$, $\sum_{i} \zeta_i = 1$.  Nhận thấy rằng với $\bmm{\gamma}' = [\m{0}^\top, \bmm{\zeta}^\top]^\top$ thì $(\bmm{\gamma}')^\top \mcl{L} = \m{0}_n^\top$, trái với giả thiết rằng $\bmm{\gamma}$ là vector riêng duy nhất của $\mcl{L}$ tương ứng với giá trị riêng $0$. Từ đây suy ra điều phải chứng minh.
\end{enumerate}
\end{proof}

Giá trị riêng $\lambda_2$ của ma trận Laplace cho biết liệu đồ thị có gốc ra hay không, và do đó còn có tên gọi là \emph{giá trị riêng liên kết}. \index{giá trị riêng!liên kết}

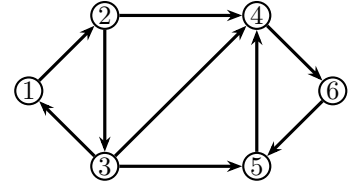
\begin{SCfigure}[.8][t!]
\caption{Đồ thị $G$ tương ứng với ma trận Laplace ở Ví dụ~\ref{eg2.6}. 
\label{fig:VD2.6_Graph}}
\hspace{3.5cm}
\begin{tikzpicture}[
roundnode/.style={circle, draw=black, thick, minimum size=2mm,inner sep= 0.25mm},
squarednode/.style={rectangle, draw=black, thick, minimum size=3.5mm,inner sep= 0.25mm},
]
    \node[roundnode] (u1) at (0,0) { $1$}; %
    \node[roundnode] (u2) at (1,1) { $2$};%
    \node[roundnode] (u3) at (1,-1) { $3$};%
    \node[roundnode] (u6) at (4,0) { $6$};%
    \node[roundnode] (u4) at (3,1) { $4$};%
    \node[roundnode] (u5) at (3,-1) { $5$};%

    \draw [very thick,-{Stealth[length=2mm]}]
    (u1) edge [bend left=0] (u2)
    (u2) edge [bend left=0] (u3)
    (u3) edge [bend left=0] (u1)
    (u2) edge [bend left=0] (u4)
    (u3) edge [bend left=0] (u4)
    (u3) edge [bend left=0] (u5)
    (u4) edge [bend left=0] (u6)
    (u6) edge [bend left=0] (u5)
    (u5) edge [bend left=0] (u4)
    ;
\end{tikzpicture}
\end{SCfigure}

\begin{example} \label{eg2.6}
Xét đồ thị $G$ có ma trận Laplace $\mcl{L}$ cho bởi:
\begin{align*}
    \mcl{L} = \begin{bmatrix}
         1  &   0  &  -1  &   0   &  0  &   0\\
        -1  &   1  &   0  &   0  &   0  &   0\\
        0  &  -1   &  1   &  0   &  0   &  0\\
        0  &  -1  &  -1   &  3  &  -1  &   0\\
        0  &   0   & -1   &  0  &   2  &  -1\\
        0  &   0   &  0   & -1  &   0  &   1
    \end{bmatrix}.
\end{align*}
Hình~\ref{fig:VD2.6_Graph} mô tả đồ thị tương ứng với ma trận $\mcl{L}$. 

Theo định lý Gersgorin, mọi giá trị riêng của $\mcl{L}$ nằm trong miền $C = \bigcup_{i=1}^6 C_i$, trong đó \[C_1 = C_2 = C_3 = C_6 = \{s=\sigma + j \omega \in \mb{C}|~ (\sigma - 1)^2 + \omega^2 = 1 \},\] \[C_4=\{s=\sigma + j \omega \in \mb{C}|~ (\sigma - 3)^2 + \omega^2 = 9\},\] và \[C_5=\{s=\sigma + j \omega \in \mb{C}|~ (\sigma - 2)^2 + \omega^2 = 4 \}.\] Với minh họa ở Hình~\ref{fig:eg2.6}, dễ thấy rằng $\Delta = 3$ và $C \equiv C_4$ chứa mọi đĩa tròn $C_i,~i=1,\ldots, 6$.
\begin{SCfigure}[][ht]
\caption{Minh họa các đĩa tròn Gerschgorin và vị trí các giá trị riêng của ma trận Laplace ở Ví dụ~\ref{eg2.6}.\label{fig:eg2.6}}
\hspace{3cm}
\begin{tikzpicture}
\begin{axis}[
  axis equal,
  samples = 200,
  xlabel={$\sigma$}, ylabel={$\jmath \omega$},
  xmin=-1, xmax=7,
  ymin=-3, ymax=3,
  axis lines=middle,
]
\addplot [
  domain=0:2*pi,
  color = cyan,
] ({cos(deg(x))+1},{sin(deg(x))});
\addplot [
  domain=0:2*pi,
  color = green,
] ({2*cos(deg(x))+2},{2*sin(deg(x))});
\addplot [
  domain=0:2*pi,
  color = blue,
] ({3*cos(deg(x))+3},{3*sin(deg(x))});
\addplot[
  mark=x,
  color = red
] coordinates {(0,0)};
\addplot[
  mark=x,
  color = red
] coordinates {(0.6753,0)};
\addplot[
  mark=x,
  color = red
] coordinates {(1.5,0.866)};
\addplot[
  mark=x,
  color = red
] coordinates {(1.5,-0.866)};
\addplot[
  mark=x,
  color = red
] coordinates {(2.6624,0.5623)};
\addplot[
  mark=x,
  color = red
] coordinates {(2.6624,-0.5623)};
\end{axis}
\end{tikzpicture}
\end{SCfigure}
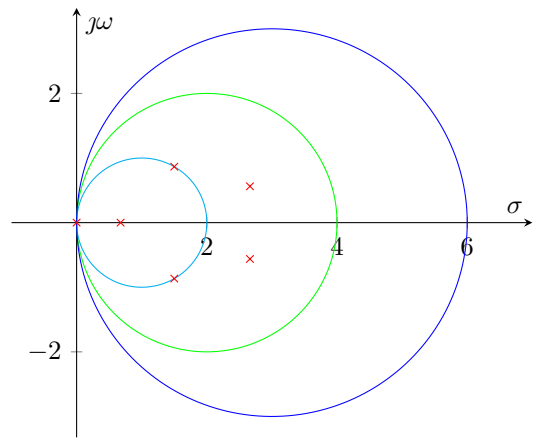
\end{example}

\section{Ghi chú và tài liệu tham khảo}
Nội dung về lý thuyết đồ thị ở chương này được trình bày ngắn gọn và được sử dụng trong phân tích hệ đa tác tử ở các chương sau. Đây cũng là cách tiếp cận chung của các giáo trình về hệ đa tác tử  \cite{MesbahiEgerstedt,Bullo2019lectures}. Những kiến thức về đồ thị trong chương này có thể tìm thấy trong hầu hết các giáo trình kinh điển về lý thuyết đồ thị \cite{Biggs1993,Godsil2001,Diestel2025graph}. Phân tích về các hệ đồng thuận ở chương \ref{chap:consensus} sẽ liên quan tới tính chất phổ của ma trận Laplace. Một số kết quả mở rộng về phổ của ma trận Laplace hữu hướng, và ma trận kề có thể tham khảo ở \cite{Horn1990,Chung1997spectral}. Những liên hệ giữa đồ thị, lý thuyết mạch (mạng điện dẫn một chiều), và bước ngẫu nhiên (random walk) có thể tham khảo tại \cite{Doyle1984random}.\index{mạch điện dẫn} \index{bước ngẫu nhiên} \index{siêu đồ thị} Một số chủ đề mở rộng trong lý thuyết đồ thị, ví dụ như bài toán ghép cặp, bài toán tô màu đồ thị, lý thuyết Ramsey, hay lý thuyết đồ thị tới hạn, siêu đồ thị (hypergraph), đồ thị ngẫu nhiên, đồ thị trọng số ma trận,\ldots có thể tham khảo những tài liệu \cite{Bollobas2004extremal,Berge1973,Frieze2015introduction,Trinh2026}. 

\section{Bài tập}
\begin{exercise} \label{Ex:2_1}
Hai đồ thị ${G}_1$ và ${G}_2$ trên Hình~\ref{fig:c2_ex21a} có đẳng cấu hay không? Cùng câu hỏi với hai đồ thị $H_1$ và $H_2$ trên Hình~\ref{fig:c2_ex21b}.
\end{exercise}

\begin{figure}[h]
\centering
\subfloat[${G}_1$]{
\begin{tikzpicture}[
roundnode/.style={circle, draw=black, thick, minimum size=1.5mm,inner sep= 0.25mm},
squarednode/.style={rectangle, draw=red!60, fill=red!5, very thick, minimum size=5mm}, scale=0.8
]
\node[roundnode]   (v1)   at   (-2,0) {};
\node[roundnode]   (v2)   at   (0,0) {};
\node[roundnode]   (v3)   at   (2,0) {};
\node[roundnode]   (v4)   at   (4,0) {};
\node[roundnode]   (v5)   at   (-2,2) {};
\node[roundnode]   (v6)   at   (0,2) {};
\node[roundnode]   (v7)   at   (2,2) {};
\node[roundnode]   (v8)   at   (4,2) {};

\draw[-, thick] (v1)--(v5);
\draw[-, thick] (v1)--(v6);
\draw[-, thick] (v1)--(v7);
\draw[-, thick] (v2)--(v5);
\draw[-, thick] (v2)--(v6);
\draw[-, thick] (v2)--(v8);
\draw[-, thick] (v3)--(v5);
\draw[-, thick] (v3)--(v7);
\draw[-, thick] (v3)--(v8);
\draw[-, thick] (v4)--(v6);
\draw[-, thick] (v4)--(v7);
\draw[-, thick] (v4)--(v8);
\end{tikzpicture}
}\qquad \qquad\qquad
\subfloat[${G}_2$]{
\begin{tikzpicture}[
roundnode/.style={circle, draw=black, thick, minimum size=1.5mm,inner sep= 0.25mm},
squarednode/.style={rectangle, draw=red!60, fill=red!5, very thick, minimum size=5mm}, scale=0.8
]
\node (A) [draw,thick,regular polygon, regular polygon sides=4, minimum size=1.2cm,outer sep=0pt] {};
\node (B) [draw,thick,regular polygon, regular polygon sides=4, minimum size=2.8cm,outer sep=0pt] {};
\foreach \n in {1,...,4} {
    \draw[-,thick] (A.corner \n)--(B.corner \n);
    \draw[fill=white,thick] (B.corner \n) circle[radius=.75mm];
    \draw[fill=white,thick] (A.corner \n) circle[radius=.75mm];
}
\end{tikzpicture}
}
\caption{Đồ thị vô hướng ${G}_1$ và ${G}_2$.}
\label{fig:c2_ex21a}
\end{figure}
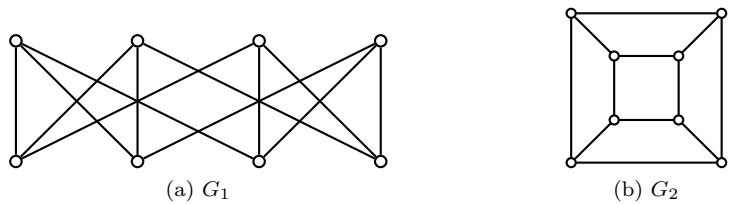

\begin{figure}
\centering
\subfloat[${H}_1$]{
\begin{tikzpicture}[scale=0.8]
\node (A) [draw=none,thick,regular polygon, regular polygon sides=5, minimum size=1.45cm,outer sep=0pt] {};
\node (B) [draw,thick,regular polygon, regular polygon sides=5, minimum size=2.8cm,outer sep=0pt] {};
\draw[-,thick] (A.corner 1)--(A.corner 3)--(A.corner 5)--(A.corner 2)--(A.corner 4)--(A.corner 1);
\foreach \n in {1,...,5} {
    \draw[-,thick] (A.corner \n)--(B.corner \n);
    \draw[fill=white,thick] (B.corner \n) circle[radius=.75mm];
    \draw[fill=white,thick] (A.corner \n) circle[radius=.75mm];
}
\end{tikzpicture}
}\qquad\qquad\qquad
\subfloat[${H}_2$]{
\begin{tikzpicture}[
roundnode/.style={circle, draw=black, thick, minimum size=1.5mm,inner sep= 0.25mm},
squarednode/.style={rectangle, draw=red!60, fill=red!5, very thick, minimum size=5mm}, scale=0.8
]
\node[roundnode]   (o)   at   (0,0) {};
\node (A) [draw=none,thick,regular polygon, regular polygon sides=3, minimum size=1.25cm,outer sep=0pt] {};
\node (B) [draw,thick,regular polygon, regular polygon sides=6, minimum size=2.8cm,outer sep=0pt] {};

\draw[-,thick] (B.corner 6)--(A.corner 1);
\draw[-,thick] (B.corner 3)--(A.corner 1);
\draw[-,thick] (B.corner 2)--(A.corner 2);
\draw[-,thick] (B.corner 5)--(A.corner 2);
\draw[-,thick] (B.corner 1)--(A.corner 3);
\draw[-,thick] (B.corner 4)--(A.corner 3);
\foreach \n in {1,...,6} {
    \draw[-,thick] (A.corner \n)--(o);
    \draw[fill=white,thick] (B.corner \n) circle[radius=.75mm];
    \draw[fill=white,thick] (A.corner \n) circle[radius=.75mm];
}
\end{tikzpicture}
}
\caption{Đồ thị vô hướng ${H}_1$ và ${H}_2$.}
\label{fig:c2_ex21b}
\end{figure}

\begin{exercise} \label{Ex:2_0}
Xét đồ thị Petersen như ở Hình~\ref{fig:c2_ex21b}.
\begin{enumerate}
    \item[i.] Hãy xác định các ma trận $\m{A}$, $\m{H}$, $\mcl{L}$ của đồ thị.
    \item[ii.] Sử dụng MATLAB, hãy tìm các giá trị riêng và vector riêng của ma trận Laplace.
    \item[iii.] Cần xóa ít nhất bao nhiêu đỉnh/cạnh để làm mất tính liên thông của đồ thị Petersen?
\end{enumerate}
\end{exercise}

\begin{exercise} Chứng minh rằng một đồ thị đơn gồm $n$ đỉnh và $k$ thành phần liên thông có nhiều nhất $\frac{(n-k)(n-k+1)}{2}$ cạnh.
\end{exercise}

\begin{exercise} Tồn tại hay không một đồ thị với chuỗi bậc cho bởi: (a) 1, 2, 2, 3, 3, 3, 4; (b) 1, 2, 2, 3, 3, 4, 4? Trong trường hợp đồ thị tồn tại, hãy biểu diễn tất cả các đồ thị thỏa mãn chuỗi bậc đó.
\end{exercise}

\begin{exercise}
Kí hiệu số cây bao trùm đồ thị $G$ bởi $\tau(G)$. Với $e$ là một cạnh bất kỳ của $G$, chứng minh rằng:
\[\tau(G) = \tau(G-e) + \tau(G\setminus e).\]
\end{exercise}

\begin{exercise}
Kí hiệu $\m{e}_i$ là vector đơn vị với phần tử thứ $i$ bằng 1 và các phần tử khác bằng 0. Với mỗi cạnh $(v_i,v_j)$ của $G$, ta định nghĩa vector $\m{e}_{ij} = \m{e}_j - \m{e}_i$. Chứng minh rằng:
\begin{enumerate}
\item[i.] Mỗi hàng của ma trận $\m{H}^\top$ tương ứng với một vector $\m{e}_{ij} \in E$.
\item[ii.] Ma trận Laplace có thể viết dưới dạng: $\mcl{L} = \sum_{(v_i,v_j)\in E} \m{e}_{ij} \m{e}_{ij}^\top$.
\end{enumerate}
\end{exercise}

\begin{exercise}
Với mỗi ma trận $\m{A} \in \mb{R}^{n \times n}$, kí hiệu $\m{A}_{ij}$ là ma trận thu được từ $\m{A}$ sau khi xóa đi hàng $i$ và cột $j$. Phần phụ đại số của $\m{A}_{ij}$ được tính bởi ${C}_{ij} = (-1)^{i+j} \text{det}(\m{A}_{ij})$. Ma trận phụ hợp của ma trận $\m{A}$ được cho bởi ${\rm adj}(\m{A}) = [C_{ji}]$, với tính chất $\m{A} {\rm adj}(\m{A}) = {\rm det}(\m{A}) \m{I}_n$. Xét đồ thị liên thông $G=(V,E)$ với $|V|=n$ và ma trận Laplace $\mcl{L}$, chứng minh rằng:
\begin{enumerate}
\item[i.] $\text{adj}(\mcl{L}) = \tau(G) \m{J}$, với $\m{J} = \m{1}_n\m{1}_n^\top$.
\item[ii.] $\tau(G) = \frac{1}{n}\prod_{i=2}^n\lambda_i(\mcl{L})$, với $\lambda_i, i = 2, \ldots, n$ là các trị riêng dương của $\mcl{L}$.
\end{enumerate}
\end{exercise}

\begin{exercise} Xét ma trận $\m{B}=k\m{I}_n+\m{A}$, trong đó $\m{A}$ là một ma trận đối xứng, bán xác định dương với các giá trị riêng $\lambda_1,\ldots,\lambda_n$.
\begin{itemize}
\item[i.]  Chứng minh rằng các giá trị riêng của ma trận $\m{B}$ là $k+\lambda_1,\ldots,k+\lambda_n$.
\item[ii.] Tìm các giá trị riêng của ma trận Laplace ứng với đồ thị đầy đủ $K_n$.
\end{itemize}
\end{exercise}

\begin{exercise}[Đồ thị đảo ngược] \cite{Bullo2019lectures}
Đồ thị đảo ngược của một đồ thị hữu hướng $G$ là một đồ thị hữu hướng trong đó mọi cạnh có hướng của  $G$ đều được đảo ngược. Đồ thị loại bỏ hướng của $G$ là đồ thị thu được sau khi thay mỗi cạnh (có hướng) của $G$ bằng một cạnh vô hướng tương ứng. Những phát biểu sau là đúng hay sai?
\begin{enumerate}
\item[i.] Đồ thị hữu hướng $G$ là liên thông mạnh khi và chỉ khi đồ thị đảo ngược của nó là liên thông mạnh.
\item[ii.] Một đồ thị hữu hướng chứa một cây bao trùm có gốc ra khi và chỉ khi đồ thị đảo ngược của nó chứa một cây bao trùm có gốc ra.
\item[iii.] Nếu như đồ thị loại bỏ hướng của $G$ là liên thông thì phải có $G$ hoặc đồ thị đảo ngược của $G$ là có chứa một cây bao trùm có gốc ra.
\item[iv.] Đồ thị hữu hướng $G$ là cân bằng khi và chỉ khi đồ thị đảo ngược của nó là cân bằng.
\end{enumerate}
\end{exercise}

\begin{exercise}[Hằng số Cheeger]
Trong bài tập này, chúng ta xem xét một số tính chất của hằng số Cheeger của đồ thị vô hướng liên thông.
\begin{itemize}
\item[i.] Tính hằng số Cheeger của đồ thị đường thẳng $P_6$ và chu trình $C_6$.
\item[ii.] Tìm hằng số Cheeger của đồ thị đường thẳng $P_n$. Nhận xét về giá trị $h(P_n)$ khi $n\to +\infty$.
\item[iii.] Tìm hằng số Cheeger của đồ thị chu trình $C_n$. Nhận xét về giá trị $h(C_n)$ khi $n\to +\infty$.
\item[iii.] Tìm hằng số Cheeger của đồ thị đầy đủ $K_n$. Nhận xét về giá trị $h(K_n)$ khi $n\to +\infty$.
\end{itemize}
\end{exercise}

\begin{exercise}
Chứng minh rằng một đồ thị gồm $n$ đỉnh và có hơn $\frac{(n-1)(n-2)}{2}$ cạnh là liên thông.
\end{exercise}

\begin{exercise}
Chứng minh rằng một đồ thị gồm $n$ đỉnh và có hơn $n-1$ cạnh thì phải chứa một chu trình.
\end{exercise}

\begin{exercise}[Đồ thị Euler] \cite{Rosen2019discrete}
Một đồ thị $G$ là Euler nếu tồn tại một mạch chứa tất cả các cạnh của $G$.  \index{đồ thị!Euler}
Chứng minh rằng một đồ thị Euler không chứa đỉnh bậc lẻ. Ngược lại, chứng minh đồ thị liên thông $G$ không có đỉnh bậc lẻ là một chu trình Euler.
\end{exercise}

\begin{exercise}
Tìm các giá trị riêng của ma trận $\m{A}$ và ma trận $\mcl{L}$ ứng với:
\begin{enumerate}
    \item[i.] Đồ thị chu trình vô hướng gồm $n$ đỉnh.
    \item[ii.] Đồ thị chu trình hữu hướng gồm $n$ đỉnh.
    \item[iii.] Đồ thị đều $K_n$ vô hướng.
    \item[iv.] Đồ thị sao $S_n$.
    \item[v.] Đồ thị bánh xe $W_n$.
\end{enumerate}
\end{exercise}

\begin{exercise} 
Thực hiện các yêu cầu sau đây:
\begin{enumerate}
    \item[i.] Hãy vẽ đồ thị chu trình $C_3$ và đồ thị hình sao $S_4$.
    \item[ii.] Hãy viết các ma trận kề $\m{A}(C_3)$ và $\m{A}(S_4)$ tương ứng của đồ thị $C_3$ và $S_4$.
    \item[iii.] Tính $\m{A}=\m{A}(C_3)\otimes \m{A}(S_4)$ và vẽ đồ thị nhận ma trận $\m{A}$ là ma trận kề.\footnote{$\otimes$ kí hiệu tích Kronecker (xem định nghĩa ở Phụ lục \ref{append:matrix_theory}).}
\end{enumerate}
\end{exercise}

\begin{exercise}
Xét đồ thị $G$ hữu hướng với ma trận kề $\m{A}\in \mb{R}^{n\times n}$. Chứng minh rằng các phần tử $[b_{ij}]$ của ma trận $\m{B}=\m{A}^2$ tương ứng với số đường đi với độ dài bằng 2 trong $G$ giữa $i$ và $j$. Tổng quát hoá kết quả trên, chứng minh rằng đồ thị có gốc ra nếu tồn tại $k\in \mb{N}_{+}$ sao cho ma trận $\sum_{i=1}^k\m{A}^i$ có một cột gồm toàn phần tử dương. Hơn nữa, chứng mình rằng đồ thị là liên thông mạnh khi và chỉ khi tồn tại một số $k\in \mb{N}_{+}$ sao cho ma trận $\sum_{i=1}^k\m{A}^i$ có tất cả các phần tử dương.
\end{exercise}

\begin{exercise}\label{ex:2_15}
Chứng minh rằng với đồ thị vô hướng có ma trận liên thuộc $\m{H}$ và ma trận Laplace $\mcl{L}$ thì $\mcl{L}=\m{H}^\top\m{H}$. Kết quả trên sẽ thay đổi như thế nào đối với đồ thị có trọng số?
\end{exercise}

\begin{exercise}[Đồ thị bù] \index{đồ thị!bù}
Đồ thị bù của đồ thị $G=(V,E)$ được kí hiệu bởi $\bar{G}=(V,\bar{E})$ với $(i,j)\in \bar{E}$ khi và chỉ khi $(i,j)\notin E$. Chứng minh rằng:
\begin{enumerate}
    \item[i.] $\mcl{L}(G)+\mcl{L}(\bar{G})=n\m{I}_n - \m{1}_n \m{1}_n^\top$ và với $2\le j \le n$,
    \begin{align*}
        \lambda_j(\bar{G}) = n - \lambda_{n+2-j}(G).
    \end{align*}
    \item[ii.] $G$ và $\bar{G}$ không thể đồng thời là không liên thông.
\end{enumerate}
\end{exercise}

\begin{exercise}
Gọi $\mcl{L}$ là ma trận Laplace của đồ thị $G$ hữu hướng gồm $n$ đỉnh. Chứng minh rằng không tồn tại vector riêng bên phải $\m{v}$ nào của $\mcl{L}$ nằm ngoài $\text{im}(\m{1}_n)$ sao cho mọi phần tử của $\m{v}$ đều dương.
\end{exercise}

\begin{exercise}\cite{Nachmias2020} \label{ex:2_18}
Xét đồ thị vô hướng liên thông $G$ với ma trận Laplace $\mcl{L}$. Khi đưa vào đỉnh $i$ và đưa ra khỏi đỉnh $j$ của đồ thị một dòng điện cường độ $I$, các đỉnh của đồ thị có các điện áp tương ứng $v_i, i=1,\ldots,n$. Chứng minh rằng các điện áp nhận mọi giá trị nằm giữa $v_i$ và $v_j$. (Nói cách khác, chứng minh rằng $v_i \ge v_k \ge v_j, \forall k \in V \setminus \{i, j\}$).
\end{exercise}
%
\part{Hệ đồng thuận}
\chapter{Thuật toán đồng thuận}
\label{chap:consensus}
Bài toán đồng thuận\index{hệ!đồng thuận} là một trong những bài toán cơ bản nhất trong điều khiển hệ đa tác tử, trong đó các tác tử thống nhất về giá trị một (hoặc một vài) biến trạng thái được quan tâm  \cite{Baillieul2015encyclopedia}. Trong tài liệu này, thuật toán \index{thuật toán!đồng thuận} (hay một giao thức) đồng thuận là một luật cập nhật dựa trên tiếp nhận và trao đổi thông tin qua mạng sao cho các biến được quan tâm của các tác tử tiệm cận về một giá trị chung - gọi là giá trị đồng thuận  \cite{Jadbabaie2003coordination,Olfati2007consensuspieee,Ren2007distr}. Các biến được quan tâm có thể là thời gian gặp mặt, trọng tâm của đội hình, góc định hướng của các tác tử, nhiệt độ của một khu vực,\ldots 

Nghiên cứu các hệ đồng thuận mở ra mối liên hệ mật thiết giữa khả năng hội tụ của hệ về một giá trị chung và cấu trúc trao đổi thông tin giữa các tác tử. Ta sẽ xem xét bài toán đồng thuận khi biến được quan tâm của các tác tử thay đổi theo một mô hình động học bậc nhất liên tục hoặc không liên tục.

\section{Hệ đồng thuận liên tục bậc nhất}
\subsection{Vấn đề đồng thuận trong hệ đa tác tử}
Xét một hệ gồm $n$ tác tử được đánh số từ $1$ đến $n$. Giả sử tại thời điểm $t \geq 0$, mỗi tác tử có một biến trạng thái $x_i(t) \in \mb{R}$ và có thể đo được các biến tương đối $x_{ij}(t)=x_j(t) - x_i(t)$ từ một số tác tử lân cận. Các tác tử cập nhật biến trạng thái của mình dựa trên tổng có trọng số của các biến tương đối. Để thể hiện sự tương tác giữa các tác tử trong hệ, ta sử dụng một đồ thị hữu hướng, có trọng số $G=(V,E,A)$. Mỗi đỉnh của đồ thị đại diện cho một tác tử và mỗi cạnh $(j,i)$ của đồ thị mô tả rằng tác tử $i$ đo được biến tương đối $x_{ij}$. Kí hiệu tập các tác tử lân cận của tác tử $i$ bởi $N_i = \{j = 1, \ldots, n| ~(j,i) \in E\}$, mỗi tác tử cập nhật biến trạng thái của mình theo thuật toán đồng thuận (còn gọi là luật đồng thuận)\index{thuật toán!đồng thuận}\index{luật!đồng thuận}, cho bởi phương trình
\begin{align} 
\dot{x}_i(t) &= \sum_{j \in N_i} a_{ij} x_{ij}(t) \nonumber \\
&= -\sum_{j \in N_i} a_{ij} (x_i(t) - x_j(t)),\; \forall i = 1, \ldots, n. \label{eq:c3_consensus_cont}
\end{align}
Sơ đồ khối thể hiện thuật toán đồng thuận đối với tác tử $i$ được cho trên Hình~\ref{fig:c3ConsensusDiagramAgenti}.
\begin{figure*}[t]
\centering
\includegraphics[height=8cm]{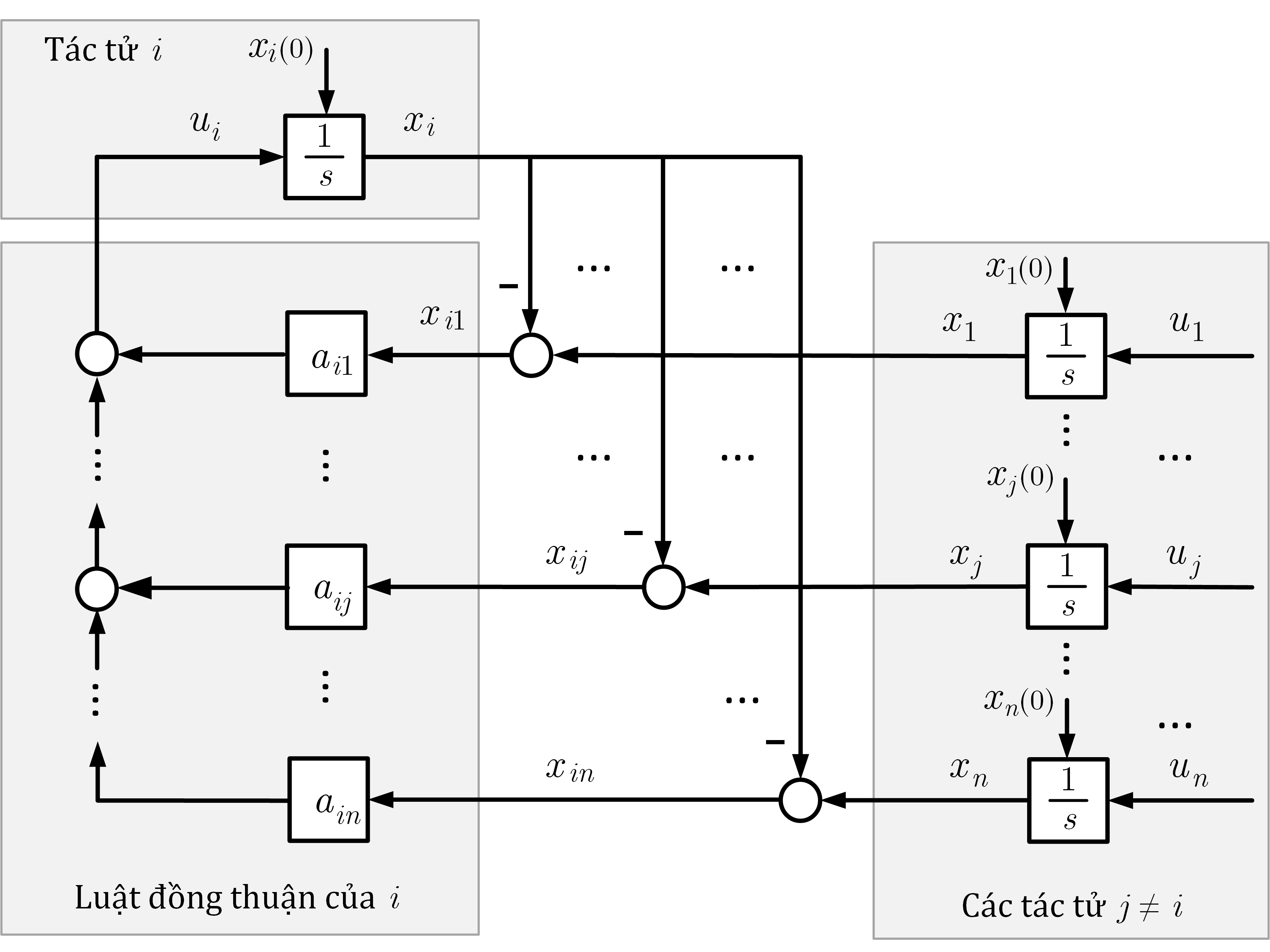}
\caption{Sơ đồ khối thuật toán đồng thuận theo góc nhìn của tác tử $i$.}
\label{fig:c3ConsensusDiagramAgenti}
\end{figure*}

Kí hiệu vector biến trạng thái\index{vector biến trạng thái} của hệ $n$ tác tử bởi $\m{x}(t) = [x_1, \ldots, x_n]^\top \in \mb{R}^n$, ta có thể biểu diễn phương trình của cả hệ như sau
\begin{equation} \label{eq:c3_consensus_cont_matrix}
\dot{\m{x}} = - \mcl{L}(G) \m{x}(t),
\end{equation}
trong đó $\mcl{L}(G)=\mcl{L}$ là ma trận Laplace (ra) của đồ thị tương tác giữa các tác tử $G$. Trong một số tài liệu, người ta cũng gọi $G$ là luồng thông tin\footnote{information flow} của hệ.\index{luồng thông tin}

Với hệ \eqref{eq:c3_consensus_cont}, nếu $x_i = x_j$, ta nói hai tác tử $i$ và $j$ đồng thuận với nhau. Nếu $x_i = x_j, \forall i, j =1, \ldots, n$, thì hệ $n$ tác tử ở trạng thái đồng thuận.\index{trạng thái đồng thuận} Định nghĩa \index{tập đồng thuận}tập đồng thuận bởi
\[\mc{A} = \{\m{x} \in \mb{R}^n|~x_1 = x_2=\ldots=x_n\},\] 
thì $\mc{A}$ là một không gian con của $\mb{R}^n$ được sinh bởi vector $\m{1}_n$. Dễ thấy rằng nếu vector $\m{x} \in \mc{A}$ thì $\m{x} = \alpha \m{1}_n$, với $\alpha \in \mb{R}$. \index{không gian con}

Trong các mục sau đây, chúng ta sẽ phân tích quá trình thay đổi theo thời gian của hệ đồng thuận \eqref{eq:c3_consensus_cont_matrix} với một số giả thiết khác nhau của đồ thị $G$.

\subsection{Trường hợp tổng quát}
Xét đồ thị $G=(V,E,A)$, điều kiện tổng quát để hệ  \eqref{eq:c3_consensus_cont_matrix} tiệm cận tới tập đồng thuận là $G$ có chứa một cây bao trùm có gốc ra. Kết quả này được cho trong định lý sau:
\begin{theorem}\label{thm:c3_consensus_rooted_tree}
Với một đồ thị $G$ có chứa một cây bao trùm có gốc ra, quỹ đạo trạng thái của hệ~\eqref{eq:c3_consensus_cont_matrix} với điều kiện đầu $\m{x}_0 = \m{x}(0)$ thỏa mãn
\begin{align*}
\lim_{t \to +\infty} \m{x}(t) = (\m{1}_n\bm{\gamma}^\top) \m{x}_0 = \m{1}_n \Big(\sum_{i=1}^n \gamma_i x_i(0)\Big),
\end{align*}
với $\m{1}_n$ và $\bm{\gamma}$ tương ứng là các vector riêng bên phải và bên trái ứng với giá trị riêng $0$ của $\mcl{L}$ đã được chuẩn hóa sao cho $\bm{\gamma}^\top \m{1}_n = 1$. Nói cách khác, $\m{x}(t) \to \mc{A}, \forall \m{x}_0 \in \mb{R}^n$, khi và chỉ khi $G$ chứa một cây bao trùm có gốc ra.
\end{theorem}

\begin{proof}
Vì đồ thị là có một gốc ra, ma trận $\mcl{L}$ có một giá trị riêng đơn tại $0$ trong khi các giá trị riêng khác của $\mcl{L}$ có phần thực không âm. Do đó, $\mcl{L}$ có thể phân tích theo dạng chuẩn Jordan như sau:
\begin{align}
\mcl{L} &= \m{P} \m{J} {\m{P}^{ - 1}} \\
&= \m{P} \left[ {\begin{array}{*{20}{c}}
  \m{J}(\lambda_1)&0& \cdots &0 \\ 
  0&\m{J}(\lambda_2)& \ddots & \vdots  \\ 
   \vdots & \ddots & \ddots &0 \\ 
  0& \cdots &0&\m{J}(\lambda_r) 
\end{array}} \right]{\m{P}^{ - 1}},
\end{align}
với $\m{J}_i(\lambda_i)$ là ma trận khối Jordan tương ứng với giá  trị riêng $\lambda_i$, và giả sử rằng có $r$ giá trị riêng khác nhau trong phân tích phổ của $\mcl{L}$. Chú ý rằng $\m{J}_1(\lambda_1) = 0$ do ma trận Laplace chỉ có duy nhất một giá trị riêng 0, và gọi $p_i$ là bội đại số của giá trị riêng $\lambda_i$ thì $n=\sum_{i=1}^r p_i$ (Xem phần Phụ lục \ref{append:matrix_theory} về dạng Jordan của ma trận). Ma trận $\m{P} = [\m{p}_1, \ldots, \m{p}_n]$ có vector cột đầu tiên là $\m{p}_1 = \m{1}_n$, và ma trận  $\m{P}^{-1} = [\m{q}_1, \ldots, \m{q}_n]^\top$ có  $\m{q}_1^\top$ là vector riêng bên trái của $\mcl{L}$ tương ứng với giá trị riêng $\lambda_1=0$. Vì $\m{P}^{-1}\m{P} = \m{I}_n$, ta có $\m{q}_1^\top \m{1}_n = 1$, nên ta đặt $\bm{\gamma} = \m{q}_1$. Chú ý rằng các phần tử của ma trận $\m{P}$ có thể là các số phức.

Theo lý thuyết về hệ tuyến tính, quỹ đạo trạng thái của hệ đồng thuận \eqref{eq:c3_consensus_cont_matrix} thỏa mãn
\begin{align*}
\lim_{t \to +\infty} \m{x}(t) &= \lim_{t \to +\infty} \mathtt{e}^{-\mcl{L}t} \m{x}_0\\
&=\m{P}\left( \lim_{t \to +\infty} \begin{bmatrix}
\mathtt{e}^0 & 0 & \cdots & 0\\
0 & \mathtt{e}^{-\m{J}(\lambda_2) t} & \ddots & \vdots\\
\vdots & \ddots& \ddots&0\\
0&\cdots & 0 & \mathtt{e}^{-\m{J}(\lambda_r)t}
\end{bmatrix} \right) \m{P}^{-1} \m{x}_0.
\end{align*}
Do các giá trị riêng khác $0$ của $\mcl{L}$ đều có phần thực dương, \[\lim_{t\to +\infty} \mathtt{e}^{-\m{J}(\lambda_i) t} = 0, \, \forall i= 2, \ldots, r.\] Bởi vậy, 
\begin{align*}
\lim_{t \to +\infty} \m{x}(t) &= \m{P} \begin{bmatrix}
1 & 0 & \cdots & 0\\
0 & 0 & \ddots & \vdots\\
\vdots & \ddots& \ddots&0\\
0&\cdots & 0 & 0
\end{bmatrix} \m{P}^{-1} \m{x}_0 \\
&= (\m{p}_1 \m{q}_1^\top)  \m{x}_0 \\
&= (\m{1}_n \bm{\gamma}^\top)\m{x}_0 \\
&=  \Big(\sum_{i=1}^n \gamma_i x_i(0)\Big) \m{1}_n,
\end{align*}
hay $x_i(t)$ tiệm cận đồng thuận tại $x^\ast = \sum_{i=1}^n \gamma_i x_i(0)$, khi $t \to +\infty$.
\end{proof}

Giá trị $x^\ast = \sum_{i=1}^n \gamma_i x_i(0)$ gọi là giá trị đồng thuận. Chú ý rằng, do $\bm{\gamma}^\top \mcl{L} = \m{0}^\top_n$, ta có\index{giá trị!đồng thuận}
\begin{align} \label{eq:c3_final_value}
\frac{d}{dt} (\bm{\gamma}^\top \m{x}(t)) = \bm{\gamma}^\top \frac{d}{dt}(\m{x}(t))= -\bm{\gamma}^\top \mcl{L} \m{x}(t) = {0}.
\end{align}
Từ đẳng thức trên, ta có định lý sau đây.
\begin{theorem}\label{thm:c3_invariant}
Giá trị trung bình cộng có trọng số \[\bm{\gamma}^\top \m{x}(t)=\bm{\gamma}^\top \m{x}(0) = \sum_{i=1}^n \gamma_i x_i(0) = x^\ast\] là bất biến với luật đồng thuận \eqref{eq:c3_consensus_cont_matrix}.
\end{theorem}

Nhận xét thêm rằng trong phân tích tiệm cận của $\m{x}(t)$, thì các thành phần tương ứng với các ma trận khối Jordan $\mathtt{e}^{-\m{J}(\lambda_k)t}$ hội tụ về 0 theo tốc độ hàm mũ. Tốc độ hội tụ của $\mathtt{e}^{-\m{J}(\lambda_k)t}$ phụ thuộc vào độ lớn của ${\rm Re}(\lambda_k)>0$, và do đó quá trình $\m{x}(t) \to \mc{A}$ phụ thuộc chủ yếu vào giá trị $\lambda_2$ của $\mcl{L}$. Vì lý do này, giá trị riêng $\lambda_2$, ngoài là một chỉ số định tính cho sự liên thông của đồ thị còn là một chỉ số định lượng cho tốc độ đồng thuận. Một số cận trên và cận dưới của $\lambda_2$ có thể được ước lượng dựa trên thông tin về đồ thị và các trọng số của ma trận Laplace. Do các kết quả này được xác định bởi Fiedler \cite{Fiedler1973AlgebraicConnectivity}, một số tài liệu gọi $\lambda_2$ là giá trị riêng Fiedler. \index{giá trị riêng!Fiedler}

\begin{story}{Sơ lược về các thuật toán đồng thuận}
Đồng thuận là vấn đề cơ bản trong lĩnh vực tính toán phân tán, được nghiên cứu từ khoảng những năm 1970. Các thuật toán đồng thuận nhằm thống nhất về các biến quyết định ở các tác tử (cơ sở dữ liệu phân tán), thông qua giao thức truyền nhận bản tin có khả năng hoạt động với lỗi trên mạng truyền thông (mất bản tin, có lỗi trong bản tin, bất đồng bộ trong đọc và xử lý bản tin, hay khi bản tin bị cố ý làm sai lệch bởi một số nút) \cite{Lynch1996distributed}. Công nghệ block-chain là một giao thức đồng thuận.

Mô hình toán mô tả quá trình đồng thuận các biến vật lý đã xuất hiện trong các lĩnh vực khác nhau: mô hình Degroot (1974) trong nghiên cứu cơ chế một nhóm người thống nhất về quan điểm qua thảo luận \cite{degroot1974reaching}, mô hình Vicsek (1995) giải thích cơ chế tụ bầy trong tự nhiên với các phân tích toán học của Ali Jadbabaie (2003) \cite{Jadbabaie2003coordination}, hay mô hình Kuramoto (1975) tổng quát hóa nhiều hiện tượng tự đồng bộ ở côn trùng, dao động cơ, máy phát điện xoay chiều điện. Thuật toán đồng thuận liên tục trong hệ đa tác tử được giới thiệu bởi lab nghiên cứu của Richard Murray tại California Institute of Technology, CA \cite{Fax2002,Olfati2004consensus}. Olfati-Saber xuất bản một loạt các công trình liên quan tới thuật toán đồng thuận và ứng dụng trong mạng cảm biến và xe tự hành trong thời gian làm Sau Tiến Sỹ tại nhóm nghiên cứu của Richard Murray \cite{Olfati2004consensus,Olfati2007consensuspieee}. Hiện tại Olfati-Saber quan tâm tới các hệ thống trí tuệ nhân tạo. Yếu tố khác biệt của thuật toán đồng thuận trong điều khiển hệ đa tác tử là sự xuất hiện của động học trong thuật toán đồng thuận. Biến đồng thuận trong hệ có thể là các biến vật lý (vị trí, góc, tần số, góc pha,...), và quá trình đồng thuận có thể được xem xét dựa trên lý thuyết ổn định. \cite{Olfati2004consensus} là một trong những bài báo được trích dẫn nhiều lần nhất về chủ đề hệ đồng thuận.

Hiện tại, Murray quan tâm tới ứng dụng của điều khiển trong các hệ nối mạng và kĩ thuật sinh học. Murray đề cao việc sử dụng kết hợp các phương pháp giải quyết vấn đề giữa hai lĩnh vực Khoa học máy tính và Điều khiển tự động. Giáo trình Điều khiển \cite{AstromMurray2008} được ông soạn cùng {\AA}str{\"o}m để phù hợp với độc giả làm việc trong lĩnh vực Khoa học máy tính. 

Một điểm đặc biệt của Murray là ông luôn để một số vấn đề ngỏ và quan trọng cần giải quyết từ khi bắt đầu là GS về Động lực học và điều khiển tại Caltech. Từ 2003 đến nay, các vấn đề và thử thách trong khắp các lĩnh vực của điều khiển được một hội đồng kĩ thuật lựa chọn công bố dưới dạng báo cáo kỹ thuật \cite{Murray2003}. Hiện tại, báo cáo có dạng một bản định hướng phát triển mở của Cộng đồng Điều khiển (IEEE Control Systems Society) tới năm 2030, với viễn kiến rằng điều khiển và AI sẽ kết hợp ngày càng chặt chẽ để giải quyết các vấn đề phức tạp (\url{https://ieeecss.org/control-societal-scale-challenges-road-map-2030}).
\end{story}

\subsection{Một số trường hợp riêng}
Trong nhiều ứng dụng, lớp các thuật toán đồng thuận cho giá trị đồng thuận tại trung bình cộng của $x_i(0)$, $i =1, \ldots, n$, được đặc biệt quan tâm. Ví dụ khi ta có $n$ cảm biến theo dõi nhiệt độ của một khu vực, các cảm biến trao đổi giá trị đo và muốn tìm giá trị nhiệt độ trung bình để đại diện cho nhiệt độ của cả khu vực. Các thuật toán này được gọi chung là các \index{thuật toán!đồng thuận trung bình cộng}thuật toán đồng thuận trung bình cộng\footnote{average consensus algorithm}.

Với thuật toán đồng thuận \eqref{eq:c3_consensus_cont_matrix}, điều kiện cần và đủ để giá trị đồng thuận là trung bình cộng phụ thuộc vào đồ thị $G$. Đầu tiên, xét $G$ là một đồ thị vô hướng, liên thông. Khi đó, do $\mcl{L} = \mcl{L}^\top$ và $\mcl{L}\m{1}_n = \m{0}_n$, ta cũng có $\m{1}^\top_n \mcl{L} = \m{0}^\top_n$. So sánh với phương trình~\eqref{eq:c3_final_value}, ta suy ra $\bm{\gamma} = \m{1}_n/n$, $\lim_{t \to +\infty} \mathtt{e}^{-\mcl{L}t} = \frac{1}{n}\m{1}_n\m{1}_n^\top$, và giá trị đồng thuận được xác định bởi $\m{1}_n^\top \m{x}_0 = \frac{1}{n}\sum_{i=1}^n x_i(0)$ -- trung bình cộng của tất cả các biến $x_i(0),\forall i\in \mc{I}$.

Từ quan sát trên, ngoài điều kiện đồng thuận là $G$ chứa một cây bao trùm có gốc ra, để giá trị đồng thuận là trung bình cộng thì ma trận Laplace phải nhận $\m{1}_n^\top$ làm một vector riêng bên trái ứng với giá trị riêng 0. Điều này tương đương với việc $l_{ii} = \sum_{i=1}^{n} a_{ij} = \sum_{j=1}^n a_{ij}$, hay $\text{deg}^+(v_i) = \text{deg}^-(v_i), \forall i = 1, \ldots, n$. Một đồ thị hữu hướng thỏa mãn tổng trọng số vào bằng tổng trọng số ra tại tất cả các đỉnh được gọi là một \index{đồ thị!cân bằng} đồ thị cân bằng (về trọng số). Ta có định lý sau:

\begin{theorem} Thuật toán~\eqref{eq:c3_consensus_cont_matrix} là một thuật toán đồng thuận trung bình cộng khi và chỉ khi đồ thị $G$ là liên thông yếu và cân bằng.
\end{theorem}

Cuối cùng, xét đồ thị $G$ là liên thông mạnh nhưng không cân bằng. Vì  $G$ liên thông mạnh nên mọi đỉnh của $G$ đều là một gốc vào. Do điều kiện đồng thuận được thỏa mãn, hệ tiến tới đồng thuận tại $x^\ast = \sum_{i=1}^n \gamma_i x_i(0)$. Theo Định lý  \ref{thm:c2_Laplace_directed}, $\bm{\gamma}$ có các phần tử  $\gamma_i>0$, $\forall i = 1, \ldots, n,$ và $\sum_{i=1}^n \gamma_i = 1$. Ta suy ra với $G$ là liên thông mạnh, giá trị đồng thuận của hệ \eqref{eq:c3_consensus_cont_matrix} là trung bình cộng có trọng số của các giá trị $x_i(0),\, i = 1, \ldots, n$.

\begin{example}\label{eg:3.4}
Mô phỏng thuật toán đồng thuận với một số đồ thị khác nhau.
Giá trị đầu của các biến trạng thái được chọn ngẫu nhiên với $-5\leq x_i(0) \leq 5$. Với luật đồng thuận \eqref{eq:c3_consensus_cont}, $x_i(t)$ tiệm cận tới không gian đồng thuận với cả ba đồ thị. Tốc độ đồng thuận phụ thuộc vào giá trị riêng liên kết $\lambda_2(\mcl{L})$ và biểu diễn trên các Hình~\ref{fig:c3_VD3.4} (b), (d), (f).
\end{example}
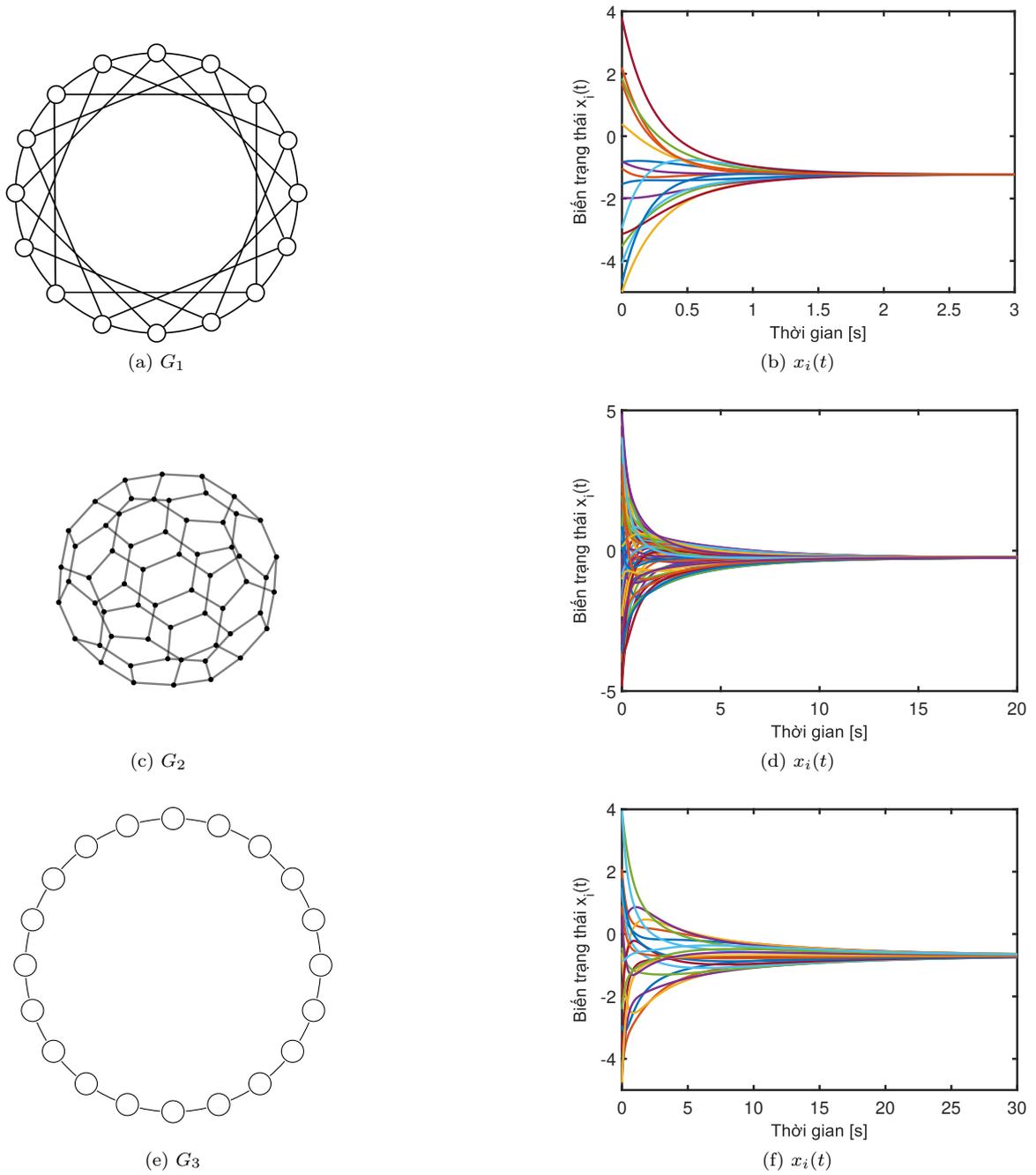
\begin{figure*}
    \centering
    \subfloat[$G_1$]{\includegraphics[height=4.5cm]{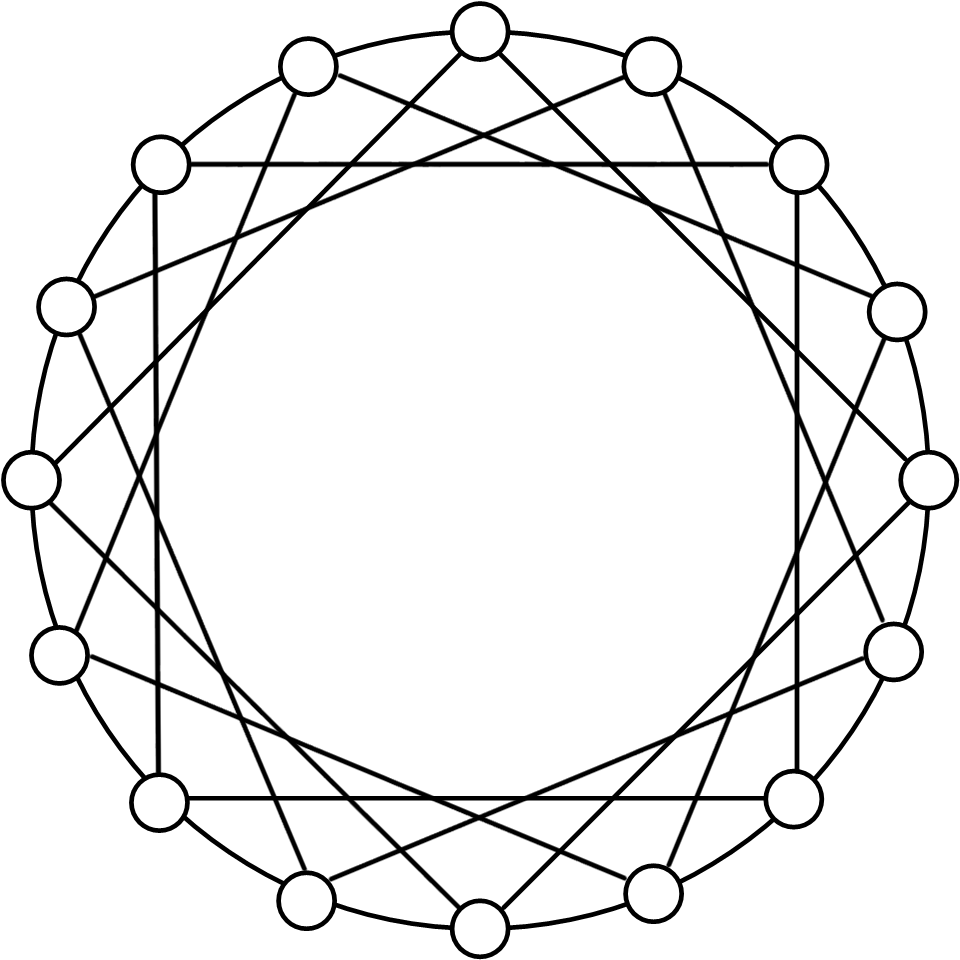}} \hfill
    \subfloat[$x_i(t)$]{\includegraphics[height=5.14cm]{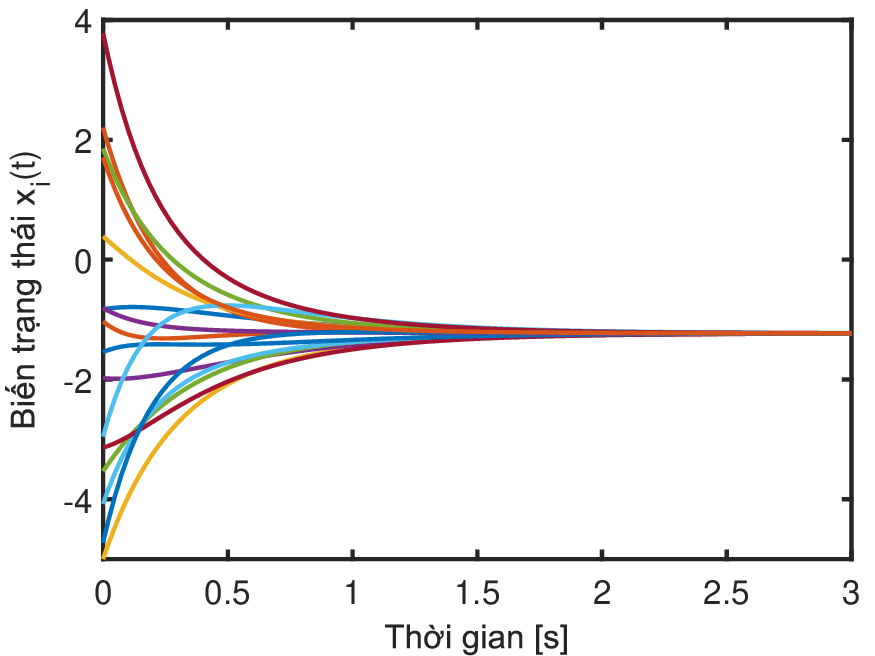}}\\
    \subfloat[$G_2$]{\includegraphics[height=4.55cm]{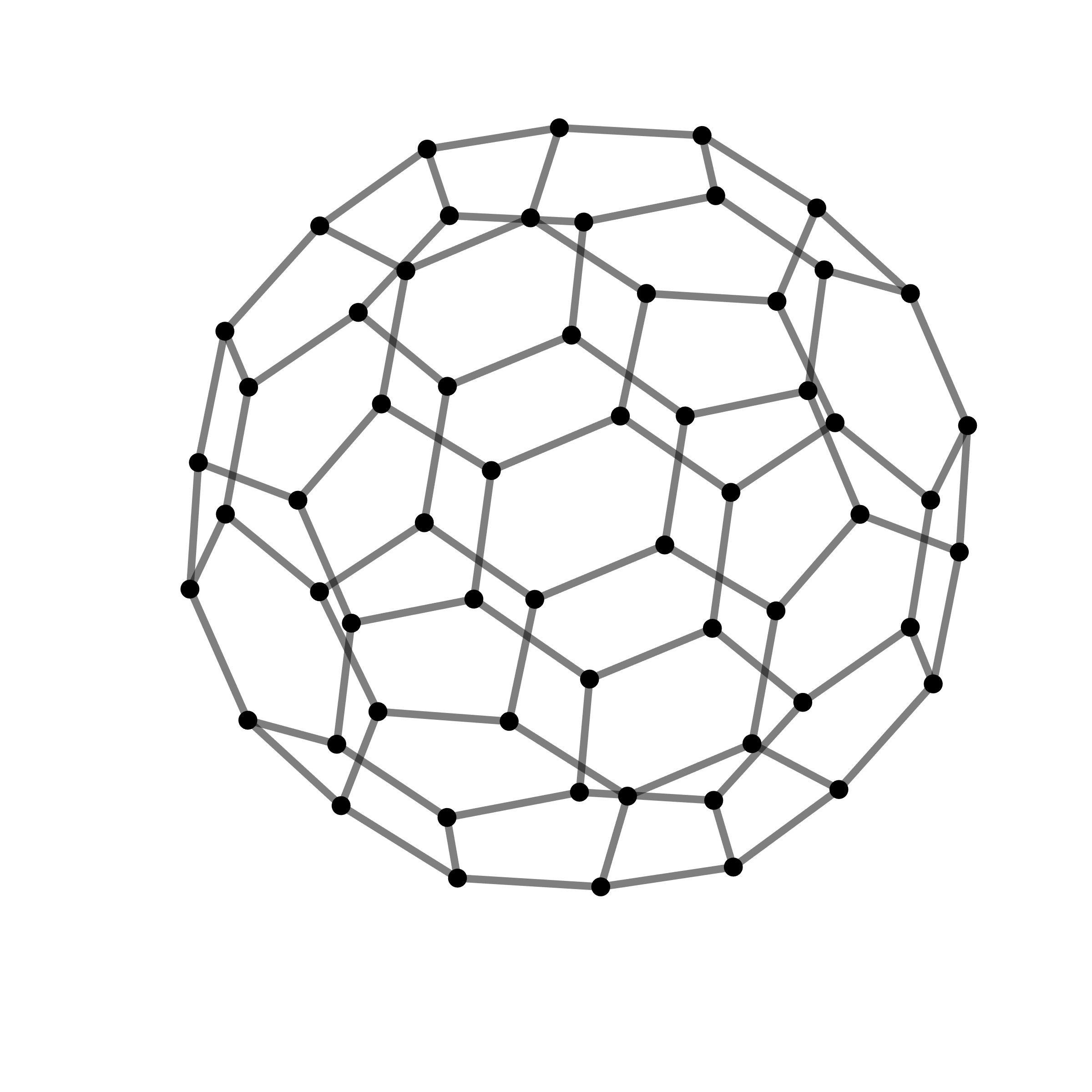}} \hfill
    \subfloat[$x_i(t)$]{\includegraphics[height=5.14cm]{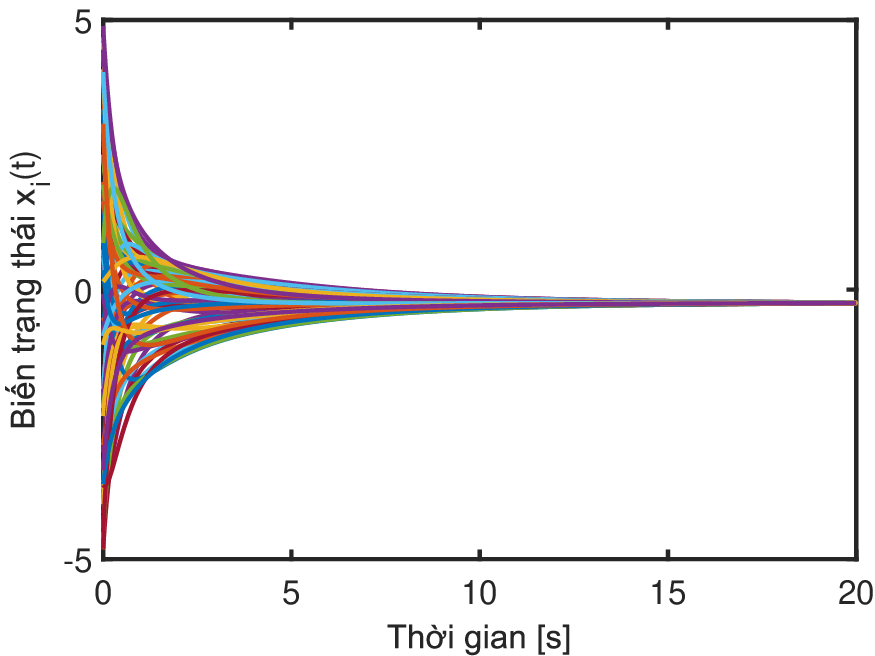}}\\
    \subfloat[$G_3$]{
        \begin{tikzpicture}[>=latex, scale=0.63]
        \def \n {20}
        \def \radius {3.5cm}
        \def \margin {5} 
        \foreach \s in {1,...,\n}
        {
        \node[draw, circle,color=black] at ({360/\n * (\s - 1)}:\radius) {};
        \draw[-, >=latex,color=black] ({360/\n * (\s - 1)+\margin}:\radius) 
        arc ({360/\n * (\s - 1)+\margin}:{360/\n * (\s)-\margin}:\radius);
        }
        \node (uu) at (0,-4) { };
        \end{tikzpicture}
    } \hfill
    \subfloat[$x_i(t)$]{\includegraphics[height=5.14cm]{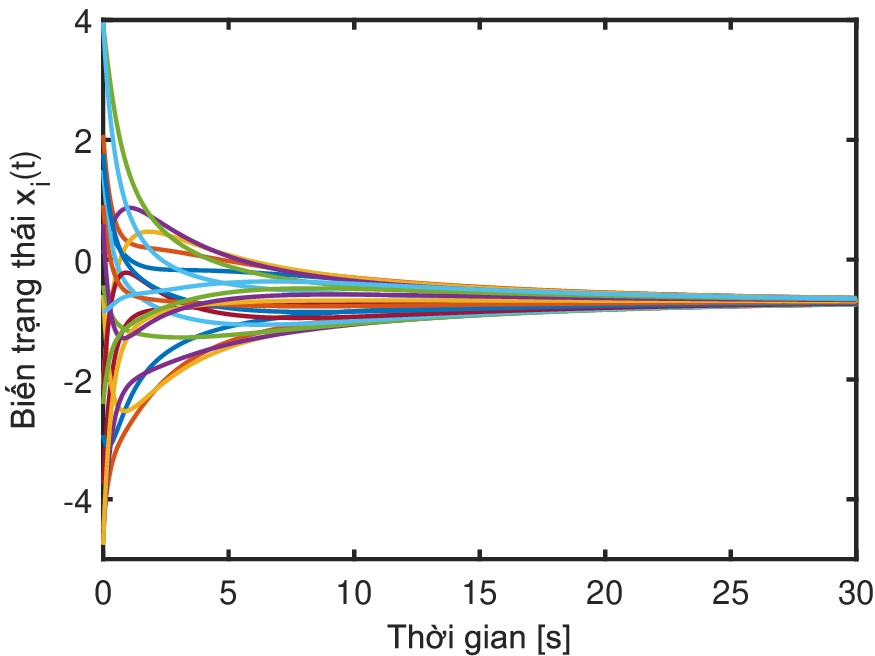}}\\
    \caption{Mô phỏng thuật toán đồng thuận với ba đồ thị khác nhau: $G_1$ là đồ thị đều gồm 16 đỉnh, mỗi đỉnh có 3 đỉnh kề, $G_2$ là đồ thị Bucky gồm 60 đỉnh và $G_3$ là chu trình gồm 20 đỉnh.}
    \label{fig:c3_VD3.4}
\end{figure*}


\section{Hệ đồng thuận liên tục bậc hai}
Ở mục này, ta xét hệ gồm các tác tử có mô hình là khâu tích phân bậc hai:
\begin{align}
    \dot{{x}}_i &= {y}_i, \nonumber\\
    \dot{{y}}_i &= {u}_i,~ i = 1, \ldots, n, \label{eq:c3_double}
\end{align}
với $x_i \in \mb{R}$, $y_i \in \mb{R}$, và $u_i \in \mb{R}$.\footnote{Thông thường, biến $x_i, y_i, u_i$ mô tả vị trí, vận tốc và gia tốc của một chất điểm trong không gian.} Luật đồng thuận được đề xuất cho hệ \eqref{eq:c3_double} là:
\begin{align}
    u_i = -\sum_{j\in {N}_i} \Big(\big(x_i - x_j) + \beta\big(y_i - y_j \big)\Big), ~ i = 1, \ldots, n, \label{eq:c3_double_u}
\end{align}
trong đó $\beta>0$ là một hằng số dương. Thuật toán đồng thuận \eqref{eq:c3_double_u} là phi tập trung theo nghĩa mỗi tác tử chỉ cần có thông tin từ một vài tác tử láng giềng. Ta sẽ chứng minh rằng với luật điều khiển \eqref{eq:c3_double_u}, hệ tiệm cận tới tập đồng thuận, thể hiện bởi $|x_i - x_j| \to 0$ và $|y_i - y_j| \to 0$, khi $t \to +\infty$. Chú ý rằng nếu $x_i$ và $y_i$ thể hiện vị trí và vận tốc của tác tử thứ $i$ thì \eqref{eq:c3_double_u} thể hiện gia tốc của tác tử này.

Đặt $\m{x} = [x_1, \ldots, x_n]^\top$ và $\m{y} = [y_1, \ldots, y_n]^\top$. Khi sử dụng luật điều khiển \eqref{eq:c3_double_u} cho hệ, ta có thể viết lại \eqref{eq:c3_double} dưới dạng ma trận như sau:
\begin{equation}
    \begin{bmatrix}\dot{\m{x}}\\
    \dot{\m{y}}
    \end{bmatrix} = \begin{bmatrix} \m{0}_{n\times n} & \m{I}_n\\
    -\mcl{L} & -\beta \mcl{L}
    \end{bmatrix} \begin{bmatrix}{\m{x}}\\
    {\m{y}}
    \end{bmatrix} = \m{M} \begin{bmatrix}{\m{x}}\\
    {\m{y}}
    \end{bmatrix}. \label{eq:c3_db}
\end{equation}
Lưu ý rằng với một ma trận khối có dạng $\m{A} = \begin{bmatrix} \m{A}_{11} & \m{A}_{12}\\
\m{A}_{21} & \m{A}_{22}
\end{bmatrix}$ thì det$(\m{M}$) = det($\m{A}_{11}\m{A}_{22} - \m{A}_{21}\m{A}_{12}$) nếu như hai ma trận $\m{A}_{11}$ và $\m{A}_{21}$ là  giao hoán được (tức là $\m{A}_{11}\m{A}_{21} = \m{A}_{21}\m{A}_{11}$)\cite{Ren2007distr}. Áp dụng công thức trên, các giá trị riêng của $\m{M}$ là nghiệm của phương trình đặc tính:
\begin{equation}
    \text{det}(s \m{I}_{2n} - \m{M}) = \text{det}\left( \begin{bmatrix} \lambda \m{I}_n & -\m{I}_n\\
    \mcl{L} & s \m{I}_n + \beta \mcl{L} 
    \end{bmatrix} \right) = \text{det}\big(s^2 \m{I}_n + (1+\beta s)\mcl{L}\big) \label{eq:c3_L}
\end{equation}
Chú ý rằng ${\rm det}(s\m{I}_n + \mcl{L})=\prod_{i=1}^n(s+\lambda_i)$, trong đó $\lambda_i$ kí hiệu giá trị riêng thứ $i$ của $\mcl{L}$. Vì vậy, so sánh với phương trình \eqref{eq:c3_L}, ta có:
\begin{align}
{\rm det}\big(s^2 \m{I}_n + (1+\beta s)\mcl{L}\big) = \prod_{i=1}^n(s^2+(1+\beta s)\lambda_i). \label{eq:c3_L2}
\end{align}
Phương trình \eqref{eq:c3_L2} chứng tỏ rằng, các giá trị riêng của $\m{M}$ đều có thể thu được từ phương trình $s^2 +\lambda_i\beta s + \lambda_i=0$.  Nghiệm của phương trình này được cho bởi:
\begin{equation}
s_{i\pm} = \frac{-\beta\lambda_i \pm \sqrt{\beta^2\lambda_i^2 - 4 \lambda_i}}{2}, \label{eq:c3_eig}
\end{equation}
với $s_{i+}$ và $s_{i-}$ là các giá trị riêng của $\m{M}$ ứng với $\lambda_i$.

Từ phương trình~\eqref{eq:c3_eig}, dễ thấy rằng ứng với mỗi giá  trị riêng $\lambda_i = 0$ của $\mcl{L}$ thì $\m{M}$ có tương ứng hai giá trị riêng bằng 0. Với giả thiết rằng đồ thị $G$ là liên thông thì  $\mcl{L}$ chỉ có một giá trị riêng duy nhất bằng 0, còn các giá trị riêng khác đều có phần thực dương (Định lý \ref{thm:c2_Laplace_directed}). Từ đây, ta suy ra $\m{M}$ có giá trị riêng 0 kép, kí hiệu là $s_{1+} = s_{1-} = 0$. Định lý sau đây cho ta một điều kiện cần và đủ để hệ đạt được đồng thuận:

\begin{theorem} \label{c3:thm_db}
Hệ~\eqref{eq:c3_db} tiệm cận tới tập ${\rm im}(\m{1}_n) \times {\rm im}(\m{1}_n)$ khi và chỉ khi ma trận $\m{M}$ có đúng hai giá trị riêng bằng 0 và các giá trị riêng khác tính theo \eqref{eq:c3_eig} đều có phần thực âm. Khi đó, $\m{x} \to \m{1}_n \bmm{\gamma}^\top \m{x}(0)+t \m{1}_n \bmm{\gamma}^\top \m{y}(0)$ khi $t \to +\infty$, trong đó $\bmm{\gamma} \in \mb{R}^n$ là một vector riêng bên phải (có các phần tử không âm) của $\mcl{L}$ ứng với giá trị riêng $0$ và $\bmm{\gamma}^\top \m{1}_n = 1$. 
\end{theorem}

\begin{proof}
(Điều kiện đủ) Đầu tiên, ta chứng minh rằng giá  trị riêng 0 của $\m{M}$ có bội hình học là 1 trong trường hợp $\m{M}$ có đúng hai giá trị riêng bằng 0. Đặt $\m{q} = [\m{q}_a^\top, \m{q}_b^\top]^\top$ là một vector riêng của $\m{M}$ ứng với giá trị riêng 0 ($\m{q}_a, \m{q}_b \in \mb{R}^n$) thì
\begin{align}
    \m{M} \m{q} = \begin{bmatrix} \m{0}_{n\times n} & \m{I}_n\\
    -\mcl{L} & -\beta \mcl{L} 
    \end{bmatrix} \begin{bmatrix} \m{q}_a\\ \m{q}_b \end{bmatrix} = \begin{bmatrix} \m{0}_n\\ \m{0}_n \end{bmatrix},
\end{align}
tức là phải có $\m{q}_b = \m{0}_n$ và $\mcl{L}\m{q}_a = \m{0}_n$. Điều này chứng tỏ rằng $\m{q}_a$ là một vector riêng của $\mcl{L}$ ứng với giá trị riêng 0. Do $\m{M}$ chỉ có hai giá trị riêng bằng 0, $\mcl{L}$ chỉ có một giá trị riêng duy nhất bằng 0. Do đó, $\text{ker}(\mcl{L})$ chỉ được sinh bởi  một vector độc lập tuyến tính duy nhất. Từ đây suy ra $\text{ker}(\m{M})$ cũng chỉ được sinh bởi duy nhất vector $\m{q} = [\m{q}_a^\top, \m{0}_n]^\top$. Điều này tương đương với việc giá  trị riêng $0$ của $\m{M}$ có bội hình học là $1$.

Chú ý rằng ta có thể viết $\m{M}$ dưới dạng sau:
\begin{align}
    \m{M} & = \m{P} \m{J} \m{P}^{-1} \nonumber\\
    &= [\m{w}_1, \ldots, \m{w}_{2n}] \begin{bmatrix} 0 & 1 & \m{0}_{1 \times (2n-2)}\\
    0 & 0 & \m{0}_{1 \times (2n-2)}\\
    \m{0}_{(2n-2)\times 1} & \m{0}_{(2n-2)\times 1} & \m{J}'
    \end{bmatrix} \begin{bmatrix} \bmm{\eta}_1^\top \\ \vdots \\ \bmm{\eta}_{2n}^\top  \end{bmatrix},
\end{align}
trong đó $\m{J}$ là dạng Jordan của $\m{M}$, $\m{w}_j \in \mb{R}^{2n}$, $j = 1, \ldots, 2n,$ là các vector riêng và vector riêng suy rộng bên phải của ma trận $\m{M}$, $\bmm{\eta}_j, i=1, \ldots, 2n,$ là các vector riêng và vector riêng suy rộng bên trái của $\m{M}$, và $\m{J}'$ là ma trận Jordan có dạng đường chéo khối (có dạng tam giác trên) ứng với các giá trị riêng khác không $s_{i+}$ và $s_{i-}$, $i=2, \ldots, n$.

Không mất tính tổng quát, ta có thể chọn $\m{w}_1 = [\m{1}_n^\top, \m{0}_n^\top]^\top$ và $\m{w}_2 = [\m{0}_n^\top, \m{1}_n^\top]^\top$ thì $\mcl{L}\m{w}_1 = \m{0}_n$ và $\mcl{L}\m{w}_2 = \m{w}_1$. Khi đó, có thể kiểm tra rằng $\bmm{\eta}_1 = [\bmm{\gamma}^\top, \m{0}^\top_n]^\top$ và $\bmm{\eta}_2 = [\m{0}^\top_n,\bmm{\gamma}^\top]^\top$ là các vector riêng suy rộng và vector riêng của $\m{M}$ ứng với giá trị riêng $0$. Chú ý rằng $\bmm{\eta}_1^\top\m{w}_1 = 1$ và $\bmm{\eta}_2^\top\m{w}_2 = 1$. Do các giá trị riêng $s_{i\pm}$ của $\m{M}$ đều có phần thực âm, ta có
\begin{align}
    \lim_{t\to+\infty} \mathtt{e}^{\m{M}t} &= \m{P} \lim_{t\to+\infty}\left( \mathtt{e}^{\m{J}t} \right) \m{P}^{-1} \nonumber\\
    &= \m{P} \lim_{t\to+\infty}\left( \begin{bmatrix}
    1 & t & \m{0}_{1 \times (2n-2)}\\
    0 & 1 & \m{0}_{1 \times (2n-2)}\\
    \m{0}_{(2n-2) \times 1} & \m{0}_{(2n-2) \times 1} & \mathtt{e}^{\m{J}'t}\\
    \end{bmatrix}
    \right) \m{P}^{-1} \nonumber\\
    &= \m{w}_1 \bmm{\eta}_1^\top + (t\m{w}_1 + \m{w}_2) \bmm{\eta}_2^\top \nonumber\\
    &= \begin{bmatrix} 
    \m{1}_n \bmm{\gamma}^\top & t \m{1}_n \bmm{\gamma}^\top\\
    \m{0}_{n\times n} & \m{1}_n \bmm{\gamma}^\top
    \end{bmatrix}.
\end{align}
Như vậy, khi $t\to +\infty$ thì $\m{x}(t) \to \m{1}_n \bmm{\gamma}^\top \m{x}(0) + t \m{1}_n \bmm{\gamma}^\top \m{y}(0)$ và $\m{y}(t) \to \m{1}_n \bmm{\gamma}^\top \m{y}(0)$. Nói cách khác, ta có $|x_i(t) - x_j(t)| \to 0$ and $|y_i(t)-y_j(0)|\to 0$ khi $t \to +\infty$, hay hệ đạt được đồng thuận.

(Điều kiện cần) Giả sử rằng điều kiện đủ ``$\m{M}$ có hai giá trị riêng bằng 0 và các giá trị riêng khác đều có phần thực âm'' không thỏa mãn. Do $\m{M}$ có ít nhất hai giá trị riêng 0, nên điều kiện đủ này không thỏa mãn khi $\m{M}$ có nhiều hơn hai giá trị riêng bằng $0$ hoặc $\m{M}$ ít nhất môt giá trị riêng với phần thực dương. Không mất tính tổng quát, giả sử rằng $\iota_1 = \iota_2 = 0$ và Re($\iota_3$)$\geq 0$, với $\iota_k, k=1, \ldots, 2n,$ kí hiệu cho giá trị riêng thứ $k$ của $\m{M}$. Với $\m{J} = [J_{kl}]$ là ma trận Jordan của $\m{M}$ thì $J_{kk} = \iota_k, k= 1, \ldots, 2n$. Khi đó ta có $\lim_{t\to+\infty} \mathtt{e}^{J_{kk}t} \neq 0, k = 1, 2, 3$, tức là ba cột đầu của $\lim_{t\to+\infty}\mathtt{e}^{\m{J}t}$ là độc lập tuyến tính. Bởi vậy,  rank($\lim_{t\to+\infty}\mathtt{e}^{\m{J}t}$)$\geq 3$. Vì hệ tiệm cận tới tập đồng thuận khi và chỉ khi $\lim_{t\to+\infty}\mathtt{e}^{\m{J}t} \begin{bmatrix}
\m{x}(0)\\
\m{y}(0)
\end{bmatrix}
 \to \begin{bmatrix}
\m{1}_n \m{p}(t)^\top \m{x}(0)\\
\m{1}_n \m{q}(t)^\top \m{y}(0)
\end{bmatrix}$, trong đó $\m{p}, \m{q} \in \mb{R}^n$, rank($\lim_{t\to+\infty}\mathtt{e}^{\m{J}t}$)$\leq 2$ và ta được một điều vô lý.
\end{proof}

Từ Định lý~\ref{c3:thm_db}, nếu như ban đầu các tác tử đứng yên ($y_i(0) = 0, \forall i = 1, \ldots, n$), thì khi $t \to +\infty$, ta có $x_i(t) \to \sum_{i=1}^n \gamma_i x_i(0)$ và  $y_i(t) \to 0, \forall i = 1, \ldots, n$. 

Nếu như vận tốc ban đầu của các tác tử khác không và ta mong muốn hệ đạt được đồng thuận về vị trí tại một điểm, thì có thể sử dụng luật đồng thuận sau:
\begin{align}
    u_i = - \alpha y_i -\sum_{j\in {N}_i} \Big(\big(x_i - x_j) + \beta\big(y_i - y_j \big)\Big), ~ i = 1, \ldots, n, \label{eq:c3_double_u1}
\end{align}
trong đó $\alpha>0$.

Ta có thể biểu diễn hệ khi áp dụng luật đồng thuận \eqref{eq:c3_double_u1} dưới dạng ma trận như sau:
\begin{align}
    \begin{bmatrix}\dot{\m{x}}\\
    \dot{\m{y}}
    \end{bmatrix} = \begin{bmatrix} \m{0}_{n\times n} & \m{I}_n\\
    -\mcl{L} & -\alpha \m{I}_n - \beta \mcl{L}
    \end{bmatrix} \begin{bmatrix}{\m{x}}\\
    {\m{y}}
    \end{bmatrix} = \m{N} \begin{bmatrix}{\m{x}}\\
    {\m{y}}
    \end{bmatrix}. \label{eq:c3_db1}
\end{align}

Các giá trị riêng của $\m{N}$ thỏa mãn
\begin{align}
 {\rm det}(s \m{I}_{2n} - \m{N}) &= \text{det}\left( \begin{bmatrix} s \m{I}_n & -\m{I}_n \nonumber\\
    \mcl{L} & (s+\alpha) \m{I}_n + \beta \mcl{L} 
    \end{bmatrix} \right) \\
    &= {\rm det}\big( (s^2 \m{I}_n + \alpha s) \m{I}_n + (1+\beta s) \mcl{L} \big) \label{eq:c3_N}
\end{align}
Các nghiệm của phương trình \eqref{eq:c3_N} có thể thu được bằng cách giải phương trình $s^2 + \alpha s+ \lambda_i(1+\beta s) = 0$, với $i=1,\ldots,n$. Nghiệm của phương trình này được cho bởi:
\begin{equation}
    s_{i\pm} = \frac{-\beta\lambda_i - \alpha \pm \sqrt{(\beta\lambda_i + \alpha)^2 - 4 \lambda_i}}{2}, \label{eq:c3_eig1}
\end{equation}
với $s_{i+}$ và $s_{i-}$ là các giá trị riêng của $\m{N}$ ứng với $\lambda_i$. Khác với $\m{M}$, mỗi giá trị riêng tại $0$ của $\mcl{L}$ sẽ cho tương ứng một giá trị riêng tại $0$ của ma trận $\m{N}$. Giả sử $\lambda_1 = 0$, ta có $s_{1+} = 0$ và $s_{1-} = - \alpha$.

\begin{theorem}
Với luật đồng thuận~\eqref{eq:c3_double_u1}, 
\begin{align*}
\m{x}(t) &\to \m{1}_n (\bmm{\gamma}^\top \m{x}(0)) + \frac{1}{\alpha} \m{1}_n (\bmm{\gamma}^\top \m{y}(0)),\\
    \m{y}(t) &\to \m{0}_{n},
\end{align*}
khi $t \to +\infty$ khi và chỉ khi $\m{N}$ có một giá trị riêng đơn bằng 0 và mọi giá trị riêng khác của $\m{N}$ có phần thực âm.
\end{theorem}

\begin{proof}
(Điều kiện đủ). Chứng minh này tương tự như chứng minh điều kiện đủ của Định lý \ref{c3:thm_db} chỉ khác rằng với ma trận $\m{N}$ thì $\bmm{\eta}_1 = [\bmm{\gamma}_1^\top, (1/\alpha)\bmm{\gamma}_1^\top]^\top$, $\m{w}_1 = [\m{1}_n^\top, \m{0}^\top_n]^\top$ và
\begin{equation*}
    \m{J} = \begin{bmatrix}
    0 & \m{0}_{2n-1}^\top\\
    \m{0}_{2n-1} & \m{J}'
    \end{bmatrix}
\end{equation*}
trong đó $\m{J}'$ là thành phần của ma trận Jordan $\m{J}$ của $\m{N}$ ứng với $(2n-1)$ giá trị riêng có phần thực âm.

(Điều kiện cần). Nếu như $\m{x}(t) \to \m{1}_n \bmm{\gamma}^\top \m{x}(0) + \frac{1}{\alpha} \m{1}_n \bmm{\gamma}^\top \m{y}(0)$ và $\m{y}(t) \to \m{0}_n$ khi $t \to +\infty$, ta biết rằng $\lim_{t \to +\infty}\m{P} \mathtt{e}^{\m{J}t}\m{P}^{-1}$ có hạng bằng 1. Điều này dẫn tới việc $\lim_{t \to +\infty} \mathtt{e}^{\m{J}t}$ cũng có hạng bằng 1. Nếu điều kiện đủ không thỏa mãn, tương tự như chứng minh điều kiện cần của Định lý \ref{c3:thm_db}, ta có thể thấy rằng $\lim_{t \to +\infty} \mathtt{e}^{\m{J}t}$ phải có hạng ít nhất là 2, và đây là một điều vô lý.
\end{proof}

\begin{example}\label{eg:3.2}
Xét hệ gồm bốn tác tử tích phân bậc hai tương tác qua một đồ thị chu trình hữu hướng (Hình~\ref{fig:VD3.2_graph}) với luật đồng thuận:
\begin{align}
    \dot{\m{p}}_i &= \m{v}_i, \nonumber\\
    \dot{\m{v}}_i &= \m{u}_i = -k_1 \sum_{j\in N_i}(\m{p}_i - \m{p}_j) - k_2 \m{v}_i,\,i=1,2,3,4, \label{eq:c3_VD3.2}
\end{align}
trong đó $k_1 = 1$, $k_2 = 3$, $\m{p}_i$ và $\m{v}_i \in \mb{R}^2$. 
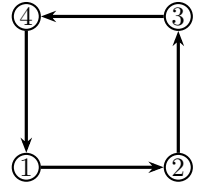
\begin{SCfigure}[][t!]
  \caption{Đồ thị chu trình hữu hướng $C_4$ mô tả luồng thông tin giữa bốn tác tử.\label{fig:VD3.2_graph}}
  \hspace{5.5cm}
\begin{tikzpicture}[
roundnode/.style={circle, draw=black, thick, minimum size=2mm,inner sep= 0.25mm},
squarednode/.style={rectangle, draw=black, thick, minimum size=3.5mm,inner sep= 0.25mm},
]
    \node[roundnode] (u1) at (0,0) { $1$}; %
    \node[roundnode] (u2) at (2,0) { $2$};%
    \node[roundnode] (u3) at (2,2) { $3$};%
    \node[roundnode] (u4) at (0,2) { $4$};%

    \draw [very thick,-{Stealth[length=2mm]}]
    (u1) edge [bend left=0] (u2)
    (u2) edge [bend left=0] (u3)
    (u3) edge [bend left=0] (u4)
    (u4) edge [bend left=0] (u1)
    ;
\end{tikzpicture}
\end{SCfigure}

Kết quả mô phỏng được cho như trong Hình~\ref{fig:VD3.2}. Dễ thấy $\m{v}_i \to \m{0}_2$ và vị trí $\m{p}_i$ các tác tử hội tụ tới cùng một điểm trên mặt phẳng (trạng thái đồng thuận). Do mô hình động học bậc nhất và bậc hai có thể mô tả đơn giản các robot di chuyển trong không gian (2 chiều hay 3 chiều), thuật toán đồng thuận là một giải pháp đơn giản cho bài toán hội ngộ (rendezvous problem) \cite{Lin2003CDC}\index{bài toán!hội ngộ}. Ở chương \ref{chap:formation}, một biến thể của thuật toán đồng thuận \eqref{eq:c3_VD3.2} sẽ được áp dụng trong điều khiển đội hình dựa trên vị trí tương đối.
\begin{figure}[th!]
    \centering
    \subfloat[Vận tốc theo trục $x$]{\includegraphics[width=.32\textwidth]{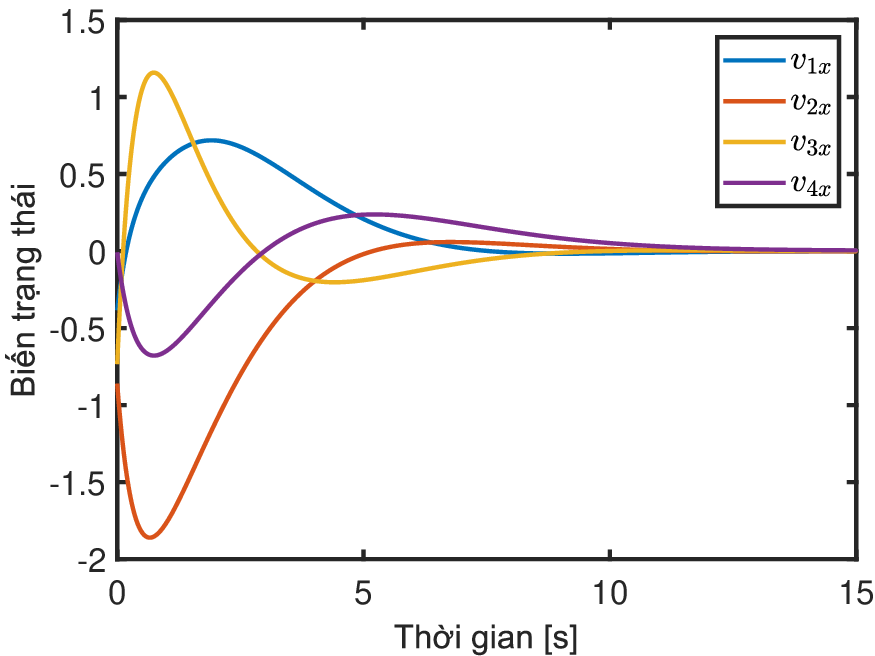}} \hfill
    \subfloat[Vận tốc theo trục $y$]{\includegraphics[width=.32\textwidth]{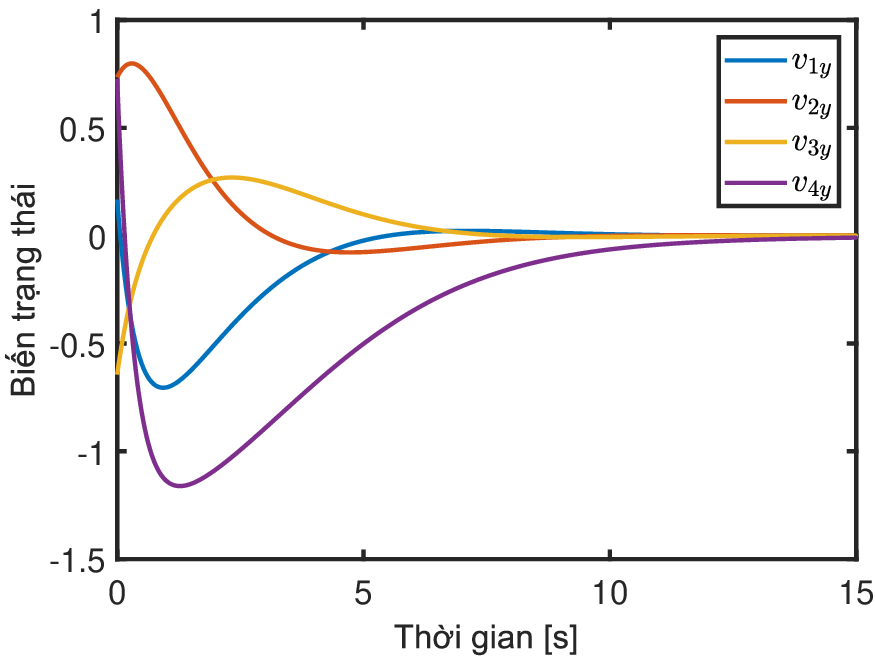}}\hfill
    \subfloat[Quỹ đạo của 4 tác tử]{\includegraphics[width=.32\textwidth]{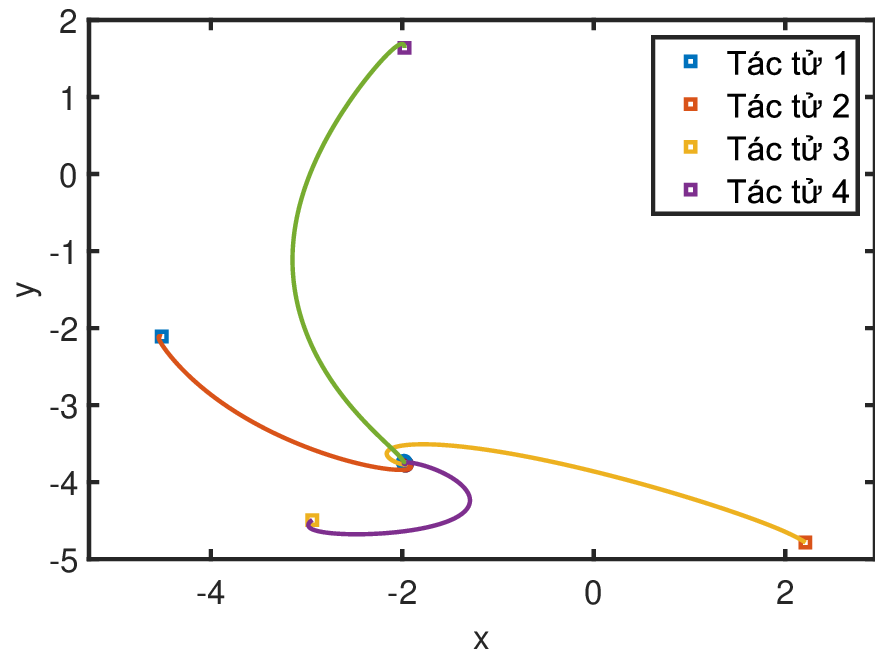}}
    \caption{Mô phỏng hệ bốn tác tử với luật đồng thuận~\eqref{eq:c3_VD3.2}.\label{fig:VD3.2}}
\end{figure}
\end{example}

\section{Hệ đồng thuận gồm các tác tử có mô hình tuyến tính liên tục}
\subsection{Thuật toán đồng bộ hóa trực tiếp}
Xét hệ gồm $n$ tác tử đồng nhất được mô tả bởi phương trình trạng thái dạng tuyến tính
\begin{align}
    \dot{\m{x}}_i = \m{A} \m{x}_i + \m{B}\m{u}_i,\, i=1,\ldots, n,
\end{align}
trong đó $\m{A}\in \mb{R}^{d\times d}$ là ma trận hệ thống, $\m{B} \in \mb{R}^{d\times p}$ là ma trận điều khiển, $\m{x}_i \in \mb{R}^d$ là vector biến trạng thái, và $\m{u}_i \in \mb{R}^p$ là tín hiệu điều khiển của tác tử $i$. Ta viết lại hệ dưới dạng ma trận như sau:
\begin{align} \label{eq:c3_linear_agent_model}
    \dot{\m{x}} = (\m{I}_n \otimes \m{A})\m{x} + (\m{I}_n \otimes \m{B})\m{u},
\end{align}
với $\m{x} = [\m{x}_1^\top, \ldots, \m{x}_n^\top]^\top = \text{vec}(\m{x}_1,\ldots,\m{x}_n) \in \mb{R}^{dn}$ và $\m{u} = [\m{u}_1^\top, \ldots, \m{u}_n^\top]^\top = \text{vec}(\m{u}_1,\ldots,\m{u}_n) \in \mb{R}^{pn}$. 

Thuật toán đồng thuận cho hệ được thiết kế có dạng:
\begin{align} \label{eq:chap3-consensus-linear-system}
    \m{u}_i = -\m{K}_1 \m{x}_i - \m{K}_2 \sum_{j\in N_i}(\m{x}_i - \m{x}_j),
\end{align}
trong đó $\m{K}_1, \m{K}_2 \in \mb{R}^{p \times d}$  là các ma trận trọng số hằng cần thiết kế. Sử dụng ma trận Laplace $\mcl{L}$, hệ gồm $n$ tác tử dưới luật đồng thuận \eqref{eq:chap3-consensus-linear-system} có thể biểu diễn dưới dạng
\begin{align}
    \dot{\m{x}} &= \big(\m{I}_n \otimes (\m{A}-\m{B}\m{K}_1)\big)\m{x} - (\mcl{L} \otimes \m{B}\m{K}_2)\m{x} \nonumber\\
    &=\big(\m{I}_n \otimes (\m{A}-\m{B}\m{K}_1) - (\mcl{L} \otimes \m{B}\m{K}_2) \big)\m{x}. \label{eq:chap3-consensus-linear-system1}
\end{align}
Để đơn giản, ta giả sử $G$ là vô hướng, liên thông. Khi đó, tồn tại $\m{P}$ là ma trận trực giao thỏa mãn \[\m{P}^{\top} \mcl{L} \m{P} = \m{\Lambda} = \text{diag}(\lambda_1(\mcl{L}),\ldots,\lambda_n(\mcl{L})).\] 
Đặt $\m{z} = (\m{P}^{\top}\otimes\m{I}_d)\m{x}$ thì hệ \eqref{eq:chap3-consensus-linear-system1} viết cho biến $\m{z} = [\m{z}_1^\top, \ldots, \m{z}_n^\top]^\top$ có dạng
\begin{align}
    \dot{\m{z}} =\big(\m{I}_n \otimes (\m{A}-\m{B}\m{K}_1) - (\m{\Lambda} \otimes \m{B}\m{K}_2) \big)\m{z}. \label{eq:chap3-consensus-linear-system2}
\end{align}
Chú ý rằng $\lambda_1(\mcl{L}) = 0$ ứng với vector riêng $\m{p}_1 = \frac{1}{\sqrt{n}} \m{1}_n$ nên $\dot{\m{z}}_1 = (\m{A}-\m{B}\m{K}_1) \m{z}_1$ không bị ảnh hưởng bởi luật đồng thuận. Với mỗi vector riêng ứng với một giá trị riêng khác 0, ta có $n-1$ hệ con
\begin{align}
\dot{\m{z}}_k = (\m{A}-\m{B}\m{K}_1 - \lambda_k \m{B}\m{K}_2 ) \m{z}_k,~k=2, \ldots, n. \label{eq:chap3-consensus-linear-system3}
\end{align}
Định lý \ref{c3:thm3.6_linear_consensus} mô tả quá trình đồng thuận của hệ \eqref{eq:chap3-consensus-linear-system1}.
\begin{theorem}\label{c3:thm3.6_linear_consensus} Xét hệ đồng thuận gồm $n$ tác tử tuyến tính \eqref{eq:c3_linear_agent_model} với đồ thị tương tác $G$ vô hướng, liên thông. Với thuật toán đồng thuận \eqref{eq:chap3-consensus-linear-system}, vector biến trạng thái $\m{x}(t)$ của hệ tiệm cận tới không gian đồng thuận khi và chỉ khi $n-1$ hệ con \eqref{eq:chap3-consensus-linear-system3} ổn định tiệm cận.
\end{theorem}

\begin{proof}
(Điều kiện đủ) Giả sử $n-1$ hệ con \eqref{eq:chap3-consensus-linear-system3} là ổn định tiệm cận, các ma trận $\m{A}-\m{B}\m{K}_1-\lambda_k\m{B}\m{K}_2$ là Hurwitz. Nghiệm của \eqref{eq:chap3-consensus-linear-system1} được cho dưới dạng:
\begin{align}
    \m{x}(t) &=  \mathtt{e}^{(\m{I}_n \otimes (\m{A}-\m{B}\m{K}_1) - (\mcl{L} \otimes \m{B}\m{K}_2))t} \m{x}(0) \nonumber\\
    &= (\m{P} \otimes \m{I}_d) \mathtt{e}^{(\m{I}_n \otimes (\m{A}-\m{B}\m{K}_1) - (\m{\Lambda} \otimes \m{B}\m{K}_2))t} (\m{P}^\top \otimes \m{I}_d) \m{x}(0) \nonumber\\
    &=\bar{\m{P}} \begin{bmatrix}
    \mathtt{e}^{(\m{A}-\m{B}\m{K}_1)t} & \m{0}\\
    \m{0} & \mathtt{e}^{(\m{I}_{n-1}\otimes (\m{A}-\m{B}\m{K}_1)-\text{diag}(\lambda_2, \ldots, \lambda_n)\otimes \m{B}\m{K}_2)t}
    \end{bmatrix}\bar{\m{P}}^\top \m{x}(0),
\end{align}
trong đó $\m{P} = [\m{p}_1, \ldots, \m{p}_n]$, $\bar{\m{P}} = \m{P}\otimes \m{I}_d$, $\m{p}_1 = \frac{1}{\sqrt{n}}\m{1}_n$. Do $\lim_{t\to +\infty }\mathtt{e}^{(\m{A}-\m{B}\m{K}_1- \lambda_i \m{B}\m{K}_2)t} = \m{0}_{d\times d}$, ta có:
\begin{align*}
(\m{P} &\otimes \m{I}_d) \mathtt{e}^{(\m{I}_n \otimes (\m{A}-\m{B}\m{K}_1) - (\m{\Lambda} \otimes \m{B}\m{K}_2))t} (\m{P}^\top \otimes \m{I}_d) = \sum_{i=1}^n (\m{p}_i\otimes \m{I}_d) \mathtt{e}^{(\m{A}-\m{B}\m{K}_1- \lambda_i \m{B}\m{K}_2)t} (\m{p}_i^\top \otimes \m{I}_d),
\end{align*}
nên
\begin{align}
&\left|\left|\sum_{i=1}^n (\m{p}_i\otimes \m{I}_d) \mathtt{e}^{(\m{A}-\m{B}\m{K}_1- \lambda_i \m{B}\m{K}_2)t} (\m{p}_i^\top \otimes \m{I}_d) - \frac{1}{n}(\m{1}_n\otimes \m{I}_d) \mathtt{e}^{(\m{A}-\m{B}\m{K}_1)t} (\m{1}_n^\top\otimes \m{I}_d)\right| \right| \nonumber\\
&=\left|\left|\sum_{i=2}^n(\m{p}_i\otimes \m{I}_d) \mathtt{e}^{(\m{A}-\m{B}\m{K}_1- \lambda_i \m{B}\m{K}_2)t} (\m{p}_i^\top \otimes \m{I}_d)\right| \right| \to 0,
\end{align}
khi $t\to +\infty$. Từ đây suy ra
\begin{align}
    \left|\left|\m{x}(t) - \frac{1}{n}(\m{1}_n\m{1}_n^\top \otimes \m{I}_d) \mathtt{e}^{(\m{A}-\m{B}\m{K}_1)t} \m{x}(0)\right|\right| = 0,
\end{align}
hay $\m{x}_i(t)$ tiệm cận tới 
\begin{align}\label{eq:c3_consensus_trajectory}
\m{x}^*(t)=\frac{1}{n} \mathtt{e}^{(\m{A}-\m{B}\m{K}_1)t} \sum_{j=1}^n\m{x}(0),\, \forall i=1,\ldots, n,
\end{align}
khi $t\to +\infty$. Chú ý rằng $\m{x}^*(t)$ ở phương trình \eqref{eq:c3_consensus_trajectory} là nghiệm của phương trình trạng thái
$\dot{\m{q}}(t)=(\m{A}-\m{B}\m{K}_1){\m{q}}(t)$, với điều kiện đầu ${\m{q}}(0)=\frac{1}{n}\sum_{j=1}^n\m{x}(0)$.

(Điều kiện cần) Giả sử hệ tiến tới đồng thuận, tồn tại quỹ đạo $\m{x}^*(t)$ sao cho $\lim_{t\to+\infty} \|\m{x}(t) - \m{1}_n \otimes \m{x}^*(t)\| = 0$. Do $ \m{1}_n^\top \m{p}_k = 0, \forall k = 2, \ldots, n$, $\m{z}_i = (\m{p}_i^\top \otimes \m{I}_d) \m{x}$ thì $\|\m{z} - (\m{p}_i^\top \m{1}_n) \otimes \m{x}^*(t) \| \to {0}$, khi $t \to +\infty, \forall i=2, \ldots, n$. Điều này chứng tỏ $n-1$ hệ con \eqref{eq:chap3-consensus-linear-system3} là ổn định tiệm cận, hay mọi ma trận $(\m{A}-\m{B}\m{K}_1 - \lambda_k \m{B}\m{K}_2 ),~k=2, \ldots, n,$ đều là các ma trận Hurwitz.
\end{proof}

Lưu ý rằng để hệ đạt đồng thuận tại một điểm trong không gian thì ta còn cần thêm điều kiện tồn tại giới hạn hữu hạn $\lim_{t\to+\infty} \mathtt{e}^{(\m{A}-\m{B}\m{K}_1)t}$. Trong trường hợp tổng quát, hệ không đồng thuận về một điểm cố định. Các biến trạng thái thỏa mãn $\m{x}_i(t)-\m{x}_j(t)\to \m{0}_d$ và cùng tiệm cận tới một nghiệm (biến thiên) của hệ $\dot{\m{q}}(t)=(\m{A}-\m{B}\m{K}_1){\m{q}}(t)$. Chúng ta sử dụng thuật ngữ đồng bộ hóa biến trạng thái để mô tả hiện tượng xảy ra với hệ đồng thuận \eqref{eq:chap3-consensus-linear-system1}.\index{đồng bộ hóa!biến trạng thái} Nếu các tác tử có các ma trận trạng thái khác nhau về giá trị và về số chiều, việc đồng bộ biến trạng thái là không thực hiện được. Khi đó, chúng ta có thể định nghĩa một số biến đầu ra của $n$ tác tử và thiết kế luật đồng thuận để các biến đầu ra của $n$ tác tử tiệm cận tới cùng một quĩ đạo đồng thuận \cite{Wieland2011internal,Su2011cooperative}. 

\begin{example} \label{VD:3.3}
Xét hệ gồm $n=6$ tác tử trong $\mb{R}^3$ với đồ thị tương tác là chu trình $C_6$ ở Hình~\ref{fig:VD3.3_graph}. 
\begin{SCfigure}[][ht]
  \caption{Đồ thị tương tác $C_6$ trong Ví dụ~\ref{VD:3.3}. \label{fig:VD3.3_graph}}
  \hspace{6cm}
  \begin{tikzpicture}[>=latex, scale=0.45] 
        \def \n {6}
        \def \radius {3.15cm}
        \def \margin {6} 
        \foreach \s in {1,...,\n}
        {
        \node[draw, circle,color=black] at ({360/\n * (\s - 1)}:\radius) {};
        \draw[-,color=black] ({360/\n * (\s - 1)+\margin}:\radius) 
        arc ({360/\n * (\s - 1)+\margin}:{360/\n * (\s)-\margin}:\radius);
        }
  \end{tikzpicture}
\end{SCfigure}
Các ma trận được chọn như sau:
\[\m{A} = \begin{bmatrix}
0 & 1 & 0\\0 & 0 & 1\\0 & 0 & 0
\end{bmatrix}, ~ \m{B} = \begin{bmatrix}
0\\0\\1
\end{bmatrix}, ~ \m{K}_1 = [1, 1, 1],~ \m{K}_2 = 400\cdot[1, 7, 1]. \]
Khi đó, ma trận $\m{A}-\m{B}\m{K}_1$ có các giá trị riêng tại $-1+0\jmath, 0\pm \jmath$ và $\m{A}-\m{B}\m{K}_1-\lambda_k \m{B}\m{K}_2,~k=2,3,4$ là các ma trận Hurwitz. 

Kết quả mô phỏng hệ được cho trên Hình~\ref{fig:VD3.3}. Các biến trạng thái $\m{x}_i$ tiệm cận về một nghiệm của phương trình vi phân $\dot{\m{r}}=(\m{A}-\m{B}\m{K}_1)\m{r}$. Do $\m{A}-\m{B}\m{K}_1$ có một cặp nghiệm phức liên hợp, quĩ đạo đồng thuận có dạng dao động điều hòa.
\begin{SCfigure}[][th!] 
\caption{Mô phỏng hệ đồng thuận ở Ví dụ~\ref{VD:3.3}. Các biến trạng thái ${x}_{ik} - {x}_{jk} \to {0}$, $i=1,\ldots,6$, $k=1,2,3$, khi $t \to +\infty$.\label{fig:VD3.3}}
\hspace{1.5cm}
\includegraphics[width=.6\linewidth]{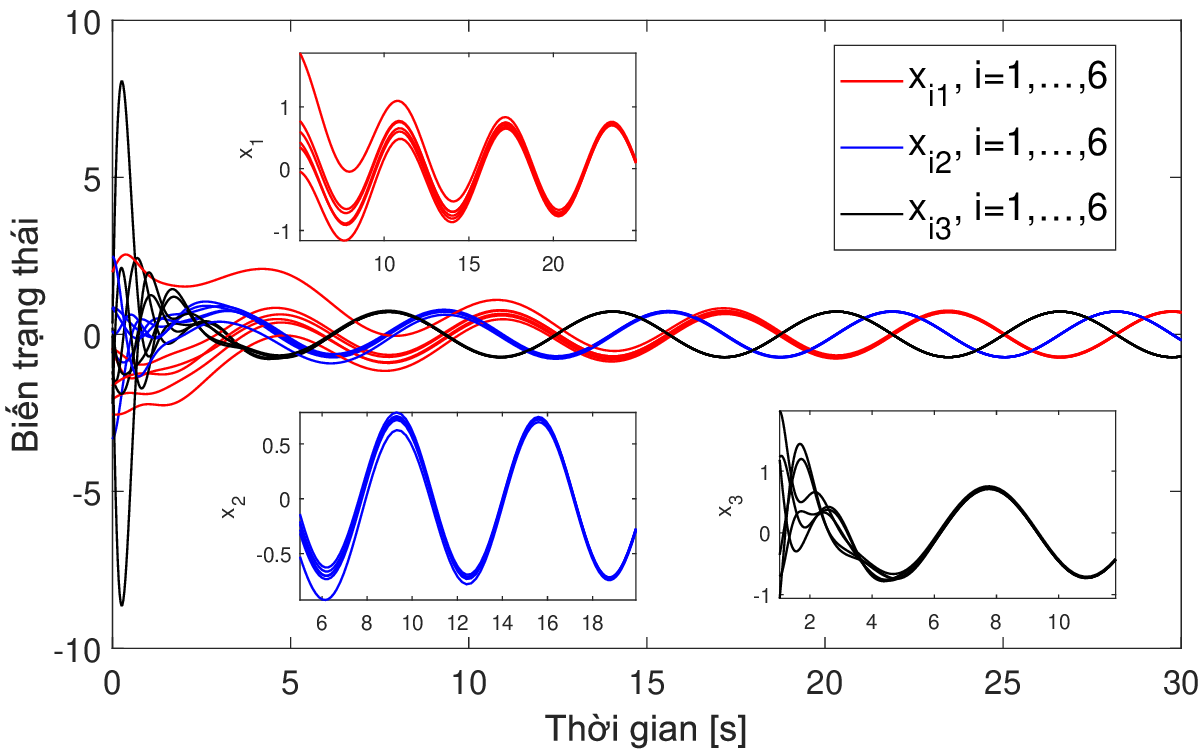}
\end{SCfigure}
\end{example}

\subsection{Thuật toán đồng bộ hóa trạng thái gián tiếp}
\label{chap3-sec3-1}
Mục này giới thiệu một phương pháp thiết kế thuật toán đồng thuận cho hệ đa tác tử tuyến tính dựa trên một số biến trung gian, thu được nhờ 
bộ quan sát trạng thái. Các giả thiết chính ở mục này bao gồm:
\begin{itemize}
    \item Mô hình của mỗi tác tử được cho bởi:
    \begin{align}
        \dot{\m{x}}_i &= \m{A} \m{x}_i + \m{B} \m{u}_i \nonumber\\
        \m{y}_i &= \m{C}\m{x}_i, \,i=1,\ldots,n, \label{eq:c3-agent-model}
    \end{align}
trong đó $\m{x}_i = [x_{i,1}, \ldots, x_{i,d}]^\top \in \mb{R}^d$, $\m{u}_i \in \mb{R}^m$, $\m{y}_i \in \mb{R}^p$ lần lượt là các vector biến trạng thái, tín hiệu điều khiển, và biến đo/biến đầu ra của tác tử $i$.\footnote{Trong bài toán đồng bộ hóa hệ đa tác tử tuyến tính, chúng ta phân biệt giữa biến đo được với biến đầu ra của mỗi tác tử. Biến đầu ra ở đây có thể là biến thứ cấp thu được từ các biến đo hoặc biến dùng để ghép nối giữa tác tử với mạng.} 
    \item $(\m{A},\m{B})$ là ổn định được\footnote{stabilizable} và $(\m{C},\m{A})$ là có thể dò được\footnote{detectable}, nghĩa là tồn tại ma trận $\m{K}\in \mb{R}^{m \times n}$ và $\m{H} \in \mb{R}^{p \times n}$ sao cho $(\m{A} + \m{B}\m{K})$ và $(\m{A} + \m{H}\m{C})$ là các ma trận Hurwitz.
    \item Đồ thị $G$ mô tả tương tác giữa các tác tử trong hệ là có gốc ra. Mỗi tác tử có thể đo được các biến đầu ra tương đối với các tác tử láng giềng.
    \item Ma trận Laplace $\mcl{L}$ của $G$ có $l$ giá trị riêng phân biệt $\lambda_i$ với bội $n_i$ ($i=1,\ldots,l$ và $\sum_{i=1}^l n_i = n$), và do đó có thể biểu diễn ma trận này ở dạng Jordan $\mcl{L} = \m{P}\m{J}\m{P}^{-1}$, trong đó $\m{J}=\text{blkdiag}(0,\m{J}(\lambda_2),\ldots,\m{J}(\lambda_l))$.
\end{itemize}
Như vậy trong bài toán này, các tác tử không có được thông tin đầy đủ về trạng thái của mình, cũng như không đo được biến trạng thái tương đối như trong phần lớn các bài toán đồng thuận đã xét ở mục trước.

\subsubsection{Sử dụng bộ quan sát Luenberger và biến trạng thái phụ}
\begin{SCfigure}[][th!]
  \caption{Sơ đồ khối mô tả hệ $n$ tác tử tuyến tính với thuật toán đồng thuận \eqref{eq:chap3_output_synch1}--\eqref{eq:chap3_output_synch4}.}
  \label{fig:chap3_output_synchronization}
  \hspace{4.5cm}
  \includegraphics[height=7.5cm]{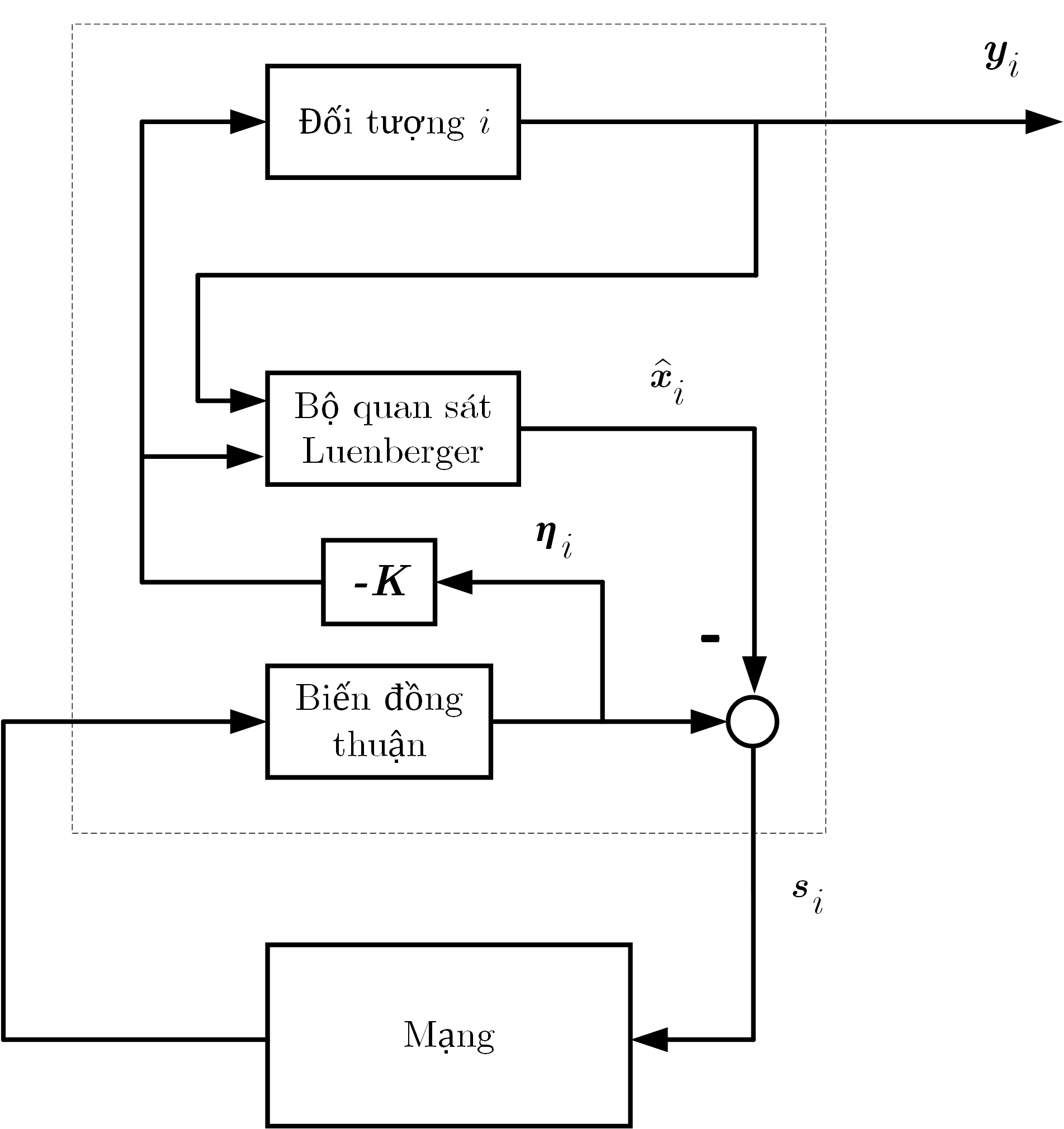}
\end{SCfigure}
Xét thuật toán đồng bộ hóa biến trạng thái sau đây \cite{Luca2009autom}:
\begin{align}
   \dot{\bmm{\eta}}_i &= \m{A}\bmm{\eta}_i + \m{B}\m{u}_i + \m{H}(\hat{\m{y}}_i - \m{y}_i) - \m{B}\m{K} \sum_{j\in N_i}(({\bmm{\eta}}_j-{\bmm{\eta}}_i)-(\hat{\m{x}}_j-\hat{\m{x}}_i)), \label{eq:chap3_output_synch1} \\
    \dot{\hat{\m{x}}}_i &= \m{A}\hat{\m{x}}_i + \m{B}\m{u}_i + \m{H}(\hat{\m{y}}_i - \m{y}_i),~\hat{\m{y}}_i=\m{C} \hat{\m{x}}_i, \label{eq:chap3_output_synch2}\\
   \dot{\m{x}}_i &= \m{A}\m{x}_i + \m{B}\m{u}_i,~\m{y}_i = \m{C}\m{x}_i, \label{eq:chap3_output_synch3}\\
    \m{u}_i & = \m{K} \bmm{\eta}_i, \label{eq:chap3_output_synch4}
\end{align}
trong đó~\eqref{eq:chap3_output_synch2} là bộ quan sát Luenberger thiết kế cho hệ~\eqref{eq:chap3_output_synch3}, \eqref{eq:chap3_output_synch4} là luật điều khiển thiết kế để làm ổn định $\bmm{\eta}_i$ trong \eqref{eq:chap3_output_synch1}, và \eqref{eq:chap3_output_synch1} là giao thức đồng thuận với biến phụ $\m{s}_i = \hat{\m{x}}_i - \bmm{\eta}_i$. Giả sử rằng $\m{A}+\m{B}\m{K}$ và $\m{A}+\m{H}\m{C}$ là các ma trận Hurwitz. Hình~\ref{fig:chap3_output_synchronization} mô tả cấu trúc đồng bộ hóa \eqref{eq:chap3_output_synch1}--\eqref{eq:chap3_output_synch4}.

Đặt $\m{x} = [\m{x}_1^\top, \ldots, \m{x}_n^\top]^\top$, $\hat{\m{x}} = [\hat{\m{x}}_1^\top, \ldots, \hat{\m{x}}_n^\top]^\top$, $\m{e} = \hat{\m{x}} - \m{x}$, ${\bmm{\eta}} = [{\bmm{\eta}}_1^\top, \ldots, {\bmm{\eta}}_n^\top]^\top$, và $\m{s} = \hat{\m{x}} - \bmm{\eta}$. Ta có thể biểu diễn hệ dưới dạng:
\begin{align}
    \dot{\bmm{\eta}}&= (\m{I}_n \otimes (\m{A}+\m{B}\m{K}))\bmm{\eta} + (\m{I}_n \otimes \m{H}\m{C}) \m{e}  - (\mcl{L} \otimes \m{B}\m{K})\m{s}\label{eq:chap3_output_synch8}\\
    \dot{\m{s}} &= (\m{I}_n \otimes \m{A})\m{s} + (\mcl{L} \otimes \m{B}\m{K})\m{s} \label{eq:chap3_output_synch6}\\
    \dot{\m{e}}&=(\m{I}_n \otimes (\m{A}+\m{H}\m{C}))\m{e}. \label{eq:chap3_output_synch7}
\end{align}
Do $\m{A}+\m{H}\m{C}$ là ma trận Hurwitz, từ \eqref{eq:chap3_output_synch7}, $\|\m{e}(t)\|$ hội tụ theo hàm mũ tới 0. Hệ~\eqref{eq:chap3_output_synch6} là một hệ đồng thuận theo biến $\m{s}$. Thực hiện phép đổi biến
\begin{align}
    \m{r} = (\m{P}^{-1}\otimes \m{I}_d)\m{s},
\end{align}
trong đó $\m{P}^{-1}\m{L}\m{P} = \m{J} = \text{blkdiag}(0, \m{J}(\lambda_2), \ldots, \m{J}(\lambda_s))$, ta thu được:
\begin{align}
    \dot{\m{r}} & = (\m{P}^{-1}\otimes \m{I}_d) (\m{I}_n \otimes \m{A} + \mcl{L} \otimes \m{B}\m{K}) (\m{P}\otimes \m{I}_d)\m{r} \\
    &= (\m{I}_n \otimes \m{A} + \m{P}^{-1}\mcl{L}\m{P} \otimes \m{B}\m{K})\m{r} \nonumber\\
    &= (\m{I}_n \otimes \m{A} + \m{J} \otimes \m{B}\m{K}) \m{r}. \label{eq:chap3_output_synch9}
\end{align}
Từ đây, ta viết lại hệ~\eqref{eq:chap3_output_synch9} thành $l$ hệ con cho bởi
\begin{align} \label{eq:chap3_output_synchx}
    \dot{\m{r}}_{[i:i+n_i]} = (\m{A} \otimes \m{I}_{n_s} + \m{J}(\lambda_i) \otimes \m{B}\m{K} )\m{r}_{[i:i+n_i]},~i=1,\ldots, l,
\end{align}
trong đó ${\m{r}}_{[i:i+n_i]}:=[\m{r}_{i}^\top,\ldots,\m{r}_{i+n_i}^\top]^\top$. Với ma trận $\m{K}$ được chọn sao cho $\m{A}+\lambda_i\m{B}\m{K},~i=2,\ldots,l$ là Hurwitz thì \eqref{eq:chap3_output_synchx} ổn định theo hàm mũ, $\m{r}_i(t) \to \m{0}_d$, $i=2, \ldots, n,$ khi $t\to +\infty$. Từ đây suy ra nghiệm $\m{s}(t)$ tiệm cận tới một nghiệm của phương trình vi phân $\dot{\m{s}}_0 = \m{A}\m{s}_0$. Cụ thể, ta có $\m{s}(t)$ tiệm cận tới
\begin{align}
    (\m{P}\otimes \m{I}_d)\begin{bmatrix}
    \mathtt{e}^{\m{A}t} \m{r}_1(0)\\
    \m{0}_d\\
    \vdots\\
    \m{0}_d
    \end{bmatrix} =  (\m{P}\otimes \m{I}_d) \begin{bmatrix}
    \mathtt{e}^{\m{A}t} \\
    \m{0}_d\\
    \vdots\\
    \m{0}_d
    \end{bmatrix} (\bm{\gamma}^\top\otimes\m{I}_d) \m{s}(0) 
    = \m{1}_n\otimes (\mathtt{e}^{\m{A}t} \bar{\m{s}}(0)),
\end{align}
khi $t\to +\infty$, trong đó $\bar{\m{s}}(0) = \sum_{i=1}^n\gamma_i \m{s}_i(0)$ là trung bình theo trọng số của các biến $\m{s}_i(0)$, $\bmm{\gamma}^\top \mcl{L} = \m{0}^\top_n$, và $\bmm{\gamma}^\top \m{1}_n = 1$. Như vậy, $(\mcl{L}\otimes \m{I}_d) \m{s}(t) \to \m{0}_{dn}$ khi $t\to +\infty$.

Xét hệ~\eqref{eq:chap3_output_synch8} với biến trạng thái $\bmm{\eta}$ và các tín hiệu bên ngoài\footnote{exogeneous input} $\m{d}_1(t) = (\m{I}_n\otimes \m{H}\m{C})\m{e}$ và $\m{d}_2(t) = (\mcl{L}\otimes \m{I}_d) \m{s}(t)$. Khi không có tác động bên ngoài $(\m{d}_1 = \m{d}_2 = \m{0}_{dn})$ thì hệ không bị kích thích\footnote{nominal system, unforced system} của \eqref{eq:chap3_output_synch9} cho bởi
\begin{equation}
    \dot{\bmm{\eta}} = (\m{I}_n \otimes (\m{A}+\m{B}\m{K})) \bmm{\eta}, \label{eq:chap3_output_synch10}
\end{equation}
với $\m{A}+\m{B}\m{K}$ Hurwitz là ổn định theo hàm mũ. Các tín hiệu bên ngoài thỏa mãn $\|\m{d}_1\| = \| (\m{I}_n\otimes \m{H}\m{C})\m{e}\| \to {0}$ và $\m{d}_2 = \|(\mcl{L}\otimes \m{I}_d) \m{s}\| \to 0$ khi $t\to +\infty$. Điều này chứng tỏ $\bmm{\eta}(t)$ tiệm cận tới một nghiệm của hệ không bị kích thích của \eqref{eq:chap3_output_synch10}, tức là $\bmm{\eta} \to \m{0}_{dn}$ khi $t\to +\infty$.

Cuối cùng, do $\m{x} = \bmm{\eta} + \m{s} - \m{e}$, mà $\m{e}\to \m{0}_{dn}$, $\bmm{\eta}\to \m{0}_{dn}$,  và $\m{s}$ tiệm cận tới không gian đồng thuận, ta có $\m{x}$ cũng tiệm cận tới không gian đồng thuận khi $t \to +\infty$,  $\m{x}_i(t) \to \mathtt{e}^{\m{A}t} \bar{\m{s}}(0) = \mathtt{e}^{\m{A}t} (\bar{\hat{\m{x}}}(0) - \bar{\bmm{\eta}}(0) )$.

\begin{example} \label{VD:3.4}
Xét hệ gồm 8 tác tử với luật đồng thuận dựa trên bộ quan sát trạng thái. Các ma trận hệ thống của mỗi tác tử $i$ được cho bởi:
\begin{align*}
\m{A} = \begin{bmatrix}
-1 & 1 & 0\\ 0 & -1.25 & 1 \\ 0 & -5.5625 & 1.25
\end{bmatrix},~\m{B} = \begin{bmatrix}
0\\ 0\\ 1
\end{bmatrix}, ~\text{ và }~ \m{C} = \begin{bmatrix}
1 & 1 & 1
\end{bmatrix}.
\end{align*}
\begin{SCfigure}[][h!]
\caption{Đồ thị mô tả tương tác giữa các tác tử trong Ví dụ~\ref{VD:3.4}.\label{fig:VD3.4_graph}}
\hspace{5cm}
\begin{tikzpicture}[
roundnode/.style={circle, draw=black, thick, minimum size=2mm,inner sep= 0.25mm},
squarednode/.style={rectangle, draw=black, thick, minimum size=3.5mm,inner sep= 0.25mm},
]
    \node[roundnode] (u1) at (2,0) { $1$}; %
    \node[roundnode] (u2) at (2.5,1.5) { $2$};%
    \node[roundnode] (u3) at (1.5,1.5) { $3$};%
    \node[roundnode] (u4) at (2.5,2.5) { $4$};%
    \node[roundnode] (u5) at (1.5,2.5) { $5$};%
    \node[roundnode] (u6) at (2,4) { $6$};%
    \node[roundnode] (u7) at (0,2) { $7$};%
    \node[roundnode] (u8) at (4,2) { $8$};%
    
    \draw [very thick,-{Stealth[length=2mm]}]
    (u1) edge [bend left=0] (u3)
    (u2) edge [bend left=0] (u1)
    (u2) edge [bend left=0] (u3)
    (u4) edge [bend left=0] (u2)
    (u5) edge [bend left=0] (u4)
    (u3) edge [bend left=0] (u5)
    (u3) edge [bend left=0] (u7)
    (u4) edge [bend left=0] (u8)
    (u5) edge [bend left=0] (u6)
    (u6) edge [bend left=0] (u4)
    (u7) edge [bend left=0] (u5)
    (u8) edge [bend left=0] (u2)
    ;
\end{tikzpicture}
\end{SCfigure}
và đồ thị tương tác của hệ là một đồ thị hữu hướng (không trọng số) biểu diễn ở Hình~\ref{fig:VD3.4_graph}.

Hai ma trận $\m{H}$ và $\m{K}$ của bộ quan sát và bộ điều khiển được thiết kế là
\begin{align*}
    \m{H} = -\begin{bmatrix}
        -0.0730 \\ 0.5255 \\ 7.5474
    \end{bmatrix},~
    \m{K} = - \begin{bmatrix}
     1 & -3.25 & 5
    \end{bmatrix}.
\end{align*}
Dễ dàng kiểm tra được $\m{A}$ có ba giá trị riêng $-1$, $0\pm \jmath 2$, $\m{A}+\m{B}\m{K}$ có giá trị riêng bội ba tại $-2+\jmath 0$, và $\m{I}_8\otimes \m{A}+ \m{L}\otimes \m{B}\m{K}$ có cặp giá trị riêng ảo liên hợp $0\pm \jmath 2$ và 22 giá trị riêng có phần thực âm.

Các điều kiện đầu $\m{x}(0)$, $\hat{\m{x}}(0)$, $\bmm{\eta}(0)$ được chọn ngẫu nhiên. Kết quả mô phỏng hệ với luật đồng thuận \eqref{eq:chap3_output_synch1}--\eqref{eq:chap3_output_synch4} được cho ở Hình~\ref{fig:VD3.4_Luenberger} và \ref{fig:VD3.4}. 

Hình~\ref{fig:VD3.4_Luenberger} biểu diễn biến trạng thái và biến ước lượng của tác tử 1. Dễ thấy $e_{1k} = \hat{x}_{1k}-x_{1k}$ hội tụ theo hàm mũ. 

\begin{SCfigure}[][th!]
\caption{Biến trạng thái và biến ước lượng (bộ quan sát Luenberger) của tác tử 1 trong Ví dụ~\ref{VD:3.4}.\label{fig:VD3.4_Luenberger}}
\hspace{3cm}
\includegraphics[width=.45\textwidth]{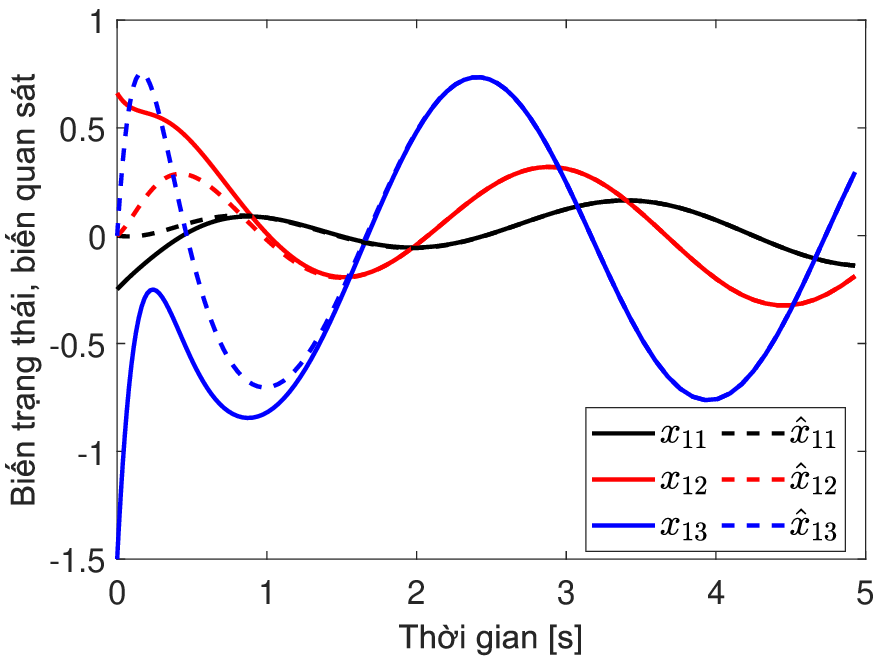}
\end{SCfigure}

Hình \ref{fig:VD3.4} (a)--(c) và (g)--(h) cho thấy các biến trạng thái $\m{x}_{ik}(t)$ và các biến trạng thái phụ ${s}_{ik}(t)$, $i=1,\ldots, 8$, dần đồng bộ tới một quĩ đạo (đồng bộ theo từng tọa độ $k=1,2,3$). Do ma trận $\m{A}$ có các giá trị riêng tại $-1+ \jmath 0, 0 \pm \jmath 2$ nên quĩ đạo đồng thuận (nghiệm của phương trình $\dot{\m{r}}=\m{A}\m{r}$) có dạng dao động điều hòa khi $t\to +\infty$.

Trong khi đó, các biến trạng thái phụ $\eta_{ik}$ hội tụ khi $t\to +\infty$ (Hình \ref{fig:VD3.4} (d)--(f)). Như vậy, kết quả mô phỏng phù hợp với phần phân tích được thực hiện ở mục này.

\begin{figure}[t!] 
\centering
\subfloat[]{\includegraphics[width=.32\textwidth]{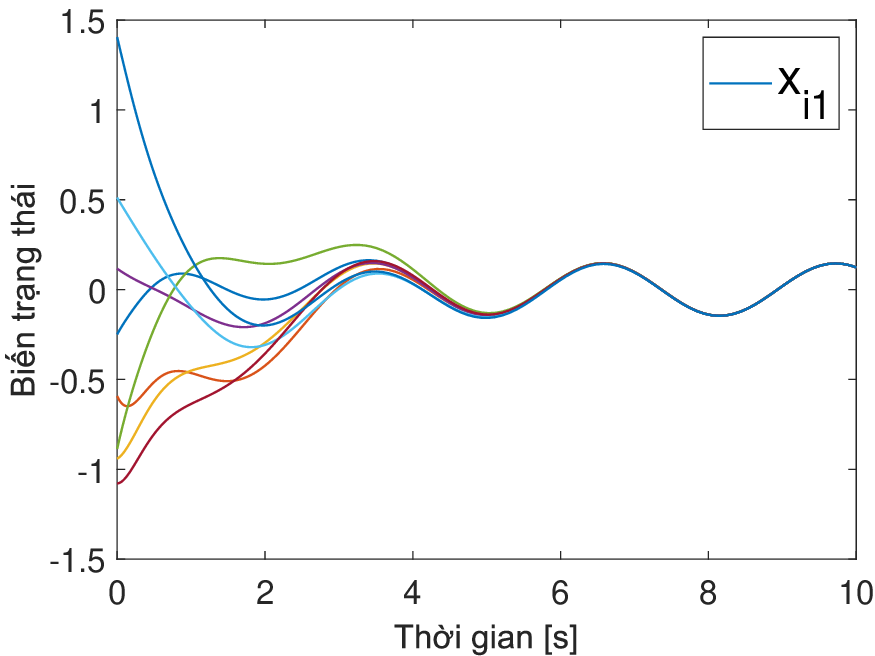}} \hfill
\subfloat[]{\includegraphics[width=.32\textwidth]{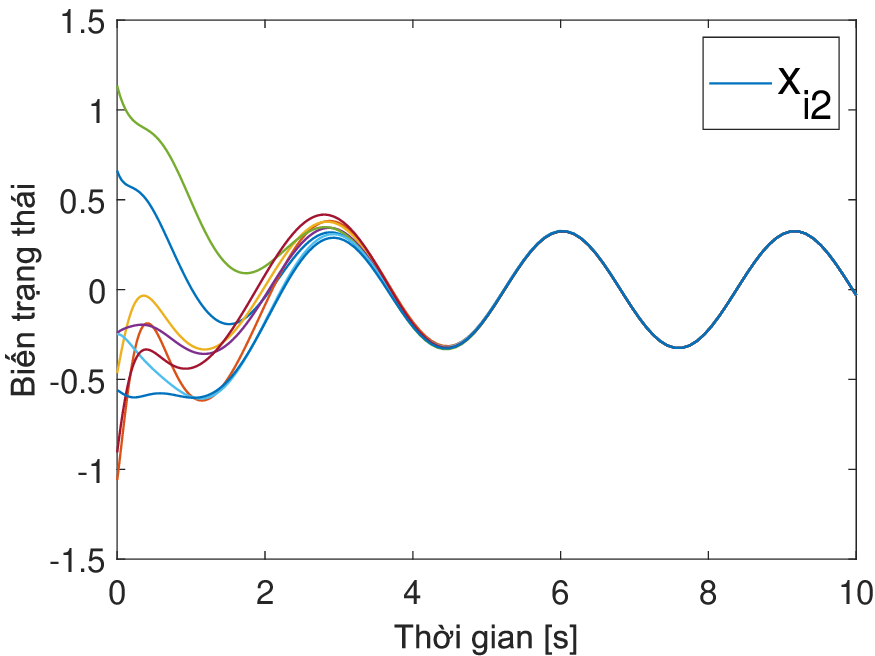}}\hfill
\subfloat[]{\includegraphics[width=.32\textwidth]{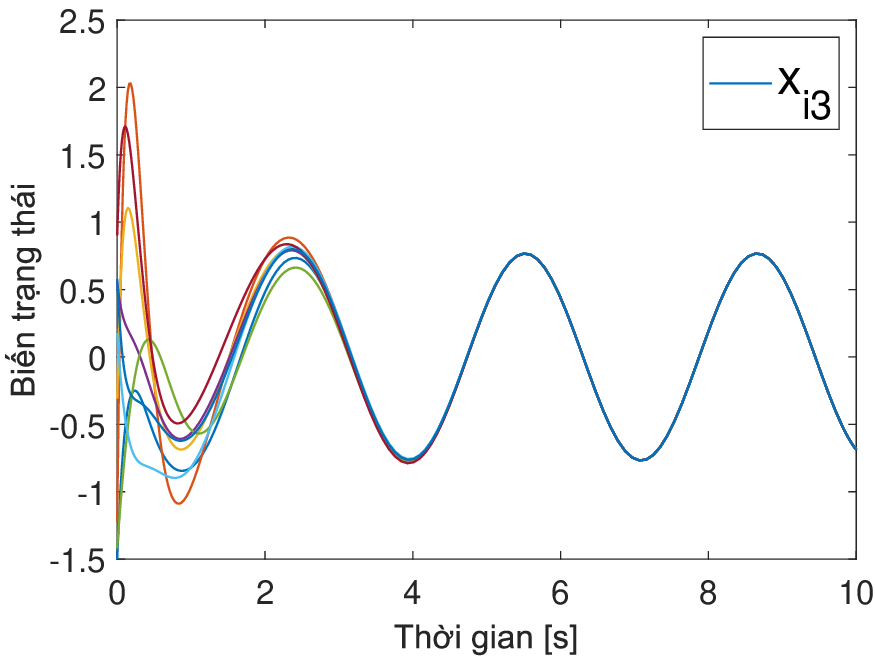}} \hfill
\subfloat[]{\includegraphics[width=.32\textwidth]{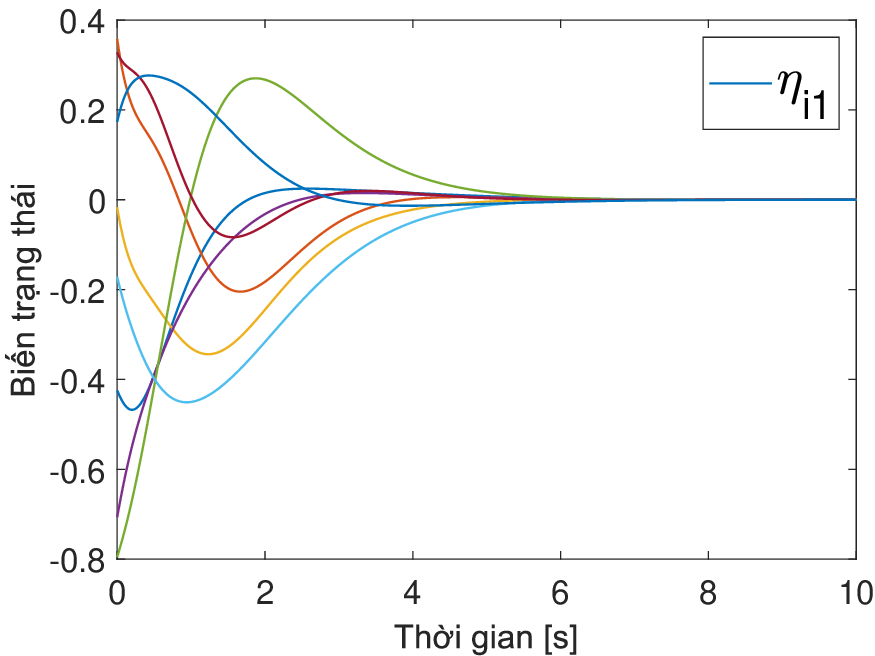}} \hfill
\subfloat[]{\includegraphics[width=.32\textwidth]{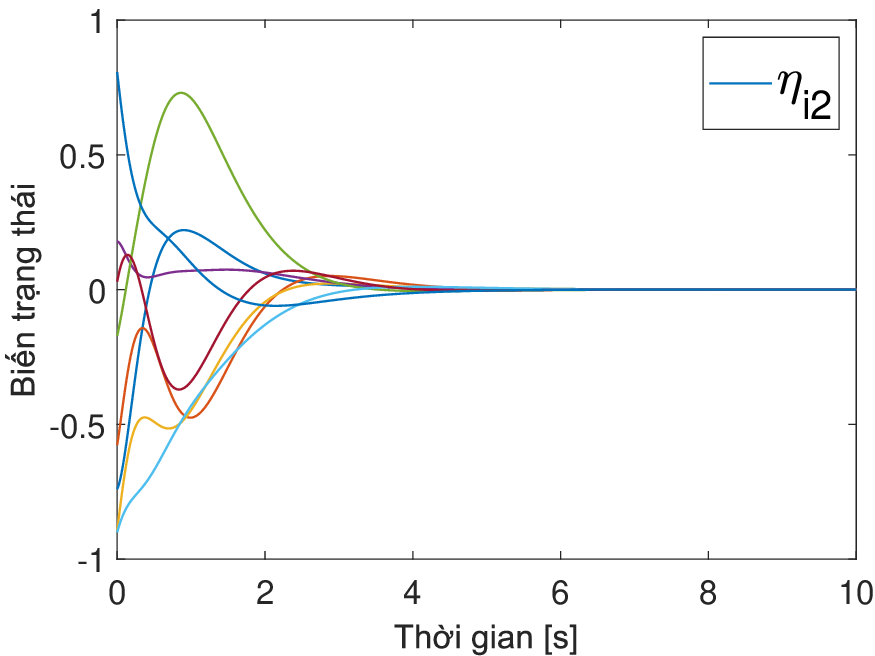}}\hfill
\subfloat[]{\includegraphics[width=.32\textwidth]{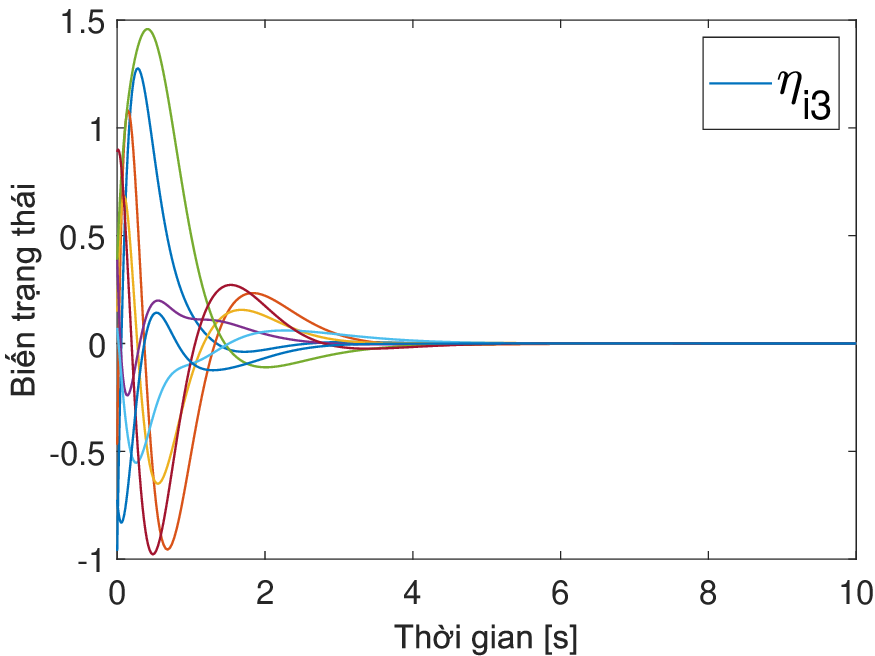}} \hfill    
\subfloat[]{\includegraphics[width=.32\textwidth]{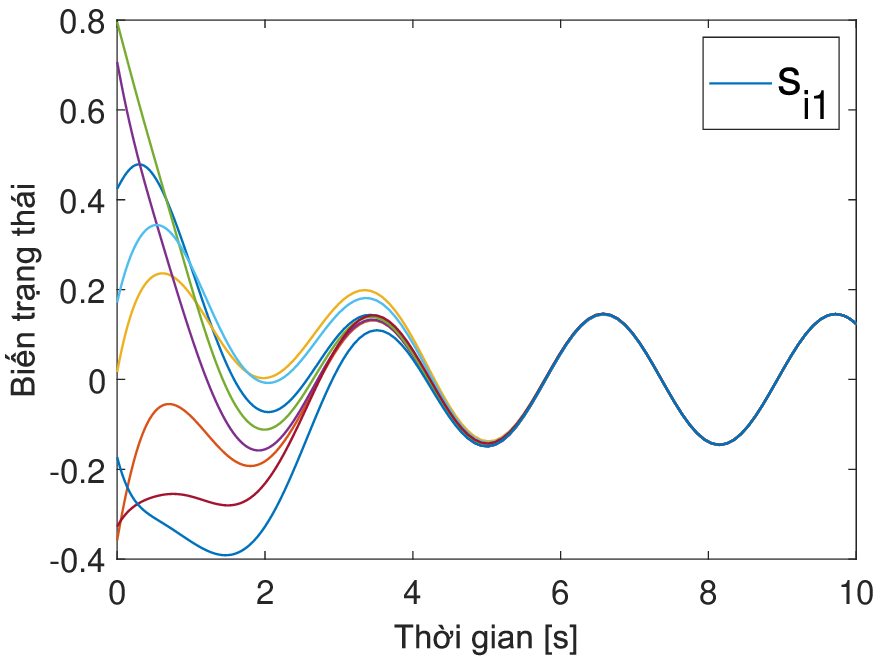}} \hfill    
\subfloat[]{\includegraphics[width=.32\textwidth]{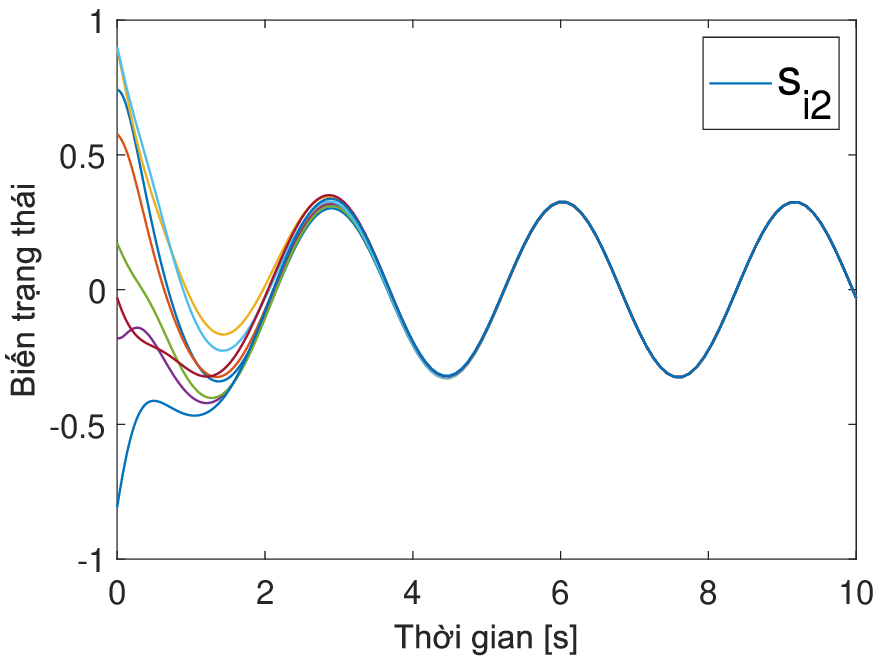}} \hfill    
\subfloat[]{\includegraphics[width=.32\textwidth]{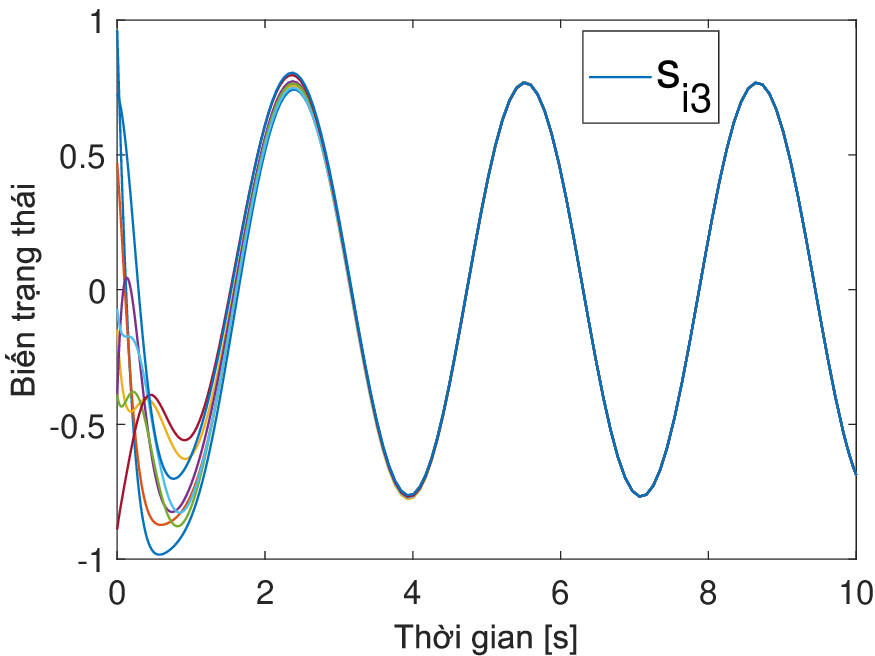}} \hfill    
\caption{Mô phỏng hệ đồng thuận gồm 8 tác tử trong Ví dụ~\ref{VD:3.4}. Các biến trạng thái ${x}_{ik}$ và các biến phụ , $i=1,\ldots,8$, $k=1,2,3,$ tiệm cận tới cùng một quĩ đạo.}
    \label{fig:VD3.4}
\end{figure}
\end{example}

\subsubsection{Bộ quan sát trạng thái dựa trên sai lệch biến đầu ra}
Trong mục này, chúng ta giả sử $\m{y}_i$ là biến đầu ra của các tác tử láng giềng được ghép nối trực tiếp với nhau và mỗi tác tử đo được sai lệch đầu ra tương đối, gộp chung lại thành vector
\begin{align}
 \bmm{\zeta}_i = \sum_{j \in {N}_i} (\m{y}_i - \m{y}_j), ~i=1,\ldots, n.
\end{align}

Do thiếu thông tin về các biến trạng thái, chúng ta thiết kế bộ quan sát trạng thái dựa trên tổng sai lệch đầu ra tương đối $\bm{\zeta}_i$ cho từng tác tử và thiết kế luật đồng thuận dựa trên biến quan sát được như sau \cite{Li2009consensus}:
\begin{align}
    \dot{\bmm{\eta}}_i &= (\m{A}+\m{B}\m{K}) \bmm{\eta}_i + \m{H}\Big( \sum_{j \in {N}_i} \m{C}(\bmm{\eta}_i - \bmm{\eta}_j) - \bmm{\zeta}_i\Big), \label{eq:c3-observer-consensus_a} \\
    \m{u}_i & = \m{K} \bmm{\eta}_i, ~i=1, \ldots, n,\label{eq:c3-observer-consensus}
\end{align}
với $\bmm{\eta}_i \in \mb{R}^d$ là các biến trạng thái phụ, $\m{F} \in \mb{R}^{d \times n}$ và $\m{K}\in \mb{R}^{m \times n}$ là các ma trận sẽ được thiết kế. Sơ đồ mô tả hệ đa tác tử với thuật toán đồng bộ hóa \eqref{eq:c3-observer-consensus_a}--\eqref{eq:c3-observer-consensus} được thể hiện ở Hình~\ref{fig:c3_output_synch1}.
\begin{SCfigure}[.8][t!]
    \caption{Sơ đồ mô tả hệ đồng thuận với thuật toán đồng thuận \eqref{eq:c3-observer-consensus_a}--\eqref{eq:c3-observer-consensus}.}
    \label{fig:c3_output_synch1}
    \hspace{5cm}
    \includegraphics[height=6.8cm]{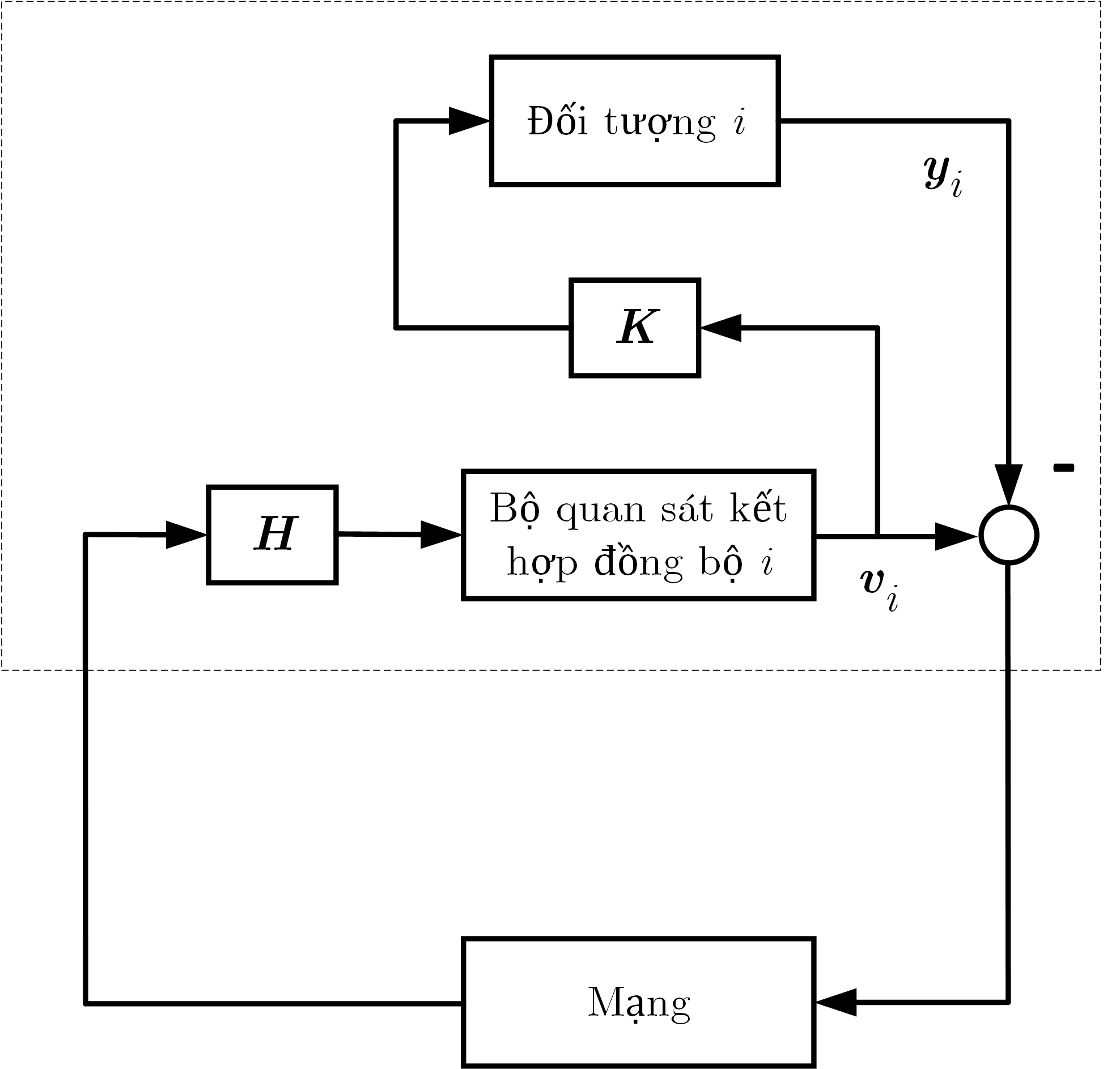}
\end{SCfigure}

Với luật đồng thuận~\eqref{eq:c3-observer-consensus_a}--\eqref{eq:c3-observer-consensus}, hệ \eqref{eq:c3-agent-model} có thể được viết lại dưới dạng ma trận:
\begin{align}
    \dot{\bmm{\eta}} &= (\m{I}_n\otimes(\m{A}+\m{B}\m{K}))\bmm{\eta} - (\mcl{L}\otimes \m{F}\m{C})(\bmm{\eta}-\m{x}),
    \label{eq:c3-system1}\\
    \dot{\m{x}} &=  (\m{I}_n\otimes\m{A})\m{x} + (\m{I}_n\otimes \m{B}\m{K})\bmm{\eta}, \label{eq:c3-system2}
\end{align}
trong đó $\bmm{\eta} = [\bmm{\eta}_1^\top,\ldots,\bmm{\eta}_n^\top]^\top$, $\m{x} = [\m{x}_1^\top,\ldots,\m{x}_n^\top]^\top$. Đặt biến phụ $\bmm{\zeta} = \bmm{\eta} - \m{x}$, ta có phương trình động học theo biến phụ:
\begin{align}
    \dot{\bmm{\zeta}} = (\m{I}_n\otimes \m{A} + \mcl{L}\otimes \m{F}\m{C}) \bmm{\zeta}.
\end{align}
Tương tự như ở mục \ref{chap3-sec3-1}, thực hiện phép đổi biến $\bmm{\delta} = (\m{P}^{-1}\otimes \m{I}_d) \bmm{\delta}$, ta có:
\begin{align}
    \dot{\bmm{\delta}} = (\m{I}_n\otimes \m{A} + \m{J}\otimes \m{F}\m{C}) \bmm{\delta}. \label{eq:c3-system3}
\end{align}
Với điều kiện $\m{A}+\lambda_i \m{F}\m{C}$ là Hurwitz, $\forall i=2,\ldots,n,$ hệ \eqref{eq:c3-system2} thỏa mãn
\begin{align}
    \bmm{\delta}(t) \to \begin{bmatrix}
     \mathtt{e}^{\m{A}t}\bmm{\delta}_1(0) \\
     \m{0}_d\\
     \vdots\\
     \m{0}_d
    \end{bmatrix},
\end{align}
khi $t\to +\infty$. Từ đây suy ra,
\begin{align}
    \bmm{\zeta}(t) \to  (\m{P}\otimes \m{I}_d)\begin{bmatrix}
     \mathtt{e}^{\m{A}t} \\
     \m{0}_d \\
     \vdots\\
     \m{0}_d \\
    \end{bmatrix} (\bmm{\gamma}^\top \otimes \m{I}_d) \bmm{\zeta}(0) = \m{1}_n \otimes (\mathtt{e}^{\m{A}t} \bar{\bmm{\zeta}}(0)),
\end{align}
với $\bar{\bmm{\zeta}}(0) = \sum_{i=1}^n \gamma_i \bmm{\zeta}_i(0)$. 

Chuyển lại với hệ động học theo biến $\bmm{\eta}$:
\begin{align}
    \dot{\bmm{\eta}} &= (\m{I}_n\otimes(\m{A}+\m{B}\m{K}))\bmm{\eta} + (\mcl{L}\otimes \m{F}\m{C}) \bmm{\zeta},
\end{align}
ta xem $(\mcl{L}\otimes \m{F}\m{C}) \bmm{\zeta}$ như một tín hiệu nhiễu tắt dần tác động vào hệ
\begin{align}
    \dot{\bmm{\eta}} = (\m{I}_n\otimes(\m{A}+\m{B}\m{K}))\bmm{\eta},
\end{align}
là một hệ ổn định tiệm cận theo hàm mũ do $\m{A}+\m{B}\m{K}$ là các ma trận Hurwitz. Do $\bmm{\zeta} $ tiệm cận tới không gian đồng thuận, ta có
\begin{align}
    (\mcl{L}\otimes \m{H}\m{C}) \bmm{\zeta} \to (\mcl{L}\otimes \m{F}\m{C}) \m{1}_n &\otimes (\mathtt{e}^{\m{A}t} \bar{\bmm{\zeta}}(0)) = \mcl{L}\m{1}_n \otimes \m{F}\m{C} (\mathtt{e}^{\m{A}t} \bar{\bmm{\zeta}}(0)) = \m{0}_{dn}.
\end{align}
Từ đây suy ra $\bmm{\eta} \to \m{0}_{dn}$ khi $t\to +\infty$. Cuối cùng, từ quan hệ $\m{x} = - \bmm{\zeta} + \bmm{\eta}$, ta có
\begin{align}
    \m{x}(t) \to - \m{1}_n \otimes (\mathtt{e}^{\m{A}t} \bar{\bmm{\zeta}}(0)) = - \m{1}_n \otimes \mathtt{e}^{\m{A}t}(\bar{\m{x}}(0) - \bar{\bmm{\eta}}(0)),
\end{align}
khi $t\to +\infty$, tức là hệ sẽ dần đạt được đồng bộ hóa.

\begin{SCfigure}[][t!]
\caption{Đồ thị mô tả tương tác giữa các tác tử trong Ví dụ~\ref{VD:3.5}.\label{fig:VD3.5_graph}}
\hspace{5cm}
\begin{tikzpicture}[
roundnode/.style={circle, draw=black, thick, minimum size=2mm,inner sep= 0.25mm},
squarednode/.style={rectangle, draw=black, thick, minimum size=3.5mm,inner sep= 0.25mm},
]
    \node[roundnode] (u1) at (2,0) { $1$}; %
    \node[roundnode] (u2) at (3,1) { $2$};%
    \node[roundnode] (u3) at (1,1) { $3$};%
    \node[roundnode] (u4) at (3,3) { $4$};%
    \node[roundnode] (u5) at (1,3) { $5$};%
    \node[roundnode] (u6) at (2,4) { $6$};%
    \node[roundnode] (u7) at (0,2) { $7$};%
    \node[roundnode] (u8) at (4,2) { $8$};%
    
    \draw [very thick,-{Stealth[length=2mm]}]
    (u1) edge [bend left=0] (u2)
    (u3) edge [bend left=0] (u2)
    (u3) edge [bend left=0] (u1)
    (u2) edge [bend left=0] (u8)
    (u2) edge [bend left=0] (u4)
    (u4) edge [bend left=0] (u5)
    (u4) edge [bend left=0] (u6)
    (u5) edge [bend left=0] (u3)
    (u5) edge [bend left=0] (u7)
    (u6) edge [bend left=0] (u5)
    (u7) edge [bend left=0] (u3)
    (u8) edge [bend left=0] (u4)
    ;
\end{tikzpicture}
\end{SCfigure}
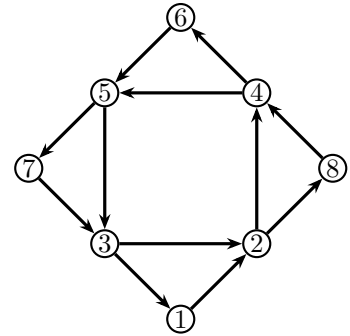

Như vậy, ta thấy rằng bài toán đồng thuận trong hệ đa tác tử với luật \eqref{eq:c3-observer-consensus} được chuyển thành việc xét tính ổn định một tập các ma trận có cùng số kích thước $d\times d$. Luật đồng thuận \eqref{eq:c3-observer-consensus_a}--\eqref{eq:c3-observer-consensus} là mở rộng của bộ điều khiển kết hợp bộ quan sát trạng thái cho hệ đa tác tử. Nguyên lý tách của bộ điều khiển dựa trên quan sát trạng thái vẫn đúng trong trường hợp hệ đa tác tử. Ảnh hưởng của đồ thị tương tác $G$ thể hiện qua các trị riêng (phức) của ma trận Laplace trong các phương trình $\m{A}+\m{B}\m{K} + \lambda_i \m{F}\m{C}$, $i = 1, \ldots, n$.

\begin{example} \label{VD:3.5}
\begin{figure}[th!] 
    \centering
    \subfloat[]{\includegraphics[width=.45\textwidth]{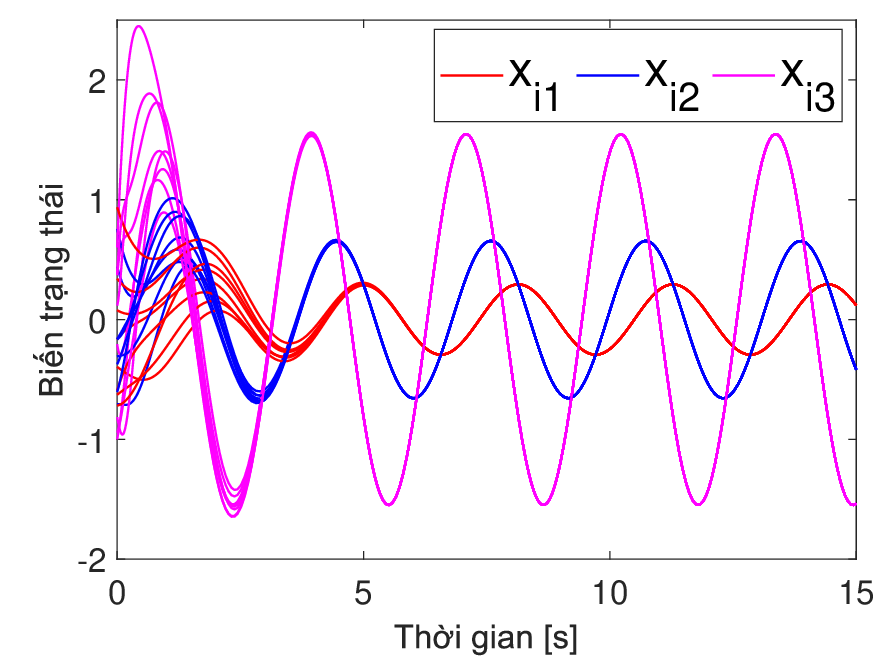}}\hfill
    \subfloat[]{\includegraphics[width=.45\textwidth]{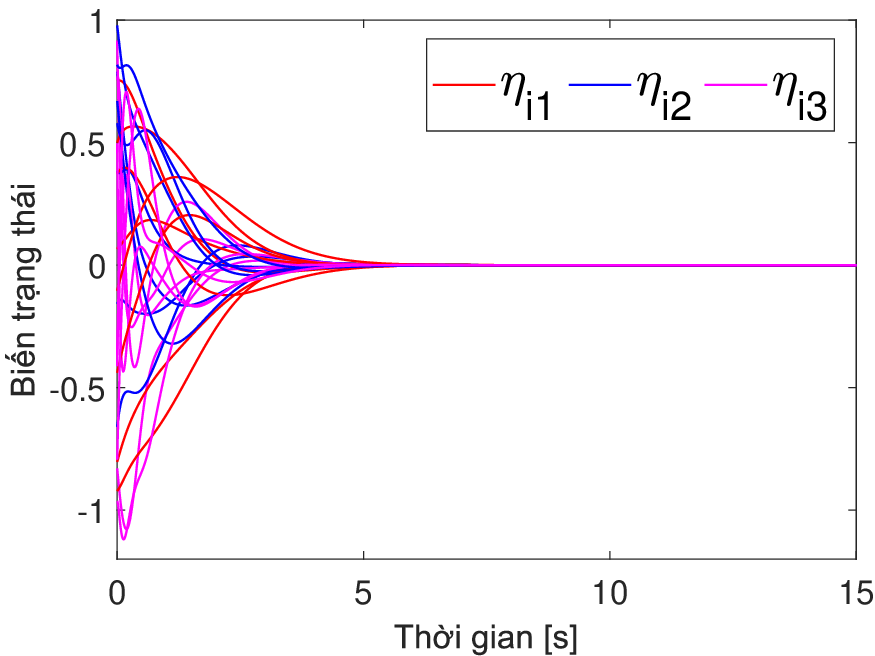}} 
    \caption{Mô phỏng hệ 8 tác tử với thuật toán đồng bộ hóa \eqref{eq:c3-observer-consensus_a}--\eqref{eq:c3-observer-consensus} ở Ví dụ~\ref{VD:3.5}.}
    \label{fig:VD3.5}
\end{figure}
Trong ví dụ này, ta mô phỏng hệ đồng thuận gồm 8 tác tử với thuật toán đồng thuận \eqref{eq:c3-observer-consensus_a}--\eqref{eq:c3-observer-consensus}. Các ma trận của hệ đa tác tử trong mô phỏng được chọn như sau:
\begin{align*}
    \m{A} &= \begin{bmatrix}
    -1  &  1      &  0\\
     0  & -1.25   &  1\\
     0  & -5.5625 & 1.25
    \end{bmatrix},~ \m{B} = \begin{bmatrix}
    0 & 0 & 1
    \end{bmatrix},~ \m{C} = \begin{bmatrix}
    1 & 1 & 1
    \end{bmatrix}, \\
    \m{F} &= \begin{bmatrix}
    -0.0730 & 0.5255 & 7.5474
    \end{bmatrix}, ~
    \m{K} = -\begin{bmatrix}
    1 & -3.25 & 5
    \end{bmatrix}.
\end{align*}
Các giá trị đầu $\m{x}_i(0), \bm{\eta}_i(0)$ được sinh ngẫu nhiên.  Hình~\ref{fig:VD3.5_graph} thể hiện đồ thị mô tả luồng thông tin $G$ giữa các tác tử. 

Các kết quả mô phỏng được cho ở Hình~\ref{fig:VD3.5}. Các biến trạng thái $\m{x}_i(t)$ tiệm cận đồng thuận tới cùng một quĩ đạo và $\bm{\eta}_i(t) \to \m{0}$, $i=1,\ldots,8$, khi $t\to +\infty$.
\end{example}

\section{Hệ đồng thuận bậc nhất không liên tục}
\subsection{Mô hình và điều kiện đồng thuận}
Trong mục này, ta xét một lớp thuật toán đồng thuận khi các tác tử chỉ cập nhật biến trạng thái tại các thời điểm $k=1, 2, \ldots, +\infty$. Xét hệ gồm $n$ tác tử, trong đó mỗi tác tử có biến trạng thái $x_i[k] \in \mb{R}, \forall k = 1, 2, \ldots$. 

Đồ thị $G$ mô tả sự trao đổi thông tin trong hệ được giả sử là hữu hướng, có trọng số và các trọng số này thỏa mãn $\sum_{j=1}^n w_{ij} = 1$. Lưu ý rằng ở đây, đồ thị $G$ có thể có các khuyên (cạnh từ đỉnh $v_i$ tới chính nó), và do đó $w_{ii}$ có thể khác 0. Định nghĩa tập $M_i = \{j\in V|~ w_{ij} > 0\}$ gồm các tác tử lân cận của tác tử $i$ (có thể bao gồm cả $i$). Tại thời điểm $k$, tác tử $i \in V$ cập nhật biến trạng thái của mình theo thuật toán đồng thuận sau:\index{thuật toán!đồng thuận rời rạc}
\begin{align}\label{eq:c3_discrete_consensus}
x_i[k+1] = \sum_{j \in M_i} w_{ij} x_j[k],
\end{align}
Thuật toán đồng thuận \eqref{eq:c3_discrete_consensus} là một thuật toán lặp tuyến tính và phân tán do chỉ yêu cầu biến trạng thái của tác tử $i$ và các tác tử láng giềng của $i$. Chúng ta có thể viết lại \eqref{eq:c3_discrete_consensus} dưới dạng sau:
\begin{equation}\label{eq:c3_discrete_consensus_matrix}
\m{x}[k+1] = \m{A}\m{x}[k],
\end{equation}
trong đó $\m{x}[k] = [x_1[k], \ldots, x_n[k]]^\top$. Điểm $\m{x}^*=[x_1^*, \ldots, x_n^*]^\top$ là một điểm cân bằng của \eqref{eq:c3_discrete_consensus_matrix} nếu $\m{x}^* = \m{A}\m{x}^*$. Hệ tiệm cận tới điểm cân bằng $\m{x}^*$ nếu $\lim_{k \to +\infty}\m{x}[k] = \m{x}^*$, khi $k \to +\infty$. \index{điểm!cố định}

Ma trận $\m{A}$ trong phương trình \eqref{eq:c3_discrete_consensus_matrix} có các phần tử không âm thỏa mãn $\m{A} \m{1}_n = \m{1}_n$ nên $\m{A}$ là một ma trận ngẫu nhiên hàng\index{ma trận!ngẫu nhiên hàng}. Một ma trận ngẫu nhiên hàng $\m{A}$ là không thể phân rã và không có chu kì (SIA)\footnote{stochastically indecomposable aperiodic matrix} nếu $\lim_{k \to +\infty} \m{A}^k = \m{1}_n \m{y}^\top$, với $\m{y}$ là một vector cột. Một đồ thị hữu hướng $G$ là có chu kỳ nếu ước chung lớn nhất của các độ dài các chu trình đơn của $G$ là lớn hơn 1. Ngược lại, đồ thị $G$ gọi là không có chu kỳ (xem Ví dụ ở Hình~\ref{fig:VD_do_thi_khong_chu_ky}).  \index{đồ thị!không có chu kỳ}
\index{ma trận!không thể phân rã và không có chu kỳ}
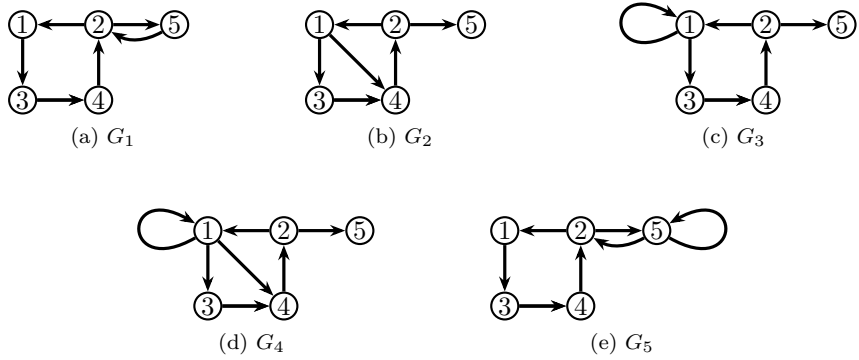
\begin{figure}[t!]
\centering
\subfloat[$G_1$]{\begin{tikzpicture}[
roundnode/.style={circle, draw=black, thick, minimum size=2mm,inner sep= 0.25mm},
squarednode/.style={rectangle, draw=black, thick, minimum size=3.5mm,inner sep= 0.25mm},
]
    \node[roundnode] (u1) at (0,1) {$1$}; %
    \node[roundnode] (u2) at (1,1) {$2$};%
    \node[roundnode] (u3) at (0,0) {$3$};%
    \node[roundnode] (u4) at (1,0) {$4$};%
    \node[roundnode] (u5) at (2,1) {$5$};%
    
    \draw [very thick,-{Stealth[length=2mm]}]
    (u1) edge [bend left=0] (u3)
    (u3) edge [bend left=0] (u4)
    (u3) edge [bend left=0] (u4)
    (u2) edge [bend left=0] (u5)
    (u2) edge [bend left=0] (u1)
    (u4) edge [bend left=0] (u2)
    (u5) edge [bend left=30] (u2)
    ;
\end{tikzpicture}} \qquad\qquad
\subfloat[$G_2$]{\begin{tikzpicture}[
roundnode/.style={circle, draw=black, thick, minimum size=2mm,inner sep= 0.25mm},
squarednode/.style={rectangle, draw=black, thick, minimum size=3.5mm,inner sep= 0.25mm},
]
    \node[roundnode] (u1) at (0,1) {$1$}; %
    \node[roundnode] (u2) at (1,1) {$2$};%
    \node[roundnode] (u3) at (0,0) {$3$};%
    \node[roundnode] (u4) at (1,0) {$4$};%
    \node[roundnode] (u5) at (2,1) {$5$};%
    
    \draw [very thick,-{Stealth[length=2mm]}]
    (u1) edge [bend left=0] (u3)
    (u3) edge [bend left=0] (u4)
    (u3) edge [bend left=0] (u4)
    (u2) edge [bend left=0] (u5)
    (u2) edge [bend left=0] (u1)
    (u4) edge [bend left=0] (u2)
    (u1) edge [bend left=0] (u4)
    ;
\end{tikzpicture}} \qquad\qquad
\subfloat[$G_3$]{\begin{tikzpicture}[
roundnode/.style={circle, draw=black, thick, minimum size=2mm,inner sep= 0.25mm},
squarednode/.style={rectangle, draw=black, thick, minimum size=3.5mm,inner sep= 0.25mm},
]
    \node[roundnode] (u1) at (0,1) {$1$}; %
    \node[roundnode] (u2) at (1,1) {$2$};%
    \node[roundnode] (u3) at (0,0) {$3$};%
    \node[roundnode] (u4) at (1,0) {$4$};%
    \node[roundnode] (u5) at (2,1) {$5$};%
    
    \draw [very thick,-{Stealth[length=2mm]}]
    (u1) edge [bend left=0] (u3)
    (u3) edge [bend left=0] (u4)
    (u3) edge [bend left=0] (u4)
    (u2) edge [bend left=0] (u5)
    (u2) edge [bend left=0] (u1)
    (u4) edge [bend left=0] (u2)
    ;
    
    \draw [draw = black, very thick, -{Stealth[length=2mm]}]
    (u1) edge [in=150, out=210,looseness=15] (u1);

\end{tikzpicture}} \\
\subfloat[$G_4$]{\begin{tikzpicture}[
roundnode/.style={circle, draw=black, thick, minimum size=2mm,inner sep= 0.25mm},
squarednode/.style={rectangle, draw=black, thick, minimum size=3.5mm,inner sep= 0.25mm},
]
    \node[roundnode] (u1) at (0,1) {$1$}; %
    \node[roundnode] (u2) at (1,1) {$2$};%
    \node[roundnode] (u3) at (0,0) {$3$};%
    \node[roundnode] (u4) at (1,0) {$4$};%
    \node[roundnode] (u5) at (2,1) {$5$};%
    
    \draw [very thick,-{Stealth[length=2mm]}]
    (u1) edge [bend left=0] (u3)
    (u3) edge [bend left=0] (u4)
    (u3) edge [bend left=0] (u4)
    (u2) edge [bend left=0] (u5)
    (u2) edge [bend left=0] (u1)
    (u4) edge [bend left=0] (u2)
    (u1) edge [bend left=0] (u4)
    ;
    
    \draw [draw = black, very thick, -{Stealth[length=2mm]}]
    (u1) edge [in=150, out=210,looseness=15] (u1);
\end{tikzpicture}} \qquad\qquad
\subfloat[$G_5$]{\begin{tikzpicture}[
roundnode/.style={circle, draw=black, thick, minimum size=2mm,inner sep= 0.25mm},
squarednode/.style={rectangle, draw=black, thick, minimum size=3.5mm,inner sep= 0.25mm},
]
    \node[roundnode] (u1) at (0,1) {$1$}; %
    \node[roundnode] (u2) at (1,1) {$2$};%
    \node[roundnode] (u3) at (0,0) {$3$};%
    \node[roundnode] (u4) at (1,0) {$4$};%
    \node[roundnode] (u5) at (2,1) {$5$};%
    \draw [very thick,-{Stealth[length=2mm]}]
    	(u1) edge [bend left=0] (u3)
    	(u3) edge [bend left=0] (u4)
    	(u3) edge [bend left=0] (u4)
    	(u2) edge [bend left=0] (u5)
    	(u2) edge [bend left=0] (u1)
    	(u4) edge [bend left=0] (u2)
    	(u5) edge [bend left=30] (u2)
    ;
    \draw [draw = black, very thick, -{Stealth[length=2mm]}]
    	(u5) edge [in=30, out=330, looseness=15] (u5);
\end{tikzpicture}}
\caption{Đồ thị $G_1$ là liên thông mạnh và có chu kỳ bằng 2; Đồ thị $G_2$ và $G_3$ là có gốc ra, thành phần liên thông mạnh chứa gốc ra là không có chu kỳ, trong đó đồ thị $G_3$ có chứa khuyên; Đồ thị $G_4, G_5$ là liên thông mạnh và không có chu kỳ.}
\label{fig:VD_do_thi_khong_chu_ky}
\end{figure}

Ta có kết quả sau:
\begin{lemma} \cite{Ren2005consensus} \label{lem:c3_discrete_G} Ma trận ngẫu nhiên $\m{A} = [w_{ij}] \in \mb{R}^{n\times n}$ có một giá trị riêng đơn $\lambda = 1$ và các giá trị riêng khác thỏa mãn $|\mu|<1$ khi và chỉ khi đồ thị tương ứng với ma trận $\m{A}$ là có gốc ra và thành phần liên thông mạnh tối đa chứa các gốc ra là không có chu kỳ. Khi đó, $\m{A}$ là SIA, $\lim_{k \to +\infty}\m{A}^k = \m{1}_n\bm{\gamma}^\top$, trong đó  $\m{A}^\top\bm{\gamma} = \bm{\gamma}$, $\m{1}_n^\top \bm{\gamma} = 1$, và $\bm{\gamma}=[\gamma_1,\ldots,\gamma_n]^\top$ có các phần tử $\gamma_i \ge 0$, $\forall i=1,\ldots, n,$ và $\gamma_i>0$ khi và chỉ khi $i$ thuộc thành phần liên thông mạnh tối đa chứa các gốc ra của đồ thị. 
\end{lemma}
Với bổ đề trên, ta có định lý sau về quá trình tiệm cận của hệ \eqref{eq:c3_discrete_consensus_matrix}:
\begin{theorem} \label{thm:c3_consensus_discrete} Với luật đồng thuận  \eqref{eq:c3_discrete_consensus_matrix}, các tác tử tiệm cận tới một điểm trong tập đồng thuận khi $k \to +\infty$ khi và chỉ khi đồ thị $G$ có gốc ra và thành phần liên thông mạnh tối đa chứa các gốc ra là không có chu kỳ.
\end{theorem}

\begin{proof}
(Điều kiện đủ) Dựa trên lý thuyết về hệ tuyến tính, ta có
\begin{align*}
\lim_{k\to +\infty} \m{x}[k+1] = \lim_{k\to +\infty} \m{A}^k \m{x}[0].
\end{align*}
Từ giả thiết $G$ có gốc ra, theo Bổ đề \ref{lem:c3_discrete_G}, ta có
\begin{align*}
\lim_{k\to +\infty} \m{x}[k+1] = (\m{1}_n \bm{\gamma}^\top) \m{x}[0].
\end{align*}
Như vậy hệ tiến tới đồng thuận tại giá trị đồng thuận $\bm{\gamma}^\top \m{x}[0] = \sum_{i=1}^n \gamma_i x_i[0]$.

(Điều kiện cần) Giả sử $G$ không có gốc ra thì từ Bổ đề \ref{lem:c3_discrete_G}, giá trị riêng $1$ là một nghiệm bội $m>1$ của đa thức đặc tính của $\m{A}$. Bởi vậy, dạng Jordan của $\m{A}^k$ có dạng $\m{A}^k = \m{M} \m{J}^k \m{M}^{-1}$, với $\m{M}$ là ma trận khả nghịch và $\m{J}^k$ là một ma trận tam giác trên với $m$ phần tử đường chéo bằng 1. Như vậy bậc của $\lim_{k \to+ \infty} \m{A}^k$ ít nhất là bằng $m>1$, khác với không gian đồng thuận. Điều này có nghĩa là tồn tại các điều kiện đầu $\m{x}[0]$ sao cho hệ không tiến tới đồng thuận, trái với giả thiết của định lý.
\end{proof}

Chú ý rằng nếu như có thêm giả thiết là đồ thị $G$ tương ứng với ma trận $\m{A}$ là liên thông mạnh và trong $G$ có ít nhất một khuyên thì ma trận $\m{A}$ là ma trận ngẫu nhiên hàng, không thể phân rã, và không có chu kì. Các kết quả ở định lý \ref{thm:c3_consensus_discrete} vẫn đúng chỉ khác là vector riêng bên trái $\bm{\gamma}^\top$ của $\m{A}$ lúc này có tất cả các phần tử dương ($\gamma_i>0, \forall i=1, \ldots, n$).

\subsection{Liên hệ với mô hình hệ đồng thuận liên tục}
Ta xét một trường hợp riêng, trong đó hệ đồng thuận không liên tục được xem như trạng thái của hệ đồng thuận liên tục $\dot{\m{x}}(t)=-\mcl{L}\m{x}(t)$ tại các thời điểm $t=k\delta$, với $k=0, 1, 2, \ldots, \infty$.

Đặt $\m{z}[k] = \m{x}(k\delta)$ thì theo công thức nghiệm của phương trình vi phân, ta có:
\begin{align}
    \m{z}[k] = \m{x}(k\delta) = \mathtt{e}^{-\mcl{L} k \delta} \m{x}(0).
\end{align}
Từ đây suy ra $\m{z}[0] = \m{x}(0)$ và 
\begin{align}
    \m{z}[k+1] = \mathtt{e}^{-\mcl{L} (k+1) \delta} \m{x}(0) = \mathtt{e}^{-\mcl{L}\delta} \mathtt{e}^{-\mcl{L} k\delta} \m{x}(0) = \m{A} \m{z}[k], \label{eq:c3_discrete_model}
\end{align}
trong đó $\m{A} = \mathtt{e}^{-\delta\mcl{L}}$. Một số tính chất của ma trận $\m{A}$ được cho trong bổ đề sau.
\begin{lemma}
Giả sử đồ thị $G$ là có gốc ra với ma trận Laplace $\mcl{L}$. Ma trận $\m{A} =[w_{ij}] = \mathtt{e}^{-\mcl{L}\delta}$ thỏa mãn:
\begin{enumerate}
    \item $\m{A}$ là ma trận ngẫu nhiên hàng.
    \item Có các phần tử không âm: $w_{ij}\ge 0, \forall i,j\in V$. 
    \item $w_{ij}>0$ khi và chỉ khi tồn tại một đường đi có hướng từ $j$ tới $i$ trong đồ thị $G$.
    \item Tồn tại ít nhất một cột của $\m{A}$ với tất cả các phần tử dương.
\end{enumerate}
\end{lemma}
\begin{proof}
Do $G$ là có gốc ra nên ker$(\mcl{L})$=im$(\m{1}_n)$. 
\begin{enumerate}
    \item Viết lại ma trận $\m{A}$ dưới dạng:
    \begin{align}
    \m{A} = \mathtt{e}^{-\delta\mcl{L}} = \sum_{k=0}^\infty \frac{(-\delta\mcl{L})^k}{k!} = \m{I}_n - \frac{\delta\mcl{L}}{1!} + \frac{(\delta\mcl{L})^2}{k!} - \ldots 
    \end{align}
    thì $\frac{(-\delta\mcl{L})^k}{k!} \m{1}_n = \m{0}$ với mọi $k=1,\ldots,n$. Do đó $\m{A}\m{1}_n = \m{I}_n \m{1}_n = \m{1}_n$, hay $\m{A}$ là ma trận ngẫu nhiên hàng. 
    \item Chọn $\mu > \max_{i}l_{ii}$ thì $\mathtt{e}^{-\delta\mcl{L}} = \mathtt{e}^{-\mu\delta} \mathtt{e}^{\delta(\mu\m{I}_n -\mcl{L})}$ nên các phần tử của hai ma trận $\m{W}$ và $\mathtt{e}^{\delta(\mu\m{I}_n -\mcl{L})}$ đều có cùng dấu. Hơn nữa, ma trận $\m{L}_+=\delta(\mu\m{I}_n -\mcl{L})$ có tất cả các phần tử là không âm nên
    \begin{align}
\mathtt{e}^{\delta(\mu\m{I}_n -\mcl{L})} = \sum_{k=0}^\infty \frac{\delta(\m{L}_+)^k}{k!} \label{eq:c3_lem3_2}
    \end{align}
    cũng là một ma trận với các phần tử không âm. 
    \item Mỗi phần tử $[\m{L}_+]_{ij}$ của ma trận $\m{L}_+=(\mu\m{I}_n -\mcl{L})$ là không âm khi và chỉ khi tồn tại một cạnh $(j,i) \in E$ (do ma trận $\m{L}_+$ có mọi phần tử ngoài đường chéo chính giống với ma trận kề của đồ thị $G$). Tương tự, $[\m{L}_+^2]_{ij}>0$ khi và chỉ khi tồn tại một đường đi có hướng độ dài bằng 2 từ $j$ tới $i$ trong đồ thị do $[\m{L}_+^2]_{ij}=\sum_{k=1}^n [\m{L}_+]_{ik}[\m{L}_+^2]_{kj}$. Tổng quát, từ phương trình \eqref{eq:c3_lem3_2}, ta suy ra mỗi phần tử $w_{ij}>0$ khi và chỉ khi tồn tại một đường đi có hướng từ $j$ tới $i$ trong $G$.
    \item Do đồ thị $G$ có gốc ra nên nếu $j$ là gốc ra thì tồn tại đường đi có hướng từ $j$ tới mọi đỉnh trong $G$. Do đó $w_{ij}>0$ với mọi $i\in V$ hay nói cách khác cột thứ $j$ của $\m{W}$ phải có mọi phần tử dương.
\end{enumerate}
\end{proof}

\begin{theorem}
Hệ đồng thuận cho bởi \eqref{eq:c3_discrete_model} tiệm cận tới một điểm cân bằng thuộc tập đồng thuận khi $k\to + \infty$.
\end{theorem}

\begin{proof}
Kết luận này có thể suy ra ngay từ tính hội tụ của mô hình đồng thuận liên tục. Một cách khác, ta thấy rằng $\m{A}$ là một ma trận ngẫu nhiên hàng, không thể phân rã và không có chu kỳ. Sự hội tụ của mô hình thu được nhờ áp dụng Định lý \ref{thm:c3_consensus_discrete}.
\end{proof}
\begin{example} \label{Ex:3.14}
Xét hệ gồm 8 tác tử với đồ thị hữu hướng $G$ như ở Hình \ref{fig:VD3.14}(a). Kết quả mô phỏng thuật toán đồng thuận liên tục và thuật toán đồng thuận không liên tục với $\delta=0.5$ được cho ở Hình~\ref{fig:VD3.14}(b). Dễ thấy sự hội tụ của mô hình không liên tục (trạng thái biểu diễn bởi `o') và mô hình liên tục (trạng thái biểu diễn bởi nét liền) là như nhau.

\end{example}
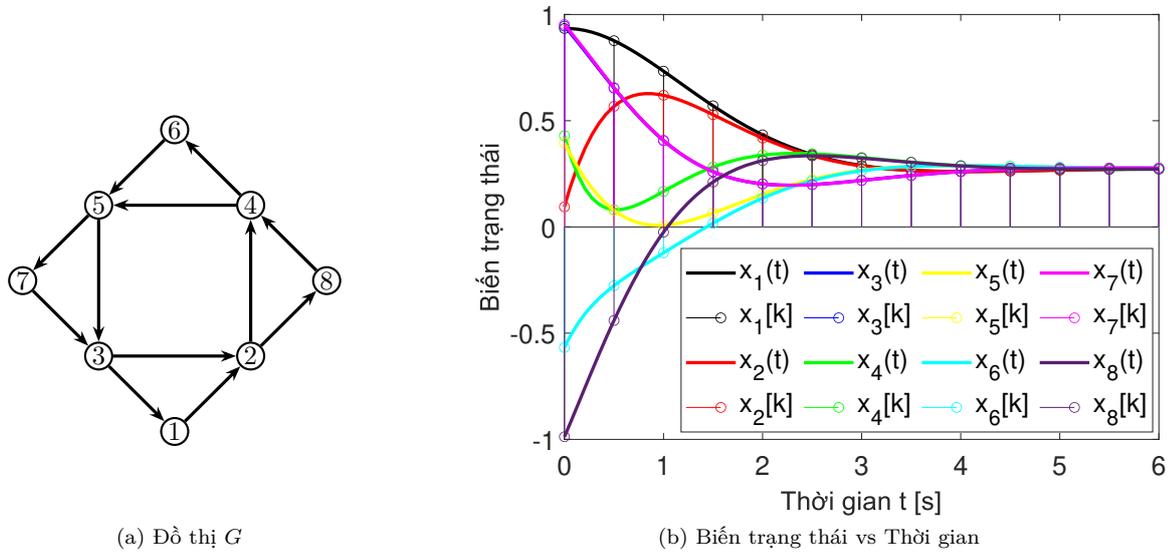
\begin{figure*}
    \centering
    \subfloat[Đồ thị $G$]{\begin{tikzpicture}[
roundnode/.style={circle, draw=black, thick, minimum size=2mm,inner sep= 0.25mm},
squarednode/.style={rectangle, draw=black, thick, minimum size=3.5mm,inner sep= 0.25mm},
]
    \node[roundnode] (u1) at (2,0) { $1$}; %
    \node[roundnode] (u2) at (3,1) { $2$};%
    \node[roundnode] (u3) at (1,1) { $3$};%
    \node[roundnode] (u4) at (3,3) { $4$};%
    \node[roundnode] (u5) at (1,3) { $5$};%
    \node[roundnode] (u6) at (2,4) { $6$};%
    \node[roundnode] (u7) at (0,2) { $7$};%
    \node[roundnode] (u8) at (4,2) { $8$};%
    \node (u9) at (0,-1) { };
    \draw [very thick,-{Stealth[length=2mm]}]
    (u1) edge [bend left=0] (u2)
    (u3) edge [bend left=0] (u2)
    (u3) edge [bend left=0] (u1)
    (u2) edge [bend left=0] (u8)
    (u2) edge [bend left=0] (u4)
    (u4) edge [bend left=0] (u5)
    (u4) edge [bend left=0] (u6)
    (u5) edge [bend left=0] (u3)
    (u5) edge [bend left=0] (u7)
    (u6) edge [bend left=0] (u5)
    (u7) edge [bend left=0] (u3)
    (u8) edge [bend left=0] (u4)
    ;
\end{tikzpicture}
}
    \hfill 
    \subfloat[Biến trạng thái vs Thời gian]{\includegraphics[width = .6 \textwidth]{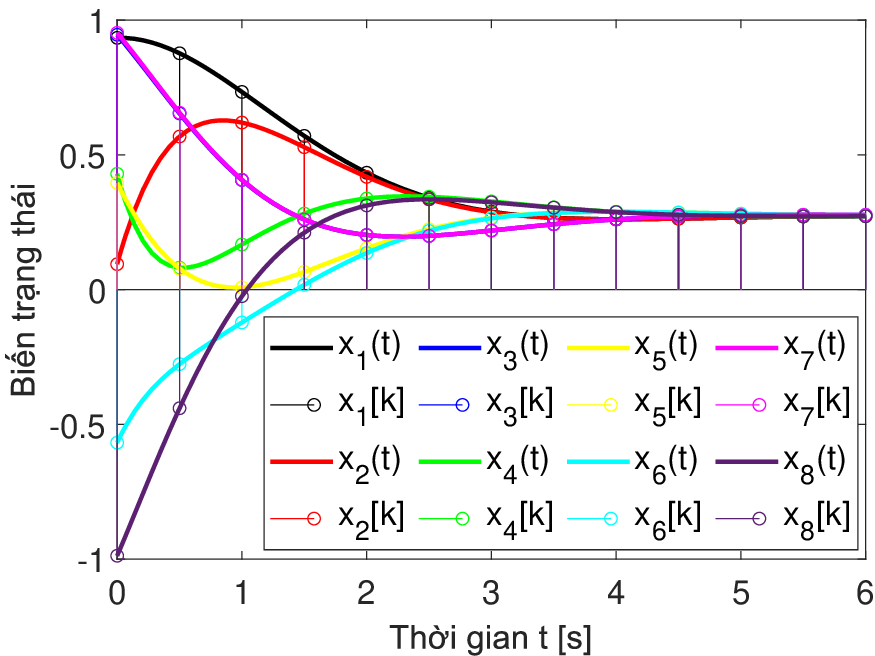}}
    \caption{Mô phỏng đối chiếu thuật toán đồng thuận liên tục và không liên tục}
    \label{fig:VD3.14}
\end{figure*}

\section{Ghi chú và tài liệu tham khảo}
Các kết quả cơ bản về hệ đồng thuận trong chương này đều dựa trên lý thuyết về phương trình vi phân/sai phân tuyến tính. Mặc dù đơn giản, các kết quả này là nền tảng cho hầu hết các mở rộng sau này. Điểm quan trọng nhất cần lưu ý là mối quan hệ giữa đồ thị $G$ và sự hội tụ của thuật toán đồng thuận. Với luật đồng thuận cho tác tử tích phân bậc nhất, điều kiện cần và đủ để hệ tiệm cận tới đồng thuận tại giá trị trung bình cộng là đồ thị tương tác là liên thông mạnh và cân bằng. Khi đồ thị là liên thông mạnh, giá trị đồng thuận là trung bình có trọng số của biến trạng thái của các tác tử. Điều kiện cần và đủ để hệ tiến tới đồng thuận là đồ thị có gốc ra. Giá trị đồng thuận lúc này chỉ phụ thuộc vào giá trị đầu của các tác tử thuộc thành phần liên thông mạnh tới hạn chứa gốc ra. 

Về tài liệu tham khảo, tổng quan, các kết quả chính, và ứng dụng của hệ đồng thuận được mô tả khá chi tiết trong \cite{Olfati2007consensuspieee} và \cite{Ren2007magazine}. Các công bố sớm nhất về thuật toán đồng thuận liên tục với tác tử tích phân bậc nhất có thể kể tới \cite{Olfati2004consensus,Moreau2005Stability}. Thuật toán đồng thuận liên tục với tác tử tích phân bậc hai, tích phân bậc cao, và tuyến tính tổng quát được đề cập ở các tài liệu  \cite{Ren2007distr,Ren2007HOConsensus} và \cite{Luca2009autom,Li2014designing,Wieland2011internal,Nguyen2016reduced}.  Trong khi đó, bài toán đồng thuận trên các mặt cong có lời giải phức tạp hơn. Điều kiện đồng thuận lúc này không chỉ liên quan cả với đồ thị mà còn cả tính chất của mặt cong \cite{Sarlette2009,Yuji2009}. Một số vấn đề liên quan như đồng thuận khi có trễ, có nhiễu được xem xét trong các tài liệu \cite{Olfati2007consensuspieee,Yu2013delay,Cao2011distributed}.

Cuối cùng, hệ đồng thuận rời rạc đã được nghiên cứu với ứng dụng trong nghiên cứu các mô hình động học ý kiến hoặc giải thích các mô hình vật lý (mô hình Vicsek) \cite{Jadbabaie2003coordination}, \cite{Ren2007magazine}, \cite{degroot1974reaching}. Tài liệu \cite{Bullo2019lectures} cung cấp phân tích khá chi tiết về thuật toán đồng thuận rời rạc cơ bản. Thuật toán đồng thuận rời rạc với đồ thị thay đổi theo thời gian được xem xét ở \cite{Ren2005consensus}. Chú ý rằng để thuật toán đồng thuận rời rạc \eqref{eq:c3_discrete_consensus} tiệm cận tới đồng thuận, một số trọng số (hằng) $w_{ij}$ có thể nhận giá trị âm \cite{Xiao2004}. 

\section{Bài tập}
\label{c3:exercise}
\begin{exercise}[Đuổi bắt theo vòng (Cyclic pursuit)]\label{ex:3_1}
Xét bốn robot trên mặt phẳng với vị trí $\m{p}_i\in \mb{R}^2$, $i=1,2,3,4,$ chuyển động với phương trình $\dot{\m{p}}_i = \m{u}_i$, với $\m{u}_i\in \mb{R}^2$ là vector vận tốc. Xét bài toán hội ngộ, trong đó các robot cần tập kết tại một điểm chung và thông tin mỗi robot có được chỉ thu được từ các cảm biến gắn trên chính robot đó. Cyclic pursuit là phương án yêu cầu cấu trúc thông tin tối giản để giải bài toán hội ngộ. Mỗi robot $i$ sẽ đuổi theo một robot thứ $i+1$ (Ở đây ta qui ước $4+1\equiv 1$.) Nói các khác, ta sử dụng luật $\m{u}_i= \m{p}_{i+1}-\m{p}_i$ và thu được hệ phương trình động học mô tả hệ:
\[\frac{d}{{dt}}\left[ {\begin{array}{*{20}{c}}
{{\m{p}_1}}\\
{{\m{p}_2}}\\
{{\m{p}_3}}\\
{{\m{p}_4}}
\end{array}} \right] = - (\mcl{L} \otimes \m{I}_2) \left[ {\begin{array}{*{20}{c}}
{{\m{p}_1}}\\
{{\m{p}_2}}\\
{{\m{p}_3}}\\
{{\m{p}_4}}
\end{array}} \right] = \left( {\left[ {\begin{array}{*{20}{c}}
{ - 1}&1&0&0\\
0&{ - 1}&1&0\\
0&0&{ - 1}&1\\
1&0&0&{ - 1}
\end{array}} \right] \otimes {\m{I}_2}} \right)\left[ {\begin{array}{*{20}{c}}
{{\m{p}_1}}\\
{{\m{p}_2}}\\
{{\m{p}_3}}\\
{{\m{p}_4}}
\end{array}} \right]\]
Chứng minh rằng:
\begin{enumerate}
    \item[i.] Trọng tâm của bốn robot $\bar{\m{p}}$ là một vector hằng với mọi $t\ge 0$.
    \item[ii.] Các robot sẽ gặp nhau tại $\bar{\m{p}}$ khi $t\to +\infty$.
    \item[iii.] Tìm các giá trị riêng và vector riêng của ma trận $\mcl{L}$.
    \item[iv.] Nếu ban đầu vị trí của 4 tác tử tạo thành một hình vuông thì tại mọi thời điểm $0\le t \le +\infty$, vị trí của 4 tác tử vẫn tạo thành một hình vuông.
    \item[v.] Mô phỏng hệ bốn robot với MATLAB với các điều kiện đầu khác nhau để kiểm chứng lại các kết luận trên.
\end{enumerate}
\end{exercise}

\begin{exercise} \cite{Bullo2019lectures} \label{Ex:3_2}
\begin{SCfigure}[][th]
    \centering
    \caption{Đồ thị ở Bài tập~\ref{Ex:3_2}.}
    \label{fig:Ex_3_2}
    \hspace{6cm}
    \begin{tikzpicture}
          \node (v1) at (0,0) {$1$};
          \node (v2) at (1.5,0) {$2$}; 
          \node (v3) at (3,0) {$3$};
          \node (v4) at (1.5,-1.5) {$4$};
          \node (v5) at (3,-1.5) {$5$};
          \draw (0,0) circle [radius=6pt];
          \draw (1.5,0) circle [radius=6pt];
          \draw (3,0) circle [radius=6pt];
          \draw (3,-1.5) circle [radius=6pt];
          \draw (1.5,-1.5) circle [radius=6pt];
          \draw[->,very thick] (v1)--(v2);
          \draw[->,very thick] (v2)--(v3);
          \draw[->,very thick] (v4)--(v2);
          \draw[->,very thick] (v2)--(v4);
          \draw[->,very thick] (v4)--(v5);
          \draw[->,very thick] (v5)--(v2);
          \draw[->,very thick] (v5)--(v3);
    \end{tikzpicture}
\end{SCfigure}
Xét giao thức đồng thuận $\dot{\m{x}}=-\mcl{L}\m{x}$ với mạng mô tả bởi đồ thị như ở Hình~\ref{fig:Ex_3_2}.
\begin{itemize}
    \item[i.] Các tác tử trong hệ có đạt được đồng thuận khi $t\to +\infty$ hay không?
    \item[ii.] Nếu như các tác tử đạt được đồng thuận thì giá trị đồng thuận có trùng với giá trị trung bình cộng hay không?
    \item[iii.] Trình bày cách thêm vào đồ thị một cạnh để hệ đạt được đồng thuận tại giá trị trung bình cộng.
\end{itemize}
\end{exercise}

\begin{exercise} \label{eq:ex3_3}
Phân tích hệ đồng thuận khi tồn tại nhiễu hằng $d_i \in \mb{R}$ tác động vào hệ:
\begin{align}
    \dot{{x}}_i = -\sum_{j\in N_i} (x_j - x_i) + d_i,~i=1,\ldots,n,
\end{align}
trong trường hợp (a) $\m{d}=[d_1,\ldots,d_n]^\top \perp \text{im}(\m{1}_n)$, và (b) $\m{d}$ bất kỳ.
\end{exercise}

\begin{exercise} Xét hệ gồm $n$ tác tử tích phân bậc nhất tương tác qua một đồ thị vô hướng, liên thông $G$. Giả sử mỗi tác tử $i$ thực hiện cập nhật biến trạng thái với tốc độ khác nhau:
\begin{align} \label{eq:ex3_4}
    \dot{\m{x}}_i = -\gamma_i \sum_{j \in N_i} (\m{x}_i - \m{x}_j),~i=1,\ldots,n,
\end{align}
trong đó $\gamma_i > 0$ là hệ số cập nhật của tác tử thứ $i$ và $\m{x}_i\in \mb{R}^d$, $d\geq 1$. Kí hiệu $\m{x}=\text{vec}(\m{x}_1,\ldots,\m{x}_n)$.
\begin{enumerate}
    \item [i.] Hãy tìm các vector riêng bên trái và bên phải của ma trận $\bm{\Gamma}\mcl{L}$, trong đó $\bm{\Gamma} = \text{diag}(\gamma_1,\ldots,\gamma_n)$ và $\mcl{L}$ là ma trận Laplace của đồ thị $G$.
    \item [ii.] Giả sử $d=1$, hãy viết phương trình mô tả hệ \eqref{eq:ex3_5} ở dạng ma trận và tìm $\lim_{t\to +\infty} \m{x}(t)$.
    \item [iii.] Giả sử $d>1$, hãy viết phương trình mô tả hệ \eqref{eq:ex3_5} ở dạng ma trận và tìm $\lim_{t\to +\infty} \m{x}(t)$.
\end{enumerate}
\end{exercise}

\begin{exercise} \cite{Olfati2004consensus} \label{eq:ex3_5}
Xét thuật toán đồng thuận trên một đồ thị có trọng số với trễ cho bởi
\begin{align}
    \dot{x}_i = \sum_{j\in N_i} a_{ij} (x_j(t-\tau) - x_i(t-\tau)),~i=1,\ldots,n,
\end{align}
với $\tau>0$. Chứng minh rằng $x_i$ này tiệm cận tới cùng một giá trị đồng thuận (là giá trị trung bình cộng của $x_i(0)$, $i=1,\ldots,n$) nếu
\begin{align*}
    \tau < \frac{\pi}{2\lambda_n(G)},
\end{align*}
trong đó $\lambda_n(G)$ là giá trị riêng lớn nhất của ma trận Laplace tương ứng của $G$. Từ đó rút ra nhận xét rằng với các giao thức đồng thuận có trễ, tồn tại một sự đánh đổi giữa tốc độ hội tụ với giới hạn ổn định đối với trễ trên các cạnh.
\end{exercise}

\begin{exercise} \cite{MesbahiEgerstedt} \label{eq:ex3_6} Tích De Cartes của hai đồ thị $G_1=(V_1, E_1)$ và $G_2=(V_2,E_2)$ được kí hiệu bởi
\begin{align*}
    G = G_1 \scalebox{.84}{$\square$} G_2
\end{align*}
là một đồ thị có tập cạnh $V_1 \times V_2$, và mỗi cặp đỉnh $(v_1,v_2)$ và $(v_1',v_2')\in V(G)$ là kề nhau khi và chỉ khi hoặc $v_1 = v_1'$ và $(v_2,v_2')$ là một cạnh trong $E_2$, hoặc $v_2 = v_2'$ và $(v_1,v_1')$ là một cạnh của $E_1$. Với $|V_1|=n$ và $|V_2|=m$ thì ta có tính chất sau của tích De Cartes của hai đồ thị
\begin{align*}
    \mcl{L}(G_1 \scalebox{.84}{$\square$} G_2) &= \mcl{L}(G_1) \otimes \m{I}_m + \m{I}_n\otimes \mcl{L}(G_2) \\
    &= \mcl{L}(G_1) \otimes \m{I}_m + \m{I}_n\otimes \mcl{L}(G_2) = \mcl{L}(G_1) \oplus \mcl{L}(G_2),
\end{align*}
trong đó $\oplus$ kí hiệu tổng Kronecker của hai ma trận.
\begin{enumerate}
    \item[i.] Tìm tích Đề Các của hai đồ thị đường thẳng $L_n$ và $L_m$.
    \item[ii.] Tìm tích Đề Các của hai đồ thị $L_n$ và $C_m$.
    \item[iii.] Chứng minh rằng với hai ma trận vuông $\m{A}$ và $\m{B}$ thì $\mathtt{e}^{\m{A}\oplus\m{B}} = \mathtt{e}^{\m{A}} \otimes e^{\m{B}}$. Từ đó chứng minh rằng nếu $\m{x}_1(t)$ và $\m{x}_2(t)$ là nghiệm của các hệ đồng thuận
    \begin{align*}
        \dot{\m{x}}_1 &= - \mcl{L}(G_1)\m{x}_1,\\
        \dot{\m{x}}_2 &= - \mcl{L}(G_2)\m{x}_2,
    \end{align*}
    thì nghiệm của hệ đồng thuận
    \begin{align*}
        \dot{\m{x}} &= - \mcl{L}(G)\m{x}
    \end{align*}
    là $\m{x}(t) = \m{x}_1(t) \otimes \m{x}_2(t)$.
\end{enumerate}
\end{exercise}

\begin{exercise}[Thuật toán đồng thuận PI] \label{eq:ex3_7}
Xét hệ với $n$ tác tử cho bởi phương trình
\begin{align*}
    \dot{x}_i = u_i + d,~i = 1,\ldots, n,
\end{align*}
trong đó $u_i \in \mb{R}$ là tín hiệu điều khiển và $d_i \in \mb{R}$ là một hằng số. Hệ phương trình trên có thể mô tả hệ $n$ đồng hồ, trong đó $x_i(t)$ là biến chỉ thời gian, $d$ là mức thay đổi về số chỉ thời gian. Ta cần thiết kế luật đồng thuận $u_i$ sao cho với mọi giá trị đầu $x_i(0) \neq x_j(0)$, và giá trị $d>0$ cho trước, $x_i(t) \to x_j(t)$ khi $t\to +\infty$ (nói cách khác, các đồng hồ dần được đồng bộ hóa).
\begin{enumerate}
    \item[i.] Chứng minh rằng luật đồng thuận cho bởi:
\begin{align*}
    {u}_i &= - \sum_{j\in N_i} a_{ij}(x_i - x_j) + y_i,\\
    \dot{y}_i &= - \sum_{j\in N_i} a_{ij}(x_i - x_j),
\end{align*}
sẽ giải quyết bài toán đồng bộ hóa các đồng hồ.
    \item[ii.] Nếu luật đồng thuận được thay bởi:
\begin{align*}
    {u}_i &= - \sum_{j\in N_i} a_{ij}(x_i - x_j) + y_i,\\
    \dot{y}_i &= - \sum_{j\in N_i} a_{ij}(x_i - x_j) - \alpha y_i,
\end{align*}
với $\alpha>0$ thì kết quả sẽ thay đổi như thế nào?
\item[iii.] Mô phỏng hệ đồng thuận với các luật đồng thuận ở trên.
\end{enumerate}
\end{exercise}

\begin{exercise} \label{eq:ex3_8}
Vẽ sơ đồ hệ đồng thuận theo động học của tác tử $i$ cho hệ đồng thuận bậc hai \eqref{eq:c3_double_u} và \eqref{eq:c3_double_u1} (Tham khảo Hình~\ref{fig:c3ConsensusDiagramAgenti}).
\end{exercise}

\begin{exercise}[Đồng thuận theo tỉ lệ \cite{Roy2015scaled}] \label{eq:ex3_9} \index{thuật toán!đồng thuận!theo tỉ lệ}
Cho hệ đồng thuận gồm $n$ tác tử tương tác qua một đồ thị hữu hướng, có gốc ra. Mỗi tác tử tác tử $i$ trong hệ có một giá trị $s_i\neq 0$ và thực hiện cập nhật biến trạng thái theo phương trình:
\begin{align}
\dot{x}_i =- \text{sgn}(s_i) \sum_{j \in N_i} (s_i x_i -s_j x_j), ~i=1,\ldots, n.
\end{align}
\begin{enumerate}
    \item[i.] Hãy viết phương trình đồng thuận dưới dạng ma trận $\dot{\m{x}} =- \m{\Theta} \m{x}$.
    \item[ii.] Chứng minh rằng $\m{\Theta}$ chỉ có duy nhất một giá trị riêng bằng 0, các giá trị riêng khác của  $\m{\Theta}$ đều nằm bên phải trục ảo.
    \item[iii.] Chứng mình rằng $s_i x_i \to s_j x_j$ khi $t\to +\infty$, hay hệ sẽ dần đạt tới đồng thuận theo tỉ lệ.
\end{enumerate}
\end{exercise}

\begin{exercise}[Đồng thuận theo tỉ lệ ma trận \cite{Trinh2022MSC}] \label{ex:3.10} \index{thuật toán!đồng thuận!theo tỉ lệ ma trận}
Cho hệ đồng thuận gồm $n$ tác tử tương tác qua một đồ thị vô hướng, liên thông. Mỗi tác tử tác tử $i$ trong hệ có một giá trị $\m{S}_i\in \mb{R}^{d\times d}$ là một ma trận xác định dương hoặc xác định âm, không nhất thiết phải đối xứng. Định nghĩa hàm dấu $\text{sign}(\m{S}_i) = 1$ nếu $\m{S}_i$ là xác định dương và $\text{sign}(\m{S}_i) = - 1$ nếu $\m{S}_i$ là xác định âm. Mỗi tác tử thực hiện cập nhật biến trạng thái theo phương trình:
\begin{align}
    \dot{\m{x}}_i =- \text{sign}(\m{S}_i) \sum_{j \in N_i} (\m{S}_i \m{x}_i - \m{S}_j \m{x}_j), ~i=1,\ldots, n,
\end{align}
trong đó $\m{x}_i \in \mb{R}^d$ là vector biến trạng thái của tác  tử thứ $i$ trong hệ.
\begin{enumerate}
    \item[i.] Hãy viết phương trình đồng thuận dưới dạng ma trận $\dot{\m{x}} =- \m{\Theta} \m{x}$.
    \item[ii.] Chứng minh rằng $\bm{\Theta}$ chỉ có $d$ giá trị riêng bằng 0, các giá trị riêng khác của  $\bm{\Theta}$ đều nằm bên phải trục ảo.
    \item[iii.] Chứng minh rằng $\m{S}_i \m{x}_i \to \m{S}_j \m{x}_j$ khi $t\to \infty$, hay hệ sẽ dần đạt tới đồng thuận theo tỉ lệ.
\end{enumerate}
\end{exercise}

\begin{exercise}[Hệ đồng thuận các tác tử tích phân bậc $p\ge 2$ \cite{Ren2007HOConsensus}] \label{ex:3.11}
\begin{SCfigure}[][h]
      \caption{Đồ thị $G$ ở Bài tập \ref{ex:3.11}.}
      \label{fig:ex:3.11}
      \hspace{6cm}
      \begin{tikzpicture}[roundnode/.style={circle, draw=black, thick, minimum size=1.5mm,inner sep= 0.25mm},
squarednode/.style={rectangle, draw=red!60, fill=red!5, very thick, minimum size=5mm},
]
                \node[roundnode] (v1) at (1,0) {$1$};
                \node[roundnode] (v2) at (0.309,0.951) {$2$}; 
                \node[roundnode] (v3) at (-0.809,0.5878) {$3$};
                \node[roundnode] (v4) at (-0.809,-0.5878) {$4$};
                \node[roundnode] (v5) at (0.309,-0.951) {$5$};
                \draw[-,very thick] (v1)--(v3);
                \draw[-,very thick] (v3)--(v5);
                \draw[-,very thick] (v5)--(v2);
                \draw[-,very thick] (v2)--(v4);
                \draw[-,very thick] (v4)--(v1);
      \end{tikzpicture}
\end{SCfigure}
Xét hệ đồng thuận gồm $n$ tác tử tương tác qua một đồ thị vô hướng, liên thông. Mỗi tác tử trong hệ cập nhật biến trạng thái theo mô hình tích phân bậc $p\ge 2$:
\begin{align*}
\dot{x}_{i,1} &= x_{i,2},\\
&\vdots \\
\dot{x}_{i,p-1} &= x_{i,p},\\
\dot{x}_{i,p} &= u_i,
\end{align*}
trong đó $\m{x}_i = [x_{i,1},\ldots,x_{i,p}]^\top \in \mb{R}^p$ và $u_i$ là vector biến trạng thái và tín hiệu điều khiển của tác tử $i=1,\ldots,n$. Đặt $\m{x} = [\m{x}_1^\top,\ldots,\m{x}_n^\top]^\top \in \mb{R}^{np}$.
\begin{enumerate}
	\item[i.] Giả sử $p=2$ và $u_i = -k\sum_{j \in N_i} (x_{i,2}-x_{j,2})$, $i=1,\ldots,n$. Hệ có tiệm cận tới giá trị đồng thuận hay không?
    \item[i.] Hãy biểu diễn hệ $n$ tác tử dưới dạng ma trận khi $u_i = -\alpha_1 x_i -\sum_{k=2}^p \alpha_k\sum_{j \in N_i} (x_{i,k}-x_{j,k})$, $i=1,\ldots,n$. 
    \item[ii.] Tìm điều kiện của các tham số $\alpha_k$, $k=1,\ldots,p$ để hệ tiệm cận tới tập đồng thuận \[\mc{A}=\{\m{x}\in \mb{R}^{np}|x_{1,1}=\ldots=x_{n,1},\,x_{1,k}=\ldots=x_{n,k}=0,\, k=2,\ldots,p\}.\]
    \item[iii.] Thực hiện mô phỏng với một hệ gồm 5 tác tử, với đồ thị tương tác cho ở Hình~\ref{fig:ex:3.11}, $p=3$, và $\alpha_1,\alpha_2,\alpha_3$ tự chọn.
\end{enumerate}
\end{exercise}

\begin{exercise}[Đồng bộ hóa hệ lò xo - vật nặng - giảm tốc] \label{ex:3.12}
\begin{SCfigure}[][h!]
\caption{Đồ thị $G$ ở Bài tập \ref{ex:3.12}.}
\label{fig:ex:3.12}
\hspace{6cm}
\begin{tikzpicture}
      \node (v1) at (0,0) {$1$};
      \node (v2) at (1.5,0) {$2$}; 
      \node (v3) at (3,0) {$3$};
      \node (v4) at (3,-1.5) {$4$};
      \node (v5) at (1.5,-1.5) {$5$};
      \node (v6) at (0,-1.5) {$6$};
      \draw (0,0) circle [radius=6pt];
      \draw (1.5,0) circle [radius=6pt];
      \draw (3,0) circle [radius=6pt];
      \draw (3,-1.5) circle [radius=6pt];
      \draw (1.5,-1.5) circle [radius=6pt];
      \draw (0,-1.5) circle [radius=6pt];
      \draw[->,very thick] (v1)--(v2);
      \draw[->,very thick] (v2)--(v3);
      \draw[->,very thick] (v4)--(v3);
      \draw[->,very thick] (v1)--(v6);
      \draw[->,very thick] (v6)--(v1);
      \draw[->,very thick] (v5)--(v6);
      \draw[->,very thick] (v4)--(v5);
      \draw[->,very thick] (v5)--(v4);
\end{tikzpicture}
\end{SCfigure}
Xét hệ các tác tử tuyến tính tổng quát, trong đó mô hình của các tác tử được cho bởi phương trình
    \begin{align}
        \dot{\m{x}}_i(t) = \begin{bmatrix}
            0 & 1\\ -\frac{k}{m} & -\frac{D}{m}
        \end{bmatrix} \m{x}_i(t) + \begin{bmatrix}
        0\\ 1    
        \end{bmatrix}u_i(t),\; i=1,\ldots,n,
    \end{align}
    với khối lượng $m=10.6$ (kg), hệ số giảm tốc $D=50.5$ ($\frac{N\cdot s}{m}$), và độ cứng lò xo $k=120$ (N/m).
    \begin{itemize}
        \item [i.] Giả sử các tác tử tương tác qua đồ thị $G$ như ở Hình~\ref{fig:ex:3.12}. Hãy viết ma trận Laplace tương ứng của đồ thị và tìm các giá trị riêng tương ứng.
        \item [ii.] Với luật đồng thuận dạng:
        \begin{align}
  u_i = \m{K}_1 \m{x}_i + \m{K}_2 \sum_{j\in N_i} (\m{x}_j - \m{x}_i),~i=1,\ldots,6,
        \end{align}
        hãy thiết kế các ma trận $\m{K}_1$ và $\m{K}_2$ để đồng bộ hóa hệ về một quĩ đạo (a) ổn định tiệm cận, hoặc (b) một quĩ đạo có dạng dao động điều hòa.
        \item [iii.] Mô phỏng hệ với MATLAB trong mỗi trường hợp.
    \end{itemize}
\end{exercise}

\begin{exercise}[Đồng thuận lệch]\cite{Ahn2019biasconsensus} \label{ex:3.13} Xét hệ đồng thuận cho bởi phương trình \index{đồng thuận!lệch}
    \begin{align}
    \dot{\m{x}}_i = - \m{R}_{i} \sum_{j \in N_i} (\m{x}_i - \m{x}_j), ~i=1,\ldots, n,
\end{align}
trong đó ${\m{x}}_i \in \mb{R}^2$, $\m{R}_i = \begin{bmatrix}
    \cos(\phi_i) & -\sin(\phi_i) \\ \sin(\phi_i)  & \cos(\phi_i)
\end{bmatrix} \in \mb{R}^{2 \times 2}$ là một ma trận quay, và đồ thị tương tác $G$ là vô hướng, liên thông. 
\begin{enumerate}
    \item [i.] Biểu diễn hệ dưới dạng ma trận.
    \item [ii.] Tìm điều kiện của các ma trận $\m{R}_i$ để hệ tiệm cận tới một giá trị đồng thuận. Xác định giá trị đồng thuận này.
    \item [iii.] Mô phỏng kiểm chứng với một số trường hợp cụ thể.
\end{enumerate}
\end{exercise}

\begin{exercise}[Hệ đồng thuận dạng leader-follower] \label{ex:3.14} Xét hệ gồm $n$ tác tử với đồ thị tương tác $G$ là vô hướng, liên thông. Giả sử các tác tử trong hệ thực hiện cập nhật biến trạng thái theo phương trình
    \begin{align}
    \dot{\m{x}}_i &= 0, ~i=1,\ldots, l,\nonumber\\
    \dot{\m{x}}_i &= - \sum_{j \in N_i} (\m{x}_i - \m{x}_j), ~i=l+1,\ldots, n,\nonumber
\end{align}
trong đó ${\m{x}}_i \in \mb{R}^d$, $d\ge 1$, và $l\ge 1$. Với thuật toán trên, $l$ tác tử $i=1,\ldots,l$ là các tác tử leader, còn các tác tử $i=l+1,\ldots,n$ là các follower. Leader giữ nguyên trạng thái ban đầu, trong khi đó follower cập nhật biến trạng thái dựa trên biến trạng thái của các tác tử láng giềng trong $G$ (bao gồm cả leader và follower).
\begin{enumerate}
    \item [i.] Đặt $\m{x}(t)=[x_1,\ldots,x_n]^\top \in \mb{R}^n$. Hãy viết phương trình động học dạng ma trận thể hiện thay đổi của $\m{x}$ theo thời gian.
    \item [ii.] Chứng minh các biến trạng thái của các tác tử follower tiệm cận tới các giá trị hằng khi $t\to +\infty$. Xác định giá trị vector $\m{x}^* = \lim_{t \to +\infty}\m{x}(t)$.
    \item [iii.] Chứng minh các giá trị $x_i^*,i=l+1,\ldots,n$ của các follower nằm trong bao lồi của $\m{x}_1(0),\ldots,\m{x}_{l}(0)$ (vị trí các tác tử leader). 

Bao lồi của các điểm $\m{x}_1(0),\ldots,\m{x}_{l}(0)$ là tập định nghĩa bởi
\begin{align*}
{\rm conv}(\m{x}_1(0),\ldots,\m{x}_{l}(0))=\{\m{q}\in \mb{R}^d|~ \m{q}= \sum_{i=1}^l \alpha_i \m{x}_i(0),\, \sum_{i=1}^l \alpha_i = 1, \alpha_i \ge 0,\, \forall i=1,\ldots,l\}
\end{align*}
    \item[iv.] Mô phỏng với một hệ cụ thể.
\end{enumerate}
\end{exercise}
\chapter[Phương pháp Lyapunov và hệ đồng thuận cạnh]{Phân tích hệ đồng thuận theo phương pháp Lyapunov và quá trình đồng thuận cạnh}
\label{chap:lyapunov}
Trong chương này, ta nghiên cứu hệ đồng thuận sử dụng công cụ là lý thuyết ổn định Lyapunov. Mặc dù các hệ đồng thuận giới thiệu ở chương trước khá đơn giản và có thể phân tích trực tiếp từ lý thuyết ma trận, phương pháp phân tích theo lý thuyết Lyapunov cho phép ta nhìn nhận sâu sắc hơn về quá trình tiến tới đồng thuận, từ đó mở rộng các kết quả của chương trước \cite{Zhang2011lyapunov}. 

Tiếp theo, thay vì quan tâm tới động học các biến trạng thái $x_i$, ta sẽ xét đến các biến tương đối $(x_i - x_j)$ ứng với các cạnh $(j,i)$ trong hệ. Khi biểu diễn động học theo các cạnh, mô hình trạng thái của hệ đồng thuận cạnh có thể được biểu diễn nhờ ma trận Laplace cạnh. Việc phân tích quá trình đồng thuận theo động học cạnh chỉ ra rằng, ta luôn có thể biểu diễn quá trình đồng thuận bởi một mô hình rút gọn gồm $(n-1)$ phương trình vi phân ứng với các cạnh của một cây bao trùm của đồ thị $G$, còn động học của $(m-n+1)$ cạnh còn lại có thể suy từ $(n-1)$ phương trình nói trên.

Mở rộng hơn, mô hình đồng thuận theo cạnh có tính thụ động. Dựa trên tính chất của hệ thụ động, ta có thể phân tích hệ đồng thuận với các tác tử và các tương tác phi tuyến. Việc đưa đầu ra của một hệ đa tác tử về đồng thuận được gọi chung là bài toán đồng thuận đầu ra. Chúng ta sẽ xét một ví dụ về đồng thuận đầu ra (pha của các dao động tử) thông qua mô hình Kuramoto đơn giản.

\section{Hàm bất đồng thuận}
\label{c4_sec1_disagreement}
Xét hệ đồng thuận gồm $n$ tác tử với mô hình là khâu tích phân bậc nhất, với đồ thị thông tin $G$ và luật đồng thuận:
\begin{align} \label{eq:c4_consensus}
\dot{\m{x}}(t) = - \mcl{L} \m{x}(t).
\end{align}
Như đã định nghĩa ở chương trước, không gian đồng thuận được định nghĩa bởi $\mc{A}=\{\m{x} \in \mb{R}^n| x_1 = \ldots = x_n\}$. Ta có thể định nghĩa mức độ bất đồng thuận giữa hai tác tử $i,j$ bất kì bởi $(x_i-x_j)^2$ với nhận xét rằng khi $(x_i-x_j)^2 = 0$, hai tác tử $i$ và $j$ đồng thuận. 

Xét đồ thị $G$ là vô hướng thì $\mcl{L} = \mcl{L}^\top$. Ta có thể định nghĩa \emph{hàm bất đồng thuận} của hệ bởi $V_L = \m{x}^\top \mcl{L} \m{x}$. Viết lại $\mcl{L} = \m{H}^\top \text{diag}(a_{ij}) \m{H}$, ta có $\m{x}^\top \m{H}^\top \text{diag}(a_{ij}) \m{H} \m{x} = \sum_{(i,j) \in E} a_{ij}(x_i - x_j)^2$. Rõ ràng, khi đồ thị $G$ là liên thông thì hệ đạt được đồng thuận khi và chỉ khi $V_L = 0$.\index{hàm bất đồng thuận}

Tiếp theo, xét $G$ là đồ thị hữu hướng và cân bằng. Khi đó $\text{deg}^+(v_i) = \text{deg}^-(v_i)$ hay $\sum_{j=1}^n a_{ij} = \sum_{i=1}^n a_{ji}$. Do đó, ta có thể viết các tổng
\begin{align*}
\m{x}^\top \mcl{L} \m{x} &= \sum_{i=1}^n \Big( x_i^2 \sum_{j=1}^n a_{ij} \Big) - \sum_{i=1}^n\sum_{j=1}^n a_{ij} x_i x_j\\
&=\sum_{i=1}^n\sum_{j=1}^n a_{ij}x_i(x_i - x_j),\\
\m{x}^\top \mcl{L}^\top \m{x} &= \sum_{i=1}^n \Big( x_i^2 \sum_{j=1}^n a_{ji} \Big) - \sum_{i=1}^n\sum_{j=1}^n a_{ij} x_i x_j\\
&=\sum_{j=1}^n \sum_{i=1}^n a_{ij} x_j^2 - \sum_{i=1}^n\sum_{j=1}^n a_{ij} x_i x_j\\
&=\sum_{i=1}^n\sum_{j=1}^n a_{ij}x_j(x_j - x_i).
\end{align*}
Như vậy, ta có thể chọn $V_L = \frac{1}{2} \m{x}^\top (\mcl{L} + \mcl{L}^\top) \m{x} = \sum_{i,j=1}^n a_{ij} (x_i - x_j)^2$. Nếu đồ thị $G$ được giả thiết là liên thông yếu thì $V_L=0$ khi và chỉ khi $x_i = x_j$, $\forall i, j = 1, \ldots, n$, $i \neq j$, hay nói cách khác, hệ~\eqref{eq:c4_consensus} ở trạng thái đồng thuận. 

Cuối cùng, xét trường hợp $G$ là hữu hướng và liên thông mạnh. Khi đó, theo Định lý \ref{thm:c2_Laplace_directed},  ma trận Laplace $\mcl{L}$ có một giá trị riêng đơn $\lambda_1 = 0$, với vector riêng bên phải $\bm{\gamma}^\top$ có tất cả các phần tử dương. Xét ma trận $\m{M} = \bm{\Gamma}\mcl{L} + \mcl{L}^\top\bm{\Gamma}$, trong đó $\bm{\Gamma} = \text{diag}(\gamma_i)$, ta có $\m{M}$ là một ma trân đối xứng. Hơn nữa, 
\begin{align*}
\m{x}^\top\bm{\Gamma} \mcl{L} \m{x} &= \sum_{i=1}^n\gamma_i x_i\sum_{j=1}^n a_{ij}(x_i -x_j)\\
&=\sum_{i=1}^n\sum_{j=1}^n \gamma_i a_{ij} x_i(x_i - x_j).
\end{align*}
Từ $\bm{\gamma}^\top\mcl{L} = \m{0}^\top$, ta suy ra  $\gamma_i \sum_{j=1}^n a_{ij} = \sum_{j=1}^n \gamma_j a_{ji}$. Do đó,
\begin{align*}
\sum_{i=1}^n \gamma_i x_i \sum_{j=1}^n a_{ij}(x_i -x_j)
&=\sum_{i=1}^n \gamma_i x_i^2 \sum_{j=1}^n a_{ij}  -\sum_{i=1}^n \sum_{j=1}^n \gamma_i a_{ij} x_i  x_j\\
&=\sum_{i=1}^n  x_i^2 \sum_{j=1}^n \gamma_j a_{ji}  -\sum_{i=1}^n \sum_{j=1}^n \gamma_i a_{ij} x_i  x_j\\
&=\sum_{j=1}^n  x_j^2 \sum_{i=1}^n \gamma_i a_{ij}  -\sum_{i=1}^n \sum_{j=1}^n \gamma_i a_{ij} x_i x_j\\
&=\sum_{i=1}^n \sum_{j=1}^n \gamma_i a_{ij} x_j (x_j - x_i).
\end{align*}
Như vậy, 
\begin{align*}
V_L &= \m{x}^\top (\bm{\Gamma}\mcl{L} +\mcl{L}^\top\bm{\Gamma})\m{x} = 2\m{x}^\top \bm{\Gamma} \mcl{L} \m{x}\\
&= \sum_{i=1}^n\sum_{j=1}^n \gamma_i a_{ij} x_i(x_i - x_j) + \sum_{i=1}^n \sum_{j=1}^n \gamma_i a_{ij} x_j (x_j - x_i)\\
&= \sum_{i=1}^n\sum_{j=1}^n \gamma_i a_{ij} (x_i - x_j)^2.
\end{align*}
Ta chứng minh rằng với $G$ là liên thông mạnh thì $\text{ker}(\m{M}) = \text{ker}(\mcl{L}) = \text{im}(\m{1}_n)$. Thật vậy, nếu $\m{x}$ là vector riêng ứng với trị riêng $0$ của $\mcl{L}$ thì $\m{x}^\top\m{M}\m{x} = 2 \m{x}^\top \bm{\Gamma}\mcl{L} \m{x} = {0}$, suy ra $\m{x} \in \text{ker}(\m{M})$. Vì $\text{ker}(\mcl{L}) = \text{im}(\m{1}_n)$, ta suy ra $\text{im}(\m{1}_n) \subseteq \text{ker}(\m{M})$. Ma trận $\m{M}$ là ma trận Laplace ứng với đồ thị $\bar{G}$ với cùng một tập đỉnh như $G$ và với trọng số của cạnh $(v_i,v_j)$ là $\bar{a}_{ij} = \gamma_i a_{ij} + \gamma_j a_{ji}$. Dễ thấy rằng $\bar{G}$ là một đồ thị vô hướng. Vì $G$ là liên thông mạnh, $\bar{G}$ là liên thông mạnh. Điều này dẫn đến $\text{rank}(\m{M}) = n-1$, và $\text{ker}(\m{M}) = \text{im}(\m{1}_n)$. 
Điều này chứng tỏ hệ đạt đồng thuận khi và chỉ khi $V_L = 0$.

\begin{example}\label{VD4.1}
Trên hình \ref{fig:VD4.1}(a), $G$ là đồ thị hữu hướng với ma trận Laplace cho bởi
\begin{align*}
\mcl{L}(G) = \begin{bmatrix}
1 & 0 & 0 & -1\\
-0.5 & 0.5 & 0 & 0\\
-0.25 & -0.5 & 0.75 & 0\\
0 & 0 & -0.75 & 0.75
\end{bmatrix}.
\end{align*}
Dễ thấy $G$ là một đồ thị liên thông mạnh và không cân bằng. Ma trận $\mcl{L}=\mcl{L}(G)$ có một vector riêng bên trái $\bm{\gamma}^\top = [\frac{3}{4},1,1,1]$ tương ứng với giá trị riêng 0. Trên hình \ref{fig:VD4.1}(b), $G'$ là đồ thị hữu hướng với ma trận Laplace không trọng số có các phần tử 0, âm, dương tương ứng với $\mcl{L}^\top$. Cuối cùng, Hình~\ref{fig:VD4.1}(c) là đồ thị vô hướng tương ứng với ma trận 
\[\m{M} = \bm{\Gamma} \mcl{L} + \mcl{L}^\top \bm{\Gamma}=\begin{bmatrix}
    1.5 &  -0.5 &  -0.25 &  -0.75\\
   -0.5 &   1   &  -0.5  &   0  \\
   -0.25&  -0.5 &   1.5  & -0.75 \\
   -0.75&   0   & -0.75  &  1.5
\end{bmatrix} = \bar{\mcl{L}}(\bar{G}).\]
\end{example}

\begin{figure}
\centering
\subfloat[${G}$]{
\centering
\begin{tikzpicture}[
roundnode/.style={circle, draw=black, thick, minimum size=1.5mm,inner sep= 0.25mm},
squarednode/.style={rectangle, draw=red!60, fill=red!5, very thick, minimum size=5mm},
]
\node[roundnode]   (v2)   at   (0,0) {};
\node[roundnode]   (v3)   at   (0,2) {};
\node[roundnode]   (v4)   at   (-2,2) {};
\node[roundnode]   (v1)   at   (-2,0) {};
\draw[->, thick] (v1)--(v2) node [midway, below] {0.5};
\draw[->, thick] (v1)--(v3) node [midway, right] {0.25};
\draw[->, thick] (v4)--(v1) node [midway, left] {1};
\draw[->, thick] (v2)--(v3) node [midway, right] {0.5};
\draw[->, thick] (v3)--(v4) node [midway, above] {0.75};
\node (a1) at (0.3,-0.3){$v_2$};
\node (a2) at (0.3,2.2){$v_3$};
\node (a3) at (-2.3,2.2){$v_4$};
\node (a4) at (-2.3,-0.3){$v_1$};
\end{tikzpicture}
}\hfill
\subfloat[${G'}$]{
\begin{tikzpicture}[
roundnode/.style={circle, draw=black, thick, minimum size=1.5mm,inner sep= 0.25mm},
squarednode/.style={rectangle, draw=red!60, fill=red!5, very thick, minimum size=5mm},
]
\node[roundnode]   (v2)   at   (0,0) {};
\node[roundnode]   (v3)   at   (0,2) {};
\node[roundnode]   (v4)   at   (-2,2) {};
\node[roundnode]   (v1)   at   (-2,0) {};
\draw[<-, thick] (v1)--(v2) ;
\draw[<-, thick] (v1)--(v3) ;
\draw[<-, thick] (v4)--(v1) ;
\draw[<-, thick] (v2)--(v3) ;
\draw[<-, thick] (v3)--(v4) ;
\node (a1) at (0.3,-0.3){$v_2$};
\node (a2) at (0.3,2.2){$v_3$};
\node (a3) at (-2.3,2.2){$v_4$};
\node (a4) at (-2.3,-0.3){$v_1$};
\end{tikzpicture}
}\hfill
\subfloat[$\bar{G}$]{
\begin{tikzpicture}[
roundnode/.style={circle, draw=black, thick, minimum size=1.5mm,inner sep= 0.25mm},
squarednode/.style={rectangle, draw=red!60, fill=red!5, very thick, minimum size=5mm},
]
\node[roundnode]   (v2)   at   (0,0) {};
\node[roundnode]   (v3)   at   (0,2) {};
\node[roundnode]   (v4)   at   (-2,2) {};
\node[roundnode]   (v1)   at   (-2,0) {};
\draw[-, thick] (v1)--(v2) node [midway, below] {$2/3$};
\draw[-, thick] (v1)--(v3) node [midway, right] {$1/3$};
\draw[-, thick] (v4)--(v1) node [midway, left] {$1$};
\draw[-, thick] (v2)--(v3) node [midway, right] {$1/3$};
\draw[-, thick] (v3)--(v4) node [midway, above] {$1$};
\node (a1) at (0.3,-0.3){$v_2$};
\node (a2) at (0.3,2.2){$v_3$};
\node (a3) at (-2.3,2.2){$v_4$};
\node (a4) at (-2.3,-0.3){$v_1$};
\end{tikzpicture}
}
\caption{(a) Đồ thị $G$ ứng với $\mcl{L}$.~(b) Đồ thị $\bar{G}$ ứng với $\bar{\mcl{L}} = \bm{\Gamma} \mcl{L} + \mcl{L}^\top \bm{\Gamma}$.}
\label{fig:VD4.1}
\end{figure}

\begin{story}{Francesco Bullo và cuốn sách ``Lectures on Network Systems''}
Francesco Bullo nhận bằng cử nhân kĩ thuật điện tại Đại học Padova, Ý, và bằng Tiến sỹ (1998) từ Caltech, Hoa Kỳ. Nghiên cứu hiện tại của Bullo về điều khiển nối mạng với các ứng dụng trong kĩ thuật, xã hội học, và các hệ vật lý. Hiện tại Bullo là giáo sư tại Khoa
Trước đó, nghiên cứu của Bullo là về điều khiển robot.

Các nghiên cứu về điều khiển nối mạng của Bullo có sự hợp tác liên ngành: nghiên cứu về mạng xã hội với Noah Friendkin, về hệ thống điện thông minh với F. D\"{o}fler, và các nghiên cứu toán ứng dụng trong điều khiển cùng J. Cortes và S. Martinez. 

Bullo tiên phong về viết giáo trình cho một số chủ đề gắn với hướng nghiên cứu của bản thân và công bố mở, miễn phí trên website cá nhân. Các giáo trình được soạn với hình thức đẹp và dựa trên triết lý của Richard Hamming: ``Thế hệ sau sẽ trân trọng những giáo trình hệ thống hóa các khái niệm cốt lõi, làm rõ các tư tưởng nền tảng, và lược bỏ những phương pháp kém hiệu quả cũng như những nội dung trùng lặp.''

``Các bài giảng về điều khiển nối mạng'' \cite{Bullo2019lectures} (viết tắt: LNS) là giáo trình được đăng công khai và dần hoàn thiện từ năm 2016 tới nay. Nội dung giáo trình gồm ba phần: Phần I đề cập tới lý thuyết đồ thị và hệ đồng thuận tuyến tính, với phần lớn nội dung đề cập tới các thuật toán đồng thuận rời rạc. Phần II đề cập tới một số chủ đề nâng cao trong hệ đồng thuận tuyến tính và phần III trình bày một số hệ nối mạng phi tuyến. So với các tài liệu hiện có về Điều khiển hệ đa tác tử, giáo trình của Bullo cung cấp một lượng bài tập khá lớn, hỗ trợ cho phần lý thuyết được trình bày khá tỉ mỉ với nhiều ví dụ minh họa. 

Giáo trình LNS của Bullo được phân phối tại \url{https://fbullo.github.io/lns/}. 
\end{story}

\section[Phân tích theo phương pháp Lyapunov]{Phân tích quá trình đồng thuận theo phương pháp Lyapunov}
\label{c4_sec2_Lyap}
Dựa vào các kết quả ở mục trước, hàm bất đồng thuận $V_L$ có thể được sử dụng như một hàm thế Lyapunov. Chúng ta sẽ phân tích quá trình đồng thuận $\dot{\m{x}}(t) = -\mcl{L}\m{x}(t)$ với các giả thiết khác nhau về đồ thị $G$ theo phương pháp Lyapunov ở mục này. 

Đầu tiên, nếu đồ thị $G$ là vô hướng và liên thông, ma trận Laplace là đối xứng và bán xác định dương, với ${\rm ker}(\m{L})=\text{im}(\m{1}_n)$. Xét hàm Lyapunov $V = \frac{1}{2}\m{x}^\top \m{x}$, khi đó,
\begin{equation*} 
\dot{V} =  \m{x}^\top \dot{\m{x}} = -\m{x}^\top \mcl{L}\m{x} \leq 0.
\end{equation*}
Theo nguyên lý bất biến LaSalle, các quĩ đạo tiệm cận tới tập bất biến lớn nhất trong $\mc{S} = \{\m{x} \in \mb{R}^n|~\mcl{L} \m{x}= \m{0}\}$ \cite{Khalil2002}. Vì đồ thị là liên thông, ta suy ra $\mc{S} = \text{im}(\m{1}_n)$. Do đó, $\m{x}(t) \to \alpha \m{1}_n$ với $\alpha \in \mb{R}$ khi $t \to +\infty$, hay hệ tiệm cận tới tập đồng đồng thuận. Dựa vào định lý~\ref{thm:c3_invariant}, ta suy ra:
\begin{itemize}
\item[i.] tập bất biến ứng với quĩ đạo $\m{x}(0)$ chỉ bao gồm phần tử ${x}^\ast = \frac{1}{n}\sum_{i=1}^nx_i(0)$.
\item[ii.] $\m{1}^\top_n(\m{x}(t) - {x}^\ast \m{1}_n) = 0$, hay $(\m{x}(t) - {x}^\ast \m{1}_n) \perp {\rm ker}(\mcl{L})$.
\end{itemize}
Từ (i), ta suy ra $\m{x} \to x^\ast \m{1}_n$ khi $t \to +\infty$. Từ (ii), chọn $W = \frac{1}{2}(\m{x}-x^\ast\m{1}_n)^\top (\m{x}-x^\ast\m{1}_n)$ ta có:
\begin{align*}
\dot{W} &= -(\m{x} - {x}^\ast \m{1}_n)^{\top} \mcl{L} \m{x}\\
&= -(\m{x} - {x}^\ast \m{1}_n)^{\top} \mcl{L} (\m{x} - {x}^\ast \m{1}_n)\\
&\leq - \lambda_2(\mcl{L}) (\m{x} - {x}^\ast \m{1}_n)^\top(\m{x} - {x}^\ast \m{1}_n)\\
&\leq -2 \lambda_2(\mcl{L}) W.
\end{align*}
Từ đó ta kết luận rằng $\m{x}(t)$ tiệm cận tới ${x}^\ast \m{1}_n$ theo tốc độ hàm mũ. Tốc độ hội tụ của $\m{x}(t) - x^*\m{1}_n$ phụ thuộc vào $\lambda_2(\mcl{L})$ - giá trị riêng dương nhỏ nhất của ma trận Laplace.

Tiếp theo, xét đồ thị $G$ là hữu hướng, cân bằng, và liên thông yếu. Lúc này, với hàm Lyapunov $V = \m{x}^\top \m{x}$, ta có 
\begin{align*}
\dot{V} = - \m{x}^\top (\mcl{L} + \mcl{L}^\top) \m{x}.
\end{align*}
Với kết quả từ mục \ref{c4_sec2_Lyap}, ta có $\dot{V} \leq 0$ và $\dot{V} = 0$ khi và chỉ khi $\m{x} \in \mc{A}$. Theo nguyên lý bất biến LaSalle, $\m{x}(t) \to \mc{A}$ khi $t \to +\infty$. 

Xét đồ thị $G$ hữu hướng và liên thông mạnh. Chọn hàm Lyapunov $V = \m{x}^\top \bm{\Gamma} \m{x}$, ta có:
\begin{align*}
\dot{V} = - \m{x}^\top (\bm{\Gamma} \mcl{L} + \mcl{L}^\top \bm{\Gamma}) \m{x}.
\end{align*}
Cũng từ kết quả ở mục \ref{c4_sec2_Lyap}, ta có $\dot{V} \leq 0$,  $\dot{V} = 0$ khi và chỉ khi $\m{x} \in \mc{A}$, và do đó $\m{x} \to \mc{A}$ khi $t \to +\infty$. 

Như vậy, vector riêng bên trái ứng với giá trị riêng $0$ đóng vai trò quan trọng trong việc chọn hàm Lyapunov. Nói cách khác, phương trình Lyapunov ứng với hệ~\eqref{eq:c4_consensus} có nghiệm là $\bm{\Gamma} = \text{diag}(\bmm{\gamma})$. 

Đối với đồ thị có một cây có gốc-vào nhưng không liên thông mạnh, vector riêng bên trái $\bmm{\gamma}^\top$ của $\mcl{L}$ có các phần tử bằng $0$ ứng với các đỉnh không thuộc thành phần liên mạnh tới hạn chứa gốc ra của $G$. Ta có thể đánh dấu các đỉnh nằm trong thành phần liên thông mạnh chứa gốc vào bởi $v_1, \ldots, v_l$. Các đỉnh còn lại là $v_{l+1}, \ldots, v_n$. Từ đó, ta có thể viết ma trận $\mcl{L}$ dưới dạng:
\begin{align*}
\mcl{L} = \begin{bmatrix}
\mcl{L}_{11} & \m{0}\\
\mcl{L}_{21} & \mcl{L}_{22}
\end{bmatrix},
\end{align*}
trong đó $\mcl{L}_{11} \in \mb{R}^{l \times l}$, $\mcl{L}_{21} \in \mb{R}^{(n-l)\times l}$, $\mcl{L}_{22} \in \mb{R}^{(n-l)\times(n-l)}$. Ma trận $\m{0}$ ở trong biểu diễn của $\mcl{L}$ thể hiện rằng không có thông tin đi từ các đỉnh $v_{l+1}, \ldots, v_n$ tới các đỉnh thuộc thành phần liên thông mạnh chứa gốc vào. Từ đó, ta có thể phân tích quá trình đồng thuận của các tác tử thuộc thành phần liên thông mạnh chứa gốc vào riêng rẽ như đã trình bày ở trên. Không mất tính tổng quát, ta có thể giả thiết đồ thị tương ứng với $G-\{v_1, \ldots, v_l\}$ là liên thông mạnh, các trường hợp khác đều có thể phân tích tương tự trường hợp này. Rõ ràng, $\text{det}(\lambda\m{I}_n - \mcl{L}) = \text{det}(\lambda\m{I}_l - \mcl{L}_{11})\text{det}(\lambda\m{I}_{n-l} - \mcl{L}_{22})=\prod_{i=1}^n(\lambda-\lambda_i)$. Vì $G$ là một đồ thị có gốc vào, $\mcl{L}$ có một giá trị riêng đơn $\lambda_1= 0$, và $\text{Re}(\lambda_i)>0$, $\forall i = 2, \ldots, n$. Do $\mcl{L}_{11}$ là một ma trận Laplace tương ứng với đồ thị liên thông mạnh $G - \{v_{l+1}, \ldots, v_n\}$ nên nghiệm đơn $\lambda_1 = 0$ là nghiệm của đa thức $\text{det}(\lambda\m{I}_l - \mcl{L}_{11})$. Do đó, các nghiệm của đa thức $\text{det}(\lambda\m{I}_{n-l} - \mcl{L}_{22})$ đều có phần thực dương, hay $-\mcl{L}_{22}$ là Hurwitz.

Đặt $\m{y} = [x_1, \ldots, x_l]^\top$ và $\m{z} = [x_{l+1}, \ldots, x_{n}]^\top$ thì
\begin{align}
\dot{\m{y}} &= - \mcl{L}_{11} \m{y}, \label{eq:c4_y_dynamic}\\
\dot{\m{z}} &= -\mcl{L}_{22} \m{z} - \mcl{L}_{21} \m{y}. \label{eq:c4_z_dynamic}
\end{align}
Ta xét hệ khi các tác tử $1, \ldots, l,$ đã ở trạng thái đồng thuận, tức là lúc $\m{y} = x^\ast \m{1}_l$. Kí hiệu $\mcl{L}'$ là ma trận Laplace ứng với đồ thị $G' = G - \{v_{1}, \ldots, v_l\}$, ta có thể viết $\mcl{L}_{22} = \mcl{L}' + \text{diag}(-\mcl{L}_{21}\m{1}_{n-l})$. Vì $-\mcl{L}_{22}$ là Hurwitz, hệ tuyến tính $\dot{\m{z}} = -\mcl{L}_{22} \m{z} - \mcl{L}_{21} \m{1}_{n-l} x^\ast$ có \[\lim_{t \to +\infty} \m{z}(t) =- \mcl{L}_{22}^{-1}\mcl{L}_{21} \m{1}_{n-l} x^\ast.\]
Chú ý rằng $\mcl{L}_{22} \m{1}_{n-l} = (\mcl{L}' + \text{diag}(-\mcl{L}_{21}\m{1}_{n-l})) \m{1}_{n-l} = -\mcl{L}_{21}\m{1}_{n-l}$. Do đó, $\lim_{t \to +\infty} \m{z}(t) = \m{1}_{n-l} x^\ast$.

Cuối cùng, ta có thể định nghĩa $\m{q} = [q_1, \ldots, q_{n-l}]^\top = \mcl{L}_{22}^{-1} \m{1}_{n-l}$, $\m{P} = \text{diag}(p_i) = \text{diag}(1/q_i)$, $\m{Q} = \m{P}\mcl{L}_{22} + \mcl{L}_{22}^\top \m{P}$, thì $\m{P}$ và $\m{Q}$ là các ma trận đối xứng xác định dương và ta có thể chọn hàm Lyapunov cho \eqref{eq:c4_z_dynamic} bởi $V_z = \m{z}^\top \m{P} \m{z}$.

\begin{example}[Mô phỏng hệ đồng thuận với một số đồ thị hữu hướng khác nhau] \label{VD_4.2} Trong ví dụ này, ta mô phỏng thuật toán đồng thuận với một số đồ thị có hướng khác nhau. Các kết quả mô phỏng được cho trong Hình~\ref{fig:VD4.2}.

Trong mô phỏng thứ nhất, giá trị đầu được cho bởi $\m{x}(0)=[1, 2, 3, 4, 5]^\top$. Đồ thị $G_1$ là chu trình có hướng gồm 5 đỉnh. Dễ thấy $G_1$ là đồ thị cân bằng. Do đó, các biến trạng thái của các tác tử dần đồng thuận tại giá trị trung bình cộng $\frac{1}{5}\sum_{i=1}^5{x}_i(0)=3$.

Trong mô phỏng thứ hai và thứ ba, giá trị đầu của hệ được cho bởi $\m{x}(0)=[1, 2, 3, 4, 5, 6]^\top$. Đồ thị $G_2$ là đồ thị có gốc ra. Thành phần liên thông mạnh chứa gốc ra là chu trình có hướng gồm 3 đỉnh 1, 2, 3. Do đó, các tác tử tiệm cận tới giá trị trung bình cộng $\frac{1}{3}(x_1(0)+x_2(0)+x_3(0))=2$. Cuối cùng, đồ thị $G_3$ có gốc ra duy nhất là đỉnh $v_1$ nên các tác tử dần đạt tới đồng thuận tại $x_1(0)=1$.
\end{example}
\begin{figure*}
    \centering
    \subfloat[$G_1$]{
    \begin{tikzpicture}[
    roundnode/.style={circle, draw=black, thick, minimum size=2mm,inner sep= 0.25mm},
    squarednode/.style={rectangle, draw=black, thick, minimum size=3.5mm,inner sep= 0.25mm},
    ]
    \node[roundnode] (u1) at (0,0) {$1$}; %
    \node[roundnode] (u2) at (1,1) {$2$};%
    \node[roundnode] (u3) at (2,1) {$3$};%
    \node[roundnode] (u4) at (2,-1) {$4$};%
    \node[roundnode] (u5) at (1,-1) {$5$};%
    
    \draw [very thick,-{Stealth[length=2mm]}]
    (u1) edge [bend left=0] (u2)
    (u2) edge [bend left=0] (u3)
    (u3) edge [bend left=0] (u4)
    (u4) edge [bend left=0] (u5)
    (u5) edge [bend left=0] (u1)
    ;
    \node (u) at (-1,-1.5) { }; %
\end{tikzpicture}    
    }\hfill 
    \subfloat[]{\includegraphics[width=.45 \textwidth]{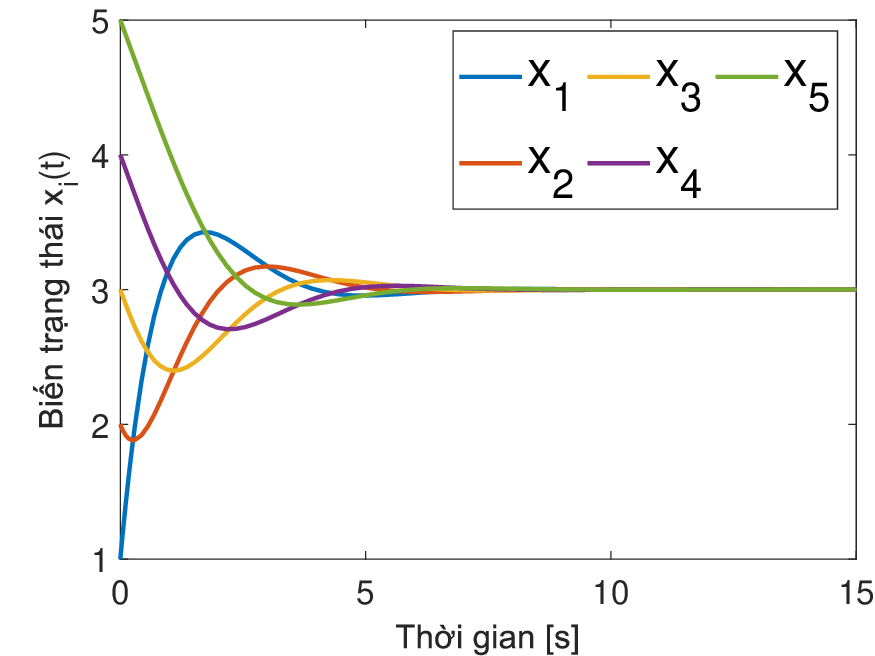}}\\
    \subfloat[$G_2$]{    
    \begin{tikzpicture}[
    roundnode/.style={circle, draw=black, thick, minimum size=2mm,inner sep= 0.25mm},
    squarednode/.style={rectangle, draw=black, thick, minimum size=3.5mm,inner sep= 0.25mm},
    ]
    \node[roundnode] (u1) at (0,0) {$1$}; %
    \node[roundnode] (u2) at (1,1) {$2$};%
    \node[roundnode] (u4) at (2,1) {$3$};%
    \node[roundnode] (u5) at (2,-1) {$4$};%
    \node[roundnode] (u3) at (1,-1) {$5$};%
    \node[roundnode] (u6) at (3,0) {$6$};%
    
    \draw [very thick,-{Stealth[length=2mm]}]
    (u1) edge [bend left=0] (u2)
    (u2) edge [bend left=0] (u3)
    (u3) edge [bend left=0] (u1)
    (u3) edge [bend left=0] (u4)
    (u2) edge [bend left=0] (u4)
    (u5) edge [bend left=0] (u4)
    (u4) edge [bend left=0] (u6)
    (u6) edge [bend left=0] (u5)
    (u3) edge [bend left=0] (u5)
    ;
    \node (u) at (-1,-1.5) { }; %
\end{tikzpicture}} \hfill
    \subfloat[]{\includegraphics[width=.45 \textwidth]{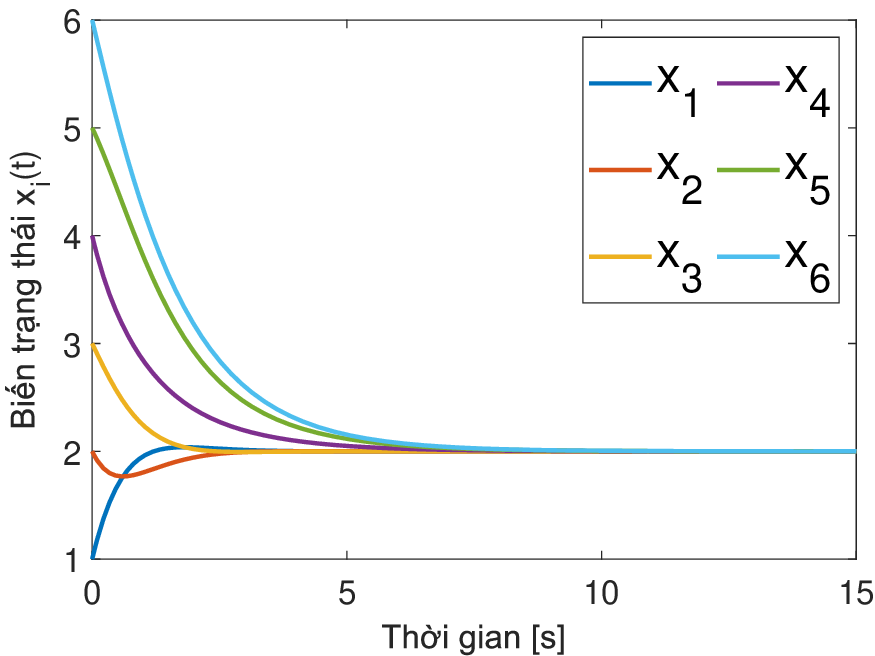}}\\
    \subfloat[$G_3$]{
    \begin{tikzpicture}[
    roundnode/.style={circle, draw=black, thick, minimum size=2mm,inner sep= 0.25mm},
    squarednode/.style={rectangle, draw=black, thick, minimum size=3.5mm,inner sep= 0.25mm},
    ]
    \node[roundnode] (u1) at (0,0) {$1$}; %
    \node[roundnode] (u2) at (1,1) {$2$};%
    \node[roundnode] (u4) at (2,1) {$3$};%
    \node[roundnode] (u5) at (2,-1) {$4$};%
    \node[roundnode] (u3) at (1,-1) {$5$};%
    \node[roundnode] (u6) at (3,0) {$6$};%
    
    \draw [very thick,-{Stealth[length=2mm]}]
    (u1) edge [bend left=0] (u2)
    (u2) edge [bend left=0] (u3)
    (u3) edge [bend left=0] (u4)
    (u2) edge [bend left=0] (u4)
    (u5) edge [bend left=0] (u4)
    (u4) edge [bend left=0] (u6)
    (u6) edge [bend left=0] (u5)
    (u3) edge [bend left=0] (u5)
    ;
    \node (u) at (-1,-1.5) { }; %
    \end{tikzpicture}
    }
    \hfill
    \subfloat[]{\includegraphics[width=.45 \textwidth]{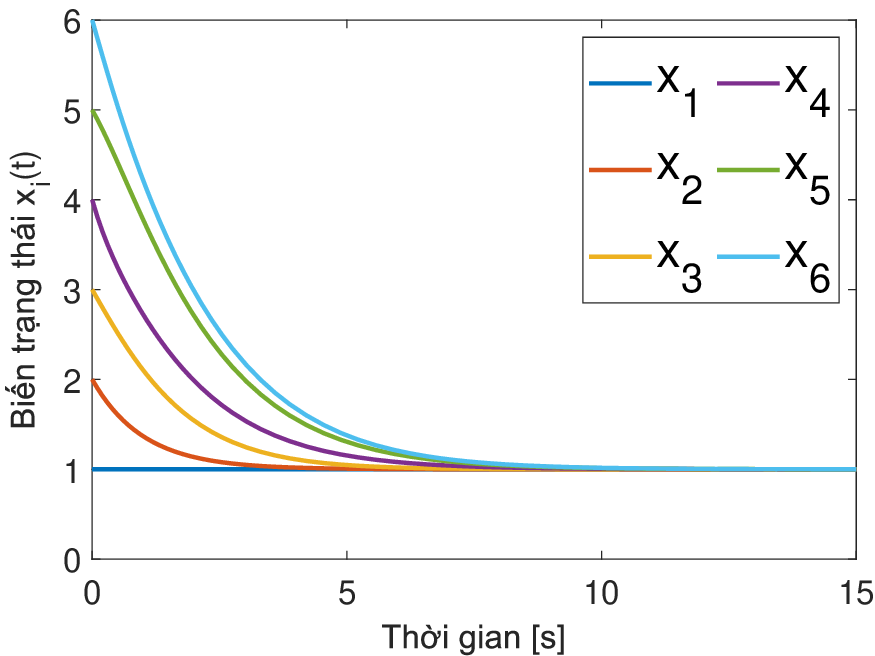}}
    \caption{Mô phỏng thuật toán đồng thuận với ba đồ thị hữu hướng khác nhau ở Ví dụ~\ref{VD_4.2}.}
    \label{fig:VD4.2}
\end{figure*}

\section{Quá trình đồng thuận cạnh}
Xét hệ đồng thuận tuyến tính mô tả bởi đồ thị vô hướng, liên thông $G=(V,E)$, trong đó các tác tử trong hệ được mô tả bởi khâu tích phân bậc nhất. Phương trình mô tả hệ đồng thuận được cho bởi:
\begin{align} \label{eq:c4_vertex_consensus}
    \dot{\m{x}} = -\mcl{L}\m{x},
\end{align}
trong đó $\m{x} = [x_1, \ldots, x_n]^\top\in\mb{R}^n$ là vector trạng thái của $n$ tác tử trong hệ. Đặt $\bmm{\zeta} = \m{H}\m{x} \in \mb{R}^m$, trong đó $\m{H}$ là ma trận liên thuộc của $G$ thì $\bmm{\zeta}$ chứa $m$ phần tử có dạng $(x_j - x_i)$ ứng với các cạnh $(i,j)$ trong đồ thị. Nhân từ phía trái cả hai vế của \eqref{eq:c4_vertex_consensus} với ma trận $\m{H}$ và chú ý rằng với $G$ vô hướng thì $\mcl{L}=\m{H}^\top\m{H}$, ta thu được \index{đồng thuận!cạnh}mô hình đồng thuận cạnh như sau:
\begin{align} \label{eq:c4_edge_consensus}
\dot{\bmm{\zeta}} =- \m{H}\m{H}^\top(\m{H}\m{x}) = -\mcl{L}_e\bmm{\zeta},
\end{align}
trong đó $\mcl{L}_e \triangleq \m{H}\m{H}^\top \in \mb{R}^{m \times m}$ được gọi là \emph{ma trận Laplace cạnh}\index{ma trận!Laplace cạnh}\footnote{edge Laplacian}. Rõ ràng, $\mcl{L}_e$ là một ma trận đối xứng, bán xác định dương. Do $G$ là liên thông, $m \geq n-1$. Gọi $\lambda$ là một giá trị riêng bất kỳ của $\mcl{L}$ và $\m{v}$ là một vector riêng tương ứng thì ta có:
\begin{align}
    \mcl{L}_e (\m{H}\m{v}) = \m{H}\m{H}^\top\m{H}\m{v} = \m{H} \mcl{L} \m{v} = \lambda (\m{H}\m{v}).
\end{align}
Do đó, $\lambda$ cũng là một giá trị riêng của $\mcl{L}_e$ với $\m{H} \m{v}$ là một vector riêng ứng. Ta lại có 
\[\text{rank}(\mcl{L})=\text{rank}(\m{H})=\text{rank}(\mcl{L}_e)=n-1.\] 
Bởi vậy, $\mcl{L}_e$ có đúng $n-1$  giá trị riêng dương (cũng chính là các giá trị riêng dương của $\mcl{L}$) và $(m-n+1)$ giá trị riêng $0$ với các vector riêng là các vector nằm trong không gian chu trình của $G$ (Chương \ref{chap:graphTheory}).

Đặt $\bmm{\delta} = \m{x} - \m{1}_n \bar{x}$, với $\bar{x} = \frac{1}{n}\m{1}_n^\top \m{x}$ thì ta biết rằng $\|\bmm{\delta}\| \to 0$ khi $t \to +\infty$. Bởi vậy, \[\|\bmm{\zeta} \| = \|\m{H}\m{x}\| = \|\m{H}\bmm{\delta}\| \leq \|\m{H}\|\|\bmm{\delta}\| \to 0,\] khi $t \to +\infty$. 
Điều này chứng tỏ rằng khi các biến trạng thái tương ứng với các đỉnh tiệm cận tới tập đồng thuận thì các biến trạng thái tương ứng với các cạnh hội tụ tiệm cận.

Giả sử chúng ta đã đánh số các cạnh của đồ thị $G$ sao cho các cạnh của một cây bao trùm $T$ lần lượt là $e_1, \ldots, e_{n-1}$, ta có thể viết lại ma trận $\m{H}$ dưới dạng:
\begin{align}
    \m{H} = \begin{bmatrix}
        \m{H}_{E(T)}\\
        \m{H}_{E\setminus E(T)}
    \end{bmatrix},
\end{align}
trong đó $\m{H}_{E(T)} \in \mb{R}^{(n-1) \times n}$ là ma trận liên thuộc của cây bao trùm $T$ và $\m{H}_{E\setminus E(T)} \in \mb{R}^{(m-n+1) \times n}$ là ma trận liên thuộc ứng với phần đồ thị còn lại của $G$ sau khi xóa đi các cạnh trong $E(T)$. Do $T$ liên thông, $\m{H}_{E(T)}$ thỏa mãn $\text{ker}(\m{H}_{E(T)}) = \text{im}(\m{1}_n)$. Do mỗi hàng trong $\m{H}_{E\setminus E(T)}$ đều có thể được biểu diễn bởi một tổ hợp tuyến tính của các hàng trong $\m{H}_{E(T)}$, tồn tại ma trận $\m{T} \in \mb{R}^{ (m-n+1)\times (n-1)}$ sao cho $\m{H}_{E\setminus E(T)} = \m{T} \m{H}_{E(T)}$. Chú ý rằng $\m{H}_{E(T)}$ là đủ hạng hàng nên $\mcl{L}_e(T) = \m{H}_{E(T)}\m{H}_{E(T)}^\top$ là khả nghịch và ta có thể viết
\begin{align}
    \m{T} = \m{H}_{E\setminus E(T)}\m{H}_{E(T)} \big(\m{H}_{E(T)}\m{H}_{E(T)}^\top\big)^{-1}.
\end{align}
Đặt $\m{R} = \begin{bmatrix}
\m{I}_{n-1}\\
\m{T}
\end{bmatrix}$, ta có $\m{H} = \m{R}\m{H}_{E(T)}$ và $\mcl{L}_e = \m{H}\m{H}^\top = \m{R}\m{H}_{E(T)}\m{H}_{E(T)}^\top\m{R}^\top = \m{R}\mcl{L}_{e}(T)\m{R}^\top$. 

Trở lại với phương trình đồng thuận cạnh \eqref{eq:c4_edge_consensus},
ta có thể biểu diễn \[\m{y} = \begin{bmatrix}
\bmm{\zeta}_t^\top,~\bmm{\zeta}_c^\top\end{bmatrix}^\top = \m{R} \bmm{\zeta}_t,\] trong đó $\bmm{\zeta}_t$ là vector biến trạng thái cạnh ứng với cây bao trùm $T$, còn $\bmm{\zeta}_c$ ứng với các cạnh còn lại. Bởi vậy, từ \eqref{eq:c4_edge_consensus}, ta có
\begin{equation} \label{eq:c4_ydot}
    \m{R} \dot{\bmm{\zeta}}_t = - \mcl{L}_e\m{R} {\bmm{\zeta}}_t.
\end{equation}
Thay $\mcl{L}_e = \m{R}\mcl{L}_{e}(T)\m{R}^\top$ vào \eqref{eq:c4_ydot}, ta thu được:
\begin{equation} \label{eq:c4_ytdot}
    \m{R} \dot{\bmm{\zeta}}_t = - \m{R} \mcl{L}_e(T) \m{R}^\top\m{R} \bmm{\zeta}_t.
\end{equation}
Từ phương trình \eqref{eq:c4_ytdot} và cấu trúc ma trận $\m{R}$, ta thấy rằng hệ rút gọn bậc mô tả bởi
\begin{equation}
    \dot{\bmm{\zeta}}_t = - \mcl{L}_e(T) \m{R}^\top\m{R} \bmm{\zeta}_t
\end{equation}
mô tả được hoàn toàn hệ đồng thuận cạnh \eqref{eq:c4_edge_consensus}. Khi động học của các biến trạng thái ứng với một cây bất kỳ được xác định, các biến trạng thái cạnh $\bmm{\zeta}_c$ còn lại có thể suy ra ngay theo công thức $\bmm{\zeta}_c = \m{T}\bmm{\zeta}_t$.

\begin{figure*}[t]
\centering
\subfloat[]{\includegraphics[height = 4.15cm]{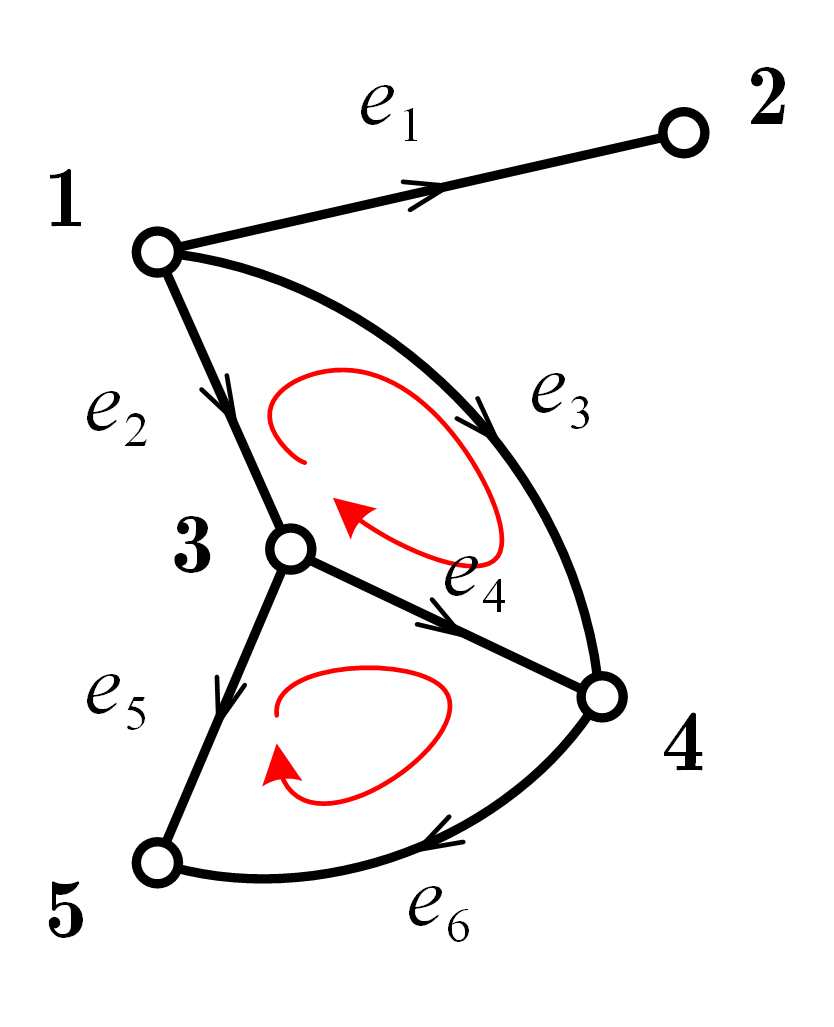}}\qquad\qquad
\subfloat[]{\includegraphics[height = 4.15cm]{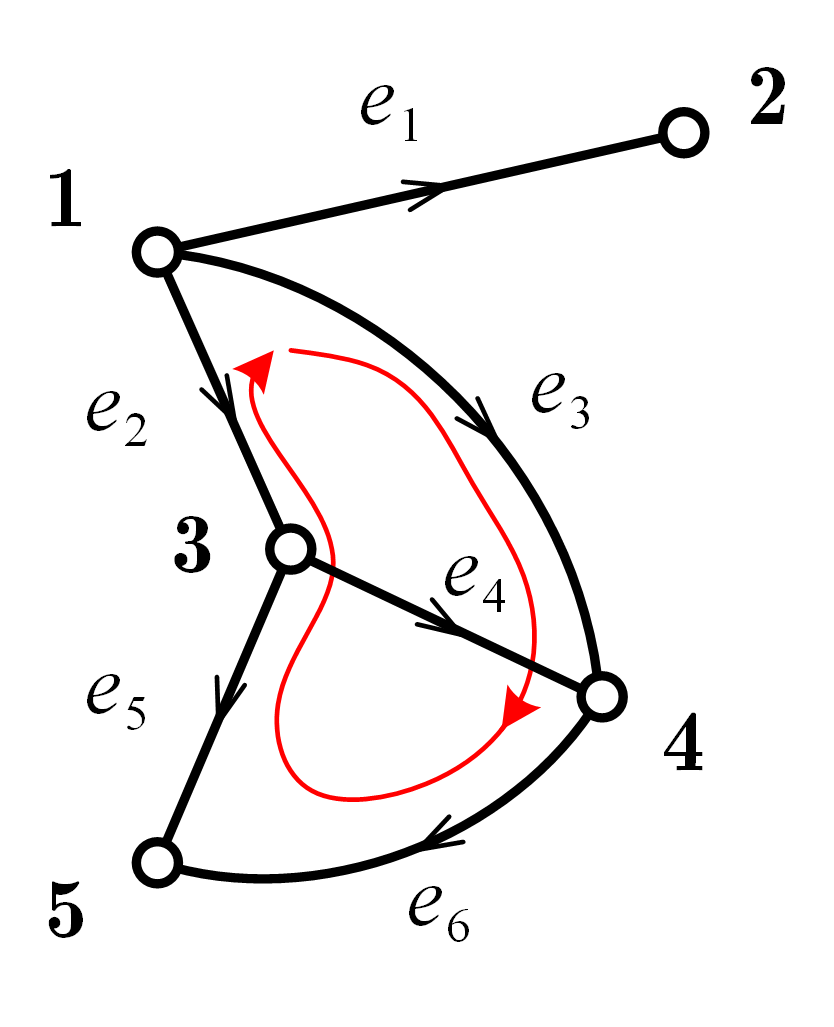}} \qquad\qquad
\subfloat[]{\includegraphics[height = 4.15cm]{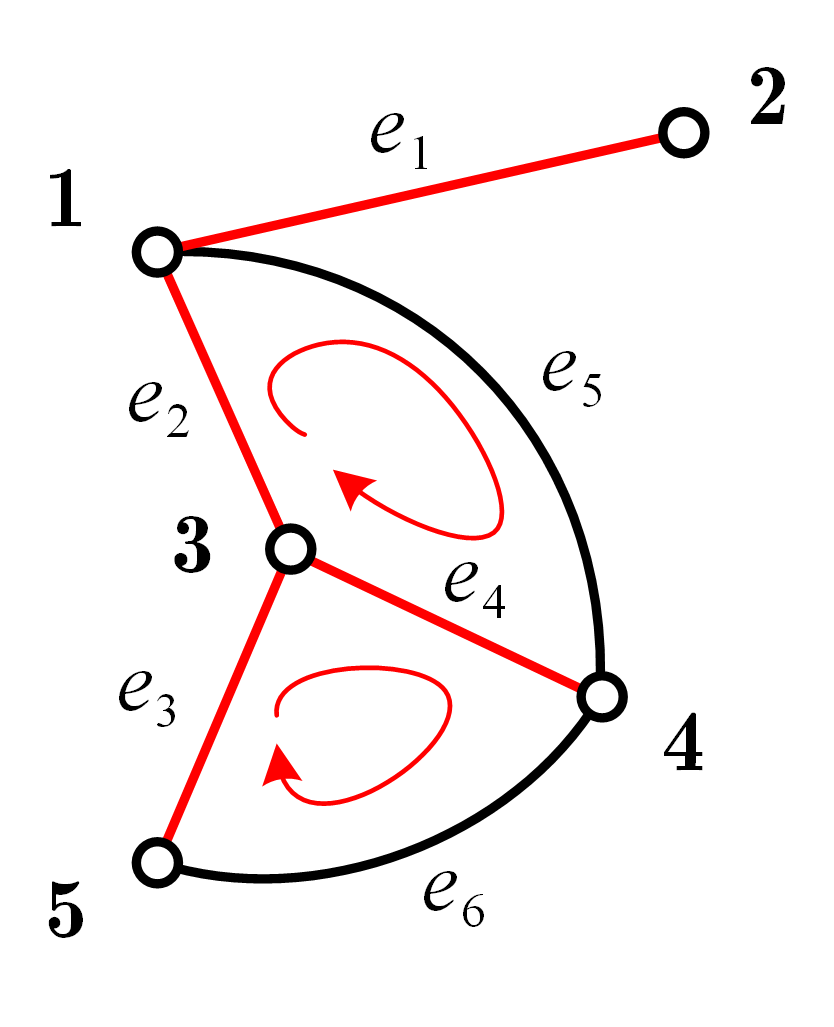}}
\caption{Đồ thị $G$ có ba chu trình, trong đó hai chu trình là độc lập.\label{fig:c4-cyclespace}}
\end{figure*}
\begin{example}
Xét đồ thị $G$ như trên Hình~\ref{fig:c4-cyclespace}. Với cây bao trùm gồm các cạnh được tô màu đỏ và các chu trình định hướng dương, ta có thể viết 
\begin{align}
    \m{H} = \begin{bmatrix}
        \m{H}_{E(T)} \\[.2em] \hline\\[-1em]
        \m{H}_{E\setminus E(T)}
    \end{bmatrix} = \left[ {\begin{array}{*{20}{c}}
{ - 1}&1&0&0&0\\
{ - 1}&0&1&0&0\\
0&0&{ - 1}&0&1\\
0&0&{ - 1}&1&0
\\[.2em] \hline\\[-1em]
{ - 1}&0&0&1&0\\
0&0&0&{ - 1}&1
\end{array}} \right],~\text{và }~ \m{T}=\begin{bmatrix} 
0 & 1 & 0 & 1\\
0 & 0 & 1 & -1
\end{bmatrix}.
\end{align}
\end{example}

\begin{figure}
    \centering
    \subfloat[Đồ thị $C_{20}$]{\centering 
    \begin{tikzpicture}[>=latex, scale=0.54] 
        \def \n {20}
        \def \radius {3.6cm}
        \def \margin {5.5} 
        \foreach \s in {1,...,\n}
        {
        \node[draw, circle,color=black] at ({360/\n * (\s - 1)}:\radius) {};
        \draw[-, >=latex,color=red, very thick] ({360/\n * (\s - 1)+\margin}:\radius) 
        arc ({360/\n * (\s - 1)+\margin}:{360/\n * (\s)-\margin}:\radius);
        }
        \draw[-, >=latex,color=black, very thick] ({360/\n * (\n - 1)+\margin}:\radius) 
        arc ({360/\n * (\n - 1)+\margin}:{360/\n * (\n)-\margin}:\radius);
        \node (u) at (0,-5) {};
    \end{tikzpicture}} \hfill
    \subfloat[$\m{\zeta}_t(t)$ ]{\includegraphics[width=0.5\linewidth]{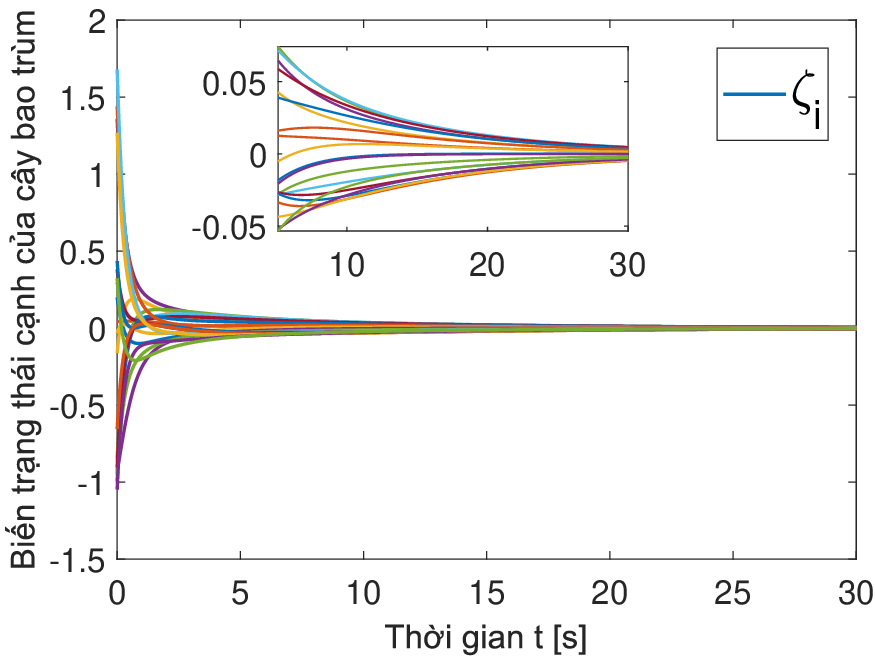}}
    \caption{Mô phỏng minh họa Ví dụ~\ref{VD4.4}. Những cạnh màu đỏ tạo thành một cây bao trùm của đồ thị. Các biến tương đối $\bmm{\zeta}(t)\to \m{0}$ khi $t\to +\infty$.}
    \label{fig:VD4.4}
\end{figure}

\begin{example} \label{VD4.4}
Trong ví dụ này, chúng ta mô phỏng quá trình đồng thuận cạnh trong một đồ thị chu trình gồm 20 đỉnh. Với đồ thị $C_{20}$, bất kỳ 19 cạnh nào cũng lập thành một cây bao trùm của đồ thị. Kết quả mô phỏng trên Hình~\ref{fig:VD4.4} cho thấy $\bmm{\zeta}_t(t) \to \m{0}$ và do đó $\bmm{\zeta}(t) \to \m{0}$ khi $t\to+\infty$.
\end{example}

\section{Đồng bộ hóa đầu ra các hệ thụ động}
Trong mục này, ta xét hệ đồng thuận phi tuyến. Quan hệ phi tuyến có thể nằm ở mô hình tác tử hoặc ở quan hệ  tương tác giữa các tác tử. Công cụ để phân tích ở đây là lý thuyết về hệ thụ động \cite{Arcak2007passivity,Zelazo2010TAC}. 

Xét hệ phi tuyến
\begin{align} \label{eq:c4_nonl_system}
    \dot{\m{z}}(t) &= \m{f}(\m{z}(t),\m{u}(t)), \\
    \m{y}(t) &= \m{z}(t),
\end{align}
với $\m{f}$ thỏa mãn điều kiện Lipschitz địa phương với biến  $\m{z}$ (tức là với mọi $\m{z}_1, \m{z}_2 \in D$, luôn tồn tại $L>0$ sao cho $\|\m{f}(\m{z}_1) - \m{f}(\m{z}_2)\| \leq L \|\m{z}_1 - \m{z}_2\|$) và $\m{f}(\m{0},\m{0}) = \m{0}$. Khi đó hệ \eqref{eq:c4_nonl_system} là thụ động nếu tồn tại một hàm khả vi liên tục, bán xác định dương $V$, gọi là hàm trữ năng thỏa mãn
\begin{equation} \label{eq:c4_passive}
    \m{u}(t)^\top \m{y}(t) \geq \dot{V}(t),~\forall t\ge 0.
\end{equation}
Nếu ở phương trình \eqref{eq:c4_passive}, $\dot{V}$ được thay thế bởi $\dot{V} + \psi(\m{z})$ với hàm $\psi$ là xác định dương, thì hệ gọi là thụ động chặt. Hơn nữa, do biến đầu ra trùng với biến trạng thái, $\m{u}^\top(t) \m{y}(t) \geq \dot{V}(t) + \psi(\m{y}(t)), \forall t,$ và hệ \eqref{eq:c4_nonl_system} là thụ động chặt đầu ra. Chúng ta có kết quả sau về hệ thụ động \eqref{eq:c4_nonl_system}:
\begin{theorem} \cite{Khalil2002}
Giả sử hệ \eqref{eq:c4_nonl_system} là thụ động thì gốc tọa độ là ổn định Lyapunov. Nếu hệ \eqref{eq:c4_nonl_system} là thụ động chặt thì gốc tọa độ sẽ là ổn định tiệm cận.
\end{theorem}

Tiếp theo, chúng ra xét bài toán thiết kế luật đồng thuận cho hệ gồm $n$ tác tử thụ động. Giả sử mỗi tác tử trong hệ có mô hình động học:
\begin{align}
    \dot{\m{x}}_i &= \m{f}_i(\m{x}_i)+\m{g}_i(\m{x}_i)\m{u}_i,\\
    \m{y}_i &= \m{h}_i(\m{x}_i),
\end{align}
với $\m{x}_i \in \mb{R}^d$, $\m{u}_i \in \mb{R}^l$, $\m{f}_i: \mb{R}^d \to \mb{R}^d$, $\m{g}_i: \mb{R}^d \to \mb{R}^{d\times l}$, $\m{h}_i: \mb{R}^d \to \mb{R}^q$, và $i=1, \ldots, n$. Giả sử các tác tử được kết nối qua đồ thị vô hướng $G=(V,E,W)$ và mỗi hệ con là thụ động với hàm trữ năng $V_i(\m{x}_i)$ thỏa mãn:
\begin{align} \label{eq:c4_passive-system}
    \dot{V}_i \le \m{y}_i^\top\m{u}_i.
\end{align}
Giả sử hàm $\bm{\phi}: \mb{R}^p \to \mb{R}^p$ là Lipschitz địa phương (tức là trong tập $D \subset \mb{R}^d$  cho trước, luôn tồn tại $L>0$ sao cho $\|\bm{\phi}(\m{r}_1)-\bm{\phi}(\m{r}_2)\| \leq L \|\m{r}_1 - \m{r}_2\|$, $\forall \m{r}_1, \m{r}_2 \in D$) và thỏa mãn: (i) $\bm{\phi}(-\m{r}_1)=-\bm{\phi}(\m{r}_1)$, (ii) $\m{r}^\top_1\bm{\phi}(\m{r}_1)>0, \forall \m{r}_1 \ne \m{0}_d$. 
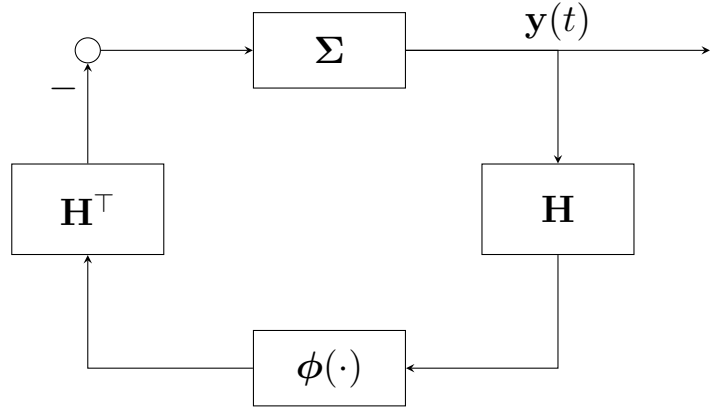
\begin{figure}
    \centering
\begin{tikzpicture}
\node[draw,
	circle,
	minimum size=0.3cm,
] (sum) at (0,0){};

\node [draw,
	minimum width=2cm,
	minimum height=1cm,
	right = 2cm of sum
]  (agents) {\Large $\boldsymbol{\Sigma}$};

\coordinate [right = 2cm of agents] (y1) {};

\node [draw,
	minimum width=2cm, 
	minimum height=1.2cm,
	below = 1.5cm of y1
] (incident) {\Large $\mathbf{H}$};

\coordinate [below = 1.5cm of incident] (y2) {};

\node [draw,
	minimum width = 2cm,
	minimum height = 1cm,
	left = 2cm of y2
]  (netw) {\Large $\boldsymbol{\phi}(\cdot)$};

\coordinate [left = 2cm of netw] (y3) {};

\node [draw,
	minimum width=2cm, 
	minimum height=1.2cm,
	below = 1.5cm of sum.center
] (collector) {\Large $\mathbf{H}^\top$};

\draw[-stealth] (agents.east) -- ++ (4,0) 
	node[midway](output){}node[midway,above] {\Large $\mathbf{y}(t)$};

\draw[-stealth] (sum.east) -- (agents.west);
\draw[-stealth] (agents.east) -| (incident.north);
\draw[-stealth] (incident.south) |- (netw.east);
\draw[-stealth] (netw.west)-| (collector.south);
\draw[-stealth] (collector.north) -- (sum.south) node[near end,left]{\Large $-$};
\end{tikzpicture}
    \caption{Hệ $\Sigma$ gồm $n$ hệ con thụ động với hàm kết nối $\bm{\phi}(\cdot)$.}
    \label{fig:c4_passive-system}
\end{figure}
Thuật toán đồng thuận được thiết kế như sau:
\begin{align} \label{eq:c4-passive-consensus-law}
    \m{u}_i = \sum_{j\in {N}_i} \bmm{\phi}(\m{y}_j - \m{y}_i).
\end{align}
Ta có sơ đồ mô tả hệ đồng thuận với luật đồng thuận \eqref{eq:c4-passive-consensus-law} như trên Hình \ref{fig:c4_passive-system}. 

Chọn hàm Lyapunov $V=\sum_{i=1}^n V_i$, ta có:
\begin{align}
\dot{V} = \sum_{i=1}^n\m{y}_i^\top \m{u}_i = - \m{y}^\top\bar{\m{H}}^\top\bm{\phi}(\bar{\m{H}}\m{y})\leq 0.
\end{align}
Như vậy, $\m{x}(t)$ bị chặn và $\dot{V} = 0$ khi và chỉ khi $\bar{\m{H}}\m{y} = \m{0}_{nq}$. Theo nguyên lý bất biến LaSalle, 
\begin{align}
    \m{x}(t) \to \Omega=\{ \m{x} \in \mb{R}^{dn}|~\m{y}_i = \m{y}_j, \forall i,j =1,\ldots,n, (i,j) \in E\}.
\end{align}
Do đồ thị liên thông, tập bất biến chỉ chứa các giá trị thỏa mãn $\m{y}_i = \m{y}_j$. Do đó, với luật đồng thuận \eqref{eq:c4-passive-consensus-law}, biến đầu ra của các tác tử tiệm cận tới không gian đồng thuận. Nếu có thêm giả thiết $\m{h}_i(\cdot)=\m{h}_j(\cdot)= \m{h}(\cdot) \forall i,j \in V$ và $\m{h}(\cdot)$ là đơn ánh thì từ $\m{h}_i(\m{x}_i) = \m{h}_j(\m{x}_j),\forall i,j =1,\ldots,n$, ta suy ra thêm được hệ đồng thuận về các biến trạng thái ($\m{x}_i \to \m{x}_j \forall i,j =1,\ldots,n$).

Ví dụ, có thể chọn hàm $\bmm{\phi}(\m{x})=\m{W}\m{x}$, trong đó $\m{W} = \text{diag}(\omega_k) \in \mb{R}^{m\times m}$ là một ma trận đường chéo, xác định dương.

\begin{example} \label{VD4.5}
[Mô hình Kuramoto đơn giản] Xét hệ $n$ tác tử dao động với tần số góc $\omega_i>0$ và có góc pha cho bởi $\theta_i(t)$. Sự tương tác giữa các tác tử được mô tả bởi một đồ thị vô hướng $G$ và cho bởi phương trình:
\begin{align}
    \dot{\theta}_i(t) = \omega_i - \frac{k}{n}\sum_{j\in N_i} \sin(\theta_i - \theta_j),~i=1,\ldots, n.
\end{align}
Mô hình Kuramoto được nghiên cứu rộng rãi trong vật lý và có ứng dụng trong giải thích các hiện tượng đom đóm nhấp nháy đồng loạt, hoặc mô hình đồng bộ hóa các nguồn tạo dao động.

Xét mô hình Kuramoto đơn giản, trong đó các tần số góc $\omega_i = \omega_j = \omega$, đồng thời liên kết giữa các tác tử được mô tả bởi đồ thị vô hướng, liên thông $G$. 

Phương trình vi phân mô tả hệ được cho bởi
\begin{align} \label{eq:c4-kuramoto}
    \dot{\bmm{\theta}}(t) = {\omega} \m{1}_n - \frac{k}{n} \m{H}^\top \sin(\m{H} \bmm{\theta}).
\end{align}
Thực hiện phép đổi biến $\bmm{\zeta}(t) = \bmm{\theta}(t) - {\omega}t\m{1}_n$ thì
\begin{align} \label{eq:c4-kuramoto1}
    \dot{\bmm{\zeta}} = - \frac{k}{n} \m{H}^\top \sin(\m{H} \bmm{\zeta}).
\end{align}
Có thể kiểm tra rằng hệ \eqref{eq:c4-kuramoto1} là một hệ thụ động. Đồng thời, hàm $\sin(x)$ thỏa mãn các điều kiện (i) và (ii) khi  $|x|<\frac{\pi}{2}$. Như vậy, với $\max_{i,j \in V}|\theta_i(0)-\theta_j(0)|<\frac{\pi}{2}$, ta có $\bmm{\zeta}(t) \to \text{im}(\m{1}_n)$ khi $t \to +\infty$. Điều này kéo theo việc $\theta_i \to \theta_j$, khi $t \to +\infty$.

Mô phỏng hệ \eqref{eq:c4-kuramoto} gồm 6 tác tử được cho như trong Hình~\ref{fig:VD4.5}. Trong ví dụ này, đồ thị $G$ được chọn có ma trận liên thuộc 
\begin{align*}
    \m{H} = \begin{bmatrix}
     -1 & 1 & 0 & 0 &0 &0\\
     -1 & 0 & 1 & 0 &0 &0\\
      0 &-1 & 0 & 1 &0 &0\\
      0 & 0 &-1 & 1 &0 &0\\
      0 & 0 &-1 & 0 &1 &0\\
      0 & 0 &-1 & 0 &0 &1\\
      0 & 0 & 0 &-1 &0 &1
    \end{bmatrix},
\end{align*}
$k=n$, và $\omega_i = 1$. Các điều kiện đầu $\theta_i(0)$ được chọn ngẫu nhiên trong khoảng $[-\frac{\pi}{4}, \frac{\pi}{4}]$, do đó $|\theta_i(0) - \theta_j(0)|<\frac{\pi}{2}$ với mọi $i\ne j$. Hình \ref{fig:VD4.5} thể hiện rằng các góc pha dần đạt tới đồng thuận khi $t\to +\infty$. (Chú ý: trong mô phỏng $\theta_i$ có xu hướng tăng tới vô cùng khi $t\to +\infty$, tuy nhiên thực tế $\theta_i$ nhận giá trị trong khoảng $[0,2\pi]$.)
\end{example}
\begin{SCfigure}[][h!]
    \label{fig:VD4.5}
    \hspace{2cm}
    \caption{Mô phỏng mô hình Kuramoto đơn giản cho hệ 6 tác tử ở Ví dụ \ref{VD4.5}.}
    \includegraphics[height=6cm]{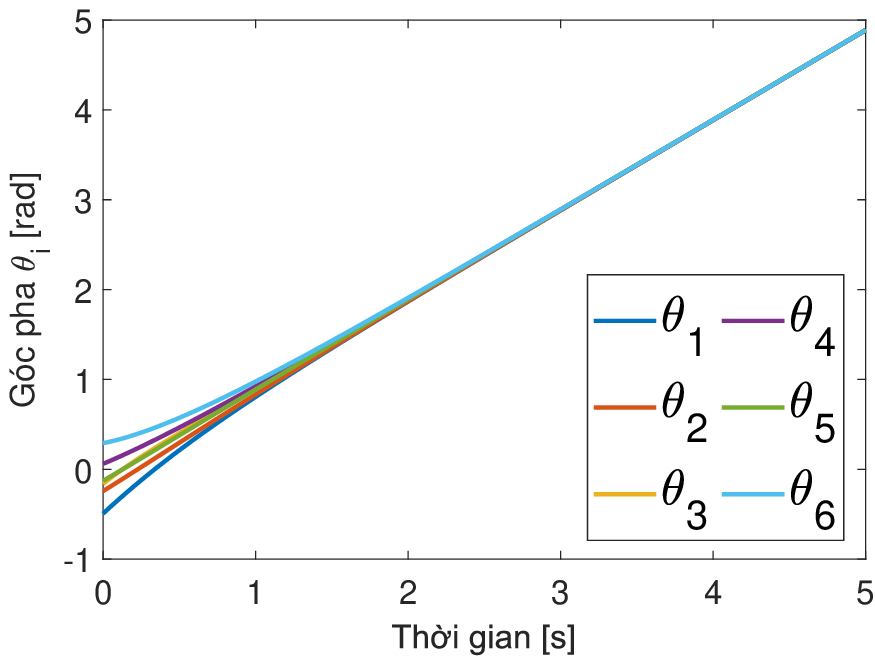}
\end{SCfigure}
%
%
%

\section{Ghi chú và tài liệu tham khảo}
Một số kết quả về chọn hàm Lyapunov cho hệ đồng thuận trong chương này được tham khảo từ \cite{Zhang2011lyapunov}. Phương pháp Lyapunov cho ta một cách trực tiếp để phân tích quá trình đồng thuận, không chỉ cho hệ đồng thuận tuyến tính mà còn cho các hệ đồng thuận với tác tử có mô hình phức tạp hơn, các hệ đồng thuận có trễ, có đồ thị thay đổi theo thời gian \cite{Olfati2007consensuspieee},... Một số nghiên cứu khác tập trung vào thiết kế luật đồng thuận trong thời gian hữu hạn, dựa trên một số định lý mở rộng của định lý Lyapunov và nguyên lý bất biến LaSalle  \cite{Wang2010,Cortes2006finite,Parsegov2013fixed,Nguyen2021}. Các thuật toán đồng thuận với tín hiệu đầu vào thay đổi theo thời gian được giới thiệu ở \cite{Spanos2005dynamic,Freeman2006stability,Kia2019tutorial,Bai2010robust}. 

Các kết quả về đồng thuận cạnh, mô hình đồng thuận rút gọn và phân tích hệ đồng thuận theo phương pháp thụ động hóa ở chương này được trình bày từ tài liệu \cite{Zelazo2010TAC}. Một số tài liệu khác về phương pháp thụ động cho hệ đồng thuận bao gồm \cite{Arcak2007passivity,Chopra2012,Burger2014duality}

\section{Bài tập}
\begin{exercise} \label{ex:4.1}
Xét hệ đồng thuận của các tác tử dạng tích phân bậc hai cho bởi
\begin{align*}
    \dot{x}_{i1} &= {x}_{i2},\\
    \dot{x}_{i2} &= -k_1 \sum_{j\in N_i}(x_{i1} - x_{j1}) - k_2{x}_{i2}, ~i=1, \ldots, n,
\end{align*}
trong đó $\m{x}_i  = [x_{i1}, x_{i2}]^\top$ là biến trạng thái của tác tử thứ $i$,  $k_1, k_2 >0$ là các hằng số dương và đồ thị tương tác $G$ là vô hướng. Hãy phân tích sự hội tụ của hệ đồng thuận trên theo phương pháp Lyapunov.
\end{exercise}

\begin{exercise}
Câu hỏi giống như bài tập \ref{ex:4.1} nhưng với hệ đồng thuận cho bởi:
\begin{align*}
    \dot{x}_{i1} &= {x}_{i2},\\
    \dot{x}_{i2} &= -k_1 \sum_{j\in N_i}(x_{i1} - x_{j1}) - k_2 {x}_{i2} - k_3 \sum_{j\in N_i}(x_{i2} - x_{j2}), ~i=1, \ldots, n.
\end{align*}
\end{exercise}

\begin{exercise}[Ảnh hưởng của nhiễu tắt dần tới hệ đồng thuận] \index{hệ đồng thuận!có nhiễu}
Xét hệ đồng thuận bị ảnh hưởng bởi nhiễu:
\begin{align} \label{eq:ex_4.3}
    \dot{x}_{i} =  - \sum_{j\in N_i}(x_{i} - x_{j}) + \omega_i(t), ~i=1, \ldots, n,
\end{align}
trong đó $\omega_i(t)$ thỏa mãn $|\omega_i (t)| \leq k_i e^{-\rho_i t}$. Chứng minh rằng $\m{x}$ tiến tới  tập đồng thuận khi $t\to +\infty$. 
(Gợi ý: Sử dụng bổ đề: ``Nếu $\dot{\m{x}}=\m{A}\m{x}+\bm{\omega}$ là hệ thỏa mãn $\m{A}$ là ma trận Hurwitz, $\bm{\omega} \to \m{0}$ khi $t \to +\infty$ thì $\m{x} = \m{0}$  là ổn định tiệm cận.'' kết hợp với việc viết lại hệ \eqref{eq:ex_4.3} dưới dạng ma trận, phân tích $\mc{L} = \m{P} \bm{\Lambda} \m{P}^{-1}$ và đổi biến $\m{z} = \m{P}^{-1} \m{x}$.)
\end{exercise}

\begin{exercise}[Thuật toán đồng thuận với thời gian hữu hạn] \index{thuật toán!đồng thuận thời gian hữu hạn}
\cite{Bhat2000,Trinh2020Pointing} Với $\mc{D} \subseteq \mb{R}^d$ và $\m{x} \in \mc{D}$. Nếu tồn tại một hàm liên tục  $V(\m{x}): \mc{D} \to \mb{R}$ sao cho các điều kiện sau thỏa mãn
\begin{itemize}
\item[(i)] $V(\m{x})$ là xác định dương,
\item[(ii)] Tồn tại $\kappa>0$, $\alpha \in (0,1)$, và một tập mở $\mc{U}_0 \in \mc{D}$ chứa gốc tọa độ mà tại đó 
$$\dot{V}(\m{x}) + \kappa (V(\m{x}))^{\alpha } \leq 0, \forall \m{x} \in \mc{U}_0 \setminus \{\m{0}_d \},$$
\end{itemize}
thì $V(\m{x})\to 0$ trong thời gian hữu hạn bị chặn trên bởi $T \leq V({0})^{1-\alpha}/(\kappa(1-\alpha))$.

Sử dụng kết quả trên, xét hệ đồng thuận với đồ thị $G$ vô hướng. Chứng minh rằng luật đồng thuận cho bởi
\begin{align}
\dot{x}_i = - k \sum_{j\in N_i} \text{sig}^\alpha (x_i - x_j), ~i=1, \ldots, n,
\end{align}
trong đó $k>0$, $\alpha \in (0,1)$ và $\text{sig}^\alpha(x) = |x|^\alpha \text{sign}(x)$, sẽ giúp hệ đạt được đồng thuận tại giá trị trung bình cộng trong thời gian hữu hạn. Tìm cận trên của thời gian hội tụ và nhận xét sự phụ thuộc của giá trị này với tham số $k$, $\alpha$, và $\lambda_2(\mcl{L})$ của hệ.
\end{exercise}

\begin{exercise} \cite{Ren2009distributed,Ye2015model} \index{hệ Euler-Lagrange}
Xét mô hình của một mạng robot tương tác qua một đồ thị vô hướng, liên thông $G$. Mỗi robot trong mạng được mô tả bởi phương trình:
\begin{equation}
    \m{M}(\m{q}_i) \ddot{\m{q}}_i + \m{C}(\m{q}_i,\dot{\m{q}}_i) \dot{\m{q}}_i + \m{g}(\m{q}_i) = \m{u}_i,~i=1,\ldots,n,
\end{equation}
trong đó:
\begin{itemize}
    \item $\m{q}_i=[{q}_{i1},\ldots,{q}_{id}]^\top$: vector các tọa độ suy rộng của các biến khớp,
    \item $\m{u}_i=[{u}_{i1},\ldots,{u}_{id}]^\top$: vector điều khiển đầu vào (ngoại lực),
    \item $\m{M}(\m{q}_i)=\m{M}(\m{q}_i)^\top>0,~\forall \m{q}_i$: ma trận quán tính, thỏa mãn $0<k_{\underline{m}}\le \|\m{M}(\m{q}_i)\|\leq k_{\bar{m}}$,
    \item $\m{C}(\m{q}_i,\dot{\m{q}}_i)\dot{\m{q}}_i$: thành phần lực hướng tâm và lực Coriolis, $\dot{\m{M}}(\m{q}_i) = \m{C}(\m{q}_i,\dot{\m{q}}_i)+\m{C}(\m{q}_i,\dot{\m{q}}_i)^\top$,
    \item $\m{g}(\m{q}_i)=\big( \frac{\partial P}{\partial \m{q}_i} \big)^\top$: trọng lực, ${P}(\m{q}_i)$: hàm thế năng, $\m{g}(\m{q}_i)=\m{0}_d \Longleftrightarrow \m{q}_i = \m{0}_d$.
\end{itemize}
\begin{enumerate}
    \item[i.] Xét robot $i$, sử dụng hàm Lyapunov $V=\frac{1}{2}(\dot{\m{q}}_i)^\top\m{M}_i(\m{q}_i) \dot{\m{q}}_i + {\rm P}(\m{q}_i)$, chứng minh rằng với luật điều khiển $\m{u}_i = -\m{K}_d \dot{\m{q}}_i$, trong đó $\m{K}_d=\m{K}_d^\top>0$ thì $\m{q}_i=\m{0}_d$, $\dot{\m{q}}_i=\m{0}_d$ là ổn định tiệm cận toàn cục.
    \item[ii.] Xét luật đồng thuận cho mỗi tác tử dạng: \[\m{u}_i = \m{M}(\m{q}_i) \Big(-\sum_{j\in N_i}a_{ij}(\m{q}_i-\m{q}_j) - k \dot{\m{q}}_i \Big) + \m{C}(\m{q}_i,\dot{\m{q}}_i) \dot{\m{q}}_i+\m{g}(\m{q}_i),\] 
    trong đó $i=1,\ldots,n$. Viết phương trình dạng ma trận mô tả hệ với luật đồng thuận trên với các biến trạng thái là $\m{q}=[\m{q}_1^\top,\ldots,\m{q}_n^\top]^\top$ và $\dot{\m{q}}=[\dot{\m{q}}_1^\top,\ldots,\dot{\m{q}}_n^\top]^\top$.
    \item[iii.] Với luật đồng thuận ở ý (ii), xét hàm Lyapunov $V_1 = \frac{1}{2} \m{q}^\top(\mc{L}\otimes \m{I}_d + 2 k_1 \m{I}_{dn})\m{q} + 2\mu \m{q}^\top \dot{\m{q}}+\frac{1}{2} \dot{\m{q}}^\top\dot{\m{q}}$, trong đó $\mcl{L}$ là ma trận Laplace của đồ thị $G$, $k_1 = \mu k$ và $k>2\mu>0$. Chứng minh $V_1$ xác định dương và $\dot{V}_1$ là bán xác định âm. Sử dụng nguyên lý bất biến LaSalle, chứng minh hệ sẽ đạt đồng thuận với mọi điều kiện đầu của $\m{q}(0)$, $\dot{\m{q}}(0)$.
    \item[iv.] Với luật đồng thuận ở ý (ii), xét hàm Lyapunov $V_1 = \frac{1}{2} \m{q}^\top(\mc{L}\otimes \m{I}_d) \m{q} + \frac{1}{2} \dot{\m{q}}^\top\dot{\m{q}}$. Chứng minh $(\mc{L}\otimes \m{I}_d) \m{q}$ và $\dot{\m{q}}$ là bị chặn. Sử dụng bổ đề Barbalat (xem phụ lục \ref{append:control_theory}), chứng minh rằng $\m{q}(t) \to \text{ker}(\mcl{L})$ và $\dot{\m{q}}\to \m{0}_{dn}$, khi $t\to +\infty$, hay các biến tọa độ suy rộng sẽ dần đạt tới đồng thuận.
\end{enumerate}
\end{exercise}

\begin{exercise} Xét đồ thị $G$ vô hướng, liên thông, gồm $n$ đỉnh và $m$ cạnh với ma trận Laplace $\mcl{L}$ và ma trận Laplace cạnh $\mcl{L}_e$. Với $T$ là một cây bao trùm của $G$ và giả sử $\lambda_i(\m{A})$ kí hiệu giá trị riêng nhỏ thứ $i$ của ma trận $\m{A}$. Chứng minh rằng khi đó
\[\lambda_i(\mcl{L}) \geq \lambda_i(\mcl{L}_e(T)).\]
\end{exercise}

\begin{exercise} Với $G=(V,E)$ là một cây với ma trận Laplace cạnh $\mcl{L}_e$ có phổ là $\lambda_1 \leq \ldots \leq \lambda_n$. Gọi $G'$ là đồ thị thu được nhờ thêm một cạnh mới vào $G$, và $\mcl{L}_e'$ là ma trận Laplace cạnh tương ứng với các giá  trị riêng $\hat{\lambda}_1 \leq \ldots \leq \hat{\lambda}_{n+1}$. Chứng minh rằng:
\begin{equation}
    0 \leq \hat{\lambda}_1 \leq \lambda_1 \leq \hat{\lambda}_2 \leq \ldots \lambda_n \leq \hat{\lambda}_{n+1}.
\end{equation}
Ghi chú: Kết quả này được suy ra từ định lý đan chéo giá trị riêng (eigenvalue interlacing theorem).
\end{exercise}

\begin{exercise}
Với đồ thị $G$ vô hướng, liên thông, gồm $n$ đỉnh và $m$ cạnh, chứng minh các quan hệ sau:
\begin{enumerate}
    \item[i.] Với $\m{S}_v = \begin{bmatrix} \m{H}_{E(T)}^\top\big(\m{H}_{E(T)}\m{H}_{E(T)}^\top\big)^{-1}~\quad \m{1}_{n}
    \end{bmatrix}$ và $(\m{S}_v)^{-1} = \begin{bmatrix} \m{H}_{E(T)}\\ \frac{1}{n} \m{1}_n^\top
    \end{bmatrix}$ thì
    \begin{align}
        (\m{S}_v)^{-1} \mcl{L} \m{S}_v = \begin{bmatrix} \mcl{L}_e(T)\big(\m{H}_{E(T)}\m{H}_{E(T)}^\top\big)^{-1} & \m{0}_{n-1}\\
        \m{0}_{n-1}^\top & 0
    \end{bmatrix}.
    \end{align}
    \item[ii.] Với $\m{S}_e = \begin{bmatrix} \m{R}~\quad \m{V}_e
    \end{bmatrix}$, $(\m{S}_e)^{-1} = \begin{bmatrix} (\m{R}^\top\m{R})^{-1}\m{R}^\top\\ \m{V}_e^\top
    \end{bmatrix}$, trong đó $\m{V}_e \in \mb{R}^{n \times (m-n+1)}$ là một ma trận với các cột tạo thành một cơ sở của không gian chu trình của $G$ thì
    \begin{align}
        (\m{S}_e)^{-1} \mcl{L}_e \m{S}_e = \begin{bmatrix} \mcl{L}_e(T)\m{R}^\top\m{R} & \m{0}\\
        \m{0} & \m{0}_{m-n+1,m-n+1}
        \end{bmatrix}.
    \end{align}
    \item[iii.] Với $\m{S} = \m{S}_e (\bar{\m{S}}_v)^{-1}$, trong đó  $\bar{\m{S}}_v = \begin{bmatrix} {\m{S}}_v & \m{0}\\
    \m{0}^\top & \m{I}_{n-m}
    \end{bmatrix}$ thì
    \begin{align}
        (\m{S})^{-1} \mcl{L}_e \m{S} = \begin{bmatrix} \mcl{L}& \m{0}\\
        \m{0} & \m{0}_{m-n,m-n}
        \end{bmatrix}.
    \end{align}
\end{enumerate}
\end{exercise}

\begin{exercise} Chứng minh rằng ma trận Laplace cạnh của một đồ thị đường thẳng $P_n$ và ma trận nghịch đảo của nó được cho bởi
\begin{align*}
    \mcl{L}_e(P_n) &= \begin{bmatrix}
    2 & -1 &    &       &\\
    -1 & 2 & -1 &       &\\
       & \ddots &\ddots & \ddots &\\
       &        &-1 & 2      &-1\\
       &        &       & -1 & 2
    \end{bmatrix},\\
    \text{và}~ [(\mcl{L}_e(P_n))^{-1}]_{ij} &= \frac{\min(i,j)(n-\max(i,j))}{n}.
\end{align*}
\end{exercise}

\begin{exercise} \label{ex:4.10}
Tìm các ma trận $\m{H}$, $\mcl{L}_e$, $\m{T}$ tương ứng của các đồ thị trong Hình \ref{fig:c4_ex4.10}. Tìm một cơ sở của không gian chu trình của các đồ thị đó.
\end{exercise}
\begin{figure}
    \centering
    \subfloat[]{
\begin{tikzpicture}[
roundnode/.style={circle, draw=black, thick, minimum size=3mm,inner sep= 0.25mm}
]
\node[roundnode] (v1) at (0,0) {};
\node[roundnode] (v2) at (1,-1) {};
\node[roundnode] (v3) at (2,1) {};
\node[roundnode] (v4) at (2.7,-1) {};
\node[roundnode] (v5) at (3.2,0.5) {};

\draw[-, thick] (v1)--(v2)--(v4)--(v5)--(v3)--(v1);
\draw[-, thick] (v1)--(v4);
\draw[-, thick] (v5)--(v2);
\draw[-, thick] (v4)--(v3);
\end{tikzpicture}
}
    \hfill
    \subfloat[]{
\begin{tikzpicture}[
roundnode/.style={circle, draw=black, thick, minimum size=3mm,inner sep= 0.25mm}
]
\node[roundnode]   (v1)   at   (1,1.732) {};
\node[roundnode]   (v2)   at   (1,0.5773) {};
\node[roundnode]   (v3)   at   (2,0) {};
\node[roundnode]   (v4)   at   (0,0) {};
\node[roundnode]   (v5)   at   (3,0.5773) {};
\node[roundnode]   (v6)   at   (3,1.732) {};
\node[roundnode]   (v7)   at   (4,0) {};
\draw[-, thick] (v2)--(v4);
\draw[-, thick] (v5)--(v7);
\draw[-, thick] (v1)--(v6);
\draw[-, thick] (v3)--(v7)--(v6)--(v5)--(v3);
\draw[-, thick] (v1)--(v2)--(v3)--(v4)--(v1);
\end{tikzpicture}
}
    \hfill
    \subfloat[]{
\begin{tikzpicture}[
roundnode/.style={circle, draw=black, thick, minimum size=3mm,inner sep= 0.25mm}
]
\node[roundnode]   (v1)   at   (0,0) {};
\node[roundnode]   (v2)   at   (1,0) {};
\node[roundnode]   (v3)   at   (.5,0.866) {};
\node[roundnode]   (v4)   at   (1.866,0.5) {};
\node[roundnode]   (v5)   at   (1.366,1.366) {};
\node[roundnode]   (v6)   at   (-0.366,1.366) {};
\node[roundnode]   (v7)   at   (-0.866,0.5) {};
\draw[-, thick] (v1)--(v2)--(v3)--(v1);
\draw[-, thick] (v2)--(v4)--(v5)--(v3);
\draw[-, thick] (v3)--(v6)--(v7)--(v1);
\end{tikzpicture}    
    }
    \caption{Các đồ thị trong Bài tập \ref{ex:4.10}.}
    \label{fig:c4_ex4.10}
\end{figure}

\begin{exercise} \label{ex:4.11}
Xét đồ thị vô hướng $G=(V,E)$ với $n$ đỉnh và $m$ cạnh và có ma trận Laplace $\mcl{L}$. Chứng minh rằng:
\begin{enumerate}
    \item[i.] Với mọi $\m{x}, \m{y}\in \mb{R}^n$, ta đều có
    \begin{align*}
        \m{y}^\top \mcl{L} \m{x} = \sum_{(i,j)\in E} a_{ij}(x_i-x_j)(y_i-y_j).
    \end{align*}
    \item[ii.] Từ đó chứng minh rằng với mọi $\m{x}\in \mb{R}^n$ thì 
    \begin{align*}
        \text{sign}(\m{x})^\top \mcl{L}\m{x}\le 0,
    \end{align*}
    và bất đẳng thức này là chặt khi và chỉ khi tồn tại một cạnh $(i,j)$ sao cho \text{sign}$(x_i) \neq \text{sign}(x_j)$.
\end{enumerate}
\end{exercise}

\begin{exercise} \label{ex:4.12}
Xét đồ thị vô hướng, liên thông $G$ gồm $n$ đỉnh và $m$ cạnh. Gọi $T(G)$ là một cây bao trùm của $G$. Chứng minh rằng $\lambda_i(\mcl{L}) \ge \lambda_i(\mcl{L}_e(T))$.
\end{exercise}

\begin{exercise} \label{ex:4.13}
\begin{figure}[h]
\centering
\subfloat[$y={\rm sgn}(x)$]{
\resizebox{0.32\linewidth}{!}{
\begin{tikzpicture}[scale=1.2]
  \draw[->] (-2.5,0) -- (2.5,0) node[right] {$x$};
  \draw[->] (0,-1.4) -- (0,1.4) node[above] {$y$};

  \draw[thick] (-2.4,-1) -- (0,-1);
  \draw[thick] (0,1) -- (2.4,1);

  \draw[thick, fill=white] (0,1) circle (2pt);
  \draw[thick, fill=white] (0,-1) circle (2pt);

  \fill (0,0) circle (2pt);
\end{tikzpicture}
}}
\hfill
\resizebox{0.32\linewidth}{!}{
\subfloat[$y={\rm sat}(x)$]{
\begin{tikzpicture}[scale=1.2]
  \draw[->] (-2.5,0) -- (2.5,0) node[right] {$x$};
  \draw[->] (0,-1.4) -- (0,1.4) node[above] {$y$};

  \draw[thick] (-2.5,-1) -- (-1,-1);
  \draw[thick] (-1,-1) -- (1,1);
  \draw[thick] (1,1) -- (2.5,1);

  \draw[dashed] (-2.5,1) -- (2.5,1);
  \draw[dashed] (-2.5,-1) -- (2.5,-1);

  \node at (1.1,-0.2) {$1$};
  \node at (-1.3,-0.2) {$-1$};
\end{tikzpicture}
}}
\hfill
\subfloat[$y={\rm tanh}(x)$]{
\resizebox{0.32\linewidth}{!}{
\begin{tikzpicture}[scale=1.2]
  \draw[->] (-2.5,0) -- (2.5,0) node[right] {$x$};
  \draw[->] (0,-1.3) -- (0,1.3) node[above] {$y$};

  \draw[thick, domain=-2.5:2.5, samples=200]
    plot (\x,{tanh(\x)});
\end{tikzpicture}
}}
\caption{Đồ thị các hàm ${\rm sgn}(x)$, ${\rm sat}(x)$, và ${\rm tanh}(x)$. \label{fig:c4_ex11}}
\end{figure}
Phân tích các hệ đồng thuận sau dựa trên lý thuyết ổn định Lyapunov:
\begin{enumerate}
    \item[i.] (Thuật toán đồng thuận với hàm dấu \cite{Cao2011distributed,Chen2011finite}) 
    \[\dot{{x}}_i = -\sum_{j\in {N}_i} \text{sgn}({x}_i - {x}_j), ~i=1, \ldots, n,\]
    trong đó
    \begin{align*}
    {\rm sgn}({x})=\left\lbrace\begin{array}{cc}
    0, & x=0,\\
    1, & x>0,\\
    -1,& x<0.
    \end{array}\right.
    \end{align*}
    trong trường hợp đồ thị vô hướng, hoặc trường hợp đồ thị có hướng, liên thông yếu và cân bằng.
    \item[ii.] (Thuật toán đồng thuận với hàm bão hòa) \[\dot{{x}}_i = -\sum_{j\in {N}_i} \text{sat}({x}_i - {x}_j), ~i=1, \ldots, n,\] trong đó
    \begin{align*}
    {\rm sat}({x})=\left\lbrace\begin{array}{cc}
    x, & |x|\leq 1,\\
    1, & x > 1.
    \end{array}\right.
    \end{align*}
    \item[iii.] (Thuật toán đồng thuận với hàm tang-hyperbol) \[\dot{{x}}_i = -\sum_{j\in {N}_i} \text{tanh}({x}_i - {x}_j), ~i=1, \ldots, n,\] trong đó
    \begin{align*}
    	{\rm tanh}({x})= \frac{e^x - e^{-x}}{e^x + e^{-x}}.
    \end{align*}
   \item[iv.] (Thuật toán đồng thuận với đầu ra bão hòa \cite{Lim2017consensus}) \[\dot{{x}}_i = -\sum_{j\in {N}_i}(\text{sat}({x}_i) - \text{sat}({x}_j)),~i=1, \ldots, n.\]
    \item[v.] (Thuật toán đồng thuận trọng số ma trận xác định dương \cite{Trinh2018matrix}) \index{thuật toán!đồng thuận!trọng số ma trận dương}
    \[\dot{\m{x}}_i = -\sum_{j\in {N}_i} \m{A}_{ij} (\m{x}_i - \m{x}_j), ~i=1, \ldots, n,\] trong đó $\m{x}_i \in \mb{R}^d$, $\m{A}_{ij} = \m{A}_{ji} = \m{A}_{ij}^\top \in \mb{R}^{d\times d}$ là các ma trận đối xứng, xác định dương.
\end{enumerate}
\end{exercise}

\begin{exercise}[Thuật toán đồng thuận thích nghi I]\index{đồng thuận!thích nghi} \cite{Li2009consensus}
Xét hệ các tác tử tuyến tính tổng quát tương tác qua đồ thị $G$ vô hướng, liên thông. Mô hình mô tả động học của các tác tử được cho bởi phương trình
    \begin{align}
        \dot{\m{x}}_i(t) = \m{A} \m{x}_i(t) + \m{B} \m{u}_i(t),~i=1,\ldots,n,
    \end{align}
    với $\m{x}_i \in \mb{R}^d$, $\m{u}_i \in \mb{R}^m$, $\m{A} \in \mb{R}^{d \times d}$, và $\m{B} \in \mb{R}^{d \times m}$. Giả sử rằng $(\m{A},\m{B})$ là điều khiển được.
    \begin{itemize}
        \item [i.] Với luật đồng thuận dạng:
        \begin{align}
            \m{u}_i(t) &= \m{K} \sum_{j\in N_i} c_{ij}(t) (\m{x}_j - \m{x}_i),~i=1,\ldots,n,\\
            \dot{c}_{ij}(t) &= \kappa \|\m{x}_j - \m{x}_i\|^2,~\forall (i,j) \in E,
        \end{align}
        trong đó $\m{A}+\m{B}\m{K}$ là Hurwitz, $c_{ij}(t)$ là các hệ số thích nghi, được chọn thỏa mãn $c_{ij}(0)>0$, và $\kappa>0$ là tốc độ thích nghi. Hãy biểu diễn hệ dưới dạng ma trận và chứng minh hệ dần đạt đồng bộ hóa tới một nghiệm của $\dot{\m{x}}_0=\m{A}{\m{x}}_0$, khi $t\to +\infty$. 
        \item [ii.] Với luật đồng thuận dạng:
        \begin{align}
            \m{u}_i(t) &= c_{i}(t) \m{K} \sum_{j\in N_i} (\m{x}_j - \m{x}_i),\\
            \dot{c}_{i}(t) &= \kappa \left|\left|\sum_{j\in N_i} c_{i}(t) (\m{x}_j - \m{x}_i)\right|\right|^2,~i=1,\ldots,n,
        \end{align}
        trong đó $\m{A}+\m{B}\m{K}$ là Hurwitz và $c_{i}(t)$ là các hệ số thích nghi, được chọn thỏa mãn $c_{i}(0)>0$, và $\kappa>0$ là tốc độ thích nghi. Hãy biểu diễn hệ dưới dạng ma trận và chứng minh hệ dần đạt đồng bộ hóa tới một nghiệm của $\dot{\m{x}}_0=\m{A}{\m{x}}_0$, khi $t\to +\infty$. 
        \item [iii.] Mô phỏng với MATLAB các trường hợp trên.
    \end{itemize}
\end{exercise}

\begin{exercise}[Thuật toán đồng thuận thích nghi II] \cite{Mei2018model}
Xét hệ đa tác tử mô tả bởi đồ thị $G$ vô hướng, liên thông. Giả sử mô hình tác tử được cho bởi $\dot{x}_i = u_i + f_i(x_i,t) \theta_i$, trong đó $\theta_i \in \mb{R}$ là một tham số chưa biết còn $f_i(x_i,t)$ là một hàm đã biết thỏa mãn điều kiện Lipschitz với biến $\m{x}_i$ và khả vi theo biến $t$. Xét thuật toán đồng thuận thích nghi cho bởi:
\begin{align}
u_i &= -\sum_{j\in {N}_i} ({x}_i - {x}_j) - {f}_i({x}_i,t) \hat{\theta}_i, \\
\dot{\hat{{\theta}}}_i & = \gamma_i {f}_i({x}_i,t) \sum_{j\in {N}_i} ({x}_i - {x}_j), ~i=1, \ldots, n,
\end{align}
trong đó $\gamma_i >0$ là hệ số thích nghi. 
\begin{enumerate}
    \item[i.] Hãy biểu diễn hệ dưới dạng ma trận.
    \item[ii.] Xét hàm Lyapunov $V = \frac{1}{2}\m{x}^\top \mcl{L} \m{x} + \frac{1}{2} \sum_{i=1}^n \frac{(\hat{\theta}_i - \theta_i)^2}{2 \gamma_i}$. Sử dụng bổ đề Barbalat (xem Phụ lục~\ref{append:control_theory}), hãy chứng minh rằng $x_i \to x_j$, $\forall i, j = 1, \ldots, n$ khi $t \to +\infty$.
\end{enumerate}
\end{exercise}

\begin{exercise}[Ma trận Laplace với trọng số phức] \cite{Lin2014distributed}
Xét đồ thị $G=(V,E)$ vô hướng, liên thông, không chứa khuyên gồm $n$ đỉnh và $m$ cạnh. Tương ứng với mỗi cạnh $(i,j)$ trong $E$, ta gán một trọng số phức $c_{ij} \in \mb{C}$. Định nghĩa ma trận Laplace với trọng số phức $\mcl{L}=[l_{ij}] \in \mb{C}^{n \times n}$, với các phần tử \index{đồ thị!trọng số phức}
\begin{align*}
l_{ij} = \left\lbrace \begin{array}{ll}
-c_{ij}, & (i,j) \in E,\\
0, & (i,j) \notin E, \\
\sum_{j=1,j\ne i}^n & i = j.
\end{array} \right.
\end{align*}
\begin{enumerate}
    \item[i.] Với $\m{H} \in \mb{R}^{m \times n}$ là ma trận liên thuộc của đồ thị $G$ và $\m{W} \triangleq {\rm diag}(c_{ij}) \in \mb{C}^{m \times m}$. Chứng minh rằng ma trận Laplace có thể biểu diễn dưới dạng $\mcl{L}=\m{H}^\top \m{W}\m{H}$.
    \item[ii.] Xét hệ gồm $n$ tác tử với đồ thị trọng số phức $G$, trong đó mỗi tác tử có một biến trạng thái phức $x_i \in \mb{C}$. Giả sử các tác tử trong hệ cập nhật biến trạng thái theo thuật toán đồng thuận trọng số phức:
\begin{align} \label{eq:complex_consensus}
\dot{x}_i(t) = -\sum_{j \in N_i}c_{ij}(x_i(t) - x_j(t)),\, i =1,\ldots,n.
\end{align}
Đặt $\m{x}=[x_1,\ldots,x_n]^\top \in \mb{C}^n$, hãy biểu diễn hệ dưới dạng ma trận.
    \item[iii.] Giả thiết thêm rằng  với ${\rm Re}(c_{ij})>0,\forall (i,j)\in E$. Chứng minh hệ $n$ tác tử tiệm cận tới tập đồng thuận $\mc{A}_C=\{\m{x}=[x_1,\ldots,x_n]^\top \in \mb{C}^n|~x_1 = \ldots = x_n\}$ và xác định giá trị đồng thuận. (Gợi ý: xét hàm Lyapunov $V = \m{x}^{\rm H} \m{x}$, trong đó $\m{A}^{\rm H}$ kí hiệu ma trận chuyển vị liên hợp (conjugate transpose) của ma trận $\m{A}$).
    \item[iv.] Thay vì trọng số phức, hãy định nghĩa ma trận Laplace trọng số quartenion đơn vị: \[q_{ij} = q_{ij,0} + \mathsf{i} q_{ij,1} + \mathsf{j} q_{ij,2} + \mathsf{k} q_{ij,3} \in \mb{H},\, (i,j)\in E,\] trong đó \cite{Goor2017} \index{quaternion}
\begin{align}
\mathsf{i}^2 = \mathsf{j}^2=\mathsf{k}^2 = \imath\jmath\mathsf{k} = -1,
\end{align}
$q_{ij,0},q_{ij,1},q_{ij,2},q_{ij,3} \in \mb{R}$, và $\|q_{ij}\| = \sqrt{q_{ij,0}^2+q_{ij,1}^2+q_{ij,2}^2+q_{ij,3}^2} =1$. Giả sử mỗi tác tử có biến trạng thái $x_i \in \mb{H}$, đề xuất thuật toán đồng thuận tương ứng và tìm điều kiện để hệ tiệm cận tới tập đồng thuận.
\end{enumerate}
\end{exercise}

\begin{exercise}[Phân tích thuật toán đồng bộ hóa theo phương pháp Lyapunov] \cite{Li2009consensus,Trinh2024msc}
Phân tích thuật toán đồng thuận cho hệ tuyến tính \eqref{eq:chap3-consensus-linear-system1} (trực tiếp), và \eqref{eq:chap3_output_synch1}--\eqref{eq:chap3_output_synch4} (gián tiếp) ở Chương~\ref{chap:consensus} sử dụng phương pháp Lyapunov.
\end{exercise}

\begin{exercise}[Đồ thị Cellular sheaf và hệ đa tác tử tương tác không đồng nhất] \cite{Hansen2019toward,Hanks2025heterogeneous}
Đồ thị cellular sheaf cung cấp một phương pháp mô tả hệ đa tác tử không đồng nhất về cả số biến trạng thái và số chiều của tương tác giữa các tác tử láng giềng. Một cách đơn giản hóa, xét hệ đa tác tử trong đó mỗi tác tử có vector biến trạng thái $\m{x}_i \in \mb{R}^{n_i}$, $n_i\geq 1$ tương tác qua một đồ thị vô hướng, liên thông $G=(V,E)$. Mỗi cạnh $(i,j)\in E$ mô tả một tương tác $m_{ij}\geq 1$ chiều giữa hai tác tử $i,j$ trong hệ. Do số biến trạng thái của hai tác tử láng giềng khác nhau và khác với số biến trạng thái trong liên kết $(i,j)$, định nghĩa hai ma trận $\m{F}_{ij} \in \mb{R}^{m_{ij} \times n_i}$ và $\m{F}_{ji}  \in \mb{R}^{m_{ij} \times n_j}$ và vector trạng thái
\begin{align}
\m{y}_{ij} = \m{F}_{ij}\m{x}_i - \m{F}_{ji}\m{x}_i = -\m{y}_{ji}
\end{align}
của cạnh $(i,j) \in E$. Mỗi biến trạng thái tương tác này tác động vào tác tử $i$ thông qua ma trận $\m{F}_{ij}^\top$ và tác tử $j$ thông qua ma trận $\m{F}_{ji}^\top$:
\begin{align}
\m{u}_i &= \sum_{j\in N_i} \m{F}_{ij}^\top \m{y}_{ij} = \sum_{j\in N_i} \m{F}_{ij}^\top (\m{F}_{ij}\m{x}_i - \m{F}_{ji}\m{x}_j), \\
\m{u}_j &= \sum_{i\in N_j} \m{F}_{ji}^\top \m{y}_{ji} = \sum_{i\in N_j} \m{F}_{ji}^\top (\m{F}_{ji}\m{x}_j - \m{F}_{ij}\m{x}_i).
\end{align}
\begin{itemize}
\item[i.] Đặt $\m{x} = [\m{x}_1^\top,\ldots,\m{x}_n^\top] \in \mb{R}^{\sum_{i=1}^n n_i}$, $\m{y} = [\m{y}_1^\top,\ldots,\m{y}_m^\top] \in \mb{R}^{\sum_{(i,j)\in E} m_{ij}}$, và $\m{u} = [\m{u}_1^\top,\ldots,\m{u}_n^\top] \in \mb{R}^{\sum_{i=1}^n n_i}$. Hãy tìm ma trận $\mcl{L}$ biểu diễn quan hệ tuyến tính giữa hai vector $\m{u}$ và $\m{x}$.
\item[ii.] Chứng minh rằng ma trận $\mcl{L}$ là đối xứng và bán xác định dương. (Gợi ý: Sử dụng hàm thế $V=\m{x}^\top \mcl{L} \m{x}$.)
\item[iii.] Tương ứng với một định hướng các cạnh của $G$, đặt $\tilde{\m{H}}=[\tilde{\m{H}}_{ki}]$, với
\begin{align*}
\tilde{\m{H}}_{kt} = \left\lbrace \begin{array}{ll}
-\m{F}_{ij}, & \text{cạnh thứ } k \text{ là } (i,j) \text{ xuất phát từ đỉnh } t=i, \\
\m{F}_{ji}, & \text{cạnh thứ } k \text{ là } (i,j) \text{ kết thúc tại đỉnh } t=j, \\
\m{0}_{m_{ij} \times n_t}, & \text{trường hợp khác.}
\end{array} \right.
\end{align*}
Biểu diễn ma trận $\mcl{L}$ theo $\tilde{\m{H}}$. Nếu mỗi cạnh $e_{ij} \in E$ có một trọng số $\m{W}_{ij} \in \mb{R}^{m_{ij} \times m_{ij}}$, biểu diễn ma trận $\mcl{L}$ theo $\tilde{\m{H}}$ và $\m{W}={\rm blkdiag}(\m{W}_{ij})$.
\item[iv.] Biểu diễn ma trận $\mcl{L}$ trong trường hợp $n_{i}=d,\forall (i,j)\in E$. Tìm điều kiện đủ của $d$ và $\m{F}_{ij}, \m{F}_{ji}$ để ${\rm rank}(\mcl{L}) = dn-d$.
\item[v.] (Liên hệ với ma trận Laplace cứng hướng) Trong trường hợp $\m{F}_{ij}=\m{F}_{ji} \in \mb{R}^{d\times d}$ và $\m{F}_{ij}^2=\m{F}_{ij}$ (idempotent), hãy biểu diễn ma trận $\mcl{L}$ trong trường hợp này. Tìm điều kiện của ma trận $\m{F}_{ij}$ và $G$ để $\mcl{L}$ có hạng $dn-d$, $dn-d-1$. (Gợi ý: tham khảo mục \ref{c5_s5:bearing_formation}).
\end{itemize}
\end{exercise}

%
\part{Một số ứng dụng của hệ đa tác tử}
\chapter{Điều khiển đội hình}
\label{chap:formation}
\section{Giới thiệu}
Điều khiển đội hình\footnote{formation control} là một trong những bài toán  được nghiên cứu rộng rãi nhất trong điều khiển hệ đa tác tử. Giả sử chúng ta cần điều khiển một hệ gồm $n$ tác tử (có thể là UAV, UUV, xe tự lái) di chuyển theo một đội hình mong muốn. Đội hình mong muốn, hay còn gọi là đội hình đặt, được xác định thông qua một tập các biến về khoảng cách, vector hướng, hay góc lệch giữa vị trí các tác tử trong hệ.

Các thuật toán điều khiển đội hình có thể ứng dụng trong hệ thống giao thông cao tốc thông minh, trong đó các phương tiện di chuyển gần nhau tự thiết lập và di chuyển sao cho khoảng cách giữa các phương tiện là không đổi và vận tốc của các phương tiện đồng thuận với nhau \cite{Fax2004}. Khi đội hình đã được thiết lập, việc di chuyển của đội hình với tốc độ cao được đảm bảo an toàn (các xe không bị va chạm với nhau) và việc không cần thay đổi vận tốc quá nhiều giúp tiết kiệm được nhiên liệu. Ngoài ứng dụng trong quân sự, sự phát triển gần đây của công nghệ UAV nảy sinh những ứng dụng dân sự mới như thăm dò địa chất, tìm kiếm cứu nạn, kiểm tra lỗi và giám sát (cánh đồng pin mặt trời, công trình xây dựng), \ldots Những ứng dụng này có thể được thực hiện hiệu quả và dễ dàng hơn nếu các UAV bay theo một đội hình định trước \cite{Anderson2008}. Cuối cùng, một số thuật toán điều khiển đội hình được dùng để mô phỏng, phân tích và giải thích các hiện tượng bầy đàn ở côn trùng, cá, chim trong tự nhiên \cite{Bullo2019lectures}.

Vấn đề trọng tâm trong điều khiển đội hình là \emph{tạo đội hình đặt}. Để giải quyết bài toán tạo đội hình đặt, rất nhiều phương pháp đã được đưa ra. Tất cả các phương pháp điều khiển đội hình hiện tại đều đòi hỏi các tác tử đo một số biến hình học về đội hình, hoặc kết hợp đo và truyền thông một số biến hình học về đội hình với các tác tử khác, từ đó điều khiển vị trí sao cho các biến hình học này thỏa mãn giá trị đặt. Trong quá trình thiết lập đội hình, đôi khi có một số tác tử đặc biệt được chọn làm tác tử dẫn đàn (leader), vị trí của các leader là tham chiếu để các tác tử còn lại (follower) điều khiển vị trí tương ứng.

Hiện nay, các phương pháp điều khiển đội hình thường được phân loại dựa trên giả thiết về các biến đo, biến điều khiển, và điều kiện về đồ thị mô tả luồng thông tin giữa các tác tử. Bảng~\ref{table:formation_control} phân loại các bài toán điều khiển đội hình theo cách phân loại này \cite{Oh2015,Ahn2019formation}. Dựa trên bảng~\ref{table:formation_control} một số phương pháp điều khiển chủ yếu gồm điều khiển dựa trên vị trí tuyệt đối, điều khiển dựa trên vị trí tương đối, điều khiển dựa trên vector hướng, điều khiển dựa trên khoảng cách và điều khiển dựa trên kết hợp các biến hình học khác nhau. Trong mỗi phương pháp, chúng ta lại có thể chia nhỏ hơn tùy theo giả thiết cụ thể. Ví dụ, vector hướng có thể được đo trong hệ qui chiếu riêng của mỗi tác tử hoặc đo trong hệ qui chiếu đã được đồng bộ.

\begin{longtable}{|p{0.15\linewidth}|p{0.12\linewidth}|p{0.12\linewidth}|p{0.135\linewidth}|p{0.315\linewidth}|}
\hline
Phương pháp   & Biến đo & Biến điều khiển  & Điều kiện đồ thị   & Tài liệu tham khảo \\ \hline
Dựa trên vị trí & Vị trí & Vị trí &  & \cite{Lewis1997high,Young2001control}\\ \hline
Dựa trên sai lệch vị trí & Sai lệch vị trí & Sai lệch vị trí & Liên thông & \cite{Fax2004,Lafferriere2005,Lin2014distributed}     \\ \hline
\multirow{4}{\linewidth}{Dựa trên khoảng cách}                                           & Sai lệch cục bộ, góc định hướng    & Sai lệch cục bộ và góc định hướng & Liên thông & \cite{Oh2014tac,Montijano2016,lee2016distributed,Lee2017arxiv}                            \\ \cline{2-5} 
& Dựa trên tọa độ tương đối địa phương & Tọa độ tương đối địa phương & Đồ thị cứng khoảng cách & \cite{Eren2003sensor,Krick2009,Tian2013,Oh2014distance,Mou2016TAC,Zhiyong2016,Pham2017ICCAS} \\ \cline{2-5} 
& Tọa độ tương đối địa phương & Tọa độ tương đối địa phương & Đồ thị không cứng khoảng cách & \cite{dimarogonas2009further,Park2015,Pham2018IJRNC}                                                            \\ \cline{2-5} 
& Khoảng cách & Khoảng cách & Đồ thị cứng khoảng cách   & \cite{Anderson2011range,cao2011formation,Suttner2018formation}    \\ \hline
\multirow{6}{\linewidth}{Dựa trên vector hướng} & Vector hướng địa phương & Góc & Đồ thị đơn giản $C_3$, $C_4$ hoặc $C_n$ & \cite{Basiri2010,Bishop2010,Bishop2011,Zhao2014,Zhao2014ijc}\\ \cline{2-5} 
    & Vector hướng địa phương & Góc & Đồ thị cứng yếu; đồ thị cứng góc (2D và 3D) & \cite{Buckley2017,jing2018weak,Chen2022globally,Chen2020angle,Chen2022maneuvering}                                       \\ \cline{2-5} 
    & Vector hướng & Vector hướng  & Đồ thị cứng hướng trong $\mb{R}^d$             &\cite{Bishop2011a,Eren2012,Eric2014,Trinh2014CN,zhao2015tac,Tron2016CDC,TrinhIJRNC2018}        \\ \cline{2-5} 
    &  Vector hướng địa phương và  Góc định hướng tương đối  & Vector hướng địa phương và  góc định hướng  tương đối  & Đồ thị cứng hướng trong $SE(d)$                           & \cite{Franchi2012,Zelazo2015,schiano2016rigidity,Michieletto2016}  \\ \cline{2-5} 
    & Vector hướng địa phương và góc định hướng tương đối & Vector hướng địa phương và góc định hướng  tương đối & Đồ thị cứng hướng trong $\mb{R}^d$     & \cite{zhao2015tac,Trinh2019TAC,TrinhCCTA2018}                     \\ \cline{2-5} 
    & Vị trí  tương đối & Vị trí  tương đối & Đồ thị cứng hướng trong $\mb{R}^d$ & \cite{Zhao2015CDC,Zhao2015ecc}   \\ \hline
Kết hợp vector hướng, góc lệch và khoảng cách & Vector hướng, góc lệch và khoảng cách & Vector hướng, góc lệch, khoảng cách, tỉ lệ khoảng cách & Đồ thị cứng hỗn hợp & \cite{Bishop2014,Fathian2016globally,Sun2017AutOrient,Park2017,kwon2018infinitesimal,Fang2025simultaneous} \\ \hline
\caption{Phân loại bài toán điều khiển đội hình}
\label{table:formation_control}
\end{longtable}
\begin{SCfigure}[][ht!]
\caption{Hệ qui chiếu toàn cục ($^g\Sigma$), hệ qui chiếu chung ($^c\Sigma$), và các hệ qui chiếu cục bộ ($^i\Sigma$ và $^j\Sigma$).}
\hspace{2.5cm}
\label{fig:reference_frame1}
\includegraphics[height=6cm]{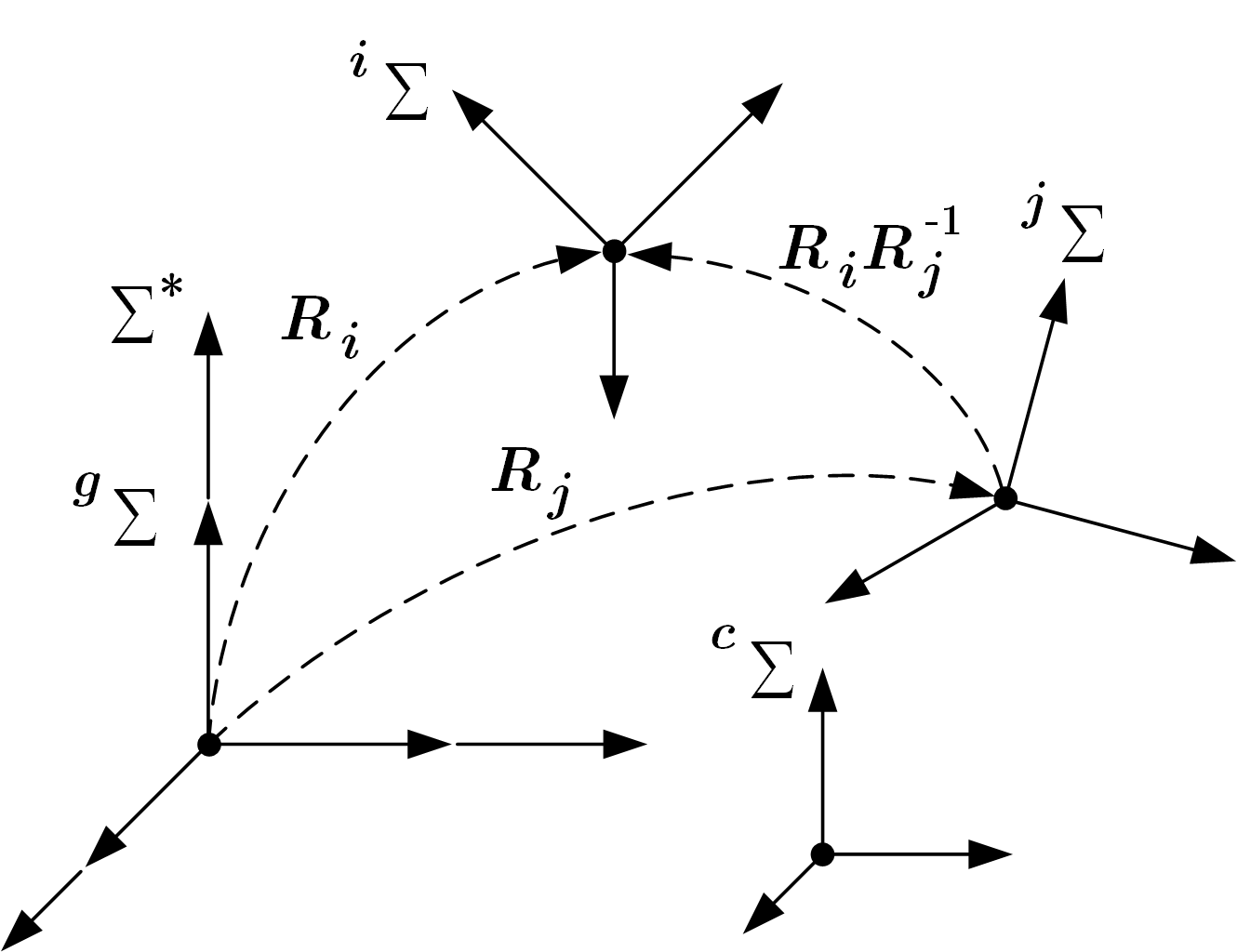}
\end{SCfigure}

Các biến trạng thái của một tác tử phân tán bao gồm một số đại lượng viết trong một hệ qui chiếu gắn với tác tử đó. Thông thường, các biến trạng thái được chọn gồm vị trí và vận tốc của tác tử cùng với hướng của các hệ tọa độ tham chiếu (gọi chung là định hướng của tác tử trong không gian). Để so sánh các giá trị của tác tử, vị trí và định hướng của tác tử cần phải được cho dưới cùng một hệ tọa độ tham chiếu. Trong không gian $\mb{R}^d$, định hướng của một tác tử $i$ được cho bởi một ma trận $\m{R}_i \in SO(d)$, trong khi định hướng và vị trí được thể hiện bởi các phần tử trong $SE(d)$. Do mỗi tác tử có một hệ tọa độ tham chiếu riêng, định hướng của các tác tử thường có sai lệch so với nhau cũng như so với hệ tham chiếu toàn cục.

Trên hình~\ref{fig:reference_frame1}, hệ qui chiếu toàn cục được kí hiệu bởi $^g\Sigma$, còn các hệ qui chiếu cục bộ/địa phương được kí hiệu bởi $^i\Sigma$ và $^j\Sigma$. Các trục tọa độ của tác tử $i$ và $j$ không trùng nhau cũng như không trùng với hệ tọa độ toàn cục. Ma trận $\m{R}_i \in SO(d)$ mô tả phép chuyển tọa độ từ hệ tọa độ $^g\Sigma$ tới $^i\Sigma$. Ma trận để chuyển từ $^j\Sigma$ tới $^i\Sigma$ được cho bởi $\m{R}_{ij} = \m{R}_i\m{R}_j^{-1}$. Như vậy, ma trận $\m{R}^{-1}_i$ mô tả phép chuyển từ hệ tọa độ địa phương $^i\Sigma$ tới hệ tọa độ toàn cục  $^g\Sigma$.

Kí hiệu $\Sigma^*$ là một hệ qui chiếu chung, không nhất thiết phải trùng với $^g\Sigma$. Một hệ tọa độ với các trục trùng với $\Sigma^*$ được gọi là một hệ qui chiếu chung $^c\Sigma$. Chú ý rằng hệ qui chiếu $\Sigma^* \neq {^c\Sigma}$ do gốc tọa độ của $\Sigma^*$ và $^c\Sigma$ có thể khác nhau. 

Trong nhiều bài toán điều khiển đội hình, ta thường mong muốn các tác tử có thể xác định vị trí và hướng dựa trên một hệ qui chiếu chung $^c\Sigma$. Ma trận $\m{R}_c \in SO(3)$ mô tả phép chuyển tọa độ từ $^g\Sigma$ sang $^c\Sigma$. Nếu như hướng của các tác tử được điều chỉnh sao cho $^i\Sigma = {^j\Sigma} = {^c\Sigma}$ thì ta nói rằng các tác tử được đồng thuận về hướng. 

Với mỗi tác tử $i$, ta kí hiệu vị trí của tác tử này trên hệ tham chiếu toàn cục là $\m{p}_i \in \mb{R}^d$. Với hai tác tử $i$ và $j$, vector sai lệch $\m{z}_{ij} = \m{p}_i - \m{p}_j$ là một vector viết trong hệ tham chiếu toàn cục $^g\Sigma$. Vector này cũng có thể biểu diễn bởi $\m{z}_{ij}^i = \m{p}_i^i - \m{p}_j^i = -\m{p}_j^i$ trong $^i\Sigma$ hoặc là $\m{z}_{ij}^j = \m{p}_j^j - \m{p}_i^j =- \m{p}_i^j$ trong $^i\Sigma$ do $\m{p}_i^i = \m{p}_j^j = \m{0}$. Chú ý rằng vector $\m{z}_{ij}$ có hướng từ $j$ tới $i$, trong khi $\m{z}_{ji}$ có hướng từ $i$ tới $j$. Chú ý rằng $\m{z}^j_{ji} = \m{p}_i^j$ (vị trí của $i$ đo bởi $j$ trong hệ qui chiếu $^j\Sigma$) hay $\m{z}_{ji}^i=-\m{p}_j^i$ (vị trí của $j$ đo bởi $i$ trong hệ qui chiếu $^i\Sigma$). Mặc dù $\m{p}_j - \m{p}_i = \m{z}_{ji}=-\m{z}_{ij}=-(\m{p}_i - \m{p}_j)$, do hệ qui chiếu của các tác tử khác nhau, ta có $\m{p}_i^j \neq -\m{p}_j^i$ do
\begin{align*}
\m{z}_{ji}^j =\m{p}_i^j=
\m{R}_j\m{R}_i^{-1}\m{z}_{ji}^i=\m{R}_{ji}(-\m{p}_j^i).
\end{align*}
Nhận xét rằng mặc dù có hướng và hệ qui chiếu khác nhau, các vector $\m{z}_{ij}^i, \m{z}_{ji}^i, \m{z}_{ij}^j$, và $\m{z}_{ji}^j$ đều có độ dài như nhau và bằng khoảng cách giữa hai tác tử $i$ và $j$. Ta kí hiệu khoảng cách từ tác tử $i$ tới tác tử $j$ bởi $d_{ij} = \|\m{p}_i - \m{p}_j\| = \|\m{z}_{ij}^i\| = \|\m{z}_{ij}^j\|=\|\m{z}_{ij}\| = d_{ji}$. 

Một đại lượng (tương đối) khác có thể được đo đạc giữa hai tác tử $i$ và $j$ là vector (đơn vị) hướng $\m{g}_{ij} = \frac{\m{p}_j - \m{p}_i}{\|\m{p}_j - \m{p}_i\|}$ với giả thiết rằng $\m{p}_i \neq \m{p}_j$. Dễ thấy $\|\m{g}_{ij}^i\| = \|\m{g}_{ij}\| = 1$. Chú ý rằng  $\m{g}_{ij}=-\m{g}_{ji}$, tuy nhiên $\m{g}_{ij}^i \neq - \m{g}_{ji}^j$ với cùng lý do như với các vector vị trí tương đối.

Với mỗi tác tử $i$, ta giả sử rằng tác tử $i$ có mô hình tích phân bậc nhất
\begin{align}\label{eq:FM_singleI}
\dot{\m{p}}_i = \m{u}_i,~i=1, \ldots, n,
\end{align}
với $\m{u}_i$ là tín hiệu điều khiển. Phương trình \eqref{eq:FM_singleI} được viết trên hệ qui chiếu $^g\Sigma$. Ta có thể biến đổi phương trình \eqref{eq:FM_singleI} như sau
\begin{align}\label{eq:FM_singleI_local}
\m{R}_i\dot{\m{p}}_i &= \m{R}_i\m{u}_i \nonumber \\
\dot{\m{p}}_i^i &= \m{u}_i^i,
\end{align}
với $\m{u}_i^i$ là tín hiệu điều khiển biểu diễn trên hệ qui chiếu địa phương $^i\Sigma$ gắn với tác tử $i$. 

Tiếp theo, ta định nghĩa khái niệm về các cấu trúc (tập cạnh) mô tả tương tác giữa các tác tử trong hệ, bao gồm: đo đạc, điều khiển, và truyền tin.

\begin{Definition}[Cấu trúc đo đạc, điều khiển, truyền thông] Nếu tác tử $i$ đo một biến tương đối nào đó với tác tử $j$ thì tác tử $j$ là một láng giềng-ra của $i$ trong đồ thị đo đạc $G^s$, kí hiệu bởi $j \in {N}_i^o$, và $(i,j)\in {E}^s$. Nếu tín hiệu điều khiển của tác tử $i$ dựa trên chuyển động của tác tử $j$ thì $j$ là một láng giềng ra của $i$ trong đồ thị điều khiển $G^a$, kí hiệu bởi  $(i,j)\in {E}^a$. Nếu tác tử $j$ gửi một biến thông tin tới tác tử $i$ thì $j$ là một láng giềng-ra của tác tử $i$ trong đồ thị thông tin $G^c$, kí hiệu bởi $(i,j) \in E^c$.
\end{Definition}

Như vậy, trong một bài toán điều khiển đội hình tổng quát, các cấu trúc về đo đạc, điều khiển, cũng như truyền thông là khác nhau: $E^s \neq E^a \neq E^c$. Khi nói đến một cấu trúc của một đội hình mà không nói gì thêm, ta ngầm nói đến cả ba cấu trúc này. Khi lượng thông tin của mỗi tác tử bị giới hạn, việc lựa chọn cấu trúc đóng vai trò quyết định liệu các tác tử có thể tạo một đội hình đặt trước hay không. Lý thuyết đồ thị cứng \index{lý thuyết!đồ thị cứng} là cơ sở để xác định cấu trúc ẩn sau các đội hình đặt.

Khi một tác tử $i$ đo vị trí tương đối với tác tử $j$, $i$ có thể không cập nhật vị trí của mình dựa trên vị trí tương đối mà chỉ dựa trên khoảng cách giữa hai tác tử, hay dựa trên vector hướng từ $i$ tới $j$. Trong tài liệu này, ta sử dụng thuật ngữ điều khiển đội hình dựa trên ``X'' nếu như các tác tử trong hệ điều khiển vị trí của mình dựa trên biến ``X'' và không tính tới việc các tác tử đo đạc hay truyền tin các biến nào khác. Với qui ước này, ta sẽ lần lượt xét các phương pháp điều khiển đội hình dựa trên vị trí tuyệt đối, vị trí tương đối, khoảng cách và vector hướng.  Đây là hai phương pháp điều khiển đội hình đã và đang được quan tâm nghiên cứu gần đây. 

\section{Điều khiển đội hình dựa trên vị trí tuyệt đối}
Khi tất cả các tác tử đều có thể nhận thông tin vị trí của mình từ hệ tham chiếu toàn cục $^g\Sigma$, bài toán điều khiển đội hình trở nên khá đơn giản. Cụ thể, khi mỗi tác tử $i$ biết được $\m{p}_i$ và mong muốn đạt tới vị trí đặt $\m{p}_i^*$  thì luật điều khiển đội hình \eqref{eq:FM_singleI} viết cho mỗi tác tử có thể thiết kế đơn giản dưới dạng
\begin{align} \label{eq:FM_positionbased}
    \dot{\m{p}}_i = \m{u}_i = -k_p (\m{p}_i - \m{p}_i^*),~ i=1, \ldots, n,
\end{align}
với $k_p > 0$ là hằng số dương. 

Kí hiệu $\m{p} = [\m{p}_1^\top, \ldots, \m{p}_n^\top]^\top$, $\m{p}^* = [(\m{p}_1^*)^\top, \ldots, (\m{p}_n^*)^\top]^\top$, và 
đặt $\m{e}_p \triangleq \m{p} - \m{p}^*$ thì ta viết chung được \eqref{eq:FM_positionbased} dưới dạng
\begin{equation} \label{eq:FM_positionbased1}
    \dot{\m{e}}_p = - k_p \m{e}_p.
\end{equation}
Rõ ràng, từ phương trình.~\eqref{eq:FM_positionbased1}, ta có ${\m{e}}_p(t) \to \m{0},~t\to +\infty$ theo hàm mũ. Điều này tương đương với việc, $\m{p}(t) \to \m{p}^*,~t\to +\infty$ theo hàm mũ.

Để cải thiện chất lượng điều khiển, ta có thể giả thiết rằng các tác tử có thể trao đổi thông tin về vị trí với nhau qua đồ thị $G$ và thêm vào luật điều khiển đội hình \eqref{eq:FM_positionbased} thành phần $\sum_{j \in N_i} a_{ij}\big((\m{p}_j - \m{p}_j^*) - (\m{p}_i - \m{p}_i^*)\big)$ để được luật điều khiển mới cho bởi:
\begin{align} 
    \m{u}_i = -k_p (\m{p}_i - \m{p}_i^*) + \sum_{j \in N_i} a_{ij}\big((\m{p}_j - \m{p}_j^*) - (\m{p}_i - \m{p}_i^*)\big),~ i=1, \ldots, n, \label{eq:FM_positionbased2}
\end{align}
trong đó $a_{ij}>0$ là trọng số ứng với $(v_j,v_i) \in E(G)$. Ta có thể viết lại phương trình \eqref{eq:FM_positionbased2} dưới dạng 
\begin{align}
    \dot{\m{p}} &= -k_p (\m{p} - \m{p}^*) - (\mcl{L} \otimes \m{I}_d) (\m{p} - \m{p}^*),
\end{align}
trong đó $\mcl{L}$ là ma trận Laplace của $G$. Lại đặt $\bmm{\delta} \triangleq \m{p} - \m{p}^*$, thì phương trình vi phân đối với sai lêch $\bmm{\delta}$ được cho dưới dạng:
\begin{align}
    \dot{\bmm{\delta}} &= - k_p \bmm{\delta} - (\mcl{L} \otimes \m{I}_d) \bmm{\delta} \nonumber\\
    &=-\big((k_p\m{I}_n +\mcl{L})\otimes \m{I}_d\big)\bmm{\delta}.
\end{align}
Giả sử rằng $G$ là một đồ thị có gốc-ra thì theo Định lý \ref{thm:c2_Laplace_directed}, các giá trị riêng của ma trận $k_p \m{I}_n + \mcl{L}$ đều lớn hơn hoặc bằng $k_p$, điều này chứng tỏ thành phần điều khiển dựa trên trao đổi thông tin có thể làm tăng tốc độ đạt được đội hình.

\begin{example} \label{eg5.2}
Mô phỏng điều khiển đội hình gồm 10 tác tử dựa trên vị trí tuyệt đối trong 2D và 3D được cho ở Hình~\ref{fig:VD5.1}. Với luật điều khiển \eqref{eq:FM_positionbased}, các tác tử di chuyển trực tiếp tới vị trí đặt trong không gian (các đỉnh của một đa giác đều gồm 10 đỉnh). Trong Phụ lục \ref{append:MATLAB} cung cấp một số code mẫu mô phỏng các thuật toán điều khiển đội hình.
\end{example}

\begin{figure}[th!]
    \centering
    \subfloat[Đội hình ban đầu trong 2D]{\includegraphics[width=0.48\linewidth]{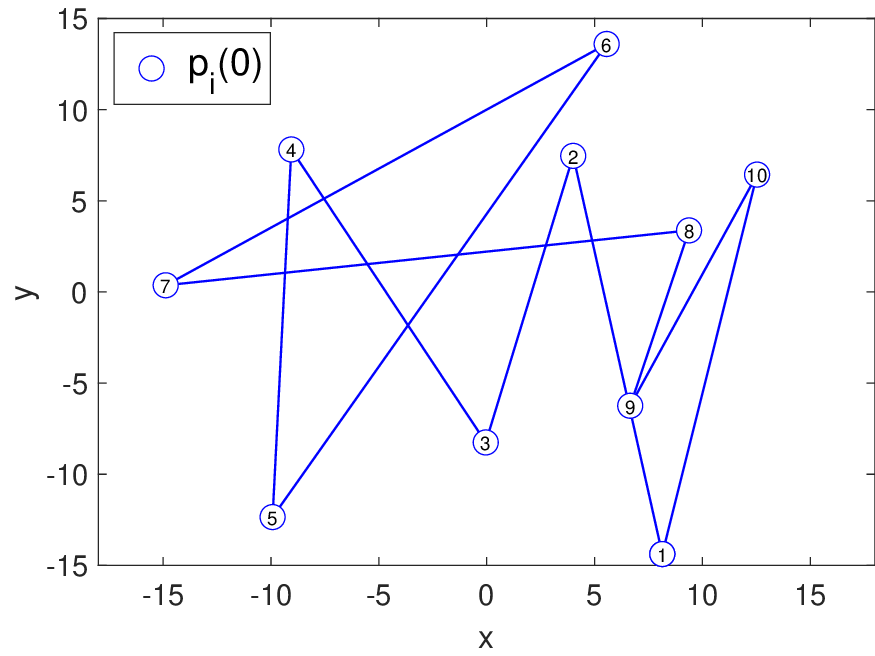}} \hfill
    \subfloat[Quá trình tạo đội hình trong 2D]{\includegraphics[width=0.48\linewidth]{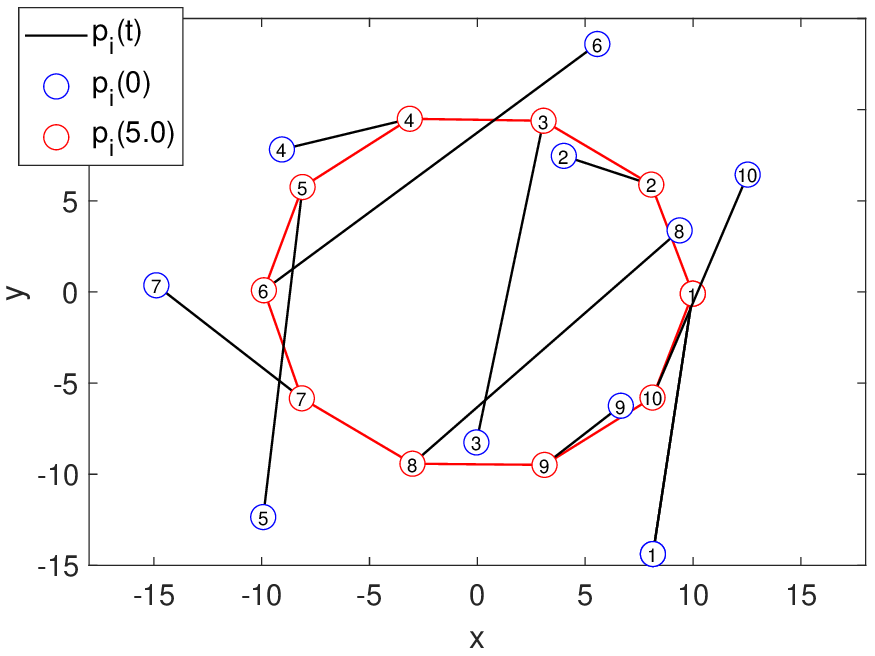}}\\
    \subfloat[Đội hình ban đầu trong 3D]{\includegraphics[width=0.45\linewidth]{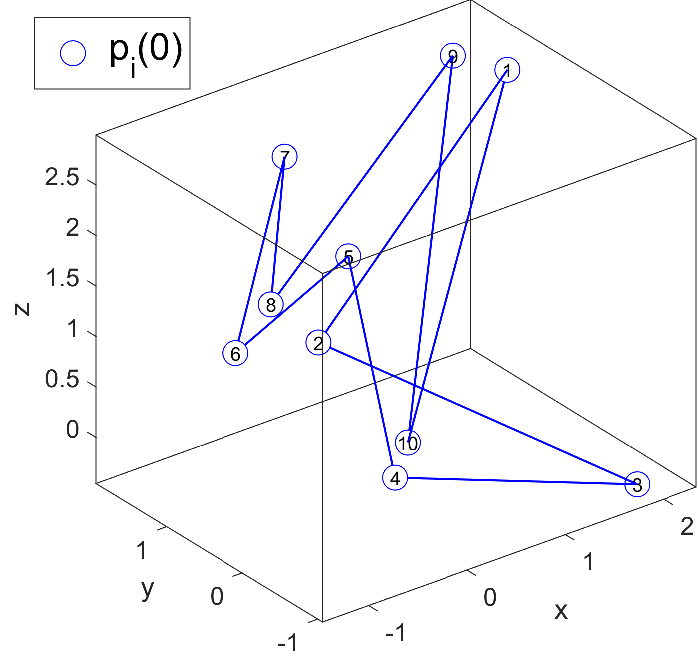}} \hfill
    \subfloat[Quá trình tạo đội hình trong 3D]{\includegraphics[width=0.45\linewidth]{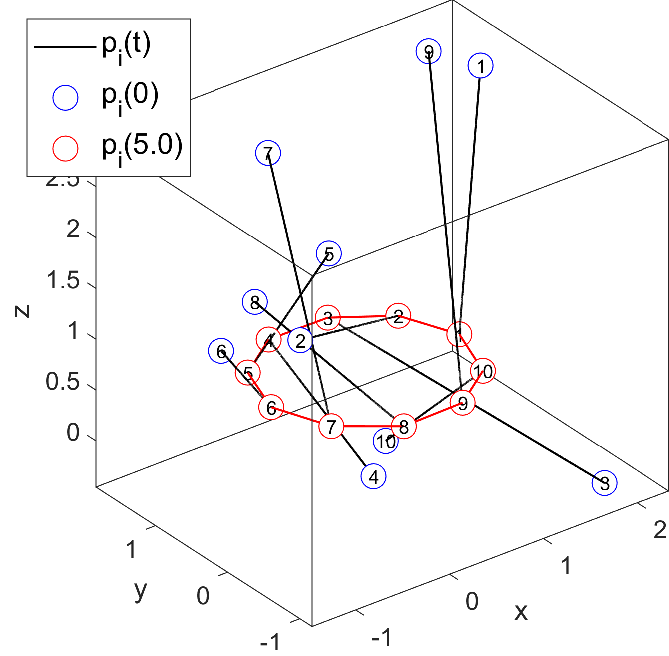}}
    \caption{Mô phỏng thuật toán điều khiển đội hình dựa trên vị trí tuyệt đối.}
    \label{fig:VD5.1}
\end{figure}
\section{Điều khiển đội hình dựa trên vị trí tương đối}
\index{điều khiển đội hình!dựa trên vị trí tương đối}
Trong phương án điều khiển dựa trên vị trí tương đối, hệ các tác tử cần thỏa mãn các giả thiết sau:
\begin{itemize}
    \item {Biến đo}: Các tác tử có các hệ qui chiếu cục bộ được định hướng như nhau và giống như hệ qui chiếu toàn cục. Tuy nhiên, các tác tử không cần biết gốc tọa độ của hệ qui chiếu toàn cục  này. Trong hệ tham chiếu địa phương, các tác tử có thể đo vị trí tương đối (vector sai lệch vị trí) và vận tốc tương đối (đối với tác tử bậc hai) của một số tác tử láng giềng. Chú ý rằng do các hệ qui chiếu địa phương là cùng hướng với hệ qui chiếu toàn cục, vector vị trí tương đối đo trong các hệ qui chiếu đều như nhau. 
    \item {Đồ thị tương tác}: Tương tác về đo đạc giữa các tác tử được cho bởi đồ thị vô hướng $G$, với giả thiết $G=(V,E)$ là một đồ thị liên thông. Đội hình đặt được cho bởi một tập các vector sai lệch vị trí mong muốn giữa các tác tử tương ứng với $G$.
\end{itemize}

\subsection{Trường hợp tác tử là khâu tích phân bậc nhất}
Xét hệ gồm $n$ tác tử trong đó mỗi tác tử trong đội hình có mô hình là khâu tích phân bậc nhất:
\begin{align}
\dot{\m{p}}_i = \m{u}_i, ~ i = 1, \ldots, n.
\end{align}
Ở đây, $\m{p}_i \in \mb{R}^d$ và $\m{u}_i \in \mb{R}^d$ lần lượt là vị trí và tín hiệu điều khiển của tác tử $i$ viết trên hệ qui chiếu toàn cục  $^g\Sigma$.
    
Từ giả thiết về đo đạc ở trên, mỗi tác tử $i$ có thể đo được $\m{z}_{ij} = \m{p}_j - \m{p}_i, \forall j \in N_i$. Đội hình đặt được cho thông qua tập ${\Gamma} = \{\m{z}_{ij}^*\}_{(i,j) \in E}$, và mục tiêu của mỗi tác tử là tạo đội hình thỏa mãn tất cả các vector sai lêch vị trí mong muốn, hay
\[\m{z}_{ij} = \m{p}_j - \m{p}_i = \m{z}_{ij}^*,~(i, j) \in E.\]
Tập ${\Gamma} = \{\m{z}_{ij}^*\}_{(i,j) \in E}$ được gọi là khả thi nếu như tập
\begin{align} \label{eq:FC_displacement1}
    \mc{E}_{\m{p}^*} \triangleq \{\m{p} \in \mb{R}^{dn}|~\m{p}_j - \m{p}_i =  \m{z}_{ij}^*, \forall (i, j) \in E\},
\end{align}
là khác tập rỗng. Ngược lại, nếu $\mc{E}_{\m{p}^*}  = \emptyset$, ta gọi ${\Gamma}$ là không khả thi. 

Trong mục này, ta giả sử ${\Gamma}$ là khả thi và xét $\m{p}^* = [(\m{p}_1^*)^\top, \ldots, (\m{p}_n^*)^\top]^\top$, là một phần tử trong $\mc{E}_{\m{p}^*}$.  Hơn nữa, ta giả sử rằng  mỗi tác tử chỉ biết được một số vector sai lêch đặt $\m{z}_{ij}^* = \m{p}_j^* - \m{p}_i^*, \forall j \in N_i$ chứ không biết được $\m{p}_i^*$ và $\m{p}_j^*$. Bài toán đặt ra là thiết kế luật điều khiển cho mỗi tác tử để $\m{p}$ tiệm cận tới một đội hình sai khác với $\m{p}^*$ bởi một phép tịnh tiến. Nói cách khác, $\m{p}(t)$ cần hội tụ tới một điểm trong tập $\mc{E}_{\m{p}^*}$ khi $t \to +\infty$. 

Với bài toán này, luật điều khiển đội hình được thiết kế như sau:
\begin{align} 
    \m{u}_i &= k_p \sum_{j \in N_i} a_{ij} (\m{z}_{ij} - \m{z}_{ij}^*) \nonumber \\
    &= k_p \sum_{j \in N_i} a_{ij} (\m{p}_{j} - \m{p}_i - (\m{p}_{j}^*-\m{p}_i^*)), \label{eq:FC_displacement2}
\end{align}
trong đó $k_p>0$. Với $\bm{\delta} = \m{p}^* - \m{p}$, ta có phương trình
\begin{align} \label{eq:FC_displacement3}
    \dot{\bmm{\delta}} = -k_p (\mcl{L} \otimes \m{I}_d) \bmm{\delta}.
\end{align}
Theo lý thuyết về hệ đồng thuận ở Chương \ref{chap:consensus}, với $G$ là một đồ thị liên thông thì $\bmm{\delta}(t) \to \bmm{\delta}^* = \m{1}_n \otimes \bar{\bmm{\delta}}$, trong đó $\bar{\bmm{\delta}} = \frac{1}{n} \sum_{i=1}^n (\m{p}^*_i(0) - \m{p}_i(0))$ là một vector hằng. Do đó, 
\begin{align}
    \m{p}^* - \m{p}(t) \to \bmm{\delta}^*, ~t \to +\infty, \text{ hay } \m{p}(t) \to \m{p}^* - \bmm{\delta}^*,~t \to +\infty,
\end{align}
tức là $\m{p}(t)$ hội tụ tới một cấu hình tĩnh thuộc tập $\mc{E}_{\m{p}^*}$.

Phân tích hệ trong trường hợp tập $\Gamma$ không khả thi sẽ được xét trong bài tập \ref{exercise:FC1}.

\begin{example} \label{VD:5.2}
Trong ví dụ này, chúng ta mô phỏng luật điều khiển đội hình dựa trên vị trí tương đối cho hệ gồm 10 tác tử với đồ thị tương tác là chu trình $C_{10}$ và đội hình đặt là các đỉnh của một đa giác đều 10 cạnh. Kết quả mô phỏng được cho trong Hình~\ref{fig:VD_5.2}. Các tác tử dần tạo thành đội hình đặt khi $t \to +\infty$. Mô phỏng được thực hiện trong hai trường hợp: đội hình định nghĩa trong không gian 2D và 3D.
\end{example}
\begin{figure}[th]
    \centering
    \subfloat[]{\includegraphics[height=6.5cm]{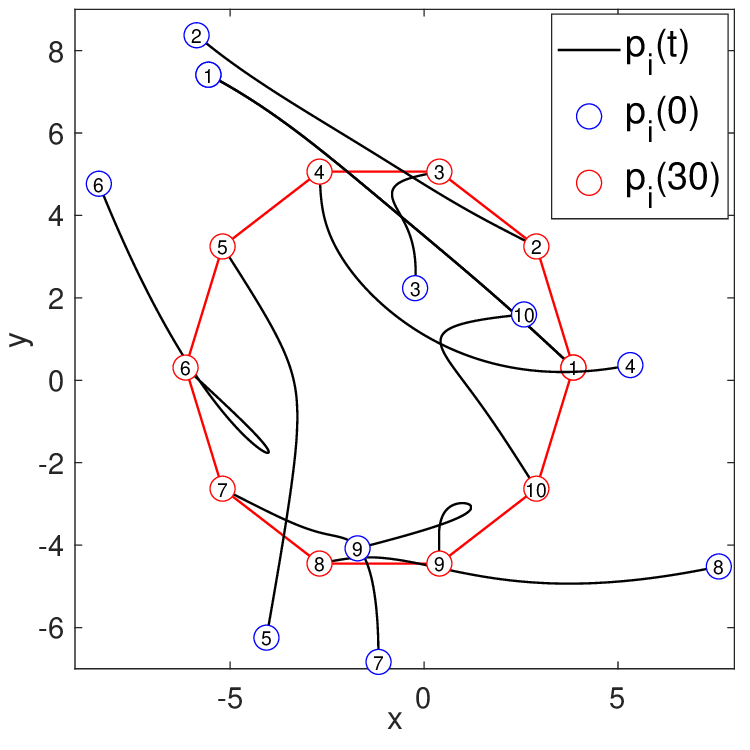}}\hfill
    \subfloat[]{\includegraphics[height=6.5cm]{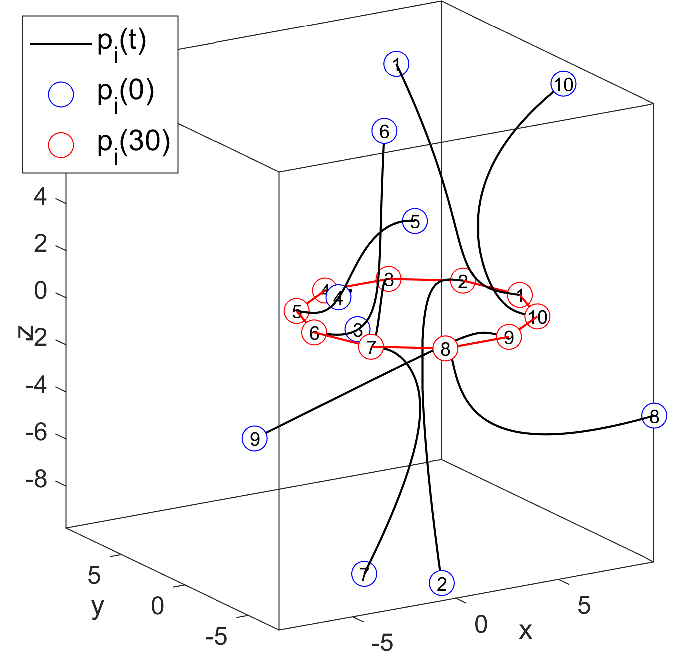}}
    \caption{Mô phỏng thuật toán điều khiển đội hình dựa trên vị trí tương đối trong 2D và 3D.}
    \label{fig:VD_5.2}
\end{figure}
%
%
%

\subsection{Trường hợp tác tử là khâu tích phân bậc hai}
Xét hệ gồm $n$ tác tử được mô hình là khâu tích phân bậc hai:
\begin{align}
\dot{\m{p}}_i &= \m{v}_i, \nonumber\\
\dot{\m{v}}_i &= \m{u}_i,~ i = 1, \ldots, n.
\end{align}
Ở đây, $\m{p}_i, \m{v}_i$, và $\m{u}_i \in \mb{R}^d$ lần lượt là vị trí, vận tốc, và tín hiệu điều khiển của tác tử $i$ viết trên $^g\Sigma$. 

Luật điều khiển viết cho từng tác tử:
\begin{align}
    \m{u}_i = -k_1 \sum_{j\in {N}_i} (\m{p}_i - \m{p}_j - (\m{p}_i^* - \m{p}_j^*)) - k_2 \m{v}_i,~i=1,\ldots, n,
\end{align}
trong đó $k_1$ và $k_2$ là các hệ số thực dương. 

Thực hiện phép đổi biến $\bmm{\delta} = \m{p}_i - \m{p}_i^*$, ta thu được phương trình
\begin{align} \label{eq:chap_formation_consensus_law}
\dot{\bmm{\delta}}_i &= \m{v}_i, \nonumber\\
\dot{\m{v}}_i &= -k_1 \sum_{j\in {N}_i} (\bmm{\delta}_i - \bmm{\delta}_j) - k_2 \m{v}_i,~i=1,\ldots, n.
\end{align}
Đối chiếu lại với phần thiết kế hệ đồng thuận với tác tử tích phân bậc hai ở mục trước, hệ \eqref{eq:chap_formation_consensus_law} tiến tới đồng thuận $\bmm{\delta}_i \to \bm{\delta}^*, \m{v}_i \to \m{0}_d$. Điều này tương đương với việc $\m{p}_i(t) \to \m{p}_i^* + \bmm{\delta}^*, t\to +\infty$.

\section{Điều khiển đội hình dựa trên khoảng cách}
\index{điều khiển đội hình!dựa trên khoảng cách}
Trong phương án điều khiển đội hình dựa trên khoảng cách (hay điều khiển đội hình dựa trên vị trí tương đối trong hệ tọa độ riêng), hệ các tác tử cần thỏa mãn các giả thiết sau:
\begin{itemize}
    \item Mô hình tác tử: Các tác tử được mô hình hóa bởi khâu tích phân bậc nhất:
    \begin{align}
    \dot{\m{p}}_i^i = \m{u}_i^i, ~ i = 1, \ldots, n,
    \end{align}
    trong đó $\m{p}_i^i \in \mb{R}^2$ và $\m{u}_i^i \in \mb{R}^2$ lần lượt là vị trí và tín hiệu điều khiển của tác tử $i$ viết trong hệ qui chiếu địa phương $^i\Sigma$.
    \item Biến đo: Các tác tử có các hệ tham chiếu riêng được định hướng khác nhau và không có thông tin về hệ tham chiếu toàn cục. Trong hệ tham chiếu riêng, các tác tử có thể đo vị trí tương đối (vector sai lệch vị trí) $\m{p}_{ij}^i = \m{p}_j^i - \m{p}_i^i$ của các tác tử láng giềng $j \in {N}_i$.
    \item Đồ thị tương tác: Tương tác về đo đạc giữa các tác tử được cho bởi đồ thị vô hướng $G$, với giả thiết $G$ là một đồ thị cứng (rigid). 
    \item Đội hình đặt: được mô tả bởi một tập các khoảng cách mong muốn $\Gamma = \{d_{ij}^* = \|\m{p}_j^* - \m{p}_i^*\|\}_{(i,j) \in E}$.
\end{itemize}

Mặc dù các tác tử trong hệ đo được vị trí tương đối của các tác tử khác trong $\mb{R}^2$, do các vector vị trí tương đối này không được biểu diễn trên cùng một hệ tọa độ chung, việc điều khiển đội hình dựa trên sai lệch vị trí là không khả thi. Thay vào đó, biến điều khiển lúc này được chọn là khoảng cách giữa các tác tử. Dễ thấy, khoảng cách giữa các tác tử có thể đo được dễ dàng từ vector sai lệch cục bộ và hoàn toàn không phụ thuộc vào các hệ qui chiếu. Do ta chỉ điều khiển tập các khoảng cách tương đối, một câu hỏi đặt ra là cần chọn bao nhiêu biến khoảng cách và cần chọn các biến này như thế nào để khi các biến này được thỏa mãn thì đội hình thu được chỉ sai khác đội hình đặt bởi một phép tịnh tiến và một phép quay. Lời giải cho câu hỏi này có ở lý thuyết cứng khoảng cách, một nhánh nghiên cứu của toán học thường được ứng dụng trong nghiên cứu về cơ học hay cấu trúc phân tử hóa học \cite{Whiteley1996,Asimow1978rigidity,Graver1993combinatorial,Connelly2005generic}. Trong tài liệu này, chúng ta sẽ giới hạn phân tích cho đội hình hai chiều.

\subsection{Lý thuyết cứng khoảng cách}
Xét đồ thị vô hướng $G = (V,E)$, trong đó $V = \{1, \ldots, n\}$ gồm $n$ đỉnh và $E \subset V \times V$ gồm $m$ cạnh. Ứng với mỗi đỉnh $i \in V$ ta có tương ứng một điểm $\m{p}_i \in \mb{R}^2$. Một \index{đội hình}đội hình (hay một mạng)\footnote{formation hoặc network} được định nghĩa bởi $(G,\m{p})$, trong đó vector $\m{p} = [\m{p}_1^\top, \ldots, \m{p}_n^\top]^\top \in \mb{R}^{2n}$ gọi là một cấu hình\footnote{configuration,  realization, hoặc embedding} của $G$ trên $\mb{R}^2$. Lưu ý rằng các điểm $\m{p}_i$ đều được cho trên một hệ qui chiếu toàn cục.

Xét một tập khoảng cách $\Gamma = \{d_{ij}>0|~{(i,j) \in {E}}\}$. Tập $\Gamma$ gọi là \emph{khả thi} \index{tập khoảng cách khả thi} nếu tồn tại ít nhất một cấu hình $\m{p}$ sao cho $\|\m{p}_j - \m{p}_i\| = d_{ij}, \forall (i,j) \in E$. Nếu không tồn tại cấu hình nào thỏa mãn tất cả các khoảng cách trong $\Gamma$ thì ta nói $\Gamma$ là không khả thi. Cấu hình $\m{p}$ thỏa mãn mọi ràng buộc khoảng cách trong $\Gamma$ được gọi là một hiện thực hóa của $\Gamma$ trên $\mb{R}^2$. Ngược lại, mỗi cấu hình $\m{p}^*$ cho ta một tập khoảng cách dẫn xuất $\Gamma = \{d_{ij}^* = \|\m{p}_j^* - \m{p}_i^*\|\}_{(i,j) \in E}$.

Xét hai cấu hình $\m{p}$ và $\m{q}$ của cùng một đồ thị $G$ trên $\mb{R}^2$. Hai cấu hình $\m{p}$ và $\m{q}$ là \index{tương đương!về khoảng cách} \emph{tương đương về khoảng cách}\footnote{distance equivalency} nếu như $\|\m{p}_i - \m{p}_j\| = \|\m{q}_i - \m{q}_j\|, \forall (i,j) \in E$, và là \index{tương đồng!về khoảng cách} \emph{tương đồng về khoảng cách}\footnote{distance congruency} nếu như  $\|\m{p}_i - \m{p}_j\| = \|\m{q}_i - \m{q}_j\|, \forall i,j \in V,~i \neq j$.

Một đội hình $(G,\m{p})$ là \emph{cứng khoảng cách}\index{cứng!khoảng cách} \footnote{rigid} nếu như tồn tại $\epsilon>0$ sao cho mọi cấu hình $\m{q}$ của $G$ thỏa mãn (i) $\in B_\epsilon \triangleq \{\m{q} \in \mb{R}^{2n}|~ \|\m{q} - \m{p}\| < \epsilon\}$ và (ii) $\m{q}$ tương đương về khoảng cách với $\m{p}$ thì $\m{q}$ cũng tương đồng về khoảng cách với $\m{p}$. Nếu không tồn tại $\epsilon>0$ như vậy thì $(G,\m{p})$ là đội hình không cứng.

Đội hình $(G,\m{p})$ là \index{cứng!khoảng cách!toàn cục} \emph{cứng khoảng cách toàn cục}\footnote{global rigidity} nếu như mọi cấu hình $\m{q}$ tương đương về khoảng cách với $\m{p}$ thì cũng tương đồng về khoảng cách với $\m{p}$ (Xem ví dụ minh họa ở Hình~\ref{fig:c5_rigid}(c)).

\begin{figure}
\centering
\subfloat[Đội hình không cứng]{
\begin{tikzpicture}[roundnode/.style={circle, draw=black, thick, minimum size=3mm,inner sep= 0.25mm}]
\node[roundnode] (v1) at (0,0) {};
\node[roundnode] (v2) at (1,0) {};
\node[roundnode] (v3) at (0,2) {};
\node[roundnode] (v4) at (1,2) {};
\node at (-1.5,0) {};
\node at (2.5,0) {};

\node[roundnode] (v5) at (1.4142,1.4142) {};
\node[roundnode] (v6) at (2.4142,1.4142) {};

\draw[-, very thick] (v1)--(v2)--(v4)--(v3)--(v1);
\draw[dashed, very thick] (v1)--(v5)--(v6)--(v2);
\draw [dashed,thick,domain=45:90] plot ({2*cos(\x)}, {2*sin(\x)});
\draw [dashed,thick,domain=45:90] plot ({1+2*cos(\x)}, {2*sin(\x)});

\draw [color = green!60!black, ultra thick,-{Stealth[length=2mm]}]
    (v3) edge [bend left=0] (0.6,2)
    (v4) edge [bend left=0] (1.6,2)
    ;
\end{tikzpicture}
}
\hfill
\subfloat[Đội hình cứng]{
\begin{tikzpicture}[roundnode/.style={circle, draw = black, thick, minimum size = 3mm,inner sep = 0.25mm}]
\node[roundnode]   (v1)   at   (0,0) {};
\node[roundnode]   (v2)   at   (1,0) {};
\node[roundnode]   (v3)   at   (0,2) {};
\node[roundnode]   (v4)   at   (1,2) {};
\node[roundnode]   (v5)   at   (-0.6,1.2) {};
\node at (-1.5,0) {};
\node at (2.5,0) {};

\draw[-, very thick] (v1)--(v2)--(v4)--(v3)--(v1);
\draw[-, very thick] (v2)--(v3);
\draw[dashed, very thick] (v2)--(v5)--(v3);
\end{tikzpicture}
}
\hfill
\subfloat[Đội hình cứng toàn cục]{
\begin{tikzpicture}[roundnode/.style={circle, draw = black, thick, minimum size = 3mm,inner sep = 0.25mm}]
\node[roundnode]   (v1)   at   (0,0) {};
\node[roundnode]   (v2)   at   (1,0) {};
\node[roundnode]   (v3)   at   (0,2) {};
\node[roundnode]   (v4)   at   (1,2) {};

\node at (-1.5,0) {};
\node at (2.5,1.4142) {};

\draw[-, very thick] (v1)--(v2)--(v4)--(v3)--(v1)--(v4);
\draw[-, very thick] (v2)--(v3);
\end{tikzpicture}
}
\caption{Một số ví dụ minh họa lý thuyết cứng khoảng cách.}
\label{fig:c5_rigid}
\end{figure}
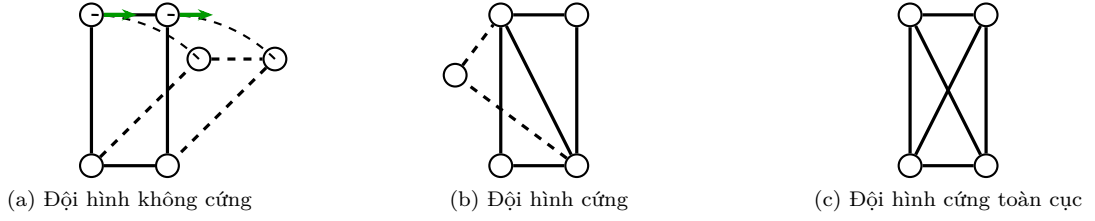

Việc xét trực tiếp liệu một đội hình $(G,\m{p})$ có là cứng khoảng cách hay không dựa trên định nghĩa là không đơn giản khi số đỉnh $n$ lớn. Để đơn giản hóa việc xác định tính cứng, ta có khái niệm \index{cứng!khoảng cách!vi phân} cứng khoảng cách vi phân\footnote{infinitesimal rigidity} - xấp xỉ bậc nhất về tính cứng của $(G,\m{p})$. Đánh số các cạnh của đồ thị từ $1$ tới $m$, với kí hiệu $e_k \equiv e_{k_{ij}} \equiv (i,j)$, ta có các vector sai lệch vị trí tương ứng $\m{z}_k \equiv \m{z}_{ij} = \m{p}_j - \m{p}_i$, $k=1, \ldots, m$. Ta định nghĩa hàm (bình phương các) khoảng cách:
\begin{align} \label{eq:c5_distance_function}
\m{f}_{{G}}(\m{p}) \triangleq \left[ \|\m{z}_1\|^2, \ldots, \|\m{z}_m\|^2 \right]^\top = [\ldots, \|\m{z}_{ij}\|^2,\ldots]^\top \in \mb{R}^m.
\end{align}
Ma trận cứng khoảng cách (rigidity matrix) của đội hình $(G,\m{p})$ được định nghĩa là:
\begin{align} \label{eq:c5_rigidity_matrix}
\m{R}(\m{p}) = \frac{1}{2} \frac{\partial \m{f}_G(\m{p})}{\partial \m{p}}.
\end{align}
Kí hiệu $\m{z} = [\m{z}_1^\top, \ldots, \m{z}_m^\top]^\top = (\m{H} \otimes \m{I}_d) \m{p} = \bar{\m{H}}\m{p}$ và $\m{D}(\m{z}) = \text{blkdiag}(\m{z}_1, \ldots, \m{z}_m)$, ta có thể viết lại $\m{f}_{{G}}(\m{p}) = \m{D}(\m{z})^\top\m{D}(\m{z})\m{1}_m = \m{D}(\m{z})^\top\m{z}$. Từ đó, ta viết được:
\begin{align} \label{eq:c5_rigidity_matrix1}
    \m{R}(\m{p}) = \m{D}^\top(\m{z})\bar{\m{H}}
\end{align}

Một cách trực quan, với cấu hình $\m{p}$ và $\bm{\delta}\m{p} = [\bm{\delta}\m{p}_1^\top,\ldots,\bm{\delta}\m{p}_n^\top]^\top \in \mb{R}^{2n}$ là một chuyển động vi phân (vô cùng nhỏ) đưa $\m{p}$ tới cấu hình $\m{q} = \m{p} + \bm{\delta}\m{q}$ thì chúng ta có xấp xỉ bậc nhất
\begin{align}
\m{f}_{{G}}(\m{q}) \approx \m{f}_{{G}}(\m{p}) + \frac{\partial \m{f}_G(\m{p})}{\partial \m{p}} \bm{\delta}\m{p} = \m{f}_{{G}}(\m{p}) + 2\m{R}(\m{p}) \bm{\delta}\m{p}.
\end{align}
Với những chuyển động vi phân $\bm{\delta}\m{p} \in {\rm ker}(\m{R}(\m{p}))$ thì $\m{f}_{{G}}(\m{q}) \approx \m{f}_{{G}}(\m{p})$, hàm khoảng cách được bảo toàn, hay hai cấu hình này tương đương về khoảng cách với nhau. 

Chú ý rằng ta luôn có $\m{R}(\m{p})(\m{1}_n \otimes \m{I}_d) = \m{Z}^\top(\m{H} \otimes \m{I}_2) (\m{1}_n \otimes \m{I}_d) = \m{Z}^\top(\m{H}\m{1}_n \otimes \m{I}_2) = \m{0}_{m\times 2}$ và $\m{R}(\m{p})(\m{I}_n \otimes \m{J})\m{p} = \m{D}^\top(\m{z})\bar{\m{H}}(\m{I}_n \otimes \m{J})\m{p} = [\ldots,\m{z}_{ij}^\top\m{J}\m{z}_{ij},\ldots]^\top = \m{0}_m$, với $\m{J} = \begin{bmatrix}
0 & -1\\ 0 & 1
\end{bmatrix}$ nên ${\rm im}([\m{1}_n \otimes \m{I}_d, (\m{I}_n \otimes \m{J})\m{p}]) \subseteq {\rm ker}(\m{R}(\m{p}))$ \cite{Trinh2016CDC}. Như vậy, ${\rm rank}(\m{R}(\m{p})) \leq 2n - 3$, ma trận cứng luôn có ít nhất $3$ vector riêng độc lập tuyến tính tương ứng với giá trị riêng 0 (chuyển động tịnh tiến và chuyển động quay toàn bộ đội hình quanh một điểm trên mặt phẳng). Những chuyển động vi phân tương ứng với phép tịnh tiến và phép quay toàn bộ đội hình theo các trục tọa độ của $^g\Sigma$ gọi là những chuyển động vi phân khoảng cách tầm thường (trivially infinitesimal distance-preserving motions).\index{chuyển động vi phân khoảng cách tầm thường}. Khi ${\rm rank}(\m{R}(\m{p})) \leq 2n - 3$, đội hình là cứng khoảng cách vi phân. Dễ thấy đội hình cứng khoảng cách vi phân chỉ có những chuyển động vi phân khoảng cách tầm thường, và khi gán vào các đỉnh của đội hình cứng khoảng cách vi phân một vận tốc trong ${\rm ker}(\m{R}(\m{q}(t)))$, ta thu được quĩ đạo $\m{q}(t),~\m{q}(0)=\m{p}$ với $\m{q}(t)$ tương đồng về khoảng cách với $\m{p}$ tại mọi thời điểm $t\geq 0$. Với đội hình không cứng khoảng cách vi phân, ${\rm ker}(\m{R}(\m{q}(t)))$ còn chứa các chuyển động bảo toàn khoảng cách không tầm thường (nontrivial). Khi gán vận tốc trong ${\rm ker}(\m{R}(\m{q}(t)))$ vào các đỉnh của đội hình, quĩ đạo $\m{q}(t),~\m{q}(0)=\m{p}$ thu được có thể chứa các cấu hình chỉ tương đương nhưng không tương đồng về khoảng cách với $\m{p}$ (Xem ví dụ ở Hình \ref{fig:c5_rigid}(a)).

\begin{example}
Ma trận cứng của đồ thị trên Hình \ref{fig:c5_rigid}(b) được cho như sau:
\[{\m{R}}\left( \m{p} \right) = \left[ {\begin{array}{*{20}{c}}
{{{\left( {{\m{p}_1} - {\m{p}_2}} \right)}^\top}}&{{{\left( {{\m{p}_2} - {\m{p}_1}} \right)}^\top}}&{{\m{0}^\top}}&{{\m{0}^\top}}\\
{{{\left( {{\m{p}_1} - {\m{p}_3}} \right)}^\top}}&{{\m{0}^\top}}&{{{\left( {{\m{p}_3} - {\m{p}_1}} \right)}^\top}}&{{\m{0}^\top}}\\
{{{\left( {{\m{p}_1} - {\m{p}_4}} \right)}^\top}}&{{\m{0}^\top}}&{{\m{0}^\top}}&{{{\left( {{\m{p}_4} - {\m{p}_1}} \right)}^\top}}\\
{{\m{0}^\top}}&{{{\left( {{\m{p}_2} - {\m{p}_3}} \right)}^\top}}&{{{\left( {{\m{p}_3} - {\m{p}_2}} \right)}^\top}}&{{\m{0}^\top}}\\
{{\m{0}^\top}}&{{\m{0}^\top}}&{{{\left( {{\m{p}_3} - {\m{p}_4}} \right)}^\top}}&{{{\left( {{\m{p}_4} - {\m{p}_3}} \right)}^\top}}
\end{array}} \right].\]
Có thể nhận xét rằng, \index{ma trận!cứng khoảng cách}ma trận cứng có cấu trúc khá tương tự như ma trận liên thuộc. Tuy nhiên, ở hàng thứ $k$ tương ứng với cạnh $(i,j)$ thì $(\m{p}_i - \m{p}_j)^\top$ được thay cho $-1$,  $(\m{p}_i - \m{p}_j)^\top$ được thay cho $1$, và $\m{0}^\top$ được thay cho $0$.
\end{example}

Đồ thị $G$ là \index{cứng!khoảng cách!phổ quát} cứng khoảng cách phổ quát (generically rigid) trong $\mb{R}^2$ nếu như với \emph{hầu hết} mọi cấu hình $\m{p}$ của $G$ trong $\mb{R}^{2n}$ thì $(G,\m{p})$ đều là cứng khoảng cách vi phân, ngoại trừ một tập con của $\mb{R}^{2n}$ với số chiều nhỏ hơn $2n$.

Dựa vào ma trận cứng khoảng cách, nếu tồn tại một cấu hình $\m{p}$ của $G$ sao cho ${\rm rank}(\m{R}(\m{p})) = 2n-3$ thì $G$ là đồ thị cứng khoảng cách phổ quát. Cấu hình $\m{p}$ của $G$ thỏa mãn ${\rm rank}(\m{R}(\m{p})) = 2n - 3$ gọi là một cấu hình với các tọa độ ở vị trí tổng quát, hay gọi tắt là một \emph{cấu hình tổng quát} (generic configuration). Ngược lại, cấu hình $\m{p}$ của $G$ mà ở đó rank$(\m{R}(\m{p})) < 2n-3$ gọi là một \emph{cấu hình suy biến} (degenerated configuration).\index{cấu hình suy biến}

Nếu đồ thị $G$ là cứng và có số cạnh đúng bằng $2n - 3$ thì $G$ được gọi là cứng khoảng cách tối thiểu. Dễ thấy, khi $d=2$, đồ thị cứng tối thiểu có $2n-3$ cạnh. Hơn nữa, trong trường hợp $d=2$, mọi đồ thị cứng tối thiểu đều là đồ thị Laman. 
\begin{Definition}[Đồ thị Laman] \label{chap8:def_Laman} Một đồ thị $G=(V,E)$ là Laman khi và chỉ khi:
\begin{itemize}
    \item $G$ có $2n-3$ cạnh,
    \item Với mỗi tập con gồm $k$ đỉnh của $V$ thì đồ thị con dẫn xuất của $G$ từ $k$ đỉnh đó có không quá $2k-3$ cạnh.
\end{itemize}
\end{Definition}
Trên không gian hai chiều, mọi đồ thị Laman \index{đồ thị!Laman} đều là đồ thị cứng tổng quát. Chứng minh cụ thể của phát biểu này có thể tham khảo từ bài báo của Laman \cite{Laman1970graphs} và sẽ không được trình bày trong tài liệu này. \index{mở rộng Henneberg} Henneberg đề xuất phương pháp xây dựng các đồ thị Laman trong hai chiều như sau: bắt đầu từ một đồ thị gồm hai đỉnh (1 và 2) và một cạnh $(1,2)$. Tại mỗi bước, chúng ta mở rộng đồ thị nhờ một trong hai toán tử sau đây:
\begin{enumerate}
    \item Cộng đỉnh: Thêm một đỉnh $k$ và hai cạnh nối $k$ với  hai đỉnh $i, j$ hiện có trong đồ thị.
    \item Chia cạnh: Xóa một cạnh $(i,j)$ hiện có trong đồ thị và thêm vào một đỉnh $k$ cùng ba cạnh $(k,i), (k,j), (k,l)$, trong đó $l$ là một đỉnh trong đồ thị hiện tại khác $i$, $j$.
\end{enumerate}
Đồ thị thu được từ xây dựng Henneberg là một đồ thị Laman. Ngược lại, mọi đồ thị Laman đều có thể xây dựng nhờ mở rộng Henneberg.

Định nghĩa \ref{chap8:def_Laman} cho ta một phương án để kiểm tra các đồ thị là cứng phổ quát với số cạnh tối thiểu. Tuy nhiên, cách kiểm tra này đòi hỏi phải xét mọi tập con của đồ thị, do đó việc kiểm tra sẽ yêu cầu khối lượng tính toán cao. Một số điều kiện tương đương khác để kiểm tra một đồ thị cứng tối thiểu với độ phức tạp thấp hơn là dựa trên định lý Crapo và định lý R\'{e}cski \cite{Crapo1990,Bereg2005certifying}. Ý nghĩa của mở rộng Henneberg là cung cấp một thuật toán đơn giản để sinh các đồ thị cứng khoảng cách phổ quát trong không gian hai chiều.

\subsection{Luật điều khiển đội hình dựa trên khoảng cách}
Với các khoảng cách được cho từ một đội hình đặt $\m{p}^* \in \mb{R}^{dn}$, ta định nghĩa các biến sai lệch khoảng cách $\sigma_{ij} = \|\m{p}_{i} - \m{p}_j\|^2 - (d_{ij}^*)^2  = (d_{ij})^2 - (d_{ij}^*)^2$, $\forall (i,j) \in {E}$ và \[\bmm{\sigma} = [\ldots, \sigma_{ij}, \ldots]^\top = [\sigma_1, \ldots, \sigma_m]^\top = \m{f}_G(\m{p}) - \m{f}_G(\m{p}^*) \in \mb{R}^m.\] 
Chọn hàm thế
\begin{align}
    V(\bm{\sigma}) = \frac{1}{4} \sum_{(i,j) \in {E}} \sigma_{ij}^2 = \frac{1}{4} \|\bmm{\sigma}\|^2 = \frac{1}{4}\|\m{f}_{G}(\m{p}) - \m{f}_{G}(\m{p}^*)\|^2,
\end{align}
là hàm tổng các sai lệch khoảng cách mong muốn giữa các tác tử trong hệ\footnote{Có rất nhiều hàm thế khác nhau có thể lựa chọn. Tương ứng với mỗi cách chọn hàm thế, ta thu được một luật điều khiển đội hình dựa trên khoảng cách tương ứng. Độc giả quan tâm có thể tham khảo thêm tại \cite{Oh2014distance,Sun2016exponential}.} thì luật điều khiển đội hình thiết kế theo phương pháp gradient được viết trên hệ tọa độ của tác tử $i$ như sau:
\begin{align} \label{eq:c5_distance_based_control_law}
    \dot{\m{p}}_i^i = - \nabla_{\m{p}_i^i}V = - \sum_{j \in {N}_i} \big( d_{ij}^2 - (d_{ij}^*)^2 \big) (\m{p}_j^i - \m{p}_i^i),~\forall i = 1, \ldots, n.
\end{align}
Dễ thấy rằng luật điều khiển \eqref{eq:c5_distance_based_control_law} là một luật phân tán, và luật này chỉ sử dụng các biến đo được bởi tác tử $i$ trong $^i\Sigma$. 

Chuyển phương trình \eqref{eq:c5_distance_based_control_law} sang hệ quy chiếu toàn cục $^g\Sigma$, ta có:
\begin{align}  \label{eq:c5_distance_based_control_law_global}
    \dot{\m{p}}_i = - \sum_{j \in {N}_i} \big(d_{ij}^2 - (d_{ij}^*)^2 \big) (\m{p}_j - \m{p}_i),~\forall i= 1, \ldots, n.
\end{align}
Ta viết lại luật điều khiển đội hình  \eqref{eq:c5_distance_based_control_law_global} dưới dạng ma trận như sau:
\begin{align}
\dot{\m{p}} &= - \nabla_{\m{p}}V \nonumber\\
    &= -\frac{1}{2} \Big( \frac{\partial \m{f}_G(\m{p})}{\partial \m{p}}\Big)^\top \big( \m{f}_G(\m{p}) - \m{f}_G(\m{p}^*) \big) \nonumber\\
    &= -\m{R}(\m{p})^\top \bmm{\sigma}. \label{eq:c5_distance_formation}
\end{align}
Dễ thấy, với luật điều khiển \eqref{eq:c5_distance_formation} thì $(\m{1}_n \otimes \m{I}_2)^\top \dot{\m{p}} = \m{0}$ nên trọng tâm của đội hình là bất biến theo thời gian. Kết hợp với quan hệ $\m{p}^\top(\m{I}_n \otimes \m{J})\dot{\m{p}} = \m{0}$, ta suy ra momen động lượng $\m{p}^\top(\m{I}_n \otimes \m{J})\m{p}$ của hệ cũng là không đổi theo thời gian \cite{Sun2018conservation}. 

Để phân tích sự hội tụ của hệ, ta định nghĩa các tập hợp:
\begin{align}
    \mc{Q} &= \{\m{p}\in\mb{R}^{dn}|~\dot{\m{p}} = \m{0}\}, \\
    \mc{D} &= \{\m{p}\in\mb{R}^{dn}|~\bmm{\sigma} = \m{0}\}, \\
    \mc{U} &= \mc{Q} \setminus \mc{D},
\end{align}
trong đó $\mc{Q}$ chứa các điểm cân bằng của \eqref{eq:c5_distance_formation}, $\mc{D}$ là tập các điểm cân bằng mong muốn (tập các đội hình thỏa mãn mọi khoảng cách đặt $d_{ij}^*$), và $\mc{U}$ là những điểm cân bằng không mong muốn. Sự tồn tại của tập $\mc{U} = \{\m{p} \in \mb{R}^{dn}|~\m{R}(\m{p})^\top \bmm{\sigma} = \m{0}_{dn}, \bmm{\sigma} \neq \m{0}_m\}$ gây khó khăn cho việc xét tính ổn định của các đội hình  tương ứng với một đồ thị cứng bất kỳ.

Ta xét trường hợp $(G,\m{p}^*)$ là cứng khoảng cách vi phân ($\text{rank}(\m{R}(\m{p}^*)) = 2n - 3 = m$).

Chuyển phương trình \eqref{eq:c5_distance_formation} sang dạng phương trình theo $\bmm{\sigma}$, với lưu ý rằng  $\m{D}(\m{z})^\top\bar{\m{H}} = \m{R}(\m{p})$ ta thu được:
\begin{align} \label{c5:distance_sigma_dynam}
\dot{\bmm{\sigma}} = - {2}\m{D}(\m{z})^\top \bar{\m{H}}\bar{\m{H}}^\top\m{D}(\m{z})\bmm{\sigma} = - 2 \m{R}(\m{p}) \m{R}(\m{p})^{\top} \bmm{\sigma}
\end{align}

Với hàm Lyapunov được cho bởi $V(\bmm{\sigma}) = \frac{1}{4}\bmm{\sigma}^\top\bmm{\sigma}$ thì dọc theo một quỹ đạo của \eqref{c5:distance_sigma_dynam}, ta có:
\begin{align} \label{eq:distance_based_boundedness}
    \dot{V} = - \bmm{\sigma}^\top \m{R}(\m{p}) \m{R}(\m{p})^\top \bmm{\sigma} = - \|\m{R}(\m{p})^\top \bmm{\sigma}  \|^2 \leq 0.
\end{align}

Từ bất đẳng thức \eqref{eq:distance_based_boundedness}, ta  suy ra $\bmm{\sigma}$ là bị chặn và tồn tại $\lim_{t \to +\infty}V$. Do $\sigma_k =\|\m{z}_k\|^2 - d_k^*$ nên suy ra $\|\m{z}_k\|=\|\m{p}_j - \m{p}_i\|,~\forall (i,j)\in E$ cũng bị chặn. Từ bất đẳng thức tam giác, ta suy ra $\|\m{p}_j - \m{p}_i\|, ~\forall i,j \in V$ là bị chặn, tức là $d_{\max}=\max_{i\ne j} \|\m{p}_i - \m{p}_j\|$ bị chặn. Bất đẳng thức sau luôn thỏa mãn:
\begin{align}
   n^2\|\bar{\m{p}} - \m{p}_i\|^2 &= \|\m{p}_1  - \m{p}_i + \ldots + \m{p}_n -\m{p}_i \|^2 \nonumber \\
   & \leq \left(\|\m{p}_1  - \m{p}_i\| + \ldots + \| \m{p}_n -\m{p}_i \|\right)^2 \nonumber\\
   & \leq n^2 \max_{j}\|\m{p}_j - \m{p}_i\|^2.\label{eq:distance_based_distance_boundedness}
\end{align}
Lấy tổng các bất phương trình \eqref{eq:distance_based_distance_boundedness} theo $i=1,\ldots,n$, ta thu được:
\begin{align}
   \sum_{i=1}^n \|\bar{\m{p}} - \m{p}_i\|^2 &\leq \sum_{i=1}^n \max_{j}\|\m{p}_j - \m{p}_i\|^2 \nonumber\\
   \|\m{p} - \m{1}_n \otimes \bar{\m{p}} \|^2 &\leq n d_{\max}^2 \le n d_{\max}^2(0).\label{eq:distance_based_ineq}
\end{align}
Bất phương trình \eqref{eq:distance_based_ineq} chứng tỏ rằng $\|\m{p} - \m{1}_n \otimes \bar{\m{p}} \|$ là bị chặn. Do $\bar{\m{p}}$ là không thay đổi và chỉ phụ thuộc vào $\m{p}(0)$, ta suy ra $\m{p}$ là bị chặn. Do đó, tồn tại $R=R(\bar{\m{p}}(0))$ sao cho tập compact $B(\bar{\m{p}}(0),R) = \{\m{p} \in \mb{R}^{2n}| ~\|\m{p}-\m{1}_n\otimes \bar{\m{p}}\|\leq R\}$ chứa $\m{p}(t),~\forall t\ge 0$. Giao của ${B}(\bar{\m{p}}(0),R(\bar{\m{p}}(0)))$ với các tập $\mc{Q}$, $\mc{D}$, $\mc{Q}$ do đó cũng là các tập bị chặn.

Đạo hàm cấp hai của $V$ là một hàm phụ thuộc vào $\m{p}$, $\m{z}$, $\bmm{\sigma}$ nên cũng là một hàm bị chặn. Từ đây, theo bổ đề Barbalat, ta có $\dot{V} \to 0$, hay $\m{R}^\top\bmm{\sigma} \to \m{0}_m$. Điều này chứng tỏ  $\m{p} \to \mc{Q}$, khi $t \to +\infty$. 

Định nghĩa $\gamma = \min_{\m{p} \in \mc{U}} V(\bmm{\sigma}(t)) >0$. Số họ cấu hình trong $\mc{U}$ (không tính tới phép quay và tịnh tiến) là một số hữu hạn nên $\gamma$ là hoàn toàn được định nghĩa. Từ đây, nếu đội hình ban đầu được chọn thỏa mãn $V(\bmm{\sigma}(0)) < \gamma$ thì rõ ràng, do $V$ không tăng, $\m{p}$ không thể tiệm cận tới tập $\mc{U}$. Do $\m{p} \to \mc{Q} = \mc{D} \cup \mc{U}$, $\mc{U} \cap \mc{D} = \emptyset$, nên phải có $\m{p} \to \mc{D}\cap B(\bar{\m{p}}(0),R)$. Từ đây suy ra $\bm{\sigma} \to \m{0}_m$.

Chú ý rằng tập $\bar{\mc{D}}=\mc{D}\cap B(\bar{\m{p}}(0),R)$ là tập compact chỉ chứa một số hữu hạn họ các cấu hình tương đồng về khoảng cách khác nhau thỏa mãn $\bm{\sigma} = \m{0}_{dm}$. Giả sử có $N\ge 1$ họ cấu hình như vậy, kí hiệu $\bar{\mc{D}}_{q} \subseteq \bar{\mc{D}}$, với $q=1,\ldots,N$, trong đó hai tập bất kỳ trong $\{\bar{\mc{D}}_q\}_{q=1}^N$ là rời nhau và $\bigcup_{q=1}^N \bar{\mc{D}}_q = \bar{\mc{D}}$. Khi đó, mỗi tập con $\bar{\mc{D}}_{q}$ được đặc trưng bởi một cấu hình $\m{p}^d \in \bar{\mc{D}}_q$, $\bar{\mc{D}}_{q}$ chứa mọi cấu hình có cùng trọng tâm với $\m{p}^d$ và sai khác với $\m{p}^d$ bởi một phép quay hoặc một phép lật. Tiếp tục chia nhỏ $\bar{\mc{D}}_{q}$ thành các tập con rời nhau $\bar{\mc{D}}_{qr}, r=1,\ldots,M_q$ thỏa mãn $\bigcup_{r=1}^{M_q} \bar{\mc{D}}_{qr} = \bar{\mc{D}}_q$ thì mọi cấu hình trong tập con $\bar{\mc{D}}_{qr}, r=1,\ldots,M_q$ chỉ sai khác nhau một phép quay quanh tâm $\bar{\m{p}}(0)$.

Do $\m{p} \to \mc{D}$, đặt $d(\m{p},\bar{\mc{D}})=\min_{\m{p}^d \in \bar{\mc{D}}} \|\m{p} - \m{p}^d\|$ là khoảng cách từ $\m{p}$ tới tập $\bar{\mc{D}}$ thì với mỗi $\varepsilon>0$ đủ nhỏ, tồn tại $T>0$ đủ lớn để $d(\m{p}(t),\bar{\mc{D}})<\varepsilon$, hay tồn tại cấu hình $\m{p}^d(t) \in \bar{\mc{D}}$ đủ gần $\m{p}(t)$. Rõ ràng $\m{p}^d$ phải nằm trong duy nhất một trong các tập hợp 
\[\bar{\mc{D}}_{11},\ldots,\bar{\mc{D}}_{1M_1},\ldots,\bar{\mc{D}}_{N1},\ldots,\bar{\mc{D}}_{NM_N}.\]
Như vậy, nếu $\m{p}(t)$ tiệm cận tới một chu trình khép kín\footnote{limit cycle} trong $\mc{D}$ thì chu trình khép kín này chỉ chứa các cấu hình sai khác nhau một phép quay quanh tâm $\bar{\m{p}}(0)$. Nhưng điều này không thể xảy ra, do $\dot{\m{p}}=\m{R}^\top\bm{\sigma} \perp {\rm im}\left((\m{I}_n \otimes \m{J})\m{p}\right)$, hay $\m{p}(t)$ không có chuyển động quay quanh tâm $\bar{\m{p}}(0)$. Do đó, tồn tại một cấu hình tĩnh $\m{p}^d \in \bar{\mc{D}}$ sao cho $\lim_{t\to +\infty}\m{p}(t) = \m{p}^d$.

\begin{story}{Bài toán Điều khiển đội hình dựa trên khoảng cách}
Bài toán điều khiển đội hình dựa trên khoảng cách là một vấn đề lý thuyết trong hệ đa tác tử thu hút sự quan tâm từ nhiều nhóm nghiên cứu khắp thế giới. Phát biểu bài toán điều khiển đội hình ở chương này dựa trên \cite{Oh2015}. Các yếu tố làm bài toán được nghiên cứu rộng rãi bao gồm: (a) bài toán được phát biểu khá đơn giản, chỉ cần kiến thức hình học là có thể hiểu phát biểu bài toán; (b) việc tìm một lời giải \emph{hoàn toàn phân tán} để \emph{ổn định toàn cục} đội hình đặt cho bởi một tập khoảng cách là vấn đề khó; (c) phép đo đạc khoảng cách có thể thực hiện tại hệ qui chiếu riêng, do đó có ý nghĩa với hệ xe tự hành; và (d) bài toán được quan tâm bởi nhiều nhóm nghiên cứu tại các đại học danh tiếng trên thế giới.

Trong \cite{Egerstedt2001}, Magnus Egerstedt và Xiaoming Hu thiết kế luật điều khiển mobile robot dựa trên hàm thế khoảng cách. Bài toán điều khiển đội hình dựa trên khoảng cách có lẽ được phát biểu đầu tiên ở \cite{Olfati2002distributed} tại IFAC 2002, trong đó Olfati-Saber và Murray xây dựng hàm thế dựa trên một tập $2n-3$ khoảng cách để thiết kế luật điều khiển cho $n$ tác tử tích phân bậc hai. Cùng thời điểm, Eren Tolga, Belhumeur, Anderson và Morse đề xuất những ý tưởng sử dụng lý thuyết cứng trong điều khiển đội hình \cite{Eren2002framework}. Nghiên cứu sử dụng lý thuyết cứng khoảng cách trong điều khiển và thay đổi cấu trúc đội hình tiếp tục được phát triển bởi cả hai nhóm nghiên cứu \cite{Olfati2002graph,Eren2002closing}, và dần nhận được quan tâm rộng rãi một phần nhờ khả năng ứng dụng trong định vị mạng cảm biến \cite{Aspnes2006}. 

Từ khoảng 2005, điều khiển đội hình với đồ thị hữu hướng được phát triển bởi Cao \cite{Cao2007controlling,Cao2011}, Yu \cite{Yu2007three}, và Hendrickx \cite{Hendrickx2007}. Dimarogonas chứng minh khả năng ổn định các đội hình tác tử tích phân bậc nhất có cấu trúc cây dựa trên chọn hàm thế theo sai lệch khoảng cách trong \cite{Dimarogonas2008}. Chứng minh ổn định tiệm cận đội hình đặt cho tác tử tích phân bậc nhất được giải quyết trong các luận văn của D\"{o}rfler \cite{Dorfler2010} và Krick (2009) \cite{Krick2009}, với giả thiết đội hình đặt hai chiều thỏa mãn tính cứng vi phân. Công cụ được sử dụng là lý thuyết đa tạp trung tâm với đồ thị là chu trình hữu hướng ba đỉnh, vô hướng, hoặc leader-follower. Oh và Ahn đề xuất luật điều khiển ổn định tiệm cận đội hình tích phân bậc nhất, bậc hai và phân tích ổn định tiệm cận địa phương dựa sử dụng bổ đề Lojasiewicz \cite{Oh2014distance}. \cite{Cai2014adaptive,DeQueiroz2019formation} tập trung vào các đội hình cứng vi phân và chứng minh ổn định tiệm cận địa phương dựa trên tính chất hạng của ma trận $\m{R}(\m{p})=2n-3$ khi $\m{p}$ nằm đủ gần một cấu hình $\m{p}^d \in \mc{D}$. Vu và cộng sự \cite{Vu2020LCSS,Vu2023distance} phân tích ổn định hệ dựa trên bổ đề Barbalat và xác định điều kiện đủ cho đội hình tiệm cận tới tập cấu hình đặt. Sun chứng minh đội hình đặt ổn định địa phương theo hàm mũ \cite{Sun2016exponential}, tuy nhiên không chỉ ra được lân cận thỏa mãn tính chất này.

Tuy nhiên, việc ổn định toàn cục hoặc gần như toàn cục các đội hình dựa trên luật điều khiển khoảng cách là không khả thi, ngoại trừ một số trường hợp đặc biệt. Lý do là sự tồn tại của các điểm cân bằng đối với $\bm{\sigma}$ và các điểm cân bằng này có khả năng là ổn định \cite{Trinh2017comments}. Nếu giả thiết không tồn tại các điểm cân bằng sai ổn định, luật điều khiển ở \cite{Tian2013} ổn định tiệm cận đội hình đặt.

Một số nhánh nghiên cứu từ nhận xét này đã được phát triển: 
\begin{itemize}
\item[i.] Chứng minh ổn định tiệm cận gần như toàn cục của một số đội hình cụ thể: acyclic leader-follower \cite{Summers2011,Pham2017ICCAS,Pham2018finite}, ${K}_4$ hai chiều \cite{Summers2009formation,Dasgupta2011controlling} và ${K}_4$ ba chiều  \cite{Park2013CDC,Park2014stability}, chu trình hữu hướng  \cite{Park2015stabilisation}, $K_4$ loại bỏ một cạnh \cite{Pham2018IJRNC}.
\item[ii.] phát triển công cụ lý thuyết để xác định số lượng điểm cân bằng sai và tính chất của các điểm cân bằng này \cite{Anderson2011,AndersonHelmke2014Counting}, Belabas \cite{Belabbas2013global}. Tuy nhiên, hướng tiếp cận này sử dụng công cụ toán khá phức tạp và chỉ cho kết luận cụ thể với một số đội hình nhỏ. Các phân tích đòi hỏi giải chính xác phương trình có bậc lớn hơn 4. Phương pháp số đã được đề xuất tại \cite{Summers2013}, giúp kiểm chứng việc tồn tại điểm cân bằng ổn định không mong muốn một cách gần đúng. Hiện nay vẫn chưa có kết luận cuối về điểm cân bằng ổn định không mong muốn, do chưa tìm được điểm cân bằng này chính xác.
\item[iii.] Tìm các phương án tránh các hạn chế của thuật toán gốc, từ đó ổn định toàn cục đội hình đặt. Oh và Ahn \cite{Oh2014tac} sử dụng thêm sai lệch hệ qui chiếu, chuyển bài toán điều khiển dựa trên khoảng cách về bài toán điều khiển dựa trên vị trí tương đối. \cite{Aranda2015coordinate,Aranda2016} giả thiết các tác tử giải các bài toán tối ưu theo từng clique trong đồ thị nhằm xác định góc định hướng phù hợp cho luật điều khiển tại mỗi clique, từ đó ổn định toàn cục đội hình đặt. Kết quả tổng quát nhất hiện nay có lẽ là ở \cite{Sakurama2015distributed,Sakurama2021}, trong đó Sakurama, Shun-ichi Azuma và Toshiharu Sugie chứng minh rằng nếu các tác tử có thể đo vị trí tương đối trong hệ qui chiếu địa phương, một luật điều khiển gradient ổn định tiệm cận toàn cục đội hình đặt có thể được xác định dựa trên giải các bài toán tối ưu trên mỗi clique của đồ thị;
\item[iv.] Ngược với (iii), một số tác giả xét bài toán ổn định đội hình đặt dựa trên ít thông tin đo đạc hơn. \cite{Anderson2011range,cao2011formation,Suttner2018formation} chỉ dùng khoảng cách chứ không có thông tin hoàn toàn về vector sai lệch địa phương.
\end{itemize}
Áp dụng của các phương pháp điều khiển nâng cao vào bài toán cũng được xem xét trong khá nhiều tài liệu đã giúp bài toán gần hơn với các điều kiện và yêu cầu thực tế, ví dụ như xét ảnh hưởng của sai lệch đo, sai lệch khoảng cách đặt \cite{Mou2016TAC,Hector2014controlling,de2015controlling}, chống nhiễu \cite{Vu2020LCSS,Trinh2022adaptive}, bám mục tiêu thay đổi thời gian Kang \cite{Kang2016,Rozenheck2015proportional,Vu2023distance}, điều khiển thời gian xác lập \cite{Park2014finite,Sun2014IFAC,Mehdifar2020prescribed}, mô hình đối tượng thực tế \cite{Cai2014adaptive,Park2015stabilisation,DeQueiroz2019formation}, điều khiển kết hợp định hướng đội hình \cite{Sun2017AutOrient}, \ldots %
Các sách chuyên khảo về điều khiển đội hình dựa trên khoảng cách bao gồm \cite{DeQueiroz2019formation,Sun2018cooperative,Ahn2019formation}.
\end{story}

Chúng ta có các nhận xét sau đây:
\begin{itemize}
\item[i.] Nếu có thể xác định được $\gamma = \min_{\m{p} \in \mc{U}} V(\bmm{\sigma}(t)) >0$ thì do $\sigma_k = \|\m{z}_k\|^2 - \|\m{z}_k^*\|^2$ nên ta có
\begin{align}
\sum_{k=1}^m \sigma_k = \sum_{k=1}^m \|\m{z}_k\|^2 - \sum_{k=1}^m \|\m{z}_k^*\|^2 = \|\m{z}\| - \|\m{z}^*\|^2.
\end{align}
Do 
\begin{align}
\left|\sum_{k=1}^m \sigma_k \right|^2 \leq m \sum_{i=1}^m \sigma_k^2 = 4m V(\bmm{\sigma}(t)),
\end{align}
ta suy ra điều kiện đủ để $\m{p}$ tiệm cận tới một cấu hình trong $\mc{D}$ là
\begin{align}
 \|\m{z}^*\|^2 - 2 \sqrt{m\gamma} \leq \|\m{z}(0)\|^2 \leq \|\m{z}^*\|^2 + 2 \sqrt{m\gamma}.
\end{align}

\item[ii.] Chú ý rằng ma trận $\mcl{R}(\bmm{\sigma}) = \m{R}(\bmm{\sigma})^\top\m{R}(\bmm{\sigma}) \in \mb{R}^{2n\times 2n}$ có cấu trúc tương tự một ma trận Laplace (với các ma trận khối $\m{z}_{ij}\m{z}_{ij}^\top$ đối xứng bán xác định dương thay cho các phần tử vô hướng). Do đó, ngoài tên gọi ma trận độ cứng (stiffness matrix), trong một số tài liệu, $\mcl{R}(\bmm{\sigma})$ còn được gọi là ma trận Laplace cứng\index{ma trận!Laplace cứng}. Giá trị riêng dương nhỏ nhất của ma trận này cũng là một chỉ số để xác định tính cứng khoảng cách của đồ thị \cite{Trinh2016CDC}.

Ta có thể biến đổi luật điều khiển \eqref{eq:c5_distance_formation} như sau:
\begin{align}
\m{u}&= - \m{R}^\top\bmm{\sigma} \nonumber \\
  &= -\bar{\m{H}}^\top \text{diag}(\m{z}_k) \bmm{\sigma} \nonumber \\
  &= - \bar{\m{H}}^\top (\text{diag}(\bmm{\sigma}) \otimes \m{I}_2) \bar{\m{H}} \m{p} \nonumber\\
  &= - \bar{\m{H}}^\top (\text{diag}(\m{f}_G(\m{p})) \otimes \m{I}_2) \bar{\m{H}} \m{p} + \bar{\m{H}}^\top (\text{diag}(\m{f}_G(\m{p}^*)) \otimes \m{I}_2) \bar{\m{H}} \m{p}.\label{eq:c5_distance_formation_1}
\end{align}
Với cách biểu diễn này, \eqref{eq:c5_distance_formation} có dạng một giao thức đồng thuận với các trọng số ${\sigma}_k, k=1, \ldots, m$, phụ thuộc vào biến trạng thái $\m{p}$. Do các trọng số ${\sigma}_k, k=1, \ldots, m$ có thể nhận giá trị âm và phụ thuộc vào cấu hình, ta không áp dụng được phương pháp phân tích như cho hệ đồng thuận thông thường. Mặt khác, \eqref{eq:c5_distance_formation_1} mô tả một hệ đồng thuận trọng số phụ thuộc trạng thái $\m{p}$ bị tác động bởi một tín hiệu đầu vào ở dạng phản hồi trạng thái $\m{u}(\m{p}) = \bar{\m{H}}^\top ({\rm diag}(\m{f}_G(\m{p}^*)) \otimes \m{I}_2) \bar{\m{H}} \m{p}$. 
\end{itemize}

\begin{example} \label{VD:5.4}
Ở ví dụ này, ta sẽ minh họa giá trị $\gamma$ trong điều kiện hội tụ trong chứng minh ở trên (Xem Hình \ref{fig:VD_5.4}). Xét đội hình gồm 3 tác tử, trong đó $G=C_3$ và cấu hình đặt là một tam giác đều cạnh bằng 1. Phương trình viết cho ba tác tử được cho lần lượt bởi:
\begin{align}
  \dot{\m{p}}_1 &= -(d_{12}^2 - 1) (\m{p}_1 - \m{p}_2) -(d_{13}^2 - 1) (\m{p}_1 - \m{p}_3) \\
  \dot{\m{p}}_2 &= -(d_{21}^2 - 1) (\m{p}_2 - \m{p}_1) -(d_{23}^2 - 1) (\m{p}_2 - \m{p}_3) \\
  \dot{\m{p}}_3 &= -(d_{31}^2 - 1) (\m{p}_1 - \m{p}_3) -(d_{23}^2 - 1) (\m{p}_3 - \m{p}_2).
\end{align}
Chúng ta sẽ xác định tập các cấu hình trong $\mc{Q}$. Rõ ràng, nếu 3 điểm $\m{p}_1$,  $\m{p}_2$,  $\m{p}_3$ không thẳng hàng thì hệ các họ các cấu hình thỏa mãn $\dot{\m{p}}_1 = \dot{\m{p}}_2 = \dot{\m{p}}_3 = 0$ phải thỏa mãn $d_{12} = d_{13} = d_{31} = 1$. Những cấu hình này thuộc tập $\mc{D}$.

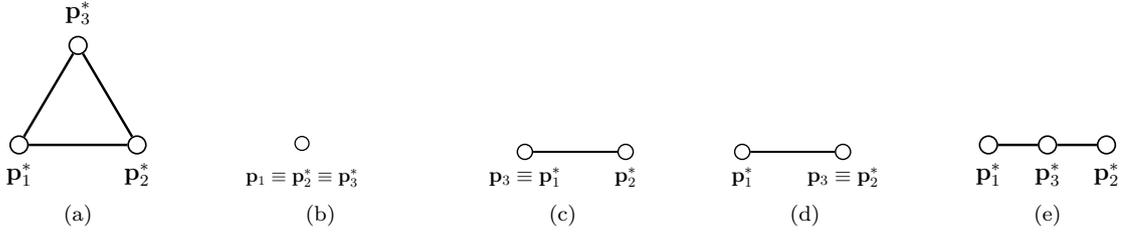
\begin{figure}[t]
\centering
\subfloat[]{
\resizebox{2.25cm}{!}{
\begin{tikzpicture}[
roundnode/.style={circle, draw=black, thick, minimum size=3mm,inner sep= 0.25mm},
squarednode/.style={rectangle, draw=red!60, fill=red!5, very thick, minimum size=5mm},
roundnode1/.style={rectangle, draw=white,},
]
\node[roundnode]   (v1)   at   (0,0) {};
\node[roundnode]   (v2)   at   (2,0) {};
\node[roundnode]   (v3)   at   (1,1.7) {};
\node  at (0,-0.5) {\large $\m{p}_1^*$};
\node  at (2,-0.5) {\large $\m{p}_2^*$};
\node  at (1,2.25) {\large $\m{p}_3^*$};
\draw[-, very thick] (v1)--(v2)--(v3)--(v1);
\end{tikzpicture}
}
}
\hfill
\subfloat[]{
\resizebox{2.25cm}{!}{
\begin{tikzpicture}[
roundnode/.style={circle, draw=black, thick, minimum size=3mm,inner sep= 0.25mm},
squarednode/.style={rectangle, draw=red!60, fill=red!5, very thick, minimum size=5mm},
roundnode1/.style={rectangle, draw=white},
]
\node[roundnode] (v1) at (0,0) {};
\node at (2,0) {};
\node at (1,1.7){};
\node at (0,-0.75) {{\large $\m{p}_1 \equiv \m{p}_2^* \equiv \m{p}_3^*$}};
\end{tikzpicture}
}
}
\hfill
\subfloat[]{
\resizebox{2.25cm}{!}{
\begin{tikzpicture}[
roundnode/.style={circle, draw=black, thick, minimum size=3mm,inner sep= 0.25mm},
squarednode/.style={rectangle, draw=red!60, fill=red!5, very thick, minimum size=5mm},
roundnode1/.style={rectangle, draw=white},
]
\node[roundnode] (v1) at (0,0) {};
\node[roundnode] (v2) at (2,0) {};
\node at (1,1.7){};
\node at (0,-0.5) {{\large $\m{p}_3 \equiv \m{p}_1^*$}};
\node at (2,-0.5) {{\large $\m{p}_2^*$}};
\draw[-, very thick] (v1)--(v2);
\end{tikzpicture}
}
}
\hfill
\subfloat[]{
\resizebox{2.25cm}{!}{
\begin{tikzpicture}[
roundnode/.style={circle, draw=black, thick, minimum size=3mm,inner sep= 0.25mm},
squarednode/.style={rectangle, draw=red!60, fill=red!5, very thick, minimum size=5mm},
roundnode1/.style={rectangle, draw=white, thick},
]
\node[roundnode]  (v1)   at   (0,0) {};
\node[roundnode]  (v2)   at   (2,0) {};
\node at (1,1.7) {};
\node at (0,-0.5){\large $\m{p}_1^*$};
\node at (2,-0.5){\large $\m{p}_3 \equiv \m{p}_2^*$};
\draw[-, very thick] (v1)--(v2);
\end{tikzpicture}
}
}
\hfill
\subfloat[]{
\resizebox{2.25cm}{!}{
\begin{tikzpicture}[
roundnode/.style={circle, draw=black, thick, minimum size=3mm,inner sep= 0.25mm},
squarednode/.style={rectangle, draw=red!60, fill=red!5, very thick, minimum size=5mm},
roundnode1/.style={rectangle, draw=white, thick},
]
\node[roundnode]  (v1)   at   (0,0) {};
\node[roundnode]  (v2)   at   (2,0) {};
\node at (1,1.7) {};
\node[roundnode]  (v3)   at   (1,0) {};
\node at (0,-0.5){\large $\m{p}_1^*$};
\node at (2,-0.5){\large $\m{p}_2^*$};
\node at (1,-0.5){\large $\m{p}_3^*$};
\draw[-, very thick] (v1)--(v3)--(v2);
\end{tikzpicture}
}
}
\caption{Minh họa Ví dụ \ref{VD:5.4}: (a) cấu hình mong muốn; (b), (c), (d), (e): bốn cấu hình khác nhau thuộc $\mc{U}^*$. \label{fig:VD_5.4}}
\end{figure}

Xét trường hợp 3 điểm $\m{p}_1$,  $\m{p}_2$,  $\m{p}_3$ trùng nhau. Khi đó, $\dot{\m{p}}_1 = \dot{\m{p}}_2 = \dot{\m{p}}_3 = \m{0}$ và $V(\bmm{\sigma})=\frac{1}{4}((d_{12}^2 - (d_{12}^*)^2)^2+(d_{13}^2 - (d_{13}^*)^2)^2+(d_{32}^2 - (d_{32}^*)^2)^2) = \frac{1}{4}(1^2+1^2+1^2)=\frac{3}{4} = \gamma_1$.

Xét trường hợp 2 trong 3 điểm trùng nhau. Do tính đối xứng, ta chỉ cần xét $\m{p}_1 = \m{p}_2 \neq \m{p}_3$. Thay vào phương trình của tác tử 1 và 2, ta có $\dot{\m{p}}_1 = -(d_{13}^2 - 1) (\m{p}_1 - \m{p}_3) = \m{0}$ nên suy ra $d_{13} = 1$ và $d_{23} = 1$. Thay vào phương trình của tác tử 3 ta thấy thỏa mãn $\dot{\m{p}}_3 = \m{0}$. Như vậy, ứng với họ cấu hình này, ta có $V(\m{\sigma})=\frac{1}{4}((1-0)^2)=\frac{1}{4} = \gamma_2$.

Xét trường hợp 3 điểm thẳng hàng và không có 2 điểm nào trùng nhau. Do tính đối xứng, ta chỉ cần xét trường hợp điểm $\m{p}_2$ nằm giữa $\m{p}_1$ và $\m{p}_3$, tức là $d_{12}+d_{23}=d_{13}$. Phương trình $\dot{\m{p}}_2 = \m{0}$ cho ta $d_{12} = d_{23} = \frac{1}{2}d_{13}$. Thay vào phương trình $\dot{\m{p}}_1 = -(d_{12}^2 - 1) (\m{p}_1 - \m{p}_2) -2(4d_{12}^2 - 1) (\m{p}_1 - \m{p}_2) =- (9d_{12}^2 - 3) (\m{p}_1 - \m{p}_2) = \m{0}$, ta suy ra $d_{12} = \frac{1}{\sqrt{3}}=d_{23}$ (vô lý).

Như vậy, $\gamma = \min_{\m{p} \in \mc{U}} V(\bmm{\sigma}) = \min \{\gamma_1, \gamma_2\} = \frac{1}{4}$. Với $V(\bmm{\sigma}(0))<\frac{1}{4}$ thì $\m{p} \to \mc{D}$.
\end{example}
\begin{SCfigure}[][th!]
\caption{Một cấu hình thỏa mãn các ràng buộc khoảng cách trong Ví dụ \ref{VD:5.5}.}
\hspace{4.5cm}
\begin{tikzpicture}[
roundnode/.style={circle, draw=black, thick, minimum size=3mm,inner sep= 0.25mm},
squarednode/.style={rectangle, draw=red!60, fill=red!5, very thick, minimum size=5mm},
roundnode1/.style={rectangle, draw=white,},
]
\node[roundnode]   (v1)   at   (0,2)  {$1$};
\node[roundnode]   (v2)   at   (-3,1) {$2$};
\node[roundnode]   (v3)   at   (0,0)  {$3$};
\node[roundnode]   (v4)   at   (2,1)  {$4$};
\node[roundnode]   (v5)   at   (1,-4) {$5$};

\node  at (0.4,1) {\large $d_{13}^*$};
\node  at (-2,1.7) {\large $d_{12}^*$};
\node  at (-1,0.7) {\large ${d}_{23}^*$};
\node  at (1.2,1.7) {\large ${d}_{23}^*$};
\node  at (1.2,0.2) {\large ${d}_{34}^*$};
\node  at (-1,-1) {\large ${d}_{25}^*$};
\node  at (2,-1) {\large ${d}_{45}^*$};

\draw[-, very thick] (v1)--(v2)--(v3)--(v4)--(v1)--(v3);
\draw[-, very thick] (v2)--(v5)--(v4);

\end{tikzpicture}
\end{SCfigure}
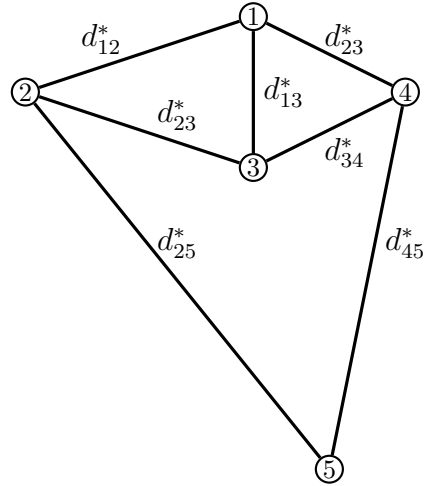

\begin{example} \label{VD:5.5}
Để minh họa sự tồn tại đồng thời của tập các điểm cân bằng mong muốn (ổn định) $\mc{D}$ và tập điểm cân bằng không mong muốn $\mc{U}$, ta xét ví dụ trình bày trong \cite{Park2016,Trinh2017comments} về đội hình gồm 5 tác tử với đồ thị $G$ cho trên Hình~\ref{fig:c5Vidu_mophong_distance_based}. Các khoảng cách đặt được cho bởi: $d_{12}^{*2}=d_{23}^{*2}=10$, $d_{13}^{*2}=4$,  $d_{14}^{*2}=d_{34}^{*2} = 5$, $d_{25}^*=41$ và $d_{45}^*=26$. Mỗi tác tử chuyển động theo luật điều khiển đội hình dựa trên khoảng cách \eqref{eq:c5_distance_based_control_law_global}, với hệ số tỉ lệ $k_p = 3$.. Tương ứng với hai cấu hình ban đầu khác nhau, hệ tiệm cận tới hai cấu hình hoàn toàn khác nhau (mong muốn và không mong muốn). 

Chú ý rằng sự tồn tại của điểm cân bằng không mong muốn và ổn định mới chỉ được kiểm chứng nhờ mô phỏng \cite{Park2016,Trinh2017comments}. Sự tồn tại của các điểm cân bằng không mong muốn và không ổn định trong $\mc{U}$ được chỉ ra cụ thể trong \cite{Summers2009formation,Dasgupta2011controlling} với một đội hình chữ nhật gồm 4 tác tử với đồ thị $K_4$.
\end{example}

\begin{figure*}[t!]
\centering
    \subfloat[]{\includegraphics[width=0.45\textwidth]{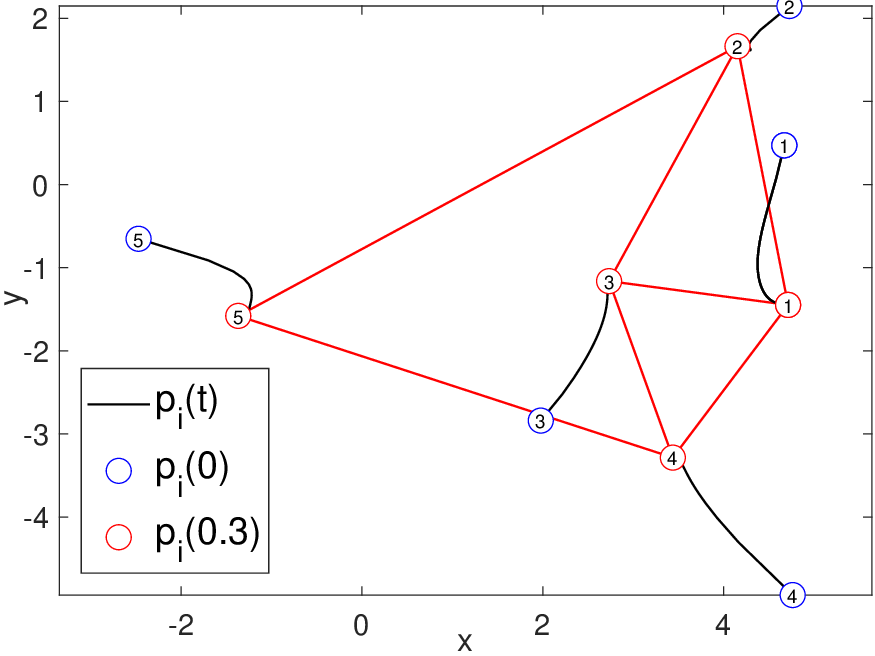}}\hfill
    \subfloat[]{\includegraphics[width=0.5\textwidth]{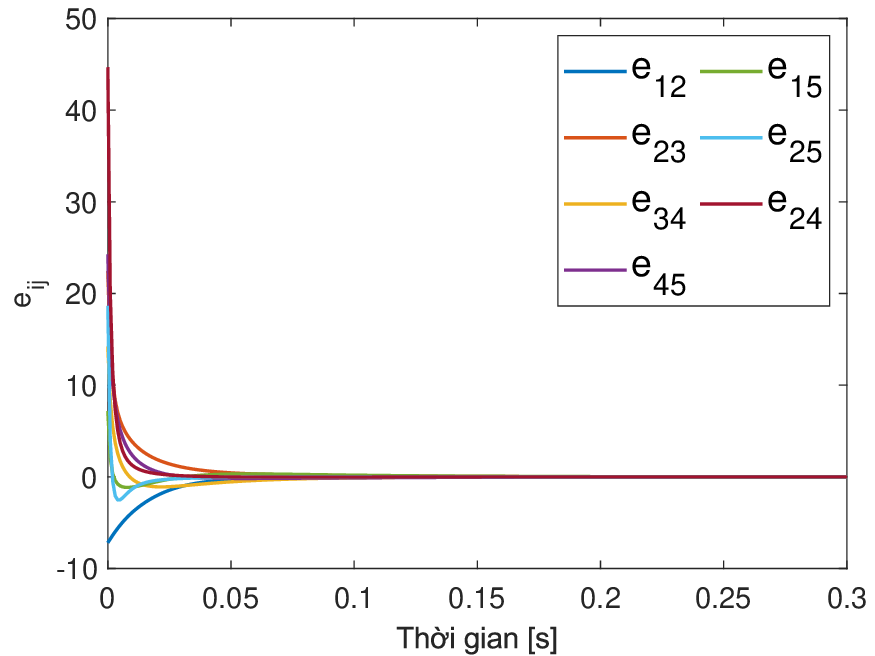}}\\
    \subfloat[]{\includegraphics[width=0.45\textwidth]{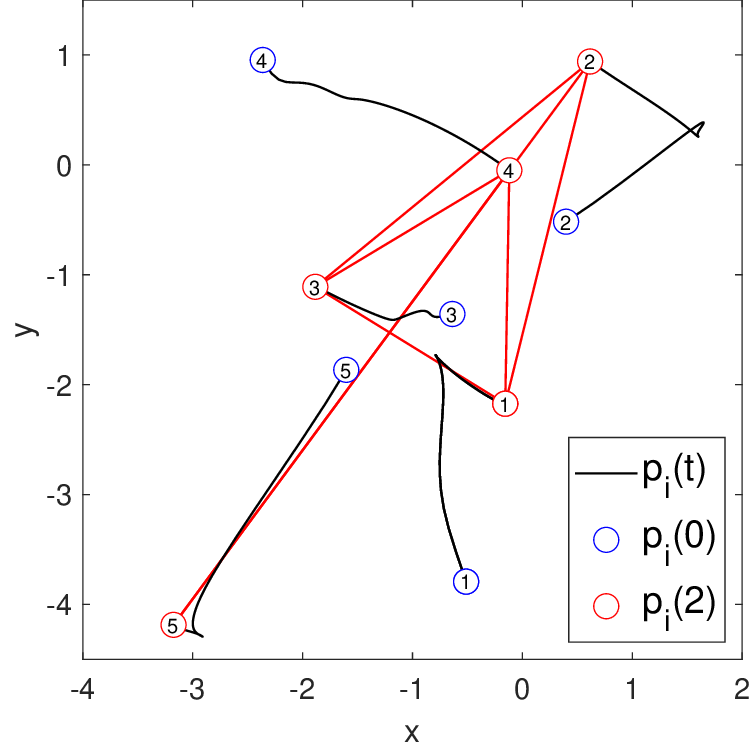}}\hfill
    \subfloat[]{\includegraphics[width=0.5\textwidth]{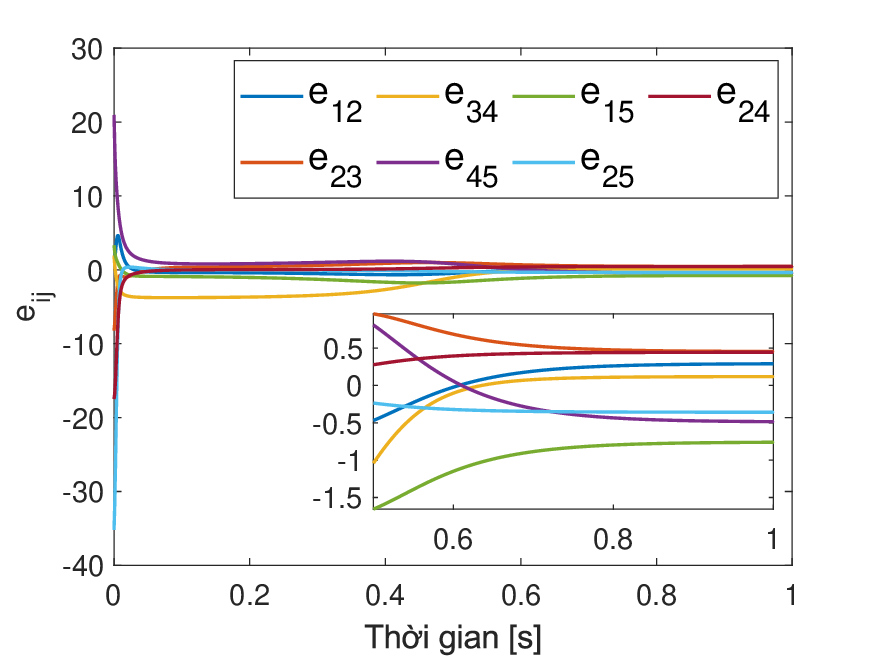}}
    \caption{Đội hình gồm 5 tác tử trong Ví dụ \ref{VD:5.5}. (a) và (b): $\m{p}$ tiệm cận tiến tới một đến một cấu hình mong muốn;~(c) và (d) $\m{p}$ tiệm cận tới một cấu hình không mong muốn.}
    \label{fig:c5Vidu_mophong_distance_based}
\end{figure*}

\section{Điều khiển đội hình dựa trên vector hướng}
\label{c5_s5:bearing_formation}
\index{điều khiển đội hình!dựa trên vector hướng}
Trong thời gian gần đây, các phương pháp điều khiển đội hình dựa trên vector hướng và góc thu hút được khá nhiều quan tâm. Sự quan tâm này một phần đến từ sự phát triển của các thiết bị bay không người lái cỡ nhỏ có gắn camera với khả năng xác định vector hướng từ ảnh thu được. Các phương án điều khiển chỉ sử dụng vector hướng hay chỉ sử dụng góc lệch không đòi hỏi hệ qui chiếu toàn cục cũng như không sử dụng các cảm biến về khoảng cách, nhờ đó mang đến một phương án thay thế khi các tác tử hoạt động ở môi trường không có GPS hay khi các cảm biến khoảng cách bị hỏng. \index{vector hướng}

Lý thuyết cứng hướng (bearing rigidity) cung cấp công cụ lý thuyết để kiểm tra và xây dựng các đội hình có thể điều khiển dựa trên vector hướng. Khi ra đời, lý thuyết cứng hướng đã được nghiên cứu với mục đích thuần toán học \cite{Whiteley1996}. Việc đưa lý thuyết cứng hướng vào các bài toán điều khiển đội hình hay định vị mạng cảm biến được được đề xuất trong một số công trình của T. Eren và các cộng sự \cite{Eren2006using,Eren2003sensor,Eren2012}. Điều khiển đội hình dựa trên vector hướng chỉ được quan tâm rộng rãi sau khi lý thuyết về cứng hướng trong $\mb{R}^d$ được trình bày trong các bài báo \cite{zhao2015tac,zhao2016aut}.\index{lý thuyết!cứng hướng trên $\mb{R}^d$} Chú ý rằng lý thuyết cứng hướng còn được mở rộng trên $\mb{R}^d\times S^{d-1}$ hay $SE(d)$ trong các tài liệu  \cite{Franchi2012,Zelazo2015,schiano2016rigidity,Michieletto2016}. Điều khiển đội hình dựa trên lý thuyết cứng hướng trên $SE(d)$ có  ưu điểm là không yêu cầu thông tin về một hệ qui chiếu chung, tuy vậy lại cần trao đổi thông tin giữa các tác tử. \index{lý thuyết!cứng hướng trên $SE(d)$}

Với phương án điều khiển đội hình dựa trên vector hướng, chúng ta xét hệ các tác tử thỏa mãn các giả thiết sau:
\begin{itemize}
\item Mô hình tác tử: Các tác tử được mô hình hóa bởi khâu tích phân bậc nhất:
\begin{align}
\dot{\m{p}}_i = \m{u}_i, ~ i = 1, \ldots, n,
\end{align}
trong đó $\m{p}_i \in \mb{R}^d$ và $\m{u}_i \in \mb{R}^d$ lần lượt là vị trí và tín hiệu điều khiển của tác tử $i$ viết trong hệ qui chiếu $^g\Sigma$.
    \item Biến đo: Các tác tử có các hệ qui chiếu riêng được định hướng giống nhau nhưng có gốc tọa độ khác nhau. Để đơn giản, chúng ta giả thiết thêm các hệ qui chiếu riêng có cùng định hướng với hệ tham chiếu toàn cục. Trong hệ qui chiếu riêng, các tác tử có thể đo vector hướng \index{vector hướng} (bearing vector) $\m{g}_{ij} = \frac{\m{p}_j - \m{p}_i}{\|\m{p}_j - \m{p}_i\|}$ của các tác tử láng giềng $j \in {N}_i$.
    \item Đồ thị tương tác: Tương tác về đo đạc giữa các tác tử được cho bởi đồ thị vô hướng $G$, với giả thiết $G$ là một đồ thị cứng hướng. 
    \item Đội hình đặt: được mô tả bởi một tập các vector hướng mong muốn $\Gamma = \{\m{g}_{ij}^* \in \mb{R}^d\}_{(i,j) \in E}$.
\end{itemize}

Lý thuyết cứng hướng trên $\mb{R}^d$ và ứng dụng trong điều khiển đội hình chỉ sử dụng vector hướng sẽ được lần lượt trình bày ở mục này.\index{lý thuyết!cứng hướng} \index{điều khiển đội hình!dựa trên khoảng cách}

\subsection{Lý thuyết cứng hướng trong không gian Euclid d chiều}
Xét một tập hợp gồm $n$ điểm trong không gian $d$ chiều ($n\geq 2, d \geq 2$) tại $\m{p}_i \in \mb{R}^d$, trong đó $\m{p}_i \neq \m{p}_j, \forall i \neq j, i, j \in \{1, \ldots, n\}$. Một đội hình trong không gian $d$ chiều $({G},\m{p})$ được cho bởi một đồ thị ${G}=(V,E)$ và một cấu hình $\m{p} = [\m{p}_1^\top, \ldots, \m{p}_n^\top]^\top$. Với mỗi cạnh $e_k=(v_i,v_j) \in {E}$, vector hướng chỉ từ $\m{p}_i$ tới $\m{p}_j$ được định nghĩa bởi
\begin{align*}
\m{g}_{ij} = \frac{\m{p}_j - \m{p}_i}{\|\m{p}_j - \m{p}_i\|} = \frac{\m{z}_{ij}}{\|\m{z}_{ij}\|},
\end{align*}
trong đó $\m{z}_{ij} = \m{p}_j - \m{p}_i$ là vector sai lệch. Chú ý rằng $\m{g}_{ij}$ là một vector đơn vị trong $\mb{R}^d$, do $\|\m{g}_{ij}\| = 1$. Với mỗi vector hướng $\m{g}_{ij}$, ta định nghĩa một ma trận chiếu vuông góc tương ứng $\m{P}_{\m{g}_{ij}} = \m{I}_d - \m{g}_{ij}\m{g}_{ij}^\top.$ \index{ma trận!chiếu vuông góc}

Hai đội hình $({G},\m{p})$ và $({G},\m{q})$ là \emph{tương đương về hướng}\footnote{bearing equivalency} \index{tương đương!về hướng}  nếu $\m{P}_{\m{g}_{ij}}(\m{q}_i-\m{q}_j) = \m{0},~\forall (v_i,v_j) \in {E}$. Hơn nữa, $({G},\m{p})$ và $({G},\m{q})$ là \emph{tương đồng về hướng}\footnote{bearing congruency} \index{tương đồng!về hướng} nếu $\m{P}_{(\m{p}_i-\m{p}_j)}(\m{q}_i-\m{q}_j) = \m{0},~\forall i \neq j,~i, j =1,\ldots, n$. Một đội hình $({G},\m{p})$ là \emph{cứng hướng toàn cục}\footnote{global bearing rigid} nếu như một đội hình bất kỳ tương đương về hướng với $({G},\m{p})$ thì cũng tương đồng về hướng với $({G},\m{p})$. \index{cứng!hướng!toàn cục}

Đặt $\m{g} = \text{vec}(\m{g}_1, \ldots, \m{g}_m) = [\ldots, \m{g}_{ij}^\top, \ldots]^\top$, \emph{ma trận cứng hướng}\index{ma trận!cứng hướng} được định nghĩa như sau:
\begin{align*}
\m{R}_b = \frac{\partial \m{g}}{\partial \m{p}} = \text{blkdiag}\Big( \frac{\m{P}_{\m{g}_k}}{\|\m{z}_k\|} \Big) \bar{\m{H}} \in \mb{R}^{dm\times dn},
\end{align*}
trong đó $\bar{\m{H}} = \m{H} \otimes \m{I}_d$ và $\m{H}$ là ma trận liên thuộc của $G$. Một đội hình là \index{cứng!hướng !vi phân} \emph{cứng hướng vi phân}\footnote{infinitesimally bearing rigid} trên $\mb{R}^d$ khi và chỉ khi ${\rm rank}(\m{R}_b) = dn-d-1$, hoặc ${\rm ker}(\m{R}_b) = {\rm im}([\m{1}_n\otimes \m{I}_d,~\m{p}])$. Một đội hình là không cứng hướng vi phân khi và chỉ khi ${\rm rank}(\m{R}_b) < dn-d-1$. Ví dụ một số đội hình cứng hướng vi phân và không cứng hướng vi phân được cho như trên Hình~\ref{fig:chap8_IBR_formations}.

\begin{figure*}[th]
    \centering
    \includegraphics[width=\linewidth]{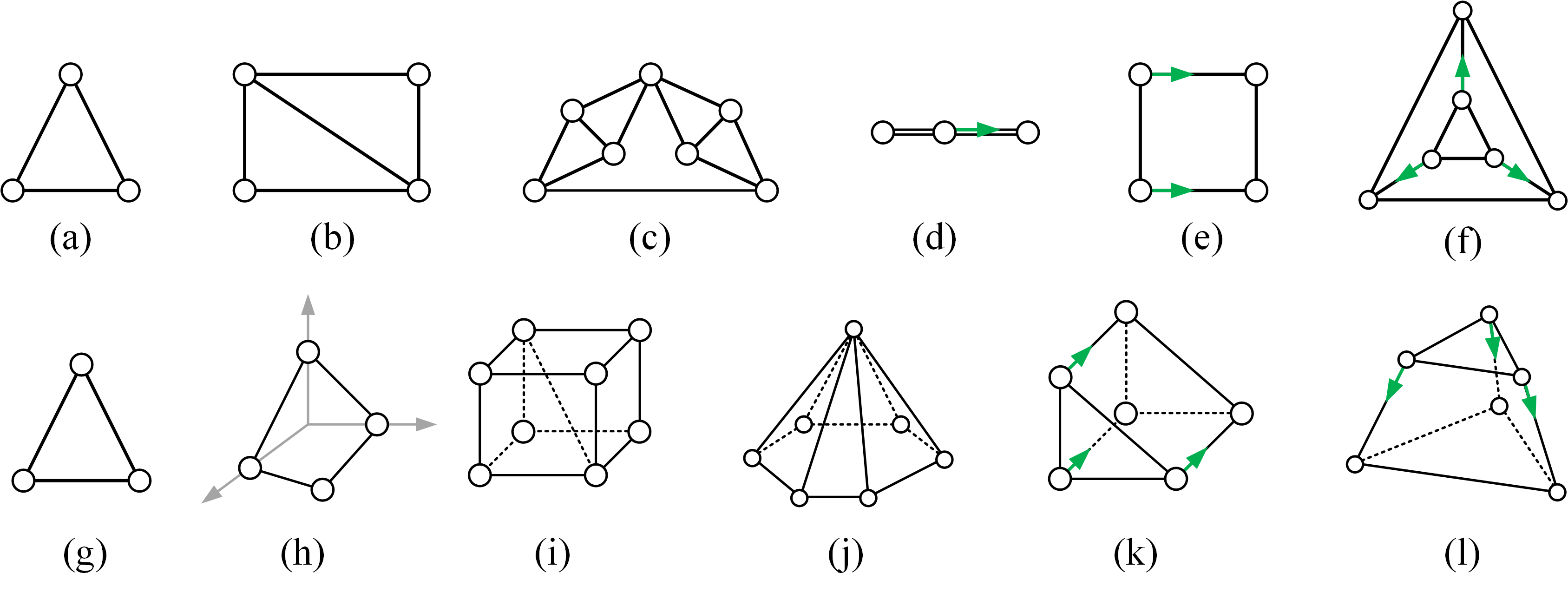}
    \caption{Ví dụ về tính cứng hướng vi phân: Trong $\mb{R}^2$, các đội hình (a), (b), (c) là cứng hướng vi phân, các đội hình (d), (e), (f) là không cứng hướng vi phân. Trong $\mb{R}^3$, các đội hình (g), (h), (i), (j) là cứng hướng vi phân, các đội hình (k), (l) là không cứng hướng vi phân.}
    \label{fig:chap8_IBR_formations}
\end{figure*}

Tính cứng hướng của một đội hình bảo toàn khi tăng số chiều $d$ của không gian. Nếu $(G,\m{p})$ là cứng hướng vi phân trong $\mb{R}^d$ thì $(G,\m{p})$ cũng cứng hướng vi phân trong $\mb{R}^{d+1}$. 

Hơn nữa, không gian sinh bởi các chuyển động vi phân\index{chuyển động vi phân bảo toàn vector hướng} làm thay đổi tỉ lệ của đội hình là  trực giao với không gian các chuyển động vi phân làm quay đội hình. Hệ quả của nhận xét này là trong không gian $d=2$ chiều, một đội hình là cứng hướng vi phân trong $\mb{R}^2$ khi và chỉ khi nó cứng khoảng cách vi phân trong $\mb{R}^2$.

Định nghĩa ma trận cứng hướng hiệu chỉnh\index{ma trận!cứng hướng hiệu chỉnh} $\tilde{\m{R}}_b = ({\rm diag}(\|\m{z}_k\|) \otimes \m{I}_{d}) \m{R}_b= {\rm blkdiag} \big(\m{P}_{\m{g}_k}\big) \bar{\m{H}}$, thì $\text{ker}(\m{R}_b) = \text{ker}(\tilde{\m{R}}_b)$. Ma trận Laplace hướng\index{ma trận!Laplace hướng} được định nghĩa bởi $\mcl{L}_b = \tilde{\m{R}}_b^\top\tilde{\m{R}}_b$ là đối xứng, bán xác định dương, và $\text{ker}(\mcl{L}_b) = \text{ker}(\m{R}_b)$. Các phần tử khối của ma trận trên được cho bởi:
\begin{align}
    [\mcl{L}_b]_{ij} =   \begin{cases}
    -\m{P}_{\m{g}_{ij}}, &  (i,j) \in {E}, \\
    \m{0}_{d\times d}, & (i,j) \notin {E}, \\
    \sum_{j\in {N}_i}\m{P}_{\m{g}_{ij}}, & i = j.
  \end{cases} 
\end{align}

\index{cứng!hướng!vi phân} Một đội hình thỏa mãn tính cứng hướng vi phân \cite{zhao2015tac} thì cũng thỏa mãn tính cứng hướng. Tính cứng hướng là một tính chất phổ quát, tức là nó được quyết định phần lớn bởi cấu trúc đồ thị hơn là bởi cấu hình cụ thể.
\begin{Definition}[Đồ thị cứng hướng phổ quát]\cite{Zhao2017cdc} Đồ thị ${G}$ là cứng hướng phổ quát\footnote{generic bearing rigidity} trong $\mb{R}^d$ nếu tồn tại ít nhất một cấu hình $\m{p} \in \mb{R}^{dn}$ sao cho $({G},\m{p})$ là cứng hướng vi phân. \index{cứng!hướng!phổ quát}
\end{Definition}
Tập các đội hình cứng hướng vi phân là trù mật trong $\mb{R}^d$, tức là nếu một đội hình $(G,\m{p})$ là cứng hướng vi phân thì ta luôn tìm được một đội hình $(G,\m{q})$ cứng hướng vi phân trong mọi lân cận $B_\epsilon=\{\m{p}' \in \mb{R}^{dn}|~\|\m{p}' - \m{p}\| < \epsilon\},$ với mọi $\epsilon>0$. Đồ thị ${G}$ là không cứng hướng (phổ quát) nếu như nó không là cứng hướng phổ quát, nói cách khác $({G},\m{p})$ không cứng hướng vi phân với mọi cấu hình $\m{p}$ của ${G}$ trong $\mb{R}^d$. 

Do tính cứng hướng vi phân là bảo toàn khi tăng số chiều của không gian nên tính cứng hướng phổ quát cũng bảo toàn khi tăng số chiều của không gian. Thật vậy, đồ thị chu trình $C_3$ là cứng hướng phổ quát trong $\mb{R}^2$ nên cũng cứng hướng phổ quát trong $\mb{R}^d$, với $d\ge 3$. Trong khi đó, đồ thị $C_4$ là không cứng hướng trong $\mb{R}^2$ nhưng là cứng hướng phổ quát trong $\mb{R}^3$. Do đó, $C_4$ cũng cứng hướng phổ quát trong $\mb{R}^d$, với $d \geq 3$.

Giả sử ${G}$ là một đồ thị gồm $n$ đỉnh và $m$ cạnh. Tương tự với lý thuyết cứng về khoảng cách, ta định nghĩa đồ thị cứng hướng tối thiểu như sau. 
\begin{Definition}[Đồ thị cứng hướng tối thiểu] Đồ thị ${G}$ là cứng hướng tối thiểu trong $\mb{R}^d$ khi và chỉ khi không tồn tại đồ thị $\mc{H}$ nào gồm $n$ đỉnh mà $\mc{H}$ là cứng hướng phổ phát trên $\mb{R}^d$ và có ít hơn $m$ cạnh. Với định nghĩa này, có thể chứng minh rằng đồ thị ${C}_n$ là cứng hướng phổ quát tối thiểu trong $\mb{R}^d$ khi và chỉ khi $n\le d+1$.
\end{Definition}

\index{đồ thị!1-thừa cứng hướng}
Đồ thị ${G}$ gồm $m$ cạnh là 1-thừa cứng hướng trong $\mb{R}^d$ khi và chỉ khi mỗi đồ thị con bao trùm của ${G}$ với $m-1$ cạnh cũng là đồ thị cứng phổ quát. Một số phương pháp xây dựng các đồ thị cứng hướng tối thiểu\footnote{minimal bearing rigidity} \index{đồ thị!cứng hướng tối thiểu} và đồ thị 1-thừa cứng hướng được trình bày tại \cite{Trinh2020Minimal}. Ý tưởng chung của phương pháp xây dựng đồ thị cứng hướng phổ quát tối thiểu với $n$ đỉnh trong $\mathbb{R}^d$ xuất phát từ toán tử cộng đỉnh trong xây dựng Henneberg. Tại mỗi bước, một chuỗi các đỉnh và cạnh\footnote{Thuật ngữ ``open ear'' được dùng ở \cite{Whitney1932}} với độ dài không quá $d$ được thêm vào hai đỉnh phân biệt của đồ thị cứng hướng phổ quát hiện tại. Phương pháp xây dựng đồ thị cứng hướng phổ quát này được mô tả ở Thuật toán~\ref{alg:MIBR}. Số cạnh tối thiểu được cho bởi hàm số $f(n,d)$ như sau: 
\begin{align}
f(n,d) = \left\lbrace\begin{array}{*{20}{l}}
n, & \text{ nếu } n\leq d+1,\\
f(n-d+1,d) + d, & \text{ nếu } n>d+1.
\end{array}\right.
\end{align}

\begin{algorithm}[t!]
\caption{Xây dựng một đồ thị cứng phổ quát tối thiểu gồm  $n~(n\geq 3)$ đỉnh trong $\mb{R}^d~(d\geq 2)$}
\label{alg:MIBR}
\begin{algorithmic}[1]
\REQUIRE{$d\geq 2$: số chiều, $n\geq 3$: số đỉnh trong đồ thị} 
\IF{$n \leq d+1$}
   \STATE ${G} \leftarrow \mc{C}_{n}$;
\ELSE
	\STATE $k \leftarrow 0$;
	\STATE ${G}[k] \leftarrow \mc{C}_{d+1}$; 
    \STATE $q\leftarrow (n-d-1)$;
\WHILE{$q \geq d-1$}
	\STATE $k \leftarrow k+1$;
    \STATE ${G}[k] \leftarrow$ Cộng một vòng hở gồm $d-1$ đỉnh và $d$ cạnh vào hai đỉnh phân biệt của ${G}$; 
    \STATE $q\leftarrow (q-d+1)$;
\ENDWHILE
\IF{$q > 0$}
   \STATE ${G} \leftarrow$ Cộng một vòng hở gồm $q$ đỉnh trong và $q+1$ cạnh vào hai đỉnh phân biệt của ${G}[k]$;
\ELSE
   \STATE ${G} \leftarrow {G}[k]$;
\ENDIF
\ENDIF
\RETURN ${G}$;
\end{algorithmic}
\end{algorithm}
\begin{figure*}[t!]
    \centering
    \includegraphics[width=0.7\linewidth]{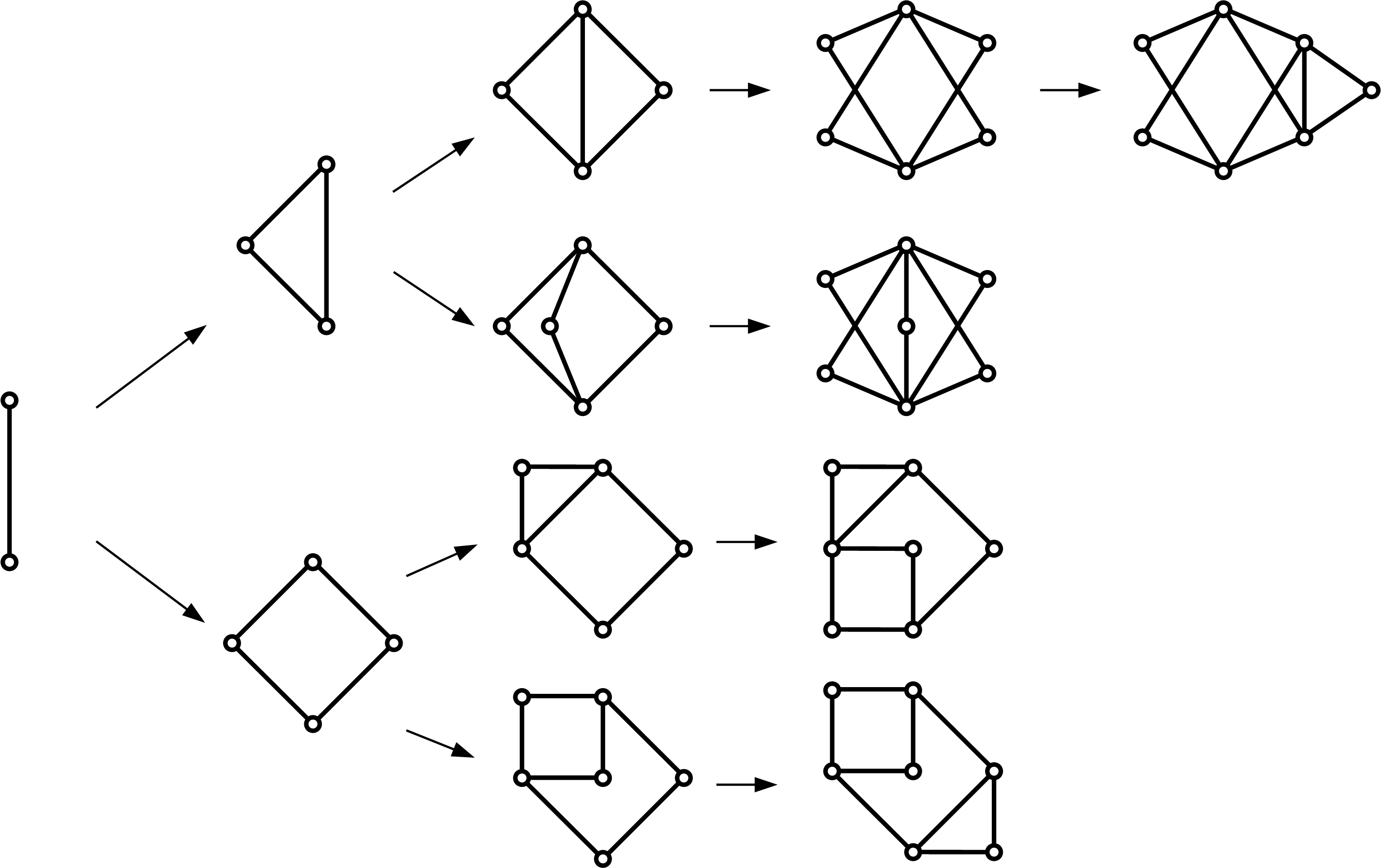}
    \caption{Xây dựng đồ thị cứng hướng phổ quát trong $\mb{R}^3$ xuất phát từ một cạnh nối hai đỉnh.}
    \label{fig:DoThiCungHuong1}
\end{figure*}
\begin{example}
Hình \ref{fig:DoThiCungHuong1} mô tả một ví dụ xây dựng một số đồ thị cứng hướng phổ quát khác nhau trong $\mb{R}^3$ xuất phát từ hai đỉnh và một cạnh. Hình \ref{fig:DoThiCungHuong2} mô tả quá trình xây dựng một đồ thị cứng hướng phổ quát tối thiểu trong $\mb{R}^3$ xuất phát từ một chu trình $C_4$.
\end{example}

\begin{figure*}[t]
    \centering
    \includegraphics[width=0.9\linewidth]{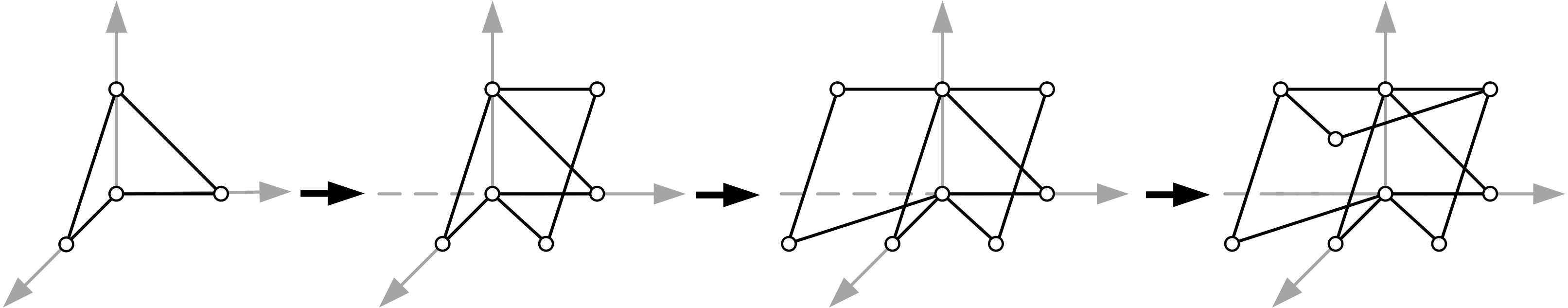}
    \caption{Xây dựng đồ thị cứng hướng phổ quát trong $\mb{R}^3$ xuất phát từ chu trình $C_4$.}
    \label{fig:DoThiCungHuong2}
\end{figure*}

\subsection{Luật điều khiển đội hình sử dụng vector hướng}
Xét đội hình gồm $n$ tác tử tích phân bậc nhất $\dot{\m{p}}_i = \m{u}_i$, $i=1,\ldots, n$, $\m{p}_i \in \mb{R}^d$, $\m{u}_i \in \mb{R}^d$ lần lượt là vị trí và tín hiệu điều khiển của tác tử thứ $i$. Mỗi tác tử $i$ có thể đo được vector hướng $\m{g}_{ij}, \forall j \in {N}_i$ và có thông tin về các vector hướng đặt $\m{g}_{ij}^*, j\in {N}_i$. Tập vector hướng mong muốn được giả thiết là khả thi, nghĩa là tồn tại một cấu hình $\m{p}^* \in \mb{R}^{dn}$ sao cho $\m{g}_{ij}^* = \frac{\m{p}_j^* - \m{p}_i^*}{\|\m{p}_j^* - \m{p}_i^*\|}$ với mọi $\m{g}_{ij}^* \in \Gamma$. Hơn nữa, đội hình đặt $(G,\m{p}^*)$ là cứng hướng vi phân và không có hai điểm nào trùng nhau. \index{điều khiển đội hình!chỉ sử dụng vector hướng}

Xét luật điều khiển chỉ sử dụng vector hướng sau đây \cite{zhao2015tac}:
\begin{align} \label{eq:chap5_bearing_only_control_law}
    \m{u}_i = - \sum_{j\in {N}_i} \m{P}_{\m{g}_{ij}} \m{g}_{ij}^*,~i = 1, \ldots, n,
\end{align}
trong đó $\m{P}_{\m{g}_{ij}} = \m{I}_d - \m{g}_{ij}\m{g}_{ij}^\top \in \mb{R}^{d\times d}$ là ma trận chiếu trực giao tính được từ vector hướng $\m{g}_{ij}$. Minh họa hình học của thuật toán này được cho trên Hình~\ref{fig:chap8_bearing_formation}(a). Với luật điều khiển \eqref{eq:chap5_bearing_only_control_law}, phương trình viết cho cả hệ có dạng
\begin{align} \label{eq:chap5_bearing_only_control_law1}
    \dot{\m{p}} = \bar{\m{H}}^\top \text{blkdiag}\left( {\m{P}_{\m{g}_k}}\right) \m{g}^* = \tilde{\m{R}}_b^\top \m{g}^*,
\end{align} 
trong đó $\m{p} = [\m{p}_1^\top, \ldots, \m{p}_n^\top]^\top \in \mb{R}^{dn}$ và $\m{g}^* = [(\m{g}_1^{*})^\top, \ldots, (\m{g}_m^{*})^\top]^\top = [\ldots, (\m{g}_{ij}^{*})^\top,\ldots]^\top \in \mb{R}^{dm}$. 

Định nghĩa $\bar{\m{p}} = \frac{1}{n}\sum_{j=1}^n \m{p}_i = \frac{1}{n} (\m{1}_n^\top \otimes \m{I}_d) \m{p}$ và $s = \frac{1}{\sqrt{n}}\|\m{p} - \m{1}_n \otimes \bar{\m{p}}\|$ lần lượt là trọng tâm và kích thước của đội hình. Do
\begin{align}
    \dot{\bar{\m{p}}} &= \frac{1}{n} (\m{1}_n^\top \otimes \m{I}_d) \bar{\m{H}}^\top \text{blkdiag}\left( \frac{\m{P}_{\m{g}_k}}{\|\m{z}_k\|} \right) \m{g}^* \nonumber\\
     &= \frac{1}{n} (\m{1}_n^\top \m{H}^\top \otimes \m{I}_d) \text{blkdiag}\left( \frac{\m{P}_{\m{g}_k}}{\|\m{z}_k\|} \right) \m{g}^*\nonumber\\
     &= \m{0}_d,\\
     \dot{s} &= \frac{(\m{p} - \m{1}_n \otimes \bar{\m{p}})^\top}{\sqrt{n}\|\m{p} - \m{1}_n \otimes \bar{\m{p}}\|} \tilde{\m{R}}_b^\top \m{g}^*= {0},
\end{align}
trọng tâm và kích thước của đội hình là không thay đổi theo thời gian.

Các điểm cân bằng của hệ \eqref{eq:chap5_bearing_only_control_law} thỏa mãn $\dot{\m{p}} =\m{0}$, hay
\begin{align*}
    0 = (\m{p}^{*})^\top \dot{\m{p}} &=(\m{p}^{*})^\top\bar{\m{H}}^\top \text{blkdiag}\left( {\m{P}_{\m{g}_k}}\right) \m{g}^* = (\m{z}^{*})^\top\text{blkdiag}\left( {\m{P}_{\m{g}_k}}\right) \m{g}^*\\
    &= \sum_{i=1}^m (\m{z}^{*}_k)^\top \m{P}_{\m{g}_k} \m{g}^*_k= \sum_{i=1}^m \|\m{z}_{k}^*\| (\m{g}^{*}_k)^\top \m{P}_{\m{g}_k} \m{g}^*_k
\end{align*}
Như vậy, tại điểm cân bằng, $\m{P}_{\m{g}_k} \m{g}^*_k = \m{0}, \forall k=1, \ldots, m$. Điều này chứng tỏ rằng điểm cân bằng là tương đương về hướng với đội hình đặt $\m{p}^*$. Mặt khác, từ giả thiết về tính cứng hướng vi phân của $(G,\m{p}^*)$, ta suy ra các điểm cân bằng đều tương đồng về hướng với đồ thị ban đầu. Kết hợp thêm điều kiện về trọng tâm $\bar{\m{p}}^* = \bar{\m{p}}(0)$ và kích thước $s^* = s(0)$, ta suy ra ứng với một điều kiện đầu, tồn tại hai điểm cân bằng: $\m{p}^*$ ứng với các vector hướng $\m{g}_{ij} = \m{g}_{ij}^*, \forall i, j \in V$; và  $\m{p}'$ ứng với các vector hướng $\m{g}_{ij} =- \m{g}_{ij}^*, \forall i, j \in V, i\ne j$ (xem Hình ~\ref{fig:chap8_bearing_formation}(b)).

Đặt $\bmm{\delta} = \m{p} - \m{p}^*$ và $\m{r} = \m{p} - \m{1}_n \otimes \bar{\m{p}}$. Khi đó, từ tính chất bảo toàn trọng tâm và tỉ lệ, ta có $\m{r} - \m{r}^* = \m{p} - \m{p}^*$ và  $\|\bmm{\delta} + \m{r}^*\| = \|\m{r}\|= \sqrt{n} s$ là hằng số. Như vậy, $\bmm{\delta}$ luôn nằm trên tập bất biến $\mc{S}=\{\bmm{\delta}\in \mb{R}^{dn}|~\|\bmm{\delta} + \m{r}^*\| = \sqrt{n} s(0)\}$ được biểu diễn trên Hình~\ref{fig:chap8_bearing_formation}(c).

\begin{figure}
    \centering
    \subfloat[]{\includegraphics[height=3.5cm]{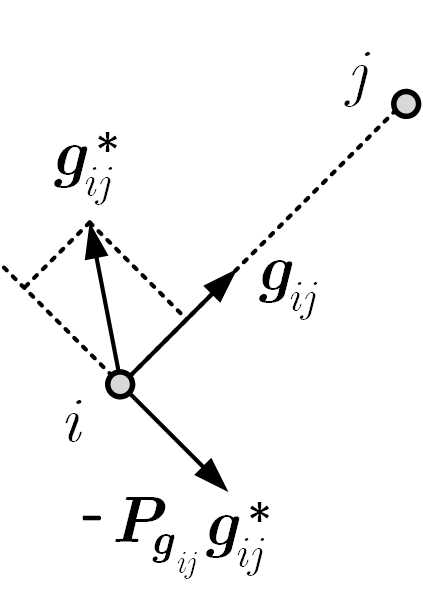}} \hfill
    \subfloat[]{\includegraphics[height=3cm]{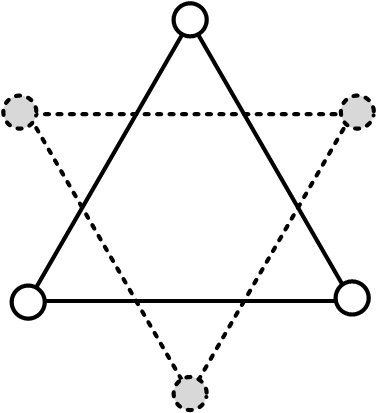}} \hfill
    \subfloat[]{\includegraphics[height=4cm]{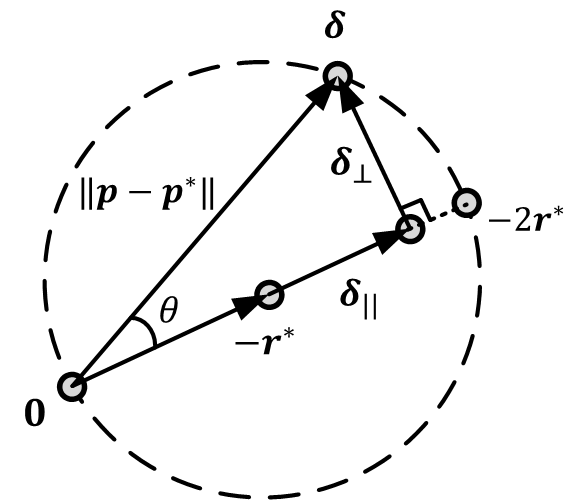}}
    \caption{Minh họa phân tích ổn định thuật toán điều khiển đội hình chỉ dựa trên vector hướng: (a) Luật điều khiển chỉ sử dụng vector hướng; (b) Ví dụ về hai điểm cân bằng đối xứng tâm và có cùng trọng tâm; (c) $\bmm{\delta}$ luôn nằm trong tập $\mc{S}$.}
    \label{fig:chap8_bearing_formation}
\end{figure}

Ta sẽ chứng minh điểm cân bằng $\m{p}=\m{p}'$ là không ổn định. Xét hàm Lyapunov $V = \frac{1}{2}\|\m{p} - \m{p}'\|^2$. Đạo hàm của $V$ dọc theo một quĩ đạo của \eqref{eq:chap5_bearing_only_control_law1} được cho bởi:
\begin{align}
    \dot{V}_1 &= (\m{p} - \m{p}')^\top \bar{\m{H}}^\top \text{blkdiag}\left( {\m{P}_{\m{g}_k}}\right) \m{g}^* \nonumber\\
    &= (\m{z}^{*})^\top\text{blkdiag}\left( {\m{P}_{\m{g}_k}}\right) \m{g}^* \nonumber\\
    &= \sum_{i=1}^m (\m{z}^{*}_k)^\top \m{P}_{\m{g}_k} \m{g}^*_k \nonumber\\
    &= \sum_{i=1}^m \|\m{z}_{k}^*\| (\m{g}^{*}_k)^\top \m{P}_{\m{g}_k} \m{g}^*_k \ge 0. \label{eq:c5_bearing_dotV1}
\end{align}
Từ phương trình \eqref{eq:c5_bearing_dotV1} suy ra $\dot{V}_1=0$ khi và chỉ khi $\m{p} = \m{p}^*$ hoặc $\m{p} = \m{p}'$. Chú ý rằng hai điểm cân bằng là rời nhau, nên khi xét một lân cận bất kỳ của $\m{p}'$ không chứa $\m{p}'$ trong $\mc{S}$, ta có $V_1>0$. Theo định lý Chetaev \cite{Khalil2002}, điểm cân bằng $\m{p}'$ là không ổn định.

Tiếp theo, ta chứng minh điểm cân bằng $\m{p}=\m{p}^*$ là ổn định. Xét hàm Lyapunov $V = \frac{1}{2} \|\bmm{\delta}\|^2$. Dọc theo một quĩ đạo của \eqref{eq:chap5_bearing_only_control_law1}, ta có
\begin{align}
    \dot{V} &= (\m{p} - \m{p}^*)^\top \bar{\m{H}}^\top \text{blkdiag}\left( {\m{P}_{\m{g}_k}}\right) \m{g}^* \nonumber\\
    &= (\m{z}^{*})^\top\text{blkdiag}\left( {\m{P}_{\m{g}_k}}\right) \m{g}^* \nonumber\\
    &= -\sum_{i=1}^m (\m{z}^{*}_k)^\top \m{P}_{\m{g}_k} \m{g}^*_k \nonumber\\
    &= -\sum_{i=1}^m \|\m{z}_{k}^*\| (\m{g}^{*}_k)^\top \m{P}_{\m{g}_k} \m{g}^*_k. \label{eq:chap5_bearing_only_control_law2}
\end{align}
Chú ý rằng $\m{g}^{*\top}_k \m{P}_{\m{g}_k} \m{g}^*_k = \m{g}^{*\top}_k (\m{I}_d - \m{g}_k \m{g}_k^\top) \m{g}^*_k = 1 - (\m{g}^{*\top}_k\m{g}_k)^2 = \m{g}_k^\top (\m{I}_d -\m{g}^{*}_k\m{g}^{*\top}_k)\m{g}_k$, nên từ bất phương trình  \eqref{eq:chap5_bearing_only_control_law2} suy ra
\begin{align}
    \dot{V} &\le -\sum_{i=1}^m \|\m{z}_{k}^*\| \m{g}^{\top}_k \m{P}_{\m{g}_k^*} \m{g}_k \nonumber\\
    &\le -\sum_{i=1}^m \frac{\|\m{z}_{k}^*\|}{\|\m{z}_{k}\|^2} \m{z}^{\top}_k \m{P}_{\m{g}_k^*} \m{z}_k \nonumber\\
    &\le - \frac{\min_k\|\m{z}_{k}^*\|}{\max_k\|\m{z}_{k}\|^2} \sum_{i=1}^m \m{z}^{\top}_k \m{P}_{\m{g}_k^*} \m{z}_k.
\end{align}
Với mọi $i\ne j$, ta có
\begin{align*}
     \|\m{z}_{ij}\| &=\|\m{p}_i - \m{p}_j\|\le \|\m{p}_i-\bar{\m{p}}\| + \|\m{p}_j-\bar{\m{p}}\| \\
     &\leq  \|\m{p}_1-\bar{\m{p}}\| + \ldots + \|\m{p}_n-\bar{\m{p}}\| \\
     &\leq \|\m{p} - \m{1}_n\otimes \bar{\m{p}}\| \\
     & = \sqrt{n} s,
\end{align*}
nên $\max_k \|\m{z}_k\| \le \sqrt{n} s$. Từ đây, đặt $\zeta = \frac{\min_k\|\m{z}_{k}^*\|}{n s^2}$ thì
 \begin{align} \label{eq:chap5_bearing_only_control_law3}
     \dot{V} &\le -\zeta \bmm{\delta}^\top\bar{\m{H}}^\top \text{blkdiag}\left( {\m{P}_{\m{g}_k^*}}\right)\bar{\m{H}}\bmm{\delta} \nonumber\\
     &= - \bmm{\delta}^\top \tilde{\m{R}}_b^\top(\m{p}^*) \tilde{\m{R}}_b(\m{p}^*) \bmm{\delta}.
 \end{align}
Với $\bmm{\delta} = \bmm{\delta}_{\parallel} +  \bmm{\delta}_{\perp}$, trong đó $\bmm{\delta}_{\perp} \perp \text{im}(\m{1}_n \otimes \m{I}_d, \m{p}^*)$ và $\bmm{\delta}_{\parallel} \in \text{im}(\m{1}_n \otimes \m{I}_d, \m{p}^*)$ thì \eqref{eq:chap5_bearing_only_control_law3} có thể viết lại thành 
\begin{align*}
    \dot{V} \le -\zeta \|\bmm{\delta}_{\perp}\|^2 = -\zeta (\sin\theta)^2 \|\bmm{\delta}\|^2,
\end{align*}
trong đó $\theta \in [0,\frac{\pi}{2})$ là góc giữa $\bmm{\delta}$ và $-\m{r}^*$. Do $\dot{V}\le 0$ nên $\bmm{\delta} \to \m{0}$ khi $t\to +\infty$ nếu $\bmm{\delta}(0) \ne -2\m{r}^*$ (tương ứng với $\m{p}(0) \ne \m{p}'$). Điều này dẫn tới $\theta(t) \ge \theta(0)$ và $\sin\theta(t) \ge \sin\theta(0)$. Như vậy, đặt $\xi = 2\zeta (\sin\theta(0))^2>0$, thì $\dot{V} \le -\xi V$, hay $\bmm{\delta} = \m{0}$ là ổn định theo hàm mũ \cite{Khalil2002}. 

Như vậy, điểm cân bằng $\m{p}=\m{p}^*$ là ổn định tiệm cận theo hàm mũ. Với mọi điều kiện đầu $\m{p}(0) \ne \m{p}'$ thì $\m{p}(t) \to \m{p}^*$, khi $t \to +\infty$.

Lưu ý rằng trong phân tích ổn định, chúng ta ngầm sử dụng giả thiết các tác tử không va chạm trong quá trình tạo đội hình đặt. Để tránh va chạm, tác tử cần thêm thông tin về khoảng cách. Trong nhiều tài liệu về điều khiển dựa trên vector hướng, các tác giả giới hạn vị trí ban đầu của các tác tử so với đội hình đặt như một điều kiện đủ để không có va chạm xảy ra.

\begin{example} \label{VD5.7}
Để minh họa thuật toán điều khiển đội hình chỉ dựa trên vector hướng, ta mô phỏng đội hình gồm 20 tác tử như trên Hình~\ref{fig:sim_bearing_formation}. Cấu hình đặt là các đỉnh của một khối dodecahedron trong không gian ba chiều gồm 39 cạnh. Nhìn chung, thời gian hội tụ của thuật toán chỉ sử dụng vector hướng khá dài do khi góc sai lệch là nhỏ thì tín hiệu điều khiển $\m{u}_i$ cũng gần như bằng 0. Việc tăng tốc luật điều khiển đội hình sử dụng vector hướng có thể thực hiện bằng cách thêm một hệ số khuếch đại $k_p$ trong luật điều khiển (trong mô phỏng $k_p=2$) hoặc biến đổi luật điều khiển để hệ đạt tới cấu hình đặt trong thời gian hữu hạn \cite{Tran2018TCNS,Nguyen2021} (Xem Bài tập ~\ref{Ex:5.12}).
\end{example}

\begin{figure*}[t]
    \centering
    \subfloat[Quá trình tiệm cận tới đội hình đặt]{\includegraphics[width = .45 \linewidth]{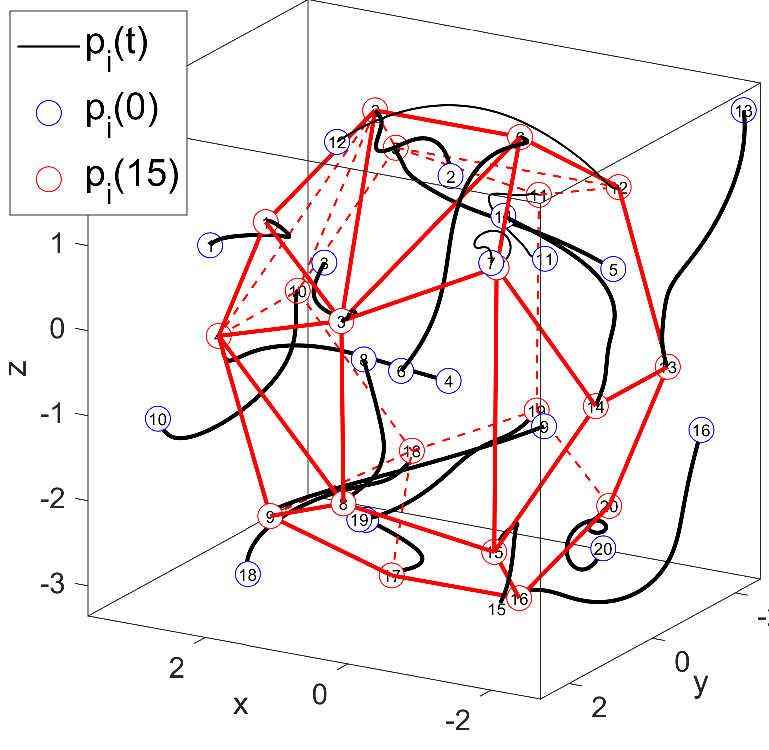}}\hfill 
    \subfloat[Đội hình ban đầu]{\includegraphics[width = .45 \linewidth]{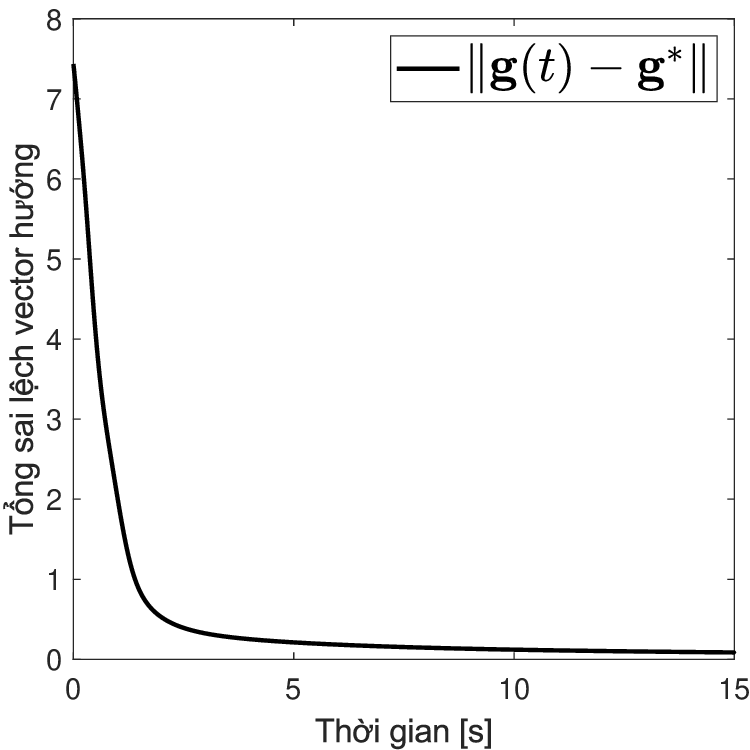}}
    \caption{Mô phỏng đội hình gồm 20 tác tử với luật điều khiển \eqref{eq:chap5_bearing_only_control_law} trong không gian ba chiều. }
    \label{fig:sim_bearing_formation}
\end{figure*}

\section{Ghi chú và tài liệu tham khảo}
Tạo đội hình đặt là bài toán cơ bản trong điều khiển đội hình. Ở chương này, chúng ta đã tìm hiểu một số phương pháp điều khiển đội hình dựa trên các biến đo, biến điều khiển khác nhau, đồng thời giới thiệu những phương pháp đã và đang được nghiên cứu khác. Việc phân chia bài toán điều khiển đội hình dựa trên biến đo ở chương này được mở rộng từ khảo cứu tổng quan \cite{Oh2015}. Nội dung phân tích điều khiển đội hình dựa trên khoảng cách sử dụng phân tích trong \cite{Vu2020LCSS,Vu2023distance}. Với luật điều khiển dựa trên khoảng cách thông thường, đội hình đặt là ổn định tiệm cận địa phương. Lý thuyết cứng hướng và phân tích hội tụ gần như toàn cục của đội hình đặt được trình bày lại từ \cite{zhao2015tac}. Lưu ý rằng điều khiển đội hình ổn định toàn cục có thể tham khảo tại \cite{Trinh2019pointing}.

Trong khuôn khổ tài liệu, chúng ta chỉ xem xét bài toán điều khiển đội hình với đồ thị tương tác vô hướng. Khi hiện thực hóa các đội hình thực tế, đồ thị đo đạc và truyền thông thường bất đối xứng và có tác tử leader. Cụ thể, cấu trúc đồ thị không chu trình có leader được nghiên cứu rộng rãi \cite{Cao2008,Hendrickx2007,Yu2009,Summers2011,Park2014finite,Pham2017ICCAS,Trinh2014CN,Trinh2019TAC}. Nghiên cứu các cấu trúc đội hình hữu hướng khác mang ý nghĩa lý thuyết nhiều hơn thực tế, và ngoại trừ phương pháp sử dụng vị trí tuyệt đối và vị trí tương đối, thường khá phức tạp và chưa có lời giải hoàn toàn. Với điều khiển đội hình dựa trên khoảng cách và vector hướng, khả năng ổn định tiệm cận các đội hình hữu hướng (ngoài dạng leader-follower không chu trình) thường dừng lại ở các đồ thị chu trình hữu hướng \cite{Park2015,Ko2020bearing}. Lưu ý rằng đội hình chu trình hữu hướng kép gồm bốn tác tử đã được chứng minh là không ổn định được toàn cục \cite{Belabbas2013global}. Mặc dù không đề cập trong chương này, bài toán đồng thuận hệ qui chiếu có liên hệ chặt chẽ với điều khiển đội hình, do các luật điều khiển dựa trên vị trí tương đối và vector hướng đều sử dụng giả thiết các hệ qui chiếu riêng của các tác tử đã được định hướng như nhau \cite{Ren2007alignment,Igarashi2009passivity,Thunberg2014aut}.

Một số vấn đề mở rộng trong nghiên cứu điều khiển đội hình đã và đang được quan tâm bao gồm:
\begin{itemize}
\item[i.] Duy trì đội hình đặt dưới ảnh hưởng của bất định và nhiễu \cite{Mou2016TAC,Vu2020LCSS}, hay các ràng buộc thực tế như miền hoạt động của cảm biến\cite{Fabris2022bearing}, trễ và lỗi truyền thông, sai lệch đo đạc và hệ qui chiếu \cite{Hector2014controlling,Meng2016aut}, lỗi cơ cấu chấp hành \cite{Zuo2014adaptive}, \ldots hoặc cải thiện chất lượng điều khiển (thời gian xác lập, độ quá điều chỉnh, độ lớn của tín hiệu điều khiển) \cite{Tran2018TCNS,Chen2020leader}
\item[ii.] Điều động đội hình: di chuyển đội hình từ một vị trí ban đầu tới một vị trí khác, hoặc di chuyển đội hình theo một quĩ đạo đặt trước \cite{Zhao2019bearing,Vu2023distance,Chen2022maneuvering,Nguyen2024TCyber}.
\item[iii.] Điều khiển đội hình với mô hình tuyến tính tổng quát hay mô hình động lực học chi tiết hơn của các tác tử \cite{Dong2014time,Cai2014adaptive,Dong2016time,Li2018bearing,Liu2018distributed,Hu2019distributed,Li2020adaptive,Heshmati2020robust,Tran2023TCNS,Su2024bearing}.
\item[iv.] Lập quĩ đạo đặt để tránh va chạm, duy trì liên kết \cite{Zavlanos2007potential}, duy trì tính cứng \cite{Zelazo2015decentralized}, dựa trên tối ưu hóa phân tán.
\item[v.] Chuyển đổi giữa các đội hình mục tiêu, tự phát hiện và cô lập tác tử gặp sự cố \cite{Shames2011distributed}, tái cấu trúc đội hình khi số tác tử thay đổi,\ldots dựa trên kết hợp của điều khiển và logic hình thức \cite{Qi2023automated}
\item[vi.] Ứng dụng của lý thuyết trò chơi (game theory) \cite{Nguyen2018distributed,Shi2021energy}, học củng cố (reinforcement learning) \cite{Luy2013reinforcement,Pham2022disturbance,Nguyen2024formation}, hay các kĩ thuật học máy và khoa học dữ liệu \cite{Hyeon2025data,Sun2023NatureComm} vào điều khiển đội hình,\ldots
\end{itemize}

\section{Bài tập}
\begin{exercise} \cite{Dimarogonas2008connection} \label{exercise:FC1}
Xét bài toán điều khiển đội hình dựa trên vị trí tương đối trong đó tập các vector sai lệch vị trí đặt ${\Gamma} = \{\m{z}_{ij}^*\}_{(i,j) \in E}$ là không khả thi. Chứng minh rằng khi đó, với luật điều khiển đội hình: 
\begin{equation}
\dot{\m{p}}_i = \sum_{j \in N_i} (\m{p}_j - \m{p}_i - \m{z}_{ij}^*),~i=1, \ldots, n,
\end{equation}
 thì vận tốc của các tác tử dần hội tụ tới một giá trị chung xác định bởi $\frac{1}{n}\sum_{i=1}^n \sum_{j \in N_i} \m{z}_{ij}^*$.
\end{exercise}
\begin{exercise} \cite{Park2016distance}
Phân tích bài toán điều khiển đội hình cứng khoảng cách vi phân với đồ thị đầy đủ $K_4$ trong $\mb{R}^3$.
\end{exercise}
\begin{exercise}
Trình bày phương pháp mở rộng Henneberg cho đồ thị cứng trên 2D. Áp dụng để xây dựng một đồ thị cứng tối thiểu gồm 8 đỉnh. Chứng minh tính cứng hướng được bảo toàn khi ta giữ nguyên đội hình và tăng số chiều của không gian.
\end{exercise}
\begin{exercise}
So sánh các phương pháp điều khiển đội hình dựa trên khoảng cách và dựa trên vector hướng về các giả thiết chính, luật điều khiển, ưu nhược điểm.
\end{exercise}
\begin{exercise}[Tính chất của các ma trận chiếu trực giao] \cite{Matsaglia1974}
Với $\m{x} \neq \m{0}_d$, định nghĩa ma trận chiếu trực giao $\m{P}_{\m{x}} = \m{I}_d - \frac{\m{x}\m{x}^\top}{ \|\m{x}\|^2}$. Chứng minh rằng:
\begin{enumerate}
    \item[i.] $\m{P}_{\m{x}} = \m{P}_{\m{x}}^\top \ge 0$.
    \item[ii.] $\m{P}_{\m{x}}$ có $n-1$ giá trị riêng bằng 1 và một giá trị riêng bằng 0. 
    \item[iii.] ker$(\m{P}_{\m{x}}) = \text{im}(\m{1}_n)$.
    \item[iv.] $\m{P}_{\m{x}} \m{y} = \m{0}_d$ khi và chỉ khi $\m{y} = k \m{x}$, $k\in \mb{R}$.
    \item[v.] Nếu $\m{x} \neq k \m{y}$ (không cùng phương) thì $\m{M} = \m{P}_{\m{x}} + \m{P}_{\m{x}}$ là một ma trận đối xứng, xác định dương. Tìm các giá trị riêng của $\m{M}$.
    \item[vi.] Nếu tồn tại ít nhất hai trong số các vector $\m{x}_1, \ldots, \m{x}_m$ không cùng phương thì ma trận $\m{M} = \sum_{i=1}^m \m{P}_{\m{x}_i}$ là đối xứng và xác định dương.
    \item[vii.] Với vector hướng $\m{g}_{ij} = \frac{\m{p}_j - \m{p}_i}{\|\m{p}_j - \m{p}_i\|}$ thì $\frac{\partial \m{g}_{ij}}{\partial \m{p}_i} = -\frac{\m{P}_{\m{g}_{ij}}}{\|\m{p}_i - \m{p}_j\|}$.
\end{enumerate}
\end{exercise}
\begin{exercise}[Tổng của hai ma trận chiếu trực giao]
Cho hai vector hướng không cùng phương $\m{g}_1$ và $\m{g}_2 \in \mb{R}^d$, $d\geq 3$. Chứng mình rằng ma trận $\m{M} = \m{P}_{\m{g}_1} + \m{P}_{\m{g}_2}$ là đối xứng, xác định dương. Tìm các giá trị riêng và vector riêng tương ứng của $\m{M}$. (Gợi ý: chứng minh $\m{M}$ nhận $\m{P}_{\m{g}_1} \m{g}_2 \pm \m{P}_{\m{g}_2} \m{g}_1$ là các vector riêng tương ứng với các giá trị riêng $1 \pm \m{g}_1^\top \m{g}_2$, và các vector riêng $\m{v}_k \perp \text{span}(\m{g}_1, \m{g}_2)$, $k=3, \ldots, d$, tương ứng với các giá trị riêng 2. 
\end{exercise}
\begin{exercise} \label{chap5_exercise:henneberg}
Xác định tính cứng khoảng cách phổ quát của các đồ thị $G_1$ và $G_2$  như trên Hình~\ref{fig:chap5_henneberg}. Tìm xây dựng Henneberg để xây dựng các đồ thị đó.
\end{exercise}
\begin{figure}[h]
    \centering
    \subfloat[]{\includegraphics[height=3cm]{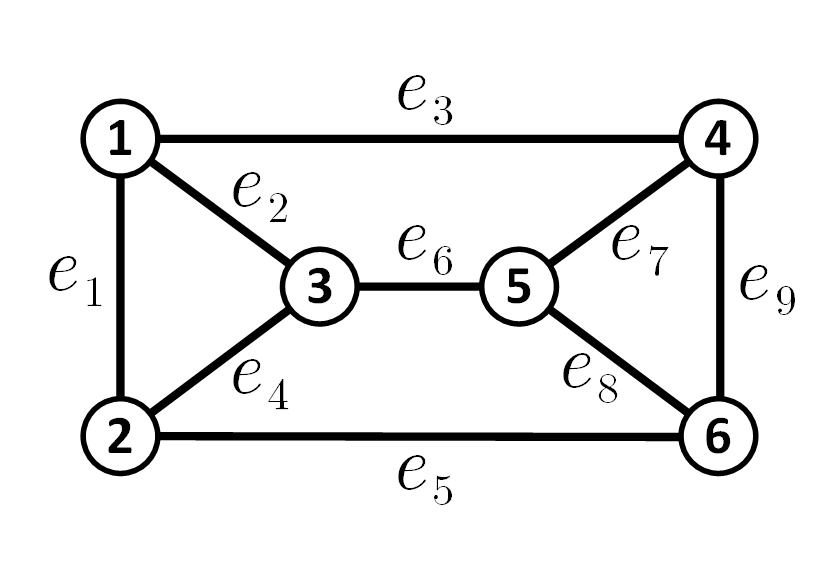}}
    \qquad\qquad\qquad
    \subfloat[]{\includegraphics[height=4.15cm]{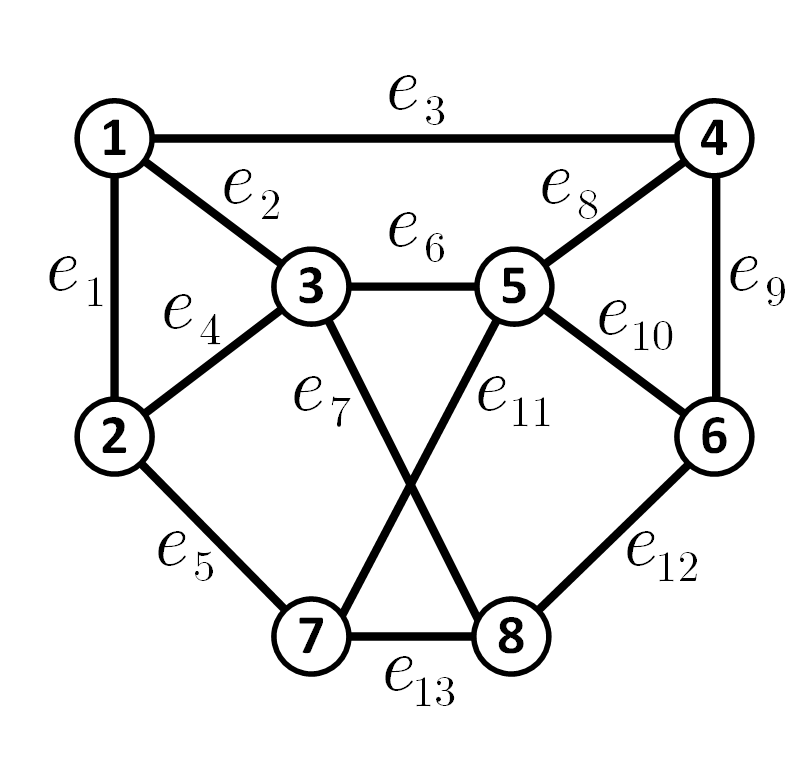}}
    \caption{Các đồ thị $G_1$ và $G_2$.}
    \label{fig:chap5_henneberg}
\end{figure}
\begin{exercise} \cite{Zhao2017cdc}
Giả sử đồ thị $G=(V,E)$ là cứng hướng phổ quát trên $\mb{R}^d$ thì khi thêm vào $G$ một số cạnh bất kỳ để được đồ thị $G'$ thì đồ thị mới này cũng cứng hướng phổ quát.
\end{exercise}
\begin{exercise}[Bảo toàn tính cứng hướng khi mở rộng số chiều] \cite{zhao2015tac}
Giả sử đội hình $(G, \m{p})$ là cứng hướng trên $\mb{R}^d$ thì đội hình $(G, \m{p}')$ với $\m{p}' = [\m{p}^\top,0]^\top \in \mb{R}^{d+1}$ cũng cứng hướng vi phân trên $\mb{R}^{d+1}$.
\end{exercise}
\begin{exercise} \cite{Trinh2020Minimal}
 Giả sử hai đồ thị $G_1$ và $G_2$ là cứng hướng phổ quát trên $\mb{R}^d$. Chứng minh rằng:
 \begin{itemize}
     \item[i.] Khi $d = 2$, ta cần thêm ít nhất 3 cạnh để ghép hai đồ thị $G_1$, $G_2$ thành một đồ thị cứng hướng phổ quát.
     \item[ii.] Nếu $d\geq 3$, ta cần thêm ít nhất 2 cạnh để ghép hai đồ thị $G_1$, $G_2$ thành một đồ thị cứng hướng phổ quát.
 \end{itemize}
\end{exercise}
\begin{exercise}[Đồ thị chu trình cứng hướng vi phân] \cite{Trinh2020Minimal}
Chứng minh đồ thị $C_{n}$ với $n\le d+1$ là cứng hướng phổ quát trong $\mb{R}^d$.
\end{exercise}
\begin{exercise}[Tương đương giữa cứng khoảng cách và cứng hướng vi phân trong $\mb{R}^2$] \cite{zhao2015tac}
Chứng minh nếu đội hình $(G,\m{p})$ là cứng khoảng cách vi phân trong $\mb{R}^2$ thì nó cũng cứng hướng vi phân trong $\mb{R}^2$.
\end{exercise}
\begin{exercise} \cite{Trinh2020Minimal}
 Chứng minh rằng một đồ thị $G=(V,E)$ với $|V|=n$ là cứng hướng trên $\mb{R}^d$ thì có tối thiểu là 
 \begin{equation}
     f(n,d)=1+ \left\lfloor \frac{n-2}{d-1}  \right \rfloor \times d + \text{mod}(n-2,d-1) + \text{sign}(\text{mod}(n-2,d-1))
 \end{equation}
 cạnh.
\end{exercise}
\begin{exercise}[Không gian nhân của ma trận cứng khoảng cách] Hãy mô tả không gian nhân của ma trận cứng khoảng cách của một đội hình cứng khoảng cách vi phân trong $\mb{R}^3$.
\end{exercise}
\begin{exercise}[Điều khiển đội hình sử dụng vị trí tương đối kháng nhiễu] Xét bài toán điều khiển đội hình trong không gian $\mb{R}^d$ trong đó $G$ là đồ thị vô hướng, liên thông. 
\begin{itemize}
\item[i.] Giả sử các tác tử di chuyển theo luật điều khiển:
\[\dot{\m{p}}_i = -\sum_{j\in {N}_i} a_{ij} (\m{p}_i - \m{p}_j - \m{p}_{ij}^*) + \m{d}_i,~i=1,\ldots,n, \]
trong đó $\m{p}_{ij}^*$ là các vector vị trí tương đối của đội hình đặt và $\m{d}_i \in \mb{R}^d$ là vector nhiễu hằng. Chứng minh rằng với luật điều khiển trên, nếu $(\m{1}_n^\top \otimes \m{I}_d)\m{d}$ $\m{p} \to {\Omega} = \{\m{p} \in \mb{R}^{dn}|~ \m{p}_i - \m{p}_j - \m{p}_{ij}^*) = \m{0}, \forall i \ne j,~i, j \in {V}\}$, khi $t \to +\infty$.
\item[ii.] Giả sử trong đội hình có một vài tác tử đánh số từ 1 tới $l \ge 2$ di chuyển với vận tốc hằng $\m{h} \in \mb{R}^d$, tức là $\dot{\m{p}}_i = \m{h},~\dot{\m{h}}=\m{0}_d, ~i=1, \ldots, l$. Các tác tử này (gọi là các tác tử dẫn đàn) ban đầu đã ở vị trí thỏa mãn các vector hướng đặt. Các tác tử follower $i=l+1,\ldots,n$ di chuyển với luật điều khiển cho bởi:
\begin{align*}
    \dot{\m{p}}_i &= -\sum_{j\in {N}_i} a_{ij} (\m{p}_i - \m{p}_j - \m{p}_{ij}^*) + \bmm{\xi}_i, \\
    \dot{\bmm{\xi}}_i &= -\sum_{j\in {N}_i} a_{ij} (\m{p}_i - \m{p}_j - \m{p}_{ij}^*),~i=l+1, \ldots, n.
\end{align*}
Chứng minh rằng khi $t \to +\infty$ hệ gồm $n$ tác tử thỏa mãn $\m{p} \to {\Omega}$ và ${\dot{\m{p}}}_i \to \m{h}$, $\forall i = l+1, \ldots, n$. 
\item[iii.] Giả sử các tác tử sử dụng luật điều khiển đội hình cho bởi
\[\dot{\m{p}}_i = -\sum_{j\in {N}_i} (k{\rm sgn}(\m{p}_i - \m{p}_j - \m{p}_{ij}^*) + \m{d}_{ij}),~i=1,\ldots,n, \] 
Hãy phân tích hệ trong hai trường hợp: (a) $\|\m{d}_{ij}\|_{\infty}<k, \forall (i,j)\in E$, và (b) tồn tại $(i,j)\in E$ với $\|\m{d}_{ij}\|_{\infty} > k$.
\end{itemize}
\end{exercise}
\begin{exercise} \cite{Cao2007controlling} Xét một hệ gồm ba tác tử, trong đó tác tử số 1 đứng yên, tác tử thứ hai cần đạt được khoảng cách đặt $d_{12}^*$ với tác tử 1, và tác tử số 3 cần đạt được hai khoảng cách đặt $d_{31}^*$ và $d_{32}^*$. Giả sử hai tác tử 2 và 3 sử dụng luật điều khiển đội hình dựa trên khoảng cách, tức là
\begin{align*}
\dot{\m{p}}_1 &= \m{0},\\
\dot{\m{p}}_2 &=-((d_{21})^2-(d_{21}^{*})^2 )(\m{p}_1-\m{p}_2),\\
\dot{\m{p}}_3 &=-((d_{31})^2-(d_{31}^*)^2 )(\m{p}_1-\m{p}_3)-((d_{32} )^2-(d_{32}^*)^2 )(\m{p}_2-\m{p}_3).
\end{align*}
Xác định các tập điểm cân bằng ứng với tác tử 2 và tác tử 3. Chứng minh rằng với luật điều khiển cho như trên, trong hầu hết mọi trường hợp, ba tác tử sẽ đạt được đội hình đặt khi $t\to +\infty$. Xác định tập các điều kiện đầu mà các tác tử sẽ không đạt được đội hình đặt. Kết quả sẽ thay đổi như thế nào nếu cả ba tác tử đều di chuyển, đồ thị tương tác là chu trình hữu hướng $C_3$, và đội hình đặt là tam giác với ba kích thước $d_{12}^*, d_{23}^*,$ và $d_{31}^*$?
\end{exercise}
\begin{exercise} \cite{Trinh2019TAC}
Xét đội hình gồm 3 tác tử trong không gian 2 chiều. Hai tác tử 1 và 2 đứng yên, còn tác tử thứ ba di chuyển theo luật điều khiển dựa trên vector hướng cho bởi
$\dot{\m{p}}_3=-\m{P}_{\m{g}_31} \m{g}_{31}^*-\m{P}_{\m{g}_{32}} \m{g}_{32}^*$ với $\m{g}_{31}^*$ và $\m{g}_{32}^*$ là hai vector hướng đặt cho trước. 
\begin{enumerate}
    \item Tìm điều kiện của $\m{p}_1$ và $\m{p}_2$ để tồn tại điểm cân bằng $\m{p}_3^*$ thỏa mãn cả hai vector hướng cho trước. Khi đó, xác định công thức tìm $\m{p}_3^*$ theo $\m{p}_1$, $\m{p}_2$, $\m{g}_{31}^*$ và  $\m{g}_{32}^*$.
    \item Chứng minh rằng với điều kiện tìm được ở ý 1 thì điểm cân bằng của hệ (nghiệm của $\dot{\m{p}}_3=\m{0}$) là duy nhất và là $\m{p}_3=\m{p}_3^*$. Sử dụng hàm Lyapunov $V=\frac{1}{2} \|\m{p}_3-\m{p}_3^*\|^2$ để chứng minh $\m{p}_3(t)\to \m{p}_3^*$ khi $t\to +\infty$.
    \item Với cùng phương án điều khiển trên, áp dụng cho đồ thị thu được  từ xây dựng Henneberg. Chứng minh tính ổn định tiệm cận của đội hình gồm $n$ tác tử dựa trên lý thuyết ổn định ISS \cite{Khalil2002}.
\end{enumerate}
\end{exercise}
\begin{exercise}[Điều khiển đội hình chỉ sử dụng vector hướng trong thời gian hữu hạn \cite{Tran2018TCNS}] \label{Ex:5.12}Xét bài toán điều khiển đội hình dựa trên vector hướng với $G$ là đồ thị vô hướng, cứng hướng phổ quát theo luật điều khiển:
\[\dot{\m{p}}_i = -\sum_{j\in {N}_i} \m{P}_{\m{g}_{ij}} \big(\text{sig}\big(\m{P}_{\m{g}_{ij}}\m{g}_{ij}^*\big)\big)^\alpha,~i=1,\ldots,n,\]
trong đó hàm $\text{sig}(\cdot)$ được định nghĩa như sau: với $\m{x}=[x_1,\ldots,x_d ]^\top\in \mb{R}^d$ thì 
\[\text{sig}(\m{x})^\alpha=[\text{sign}(x_1 ) |x_1 |^\alpha,\ldots, \text{sign}(x_d )|x_d |^\alpha]^\top,\] với sign$(\cdot)$ là hàm dấu cho bởi $\text{sign}(x_k)=1$ nếu $ x_k>0,$ $\text{sign}(x_k )=-1$, nếu $x_k<0$ và sign$(x_k )=0$ nếu $x_k=0.$
\end{exercise}

\begin{exercise}[Điều động đội hình dựa trên vector hướng \cite{Zhao2015MSC}] \label{Ex:FC14} 
Xét bài toán điều khiển đội hình trong đó $G$ là đồ thị vô hướng, cứng hướng vi phân trong $\mb{R}^d$. 
\begin{itemize}
\item[i.] Giả sử các tác tử di chuyển theo luật điều khiển:
\[\dot{\m{p}}_i = -\sum_{j\in {N}_i} \m{P}_{\m{g}_{ij}^*} (\m{p}_i - \m{p}_j),~i=1,\ldots,n, \]
trong đó $\m{g}_{ij}^*$ là các vector hướng đặt. Chứng minh rằng với luật điều khiển trên, $\m{p} \to {\Omega} = \{\m{p} \in \mb{R}^{dn}|~ \m{P}_{\m{g}_{ij}^*}\m{g}_{ij} = \m{0}, \forall i \ne j,~i, j \in {V}\}$, khi $t \to +\infty$.
\item[ii.] Giả sử trong đội hình có một vài tác tử đánh số từ 1 tới $l \ge 2$ di chuyển với vận tốc hằng $\m{h} \in \mb{R}^d$, tức là $\dot{\m{p}}_i = \m{h},~\dot{\m{h}}=\m{0}_d, ~i=1, \ldots, l$. Các tác tử này (gọi là các tác tử dẫn đàn) ban đầu đã ở vị trí thỏa mãn các vector hướng đặt. Các tác tử còn lại di chuyển với luật điều khiển cho bởi:
\begin{align*}
    \dot{\m{p}}_i &= -\sum_{j\in {N}_i} \m{P}_{\m{g}_{ij}^*} (\m{p}_i - \m{p}_j) + \bmm{\xi}_i, \\
    \dot{\bmm{\xi}}_i &= -\sum_{j\in {N}_i} \m{P}_{\m{g}_{ij}^*} (\m{p}_i - \m{p}_j),~i=l+1, \ldots, n.
\end{align*}
Chứng minh rằng khi $t \to +\infty$ hệ gồm $n$ tác tử dần đạt được $\m{g}_{ij} \to \m{g}_{ij}^*, \forall i, j \in {V}, i\ne j$ và ${\dot{\m{p}}}_i \to \m{h}$, $\forall i = l+1, \ldots, n$. 
\end{itemize}
\end{exercise}

\begin{exercise}[Điều động đội hình chỉ sử dụng vector hướng \cite{Zhao2019bearing}] \label{Ex:FC15} 
Xét bài toán điều khiển đội hình trong đó $G$ là đồ thị vô hướng, cứng hướng vi phân trong $\mb{R}^d$. 
\begin{itemize}
\item[i.] Giả sử các tác tử di chuyển theo luật điều khiển:
\[\dot{\m{p}}_i = \sum_{j\in {N}_i} (\m{g}_{ij} - \m{g}_{ij}^*),~i=1, \ldots, n,\]
trong đó $\m{g}_{ij}^*$ là các vector hướng đặt. Viết lại hệ dưới dạng ma trận và chứng minh rằng với luật điều khiển trên, $\m{p} \to {\Omega} = \{\m{p} \in \mb{R}^{dn}|~\m{P}_{\m{g}_{ij}^*}\m{g}_{ij} = \m{0}, \forall i \ne j,~i, j \in {V}\}$, khi $t \to +\infty$.
\item[ii.] Giả sử trong đội hình có một vài tác tử đánh số từ 1 tới $l \ge 2$ di chuyển với vận tốc hằng $\m{h} \in \mb{R}^d$, tức là $\dot{\m{p}}_i = \m{h},~\dot{\m{h}}=\m{0}_d, ~i=1, \ldots, l$. Các tác tử này (gọi là các tác tử dẫn đàn) ban đầu đã ở vị trí thỏa mãn các vector hướng đặt. Các tác tử còn lại di chuyển với luật điều khiển cho bởi:
\begin{align*}
    \dot{\m{p}}_i &= \sum_{j\in {N}_i} (\m{g}_{ij} - \m{g}_{ij}^*) + \bmm{\xi}_i, \\
    \dot{\bmm{\xi}}_i &= \sum_{j\in N_i} (\m{g}_{ij} - \m{g}_{ij}^*),~i=l+1, \ldots, n.
\end{align*}
Chứng minh rằng khi $t \to +\infty$, hệ gồm $n$ tác tử dần đạt được $\m{g}_{ij} \to \m{g}_{ij}^*, \forall i, j \in {V}, i\ne j$ và ${\dot{\m{p}}}_i \to \m{h}$, $\forall i = l+1, \ldots, n$. 
\end{itemize}
\end{exercise}


%
\chapter{Định vị mạng cảm biến}
\label{chap:network_localization}
\section{Bài toán định vị mạng cảm biến}
\index{định vị mạng cảm biến}
Đối tượng trong chương này là mạng cảm biến không dây có khả năng đo đạc, tính toán và truyền thông. Theo tài liệu \cite{Aspnes2006}, định vị mạng cảm biến là quá trình các tác tử (ở đây là các nút mạng) xác định vị trí của mình trong không gian dựa trên thông tin thu được nhờ đo đạc và trao đổi  với các nút mạng khác. Ở tài liệu này, phương pháp tiếp cận bài toán định vị mạng là dựa trên tính đối ngẫu giữa bài toán định vị mạng và bài toán điều khiển đội hình. Ngoài ra, trong trường hợp biến đo là khoảng cách, phương pháp xây dựng mạng thỏa mãn tính cứng toàn cục cũng như tính định vị được sẽ được trình bày. \index{định vị mạng cảm biến} \index{bài toán đối ngẫu}

Một mạng\footnote{network} được định nghĩa bởi $(G,\m{p})$, trong đó $G=(V,E)$ là một đồ thị vô hướng và $\m{p} = \text{vec}(\m{p}_1, \ldots, \m{p}_n) \in \mb{R}^{dn}$ là vector vị trí tuyệt đối của các nút mạng, hay còn gọi là một cấu hình. Nếu như các nút mạng có thông tin về vị trí tuyệt đối của mình, có thể nhờ cài đặt bên ngoài, hoặc có thể thông qua GPS, bài toán định vị mạng trở nên tầm thường. Do đó, trong chương này, ta sẽ giả sử mọi tác tử không biết vị trí tuyệt đối, hoặc chỉ một vài tác tử trong mạng (gọi là các nút tham chiếu), được đánh số từ $1$ tới $l$ ($0\leq l <n$), là có thông tin về vị trí tuyệt đối của mình. Các tác tử khác trong mạng, đánh số từ $l+1$ tới $n$, lưu biến vị trí ước lượng $\hat{\m{p}}_i\in \mb{R}^d$ và cập nhật biến này dựa trên các biến đo được. Hình \ref{fig:netwkLocwSensors} minh họa một mạng cảm biến gồm ba nút tham chiếu và tám nút thường.\footnote{Trong một số tài liệu, các nút tham chiếu có tên gọi khác là các nút mỏ neo/nút mốc (beacon). Tên gọi này bắt nguồn từ nhận xét rằng khi không có vị trí tham chiếu, cấu hình định vị được sẽ bị sai khác với cấu hình thực và không có phương pháp nào tìm được cấu hình thực này. Hiện tượng này có thể quan sát trong các phương pháp định vị mạng được trình bày ở các phần tiếp theo của chương này, và có tên gọi là hiện tượng trôi. Sự tồn tại các nút tham chiếu trong mạng giúp các nút định vị được vị trí thực trong không gian, loại bỏ hiện tượng trôi. Do đó, các nút tham chiếu có thể coi như những nút mốc/nút mỏ neo (beacon) để cố định cấu hình ước lượng của mạng.} Tương tự như trong bài toán điều khiển đội hình, bài toán định vị mạng có thể được phân chia dựa trên biến đo ở mỗi nút mạng. Ở chương này sẽ giới thiệu các phương pháp định vị mạng dựa trên vị trí tương đối, dựa trên khoảng cách và dựa trên vector hướng.\index{nút tham chiếu}
\begin{SCfigure}[][t!]
    \centering
	\begin{tikzpicture}[
roundnode/.style={circle, draw=black, thick, minimum size=2mm,inner sep= 0.25mm}
]
    \node[roundnode,fill=blue!30] (u1) at (-1,0) { }; %
    \node[roundnode,fill=red] (u2) at (0,0) { };%
    \node[roundnode,fill=blue!30] (u3) at (.5,-2) { };%
    \node[roundnode,fill=red] (u4) at (1,1) { };%
    \node[roundnode,fill=red] (u5) at (2,-1) { };%
    \node[roundnode,fill=blue!30] (u6) at (2,-3) { };%
    \node[roundnode,fill=blue!30] (u7) at (3,1) { };%
    \node[roundnode,fill=blue!30] (u8) at (3.5,-1.5) { };%
    \node[roundnode,fill=blue!30] (u9) at (4,2) { };%
    \node[roundnode,fill=blue!30] (u10) at (5,0) { };%
    \node[roundnode,fill=blue!30] (u11) at (7,-.5) { };%

    \node (u12) at (-1,2) {Nút tham chiếu};
    \node (u13) at (-2,-1.5) {Nút thường};
    \draw [very thick]
    (u1) edge [bend left=0] (u2)
    (u1) edge [bend left=0] (u3)
    (u1) edge [bend left=0] (u4)
    (u2) edge [bend left=0] (u3)
    (u2) edge [bend left=0] (u4)
    (u2) edge [bend left=0] (u5)
    (u3) edge [bend left=0] (u5)
    (u3) edge [bend left=0] (u6)
    (u4) edge [bend left=0] (u5)
    (u4) edge [bend left=0] (u7)
    (u4) edge [bend left=0] (u8)
    (u5) edge [bend left=0] (u7)
    (u5) edge [bend left=0] (u8)
    (u6) edge [bend left=0] (u8)
    (u7) edge [bend left=0] (u9)
    (u7) edge [bend left=0] (u10)
    (u8) edge [bend left=0] (u9)
    (u8) edge [bend left=0] (u10)
    (u8) edge [bend left=0] (u11)
    (u9) edge [bend left=0] (u10)
    (u9) edge [bend left=0] (u11)
    (u10) edge [bend left=0] (u11)
    ;
    \draw [very thick, color = teal,-stealth]
    (u12) edge [bend left=0] (0.8,1.1)
    (u13) edge [bend left=0] (-1.1,-.2)
    ;
\end{tikzpicture}    
    \caption{Minh họa một mạng cảm biến với các nút tham chiếu và các nút thường.}
    \label{fig:netwkLocwSensors}
\end{SCfigure}
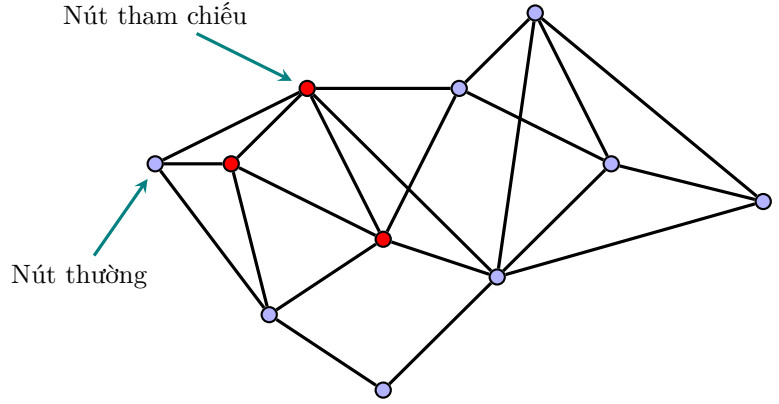

\section{Định vị mạng dựa trên vị trí tương đối}
\label{chap6:relative_position}
\subsection{Trường hợp mạng không có nút tham chiếu}
Giả sử đồ thị $G$ mô tả mạng là đồ thị có gốc ra. Do không có nút mạng nào có thông tin về vị trí tuyệt đối của mình, mỗi tác tử $i\in V$ lưu ước lượng vị trí của mình tại biến $\hat{\m{p}}_i \in \mb{R}^d$. Mỗi cạnh $(j,i) \in E$ mô tả việc nút $i$ có thể đo được $\m{z}_{ij}$ và nhận được $\hat{\m{p}}_j$ từ tác tử $j$. Tác tử $i$ cập nhật biến  $\hat{\m{p}}_i$ theo luật định vị mạng $\dot{\hat{\m{p}}}_i = \m{u}_i$.\index{thuật toán!định vị mạng!dựa trên sai lệch vị trí}

Với luật định vị mạng được thiết kế dưới dạng:
\begin{align} \label{eq:c6_network_loc_displacement}
    \m{u}_i &= - \sum_{j\in {N}_i} \big( (\hat{\m{p}}_i -\hat{\m{p}}_j) - (\m{p}_i -\m{p}_j)\big) \nonumber \\
    &= \sum_{j\in {N}_i} \big( \hat{\m{z}}_{ij} - {\m{z}}_{ij}\big),~i=1,\ldots,n,
\end{align}
trong đó ${\m{z}}_{ij}$ thu được từ đo đạc, $\hat{\m{p}}_j$ nhận được từ tác tử $j$, và $\hat{\m{z}}_{ij} = \hat{\m{p}}_j -\hat{\m{p}}_i$, phương trình viết cho hệ với $n$ nút mạng có dạng:
\begin{align}
    \dot{\hat{\m{p}}} = -\big(\mcl{L}\otimes \m{I}_d\big)(\hat{\m{p}} - \m{p}), \label{eq:c6_matrix_form_displacement}
\end{align}
trong đó  $\m{p} = \text{vec}(\hat{\m{p}}_1, \ldots, \hat{\m{p}}_n) \in \mb{R}^{dn}$ là cấu hình ước lượng tại thời điểm $t\ge 0$. 

Đặt $\m{r} = \hat{\m{p}} - \m{p}$, ta có thể viết lại hệ \eqref{eq:c6_matrix_form_displacement} 
\begin{align}
    \dot{\m{r}} = -\big(\mcl{L}\otimes \m{I}_d\big) \m{r},
\end{align}
là một hệ đồng thuận đối với biến $\m{r}$. Theo lý thuyết về hệ đồng thuận, ta có 
\begin{align}
    \m{r}(t) \to \m{1}_n \otimes \m{r}^* \in \text{im}(\m{1}_n \otimes \m{I}_d),
\end{align}
hay nói cách khác, $\m{r}_i(t) \to \m{r}^*, \forall i=1,\ldots, n$, trong đó $\m{r}^*$ là giá trị đồng thuận. Điều này tương đương với việc, $\hat{\m{p}}_i(t) \to \m{p}_i+\m{r}^*_i,~ \forall i =1, \ldots, n$ khi $t \to +\infty$. Như vậy, sau quá trình định vị mạng, các nút mạng xác định được một cấu hình $\hat{\m{p}}(\infty)$ sai khác với cấu hình thực tế bởi một phép tịnh tiến (với vector tịnh tiến $\m{r}^*$).

\begin{story}{Wei Ren và Điều khiển hệ đa tác tử}
Wei Ren là giáo sư tại Khoa Điện và Khoa học máy tính, Đại học California, Riverside. Nghiên cứu của Ren hầu như chỉ tập trung vào điều khiển hệ đa tác tử, ông có đóng góp trong hầu hết các chủ đề được đề cập của giáo trình này. Các nghiên cứu sau đó của Ren liên quan tới thuật toán đồng thuận với những giả thiết khác nhau về truyền thông và mô hình đối tượng với áp dụng ở các ứng dụng như đồng bộ hóa, điều khiển đội hình, tối ưu phân tán.

Ren là tác giả của ba sách chuyên khảo của Springer về điều khiển hệ đa tác tử \cite{Ren2008springer,Ren2010Springer,Chen2020distributed}, trong đó nội dung trong sách chuyên khảo đầu tiên \cite{Ren2008springer} dựa trên nghiên cứu của Ren về hệ đa tác tử khi làm Ph.D. tại Bringham Young University, Provo, UT (2004). Đây cũng là một sách chuyên khảo đầu tiên về điều khiển hệ đa tác tử.
\end{story}

\subsection{Trường hợp có nút tham chiếu trong mạng}
Giả sử các nút $1, \ldots, l$ biết được vị trí tuyệt đối của mình trong không gian. Khi đó, các nút mạng này, gọi là các nút tham chiếu, không cần cập nhật biến ước lượng $\hat{\m{p}}_i$ của mình mà chỉ cần gửi thông tin về vị trí tuyệt đối của mình cho các nút khác trong mạng. Nói cách khác, phương trình định vị mạng viết cho nút $i=1,\ldots, l$ được cho bởi:
\begin{align}
\dot{\hat{\m{p}}}_i = \m{0},~{\hat{\m{p}}}_i(0) = \m{p}_i,~ i=1,\ldots, l.
\end{align}
Luật định vị mạng viết cho các nút mạng $l+1,\ldots, n$ cũng được cho như ở phương trình  \eqref{eq:c6_network_loc_displacement} (Xem Hình~\ref{fig:c6_network_localization_beacon}). 
\begin{SCfigure}[][t!]
    \caption{Mô tả mạng cảm biến với các nút tham chiếu và các nút mạng thường. Mỗi cạnh của đồ thị thể hiện luồng thông tin (đo đạc hoặc truyền thông) giữa các nút mạng. Nhiễu $\bm{\varepsilon}_{ij}$ có thể xuất hiện trong từng cạnh của đồ thị.}
    \label{fig:c6_network_localization_beacon}
    \hspace{3cm}
\begin{tikzpicture}[
roundnode/.style={circle, draw=black, thick, minimum size=2mm,inner sep= 0.25mm}
]
	\draw[rotate around={30:(.5,.25)}, fill=red!5, dashed] (.5,.25) ellipse (.9cm and .35cm);
	\draw[rotate around={30:(1.9,-1.5)}, fill=blue!5, dashed] (1.9,-1.5) ellipse (2cm and 1.3cm);
	
    \node[roundnode,fill=red!50] (u2) at (1,0.5) { }; %
    \node[roundnode,fill=red!50] (u1) at (0,0) { };%
    \node[roundnode,fill=blue!50] (u3) at (.5,-2) { };%
    \node[roundnode,fill=blue!50] (u4) at (1.2,-1.2) { };%
    \node[roundnode,fill=blue!50] (u5) at (2,-.7) { };%
    \node[roundnode,fill=blue!50] (u6) at (2.7,-.5) { };%
    \node[roundnode,fill=blue!50] (u7) at (1.5,-2.5) { };%
    \node[roundnode,fill=blue!50] (u8) at (3,-1.8) { };%

    \node at (-.8,1) {Nút tham chiếu};
    \node at (-1.2,-1.5) {Nút thường};
    \node at (3,-0.3) {\large $i$};
    \node at (3.3,-1.8) {\large $j$};
    \node at (4.2,-1) {\large{\color{teal} $\bm{\varepsilon}_{ij}$}};
    \draw [very thick]
    (u3) edge [bend left=0] (u7)
    (u3) edge [bend left=0] (u8)
    (u4) edge [bend left=0] (u5)
    (u4) edge [bend left=0] (u8)
    (u5) edge [bend left=0] (u7)
    (u6) edge [bend left=0] (u8)
    (u7) edge [bend left=0] (u8)
    ;
    \draw [very thick,-stealth]
    (u1) edge [bend left=0] (u3)
    (u1) edge [bend left=0] (u5)
    (u2) edge [bend left=0] (u4)
    (u2) edge [bend left=0] (u6)
    ;
    \draw[very thick,-stealth] 
    (3.85,-1) -- (3,-1.2)
    ;
\end{tikzpicture}
\end{SCfigure}
Đặt $\m{p}^l = [\m{p}_1^\top,\ldots,\m{p}_l^\top]^\top$, $\m{p}^f = [\m{p}_{l+1}^\top,\ldots,\m{p}_n^\top]^\top$, $\hat{\m{p}}^l = [\hat{\m{p}}_{1}^\top, \ldots, \m{p}_{l}^\top]^\top$, và $\hat{\m{p}}^f = [\m{p}_{l+1}^\top, \ldots, \m{p}_{n}^\top]^\top$. Ta viết lại phương trình cho các tác tử $i = l+1, \ldots, n$ dưới dạng
\begin{align} \label{eq:c6_network_localization1}
    \dot{\hat{\m{p}}}^f = \big(\text{blkdiag}(\mcl{L}_{fl}\m{1}_l) \otimes \m{I}_d)\big) (\hat{\m{p}}^f - \m{p}^f) - (\mcl{L}(G') \otimes \m{I}_d)(\hat{\m{p}}^f - \m{p}^f)
\end{align}
trong đó $\mcl{L}_{fl} \in \mb{R}^{df \times dl}$, $\mcl{L}(G')$ là các ma trận khối tương ứng của ma trận Laplace của đồ thị $G$, còn $G'$ là ma trận Laplace của đồ thị $G'$ thu được từ $G$ sau khi xóa đi các nút tham chiếu. Chú ý rằng ma trận Laplace của $G$ có thể viết dưới dạng
\begin{align*}
    \mcl{L}(G)=\begin{bmatrix} 
        \m{0}_{l \times l} & \m{0}_{l\times f} \\
        \mcl{L}_{fl} & \mcl{L}_{ff}
    \end{bmatrix},
\end{align*}
trong đó $\mcl{L}_{ff} \in \mb{R}^{df \times df}$, với $f=n-l$, còn có tên gọi là \index{ma trận!Laplace nối đất} ma trận Laplace nối đất khi $l=1$. Với $G$ là đồ thị vô hướng và liên thông thì ma trận $\mcl{L}_{ff}$ có một số tính chất như sau (xem Bài tập \ref{ex:6.2}):
\begin{itemize}
    \item Đối xứng, xác định dương
    \item $\mcl{L}_{ff} = \mcl{L}(G') - \text{diag}(\mcl{L}_{fl}(\m{1}_l))$, trong đó $G'$ là đồ thị thu được sau khi xóa đi các nút tham chiếu.
\end{itemize}
Từ phương trình \eqref{eq:c6_network_localization1}, ta có
\begin{align} \label{eq:c6_network_localization2}
    \dot{\hat{\m{p}}}^f = - (\mcl{L}_{ff} \otimes \m{I}_d) (\hat{\m{p}}^f - \m{p}^f).
\end{align}
Do $\mcl{L}_{ff}$ là ma trận xác định dương, hệ \eqref{eq:c6_network_localization2} chỉ có điểm cân bằng duy nhất ${\hat{\m{p}}}^f = \m{p}^f$ là ổn định tiệm cận theo hàm mũ. Điều này tương đương với việc các giá trị ước lượng sẽ hội tụ tới vị trí thực của các nút khi $t\to +\infty$.

Nếu ta xét thêm sự ảnh hưởng của nhiễu hằng $\bmm{\epsilon}_{ij}=-\bmm{\epsilon}_{ij}$ tới từng cạnh $(j,i)$ với $j\notin \{1, \ldots, l\}$ thì
\begin{align*}
    \dot{\hat{\m{p}}}_i = - \sum_{j =1}^l a_{ij} ({\hat{\m{p}}}_i - \m{p}_j -(\m{p}_i-\m{p}_j)) - \sum_{j =l+1}^n a_{ij} ({\hat{\m{p}}}_i - {\hat{\m{p}}}_j -(\m{p}_i-\m{p}_j) + \bmm{\epsilon}_{ij})
\end{align*}
phương trình định vị mạng lúc này được tương ứng cho bởi
\begin{align} \label{eq:c6_network_localization3}
    \dot{\hat{\m{p}}}^f = - (\mcl{L}_{ff} \otimes \m{I}_d)(\hat{\m{p}}^f - \m{p}^f) + \m{d},
\end{align}
trong đó $\m{d} = (\m{H}(G')^\top \otimes \m{I}_d)\bmm{\epsilon}$ và $\bmm{\epsilon} = \text{vec}(\ldots, \bmm{\epsilon}_{ij}, \ldots) = \text{vec}(\bmm{\epsilon}_1, \ldots, \bmm{\epsilon}_m) \in \mb{R}^{dm}$. Hệ \eqref{eq:c6_network_localization3} có điểm cân bằng thỏa mãn $\m{0}= - (\mcl{L}_{ff} \otimes \m{I}_d)(\hat{\m{p}}^{f*} - \m{p}^f) + \m{d}$, hay $\hat{\m{p}}^{f*} = \m{p}^f + (\mcl{L}_{ff}^{-1} \otimes \m{I}_d) \m{d}$. Với phép đổi biến $\m{r}= \hat{\m{p}}^{f} - \hat{\m{p}}^{f*}$, ta thu được hệ
\begin{align*}
    \dot{\m{r}} = - (\mcl{L}_{ff} \otimes \m{I}_d)(\hat{\m{p}}^f - \m{p}^f- (\mcl{L}_{ff}^{-1} \otimes \m{I}_d)\m{d}) = - (\mcl{L}_{ff} \otimes \m{I}_d) \m{r}.
\end{align*}
Từ đây suy ra $\m{r}(t) = \m{0}$ ổn định tiệm cận theo hàm mũ, hay $\hat{\m{p}}^f(t) \to \hat{\m{p}}^{f*}$, khi $t \to +\infty$. Như vậy, nhiễu hằng số làm cho kết quả định vị mạng bị sai lệch so với vị trí thực một khoảng giới hạn bởi $\|(\mcl{L}_{ff}^{-1} \otimes \m{I}_d) \m{d}\| \leq \|\mcl{L}_{ff}^{-1} \otimes \m{I}_d\| \|\m{d}\| \leq \sigma(\mcl{L}_{ff}^{-1})\|\m{d}\|$, trong đó $\sigma(\mcl{L}_{ff}^{-1})=\lambda_{\min}(\mcl{L}_{ff})^{-1}$ là giá trị suy biến lớn nhất của ma trận đối xứng, xác định dương $\mcl{L}_{ff}^{-1}$.

\subsection{Phương pháp dựa trên vector hướng}
\label{subsec10:bearing_basedNL}
Xét mạng gồm $n$ nút $(G,\m{p})$ cứng hướng vi phân trong $\mb{R}^d$ và giả sử rằng mỗi nút mạng có thể đo vector hướng $\m{g}_{ij}$ và nhận/gửi thông tin về $\hat{\m{p}}_j$ từ một số nút mạng $j\in {N}_i$. Tương tự như trong mục \ref{eq:c6_network_loc_displacement}, chúng ta xét hai trường hợp mạng có nút tham chiếu và mạng không có nút tham chiếu. 
\index{thuật toán!định vị mạng!dựa trên vector hướng}

\subsubsection{Trường hợp mạng không có nút tham chiếu}
Luật định vị mạng trong trường hợp này được cho bởi:
\begin{align} \label{eq:c6_bearing_NL_no_beacon}
    \dot{\hat{\m{p}}}_i = - \sum_{j \in {N}_i} \m{P}_{\m{g}_{ij}} ({\hat{\m{p}}}_i - {\hat{\m{p}}}_j),~\forall i = 1,\ldots, n,
\end{align}
trong đó $\m{P}_{\m{g}_{ij}}$ tính toán được từ vector hướng $\m{g}_{ij}$ là một ma trận chiếu trực giao. Sử dụng ma trận Laplace cứng hướng $\mcl{L}_b = \tilde{\m{R}}_b^\top\tilde{\m{R}}_b$, phương trình \eqref{eq:c6_bearing_NL_no_beacon} có thể biểu diễn lại dưới dạng
\begin{align}  \label{eq:c6_bearing_NL_no_beacon1}
    \dot{\hat{\m{p}}} = - \mcl{L}_b \hat{\m{p}}.
\end{align}
Từ giả thiết mạng là cứng vi phân, rank$(\mcl{L}_b) = dn-d-1$,  $\text{ker}(\mcl{L}_b) = \text{im}(\m{1}_n\otimes \m{I}_d, \m{p})$ và các giá trị riêng khác 0 của $\mcl{L}_b$ đều là các số dương. Dễ thấy trọng tâm $\bar{\hat{\m{p}}}(t)$ là không thay đổi theo thời gian. Theo lý thuyết về hệ tuyến tính, \eqref{eq:c6_bearing_NL_no_beacon1} hội tụ tới 
\begin{align}
    \hat{\m{p}}(\infty) &= \lim_{t\to +\infty} \mathtt{e}^{-\mcl{L}_b t} \hat{\m{p}}(0) \nonumber\\
    &= \begin{bmatrix}\frac{\m{1}_n\otimes \m{I}_d}{\sqrt{n}}, \frac{\m{p}-\m{1}_n\otimes \bar{\m{p}}}{\|\m{p}-\m{1}_n\otimes \bar{\m{p}\|}}\end{bmatrix}
    \begin{bmatrix}\frac{\m{1}_n\otimes \m{I}_d}{\sqrt{n}}, \frac{\m{p}-\m{1}_n\otimes \bar{\m{p}}}{\|\m{p}-\m{1}_n\otimes \bar{\m{p}\|}}\end{bmatrix}^\top \hat{\m{p}}(0) \nonumber\\
    &= \m{1}_n \otimes \bar{\hat{\m{p}}}(0) + \frac{\m{p}-\m{1}_n\otimes \bar{\m{p}}}{\|\m{p}-\m{1}_n\otimes \bar{\m{p}}\|^2} (\m{p}-\m{1}_n\otimes \bar{\m{p}})^\top\hat{\m{p}}(0)
\end{align}
là một điểm thuộc $\text{ker}(\mcl{L}_b)$ khi $t \to +\infty$. 

Như vậy, cấu hình mạng ước lượng sẽ hội tụ tới một cấu hình tương đồng về hướng với cấu hình mạng thực tế $\m{p}$. Do không có các nút tham chiếu, tồn tại khả năng các tác tử hội tụ tới một cấu hình đối xứng qua gốc tọa độ với cấu hình gốc với các vector hướng thỏa mãn $\hat{\m{g}}_{ij} = -\m{g}_{ij}, \forall i, j \in V, i\ne j$. Các nút mạng có thể kiểm tra vector hướng ứng với cấu hình cuối, từ đó khởi tạo lại giá trị $\hat{\m{p}}(0)$ của mình để định vị lại mạng. Tuy nhiên, cách này không đảm bảo rằng cấu hình cuối sau khi định vị mạng lại sẽ thỏa mãn $\hat{\m{g}}_{ij} = \m{g}_{ij}, \forall i,j \in V, i\ne j$.

\subsubsection{Trường hợp mạng có nút tham chiếu}
Giả sử tồn tại các nút tham chiếu $1,\ldots, l$ trong mạng, với $l\ge 2$. Vị trí các nút mạng cần thỏa mãn điều kiện 
\begin{align*}
    \mcl{L}_b \m{p} = \begin{bmatrix}
    \mcl{L}_{bll} & \mcl{L}_{blf} \\
    \mcl{L}_{bfl} & \mcl{L}_{bff}
    \end{bmatrix} \m{p} = \m{0},
\end{align*}
trong đó $\mcl{L}_{bll} \in \mb{R}^{dl\times dl}$, $\mcl{L}_{blf} = \mcl{L}_{bfl}^\top \in \mb{R}^{dl\times df} $, và $\mcl{L}_{bff} \in \mb{R}^{df\times df}$. Từ đây, kí hiệu $\m{p}^l = [\m{p}_1^\top,\ldots,\m{p}_l^\top]^\top$, $\m{p}^f = [\m{p}_{l+1}^\top,\ldots,\m{p}_n^\top]^\top$ thì $\mcl{L}_{bfl} \m{p}^l + \mcl{L}_{bff} \m{p}^f = \m{0}$. Nếu $l\ge 2$ và $(G,\m{p})$ là cứng hướng vi phân thì ma trận $\mcl{L}_{bff}$ là đối xứng, xác định dương, đồng thời $\mcl{L}_{bff} = \mcl{L}_{b}(G',\m{p}^f) - \text{blkdiag}(\mcl{L}_{blf}(\m{1}_l \otimes \m{I}_d))$, trong đó $G'$ là đồ thị thu được từ $G$ sau khi bỏ đi các nút tham chiếu và $\mcl{L}_{b}(G',\m{p}^f)$ là ma trận cứng hướng tương ứng viết cho mạng $(G',\m{p}^f)$. Từ đây, ta suy ra vị trí các nút còn lại có thể xác định duy nhất bởi $\m{p}^f = -\mcl{L}_{bff}^{-1}\mcl{L}_{bfl} \m{p}^l$. 

Do các nút tham chiếu có thông tin về vị trí tuyệt đối của mình, chúng không cần cập nhật giá trị ước lượng mà chỉ truyền các giá trị này tới các nút mạng khác. Luật định vị mạng viết cho các nút $l+1, \ldots, n$ có dạng giống như \eqref{eq:c6_bearing_NL_no_beacon}, và có thể biểu diễn chung lại như sau \cite{zhao2016aut}:
\begin{align}
    \dot{\hat{\m{p}}}^f &= - \mcl{L}_{bfl} \m{p}^l - \mcl{L}_{bff} \hat{\m{p}}^f \nonumber\\
    &= -\mcl{L}_{bff} (\hat{\m{p}}^f - \m{p}^f).
\end{align}
Từ tính đối xứng, xác định dương của $\mcl{L}_{bff}$, dễ thấy ${\hat{\m{p}}}^f(t) \to \m{p}^f$ khi $t\to +\infty$. Nói cách khác, vị trí ước lượng các nút mạng sẽ dần tiến tới vị trí tuyệt đối của nút trong không gian. Sự ảnh hưởng của nhiễu hằng tới luật định vị mạng lúc này cũng có thể phân tích tương tự như với định vị mạng sử dụng vị trí tương đối.

\begin{figure*}
    \centering
    \includegraphics[width=1\linewidth]{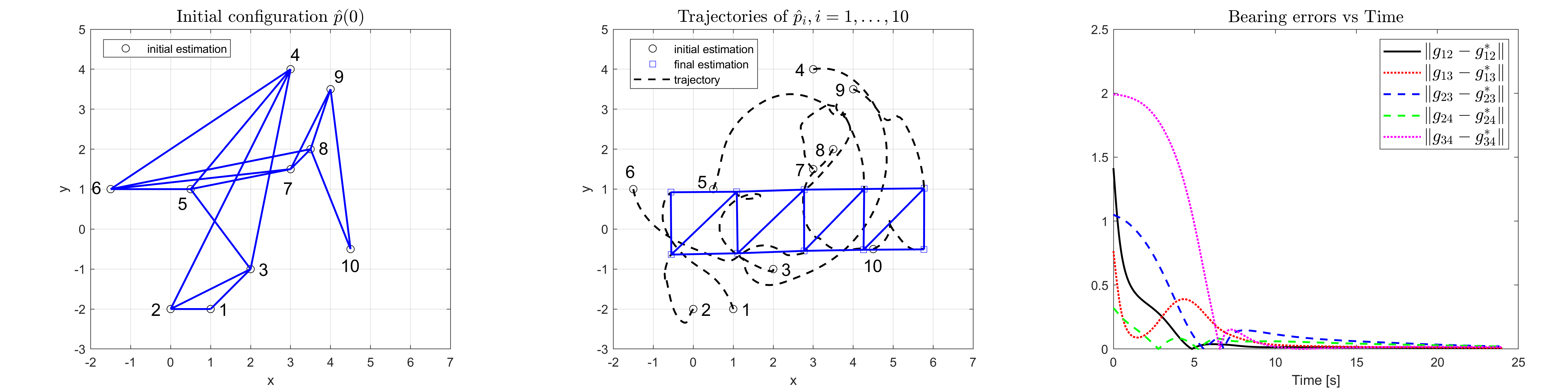}
    \caption{Minh họa định vị mạng cảm biến gồm 10 nút với luật định vị mạng \eqref{eq:c6_bearing_NL_no_beacon}}
    \label{fig:c6_NL_no_beacon}
\end{figure*}

Ý nghĩa của luật định vị mạng \eqref{eq:c6_bearing_NL_no_beacon1}: do mục tiêu của bài toán là định vị các nút mạng sao cho cấu hình cuối là tương đồng về hướng đối với cấu hình thực tế, ${\hat{\m{p}}}^f$ có thể coi là nghiệm của bài toán tối ưu
\begin{align}
\begin{array}{*{20}{cl}}
\text{minimize}&{\frac{1}{2}\|\m{P}_{\m{g}_{ij}}\bar{\m{H}}(\hat{\m{p}} -{\m{p}}) \|^2}\\
\text{s.t.}&{\hat{\m{p}}^l = {\m{p}}^l}
\end{array}
\end{align}
Luật định vị mạng khi đó thiết kế cho các nút $i=l+1,\ldots, n$ có hướng ngược với hướng gradient của hàm mục tiêu $V = \frac{1}{2} \|\m{P}_{\m{g}_{ij}}\bar{\m{H}}(\hat{\m{p}} -{\m{p}})\|^2$, tức là $\hat{\m{p}}^f = - \nabla_{\hat{\m{p}}^f} V$. Chú ý rằng luật định vị chỉ giúp đưa hệ về cấu hình tương đương về hướng với cấu hình thực. Ta còn cần thêm điều kiện về tính cứng hướng vi phân của mạng để đảm bảo rằng cấu hình cuối thu được là tương đồng về hướng với cấu hình ban đầu.

\begin{example}
Ví dụ định vị mạng dựa trên vector hướng: xét hệ gồm 10 cảm biến phân bố trong không gian hai chiều với luật định vị mạng dựa trên vector hướng. Ước lượng vị trí ban đầu của các nút được cho như trong Hình~\ref{fig:c6_NL_no_beacon}~(trái). Sử dụng luật định vị mạng, các giá trị ước lượng vị trí có quỹ đạo như ở Hình~ \ref{fig:c6_NL_no_beacon}~(giữa). Ta có thể thấy rằng chúng tạo thành một cấu hình tương đồng về hướng như với cấu hình thực. Các sai lệch vector hướng $\|\m{g}_{ij} - \m{g}_{ij}^*\|$ được cho như trên Hình~\ref{fig:c6_NL_no_beacon}~(phải). Dễ thấy  $\|\m{g}_{ij} - \m{g}_{ij}^*\| \to 0$, hay các tác tử dần định vị được vị trí phù hợp với các vector hướng đo được.
\end{example}

\subsection{Phương pháp dựa trên khoảng cách}
\index{thuật toán!định vị mạng!dựa trên khoảng cách}
Xét bài toán định vị mạng cảm biến dựa trên khoảng cách trong không gian hai chiều. Để xác định duy nhất một vị trí $\m{p}_4$ trong không gian 2 chiều dựa trên khoảng cách, ta cần vị trí của ba nút tham chiếu $\m{p}_1, \m{p}_2, \m{p}_3$ trong không gian và ba khoảng cách $d_{4i} = \|\m{p}_4 - \m{p}_i\|, i=1, 2, 3$. Vị trí của nút 4 là giao của ba đường tròn $(\m{p}_1,d_{41})$, $(\m{p}_2,d_{42})$ và $(\m{p}_3,d_{43})$.

\begin{SCfigure}[][h!]
    \caption{Định vị nút 4 dựa vào 3 nút tham chiếu và 3 biến đo khoảng cách}
    \label{fig:c10_Distance_based_NL}
    \hspace{3.5cm}
    \begin{tikzpicture}[scale=1]

  \coordinate (A) at (0,1.5);
  \coordinate (B) at (2,1.5);
  \coordinate (C) at (1.5,0);

  \coordinate (P) at (-1,0);

  \fill[color=green!50!black] (A) circle (3pt) node[right] {\color{black} $\m{p}_1$};
  \fill[color=black!50] (B) circle (3pt) node[right] {\color{black} $\m{p}_2$};
  \fill[color=blue!50]  (C) circle (3pt) node[right] {\color{black} $\m{p}_3$};
  
  \draw[-,color=green!50!black] (A) -- (P);
  \draw[-,color=black!50] (B) -- (P);
  \draw[-,color=blue!50]  (C) -- (P);

  \draw[color=green!50!black] (A) circle ({sqrt(1 + 2.25)});
  \draw[color=black!50] (B) circle ({sqrt(9 + 2.25)});
  \draw[color=blue!50] (C) circle ({sqrt(6.25 + 0)});
  \fill[color=yellow!80] (P) circle (3pt) node[left=.3cm] {\color{black} $\m{p}_4$};

\end{tikzpicture}
\end{SCfigure}
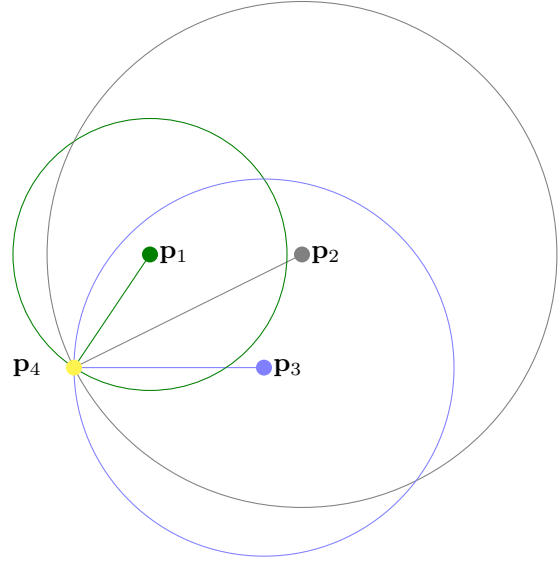

Ta sẽ xây dựng một công thức để tìm $\m{p}_4$ dựa trên các thông tin thu nhận được (ba biến đo  khoảng cách và ba biến vị trí nhận được từ các nút tham chiếu $1, 2, 3$). Kí hiệu $\m{p}_i = [x_i,y_i]^\top \in \mb{R}^2$, ta có các phương trình
\begin{align} \label{eq:c6_distance}
    d_{4i}^2 = (x_4-x_i)^2+(y_4-y_i)^2,\, i=1, 2, 3
\end{align}
Trừ từng vế các phương trình với $d_{42}^2$ và $d_{43}^2$ cho $d_{41}^2$, ta thu được
\begin{align} \label{eq:c6_distance1}
    d_{4i}^2-d_{41}^2 &= (x_i^2+y_i^2-x_1^2-y_1^2)-2(x_i-x_1)x_4-2(y_i-y_1)y_4,~i=2,3
\end{align}
Như vậy, ta thu được phương trình
\begin{equation}  \label{eq:c6_distanceA}
    \m{A}\m{p}_4 = \m{b},
\end{equation}
trong đó $\m{A}=\begin{bmatrix}
    2(x_2-x_1) & 2(y_2-y_1)\\ 2(x_3-x_1) & 2(y_3-y_1)
\end{bmatrix}$ và $\m{b}=\begin{bmatrix}
x_2^2+y_2^2-x_1^2 - y_1^2 - d_{42}^2 + d_{41}^2\\ x_3^2+y_3^2-x_1^2 - y_1^2 - d_{43}^2 + d_{41}^2
\end{bmatrix}$. Nếu các nút tham chiếu không thẳng hàng, ma trận $\m{A}$ khả nghịch, ta xác định được nghiệm là vị trí chính xác của nút 4 theo công thức $\m{p}=\m{A}^{-1}\m{b}$. 

Công thức trên có thể mở rộng cho trường hợp định vị mạng khi cần định vị nút mạng tại $\m{p}_{N+1}=[x_{N+1},y_{N+1}]^\top$ dựa trên $N\geq 3$ nút tham chiếu \cite{Sayed2005network}. Thực hiện các bước tương tự như trên, vị trí của nút $(N+1)$ thỏa mãn phương trình \eqref{eq:c6_distanceA}, với 
\[\m{A}=\begin{bmatrix}
    2(x_2-x_1) & 2(y_2-y_1)\\ \vdots \\ 2(x_3-x_1) & 2(y_3-y_1)
\end{bmatrix} \text{ và } \m{b}=\begin{bmatrix}
x_2^2 + y_2^2-x_1^2 - y_1^2 - d_{N+1,2}^2 + d_{N+1,1}^2\\ \vdots \\x_N^2+y_N^2-x_1^2 - y_1^2 - d_{N+1,N}^2 + d_{N+1,1}^2
\end{bmatrix}.\]
Nếu tồn tại ít nhất ba nút tham chiếu không thẳng hàng, ma trận $\m{A}$ đủ hạng cột, ta xác định được vị trí nút $N+1$ theo phương trình 
\begin{align}
    \m{p}_{N+1} = (\m{A}^\top\m{A})^{-1}\m{A}^\top\m{b}.
\end{align}

Từ kết quả trong trường hợp định vị một nút nêu trên, ta sẽ xây dựng một đồ thị có thể định vị được bằng khoảng cách.\index{đồ thị!định vị được bằng khoảng cách}

Một đồ thị là $k$-liên thông cạnh nếu xóa đi $k$ cạnh bất kỳ trong $G$ thì đồ thị thu được vẫn là liên thông. Tương tự, một đồ thị $G$ là $k-$thừa cứng nếu xóa đi $k$ cạnh bất kỳ trong $G$ thì ta vẫn được một đồ thị cứng khoảng cách phổ quát. Ví dụ về đồ thị \index{1-thừa cứng} 1-thừa cứng khoảng cách trong $\mb{R}^2$ là đồ thị đầy đủ gồm 4 đỉnh ${K}_4$. Ví dụ về đồ thị 3-liên thông nhưng không 1-thừa cứng khoảng cách được cho trên Hình~\ref{fig:c6_k_Redundant_Edges}. \index{đồ thị!$k$-liên thông cạnh}
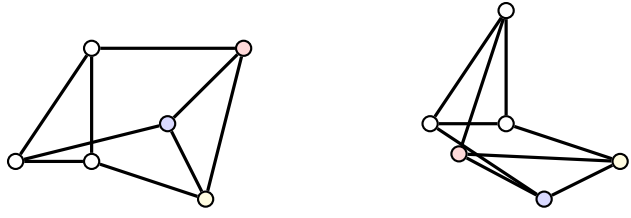
\begin{figure}[t!]
    \centering
    \begin{tikzpicture}[
roundnode/.style={circle, draw=black, thick, minimum size=2mm,inner sep= 0.25mm}
]
	
    \node[roundnode] (u1) at (-1,0) { }; %
	\node[roundnode] (u2) at (0,0) { }; %
	\node[roundnode] (u3) at (0,1.5) { }; %
	\node[roundnode,fill = yellow!15] (u4) at (1.5,-0.5) { }; %
	\node[roundnode,fill = blue!15] (u5) at (1.0,0.5) { }; %
	\node[roundnode,fill = red!15] (u6) at (2.0,1.5) { }; %
	
    \draw [very thick]
    (u1) edge [bend left=0] (u2)
    (u1) edge [bend left=0] (u3)
    (u2) edge [bend left=0] (u3)
    (u1) edge [bend left=0] (u5)
    (u2) edge [bend left=0] (u4)
    (u4) edge [bend left=0] (u5)
    (u4) edge [bend left=0] (u6)
    (u3) edge [bend left=0] (u6)
    (u5) edge [bend left=0] (u6)
    ;
\end{tikzpicture}\qquad\qquad\qquad
    \begin{tikzpicture}[
roundnode/.style={circle, draw=black, thick, minimum size=2mm,inner sep= 0.25mm}
]
    \node[roundnode] (u1) at (-1,0) { }; %
	\node[roundnode] (u2) at (0,0) { }; %
	\node[roundnode] (u3) at (0,1.5) { }; %
	\node[roundnode,fill=yellow!15] (u4) at (1.5,-0.5) { }; %
	\node[roundnode,fill=blue!15] (u5) at (0.5,-1) { }; %
	\node[roundnode,fill=red!15] (u6) at (-0.62,-0.4) { }; %
	
    \draw [very thick]
    (u1) edge [bend left=0] (u2)
    (u1) edge [bend left=0] (u3)
    (u2) edge [bend left=0] (u3)
    (u1) edge [bend left=0] (u5)
    (u2) edge [bend left=0] (u4)
    (u4) edge [bend left=0] (u5)
    (u4) edge [bend left=0] (u6)
    (u3) edge [bend left=0] (u6)
    (u5) edge [bend left=0] (u6)
    ;
\end{tikzpicture}
    \caption{Ví dụ đồ thị cứng khoảng cách 3-liên thông có 2 hiên thực hóa trong 2D. Đồ thị này không 1-thừa cứng khoảng cách.}
    \label{fig:c6_k_Redundant_Edges}
\end{figure}

Một đồ thị $G$ với $n \ge 4$ đỉnh là \index{cứng!khoảng cách!toàn cục phổ quát}cứng toàn cục phổ quát trong $\mb{R}^2$ khi và chỉ khi nó là $3-$liên thông và $1-$thừa cứng trong $\mb{R}^2$ \cite{Aspnes2006}. 

Một phương án xây dựng các đồ thị cứng toàn cục phổ quát có thể định vị được bằng khoảng cách trong $\mb{R}^d$:
\begin{itemize}
    \item Xuất phát từ một đồ thị đầy đủ gồm $d+1$ đỉnh ${K}_{d+1}$ tương ứng với $d+1$ nút tham chiếu.
    \item Tại mỗi bước $k=d+2, \ldots, n$, ta thêm một đỉnh mới và $d+1$ cạnh với các đỉnh hiện có của đồ thị.
\end{itemize}
Khi đó, đồ thị thu được thỏa mãn tính cứng hướng phổ quát và có thể định vị được dựa trên khoảng cách khi $d+1$ nút $1,\ldots, d+1$ là các nút tham chiếu trong mạng. Mỗi nút $k=d+2, \ldots, n$ định vị vị trí của mình dựa trên $\{d_{ij}, \m{p}_j\}_{j\in N_i \cap \{1, \ldots, k-1\}}$ sau đó gửi vị trí định vị được cho các nút $j\in N_i \cap \{k+1, \ldots, n\}$.

\section{Bài tập}
\label{c6:exercise}
\begin{exercise} \label{ex:6.1}
Xét mạng cảm biến với đồ thị $G$ được cho như ở Hình~\ref{fig:c6_ex1}. Chọn một nút bất kỳ trong mạng là nút tham chiếu, hãy viết ma trận Laplace nối đất tương ứng của mạng.
\end{exercise}
\begin{SCfigure}[][t!]
\caption{Đồ thị trong Bài tập \ref{ex:6.1}.}
\label{fig:c6_ex1}
\hspace{4cm}
\begin{tikzpicture}[
roundnode/.style={circle, draw=black, thick, minimum size=2mm,inner sep= 0.25mm}
]
    \node[roundnode] (u1) at (0,0) { }; %
    \node[roundnode] (u2) at (3,0) { };%
    \node[roundnode] (u3) at (3,3) { };%
    \node[roundnode] (u4) at (0,3) { };%
    \node[roundnode] (u5) at (1,1) { };%
    \node[roundnode] (u6) at (2,1) { };%
    \node[roundnode] (u7) at (2,2) { };%
    \node[roundnode] (u8) at (1,2) { };%

    \draw [very thick,-]
    (u1) -- (u2) -- (u3) -- (u4) -- (u1) -- (u5) -- (u6) -- (u7) -- (u8) -- (u5);
    
    \draw [very thick]
    (u2) edge [bend left=0] (u6)
    (u3) edge [bend left=0] (u7)
    (u4) edge [bend left=0] (u8)
    ;
\end{tikzpicture}
\end{SCfigure}

\begin{exercise} \label{ex:6.2}
Xét mạng cảm biến với đồ thị $G$ là vô hướng, liên thông và chọn $l\ge 1$ là số nút tham chiếu trong mạng. Giả sử ma trận $\mcl{L}$ được phân hoạch tương ứng dưới dạng
\begin{align*}
    \mcl{L} = \begin{bmatrix} \mcl{L}_{ll} & \mcl{L}_{lf}\\ \mcl{L}_{fl} & \mcl{L}_{ff} \end{bmatrix},
\end{align*}
trong đó $\mcl{L}_{ff} \in \mb{R}^{(n-l) \times (n-l)}$ là ma trận Laplace nối đất. Chứng minh rằng:
\begin{itemize}
    \item[i.] $\mcl{L}_{ff} = \mcl{L}(G') - \text{diag}(\mcl{L}_{fl} \m{1}_l)$, trong đó $G'$ là đồ thị thu được sau khi xóa đi các nút tham chiếu.
    \item[ii.] $\mcl{L}_{ff}$ là đối xứng và xác định dương.
\end{itemize}
\end{exercise}

\begin{exercise} \label{ex:6.3}
Trình bày cách lập một đồ thị 3-thừa cứng khoảng cách gồm 10 đỉnh trong 2D phục vụ cho định vị mạng dựa trên  khoảng cách.
\end{exercise}

\begin{exercise} \label{ex:6.4}
Xét mạng cảm biến với $(G,\m{p})$ được cho như ở Hình~\ref{fig:c6_ex42}, trong đó mỗi trục tọa độ có đơn vị là 1. Viết ma trận cứng hướng  và kiểm tra tính cứng hướng vi phân của mạng. Chọn nút 1 và 2 là nút tham chiếu, hãy viết ma trận $\mcl{L}_{bff}$ tương ứng của mạng.
\end{exercise}
\begin{SCfigure}[][h!]
    \caption{Mạng cảm biến $(G,\m{p})$ trong Bài tập \ref{ex:6.4} }
    \label{fig:c6_ex42}
    \hspace{3cm}
	\begin{tikzpicture}[scale=.8]
  \draw[step=1cm, gray!30, very thin] (-1,-3) grid (6,2);

  \draw[->,color=black!50] (-1.5,0) -- (6.4,0) node[right] {$x$};
  \draw[->,color=black!50] (0,-3) -- (0,2.4) node[above] {$y$};

  \foreach \x in {-1,1,2,3,4,5,6} {
    \draw (\x,0.05) -- (\x,-0.05);
  }

  \foreach \y in {-3,-2,-1,1,2} {
    \draw (0.05,\y) -- (-0.05,\y);
  }
  
  \draw[very thick,-] (0,1) -- (0,-2) -- (1,0)-- (0,1) -- (5,0) -- (0,-2);
  \draw[very thick,-] (0,1) -- (0,-2);
  \draw[fill=white] (0,1) circle (3pt) node[left]  {\color{black} $\m{p}_1$};
  \draw[fill=white] (0,-2) circle (3pt) node[left] {\color{black} $\m{p}_3$};
  \draw[fill=white] (1,0) circle (3pt) node[right] {\color{black} $\m{p}_2$};
  \draw[fill=white] (5,0) circle (3pt) node[below] {\color{black} $\m{p}_4$};
  
\end{tikzpicture}
\end{SCfigure}

\begin{exercise} \label{ex:6.5}
Xét bài toán định vị mạng cảm biến dựa trên vector hướng với $(G,\m{p})$ là cứng hướng vi phân với $\m{p} = \text{vec}(\m{p}_1,\ldots, \m{p}_n)$ là vị trí tuyệt đối của các nút mạng, $\m{p}_i \in \mb{R}^d$.  Giả sử trong mạng có $l\ge 2$ nút tham chiếu, đồng thời ma trận $\mcl{L}_b$ được phân hoạch tương ứng dưới dạng
\begin{align*}
    \mcl{L}_b = \begin{bmatrix} \mcl{L}_{bll} & \mcl{L}_{blf}\\ \mcl{L}_{bfl} & \mcl{L}_{bff} \end{bmatrix}
\end{align*}
với $\mcl{L}_{bff} \in \mb{R}^{d(n-l) \times d(n-l)}$. Chứng minh rằng:
\begin{item}
    \item[i.] $\mcl{L}_{bff} = \mcl{L}_b(G') - \text{blkdiag}(\mcl{L}_{fl} (\m{1}_l \otimes \m{I}_d))$, trong đó $G'$ là đồ thị thu được sau khi xóa đi các nút tham chiếu.
    \item[ii.] $\mcl{L}_{ff}$ là đối xứng và xác định dương.
\end{item}
\end{exercise}

\begin{exercise}\cite{Bullo2019lectures} \label{ex:6.6}
Với $\mcl{L}$ là ma trận Laplace của một đồ thị vô hướng, liên thông, có trọng số gồm $n$ nút. Định nghĩa ma trận $\m{X} = \mcl{L} + \frac{k}{n}\m{1}_n\m{1}_n^\top$, với $k\in \mb{R}$. Chứng minh rằng:
\begin{item}
    \item[i.] $\m{X}$ là xác định dương nếu $k>0$.
    \item[ii.] $\m{X}^{-1} = \left(\mcl{L} +  \frac{k}{n}\m{1}_n\m{1}_n^\top \right) ^{-1} = \mcl{L}^+ +\frac{1}{kn}\m{1}_n\m{1}_n^\top$, trong đó  $\mcl{L}^+$ là ma trận giả nghịch đảo của $\mcl{L}$ được thỏa mãn (i) $\mcl{L}\mcl{L}^+ \mcl{L} = \mcl{L}$, (ii) $\mcl{L}^+ \mcl{L}\mcl{L}^+ = \mcl{L}^+$, và (iii) $\mcl{L} \mcl{L}^+ = \mcl{L}^+ \mcl{L} = \m{I}_n - \frac{1}{n} \m{1}_n\m{1}_n^\top$.
\end{item}
\end{exercise}

\begin{exercise}[Rút gọn Kron \cite{Dorfler2012kron}] \label{ex:6.7}
Với $\mcl{L}$ là ma trận Laplace của một đồ thị vô hướng, liên thông, có trọng số gồm $n$ nút. Xét phương trình $\m{y} = \mcl{L}\m{x}$, với $\m{x} \in \mb{R}^n$ là vector các biến chưa biết, $\m{y}\in \mb{R}^n$ là một vector. 
\begin{itemize}
\item[i.] Giả sử nút thứ $n$ trong đồ thị được chọn là nút tham chiếu. Rút gọn $x_n$ từ phương trình  $[\m{y}]_n = [\mcl{L}\m{x}]_n$ và thế vào $n-1$ phương trình đối với $\m{y}_{[1:n-1]}=[y_1, \ldots, y_{n-1}]^\top$ và $\m{x}_{[1:n-1]}=[x_1, \ldots, x_{n-1}]^\top$, ta thu được phương trình dạng:
\begin{equation}
    \m{y}_{[1:n-1]} + \m{b} y_n = \mcl{L}_{\rm red} \m{x}_{[1:n-1]}.
\end{equation}
Chứng minh rằng: (i) các phần tử của ma trận rút gọn Kron của $G$ ứng với nút $n$,  $\mcl{L}_{red}$, được cho bởi $l_{ij}- l_{in}l_{jn}/l_{nn}$; (ii) ma trận  $\mcl{L}_{red}$ là ma trận Laplace của một đồ thị vô hướng, liên thông; (iii) tìm công thức tính các giá trị của $\mcl{L}_{red}$ cho trường hợp đồ thị hình sao $S_4$ (biến đổi $\Delta - Y$).
\item[ii.] Tổng quát hóa kết quả trên cho trường hợp $2\le l \le n-1$ nút trong mạng được chọn là nút tham chiếu.
\end{itemize}
\end{exercise}

\begin{exercise}[Định vị mạng khi có trễ] \cite{Olfati2004consensus,Pham2025leader} Giả sử mạng cảm biến được mô tả bởi đồ thị $G$ vô hướng, liên thông gồm $n$ nút mạng. 
Mỗi tác tử trong hệ cập nhật biến ước lượng vị trí của mình dựa trên luật định vị mạng dựa trên sai lệch vị trí.
\begin{itemize}
    \item[i.] Giả sử mỗi nút cần một khoảng thời gian $\tau >0$ để xử lý thông tin thu về và tính toán luật cập nhật, phương trình cập nhật ước lượng vị trí của nút $i$ được cho bởi:
    \begin{align}
        \dot{\hat{\m{p}}} _i = \sum_{j\in {N}_i} \big( (\hat{\m{p}}_i(t-\tau) - \hat{\m{p}}_j(t-\tau) )- (\m{p}_i - \m{p}_j) \big),~i=1,\ldots, n.
    \end{align}
    Chứng minh rằng nếu $\tau < \frac{\pi}{2\lambda_n(\mcl{L})}$ thì các nút sẽ dần định vị được vị trí thực của mình. Mô phỏng cho trường hợp đồ thị được cho ở Hình~\ref{fig:c6_ex42}.
    \item[ii.] Giả sử trong mạng có trễ truyền thông $\tau>0$, phương trình cập nhật ước lượng vị trí của nút $i$ được cho bởi \cite{Seuret2008consensus}:
        \begin{align}
        \dot{\hat{\m{p}}}_i = \sum_{j\in {N}_i} \big( (\hat{\m{p}}_i(t) - \hat{\m{p}}_j(t-\tau) )- (\m{p}_i - \m{p}_j) \big),~i=1,\ldots, n.
    \end{align}
    Phân tích tính ổn định của mạng trong trường hợp này.
\end{itemize}
Thực hiện các yêu cầu trên khi các tác tử định vị mạng dựa trên vector hướng.
\end{exercise}

\begin{exercise}\cite{zhao2016aut} \label{ex:6.9} Giả sử mạng cảm biến được mô tả bởi đồ thị $G$ vô hướng, liên thông gồm $n$ nút mạng. Mỗi tác tử trong hệ cập nhật biến ước lượng vị trí của mình dựa trên luật định vị mạng dựa trên sai lệch vị trí.
\begin{align} \label{eq:ex_6.9}
    \dot{\hat{\m{p}}} _i = \sum_{j\in {N}_i} \big( (\hat{\m{p}}_i- \hat{\m{p}}_j)- (\m{p}_i - \m{p}_j) \big),~i=1,\ldots, n.
\end{align}
Chứng minh rằng luật định vị mạng \eqref{eq:ex_6.9} cho nghiệm của bài toán tối ưu:
\begin{mini}
{\hat{\m{p}}\in \mb{R}^{dn}}{\frac{1}{2}\|\bar{\m{H}}\hat{\m{p}} -\bar{\m{H}}{\m{p}}\|^2}
{\label{eq:c6_network_localization_optim}}{}
\end{mini}
trong đó $\m{H}$ là ma trận liên thuộc của $G$ và $\bar{\m{H}} = \m{H} \otimes \m{I}_d$.
\end{exercise}

\begin{exercise}[Điều khiển đội hình dựa trên định vị mạng] \cite{Oh2012formation,Li2009consensus} \label{ex:6.10} Xét đội hình gồm $n$ tác tử trong không gian $d$ chiều, vị trí của các tác tử được cho bởi vector $\m{p}_i \in \mb{R}^d$, $i=1,\ldots,n$.  Giả sử đồ thị $G$ mô tả tương tác đo, điều khiển, và trao đổi thông tin giữa các tác tử trong đội hình là có gốc ra. Mỗi tác tử $i$ có thông tin về đội hình đặt $\m{p}^*$ thông qua các vector sai lệch đặt $\m{p}_i^* - \m{p}_j^*$, $(i,j) \in E$. Giả thiết rằng mỗi tác tử $i$ ngoài đo vị trí tương đối $\m{p}_i - \m{p}_j,~j\in {N}_i$ còn có thể trao đổi thông tin với các tác tử khác trong đội hình. Mỗi tác tử $i$ trong đội hình thực hiện ước lượng vị trí $\hat{\m{p}}_i \in \mb{R}^d$ và trao đổi biến ước lượng này với các tác tử $j \in N_i$. Dựa trên biến ước lượng, các tác tử thực hiện luật điều khiển đội hình cho bởi
\begin{subequations}
\begin{align}
\dot{\hat{\m{p}}}_i &= - \sum_{j \in N_i} (\hat{\m{p}}_i - \hat{\m{p}}_j - (\m{p}_i - \m{p}_j) + (\m{p}_i^* - \m{p}_j^*)) + \m{u}_i,\label{eq:ex_6.10a}\\
\dot{\m{p}}_i &= \m{u}_i, \label{eq:ex_6.10b}\\
\m{u}_i &= - k \hat{\m{p}}_i, \label{eq:ex_6.10c}
\end{align}
\end{subequations}
trong đó $k>0$. Hãy thực hiện các yêu cầu sau:
\begin{enumerate}
\item[i.] Biểu diễn hệ \eqref{eq:ex_6.10a}--\eqref{eq:ex_6.10c} ở dạng ma trận.
\item[ii.] Đặt $\m{e} = \hat{\m{p}} - \m{p} + \m{p}^*$, hãy chứng minh $\m{e} \to \m{1}_n \otimes \bar{\m{e}}(0)$, trong đó $\bar{\m{e}}(0) = \sum_{i=1}^n \gamma_i \m{e}_i(0)$, $\bm{\gamma} = [\gamma_1,\ldots,\gamma_n]^\top$ là vector riêng bên trái đã chuẩn hóa ứng với ma trận Laplace của $G$.
\item[iii.] Chứng minh rằng $\m{p}$ dần đạt tới một đội hình cuối chỉ sai khác với đội hình đặt bởi một phép tịnh tiến.
\end{enumerate}
\end{exercise}

\begin{exercise}[Điều khiển đội hình dựa trên định vị mạng theo lý thuyết cứng hướng] \cite{Tran2018bearing} \label{ex:6.11} Xét đội hình gồm $n$ tác tử trong không gian $d$ chiều, vị trí của các tác tử được cho bởi vector $\m{p}_i \in \mb{R}^d$, $i=1,\ldots,n$. Đồ thị $G$ mô tả tương tác đo, điều khiển, và trao đổi thông tin giữa các tác tử trong đội hình. Mỗi tác tử $i$ có thông tin về đội hình đặt $\m{p}^*$ thông qua các vector sai lệch đặt $\m{p}_i^* - \m{p}_j^*$, $(i,j) \in E$. Đội hình đặt $(G,\m{p}^*)$ là cứng hướng vi phân. Giả thiết rằng mỗi tác tử $i$ ngoài đo vị trí tương đối $\m{p}_i - \m{p}_j,~j\in {N}_i$ còn có thể trao đổi thông tin với các tác tử khác trong đội hình. Mỗi tác tử $i$ trong đội hình thực hiện ước lượng vị trí $\hat{\m{p}}_i \in \mb{R}^d$ và trao đổi biến ước lượng này với các tác tử $j \in N_i$. Dựa trên biến ước lượng, các tác tử thực hiện luật điều khiển đội hình cho bởi
\begin{subequations}
\begin{align}
\dot{\hat{\m{p}}}_i &= - \sum_{j \in N_i} \m{P}_{\m{g}_{ij}^*}(\hat{\m{p}}_i - \hat{\m{p}}_j - (\m{p}_i - \m{p}_j)) + \m{u}_i,\label{eq:ex_6.11a}\\
\dot{\m{p}}_i &= \m{u}_i, \label{eq:ex_6.11b}\\
\m{u}_i &= - k \hat{\m{p}}_i, \label{eq:ex_6.11c}
\end{align}
\end{subequations}
với $k>0$ là một hệ số dương và $i=1,\ldots,n$. Hãy thực hiện các yêu cầu sau:
\begin{enumerate}
\item[i.] Biểu diễn hệ \eqref{eq:ex_6.11a}--\eqref{eq:ex_6.11c} ở dạng ma trận.
\item[ii.] Đặt $\m{e} = \hat{\m{p}} - \m{p} + \m{p}^*$, hãy chứng minh $ \mcl{L}_b \m{e} \to \m{0}_n$ khi $t \to +\infty$.
\item[iii.] Chứng minh rằng $\m{p}$ dần đạt tới một đội hình cuối tương đồng về hướng với đội hình đặt.
\end{enumerate}
\end{exercise}


\begin{exercise}[Định vị mạng dựa trên vector hướng] \cite{Trinh2019pointing} \label{ex:6.12} Xét mạng gồm $n$ tác tử trong không gian $d$ chiều, vị trí của các tác tử được cho bởi các vector $\m{p}_i \in \mb{R}^d$, $i=1,\ldots,n$. Giả sử mạng $(G,\m{p})$ mô tả tương tác đo và trao đổi thông tin giữa các tác tử trong đội hình là cứng hướng phổ quát. Mạng $(G,\m{p}^*)$ là cứng hướng vi phân. Giả thiết rằng mỗi tác tử $i$ ngoài đo vector hướng $\m{g}_{ij},~j\in {N}_i$ còn có thể trao đổi thông tin với các tác tử khác trong đội hình. Mỗi tác tử $i$ trong đội hình thực hiện ước lượng vị trí $\hat{\m{p}}_i \in \mb{R}^d$ và trao đổi biến ước lượng này với các tác tử $j \in N_i$. Các tác tử thực hiện cập nhật biến ước lượng theo luật định vị mạng:
\begin{align}
\dot{\hat{\m{p}}}_i &= - \sum_{j \in N_i} \m{P}_{\hat{\m{g}}_{ij}}\m{g}_{ij}^*, \label{eq:ex_6.12}
\end{align}
trong đó $k>0$, $\hat{\m{g}}_{ij} = \frac{\hat{\m{p}}_{j}-\hat{\m{p}}_{i}}{\|\hat{\m{p}}_{j}-\hat{\m{p}}_{i}\|}$, và $\m{P}_{\hat{\m{g}}_{ij}}=\m{I}_d - \hat{\m{g}}_{ij}\hat{\m{g}}_{ij}^\top$ được tính toán dựa trên các biến trao đổi. Hãy thực hiện các yêu cầu sau:
\begin{enumerate}
\item[i.] Đặt $\hat{\m{p}}=[\m{p}_1^\top,\ldots,\m{p}_n^\top]^\top \in \mb{R}^{dn}$, hãy biểu diễn mạng $n$ nút với thuật toán định vị mạng \eqref{eq:ex_6.12} ở dạng ma trận.
\item[ii.] Chứng minh rằng $\bar{\hat{\m{p}}}=\frac{1}{n}\sum_{i=1}^n \hat{\m{p}}_i$ và $\hat{s}=\|\hat{\m{p}}-\m{1}_n\otimes \bar{\hat{\m{p}}}\|$ là bất biến theo thời gian. Từ đó chứng minh sự tồn tại của điểm cân bằng $\m{p}^*$ của \eqref{eq:ex_6.12} tương đồng về hướng với cấu hình thực $\m{p}$.
\item[iii.] Chứng minh rằng $\hat{\m{p}}$ tiệm cận tới $\m{p}^*$ với hầu hết điều kiện đầu $\hat{\m{p}}(0)$.
\item[iv.] Giả sử trong mạng có một số nút tham chiếu $i=1,\ldots,l$ với $l\ge 2$. Các nút tham chiếu có thông tin về vị trí tuyệt đối trong hệ qui chiếu, không cập nhật biến trạng thái $\hat{\m{p}}_i = \m{p}_i \forall t\geq 0$, và chỉ gửi thông tin về biến trạng thái của mình cho các nút mạng còn lại (follower). Các nút mạng follower cập nhật ước lượng vị trí theo phương trình \eqref{eq:ex_6.12}. Chứng minh rằng $\hat{\m{p}}$ tiệm cận tới $\m{p}$.
\end{enumerate}
\end{exercise}

\begin{exercise}[Định vị mạng dựa trên vector hướng với thời gian hữu hạn không phụ thuộc điều kiện đầu] \cite{Trinh2020fixed} \label{ex:6.13} Xét mạng gồm $n$ tác tử trong không gian $d$ chiều, vị trí của các tác tử được cho bởi các vector $\m{p}_i \in \mb{R}^d$, $i=1,\ldots,n$. Giả sử trong mạng có $3\le l \le n-1$ nút tham chiếu, các nút tham chiếu có thông tin về vị trí trong hệ qui chiếu toàn cục. Các nút thường $i=l+1,\ldots,n$ không có thông tin về vị trí, và cần ước lượng $\m{p}_i$ thông qua đo đạc vector hướng $\m{g}_{ij}$ và trao đổi thông tin ước lượng $\hat{\m{p}}_i\in \mb{R}^d$ với các nút láng giềng. Đồ thị đo đạc $G$ có dạng một đồ thị hữu hướng dạng leader-follower không chu trình, được xây dựng dựa trên cộng dần các nút mạng như sau:
\begin{itemize}
\item nút $l+1$ có $3\le |N_{l+1}| \le l$ cạnh $(l+1,j)$, $j \in \{1,\ldots,l\}$,
\item nút $l+2$ có $3\le |N_{l+2}| \le l+1$ cạnh $(l+2,j)$, $j \in \{1,\ldots,l+1\}$,
\item \ldots 
\item nút $n$ có $3\le |N_{n}| \le n-1$ cạnh $(n,j)$, $j \in \{1,\ldots,n-1\}$.
\end{itemize}
Giả sử các nút mạng thông thường cập nhật biến ước lượng theo phương trình
\begin{align}
\dot{\hat{\m{p}}}_i &= \sum_{j \in N_i} \m{P}_{\hat{\m{g}}_{ij}} \left(\left(\m{P}_{\hat{\m{g}}_{ij}}(\hat{\m{p}}_{j}-\hat{\m{p}}_{i}\right)^\alpha + \left(\m{P}_{\hat{\m{g}}_{ij}}(\hat{\m{p}}_{j}-\hat{\m{p}}_{i}\right)^\beta\right), \label{eq:ex_6.13}
\end{align}
trong đó $i=l+1,\ldots,n$, $\m{P}_{{\m{g}}_{ij}}=\m{I}_d - {\m{g}}_{ij}{\m{g}}_{ij}^\top$, $\alpha=1-\frac{1}{2\mu}$, $\beta=1+\frac{1}{2\mu}$ với $\mu>1$, và 
\begin{align*}
{\rm sig}^{\alpha}\left( \m{x} \right) = [{\rm sig}^{\alpha}\left(x_1\right),\ldots,{\rm sig}^{\alpha}\left(x_d\right)]^\top = [{\rm sgn}(x_1)|x_1|^{\alpha},\ldots,{\rm sgn}(x_1)|x_1|^{\alpha}]^\top.
\end{align*}
Hãy thực hiện các yêu cầu sau:
\begin{enumerate}
\item[i.] Chứng minh rằng \eqref{eq:ex_6.13} với $i=l+1$ có điểm cân bằng duy nhất $\hat{\m{p}}_{l+1} = \m{p}_{l+1}$.
\item[ii.] Chứng minh rằng điểm cân bằng duy nhất $\hat{\m{p}}_{l+1} = \m{p}_{l+1}$ là ổn định với thời gian hữu hạn và thời gian hữu hạn này không phụ thuộc vào $\hat{\m{p}}_{l+1}(0)$.
\item[iii.] Sử dụng qui nạp toán học, hãy chứng minh các nút mạng sẽ định vị được vị trí toàn cục trong một thời gian hữu hạn.
\end{enumerate}
(Gợi ý: để chứng minh một điểm cân bằng là ổn định với thời gian hữu hạn, sử dụng kết quả sau: ``Xét hệ $\dot{\m{x}} = \m{f}(\m{x}),~\m{x}(0)=\m{x}_0$, trong đó $\m{x}\in \mb{R}^d$ là vector biến trạng thái, $\m{f}:\mb{R}^d \to \mb{R}^d$ là một hàm phi tuyến, và $\m{x}=\m{0}_d$ là một điểm cân bằng của hệ. Nếu tồn tại $V:\m{x} \mapsto \mb{R}_+\cup \{{0}\}$ là hàm liên tục thỏa mãn: (i) $V(\m{x})=0$ khi và chỉ khi $\m{x}=\m{0}_d$, và (ii) $\dot{V}\le -\kappa_1 V^p(\m{x})-\kappa_2V^q(\m{x})$ với $\kappa_1,\kappa_2>0$, $p=1-(2\mu)^{-1}$, $q=1+(2\mu)^{-1}$, $\mu>1$, thì $\m{x}=\m{0}_d$ là ổn định toàn cục trong thời gian hữu hạn với thời gian xác lập bị chặn trên bởi $T_{\max} = \frac{\pi\mu}{\sqrt{\kappa_1\kappa_2}},~\forall \m{x}_0 \in \mb{R}^d$.'')
\end{exercise}

\begin{exercise}[Định vị mạng chu trình hữu hướng dựa trên vector hướng] \cite{Ko2020bearing} \label{ex:6.14} Xét mạng gồm $n$ tác tử trong không gian $n-1$ chiều, vị trí của các tác tử được cho bởi các vector $\m{p}_i \in \mb{R}^d$, $i=1,\ldots,n$. Đồ thị đo đạc $G$ là một chu trình hữu hướng, và $(H,\m{p})$ là cứng hướng vi phân, với $H$ là đồ thị vô hướng tương ứng với $G$. 

Giả sử các nút mạng cập nhật biến ước lượng theo phương trình
\begin{align}
\dot{\hat{\m{p}}}_i = \m{P}_{{\m{g}}_{i,i+1}}(\hat{\m{p}}_{i}-\hat{\m{p}}_{i+1}), \label{eq:ex_6.14}
\end{align}
trong đó $i=1,\ldots,n$, $\m{P}_{{\m{g}}_{ij}}=\m{I}_d - {\m{g}}_{ij}{\m{g}}_{ij}^\top$ tính được từ phép đo vector hướng $\m{g}_{i,i+1}$ bởi tác tử $i$ tới tác tử $i+1$, và $\hat{\m{p}}_{i+1}$ gửi thông tin về vị trí cho tác tử $i$. Hãy thực hiện các yêu cầu sau:
\begin{itemize}
\item[i.] Chứng minh rằng \eqref{eq:ex_6.14} có điểm cân bằng duy nhất $\m{p}^*$ có giá trị chỉ phụ thuộc vào $\hat{\m{p}}_i(0)$.
\item[ii.] Chứng minh rằng điểm cân bằng duy nhất $\hat{\m{p}} = \m{p}^*$ là ổn định tiệm cận.
\end{itemize}
\end{exercise}

\begin{exercise}[Định thức Cayley-Menger] \cite{Blumenthal1953} \label{ex:6.15} Với $\m{p}_0,\m{p}_1,\ldots,\m{p}_n$ là $n+1$ điểm trên $\mb{R}^d$ với $d \geq n$. Các điểm này tạo thành một đa diện đơn giản trong không gian $n$ chiều: đoạn thẳng với $n=1$, tam giác với $n=2$, tứ diện với $n=3$,\ldots Kí hiệu $d_{ij}=\|\m{p}_i-\m{p}_j\|$ và
\begin{align}
\m{M}(\m{p}_1,\ldots,\m{p}_n) = \begin{bmatrix}
0 & d_{01}^2 & d_{02}^2 & \ldots & d_{0n}^2 & 1 \\
d_{01}^2 & 0 & d_{12}^2 & \ldots & d_{1n}^2 & 1 \\
d_{02}^2 & d_{12}^2 & 0 & \ldots & d_{2n}^2 & 1 \\
\vdots & \vdots & \vdots & \ddots & \vdots & \vdots \\
d_{0n}^2 & d_{1n}^2 & d_{2n}^2 & \ldots & 0 & 1 \\
1 & 1 & 1 & \ldots & 1 & 0
\end{bmatrix}.
\end{align}
Định thức Cayley-Menger được định nghĩa là ${D}(\m{p}_1,\ldots,\m{p}_n) \triangleq {\rm det}(\m{M}(\m{p}_1,\ldots,\m{p}_n))$.
\begin{itemize}
\item[i.] Chứng minh rằng khi $n=1$ thì $D(\m{p}_0,\m{p}_1)=2d_{01}^2$;
\item[ii.] Chứng minh rằng khi $n=2$ thì $D(\m{p}_0,\m{p}_1,\m{p}_2)=-16 S^2$, với $S$ là diện tích tam giác tạo bởi $\m{p}_0,\m{p}_1,\m{p}_2$;
\item[iii.] Chứng minh rằng $\m{M}(\m{p}_0,\m{p}_1,\ldots,\m{p}_n)=\m{P}^\top\m{Q}\m{P}$, với
\begin{align*}
\m{P}&= \begin{bmatrix}
\m{p}_0 & \m{p}_1 & \ldots & \m{p}_n & 0\\
1 & 1 & \ldots & 1 & 0 \\
\|\m{p}_0\| & \|\m{p}_1\|^2 & \ldots & \|\m{p}_n\|^2 & 1
\end{bmatrix},\quad \m{Q} = \begin{bmatrix}
-2\m{I}_n & \m{0}_{n\times 2} \\
\m{0}_{2\times n} & \m{T}
\end{bmatrix},\quad \m{T} = \begin{bmatrix}
0 & 1 \\ 1 & 0
\end{bmatrix}.
\end{align*}
\item[iv.] Biết công thức tính thể tích của đa diện đơn giản tạo bởi $n+1$ điểm $\m{p}_0,\m{p}_1,\ldots,\m{p}_n$ là
\begin{align}
V_n = \frac{1}{n!} \left| 
\begin{bmatrix}
\m{p}_0 & \m{p}_1 & \ldots & \m{p}_n \\
1 & 1 & \ldots & 1
\end{bmatrix}
\right| .
\end{align}
Chứng minh ${D}(\m{p}_1,\ldots,\m{p}_n)={\rm det}(\m{P}^\top\m{Q}\m{P})=(-1)^{n+1}2^n(n!)^2 V_n^2$.
\item[v.] Sử dụng tính chất của định thức, chứng minh rằng:
\begin{align*}
V_n^2 & = \frac{1}{(n!)^22^n} {\rm det}\left( \begin{bmatrix}
2{d}_{01}^2 & {d}_{01}^2+{d}_{02}^2-{d}_{12}^2 & \ldots & {d}_{01}^2 + {d}_{02}^2 - {d}_{1n}^2 \\
{d}_{01}^2 + {d}_{02}^2 - {d}_{12}^2 & 2d_{02}^2 & \ldots & {d}_{02}^2+{d}_{0n}^2-{d}_{2n}^2 \\
\vdots & \vdots & \ddots & \vdots \\
{d}_{01}^2+{d}_{0n}^2-{d}_{1n}^2 & d_{02}^2 + d_{0n}^2-d_{2n}^2 & \ldots & 2{d}_{0n}^2
\end{bmatrix} \right)\\
& = \frac{(-1)^{n+1}}{(n!)^22^n} {D}(\m{p}_1,\ldots,\m{p}_n).
\end{align*}
\end{itemize}
\end{exercise}

\begin{exercise}\cite{Blumenthal1953}
Chứng minh rằng với $4$ điểm $\m{p}_0,\m{p}_1,\m{p}_{2},\m{p}_3 \in \mb{R}^2$ bất kỳ thì ${\rm rank}(\m{M}(\m{p}_0,\m{p}_1,\m{p}_{2})) \leq 3$ và $D(\m{p}_0,\m{p}_1,\m{p}_{2},\m{p}_3) = 0$. 
\end{exercise}

\begin{exercise}\cite{Cao2006sensor}
Xét mạng cảm biến với $3$ nút tham chiếu tại $\m{p}_1,\m{p}_{2},\m{p}_3 \in \mb{R}^2$ không thẳng hàng và nút mạng $\m{p}_0$ tại vị trí chưa xác định. Giả sử nút 0 đo được ba khoảng cách $\bar{d}_{0i}$, $i=1,2,3$ có sai lệch, và biết chính xác các khoảng cách $d_{ij}$ với $i,j=1,2,3$. 
Giả sử $\bar{d}_{0i}^2 = d_{0i}^2 - \varepsilon_i$ với $\varepsilon_i$ là sai lệch đo.
\begin{itemize}
\item[i.] Đặt 
\[\m{E} = \begin{bmatrix}
0 & d_{12}^2 & d_{13}^2 & 1\\
d_{12}^2 & 0 & d_{23}^2 & 1 \\
d_{13}^2 & d_{23}^2 & 0 & 1 \\
1 & 1 & 1 & 0
\end{bmatrix}^{-1},\]
chứng mimh rằng
\begin{align}
\begin{bmatrix}
d_{01}^2 & d_{02}^2 & d_{03}^2 & 1
\end{bmatrix} \m{E} \begin{bmatrix}
d_{01}^2 & d_{02}^2 & d_{03}^2 & 1
\end{bmatrix}^\top = 0.
\end{align}
\item[ii.] Với $\bm{\varepsilon}=[\varepsilon_1,\varepsilon_2,\varepsilon_3]^\top$, tìm $\m{A}_3\in \mb{R}^{3\times 3}$, $\m{b}_3\in \mb{R}^3$, ${c}_3\in \mb{R}$ thỏa mãn 
\begin{align} \label{eq:ex_6.17}
f_3(\varepsilon_1,\varepsilon_2,\varepsilon_3)=\bm{\varepsilon}^\top\m{A}_3\bm{\varepsilon}+\bm{\varepsilon}^\top\m{b}_3+c_3= 0.
\end{align}
Ràng buộc \eqref{eq:ex_6.17} gọi là ràng buộc đồng phẳng của bốn điểm $\m{p}_0,\ldots,\m{p}_3$.
\item[iii.] Giả sử mạng có $n \geq 3$ nút tham chiếu và nút mạng 0 có thể đo các khoảng cách (bị lệch bởi nhiễu) $\bar{d}_{0i}^2 = d_{0i}^2 - \varepsilon_i$ tới các nút tham chiếu. Ảnh hưởng của sai lệch làm các ràng buộc \eqref{eq:ex_6.17} không nhất quán với nhau. Để xác định khoảng cách chính xác tới các nút tham chiếu, chúng ta tối thiểu hóa hàm tổng bình phương sai lệch $J=\sum_{i=1}^n \varepsilon_i^2$, với các ràng buộc đồng phẳng: $f(\varepsilon_1,\varepsilon_2,\varepsilon_i)=0$, với $i=3,\ldots,n$. Hãy tìm giá trị tối ưu $\varepsilon_1^*,\ldots,\varepsilon_n^*$ của bài toán.
\end{itemize}
\end{exercise}

\chapter{Tối ưu phân tán}
\label{chap:distributed_optimization}
\section{Bài toán tối ưu hóa trong hệ đa tác tử}

Một bài toán tối ưu hóa thường được cho dưới dạng
\begin{mini!}
{\m{x}\in \mb{R}^n}{f(\m{x})}
{\label{eq:toi_uu_phan_tan}}{}
\addConstraint{g_k(\m{x})}{\leq 0,\,k=1,\ldots,p}
\addConstraint{h_r(\m{x})}{=0,\, r=1,\ldots, q,}
\end{mini!}
trong đó $\m{x} \in \mb{R}^n$ là biến tối ưu, $f:\mb{R}^n \to \mb{R}$ là hàm mục tiêu, $g_k(\m{x}) \leq 0$, $h_r(\m{x})=0$ lần lượt là các ràng buộc bất đẳng thức và đẳng thức mà nghiệm của bài toán phải thỏa mãn. Nếu bài toán \eqref{eq:toi_uu_phan_tan} khuyết các ràng buộc $g_k(\m{x}) \leq 0$, $h_r(\m{x})=0$, chúng ta có một bài toán tối ưu không ràng buộc. 

Xét bài toán tối ưu \eqref{eq:toi_uu_phan_tan} trong đó tập chấp nhận được $D\subseteq \mb{R}^n$ xác định bởi
\begin{align}
    D = \{\m{x}\in \mb{R}^n|~g_k(\m{x})\leq 0,\,k=1,\ldots,p,~h_r(\m{x})=0,\, r=1,\ldots, q \}
\end{align}
thỏa mãn $D \neq \emptyset$. Khi đó, bài toán \eqref{eq:toi_uu_phan_tan} có nghiệm tối ưu nếu tồn tại $\m{x}^* \in D$ sao cho $f(\m{x}^*)\leq f(\m{x}) \forall \m{x} \in D$. Tùy theo các giả thiết về hàm mục tiêu và tập các ràng buộc mà bài toán tối ưu có thể được phân chia thành các lớp nhỏ hơn. Trong tài liệu này, chúng ta chỉ xét trường hợp hàm mục tiêu $f$ là hàm lồi và tập $D$ xác định bởi các ràng buộc là tập lồi \cite{Nguyen2023,Luenberger2008linear}.

Với định nghĩa trên, bài toán định vị mạng cảm biến ở Chương \ref{chap:network_localization} là một bài toán tối ưu không ràng buộc, trong đó hàm mục tiêu là tổng bình phương sai lệch giữa các biến đo và các biến ước lượng. Trong các ứng dụng của hệ đa tác tử, việc tìm nghiệm của bài toán \eqref{eq:toi_uu_phan_tan} cần được thực hiện một cách phân tán. Cụ thể hơn, mỗi tác tử chỉ có thông tin về biến trạng thái riêng $\m{x}_i$, hàm mục tiêu riêng $f_i(\m{x}_i)$, các ràng buộc riêng $g_i(\m{x}_i), h_i(\m{x}_i)$, và một số biến trạng thái tương đối $\m{x}_i - \m{x}_j$ thu được qua mạng truyền thông hoặc đo đạc với các tác tử láng giềng $j \in {N}_i$. Nhờ thiết kế hàm mục tiêu riêng và các ràng buộc một cách hợp lý, việc giải bài toán tối ưu \eqref{eq:toi_uu_phan_tan} được thực hiện thông qua giải các bài toán tối ưu riêng, có độ phức tạp thấp hơn, của mỗi tác tử trong hệ.

\section{Tối ưu phân tán với ràng buộc đẳng thức}
\label{sec:dist_optim_consensus}
Xét hệ gồm $n$ tác tử với đồ thị trao đổi thông tin $G=(V,E)$ vô hướng, liên thông với ma trận Laplace $\mcl{L}\in \mb{R}^{n \times n}$. Mỗi tác tử $i\in V$ có hàm mục tiêu riêng ${f}_i:\mb{R}^d \to \mb{R}$, với $f_i$ được giả thiết là hàm lồi và khả vi liên tục. Mục tiêu của hệ là giải bài toán tối ưu
\begin{mini!}
{\m{z}\in \mb{R}^d}{f(\m{z})=\sum_{i=1}^nf_i(\m{z})}
{\label{eq:OP1}}{}
\end{mini!}
một cách phân tán, với giả thiết hàm mục tiêu riêng là thông tin không được chia sẻ với các tác tử khác trong hệ.

Mỗi tác tử sử dụng biến trạng thái $\m{x}_i(t)$ để ước lượng của nghiệm tối ưu của \eqref{eq:OP1}. Việc cập nhật biến $\m{x}_i(t)$ dựa trên tối ưu hàm mục tiêu riêng và trao đổi thông tin về biến cập nhật với các tác tử láng giềng. Do $f$ là tổng của các hàm lồi khả vi liên tục, ${f}$ cũng là hàm lồi khả vi liên tục. Đặt $\m{x}=[\m{x}_1^\top,\ldots,\m{x}_n^\top]^\top\in\mb{R}^{dn}$, ${\tilde{f}(\m{x})=\sum_{i=1}^n f_i(\m{x}_i)}$ thì do $\tilde{f}(\m{1}_n\otimes \m{z}^*)=f(\m{z}^*)$, nghiệm của \eqref{eq:OP1} đạt được tại một điểm $\m{z}^*\in \mb{R}^d$ khi và chỉ khi $\m{x}_1 = \ldots = \m{x}_n = \m{z}^*$. Do đó, bài toán \eqref{eq:OP1} là tương đương với bài toán
\begin{mini!}
{\m{x}\in \mb{R}^{dn}}{\tilde{f}(\m{x})=\sum_{i=1}^nf_i(\m{x}_i)}
{\label{eq:OP2}}{}
\addConstraint{\bar{\mcl{L}}\m{x}}{=\m{0}_{dn}}
\end{mini!}
trong đó $\bar{\mcl{L}}=\mcl{L}\otimes \m{I}_d \in \mb{R}^{dn \times dn}$. 

Để giải bài toán \eqref{eq:OP2}, chúng ta định nghĩa hàm Lagrange mở rộng (augmented Lagrangian) 
\begin{align}\label{eq:Augmented_Lagrangian}
L(\m{x},\bm{\lambda}) = \tilde{f}(\m{x}) + \bm{\lambda}^\top\bar{\mcl{L}}\m{x} + \frac{1}{2}\m{x}^\top\bar{\mcl{L}}\m{x},
\end{align}
với $\bm{\lambda} = [\bm{\lambda}_1^\top,\ldots,\bm{\lambda}_n^\top]^\top \in \mb{R}^{dn}$ là nhân tử Lagrange và thành phần $\frac{1}{2}\m{x}^\top\bar{\mcl{L}}\m{x}$ là hàm bổ sung để phạt mức độ bất đồng thuận của các biến ước lượng của các tác tử trong hệ.

Bài toán đối ngẫu của \eqref{eq:OP2} là
\begin{equation}
\max_{\bm{\lambda} \min_{\m{x} \in \mb{R}^{dn}} \in \mb{R}^{dn}} L(\m{x},\bm{\lambda})
\end{equation}

Điểm $(\m{x}^*,\bm{\lambda}^*)\in \mb{R}^{dn} \times \mb{R}^{dn}$ là điểm yên ngựa địa phương của $L$ nếu tồn tại một lân cận mở $B_{\m{x}^*} \subset \mb{R}^{dn}$ và $B_{\bm{\lambda}^*} \subset \mb{R}^{dn}$ của $\m{x}^*$ và $\bm{\lambda}^*$ sao cho
\begin{align}\label{eq:Saddle_point}
L(\m{x}^*,\bm{\lambda}) \leq L(\m{x}^*,\bm{\lambda}^*) \leq L(\m{x},\bm{\lambda}^*),\, \forall \m{x} \in B_{\m{x}^*}, \forall \bm{\lambda}^* \in B_{\bm{\lambda}^*}.
\end{align}
Khi $B_{\m{x}^*} = \mb{R}^{dn}$ và $B_{\bm{\lambda}^*} = \mb{R}^{dn}$, $(\m{x}^*,\bm{\lambda}^*)$ là điểm yên ngựa toàn cục của $L$.

Nếu $(\m{x}^*,\bm{\lambda}^*)$ là một điểm yên ngựa của $L$, ta có $L(\m{x}^*,\bm{\lambda}^*+\m{1}_n \otimes \m{a}) = L(\m{x}^*,\bm{\lambda}^*)$. Do đó, $(\m{x}^*,\bm{\lambda}^*+\m{1}_n \otimes \m{a})$ cũng là một điểm yên ngựa của $L$. \index{điểm!yên ngựa} 

Các điểm tới hạn $(\m{x}^*,\bm{\lambda}^*)$ của hàm Lagrange $L$ thỏa mãn\index{điểm!tới hạn}
\begin{subequations} \label{eq:Critical_point}
\begin{align}
\frac{\partial L(\m{x}^*,\bm{\lambda}^*)}{\partial \m{x}} &= \frac{\partial \tilde{f}(\m{x}^*)}{\partial \m{x}} + (\bm{\lambda}^*)^\top \bar{\mcl{L}} + (\m{x}^*)^\top \bar{\mcl{L}} = \m{0}_{dn}^\top, \label{eq:Critical_point1} \\
\frac{\partial L(\m{x}^*,\bm{\lambda}^*)}{\partial \bm{\lambda}} &= (\m{x}^*)^\top \bar{\mcl{L}} = \m{0}_{dn}^\top. \label{eq:Critical_point2}
\end{align}
\end{subequations}
Từ phương trình \eqref{eq:Critical_point2}, ta có $\m{x}^* \in {\rm ker}(\bar{\mcl{L}})$ nên $\m{x}^* = \m{1}_n \otimes \m{a}$, với $\m{a}\in \mb{R}^d$. Thay vào \eqref{eq:Critical_point1}, ta có
\begin{subequations}
\begin{align}
\frac{\partial \tilde{f}(\m{x}^*)}{\partial \m{x}} = \frac{\partial \tilde{f}(\m{1}_n \otimes \m{a})}{\partial \m{x}} &= -(\bm{\lambda}^*)^\top \bar{\mcl{L}} \label{eq:Critical_point3} \\
\frac{\partial \tilde{f}(\m{1}_n \otimes \m{a})}{\partial \m{x}} (\m{1}_n\otimes \m{I}_d) = \frac{\partial f(\m{a})}{\partial \m{z}} = \sum_{i=1}^n \frac{\partial {f}_i(\m{a})}{\partial \m{z}} &= -(\bm{\lambda}^*)^\top \bar{\mcl{L}}(\m{1}_n\otimes \m{I}_d) = \m{0}_{d}^\top. \label{eq:Critical_point4}
\end{align}
\end{subequations}
Như vậy, $\m{x}^* = \m{1}_n \otimes \m{a}$ là nghiệm của bài toán \eqref{eq:OP2}. Đồng thời, phương trình \eqref{eq:Critical_point3} chứng tỏ sự tồn tại của $\bm{\lambda}^*$ thỏa mãn $\bar{\mcl{L}} \bm{\lambda}^* = -\left( \frac{\partial \tilde{f}(\m{x}^*)}{\partial \m{x}} \right)^\top = -\nabla_{\m{x}}\tilde{f}(\m{x}^*)$. Chú ý rằng $L:\mb{R}^{dn} \times \mb{R}^{dn} \to \mb{R}$ là hàm lồi theo biến $\m{x}$ và là hàm affine (do đó là hàm lõm) theo biến $\bm{\lambda}$, nên các điểm tới hạn $(\m{x}^*,\bm{\lambda}^*)$ cũng là các điểm yên ngựa của $L$.

Việc tìm nghiệm bài toán tối ưu \eqref{eq:OP2} được qui về tìm một điểm yên ngựa của $L$. Do đó, thuật toán tối ưu thực hiện đồng thời hai quá trình: tìm cực tiểu theo $\m{x}$ và tìm cực đại theo $\bm{\lambda}$. Xét thuật toán tối ưu phân tán: \index{thuật toán!tối ưu phân tán}
\begin{subequations} \label{eq:Dist_optim_alg}
\begin{align}
\dot{\m{x}}_i &= -\sum_{j \in N_i} (\m{x}_i - \m{x}_j) - \sum_{j \in N_i} (\bm{\lambda}_i - \bm{\lambda}_j) - \nabla {f}_i(\m{x}_i),\label{eq:Dist_optim_alg1}\\
\dot{\bm{\lambda}}_i &= \sum_{j \in N_i} (\m{x}_i - \m{x}_j),\label{eq:Dist_optim_alg2}
\end{align}
\end{subequations}
với $i = 1,\ldots,n$. Như vậy, ngoài biến ước lượng cực tiểu $\m{x}_i$, mỗi tác tử duy trì thêm biến $\bm{\lambda}_i \in \mb{R}^d$. Cả hai biến $\m{x}_i$ và $\bm{\lambda}_i$ đều được trao đổi với các tác tử láng giềng nhằm xác định thuật toán \eqref{eq:Dist_optim_alg}. Ở phương trình \eqref{eq:Dist_optim_alg}, $\bm{\lambda}_i$ có dạng một khâu tích phân của tổng sai lệch biến trạng thái $\m{x}_i-\m{x}_j$. 
Thuật toán \eqref{eq:Dist_optim_alg} có thể viết lại dưới dạng ma trận như sau:
\begin{subequations} \label{eq:Dist_optim_alg_matrix_form}
\begin{align}
\dot{\m{x}} &= -\left(\frac{\partial L}{\partial \m{x}} \right)^\top =-\bar{\mcl{L}}\m{x} - \bar{\mcl{L}}\bm{\lambda} - \nabla \tilde{f}(\m{x}),\label{eq:Dist_optim_alg_matrix_form1}\\
\dot{\bm{\lambda}} &= \left(\frac{\partial L}{\partial \bm{\lambda}} \right)^\top = \bar{\mcl{L}}\m{x},\label{eq:Dist_optim_alg_matrix_form2}
\end{align}
\end{subequations}
trong đó $\nabla \tilde{f}(\m{x}) = [\nabla f_1(\m{x}_1)^\top,\ldots,\nabla f_n(\m{x}_n)^\top]^\top$. Với biểu diễn dạng ma trận, $\dot{\m{x}}$ ngược hướng gradient của $L$ theo $\m{x}$ và $\dot{\bm{\lambda}}$ cùng hướng với gradient của $L$ theo $\bm{\lambda}$. Các điểm cân bằng của \eqref{eq:Dist_optim_alg_matrix_form} cũng chính là các điểm yên ngựa của hàm Lagrange $L$.

Gọi $(\m{x}^*,\bm{\lambda}^*)$ là một điểm yên ngựa của hàm Lagrange $L$, với hàm Lyapunov $V = \frac{1}{2}\|\m{x}-\m{x}^*\|^2 + \frac{1}{2}\|\bm{\lambda}-\bm{\lambda}^*\|^2$, ta có
\begin{align} \label{eq:DistOpt_dotV}
\dot{V} = -(\m{x}-\m{x}^*)^\top \nabla_{\m{x}}L(\m{x},\bm{\lambda}) + (\bm{\lambda}-\bm{\lambda}^*)^\top \nabla_{\bm{\lambda}}L(\m{x},\bm{\lambda}).
\end{align}
Vì $L$ là hàm lồi theo $\m{x}$ và affine theo $\bm{\lambda}$ nên
\begin{subequations} \label{eq:Convex_property_L}
\begin{align}
L(\m{x},\bm{\lambda}) - L(\m{x}^*,\bm{\lambda}) & \leq   \frac{\partial L(\m{x},\bm{\lambda})}{\partial \m{x}} (\m{x} - \m{x}^*) = (\nabla_{\m{x}}L(\m{x},\bm{\lambda}))^\top (\m{x} - \m{x}^*) \\
L(\m{x},\bm{\lambda}) - L(\m{x},\bm{\lambda}^*) & =  \frac{\partial L(\m{x},\bm{\lambda})}{\partial \bm{\lambda}} (\bm{\lambda} - \bm{\lambda}^*) = (\nabla_{\bm{\lambda}}L(\m{x},\bm{\lambda}))^\top(\bm{\lambda} - \bm{\lambda}^*).
\end{align}
\end{subequations}
Thế phương trình \eqref{eq:Convex_property_L} vào $\dot{V}$ (phương trình \eqref{eq:DistOpt_dotV}, ta thu được
\begin{align}
\dot{V} \leq [L(\m{x}^*,\bm{\lambda}) -  L(\m{x}^*,\bm{\lambda}^*)] - [L(\m{x},\bm{\lambda}^*) - L(\m{x}^*,\bm{\lambda}^*)] \leq 0,
\end{align}
theo tính chất của điểm yên ngựa. Do đó, nghiệm $(\m{x}(t),\bm{\lambda}(t))$ của \eqref{eq:Dist_optim_alg_matrix_form} là bị chặn. Hơn nữa, do 
\[(\m{1}_n^\top\otimes \m{I}_d)\dot{\bm{\lambda}}=\m{0}_{d},\]
nên $\sum_{i=1}^n\bm{\lambda}_i(t) = \sum_{i=1}^n\bm{\lambda}_i(0),~ \forall t \geq 0$. 

Theo nguyên lý bất biến LaSalle, mọi quĩ đạo xuất phát từ $(\m{x}_0,\bm{\lambda}_0)$ tiệm cận tới tập bất biến lớn nhất trong $\{(\m{x},\bm{\lambda})\in \mb{R}^{dn}\times \mb{R}^{dn}|~\dot{V} \equiv 0,~ \sum_{i=1}^n\bm{\lambda}_i(t) = \sum_{i=1}^n\bm{\lambda}_i(0)\}$. Điều kiện để $\dot{V} \equiv 0$ là
\begin{subequations}\label{eq:DistOpt_Lasalle}
\begin{align}
L(\m{x}^*,\bm{\lambda}) -  L(\m{x}^*,\bm{\lambda}^*) &= (\m{x}^*)^\top\bar{\mcl{L}}(\bm{\lambda}-\bm{\lambda}^*) \equiv 0, \label{eq:DistOpt_Lasalle1}\\
L(\m{x},\bm{\lambda}^*) - L(\m{x}^*,\bm{\lambda}^*) &\equiv 0.\label{eq:DistOpt_Lasalle2}
\end{align}
\end{subequations}
Do $\m{x}^*$ thuộc tập điểm yên ngựa của $L$, điều kiện \eqref{eq:DistOpt_Lasalle1} luôn thỏa mãn. Điều kiện \eqref{eq:DistOpt_Lasalle2} tương đương với
\begin{align}
\tilde{f}(\m{x}) - \tilde{f}(\m{x}^*) + (\bm{\lambda}^*)^\top \bar{\mcl{L}} \m{x} + \frac{1}{2}\m{x}^\top\bar{\mcl{L}}\m{x} \equiv 0. \label{eq:DistOpt_Lasalle3}
\end{align}
Theo tính chất của hàm lồi và tính chất của điểm yên ngựa \eqref{eq:Critical_point3}, 
\begin{align} \label{eq:DistOpt_Lasalle4}
\tilde{f}(\m{x}) - \tilde{f}(\m{x}^*) \geq (\nabla \tilde{f}(\m{x}))^\top (\m{x}-\m{x}^*)= -(\bm{\lambda}^*)^\top \bar{\mcl{L}}\m{x}
\end{align}
nên để \eqref{eq:DistOpt_Lasalle3} thỏa mãn thì $\frac{1}{2} \m{x}^\top\bar{\mcl{L}}\m{x} \equiv 0$, hay $\m{x}\equiv \m{1}_n \otimes \m{b}(t)$, với $\m{b}(t) \in \mb{R}^{d}$ thỏa mãn \eqref{eq:DistOpt_Lasalle4}, tức là 
\begin{align} \label{eq:DistOpt_Lasalle5}
f(\m{b})=\tilde{f}(\m{1}_n \otimes \m{b}(t)) = \tilde{f}(\m{x}^*),~\forall t\geq 0.
\end{align}
Thế vào phương trình \eqref{eq:Dist_optim_alg_matrix_form1}, ta suy ra
\begin{align}
\m{1}_n\otimes\dot{\m{b}} &\equiv -\bar{\mcl{L}}\bm{\lambda}^* - \nabla \tilde{f}(\m{1}_n\otimes \m{b}) \label{eq:DistOpt_Lasalle6}\\
n\dot{\m{b}} &\equiv -(\m{1}_n^\top\otimes \m{I}_d)\bar{\mcl{L}}\bm{\lambda}^* - (\m{1}_n^\top\otimes \m{I}_d) \nabla \tilde{f} (\m{1}_n \otimes \m{b}(t))=-\sum_{i=1}^n \nabla {f}_i(\m{b}) = -\nabla f(\m{b}), \label{eq:DistOpt_Lasalle7}
\end{align}
Hệ $\dot{\m{b}} = -\frac{1}{n}\nabla f(\m{b})$ là một hệ gradient với hàm thế $f(\m{z})$, có tập điểm cân bằng là các điểm cực tiểu của $f(\m{z})$. Do \eqref{eq:DistOpt_Lasalle5}, $\m{b}(0)$ thuộc tập các cực tiểu của $f(\m{z})$. Nếu tồn tại thời điểm $t_1>0$ nhỏ nhất để $\m{b}(t_1)\neq \m{b}(0)$ thì phải có $\dot{f}(\m{b}(t_1))=-\|\nabla f(\m{b}_1)\|^2<0$. Tuy nhiên, do $\m{b}(t)$ thuộc tập các điểm cực tiểu của $f$, $\forall t\ge 0$, nên $\nabla f(\m{b}_1)=\m{0}_d$, ta có một mâu thuẫn. Như vậy, ta có $\dot{\m{b}} = \m{0}_d$. Thay vào \eqref{eq:DistOpt_Lasalle6}, đẳng thức $\bar{\mcl{L}}\bm{\lambda}^*= - \nabla \tilde{f}(\m{1}_n\otimes \m{b})$ thỏa mãn, nên $(\m{1}_n\otimes \m{b},\bm{\lambda}^*)$ là một điểm yên ngựa của $L$, với $\sum_{i=1}^n\bm{\lambda}_i^* = \sum_{i=1}^n\bm{\lambda}_i(0)$. 

Như vậy, với thuật toán \eqref{eq:Dist_optim_alg_matrix_form}, $(\m{x}(t),\bm{\lambda}(t))$ tiệm cận tới một điểm yên ngựa của $L$. Nếu $f(\m{z})$ có hữu hạn điểm cực tiểu, $\m{x}(t)$ tiệm cận tới một nghiệm của \eqref{eq:OP2}.
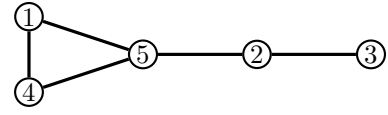
\begin{SCfigure}[][t!]
\caption{Đồ thị mô tả tương tác của năm tác tử trong Ví dụ~\ref{VD:7.1}.\label{fig:VD_7.1_graph}}
\hspace{5cm}
\begin{tikzpicture}[
roundnode/.style={circle, draw=black, thick, minimum size=3.5mm,inner sep= 0.25mm}
]
	
    \node[roundnode] (u1) at (0,1) {1}; %
	\node[roundnode] (u4) at (0,0) {4 }; %
	\node[roundnode] (u5) at (1.5,0.5) {5 }; %
	\node[roundnode] (u2) at (3,0.5) { 2}; %
	\node[roundnode] (u3) at (4.5,0.5) { 3}; %
	\node (u) at (5.5,0) { };
    \draw [very thick,-] (u1)--(u5)--(u4)--(u1);
    \draw [very thick,-] (u5)--(u2)--(u3);
    ;
\end{tikzpicture}
\end{SCfigure}
\begin{example} \label{VD:7.1}
Xét hệ gồm năm tác tử với đồ thị tương tác cho trên Hình \ref{fig:VD_7.1_graph}. Các hàm mục tiêu của mỗi tác tử là các hàm lồi chặt được chọn lần lượt bởi:
\begin{align*}
f_i(z) = z^2 + b_i z + c_i, i=1,\ldots, 4,\,\text{ và }
f_5(z) = \mathtt{e}^{0.25z},
\end{align*}
với $b_1=-1, b_2 = -9, b_3 = -9, b_4 = -8$ và $c_1=5, c_2=1, c_3=3, c_4 = 5$. Đồ thị $f_i(z),i=1,\ldots,5$ và $f(z)=\sum_{i=1}^5f_i(z)$ trong khoảng $[-5,5]$ được biểu diễn trên Hình \ref{fig:c7_VD7.1} (a). Điểm cực tiểu toàn cục của $f(z)$ được xấp xỉ tại $(z^*,f(z^*))=(3.3036,-29.2582)$.

Mô phỏng hệ với thuật toán tối ưu phân tán \eqref{eq:Dist_optim_alg} trong khoảng thời gian $0\leq t \leq 60$ (s), chúng ta thu được kết quả ở Hình \ref{fig:c7_VD7.1} (b)--(d). Nhận xét rằng các biến $x_i$ tiệm cận tới một giá trị đồng thuận và $\lambda_i(t)$ tiệm cận tới các giá trị hằng. Tại $t=60$, 
\begin{align*}
\m{x}(t=60) &= [3.3028,3.3041,3.3048,3.3028,3.3031]^\top, \\ \bm{\lambda}(t=60) &=[-4.2782,    3.7753,    6.1653,   -1.9449,   -1.0062]^\top,
\end{align*}
và $\tilde{f}(\m{x}) = -29.2658$, tương đối gần với cực tiểu toàn cục của $f(z)$ theo tính toán.
\begin{figure}[th!]
    \subfloat[]{\includegraphics[width=0.45\textwidth]{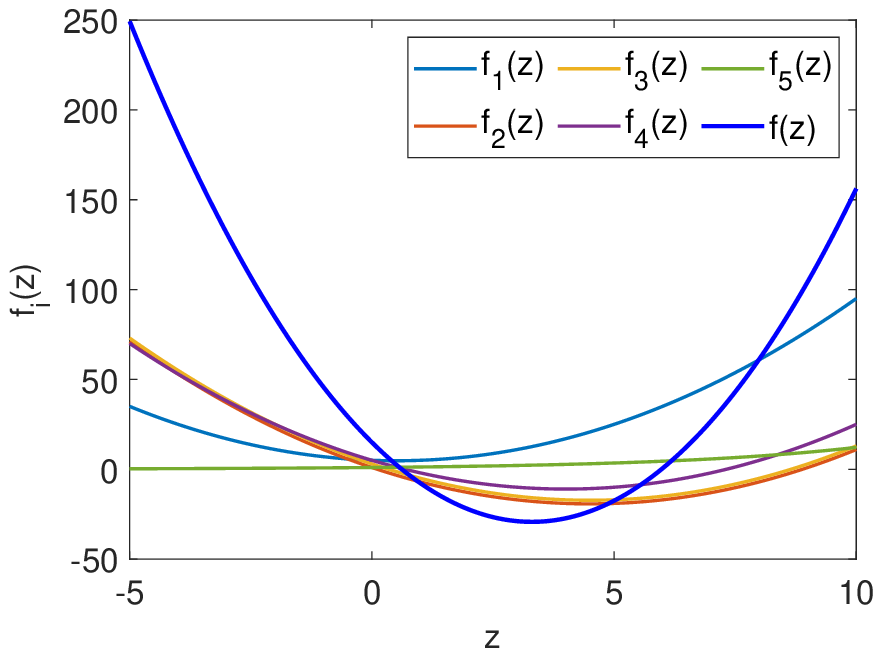}}\hfill
    \subfloat[]{\includegraphics[width=0.45\textwidth]{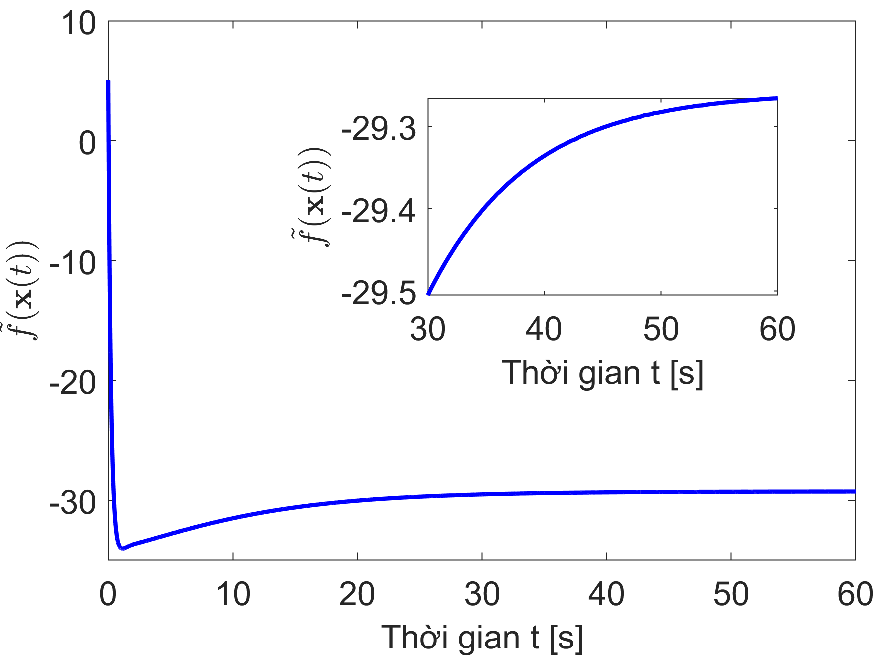}}\\
    \subfloat[]{\includegraphics[width=0.45\textwidth]{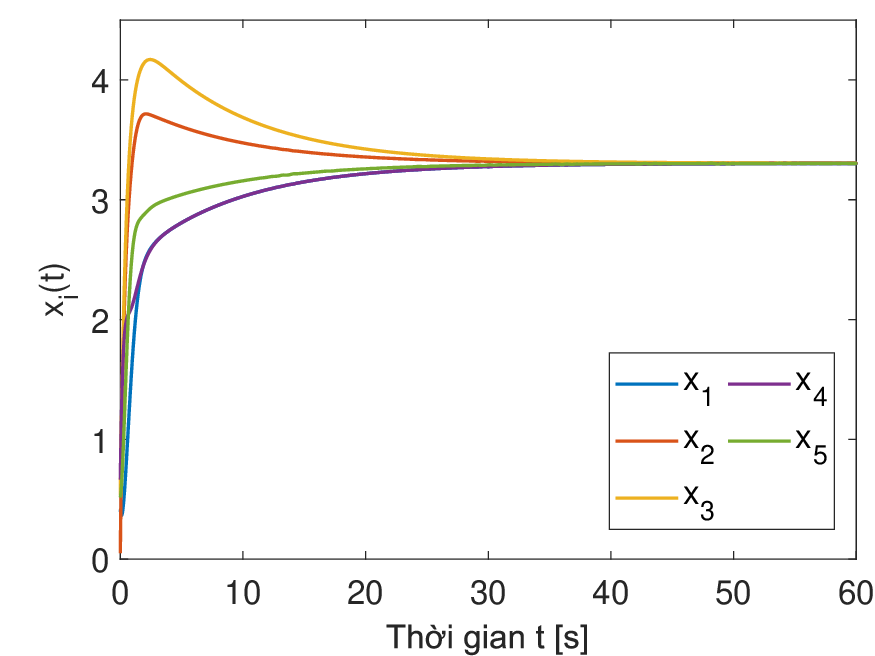}}\hfill
    \subfloat[]{\includegraphics[width=0.45\textwidth]{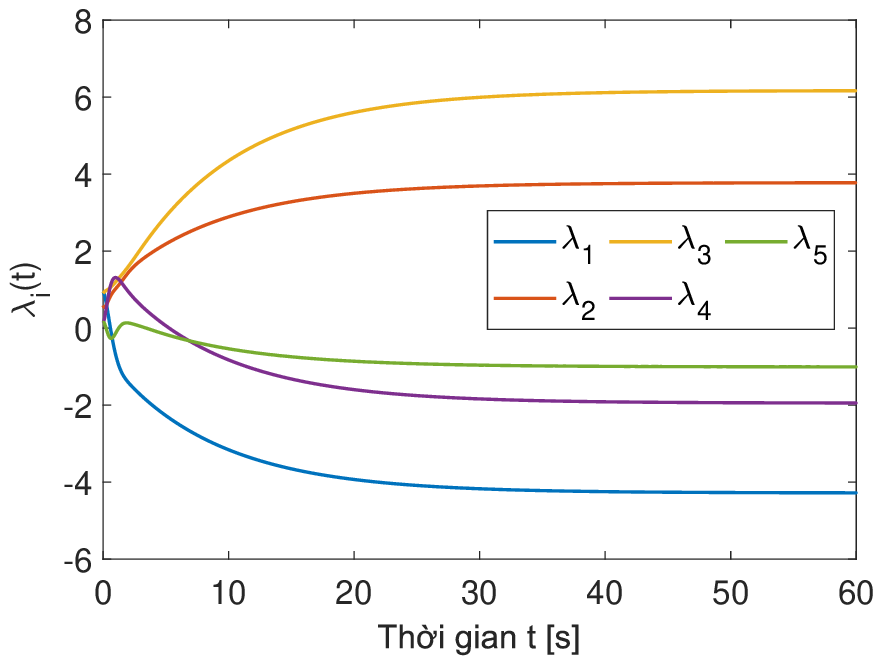}}
    \caption{Mô phỏng hệ 5 tác tử trong Ví dụ \ref{VD:7.1}. (a) Giá trị các hàm mục tiêu $f_i(z)$ và $f(z)$ (b): Giá trị hàm mục tiêu $\tilde{f}(\m{x}(t))$; (c) Các biến trạng thái ước lượng $x_i(t)$; và (d) Các biến $\lambda_i(t)$.    \label{fig:c7_VD7.1}}
\end{figure}
\end{example}

Tiếp theo, chúng ta xét một ví dụ sử dụng tối ưu phân tán với ràng buộc đẳng thức để giải hệ phương trình tuyến tính.

\begin{example}[Giải phân tán hệ phương trình tuyến tính] \label{VD:7.2} Xét hệ $n$ tác tử với các hàm mục tiêu riêng $f_i(\m{z}) = \frac{1}{2} \|\m{A}_i \m{z} - \m{b}_i\|^2$, trong đó $\m{z} \in \mb{R}^d$, $\m{A}_i \in \mb{R}^{d_i\times d}$, $\m{b}_i \in \mb{R}^{d_i}$. Nghiệm tối ưu $\m{z}^*\in\mb{R}^{d}$ của bài toán
\begin{mini!}
{\m{z}\in \mb{R}^{dn}}{{f}(\m{z})=\sum_{i=1}^nf_i(\m{z})=\frac{1}{2} \sum_{i=1}^n \|\m{A}_i \m{z} - \m{b}_i\|^2}
{\label{eq:OP_VD7.2}}{}
\end{mini!}
là nghiệm của đồng thời $n$ phương trình $\m{A}_i\m{x}_i=\m{b}_i,i=1,\ldots,n,$ khi $n$ phương trình có nghiệm chung. Trong trường hợp $n$ phương trình không có nghiệm chung, nghiệm tìm được tối thiểu hóa tổng bình phương sai lệch của $n$ ràng buộc.

Để giải phân tán \eqref{eq:OP_VD7.2}, chúng ta viết bài toán tối ưu dưới dạng \eqref{eq:OP2} và thuật toán tối ưu phân tán được cho tương ứng là:
\begin{subequations}
\begin{align}
\dot{\m{x}}_i &= -\sum_{j\in N_i}(\m{x}_i - \m{x}_j) - \sum_{j\in N_i}(\bm{\lambda}_i - \bm{\lambda}_j) - \m{A}_i^\top(\m{A}_i\m{x}_i - \m{b}_i),\\
\dot{\bm{\lambda}}_i &= \sum_{j\in N_i}(\m{x}_i - \m{x}_j),
\end{align}
\end{subequations}
với $\bm{\lambda}_i \in \mb{R}^d$, $i=1,\ldots,n$.
\end{example}

\section{Tối ưu phân tán có ràng buộc bất đẳng thức}\label{sec:dist_optim_consensus_constraints}
Ở mục này, chúng ta xét bài toán tối ưu phân tán có thêm các ràng buộc bất đẳng thức:
\begin{mini!}
{\m{x}\in \mb{R}^{dn}}{\tilde{f}(\m{x})=\sum_{i=1}^nf_i(\m{x}_i)}
{\label{eq:OP3}}{}
\addConstraint{\bar{\mcl{L}}\m{x}}{=\m{0}_{dn}}
\addConstraint{\m{g}_i(\m{x}_i)}{\leq \m{0}_{m_i},\, i=1,\ldots,n}
\end{mini!}
trong đó $f_i:\mb{R}^{d}\to \mb{R},i=1,\ldots,n$ là các hàm lồi, có đạo hàm liên tục đến cấp hai, $\m{g}_i(\m{x}_i)= [g_{i1}(\m{x}_i),\ldots,\m{g}_{im_i}(\m{x}_i)]^\top$ có các hàm thành phần $g_{ik}$ là các hàm lồi, khả vi liên tục. Hơn nữa, hàm $f(\m{z})=\sum_{i=1}^n f_i(\m{z})$ là hàm lồi chặt. Như vậy, ngoài việc tối thiểu hóa phân tán hàm $f$ thông qua đồng thuận, biến trạng thái của mỗi tác tử bị giới hạn trong tập lồi định nghĩa bởi các bất đẳng thức (\ref{eq:OP3}c).

Từ các giả thiết về các hàm mục tiêu và hàm ràng buộc, định nghĩa hàm Lagrange mở rộng 
\begin{align}\label{eq:Augmented_Lagrangian_constraints}
L(\m{x},\bm{\lambda},\bm{\nu}) = \tilde{f}(\m{x}) + \bm{\lambda}^\top\bar{\mcl{L}}\m{x} + \sum_{i=1}^n \bm{\nu}^\top_i\m{g}_i(\m{x}_i) + \frac{1}{2}\m{x}^\top\bar{\mcl{L}}\m{x},
\end{align}
với $\bm{\lambda} = [\bm{\lambda}_1^\top,\ldots,\bm{\lambda}_n^\top]^\top \in \mb{R}^{dn}$ và $\bm{\nu} = [\bm{\nu}_1^\top,\ldots,\bm{\nu}_n^\top]^\top \in \mb{R}^{M}$, $M=m_1+\ldots+m_n$ là các nhân tử Lagrange. Hàm Lagrange $L(\m{x},\bm{\lambda},\bm{\nu})$ là hàm lồi theo $\m{x}$ và affine (do đó là hàm lõm) theo $(\bm{\lambda},\bm{\nu})$. 

Bài toán đối ngẫu của \eqref{eq:OP2} là
\begin{equation}\label{eq:minimax_constraints}
\max_{(\bm{\lambda},\bm{\nu})\in \mb{R}^{dn+M}} \min_{\m{x} \in \mb{R}^{dn}}  L(\m{x},\bm{\lambda},\bm{\nu})
\end{equation}
Giả thiết rằng tồn tại điểm $\m{x}^* = \m{1}\otimes \m{a} \in {\rm ker}(\m{L})$ với $\m{g}_{ik}(\m{a})<0,~\forall i=1,\ldots,n,\, \forall k=1,\ldots,m_i$ thì điều kiện Slater thỏa mãn, giá trị tối ưu của bài toán gốc và bài toán đối ngẫu là như nhau. Nghiệm $(\m{x}^*,\bm{\lambda}^*,\bm{\nu}^*)$ của bài toán \eqref{eq:OP3} cũng là một điểm yên ngựa của $L$. Hơn nữa, $(\m{x}^*,\bm{\lambda}^*,\bm{\nu}^*)$ phải thỏa mãn điều kiện Karush-Kuhn-Tucker
\begin{subequations}
\begin{align}\label{eq:KKT_constraints}
\nabla \tilde{f}(\m{x}^*) + \bar{\mcl{L}}\bm{\lambda}^* + \sum_{i=1}^n  \sum_{k=1}^{m_i} {\nu}_{ik}^* \nabla {g}_{ik}(\m{x}^*) &= \m{0}_{dn},\\
\bar{\mcl{L}}\m{x}^* &= \m{0}_{dn},\\
\m{g}(\m{x}^*) \leq \m{0}_M,\, \bm{\nu}^* \ge \m{0}_M, (\bm{\nu}^*)^\top \m{g}(\m{x}^*) &= 0.
\end{align}
\end{subequations}
Với điều kiện hàm $f$ là hàm lồi chặt, tồn tại duy nhất giá trị tối ưu $\m{x}^*\in \mb{R}^{dn}$. 

Sử dụng hàm Lagrange $L$, thuật toán giải bài toán tối ưu \eqref{eq:OP3} được đề xuất như sau:
\begin{subequations} \label{eq:Dist_Alg_constraints_Matrix}
\begin{align}
\dot{\m{x}} &= -\nabla_{\m{x}} L(\m{x},\bm{\lambda},\bm{\nu}) = - \bar{\mcl{L}}\m{x} - \bar{\mcl{L}}\bm{\lambda} - \sum_{i=1}^n  \sum_{k=1}^{m_i} {\nu}_{ik} \nabla {g}_{ik}(\m{x}) - \nabla \tilde{f}(\m{x}),  \label{eq:Dist_Alg_constraints_Matrix1} \\
\dot{\bm{\lambda}} &= \nabla_{\bm{\lambda}} L(\m{x},\bm{\lambda},\bm{\nu}) = \bar{\mcl{L}}\m{x}, \label{eq:Dist_Alg_constraints_Matrix2} \\
\dot{\bm{\nu}} &= [\nabla_{\bm{\nu}} L(\m{x},\bm{\lambda},\bm{\nu})]^{+}_{\bm{\nu}} = [\m{g}(\m{x})]^{+}_{\bm{\nu}}. \label{eq:Dist_Alg_constraints_Matrix3}
\end{align}
\end{subequations}
Ở thuật toán \eqref{eq:Dist_Alg_constraints_Matrix}, để giới hạn việc cập nhật biến trạng thái của mỗi tác tử trong miền khả thi, chúng ta sử dụng hàm chiếu được định nghĩa như sau: Với $a, b \in \mb{R}$, 
\begin{align} \label{eq:c7_projection}
[a]_b^+ = \left\lbrace \begin{array}{ll}
a, & b>0, \\
\max\{0, a\}, & b=0.
\end{array}\right.
\end{align}
Với hai vector $\m{a}, \m{b} \in \mb{R}^p$, chúng ta viết $[\m{a}]_{\m{b}}^+$ để kí hiệu vector với các phần tử $[a_{i}]_{b_i}^+, \forall i=1,\ldots, p$. Chú ý rằng hàm chiếu là không liên tục (ví dụ khi hệ chuyển từ trạng thái $a<0$, $b>0$ tới $b=0$ và $a<0$). Do đó, hệ \eqref{eq:Dist_Alg_constraints_Matrix} có đạo hàm không liên tục. Chúng ta sẽ không đi sâu vào chi tiết về hệ với đạo hàm không liên tục \cite{Cortes2008} và thừa nhận rằng, nghiệm của \eqref{eq:Dist_Alg_constraints_Matrix} ứng với mỗi điều kiện đầu $(\m{x},\bm{\lambda},\bm{\nu})\in \mb{R}^{dn}\times \mb{R}^{dn}_{+}\times \mb{R}^{M}_+$ tồn tại, duy nhất, và các quĩ đạo nghiệm là liên tục theo từng biến trạng thái.

Thuật toán \eqref{eq:Dist_Alg_constraints_Matrix} viết cho mỗi tác tử $i$ như sau:
\begin{subequations} \label{eq:Dist_Alg_constraints}
\begin{align}
\dot{\m{x}}_i &= -\sum_{j \in N_i} (\m{x}_i - \m{x}_j) - \sum_{j \in N_i} (\bm{\lambda}_i - \bm{\lambda}_j)  -  \sum_{k=1}^{m_i} {\nu}_{ik} \nabla {g}_{ik}(\m{x}_i) - \nabla {f}_i(\m{x}_i),  \label{eq:Dist_Alg_constraints1} \\
\dot{\bm{\lambda}}_i &= \sum_{j \in N_i} (\m{x}_i - \m{x}_j), \label{eq:Dist_Alg_constraints2} \\
\dot{\bm{\nu}}_i &= [\m{g}_i(\m{x}_i)]^{+}_{\bm{\nu}_i}, \label{eq:Dist_Alg_constraints3}
\end{align}
\end{subequations}
với $i=1,\ldots,n$. Dễ thấy thuật toán~\eqref{eq:Dist_Alg_constraints} là phân tán do chỉ dùng thông tin về biến trạng thái, hàm mục tiêu, các hàm ràng buộc tại $i$ và các biến trạng thái nhận được từ các tác tử láng giềng ($\m{x}_j,\bm{\lambda}_j$, với $j \in N_i$).

Tương tự với trường hợp không có ràng buộc bất đẳng thức, để phân tích ổn định hệ \eqref{eq:Dist_Alg_constraints_Matrix}, xét hàm Lyapunov $V(\m{x},\bm{\lambda},\bm{\mu})=\frac{1}{2}(\|\m{x}-\m{x}^*\|^2 + \|\bm{\lambda}-\bm{\lambda}^*\|^2 + \|\bm{\nu}-\bm{\nu}^*\|^2)$, trong đó $(\m{x}^*,\bm{\lambda}^*,\bm{\nu}^*)$ là một điểm trong tập nghiệm tối ưu. Do quĩ đạo nghiệm không trơn, đạo hàm của $V$ theo thời gian là đạo hàm suy rộng, có giá trị
\begin{align*}
\dot{V} &= -(\m{x}-\m{x}^*)^\top\nabla_{\m{x}} L(\m{x},\bm{\lambda},\bm{\nu}) + \begin{bmatrix}
\bm{\lambda}-\bm{\lambda}^*\\
\bm{\nu}-\bm{\nu}^*
\end{bmatrix}^\top \begin{bmatrix}
\nabla_{\bm{\lambda}}L(\m{x},\bm{\lambda},\bm{\nu}) \\
[\nabla_{\bm{\nu}} L(\m{x},\bm{\lambda},\bm{\nu})]^{+}_{\bm{\nu}}
\end{bmatrix} \\
&= -(\m{x}-\m{x}^*)^\top\nabla_{\m{x}} L(\m{x},\bm{\lambda},\bm{\nu}) + \begin{bmatrix}
\bm{\lambda}-\bm{\lambda}^*\\
\bm{\nu}-\bm{\nu}^*
\end{bmatrix}^\top \begin{bmatrix}
\nabla_{\bm{\lambda}}L(\m{x},\bm{\lambda},\bm{\nu}) \\
\nabla_{\bm{\nu}} L(\m{x},\bm{\lambda},\bm{\nu})
\end{bmatrix} \\
&\qquad + (\bm{\nu}-\bm{\nu}^*)^\top \left([\nabla_{\bm{\nu}} L(\m{x},\bm{\lambda},\bm{\nu})]^{+}_{\bm{\nu}} -  \nabla_{\bm{\nu}} L(\m{x},\bm{\lambda},\bm{\nu})\right).
\end{align*}
Sử dụng tính chất hàm Lagrange $L$ là lồi theo $\m{x}$ và lõm theo $(\bm{\lambda},\bm{\nu})$, ta có
\begin{align*}
L(\m{x},\bm{\lambda},\bm{\nu}) - L(\m{x}^*,\bm{\lambda},\bm{\nu}) &\leq (\nabla_{\m{x}} L(\m{x},\bm{\lambda},\bm{\nu}))^\top (\m{x}-\m{x}^*) \\
L(\m{x},\bm{\lambda},\bm{\nu}) - L(\m{x},\bm{\lambda}^*,\bm{\nu}^*) &\geq \begin{bmatrix}
\nabla_{\bm{\lambda}}L(\m{x},\bm{\lambda},\bm{\nu}) \\
\nabla_{\bm{\nu}} L(\m{x},\bm{\lambda},\bm{\nu})
\end{bmatrix}^\top \begin{bmatrix}
\bm{\lambda}-\bm{\lambda}^*\\
\bm{\nu}-\bm{\nu}^*
\end{bmatrix}.
\end{align*}
Sử dụng các bất đẳng thức này, ta có
\begin{align}
\dot{V} &\leq -[L(\m{x},\bm{\lambda},\bm{\nu}) - L(\m{x}^*,\bm{\lambda},\bm{\nu})] + [L(\m{x},\bm{\lambda},\bm{\nu}) - L(\m{x},\bm{\lambda}^*,\bm{\nu}^*)] \nonumber \\
&\qquad + (\bm{\nu}-\bm{\nu}^*)^\top \left([\nabla_{\bm{\nu}} L(\m{x},\bm{\lambda},\bm{\nu})]^{+}_{\bm{\nu}} -  \nabla_{\bm{\nu}} L(\m{x},\bm{\lambda},\bm{\nu})\right) \nonumber \\
& \leq [L(\m{x}^*,\bm{\lambda},\bm{\nu})-L(\m{x}^*,\bm{\lambda},\bm{\nu}^*)] + [L(\m{x}^*,\bm{\lambda},\bm{\nu}^*)-L(\m{x}^*,\bm{\lambda}^*,\bm{\nu}^*)] \nonumber\\
&\qquad + [L(\m{x}^*,\bm{\lambda}^*,\bm{\nu}^*) - L(\m{x},\bm{\lambda}^*,\bm{\nu}^*)] + (\bm{\nu}-\bm{\nu}^*)^\top \left([\nabla_{\bm{\nu}} L(\m{x},\bm{\lambda},\bm{\nu})]^{+}_{\bm{\nu}} -  \nabla_{\bm{\nu}} L(\m{x},\bm{\lambda},\bm{\nu})\right) \nonumber \\
&\leq M_1 + M_2 + M_3 + M_4,\label{eq:c7dotV_constraint}
\end{align}
với $M_1\leq 0$, $M_2 \leq 0$, $M_3 \leq 0$. Biểu diễn $M_4$ dưới dạng
\begin{align*}
M_4 &= (\bm{\nu}-\bm{\nu}^*)^\top \left([\nabla_{\bm{\nu}} L(\m{x},\bm{\lambda},\bm{\nu})]^{+}_{\bm{\nu}} -  \nabla_{\bm{\nu}} L(\m{x},\bm{\lambda},\bm{\nu})\right)\\
&=\sum_{i=1}^n\sum_{k=1}^{d_i}(\nu_{ik}-\nu_{ik}^*)([g_{ik}(\m{x}_i)]_{\nu_{ik}}^+ - g_{ik}(\m{x}_i)) \\
&= \sum_{i=1}^n\sum_{k=1}^{d_i} M_4(i,k),
\end{align*}
thì với mỗi $M_4(i,k)$, ta có hai trường hợp:
\begin{itemize}
\item Nếu $\nu_{ik}>0$, ta có $[g_{ik}(\m{x}_i)]_{\nu_{ik}}^+ = g_{ik}(\m{x}_i)$ nên $M_4(i,k)=0$,
\item Nếu $\nu_{ik}=0$, ta có $[g_{ik}(\m{x}_i)]_{\nu_{ik}}^+ \geq g_{ik}(\m{x}_i)$ và $\nu_{ik} - \nu_{ik}^*\leq 0$ nên $M_4(i,k)\leq 0$.
\end{itemize}
Như vậy, $M_4(i,k)\leq 0$ và do đó $M_4 \leq 0$. Từ đây suy ra
\begin{align}
\dot{V} \leq 0,
\end{align}
quĩ đạo của hệ bị chặn và giới nội trong $V^{-1}(V(\m{x}(0),\bm{\lambda}(0),\bm{\nu}(0))) \subset S = \mb{R}^{dn}\times \mb{R}^{dn}_{+}\times \mb{R}^{M}_+$. Sử dụng định lý LaSalle cho hệ có đạo hàm không liên tục và Lipschitz địa phương \cite{Shevitz1994}, mỗi nghiệm của hệ xuất phát tại $(\m{x}_0,\bm{\lambda}_0,\bm{\nu}_0)\in S$ tiệm cận tới tập bất biến lớn nhất trong $\{(\m{x},\bm{\lambda},\bm{\nu})\in V^{-1}(V(0))|~ \dot{V}=0\}$.

Vì $\m{x}^*$ là cực tiểu duy nhất, để $M_3 \equiv 0$ thì phải có $\m{x}=\m{x}^*$. Để có thêm $M_1=M_2 = 0$ thì phải có $(\bm{\lambda}-\bm{\lambda}^*)^\top\mcl{L}\m{x}=0$, và $(\bm{\nu}-\bm{\nu}^*)^\top\m{g}(\m{x}^*)=0$. Do $\nu_{ik}^*g_{ik}(\m{x}^*_i)=0$ nên ta có $\bm{\nu}^\top\m{g}(\m{x}^*)=0$.

Như vậy, để $(\m{x},\bm{\lambda},\bm{\nu})$ không rời khỏi tập bất biến $M=\{(\m{x},\bm{\lambda},\bm{\nu})| \m{x}=\m{x}^*, \bm{\nu}^\top\m{g}(\m{x}^*)=0\}$ thì phải có $\dot{\m{x}}|_{\m{x}=\m{x}^*}=\nabla_{\m{x}}L(\m{x}^*,\bm{\lambda},\bm{\nu})=\m{0}_{dn}$. Do $\m{L}\m{x}^*=\m{0}_{dn}$, $\m{g}(\m{x}^*)\le \m{0}_M$, $\bm{\lambda}\ge \m{0}_M$, và $\bm{\lambda}^\top\m{g}(\m{x}^*)=0$, điều kiện KKT thỏa mãn với $(\m{x}^*,\bm{\lambda},\bm{\nu}) \in M$. Do đó $M$ là tập con của tập các điểm yên ngựa của $L$. 

Như vậy, mọi quĩ đạo xuất phát từ một điểm trong $S$ tiệm cận tới tập các điểm yên ngựa của $L$. Cụ thể hơn, mỗi quĩ đạo tiệm cận tới một điểm yên ngựa của $L$ thỏa mãn $(\m{1}_n^\top\otimes \m{I}_d)\bm{\lambda}^*=(\m{1}_n^\top\otimes \m{I}_d)\bm{\lambda}(0)$.

\begin{example}[Điều kiên lồi chặt] \cite{Feijer2010stability}
Với bài toán \eqref{eq:OP3}, nếu chỉ có điều kiện $f(\m{x})$ là hàm lồi thì các phân tích không còn đúng. Xét bài toán qui hoạch tuyến tính
\begin{mini!}
{\m{x}\in \mb{R}^{n}}{\tilde{f}(\m{x})=\m{c}^\top\m{x}}
{\label{eq:OP4}}{}
\addConstraint{\m{G}\m{x}-\m{h}}{\leq \m{0}_{m},}
\end{mini!}
trong đó ``$\le$'' được hiểu là các phần tử của vector vế trái không lớn hơn các phần tử của vector ở vế phải theo từng tọa độ. Lập hàm Lagrange $L(\m{x},\bm{\lambda}) = \m{c}^\top\m{x} + \bm{\lambda}^\top (\m{G}\m{x}-\m{h})$, thuật toán tìm điểm yên ngựa của $L$ tương ứng có dạng:
\begin{subequations} \label{eq:c7_saddle_point}
\begin{align}
\dot{\m{x}} &= \m{c} + \m{G}^\top\bm{\lambda}, \\
\dot{\bm{\lambda}} &= [\m{G}\m{x}-\m{h}]_{\bm{\lambda}}^+.
\end{align}
\end{subequations}
Tuyến tính hóa hệ tại một lân cận của một điểm yên ngựa $(\m{x}^*,\bm{\lambda}^*) $ với $\bm{\lambda}^*>0$, thì \eqref{eq:c7_saddle_point} có dạng hệ tuyến tính
\begin{align}
\begin{bmatrix}
\dot{\tilde{\m{x}}} \\ \dot{\tilde{\bm{\lambda}}}
\end{bmatrix} = \underbrace{\begin{bmatrix}
\m{0} & -\m{G}^\top \\ \m{G} & \m{0}
\end{bmatrix}}_{\m{A}} \begin{bmatrix}
\tilde{\m{x}} \\ \tilde{\bm{\lambda}}
\end{bmatrix},
\end{align}
với $\tilde{\m{x}} = \m{x}-\m{x}^*$, $\tilde{\m{x}} = \bm{\lambda}-\bm{\lambda}^*$. Do ma trận $\m{A}$ có các cặp giá trị ảo liên hợp nên điểm cân bằng không ổn định tiệm cận.
\end{example}

\begin{story}{Angelia Nedic và thuật toán tối ưu phân tán}
Angelia Nedich nhận bằng T.S. về Toán học tính toán từ Đại học tổng hợp Moscov, Moscow, Nga (1994) và Kĩ thuật Điện và Khoa học máy tính từ Viện Công nghệ Massachusetts, Cambridge, Hoa Kỳ (2002). Hiện tại, Nedich là giáo sư tại Viện Kĩ thuật Điện, Máy tính và Năng lượng thuộc Đại học bang Arizona, Hoa Kỳ. Nghiên cứu của Nedich về tối ưu hóa, tối ưu phân tán cho hệ đa tác tử, xấp xỉ ngẫu nhiên. 

Trong các ứng dụng tối ưu hóa thực tế, các thuật toán tối ưu rời rạc có được quan tâm hơn các thuật toán liên tục. Thuật toán tối ưu phân tán subgradient \cite{Nedic2009distributed} của bà là một trong những công bố đầu tiên nghiên cứu về chủ đề tối ưu trong hệ đa tác tử.
\end{story}
\section{Ứng dụng trong hệ thống sản xuất - phân phối năng lượng} 
Xét hệ thống điện gồm $n$ tác tử, mỗi tác tử có kết nối với nguồn sản xuất điện năng và tải tiêu thụ riêng, đồng thời kết nối với các tác tử khác thông qua mạng lưới truyền tải điện năng (tầng liên kết vật lý) và mạng truyền thông (tầng liên kết thông tin). Hệ thống sản xuất - phân phối điện năng một ví dụ về hệ thống mạng - vật lý (cyber-physical system), trong đó các quyết định điều khiển, vận hành các máy phát và lượng điện truyền tải giữa các tác tử được tính toán dựa yêu cầu về tải tại từng tác tử và thông tin trao đổi giữa các tác tử.

Giả sử mạng lưới truyền tải điện năng được mô tả bởi một đồ thị $G=(V,E)$ gồm $n$ đỉnh và $m$ cạnh, trong đó mỗi đỉnh của đồ thị mô tả một tác tử và mỗi cạnh $(i,j)$ thể hiện điện năng có thể truyền tải giữa hai tác tử $i$ và $j$. Kí hiệu $x_i$ là lượng điện năng sản xuất tại nút $i$ (tác tử $i$), $y_k$ là điện năng truyền tải trong bus thứ $k$, và $x^d_i>0$ là yêu cầu năng lượng từ tải tại nút $i$. Bài toán điều phối sản xuất và tiêu thụ năng lượng có thể được mô tả dưới dạng một bài toán tối ưu phân tán với các ràng buộc đẳng thức và bất đẳng thức như sau \cite{Yi2015distributed}:
\begin{mini!}
{\m{x}\in \mb{R}^{n},\m{y}\in\mb{R}^m}{\tilde{f}(\m{x},\m{y})=\sum_{i=1}^n f_i({x}_i)}
{\label{eq:OP5}}{}
\addConstraint{\m{x}+\m{H}^\top\m{y}}{=\m{x}^d}
\addConstraint{\underline{x}_i \leq {x}_i \leq \overline{x}_i,}{~\forall i=1,\ldots,n}
\addConstraint{\underline{y}_k \leq {y}_k \leq \overline{y}_k,}{~\forall k=1,\ldots,m,}
\end{mini!}
trong đó $f_i$ là hàm mục tiêu riêng cần tối ưu hóa tại nút $i$, $\m{x}=[x_1,\ldots,x_n]^\top$, $\m{y}=[\ldots,y_{ij},\ldots]^\top = [y_1,\ldots,y_m]^\top$, $\m{x}^d = [x_1^d,\ldots,x_n^d]^\top$, $\m{H}$ là ma trận liên thuộc của đồ thị $G$, $\underline{x}_i,\overline{x}_i$ là giới hạn dưới và giới hạn trên để vận hành máy phát ở nút $i$, và $\underline{y}_i,\overline{y}_i$ là giới hạn dưới và giới hạn trên về điện năng truyền tải trên bus thứ $k$.

Hàm mục tiêu và các ràng buộc vật lý được giả sử là thông tin không được trao đổi bởi mỗi tác tử. Mặc dù $\m{y}$ không xuất hiện trực tiếp trong hàm mục tiêu, sự tồn tại của $\m{y}$ thỏa mãn các ràng buộc là cần thiết để vận hành hệ. 

Giả sử cạnh $e_k=(i,j)\in E$ liên kết hai nút $i,j\in V$. Với mỗi cạnh $(i,j)$, chúng ta định nghĩa hai biến $y_k^i,y_k^j$ tương ứng là ước lượng về điện năng truyền tải tối ưu trên bus thứ $k$ của nút $i$ và nút $j$. Hai biến này cần thỏa mãn điều kiện ràng buộc $\m{E}\begin{bmatrix}
y_k^i \\ y_k^j
\end{bmatrix} = \m{0}_2$, với $\m{E}=\begin{bmatrix}
1 & -1\\ -1 & 1
\end{bmatrix}$. Với biến bổ sung, chúng ta xét bài toán tối ưu:
\begin{mini!}
{\m{x}\in \mb{R}^{n},\m{y}\in\mb{R}^m}{\tilde{f}(\m{x},\m{y})=\sum_{i=1}^n f_i({x}_i) + \frac{1}{2}\sum_{e_k=(i,j) \in E}\begin{bmatrix}
y_k^i & y_k^j \end{bmatrix}^\top \m{E}\begin{bmatrix}
y_k^i \\ y_k^j \end{bmatrix}}
{\label{eq:OP6}}{}
\addConstraint{{x}_i+\sum_{e_k \sim i} y_{k}^i - {x}^d_i =0,}{~i=1,\ldots,n}
\addConstraint{\underline{x}_i \leq {x}_i \leq \overline{x}_i,}{~\forall i=1,\ldots,n}
\addConstraint{\underline{y}_k \leq {y}_k^i \leq \overline{y}_k,}{~\forall i=1,\ldots,n, \forall e_k \sim i}
\addConstraint{\m{E}\begin{bmatrix}
y_k^i \\ y_k^j \end{bmatrix} = \m{0}_2,}{~\forall e_k \in E,~i,j \sim e_k,}
\end{mini!}
Như vậy, các biến tối ưu tại nút $i$ bao gồm $x_i$ và $\{y_k^i\}_{e_k \sim i}$ ($1+|N_i|$ biến), và tổng cộng ở cả hệ có $|V|+2|E|$ biến tối ưu. Việc định nghĩa thêm các biến $y_k^i,y_k^j$ giúp cho bài toán \eqref{eq:OP5} có thể giải một cách phân tán. Hơn nữa, với ràng buộc (\ref{eq:OP6}e), nghiệm tối ưu của \eqref{eq:OP6} thỏa mãn $y_k^{i*}=y_k^{j*}=y_k^*~\forall e_k \in E,~i,j \sim e_k$, do đó nghiệm tối ưu $({x}^*_i,{y}^*_k)$ ở hai bài toán \eqref{eq:OP5} và \eqref{eq:OP6} là như nhau.

Hàm Lagrange tương ứng của bài toán \eqref{eq:OP6} có dạng
\begin{align}
L&\left(\{x_i\}_{i\in V},\{\theta_i\}_{i\in V},\{y_k^i\}_{\substack{e_k\in E \\ e_k\sim i}},\{\underline{\lambda}_i,\overline{\lambda}_i\}_{i \in V},\{\underline{\mu}_k^i,\overline{\mu}_k^i\}_{\substack{i \in V,\\ e_k \in E \\ e_k \sim i}},\{\nu_k^i, \nu_k^j\}_{e_k=(i,j)\in E} \right) \nonumber \\
 \triangleq & L(x_i,y_k^i, \theta_i,\underline{\lambda}_i,\overline{\lambda}_i, \underline{\mu}_k^i, \overline{\mu}_k^i, \nu_k^i, \nu_k^j) \nonumber \\
=& \sum_{i=1}^n \left(f_i({x}_i) + \underline{\lambda}_i(\underline{x}_i - x_i) + \overline{\lambda}_i (x_i - \overline{x}_i) + \sum_{e_k\sim i}(\underline{\mu}_k^i(\underline{y}_k-y_k^i) + \overline{\mu}_k^i(y_k^i - \overline{y}_k)) \right. \nonumber \\
& \left. + \theta_i \left( x_i + \sum_{e_k \sim i} h_{ki}y_{k}^i - {x}^d_i \right) \right) + \sum_{e_k=(i,j) \in E} \left( \frac{1}{2}\begin{bmatrix}
y_k^i,~y_k^j \end{bmatrix}^\top \m{E}\begin{bmatrix}
y_k^i \\ y_k^j \end{bmatrix} + \begin{bmatrix}
\nu_k^i,~\nu_k^j \end{bmatrix}^\top \m{E}\begin{bmatrix}
y_k^i\\ y_k^j \end{bmatrix} \right). \label{eq:c7_Lagrange_OP6}
\end{align}
Thuật toán tối ưu phân tán tìm điểm yên ngựa của hàm Lagrange \eqref{eq:c7_Lagrange_OP6} có dạng như sau:
\begin{subequations} \label{eq:c7_alg_OP6}
\begin{align}
\dot{x}_i &= -\nabla_{x_i} L = -\nabla f_i(x_i) - \theta_i + \underline{\lambda}_i - \overline{\lambda}_i, \, \forall i \in V,\\
\dot{y}_k^i &= -\nabla_{y_k^i} L = -(y_k^i-y_k^j) -(\nu_k^i-\nu_k^j) -h_{ki}\theta_i + \underline{\mu}_i^k - \overline{\mu}_i^k, \, \forall i \in V,\, \forall e_k \sim i, \\
\dot{\theta}_i &= \nabla_{\theta_i} L = x_i + \sum_{e_k \sim i} h_{ki}y_{k}^i - {x}^d_i, \, \forall i \in V,\\
\dot{\nu}_k^i &= \nabla_{\nu_k^i} L = y_k^i - y_k^j, \, \forall e_k = (i,j) \in E,\\
\dot{\underline{\lambda}_i} &= [\nabla_{\underline{\lambda}_i} L]_{\underline{\lambda}_i}^+ = [\underline{x}_i - x_i]_{\underline{\lambda}_i}^+, \, \forall i \in V,\\
\dot{\overline{\lambda}_i} &= [\nabla_{\overline{\lambda}_i} L]_{\overline{\lambda}_i}^+ = [x_i - \overline{x}_i]_{\overline{\lambda}_i}^+, \, \forall i \in V,\\
\dot{\underline{\mu}_k^i} &= [\nabla_{\underline{\mu}_k^i} L]_{\underline{\mu}_k^i}^+ = [\underline{y}_k - y_k^i]_{\underline{\mu}_k^i}^+, \, \forall e_k \in E, \, e_k \sim i,\\
\dot{\overline{\mu}_k^i} &= [\nabla_{\overline{\mu}_k^i} L]_{\overline{\mu}_k^i}^+ = [y_k^i - \overline{y}_k]_{\overline{\mu}_k^i}^+, \, \forall e_k \in E, \, e_k \sim i.
\end{align}
\end{subequations}
Các biến trạng thái mà tác tử $i$ cần nhận từ tác tử láng giềng bao gồm các biến $\{y_k^j, \nu_k^j\}_{\forall j \in N_i}$ liên quan tới các cạnh $e_k$ kề với đỉnh $i\in V$. Để thực hiện thuật toán \eqref{eq:c7_alg_OP6}, mỗi tác tử cần $4+4|N_i|$ biến trạng thái. Số biến trạng thái của hệ $n$ tác tử là $4|V|+8|E|$.

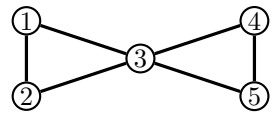
\begin{SCfigure}[][t!]
\caption{Đồ thị mô tả lưới điện gồm 5 tác tử trong Ví dụ~\ref{VD:7.4}.\label{fig:VD_7.4_graph}}
\hspace{5cm}
\begin{tikzpicture}[
roundnode/.style={circle, draw=black, thick, minimum size=3.5mm,inner sep= 0.25mm}
]
	
    \node[roundnode] (u1) at (0,1) {1}; %
	\node[roundnode] (u2) at (0,0) {2}; %
	\node[roundnode] (u3) at (1.5,0.5) {3}; %
	\node[roundnode] (u4) at (3,1) {4}; %
	\node[roundnode] (u5) at (3,0) {5}; %
	\node (u) at (4,0) { };
    \draw [very thick,-] (u1)--(u2)--(u3)--(u4)--(u5)--(u3)--(u1);
    ;
\end{tikzpicture}
\end{SCfigure}
\begin{example}[Hệ thống sản xuất - phân phối điện năng gồm 5 tác tử] 
\label{VD:7.4} Xét mạng điện gồm 5 tác tử với đồ thị như ở Hình~\ref{fig:VD_7.4_graph}. Các hàm mục tiêu riêng của các tác tử có dạng $f_i(x_i)=x^2+c_i$, với $c_1=3, c_2 = 4, c_3 = 1, c_4 = 2, c_5 =1$. Các giá trị biên về năng lượng sản xuất của mỗi tác tử được cho như ở Bảng \ref{tab:c7_xi}. Với các cạnh được đánh số và định hướng  bởi $e_1=(1,2)$, $e_2=(1,3)$, $e_3=(2,3)$, $e_4=(3,4)$, $e_5=(3,5)$, $e_6=(4,5)$, các giá trị định mức về điện năng có thể truyền tải qua các cạnh được cho trong Bảng \ref{tab:c7_yk}.
\begin{SCtable}[][h]
\caption{Các giá trị biên của $x_i$}
\label{tab:c7_xi}
\hspace{8cm}
\begin{tabular}{@{}llllll@{}}
\toprule
$i$ & $1$ & $2$ & $3$ & $4$ & $5$  \\ \midrule
$\underline{x}_i$ & $0$ & $0$ & $0$ & $0$ & $0$ \\
$\overline{x}_i$  & $9$ & $3$ & $6$ & $4$ & $5$ \\ \bottomrule
\end{tabular}
\end{SCtable}
\begin{SCtable}[][h]
\hspace{6.5cm}
\caption{Các giá trị biên của $y_k$}
\label{tab:c7_yk}
\begin{tabular}{@{}lrrrrrr@{}}
\toprule
$k$ & $1$ & $2$ & $3$ & $4$ & $5$ & $6$ \\ \midrule
$\underline{y}_k$ & $-6$ & $-4$ & $-3$ & $-2$ & $-4$ & $-4$\\
$\overline{y}_k$  &  $6$ &  $4$ &  $3$ &  $2$ & $4$  & $4$ \\ \bottomrule
\end{tabular}
\end{SCtable}

Trong thời gian $0\leq t \leq 100$ (s), tải đặt tại mỗi tác tử là $\m{x}^d = [5,2,5,3,4]^\top$. Các giá trị tải nằm trong dải điện năng sản xuất được của cả 5 tác tử. Mô phỏng trong khoảng thời gian này cho thấy  điện năng sản xuất của các tác tử tiệm cận $\m{x}^*=[4,3,4,4,4]^\top$ và điện năng trao đổi trên các bus tương ứng thỏa mãn $y_k^i \to y_k^j$ và tiệm cận tới các giá trị:
\[-0.6297,-0.3704,0.4705,-0.3648,-0.6345,0.6350.\]
Hình \ref{fig:c7_VD7.4}(c) cho thấy tổng điện năng tại mỗi nút (bao gồm điện năng tự sản xuất và điện năng nhận từ tác tử láng giềng) dần cân bằng với tải đặt.

Tại $t>100$ (s), tải đặt thay đổi $\m{x}^d = [7,4,5,3,5]^\top$. Nếu hoạt động độc lập, tác tử 2 không thể sản xuất đáp ứng tải mới do $\overline{x}_2=3 < x^d_2(t>100) = 4$ (pu). Nhờ có sự phối hợp hoạt động giữa các tác tử, lượng điện năng thiếu hụt dần được bù bởi các tác tử khác trong hệ (Hình \ref{fig:c7_VD7.4}(c)). Hình \ref{fig:c7_VD7.4}(a)--(b) cho thấy $\m{x}(t) \to \m{x}^*=[6,3,6,4,5]^\top$, phần điện năng bù này chủ yếu sản xuất bởi tác tử 3, 4. 

Cuối cùng, đồ thị biểu diễn hàm mục tiêu $\tilde{f}(\m{x})=\sum_{i=1}^n f_i(x_i)$ theo thời gian được thể hiện tương ứng ở Hình \ref{fig:c7_VD7.4}(d). 
\end{example}
\begin{figure*}[ht!]
\centering
    \subfloat[]{\includegraphics[width=0.7\textwidth]{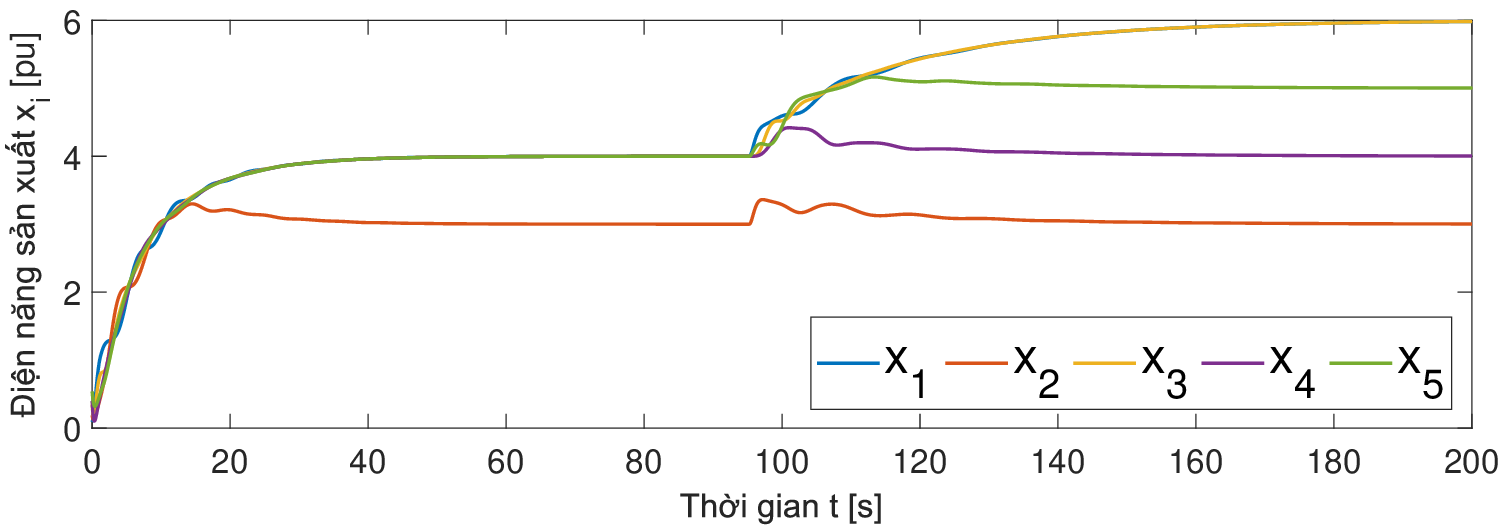}}\\
    \subfloat[]{\includegraphics[width=0.7\textwidth]{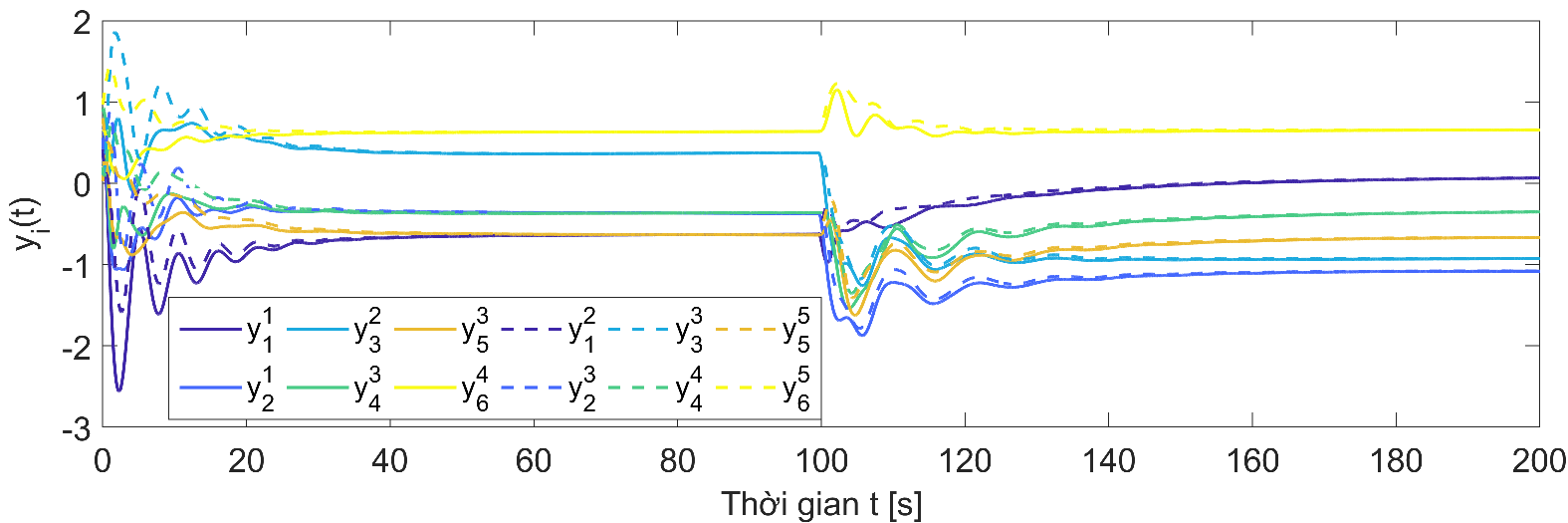}}\\
    \subfloat[]{\includegraphics[width=0.7\textwidth]{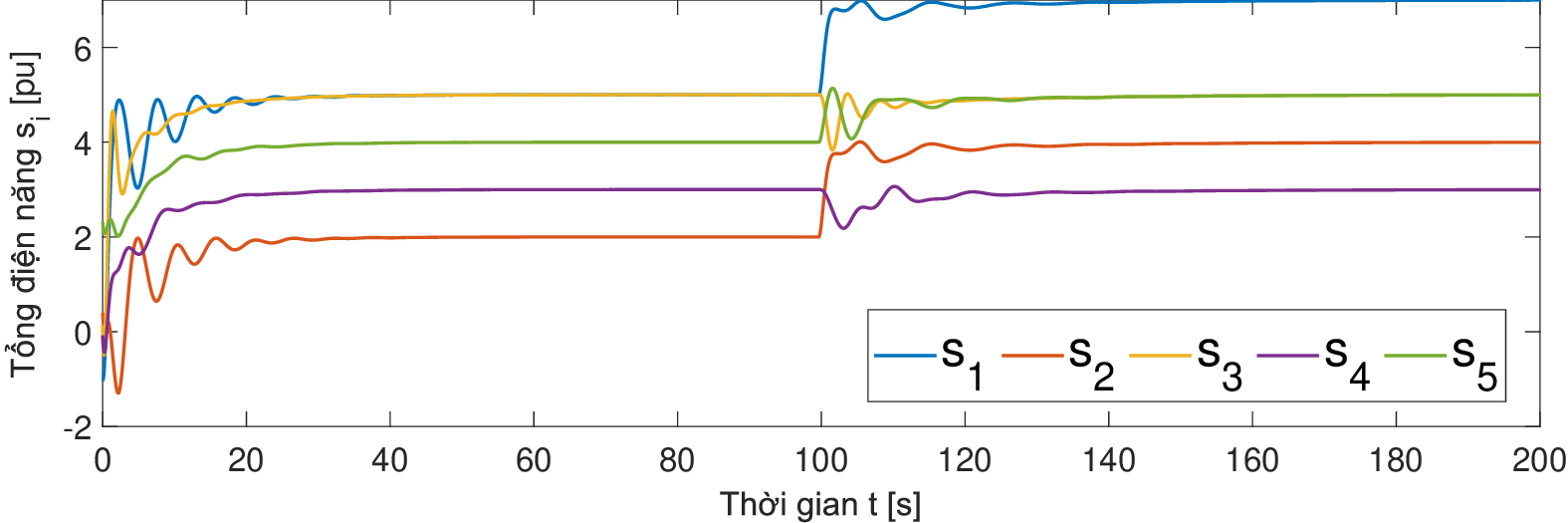}}\\
    \subfloat[]{\includegraphics[width=0.7\textwidth]{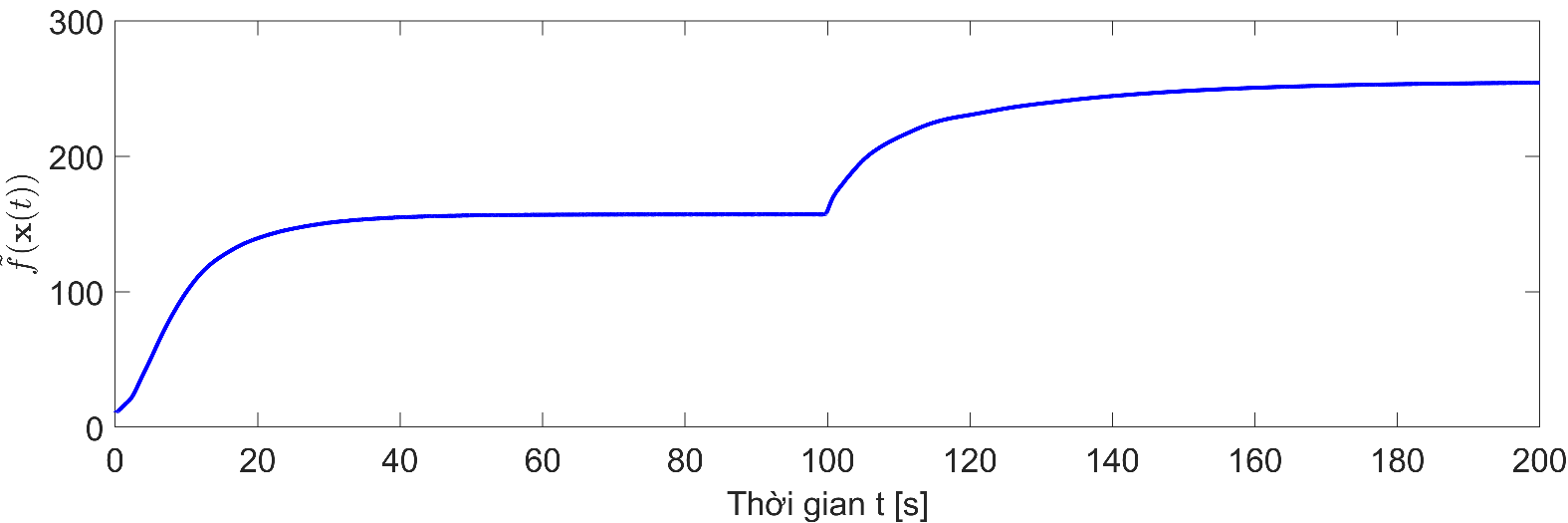}}
    \caption{Mô phỏng hệ 5 tác tử trong Ví dụ \ref{VD:7.4}. (a): Điện năng sản xuất $x_i(t)$ (pu), $i=1,\ldots,6$; (b) Ước lượng điện năng truyền tải trên các đường dây $y_k^i(t),y_k^j(t)$ (pu) với $k=1,\ldots,6$; và (c) Tổng điện năng tại mỗi nút $s_i=x_i-\sum_{e_k\sim i}h_{ki}y_k^{i}$ (pu), $i=1,\ldots,6$. (d) Giá trị hàm mục tiêu $\tilde{f}(\m{x}(t))$. \label{fig:c7_VD7.4}}
\end{figure*}

\section{Ghi chú và tài liệu tham khảo}
Phần phân tích ổn định của thuật toán tối ưu phân tán ở mục \ref{sec:dist_optim_consensus} dựa trên tài liệu \cite{Gharesifard2013distributed}. Lưu ý rằng trong \cite{Gharesifard2013distributed}, điều kiện khả vi liên tục có thể thay bởi điều kiện Lipschitz địa phương, đồng thời đồ thị $G$ có thể là liên thông mạnh và cân bằng. Các giả thiết về bài toán tối ưu có ràng buộc ở mục \eqref{sec:dist_optim_consensus_constraints} tương tự với \cite{Hatanaka2018passivity,Doan2012}, tuy nhiên phương pháp phân tích được biến đổi từ \cite{Cherukuri2016asymptotic}. Chứng minh ổn định theo hàm mũ của thuật toán có thể tham khảo tại \cite{Ding2019global}. Về một số phương pháp phân tích ổn định khác dựa trên hệ thụ động hoặc định lý LaSalle cho hệ lai, tham khảo \cite{Wen2004unifying,Hatanaka2018passivity,Feijer2010stability,Doan2012}.

Nội dung ở chương này yêu cầu một số kiến thức về tối ưu lồi và động học điểm yên ngựa. Một số nội dung tương tự có thể tham khảo ở \cite{Dofler2017}. Thuật toán tối ưu khi có ràng buộc bất đẳng thức dựa trên hàm cập nhật biến đối ngẫu không liên tục. Thuật toán cập nhật biến đối ngẫu liên tục và ứng dụng có thể tham khảo ở \cite{Dorr2012,Hoang2017ICCAS}. Ứng dụng của tối ưu phân tán trong bài toán sản xuất và phân phối điện năng được trình bày dựa trên tài liệu \cite{Yi2016initialization}. Thuật toán sản xuất - phân phối với tải thay đổi theo thời gian được đề xuất tại \cite{Cherukuri2016initialization} dựa trên thuật toán đồng thuận động. Về thuật toán tối ưu phân tán với hàm mục tiêu phụ thuộc thời gian, người đọc có thể tham khảo tại \cite{Salar2017}.

Những ý tưởng về áp dụng lý thuyết điều khiển cho bài toán tối ưu phân tán với các thuật toán liên tục được trình bày ở \cite{Wen2004unifying,Lu2010,Wang2011control}. Tối ưu phân tán với hệ rời rạc đã được ứng dụng trong truyền thông, mạng cảm biến và hệ thống điện \cite{Low1999optimization,Rabbat2004distributed,Xiao2005scheme,Madrigal2002analytical}, trước khi được  nghiên cứu sâu hơn dựa trên lý thuyết điều khiển \cite{Nedic2009distributed}. Tổng quan về các thuật toán cũng như các ứng dụng của tối ưu phân tán trong điều khiển có thể tham khảo tại \cite{Nedic2018distributed}. Một số vấn đề liên quan tới tối ưu phân tán bao gồm tìm điểm cực trị (extremum seeking), lý thuyết trò chơi (tìm điểm cân bằng Nash) \cite{Nguyen2018distributed}, phân việc \cite{Tran2022distributed}, tối ưu hóa trong các hệ ngăn (compartmental systems) như mạng phân phối nước \cite{Nguyen2023coordination}, hệ điều khiển nhiệt trong tòa nhà \cite{Kim2015consensus}, mạng lưới giao thông \cite{Kim2014distributed,Pham2022distributed}, ...

\section{Bài tập}

\begin{exercise}[Phần bù Schur] \label{ex:c7_Schur_complement}
Xét ma trận khối 
\[\m{M}=\begin{bmatrix}
\m{A} & \m{B} \\ \m{C} & \m{D}
\end{bmatrix}.\] 
\begin{itemize}
\item[i.] Giả sử $\m{A}$ khả nghịch. Chứng minh $\m{M}$ có thể biểu diễn dưới dạng:
\begin{align*}
\m{M} = \begin{bmatrix}
\m{I} & \m{0} \\ 
\m{C}\m{A}^{-1} & \m{I}
\end{bmatrix} 
\begin{bmatrix}
\m{A} & \m{0} \\ 
\m{0} & \m{D}-\m{C}\m{A}^{-1}\m{B}
\end{bmatrix} 
\begin{bmatrix}
\m{I} & \m{A}^{-1}\m{B} \\ \m{0} & \m{I}
\end{bmatrix}.
\end{align*}
Hơn nữa, $\m{M}$ khả nghịch khi và chỉ khi phần bù Schur $\m{D}-\m{C}\m{A}^{-1}\m{B}$ của $\m{A}$ là khả nghịch.
\item[ii.] Giả sử $\m{C}=\m{B}^\top$ chứng minh rằng $\m{M}$ là ma trận đối xứng xác định dương khi và chỉ khi $\m{D}$ và $\m{A}-\m{B}\m{D}^{-1}\m{B}^\top$ đối xứng và xác định dương.
\item[iii.] Giả sử $\m{C}=\m{B}^\top$ chứng minh rằng nếu $\m{D}$  là ma trận đối xứng xác định dương thì $\m{M}$ và $\m{A}-\m{B}\m{D}^{-1}\m{B}^\top$ đối xứng và bán xác định dương.
\end{itemize}
\end{exercise}

\begin{exercise}[Ma trận giả nghịch đảo] \label{ex:c7_pseudo_inverse}
Xét ma trận vuông $\m{M}\in \mb{R}^{n \times n}$ thỏa mãn ${\rm rank}(\m{M})=r$. 
\begin{itemize}
\item[i.] Chứng minh rằng ma trận $\m{M}^\top\m{M}$ và ma trận $\m{M}\m{M}^\top$ có cùng tập các giá trị riêng là $\sigma_1 \ge \ldots \ge \sigma_r>0$. Tìm liên hệ của $\sigma_i$ với các giá trị riêng của $\m{M}$.
\item[ii.] $\m{M}$ có phân tích theo giá trị suy biến (SVD):
\[\m{M}=\m{U}\bm{\Sigma}\m{V}^\top,\]
với $\m{U},\m{V}$ là các ma trận trực giao và $\bm{\Sigma}={\rm diag}(\sigma_1,\ldots,\sigma_r,0,\ldots,0)$. Định nghĩa \emph{ma trận giả nghịch đảo} của $\m{M}$ bởi $\m{M}^{\dagger}=\m{U}\bm{\Sigma}^{\dagger}\m{V}^\top$, trong đó $\bm{\Sigma}={\rm diag}(\sigma_1^{-1},\ldots,\sigma_r^{-1},0,\ldots,0)$. Chứng minh rằng:
\begin{subequations}
\begin{align}
\m{M}\m{M}^{\dagger}\m{M} &=\m{M},\\
\m{M}^{\dagger}\m{M}\m{M}^{\dagger} &=\m{M}^{\dagger},\\
{\rm ker}(\m{M}) = {\rm ker}(\m{M}^{\dagger}) &= {\rm ker}(\m{M}\m{M}^{\dagger})= {\rm ker}(\m{M}^{\dagger}\m{M})
\end{align}
\end{subequations}
\end{itemize} 
\end{exercise}

\begin{exercise}[Dạng toàn phương I] \cite{Gallier2010notes}\label{ex:c7_quad_form1}
Xét hàm $f(\m{x})=\frac{1}{2}\m{x}^\top\m{P}\m{x} + \m{x}^\top\m{b}$. 
\begin{itemize}
\item[i.] Chứng minh rằng nếu $\m{P}$ khả nghịch thì $f(\m{x})$ có cực tiểu khi và chỉ khi $\m{P}$ đối xứng xác định dương. Tìm giá trị nhỏ nhất của $f(\m{x})$ và điểm cực tiểu trong trường hợp này.
\item[ii.] Chứng minh rằng nếu $\m{P}$ đối xứng thì $f(\m{x})$ có cực tiểu khi và chỉ khi $\m{P}$ là bán xác định dương đồng thời $\m{b} \in {\rm im}(\m{P})$. Tìm giá trị nhỏ nhất của $f(\m{x})$ và tập các điểm cực tiểu trong trường hợp này.
\end{itemize}
\end{exercise}

\begin{exercise}[Dạng toàn phương II] \cite{Gallier2010notes} \label{ex:c7_quad_form2}
Xét hàm hai biến \[f(\m{x},\m{y})=\begin{bmatrix}
\m{x}^\top & \m{y}^\top
\end{bmatrix}\begin{bmatrix}
\m{A} & \m{B} \\ \m{B}^\top & \m{C}
\end{bmatrix}\begin{bmatrix}
\m{x} \\ \m{y}
\end{bmatrix}=\begin{bmatrix}
\m{x}^\top & \m{y}^\top
\end{bmatrix}\m{M}
\begin{bmatrix}
\m{x} \\ \m{y}
\end{bmatrix},\]
với $\m{M}^\top = \m{M}$. Chứng minh rằng $f(\m{x},\m{y})$ có cực tiểu khi và chỉ khi một trong các điều kiện sau thỏa mãn:
\begin{itemize}
\item[i.] $\m{M}$ là bán xác định dương;
\item[ii.] $\m{A}$ là bán xác định dương, $(\m{I}-\m{A}\m{A}^{\dagger})\m{B}=\m{0}$, và $\m{C}-\m{B}^\top\m{A}^{\dagger}\m{B}$ là bán xác định dương;
\item[iii.] $\m{C}$ là bán xác định dương, $(\m{I}-\m{C}\m{C}^{\dagger})\m{B}^\top=\m{0}$, và $\m{A}-\m{B}\m{C}^{\dagger}\m{B}^\top$ là bán xác định dương;
\end{itemize}
\end{exercise}
\begin{exercise}[Giải hệ phương trình tuyến tính phân rã được thành các hệ phương trình nhỏ hơn] \cite{Cao2017continuous}
Xét hệ $n$ tác tử tương tác qua đồ thị vô hướng liên thông $G$. Giả sử hệ cần giải bài toán tìm nghiệm của hệ phương trình tuyến tính có dạng
\begin{align}\label{ex:c7_linearEq}
\m{A}\m{x} = \sum_{i=1}^n \m{A}_i \m{x}_i = \m{b}_0,
\end{align}
trong đó $\m{x}_i \in \mb{R}^{d_i}$ và $\m{A}_i\in \mb{R}^{N \times d_i}$ là biến trạng thái và ma trận ràng buộc của tác tử thứ $i$, $\m{A}=[\m{A}_1,\ldots,\m{A}_n] \in \mb{R}^{N \times q}$, $\m{x} = [\m{x}_1^\top,\ldots,\m{x}_n^\top]^\top\in \mb{R}^{q}$, $\m{b}_0 \in \mb{R}^N$, $q=\sum_{i=1}^n d_i$. Giả sử thêm rằng thông tin về  vector $\m{b}_i$ là được biết bởi tác tử $i$, và các vector này thỏa mãn $\sum_{i=1}^n \m{b}_i = \m{b}_0$.

\begin{itemize}
\item[i.] Đặt $\bar{\m{A}}={\rm blkdiag}(\m{A}_1,\ldots,\m{A}_n)\in \mb{R}^{Nn \times q}$. Chứng minh $\m{x}^*$ là nghiệm của \eqref{ex:c7_linearEq} khi và chỉ khi tồn tại $\m{z}^*\in \mb{R}^{Nn}$ và $\m{b}_i\in\mb{R}^{N},i=1,\ldots,n$, sao cho 
\begin{align} \label{ex:c7_linearEq_constraint}
\bar{\m{A}}\m{x}^* - \m{b} - \bar{\mcl{L}}\m{z}^* &= \m{0}_{Nn},
\end{align}
trong đó $\m{b}=[\m{b}_1^\top,\ldots,\m{b}_n^\top]^\top$ và $\bar{\mcl{L}} = \mcl{L}\otimes \m{I}_N$.

\item[ii.] Để tìm nghiệm \eqref{ex:c7_linearEq}, chúng ta tìm $\m{x}^*$ và $\m{z}^*$ thỏa mãn \eqref{ex:c7_linearEq_constraint}. Điều này được thực hiện thông qua việc tìm nghiệm bài toán tối ưu
\begin{mini!}
{\m{x}\in \mb{R}^{q},\m{y}\in \mb{R}^{Nn},\m{z}\in \mb{R}^{Nn}}{\|\bar{\m{A}}\m{x}^* - \m{b} - \bar{\mcl{L}}\m{z}^*\|^2}
{\label{ex:c7_OPex1}}{}
\addConstraint{\m{y}+\bar{\mcl{L}}\m{z}}{=\m{0}_{Nn}}
\end{mini!}
trong đó $\m{y}_i \in \mb{R}^{N},i=1,\ldots,n$ là các biến trạng thái phụ, $\m{y}=[\m{y}_1^\top,\ldots,\m{y}_n^\top]^\top\in\mb{R}^{Nn}$. Định nghĩa hàm Lagrange mở rộng
\begin{align}
L(\m{x},\m{y},\m{z},\bm{\lambda}) = \|\bar{\m{A}}\m{x}^* - \m{b} - \bar{\mcl{L}}\m{z}^*\|^2 + \bm{\lambda}^\top(\m{y}+\bar{\mcl{L}}\m{z}),
\end{align}
hãy viết thuật toán tối ưu phân tán tìm nghiệm của  \eqref{ex:c7_OPex1}.
\item[iii.] Tìm tập các điểm cân bằng của hệ với thuật toán viết ở ý (ii). Phân tích ổn định của hệ khi $t\to\infty$.
\end{itemize}
\end{exercise}


\begin{exercise}[Thuật toán tìm điểm yên ngựa với ma trận tỉ lệ] \cite{Cherukuri2016asymptotic} 
Xét hệ $n$ tác tử với thuật toán gradient 
\begin{subequations} \label{ex:c7_Dist_Alg_constraints_Matrix}
\begin{align}
\dot{\m{x}} &= -\m{K}_1\nabla_{\m{x}} L(\m{x},\bm{\lambda},\bm{\nu}),  \label{ex:c7_Dist_Alg_constraints_Matrix1} \\
\dot{\bm{\lambda}} &= \m{K}_2\nabla_{\bm{\lambda}} L(\m{x},\bm{\lambda},\bm{\nu}), \label{ex:c7_Dist_Alg_constraints_Matrix2} \\
\dot{\bm{\nu}} &= \m{K}_3[\nabla_{\bm{\nu}} L(\m{x},\bm{\lambda},\bm{\nu})]^{+}_{\bm{\nu}}. \label{ex:c7_Dist_Alg_constraints_Matrix3}
\end{align}
\end{subequations}
trong đó $\m{K}_1, \m{K}_2 \in \mb{R}^{dn \times dn}$ và $\m{K}_3 \in \mb{R}^{M \times M}$ là các ma trận đường chéo, đối xứng và xác định dương.
\begin{itemize}
\item[i.] Viết phương trình thuật toán \eqref{ex:c7_Dist_Alg_constraints_Matrix} cho mỗi tác tử.
\item[ii.] Tìm tập điểm cân bằng của hệ.
\item[iii.] Chứng minh mỗi quĩ đạo xuất phát từ một điểm $(\m{x},\bm{\lambda},\bm{\nu})\in \mb{R}^{dn}\times \mb{R}^{dn} \times \mb{R}^M_+$ tiệm cận tới một điểm trong tập điểm yên ngựa của $L$.
\end{itemize}
\end{exercise}

\begin{exercise}[Thuật toán tối ưu phân tán dựa trên đồng thuận với thời gian hội tụ hữu hạn] \cite{Song2016ietCTA}
Xét hệ $n$ tác tử tương tác qua đồ thị vô hướng liên thông với hàm mục tiêu $\tilde{f}(\m{x})=\sum_{i=1}^n f_i(\m{x}_i)$, trong đó $f_i:\mb{R}^d \to \m{R},i=1,\ldots,n$ là các hàm lồi chặt, khả vi liên tục tới cấp hai, và thỏa mãn điều kiện:
\begin{align}
\forall \m{x},\m{y} \in \mb{R}^d,~\exists \theta>0:~\nabla^2f_i(\m{x}_i) \geq \theta \m{I}_d.
\end{align}
Xét thuật toán tối ưu phân tán 
\begin{subequations} \label{ex:c7_Dist_Alg_FT}
\begin{align}
\dot{\m{x}}_i &= -\gamma \left(\nabla^2 {f}_i(\m{x}_i)\right)^{-1} \sum_{j\in N_i}{\rm sig}^{\alpha}(\m{x}_j-\m{x}_i),  \label{ex:c7_Dist_Alg_FT1} \\
\m{x}_i(0) &= \m{x}_i^*,\, i=,\ldots,n. \label{ex:c7_Dist_Alg_FT2} 
\end{align}
\end{subequations}
trong đó $\m{x}_i^*$ là điểm cực tiểu của hàm mục tiêu riêng $f_i(\m{x}_i)$.
\begin{itemize}
\item[i.] Hãy viết lại hệ \eqref{ex:c7_Dist_Alg_FT} dưới dạng ma trận.
\item[ii.] Chứng minh rằng $\sum_{i=1}^n \nabla f_i(\m{x}_i)$ là bất biến theo thời gian.
\item[iii.] Chứng minh với thuật toán \eqref{ex:c7_Dist_Alg_FT}, $\m{x} \to \m{1}_n\otimes \m{x}^*$, với $\m{x}^*$ là điểm cực tiểu của hàm $f(\m{z})=\sum_{i=1}^n f_i(\m{z})$, trong thời gian hữu hạn. Hãy ước lượng cận trên của thời gian hội tụ. (Gợi ý: sử dụng hàm Lyapunov $V(\m{x})={f}(\m{x}^*)-\tilde{f}(\m{x})-(\nabla f(\m{x}))^\top (\m{1}_n\otimes \m{x}^* - \m{x})$ và tham khảo kết quả liên quan tới chứng minh hội tụ thời gian hữu hạn trong Bài tập 5.11)
\end{itemize}
\end{exercise}


\begin{exercise}[Thuật toán điểm yên ngựa với thời gian hội tụ hữu hạn] 
\cite{Shi2022finite}
Xét bài toán tối ưu có ràng buộc:
\begin{mini!}
{\m{x}\in \mb{R}^{n}}{f(\m{x})}
{\label{ex:c7_OPex2}}{}
\addConstraint{\m{A}\m{x}=\m{b}}{}
\addConstraint{\m{h}(\m{x})\leq \m{0}}{}
\end{mini!}
trong đó $f: \mb{R}^n \to \mb{R}$ là hàm mục tiêu lồi chặt, khả vi liên tục tới cấp hai, $\m{h}: \mb{R}^n \to \mb{R}^p$ là hàm lồi, khả vi liên tục tới cấp hai, $\m{A}\in \mb{R}^{m\times n}$ và $\m{b} \in \mb{R}^m$.
\begin{itemize}
\item[i.] Hãy lập hàm Lagrange mở rộng tương ứng của \eqref{ex:c7_OPex2} và viết thuật toán động học tìm điểm yên ngựa tương ứng.
\item[ii.] Xét thuật toán tìm điểm yên ngựa
\begin{subequations} \label{ex:c7_FT_constraints_Matrix}
\begin{align}
\dot{\m{x}} &= -{\rm sign}\left(\nabla_{\m{x}} L(\m{x},\bm{\lambda},\bm{\nu})\right),  \label{ex:c7_FT_constraints_Matrix1} \\
\dot{\bm{\lambda}} &= {\rm sign}\left(\nabla_{\bm{\lambda}} L(\m{x},\bm{\lambda},\bm{\nu})\right), \label{ex:c7_FT_constraints_Matrix2} \\
\dot{\bm{\nu}} &= {\rm sign}\left([\nabla_{\bm{\nu}} L(\m{x},\bm{\lambda},\bm{\nu})]^{+}_{\bm{\nu}}\right). \label{ex:c7_FT_constraints_Matrix3}
\end{align}
\end{subequations}
Hãy tìm tâp điểm cân bằng $S^*=\{\bm{\xi}=(\m{x},\bm{\lambda},\bm{\nu})|~\dot{\m{x}}=0,\dot{\bm{\lambda}}=0,\dot{\bm{\nu}}=0\}$ của \eqref{ex:c7_Dist_Alg_constraints_Matrix}. 

Giả thiết rằng tại mọi điểm $\bm{\xi}(t)\notin S^*$, các điều kiện sau thỏa mãn:
\begin{itemize}
\item[a.] $\nabla_{\m{x}}L(\bm{\xi}) \perp {\rm ker}(\nabla_{\m{x}\m{x}}L(\bm{\xi}))$
\item[b.] $\m{h}(\m{x}) \in {\rm im}(\nabla \m{h}(\m{x})$
\item[c.] tồn tại các hằng số dương $\delta_1,\delta_2,\delta_3$ để
\begin{align*}
\delta_1 \leq \lambda_2(\nabla_{\m{x}\m{x}}L(\bm{\xi})) &\leq \lambda_n(\nabla_{\m{x}\m{x}}L(\bm{\xi})) \leq \delta_2,\\
\lambda_2(\m{M}(\m{x})\m{M}(\m{x})^\top) \geq \delta_3,& \, \text{ với } \m{M}(\m{x})= \begin{bmatrix}
\m{A}^\top,\nabla^\top\m{h}(\m{x})
\end{bmatrix}^\top,
\end{align*}
\end{itemize}
thì $\bm{\xi}(t)$ hội tụ tới $S^*$ trong thời gian hữu hạn.
\item[iii.] Áp dụng thuật toán ở ý (ii) cho bài toán tối ưu phân tán
\begin{mini!}
{\m{x}\in \mb{R}^{N}}{\tilde{f}(\m{x})=\sum_{i=1}^nf_i(\m{x}_i)}
{\label{ex:c7_OPex3}}{}
\addConstraint{\sum_{i=1}^n\m{A}_i\m{x}_i}{=\sum_{i=1}^n\m{b}_i=\m{b}}
\end{mini!}
trong đó $f_i:\mb{R}^{d_i}\to \mb{R}$ là hàm mục tiêu riêng (lồi chặt, khả vi liên tục tới cấp hai) và $\m{A}_i,\m{b}_i$ là các ma trận chỉ được biết bởi tác tử $i$, $N=\sum_{i=1}^n d_i$.
\end{itemize}
\end{exercise}

\begin{exercise}
Xét bài toán tối ưu phân tán
\begin{mini!}
{\m{x}\in \mb{R}^{dn}}{\tilde{f}(\m{x})=\sum_{i=1}^nf_i(\m{x}_i)}
{\label{eq:OPex4}}{}
\addConstraint{\bar{\mcl{L}}\m{x}}{=\m{0}_{dn}}
\end{mini!}
và thuật toán \eqref{eq:Dist_optim_alg}. Sử dụng hàm thế $V(\m{x},\bm{\lambda}) = \frac{1}{2}\|\dot{\m{x}}\|^2 + \frac{1}{2}\|\dot{\bm{\lambda}}\|^2$, hãy chứng minh rằng $\dot{V}\leq 0$.
\end{exercise}

\begin{exercise}[Bài toán kiểm soát tắc nghẽn mạng (internet congestion control)] \cite{Feijer2010stability} Xét mạng gồm $n$ nguồn thông tin trong tập $S$, gửi thông tin qua tập gồm $m$ đường truyền trong tập $L$. Mỗi nguồn tin có một hàm hiệu năng $U_j(x_j)$ là hàm lõm, tăng theo lưu lượng thông tin $x_j$ được truyền tải qua mạng. Nhiệm vụ của kiểm soát tắc nghẽn băng thông là tối ưu hóa hàm hiệu năng của mạng:
\begin{maxi*}
{\m{x}\in \mb{R}^{dn}}{\tilde{U}(\m{x})=\sum_{j\in S}^nU_j(\m{x}_j)}
{\label{eq:OPex5}}{}
\addConstraint{y_i = \sum_{j \in S(i)}x_j\leq c_i}{i \in L}
\end{maxi*}
trong đó $S(i) \subset S$ là tập các nguồn tin sử dụng chung đường truyền $i$, với băng thông $y_i$
\end{exercise}

\begin{exercise}[Kiểm soát tắc nghẽn và tranh chấp băng thông chéo trong mạng không dây] \cite{Feijer2010stability}
Xét mạng không dây gồm một tập các nút nguồn, sử dụng giao thức điều khiển đa truy cập ngẫu nhiên. Mỗi nút $k$ được chọn để truyền tin với xác suất $P_k$, và khi truyền tin, nút $k$ sẽ chọn một trong các kênh truyền trong $L_{\rm out}(n)$ với xác suất $p_i$. Do đó, chúng ta có các ràng buộc sau cho mỗi nút:
\begin{align}
p_i\geq 0,\quad \sum_{i \in L_{\rm out}(k)}p_i = P_k,\quad P^k \leq 1.
\end{align}
Việc truyền tin trong một kênh truyền $i$ sẽ bị ảnh hưởng bởi một kênh truyền $j$ khác, nếu nút nhận tin của $i$ nằm trong vùng truyền tin của kênh truyền $j$. Xung đột kênh truyền sẽ dẫn đến việc mất gói tin truyền. Dung lượng của mỗi kênh $i$, do đó, là hàm phụ thuộc vào xác xuất kênh $i$ được chọn khi không có xung đột:
\begin{align}
C_i(p)=c_ip_i \prod_{k \in N_I(i)}(1-P_k),
\end{align}
với $N_I(i)$ là tập các nút có thể gây ảnh hưởng tới kênh truyền $i$. Bài toán đặt ra là tìm phân phối băng thông và xác suất truyền tin để tối ưu hóa hiệu năng mạng:
\begin{maxi!}
{\m{x}\in \mb{R}^{n}}{\tilde{U}(\m{x})=\sum_{j\in S}^n U_j({x}_j)}
{\label{eq:OPex6}}{}
\addConstraint{\sum_{j \in S(i)} x_j\leq C_i(p),}{\forall j}
\end{maxi!}
\begin{itemize}
\item[i.] Hãy thực hiện phép đổi biến $\tilde{x}_j \triangleq \log x_j$ chuyển ràng buộc bất đẳng thức trong \eqref{eq:OPex6} thành một ràng buộc lồi.
\item[ii.] Giải bài toán \eqref{eq:OPex6} với giả thiết $\nabla U_j(x_j) = x_j^{-\alpha}$ và $\alpha > 1$. 
\end{itemize}
\end{exercise}

\begin{exercise}\cite{Kar2012} 
Xét bài toán sản xuất phân phối trong đó hàm mục tiêu của mỗi máy phát dạng tam thức bậc hai $f_i(x_i)=a_ix_i^2+b_ix_i+c_i$, với $a_i$, $b_i$ , $c_i$ là các số thực, $a_i>0$, $x_i$ là sản lượng điện sản xuất bởi máy phát $i$. Điện năng sản xuất tối ưu tương ứng với bài toán tối ưu:
\begin{mini!}
{\m{x}\in \mb{R}^{dn}}{\tilde{f}(\m{x})=\sum_{i=1}^n f_i({x}_i)}
{\label{eq:OPex7}}{}
\addConstraint{\sum_{i=1}^n x_i = \sum_{i=1}^n x^l_i =x^*}{ }
\addConstraint{0 \leq x_i \leq \bar{x}_i.}{ }
\end{mini!}
trong đó $x_i^l$ là điện năng tiêu thụ bởi tải tại nút thứ $i$, $x^*$ là tổng điện năng tiêu thụ trong lưới điện, và $\bar{x}_i>0$ là giới hạn sản xuất điện năng của máy phát $i$. 
\begin{itemize}
\item[i.] Nếu các máy phát không có giới hạn trên ($\bar{x}_i = \infty$), hãy lập hàm Lagrange tương ứng của bài toán \eqref{eq:OPex7}. Tìm giá trị nghiệm tối ưu $x_i^*,\lambda^*$ tương ứng bằng phương pháp giải tích.

\item[ii.] Khi xét tới các ràng buộc, kí hiệu $\Omega_{\bar{a}}$, $\Omega_{\underline{a}}$ là tập các máy phát phát điện tại cận trên và cận dưới. Chứng minh rằng nghiệm của \eqref{eq:OPex7} thỏa mãn:
\begin{subequations}
\begin{align}
\lambda^* &= 2a_i x^* + b_i,\quad n \notin \Omega_{\bar{a}}, \\
x_i^* &= \left\lbrace \begin{array}{cl}
\frac{\lambda^* - b_i}{2a_i}, & i \notin \Omega_{\bar{a}} \cup \Omega_{\underline{a}}\\
\bar{x}_i, & i \in \Omega_{\bar{a}}\\
0,	& i \in \Omega_{\underline{a}}
\end{array} \right.
\end{align}
\end{subequations}
\item[iii.] Tiếp theo ý (ii), hãy chứng minh biểu thức liên hệ của $\lambda^*$ với $x^*$ và $\bar{x}_i$:
\begin{align}
\lambda^*=\left(\sum_{i \notin \Omega_{\bar{a}} \cup \Omega_{\underline{a}}} \frac{1}{2a_i} \right)^{-1} \left(x^* - \sum_{i \in \Omega_{\bar{a}}} \bar{x}_i +  \sum_{i \notin \Omega_{\bar{a}} \cup \Omega_{\underline{a}}} \frac{b_i}{2a_i} \right).
\end{align}
\end{itemize}
\end{exercise}

%
%

%
\chapter{Mô hình động học ý kiến}
\label{chap:social_network}
Nghiên cứu về các hiện tượng xã hội đã đưa đến những mô hình toán học liên quan tới đồ thị và hệ đồng thuận. Trong thời gian gần đây, các mô hình động học ý kiến/quan điểm cổ điển đã được xem xét lại dưới góc nhìn của lý thuyết điều khiển, đồng thời một số mô hình mới đã được đề xuất nhờ sự hợp tác giữa một số nhà xã hội học và nhà điều khiển học \cite{Jia2015opinion,Parsegov2016novel,Proskurnikov2017Tut,Proskurnikov2018Tut}.

\section{Mô hình French - Degroot}
\index{mô hình!French-Degroot}
Xét một hệ gồm $n$ tác tử, mỗi tác tử có một quan điểm/ý kiến riêng cho bởi $x_i[k]$, $i=1,\ldots,n$, và $k=0,1,2,\ldots$ là chỉ số thể hiện thời điểm tương tác (rời rạc). Mạng xã hội thể hiện tương tác giữa các tác tử, được mô tả bởi một đồ thị $G=(V,E)$ có tính đến khuyên (các cạnh có dạng $(i,i)$). Trọng số $\omega_{ji}$ tại mỗi cạnh $(j,i)$ của $G$ thể hiện ảnh hưởng xã hội của tác tử $j$ trong việc hình thành quan điểm của tác tử $i$. Tại mỗi thời điểm $k\ge 0$, mỗi tác tử tổng hợp ý kiến của bản thân và các tác tử khác trong hệ để thay đổi ý kiến của bản thân theo phương trình \cite{degroot1974reaching}:
\begin{align} \label{eq:c8_Degroot_model}
    x_i[k+1] = \sum_{j\in M_i} a_{ij} x_i[k],~i=1,\ldots,n,
\end{align}
với $M_i=N_i \cup \{i\}$ và các trọng số thỏa mãn $a_{ij} = \omega_{ji}>0$ nếu $j\in N_i$, $a_{ij}=0$ nếu $j\notin M_i$ và $\sum_{j\in M_i}a_{ij}=1$. Trọng số $a_{ii}$ thể hiện mức độ cởi mở trong tiếp thu ý kiến bên ngoài của tác tử $i$. Nếu $a_{ii}=0$, quan điểm của $i$ hoàn toàn phụ thuộc vào các tác tử láng giềng. Ngược lại, nếu $a_{ii}=1$, tác tử $i$ là một tác tử bảo thủ không tiếp thu ý kiến từ bên ngoài ($x_i[k] = x_i[0], \forall k \ge 0$). Mô hình \eqref{eq:c8_Degroot_model} với $a_{ij} = \frac{1}{|M_i|}, \forall j \in M_i$, được đề xuất bởi French để nghiên cứu về quyền lực xã hội \cite{French1956}. Degroot đề xuất mô hình \eqref{eq:c8_Degroot_model} độc lập với French nhằm nghiên cứu sự thay đổi quan điểm của một nhóm người sau một số lần họp mặt và trao đổi ý kiến. Có thể thấy rằng, mô hình French-Degroot thực chất là một mô hình đồng thuận rời rạc. Viết lại phương trình \eqref{eq:c8_Degroot_model} dưới dạng ma trận
\begin{align} \label{eq:c8_Degroot_model1}
    \m{x}[k+1] = \m{A} \m{x}[k],~k=0,1,2,\ldots
\end{align}
trong đó $\m{x}[k] = [x_1[k],\ldots,x_n[k]]^\top$ và $\m{A}$ là một ma trận có các phần tử không âm ($a_{ij}\ge 0, \forall i, j = 1,\ldots, n$), ngẫu nhiên hàng ($\m{A}\m{1}_n = \m{1}_n$). Hệ \eqref{eq:c8_Degroot_model1} hội tụ nếu tồn tại giới hạn $\m{x}^\infty = \lim_{k\to +\infty} \m{x}[k] = \lim_{k\to +\infty} \m{A}^k \m{x}[0]$. Hơn nữa, hệ \eqref{eq:c8_Degroot_model1} là tiến tới đồng thuận nếu $\m{x}^\infty = \bar{x} \m{1}_n$ với $\bar{x} = \bm{\gamma}^\top \m{x}[0] = \sum_{j=1}^n \gamma_i x_i[0]$, hay $\lim_{k\to +\infty} \m{A}^k = \m{1}_n \bm{\gamma}^\top$. Ta có một số kết luận về hệ \eqref{eq:c8_Degroot_model1} đối với các giả thiết thêm:
\begin{itemize}
    \item Nếu đồ thị $G$ là liên thông mạnh thì $\m{x}[k]$ ở mô hình \eqref{eq:c8_Degroot_model1} tiệm cận tới một điểm trong tập đồng thuận khi và chỉ khi $G$ là không có chu kỳ. Ngược lại, \eqref{eq:c8_Degroot_model1} phân kỳ với hầu hết mọi điều kiện đầu $\m{x}(0)$.
    \item Hệ \eqref{eq:c8_Degroot_model1} tiệm cận tới một điểm cân bằng khi và chỉ khi mọi thành phần liên thông mạnh tối đa của $G$ là không có chu kì. Hệ \eqref{eq:c8_Degroot_model1} tiệm cận tới một điểm cố định trong tập đồng thuận khi và chỉ khi $G$ có gốc ra và thành phần liên thông mạnh chứa gốc ra là không có chu kỳ.
    \item Nếu $a_{ii}>0$ với $\forall i = 1, \ldots, n$ thì giả thiết đồ thị không có chu kỳ được thỏa mãn, \eqref{eq:c8_Degroot_model1} tiệm cận tới một điểm cụ thể. Nếu có thêm giả thiết $G$ là có gốc ra thì điểm cân bằng của \eqref{eq:c8_Degroot_model1} thuộc tập đồng thuận.
    \item Giả sử hệ \eqref{eq:c8_Degroot_model1} tiệm cận tới một điểm trong tập đồng thuận. Khi đó, $\bm{\gamma}$ là vector riêng bên trái duy nhất của $\m{A}$ ứng với giá trị riêng 1 và thỏa mãn $\bm{\gamma}^\top\m{1}_n=1, \gamma_i \ge 0, \forall i=1, \ldots, n$. Mỗi phần tử $\gamma_i$ thể hiện mức ảnh hưởng của tác tử $i$ tới ý kiến đồng thuận và còn được gọi là chỉ số trung tâm hay chỉ số ảnh hưởng của tác tử thứ $i$ trong mạng xã hội \cite{French1956}.
\end{itemize}

\section{Mô hình Friendkin - Johnsen}
\label{c8:FJ_model}
Friendkin và Johnsen \cite{Friedkin1986,Friedkin1990} đề xuất một mở rộng của mô hình French-Degroot bằng cách xem xét thêm ảnh hưởng của định kiến tới sự thay đổi quan điểm của các tác tử. \index{mô hình!Friedkin-Johnsen}Mô hình Friedkin-Johnsen (F-J) được mô tả bởi phương trình:
\begin{align} \label{eq:c8_FJ_model1}
x_i[k+1] = \theta_{ii} \sum_{j\in M_i} a_{ij}x_j[k] + (1-\theta_{ii})u_i,
\end{align}
hay biểu diễn ở dạng ma trận:
\begin{align} \label{eq:c8_FJ_model2}
    \m{x}(k+1)= \m{\Theta}\m{A}\m{x}[k] + (\m{I}_n - \m{\Theta}) \m{u},
\end{align}
trong đó ma trận $\m{A}$ được định nghĩa giống như với mô hình French-Degroot, $\m{\Theta} = \text{diag}(\theta_{11}, \ldots, \theta_{nn})$, với $\theta_{ii} \in [0,1]$ mô tả mức ảnh hưởng từ bên ngoài tới quan điểm của tác tử thứ $i$ và $\m{u}\in \mb{R}^n$ được gọi là vector định kiến của các tác tử, thường được chọn bởi $\m{u}=\m{x}[0]$. Dễ thấy khi $\bm{\Theta}=\m{I}_n$, mô hình Friendkin-Johnsen \eqref{eq:c8_FJ_model2} trở thành mô hình French-DeGroot. 

Xét trường hợp $\bm{\Theta}\ne\m{I}_n$. Một tác tử $i$ là \textit{có định kiến} nếu $\theta_i<1$. Một tác tử $i$ là \textit{bị ảnh hưởng bởi định kiến} nếu $i$ là tác tử có định kiến hoặc $i$ bị ảnh hưởng bởi một tác tử có định kiến, hay nói cách khác, tồn tại một đường đi từ một tác tử $j$ có định kiến tới $i$. Tác tử $i$ không bị ảnh hưởng bởi định kiến gọi là một \textit{tác tử độc lập}. Với cách phân chia này, đồ thị $G$ có thể được đánh số lại sao cho các tác tử bị ảnh hưởng bởi định kiến được cho bởi $1,\ldots, r$ và các tác tử độc lập là $r+1,\ldots,n$, với $r\ge1$. Ý kiến của hai nhóm tác tử này được cho tương ứng bởi các vector $\m{x}_1\in \mb{R}^r$ và $\m{x}_2\in \mb{R}^r$, thay đổi theo phương trình:
\begin{align}
    \m{x}_1[k+1]&= \bm{\Theta}^{11}[\m{A}^{11} \m{x}_1[k] + \m{A}^{11} \m{x}_2[k]] + (\m{I}_r - \m{\Theta}^{11})\m{u}_1 \label{eq:c8_FJ_Model3}\\
    \m{x}_2[k+1]&= \m{A}^{22} \m{x}_2[k], \label{eq:c8_FJ_Model4}
\end{align}
trong đó $\m{A}^{22}$ là một ma trận ngẫu nhiên hàng (do đó hệ con \eqref{eq:c8_FJ_Model4}) có dạng mô hình French-Degroot. Một số kết quả về tính hội tụ của \eqref{eq:c8_Degroot_model1} được tóm gọn lại như sau \cite{Proskurnikov2016}:
\begin{itemize}
    \item Hệ con \eqref{eq:c8_FJ_Model3} là ổn định tiệm cận với ma trận $\bm{\Theta}^{11}\m{A}^{11}$ là ổn định Schur (tức là ma trận có bán kính quang phổ $\rho(\bm{\Theta}^{11}\m{A}^{11})<1$). Hệ \eqref{eq:c8_FJ_Model3} tiêm cận tới:
\begin{align}
\m{x}^{1\infty} = \m{V} \begin{bmatrix}
\m{u}^1\\ \m{x}^{2\infty}
\end{bmatrix}, \text{ với } \m{V} = (\m{I}_r - \bm{\Theta}^{11}\m{A}^{11})^{-1}[\m{I}_r-\bm{\Theta}^{11}~ \bm{\Theta}^{11}\m{A}^{11}]
\end{align}
khi và chỉ khi $r=n$ hoặc ma trận $\m{A}^{22}$ là hội tụ ($\lim_{k\to \infty}(\m{A}^{22})^k = \m{0}$).
\item Hệ \eqref{eq:c8_FJ_model2} ổn định tiệm cận khi và chỉ khi mọi tác tử là bị ảnh hưởng bởi định kiến. Điều này đạt được khi và chỉ khi $\bm{\Theta}<\m{I}_n$ hoặc $\bm{\Theta}<\m{I}_n$ và $G$ là liên thông mạnh. Lúc này, $\m{x}^\infty = \m{V}\m{u}$, với $\m{V} = (\m{I}_n - \bm{\Theta}\m{A})^{-1}(\m{I}_n - \bm{\Theta})$.
\end{itemize}

\begin{story}{Mô hình động học ý kiến và lý thuyết điều khiển}
Những nghiên cứu gần đây về mô hình động học ý kiến Friendkin - Johnsen đa chiều là một ví dụ khá thành công của hợp tác học thuật giữa các nhà xã hội học (Noah Friedkin - Đại học California, Santa Barbara, Hoa Kỳ) và các nhà điều khiển học (A. Proskunikov - Đại học Bách Khoa Turin, Ý; R. Tempo - Đại học Bách Khoa Torino, Ý; S. Parsegov - Viện Vật lý và Công nghệ Moscow, Nga).

Nghiên cứu các mô hình động học ý kiến giải thích về mặt toán học mối liên hệ của các cấu trúc xã hội đối với việc hình thành và thay đổi quan điểm cá nhân. Tuy nhiên, để một mô hình toán học lý thuyết được chấp nhận trên các tạp chí khoa học như Science \cite{Friedkin2016}, mô hình phải được xây dựng dựa trên các tiền đề đã được kiểm chứng qua nghiên cứu thực nghiệm từ tâm lý học hành vi và xã hội học thực nghiệm. Một số giả thiết quan trọng để bảo vệ sự hợp lý của các mô hình động học ý kiến Friendkin-Johnsen bao gồm:
\begin{itemize}
\item Nghiên cứu về một chủ đề xã hội có thể được xem xét từ các đơn vị tối thiểu là cá nhân.
\item Ý kiến, quan điểm, hay thái độ của mỗi cá nhân với một đối tượng (chủ đề xã hội, sự kiện, hay cá nhân khác) có thể được lượng hóa bởi các đại lượng vô hướng hoặc vector. Việc lượng hóa cho phép đánh giá mức độ tin cậy chủ quan của cá nhân đối với đối tượng (khi giá trị quan điểm được chuẩn hóa trong khoảng $[0,1]$, hay thái độ đối với đối tượng (khi giá trị quan điểm có thể nhận cả giá trị dương và âm).
\item Việc hình thành quan điểm cá nhân là thông qua những lần tương tác với các cá thể khác, ý kiến cá nhân là một quá trình động, và các tương tác là các quá trình ngẫu nhiên. Do mạng xã hội là chứa rất nhiều cá thể, việc lựa chọn các hệ số tĩnh trong mô hình là dựa trên việc lấy trung bình các khả năng có thể, hoặc thông qua dữ liệu thực nghiệm đã chuẩn hóa.
\item Các tương tác không có giá trị giống nhau, ảnh hưởng của các cá thể khác nhau đối với một cá thể dựa trên nhiều yếu tố. Điều này giải thích việc chuẩn hóa các trọng số và cập nhật ý kiến mới theo một tổ hợp lồi của ý kiến từ các nguồn khác nhau.
\end{itemize}
Ngay trong quá trình hình thành, điều khiển đa tác tử đã là một hướng nghiên cứu liên ngành. Các chủ đề như động học ý kiến, mô hình lan truyền dịch bệnh là các ví dụ gần đây thể hiện tính liên ngành trong nghiên cứu khoa học hiện nay.
\end{story}

\begin{example}[Mô hình Friendkin - Johnsen] \label{VD:8.1}
Xét hệ gồm 4 tác tử với các giá trị ý kiến ban đầu lấy ngẫu nhiên trong khoảng $[0,1]$. Đầu tiên, xét hệ có đồ thị tương tác mô tả bởi ma trận
\begin{align*}
    \m{A} = \begin{bmatrix} 
     0.220 & 0.120 & 0.360 & 0.300\\
     0.147 & 0.215 & 0.344 & 0.294\\
     0.000 & 0.000 & 1.000 & 0.000\\
     0.090 & 0.178 & 0.446 & 0.286
    \end{bmatrix}.
\end{align*}
Với ma trận kề này, tác tử 3 hoàn toàn không cập nhật quan điểm của mình do $\lambda_{33} = 1 - a_{33}=0$. Kết quả mô phỏng với hệ này được thể hiện trên Hình~\ref{fig:VD_8.1}a. Các tác tử không đạt được đồng thuận do ảnh hưởng của định kiến. Khi ma trận kề được lựa chọn là 
\begin{align*}
    \m{A} = \begin{bmatrix} 
     0.220 & 0.120 & 0.360 & 0.300\\
     0.147 & 0.215 & 0.344 & 0.294\\
     0.000 & 0.400 & 0.600 & 0.000\\
     0.090 & 0.178 & 0.446 & 0.286
    \end{bmatrix}.
\end{align*}
mọi tác tử cập nhật ý kiến, tuy nhiên giá trị đồng thuận không đạt được như thể hiện trên Hình \ref{fig:VD_8.1}b.
\end{example}
\begin{figure}[th]
\centering
\subfloat[]{\includegraphics[height = 5.23cm]{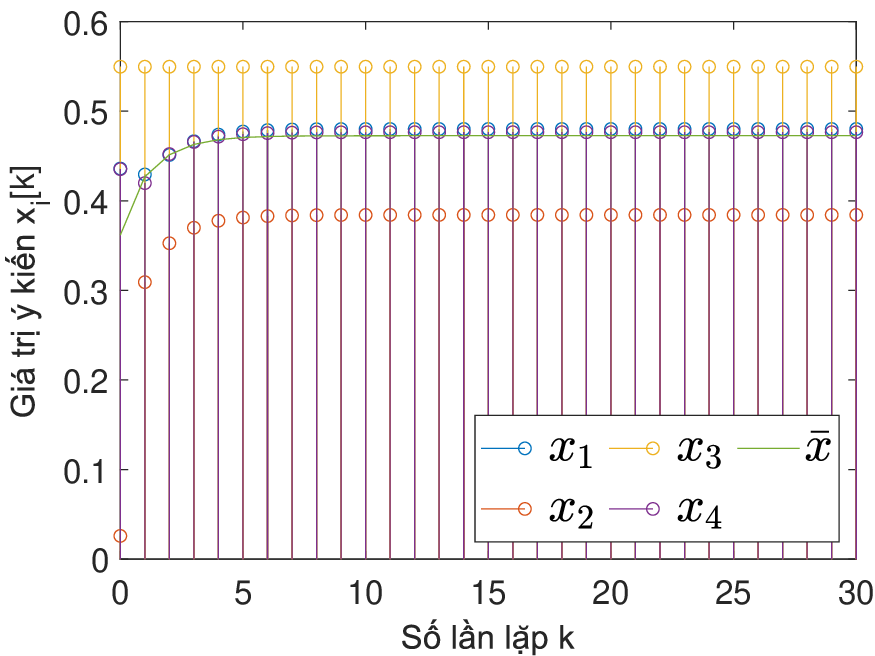}}\hfill 
\subfloat[]{\includegraphics[height = 5.23cm]{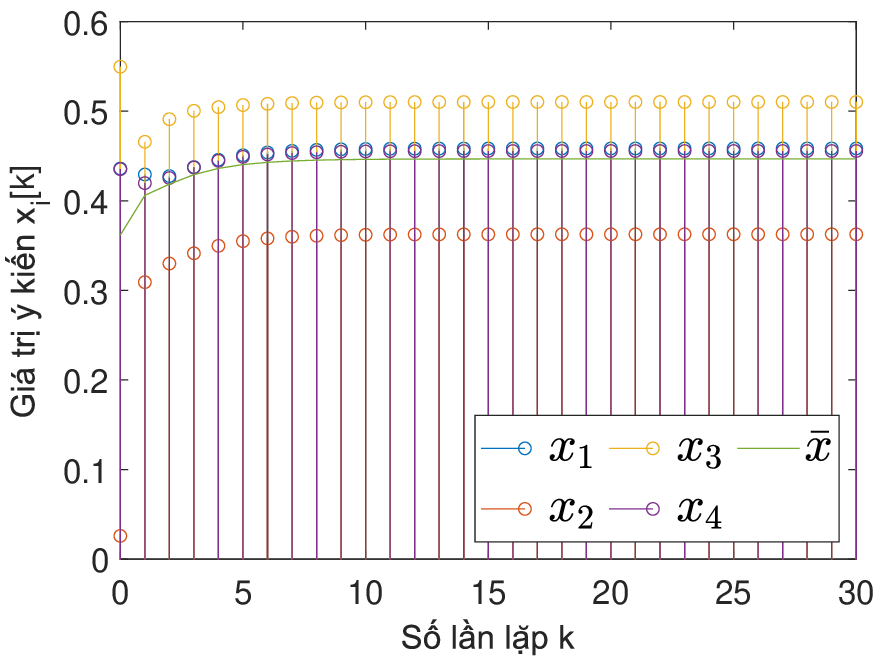}}
\caption{Mô phỏng hệ 4 tác tử với mô hình F-J trong hai trường hợp khác nhau của đồ thị tương tác.}
\label{fig:VD_8.1}
\end{figure}

\section{Mô hình Abelson và mô hình Taylor}
\label{c8:Abelson_Taylor}
\index{mô hình!Abelson}
Dạng liên tục của mô hình French-Degroot và mô hình Friedkin-Johnsen là mô hình Abelson \cite{Abelson1967} và mô hình Taylor \cite{Taylor1968}. Abelson \cite{Abelson1967} đưa ra một mô hình giải thích hiện tượng ``quan điểm của các cá thể có xu hướng tiến lại gần nhau sau quá trình thảo luận'' dưới dạng phương trình hệ đồng thuận tuyến tính liên tục:
\begin{align}
\dot{x}_i(t) = \sum_{j\in N_i} a_{ij}(x_j - x_i), ~ i=1, \ldots, n.
\end{align}
Mô hình Abelson có thể coi là dạng liên tục của mô hình Degroot. Điều kiện để mọi tác tử hội tụ tới giá trị đồng thuận là đồ thị G có gốc ra. Mô hình Taylor mở rộng mô hình Abelson, giải thích sự phân hóa về quan điểm do ảnh hưởng của các nguồn thông tin bên ngoài:
\begin{align} \label{eq:c10_taylor}
\dot{x}_i(t) = \sum_{j\in N_i} a_{ij}(x_j - x_i)+\sum_{r=1}^l b_{ir}(s_r - x_i), ~ i=1,\ldots, n.
\end{align}
trong đó $s_1,\ldots, s_l$ là các hằng số mô tả ảnh hưởng từ bên ngoài và $b_{ir} \ge 0$ thể hiện ảnh hưởng từ nguồn thông tin thứ $k$ tới tác tử $i$. Khi $G$ là có gốc ra, mô hình Taylor ổn định tiệm cận và hội tụ tới một điểm cân bằng duy nhất xác định bởi đồ thị $G$ và các giá trị $s_1,\ldots,s_l$ \cite{Taylor1968,Ren2007distr}. Chú ý rằng mô hình Taylor \eqref{eq:c10_taylor} được nghiên cứu trong \cite{Ren2007distr} với tên gọi điều khiển bao của hệ đa tác tử\footnote{containment control}\index{điều khiển!bao lồi}, trong đó các tác tử theo sau cần hội tụ tới bao lồi định nghĩa bởi tập vị trí của các tác tử dẫn đầu (leader).\index{mô hình!Taylor}

\begin{example}[Mô hình Taylor] \label{VD:8.2}
Xét ví dụ hệ gồm 3 leader tại $\m{x}_1=[0, 0]^\top$, $\m{x}_2=[1, 0]^\top$ và $\m{x}_3=[0, 1]^\top$ và 7 tác tử thường với vị trí đầu được lấy ngẫu nhiên trong khoảng $-3$ tới $3$. Vị trí đầu của các leader được kí hiệu bởi khuyên tròn nhỏ màu đỏ, còn các tác tử khác kí hiệu bởi màu đen. 

Mô phỏng cho thấy sau khoảng 10 giây, các tác tử $i=4,\ldots,10$ hội tụ về miền tam giác với ba đỉnh $\m{x}_i~,i=1, 2, 3$ như trên Hình~\ref{fig:VD8.2}. Chú ý rằng các tác tử leader không thay đổi vị trí (không cập nhật ý kiến) từ các tác tử thường.
\end{example}

\begin{SCfigure}[][t!]
\caption{Mô phỏng hệ 10 tác tử với mô hình Taylor mở rộng}
\label{fig:VD8.2}
\hspace{4cm}
\includegraphics[height=6.5cm]{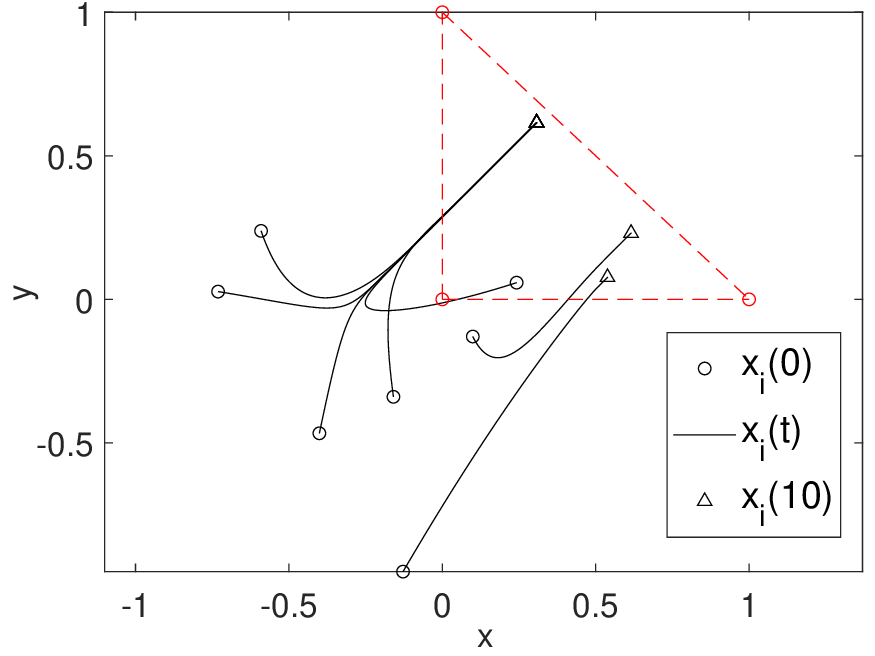}
\end{SCfigure}
\section{Mô hình Friendkin - Johnsen đa chiều và một số mở rộng}
\index{mô hình!Friendkin - Johnsen đa chiều}
Một mở rộng tự nhiên của mô hình F-J là xét tới trường hợp các tác tử có quan điểm về nhiều vấn đề khác nhau. Các tác giả ở \cite{Friedkin2016,Proskurnikov2016} đề xuất thêm một ma trận liên hệ logic vào mô hình F-J để thể hiện quá trình suy lý của các tác tử khi có nhiều hơn một chủ đề được xem xét trong quá trình thảo luận. Giả sử mỗi tác tử $i$ có một vector quan điểm $\m{x}_i[k]\in \mb{R}^d$, mô hình F-J đa chiều được cho dưới dạng:
\begin{align} \label{eq:c8_MDFJ_Model1}
\m{x}_i[k+1] = \theta_{ii} \m{C} \sum_{j \in M_i} a_{ij} \m{x}_j[k] + (1-\theta_{ii}) \m{u}_i,~i=1, \ldots, n.
\end{align}
Trong trường hợp $\m{C} = \m{I}_n$, mô hình \eqref{eq:c8_MDFJ_Model1} trở thành mô hình F-J thông thường. Định nghĩa $\m{x} = [\m{x}_1^\top, \ldots, \m{x}_n^\top]^\top$, ta viết lại mô hình \eqref{eq:c8_MDFJ_Model1} dưới dạng ma trận như sau:
\begin{align} \label{eq:c8_MDFJ_Model2}
\m{x}[k+1] = [(\bm{\Theta}\m{A})\otimes \m{C}]\m{x}[k] + [(\m{I}_n - \bm{\Theta}) \otimes \m{I}_d] \m{u}.
\end{align}
Mô hình \eqref{eq:c8_MDFJ_Model2} thể hiện rằng quá trình hình thành quan điểm của mỗi tác tử là sự tương tác giữa những ảnh hưởng từ bên ngoài (ma trận $\m{A}$), định kiến ban đầu (ma trận $\bm{\Theta}$) và hệ thống niềm tin của mỗi tác tử (ma trận $\m{C}$). Một số kết quả về mô hình \eqref{eq:c8_MDFJ_Model2} được tóm tắt lại như sau:
\begin{itemize}
    \item Mô hình \eqref{eq:c8_MDFJ_Model2} ổn định khi và chỉ khi $\rho(\bm{\Theta}\m{A})\rho(\m{C}) < 1$. Khi đó, $\m{x}^\infty = \lim_{k \to +\infty}\m{x}[k] = (\m{I}_{dn}-\bm{\Theta}\m{A} \otimes \m{C})^{-1}[(\m{I}_n - \bm{\Theta})\otimes \m{I}_d]\m{u}$.
    \item Trong trường hợp hệ có một số tác tử độc lập, ta đánh số thứ tự các tác tử như ở mô hình F-J thường. Mô hình \eqref{eq:c8_MDFJ_Model2} hội tụ khi và chỉ khi tồn tại giới hạn $\m{C}_* = \lim_{k\to +\infty}\m{C}^k$ đồng thời hoặc $\m{C}_* = \m{0}$ hoặc tồn tại $\m{A}_*^{22} = \lim_{k\to +\infty}\m{A}^{22}$. Khi đó, 
\begin{align}
\lim_{k\to +\infty} \m{x}[k] \to &\begin{bmatrix}
    (\m{I}_r - \m{\Theta}^{11}\m{A}^{11}\otimes \m{C})^{-1} & \m{0}\\ \m{0} & \m{I}_{n-r}
    \end{bmatrix} \cdot \nonumber\\
    &\qquad \cdot \begin{bmatrix} 
    (\m{I}_r - \m{\Theta}^{11})\otimes \m{I}_d & (\m{\Theta}^{11}\m{A}^{12}\m{A}^{22}_*)\otimes \m{C}\m{C}_*\\
    \m{0} & \m{A}_*^{22}\otimes \m{C}_*
    \end{bmatrix} \m{u},
    \end{align}
    với qui ước rằng khi $\m{C}_* = \m{0}$ nhưng $\lim_{k\to +\infty}\m{A}^{22}$ không tồn tại thì ta gán $\m{A}_*^{22} = \m{0}$.
\end{itemize}
\begin{figure}[t!]
\centering
\subfloat[]{\includegraphics[height=5.5cm]{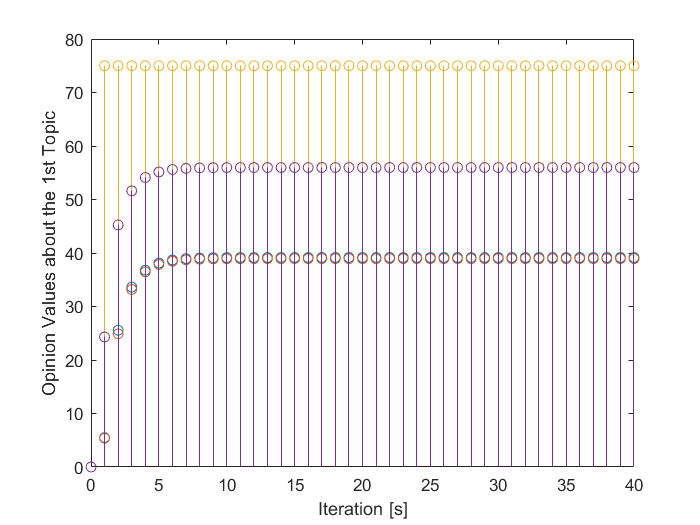}} \hfill
\subfloat[]{\includegraphics[height=5.5cm]{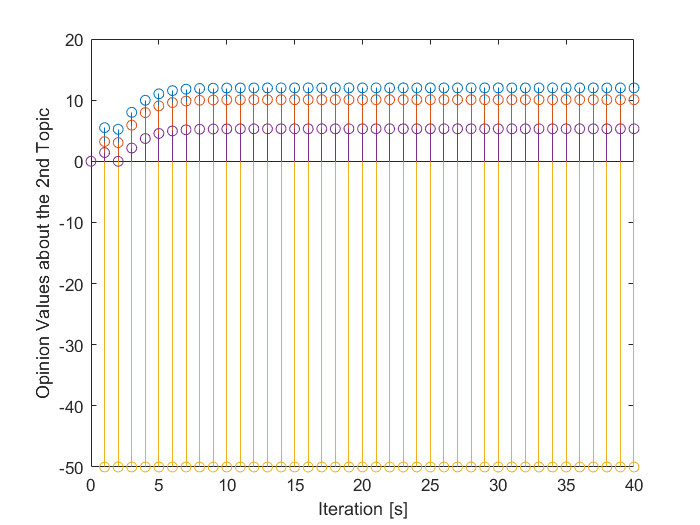}}
\caption{Mô phỏng mô hình F-J với ma trận $\m{C}=\begin{bmatrix} 0.8 & 0.2\\0.3 & 0.7 \end{bmatrix}$.}
\label{fig:VD_83_1}
\end{figure}
\begin{figure}[th!]
\centering
\subfloat[]{ \includegraphics[height=5.5cm]{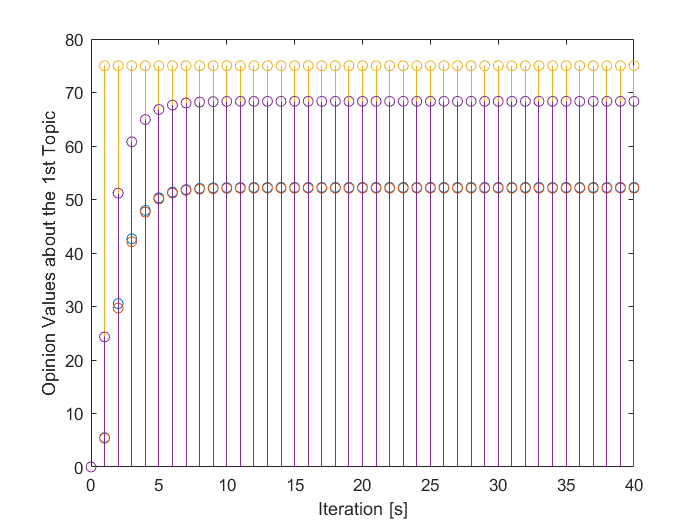}} \hfill
\subfloat[]{ \includegraphics[height=5.5cm]{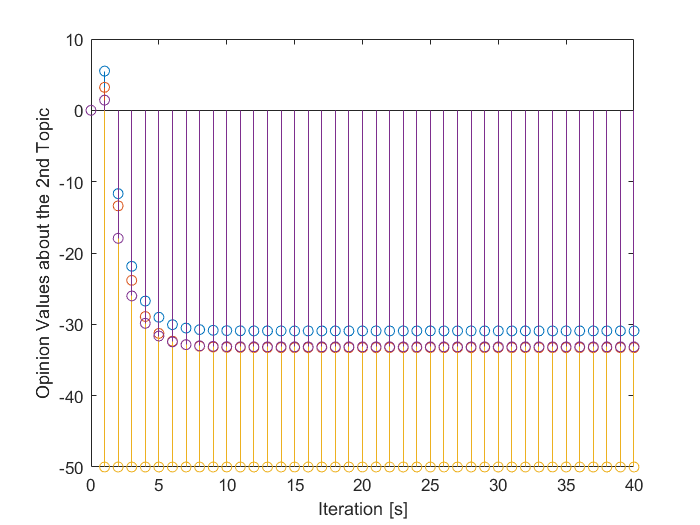}} 
\caption{Mô phỏng mô hình F-J với ma trận $\m{C}=\begin{bmatrix} 0.8 & -0.2\\-0.3 & 0.7 \end{bmatrix}$.}
    \label{fig:VD_83_2}
\end{figure}

\begin{example}[Mô hình Friedkin-Johnsen đa chiều] \cite{Proskurnikov2016} \label{VD:8.3} 
Xét một mạng xã hội gồm $n=4$ tác tử, trong đó các mối quan hệ giữa các cá nhân được cho bởi ma trận kề như ở Ví dụ \ref{VD:8.1}. Các tác tử thảo luận về hai chủ đề, (a) và (b), hai chủ đề có liên quan đến nhau nhưng đối lập nhau, ví dụ, các tác tử thảo luận về quan điểm của họ đối với chế độ ăn chay và với chế độ ăn toàn thịt. Giá trị đầu của ý kiến các tác tử được cho bởi $\m{x}[0]=[25, 25, 25, 15, 75, -50, 35, 5]^\top$. 

Khi ma trận logic được chọn là $\m{C} =\begin{bmatrix} 0.8 & 0.2\\0.3 & 0.7 \end{bmatrix},$ ta có
\[\lim_{t\to+\infty} \m{x}(t) = [39.160, ~12.001, ~38.948, ~10.071, ~75, ~-50, ~55.977,~ 5.314]^\top.\]
Có thể thấy rằng, ngoài tác tử 3 không thay đổi quan điểm thì quan điểm của các tác tử về việc ăn chay và ăn thịt đều có xu hướng tích cực hơn (tăng lên). Điều này tỏ ra mâu thuẫn do hai chủ đề đối nghịch nhau (người ăn chay thường có xu hướng không đánh giá tích cực chế độ ăn toàn thịt).

Tiếp theo, với cùng giá trị đầu $\m{x}[0]$ nhưng ma trận logic $\m{C}$ được thay bởi $\m{C} =\begin{bmatrix} 0.8 & -0.2\\-0.3 & 0.7 \end{bmatrix},$ ta có
\[\lim_{k\to+\infty} \m{x}[k] = [52.27,~-30.92,~52.13,~-33.27,~75,~
-50,~68.35,~-33.15]^\top.\]
Trong trường hợp này, giá trị hội tụ $\m{x}^\infty$ thể hiện tính nhất quán trong quan điểm của các tác tử (tích cực về chế độ ăn chay, tiêu cực về chế độ ăn toàn thịt). 

Hai ví dụ mô phỏng củng cố kết luận rằng việc lựa chọn ma trận logic $\m{C}$ ảnh hưởng tới tính hợp lý của mô hình. Một số nghiên cứu hiện nay cố gắng ước lượng giá trị của ma trận logic từ phân tích các dữ liệu thực tế.
\end{example}

Hai mô hình liên tục tương ứng của mô hình F-J đa chiều được đề xuất bởi Ye trong \cite{Ye2020Aut} có dạng như sau:\index{mô hình!Ye}
\begin{itemize}
    \item Mô hình 1:
    \begin{subequations}
    \begin{align}
    \dot{\m{x}}_i &= \sum_{j\in N_i}a_{ij}\m{C}(\m{x}_j - \m{x}_i) + (\m{C}-\m{I}_d)\m{x}_i + b_i(\m{x}_i(0)-\m{x}_i),~i=1, \ldots, n,\\
    \dot{\m{x}} &=-\left((\mcl{L}-\m{I}_n)\otimes\m{C}+(\m{B}+\m{I}_n)\otimes\m{I}_d \right)\m{x} + (\m{B}\otimes\m{I}_d)\m{x}(0).
    \end{align}
    \end{subequations}
    \item Mô hình 2:
    \begin{subequations}
    \begin{align}
     \dot{\m{x}}_i &= \sum_{j\in N_i}a_{ij}(\m{x}_j - \m{x}_i) + (\m{C}-\m{I}_d)\m{x}_i + b_i(\m{x}_i(0)-\m{x}_i),~i=1, \ldots, n,\\  
     \dot{\m{x}} &=-((\mcl{L}+\m{B})\otimes\m{I}_d - \m{I}_n \otimes (\m{C}-\m{I}_d))\m{x}+(\m{B}\otimes \m{I}_d) \m{x}(0).
    \end{align}
    \end{subequations}
\end{itemize}
Xét trường hợp $\m{B}=\text{diag}(b_i)\otimes \m{I}_d =\m{0}$ (các tác tử không có định kiến), và giả sử ma trận $\m{C}$ có $p$ giá trị riêng bằng 1 với $p$ vector riêng bên trái và bên phải trực chuẩn $\bm{\zeta}_r$ và $\bm{\xi}_r^\top$, $r=1,\ldots,p$ và các giá trị riêng khác đều có phần thực âm. Giả sử đồ thị $G$ có gốc ra, khi một số điều kiện liên hệ giữa ma trận $\m{C}$ và ma trận Laplace $\mcl{L}$ được thỏa mãn, hai mô hình động học liên tục trên hội tụ tới cùng một giá trị đồng thuận $\m{x}_i (t) \to (\sum_{r=1}^p\bm{\zeta}_r\bm{\xi}_r^\top) \sum_{j=1}^n \gamma_j \m{x}_j(0), \forall i=1, \ldots, n$.

Theo một góc nhìn khác, ta có thể coi ma trận $\m{C}$ đóng vai trò như một trọng số ma trận đồng nhất ảnh hưởng vào quá trình đồng thuận của mỗi tác tử \cite{Trinh2018matrix}. 
\begin{example}[Mô hình Ye] \cite{Ye2020Aut}
Xét hệ 8 tác tử với đồ thị tương tác giữa các tác tử có ma trận Laplace
\begin{align*}
    \mcl{L} = \begin{bmatrix}
    1 & 0 & -1 & 0 & 0 & 0 & 0 & 0\\
    -1 & 1 & 0 & 0 & 0 & 0 & 0 & 0 \\
    0 & -0.8 & 1 & -0.2 & 0 & 0 & 0 & 0\\
    0 & 0 & -1 & 1 & 0 & 0 & 0 & 0\\
    0 & 0 & 0 & -0.4 & 1 & 0 & -0.6 & 0\\
    0 & 0 & -0.2 & 0&-0.8 & 1 & 0 & 0\\
    0 & 0 & 0 & 0 & 0 & -1 & 1 & 0\\
    -0.3 & -0.7 & 0 & 0 & 0 & 0 & 0 & 1
    \end{bmatrix}.
\end{align*}
Ma trận thể hiện sự phụ thuộc logic giữa ba chủ đề được cho bởi 
\begin{equation}
\m{C} = \begin{bmatrix} 1 & 0 & 0\\ -0.1 & 0.2 & 0.7 \\ 0.1 & -0.8 & 0.1  \end{bmatrix}.
\end{equation}
Ma trận $\m{C}$ có thể mô tả 3 chủ đề như sau: 
\begin{itemize}
    \item Chủ đề 1: Triều Tiên có vũ khí hạt nhân đủ khả năng tấn công tới Mỹ.
    \item Chủ đề 2: Do Trung Quốc là đồng minh của Triều Tiên, Trung Quốc sẽ bảo vệ Bắc Hàn khi bị tấn công.
    \item Chủ đề 3: Mỹ sẽ tấn công Triều Tiên bằng vũ khí hạt nhân.
\end{itemize}

Với các giá trị quan điểm ban đầu được lấy ngẫu nhiên trong khoảng -1 tới 1, chúng ta mô phỏng mô hình Ye trong một số trường hợp. Đầu tiên, xét $b_i = 0, \forall i = 1, \ldots, 8$. Kết quả mô phỏng với cả hai mô hình Ye cho thấy các tác tử tiệm cận tới đồng thuận về cả 3 chủ đề, hơn nữa điểm đồng thuận của cả hai mô hình là như nhau (Hình~\ref{fig:c8_fig_ModelYe_Stable}). Tiếp theo, chúng ta thêm vào mô hình vector định kiến $\m{b} = [0, 0.1, 0, 0.05, 0, 0.4, 0, 0.3]^\top$. Lúc này, các tác tử không đạt được đồng thuận về cả ba chủ đề (Hình~\ref{fig:c8_ModelYe_Stubborn}). Cuối cùng, khi tăng tương tác giữa các tác tử bằng cách thay $\mcl{L}$ bởi $3\mcl{L}$, kết quả mô phỏng được cho trên Hình \ref{fig:c8_modelYe_Fast}. Mô hình 1 lúc này phân kỳ. Trong khi đó, việc tăng mức độ tương tác đẩy nhanh quá trình tiến tới đồng thuận trong mô hình 2. Như vậy, mô hình 2 không cho phép mô tả quá trình quá tải về thông tin giữa các tác tử.
\end{example}

\begin{figure}[th!]
\subfloat[]{\includegraphics[height=0.425\linewidth,angle=-90]{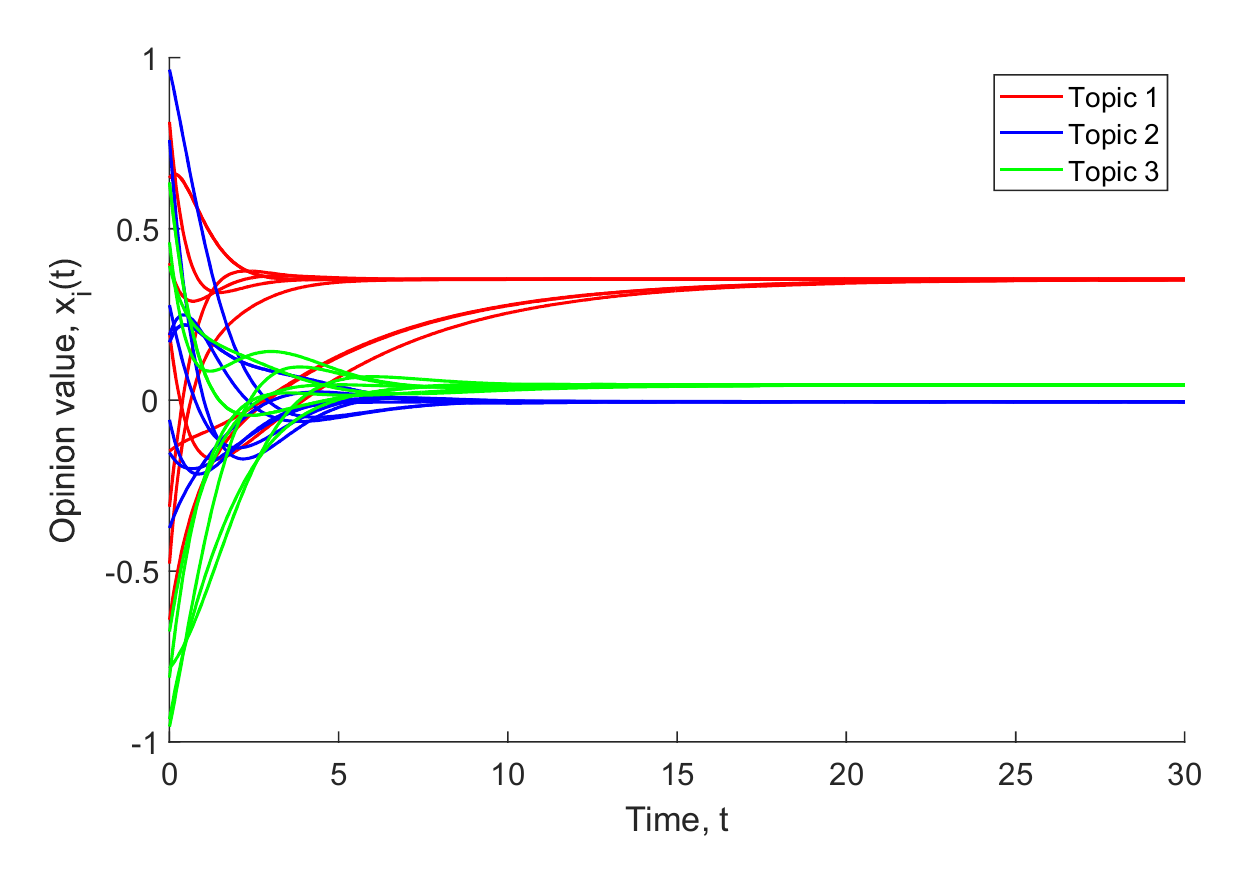}} \hfill
\subfloat{\includegraphics[height=0.425\linewidth,angle=-90]{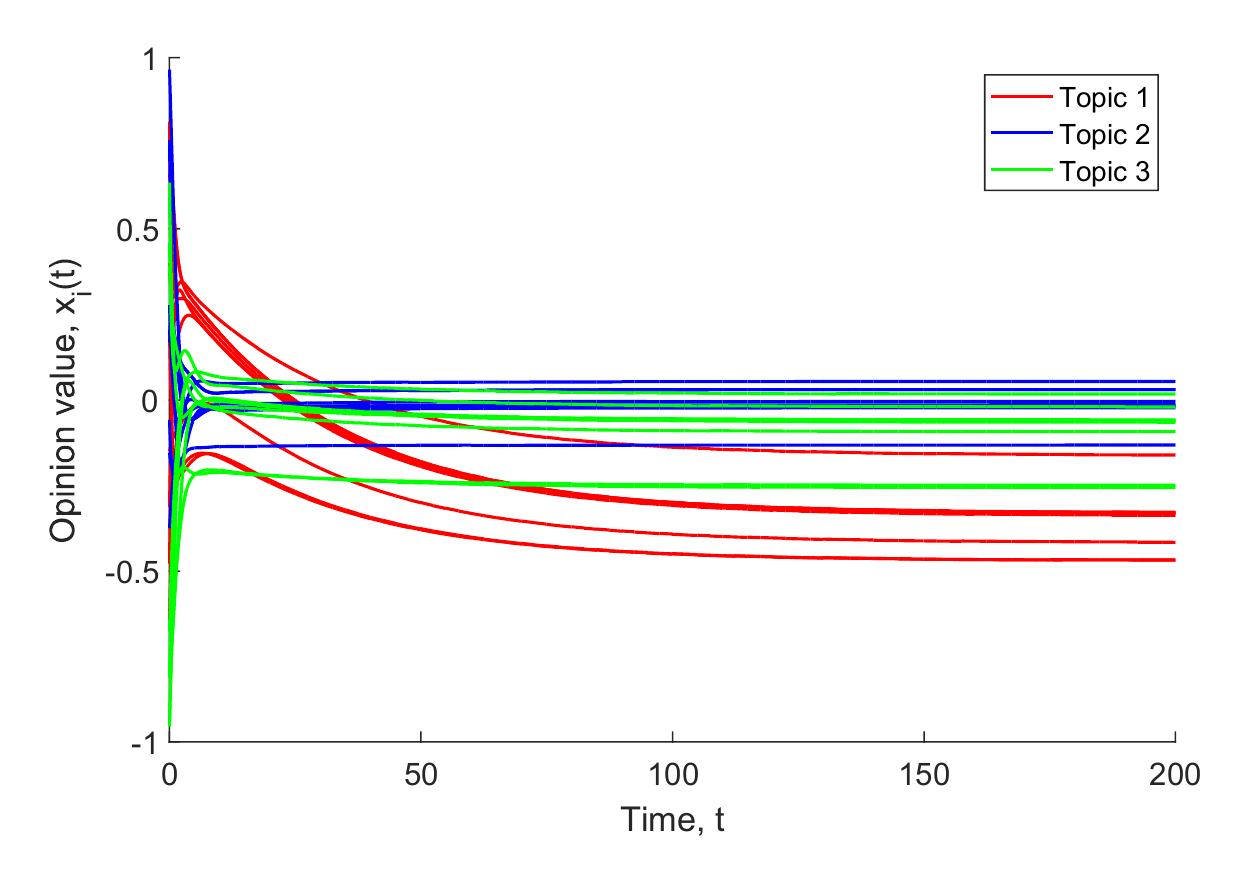}}
\caption{Mô hình Ye 1 khi hệ có và không có định kiến.}
\label{fig:c8_fig_ModelYe_Stable}
\end{figure}
\begin{figure}[th!]
\subfloat[]{\includegraphics[height=0.425\linewidth,angle=-90]{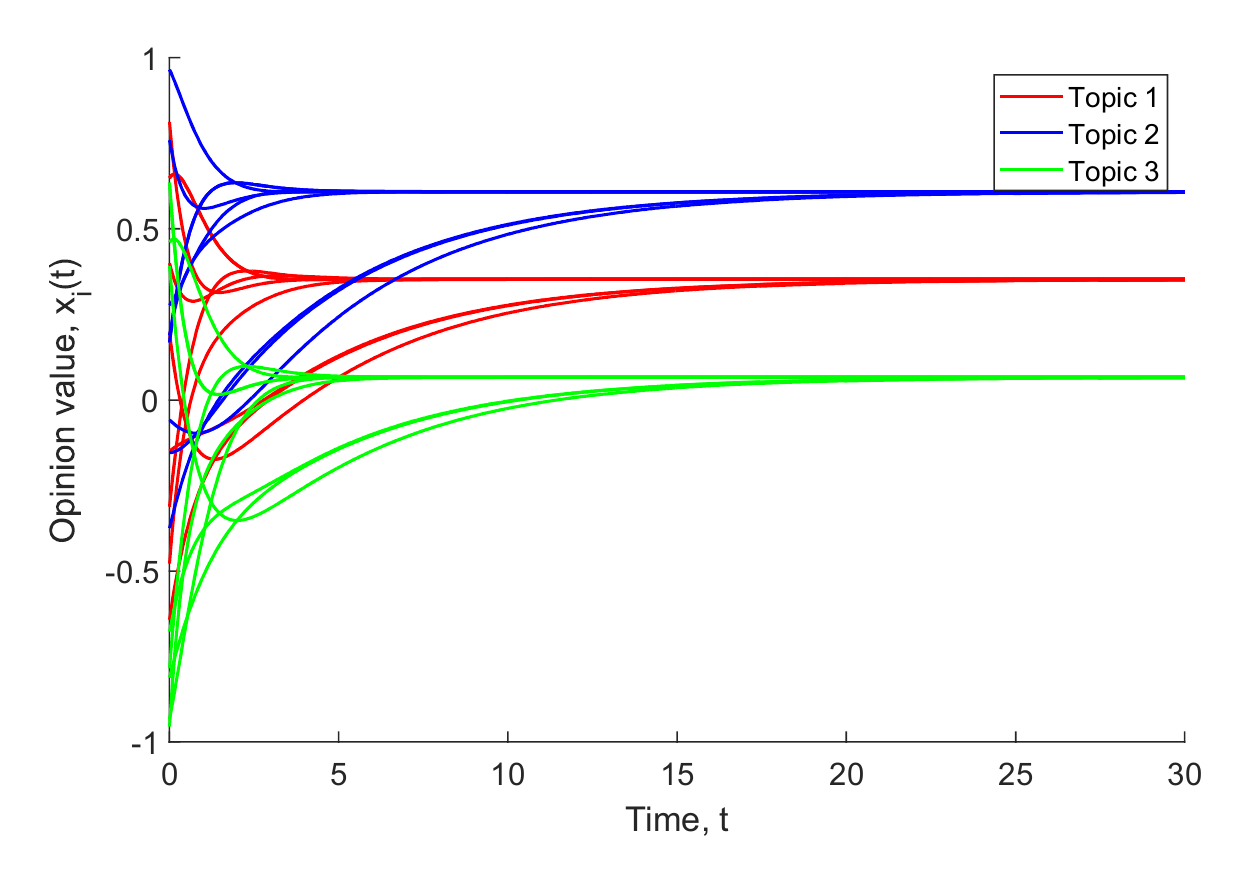}} \hfill
\subfloat[]{\includegraphics[height=0.425\linewidth,angle=-90]{Figures/c8_fig_Model1_Stubborn.pdf}}
\caption{Mô hình Ye 2 khi hệ có và không có định kiến.}
\label{fig:c8_ModelYe_Stubborn}
\end{figure}
\begin{figure}[!t]
\subfloat{\includegraphics[height=0.425\linewidth,angle=-90]{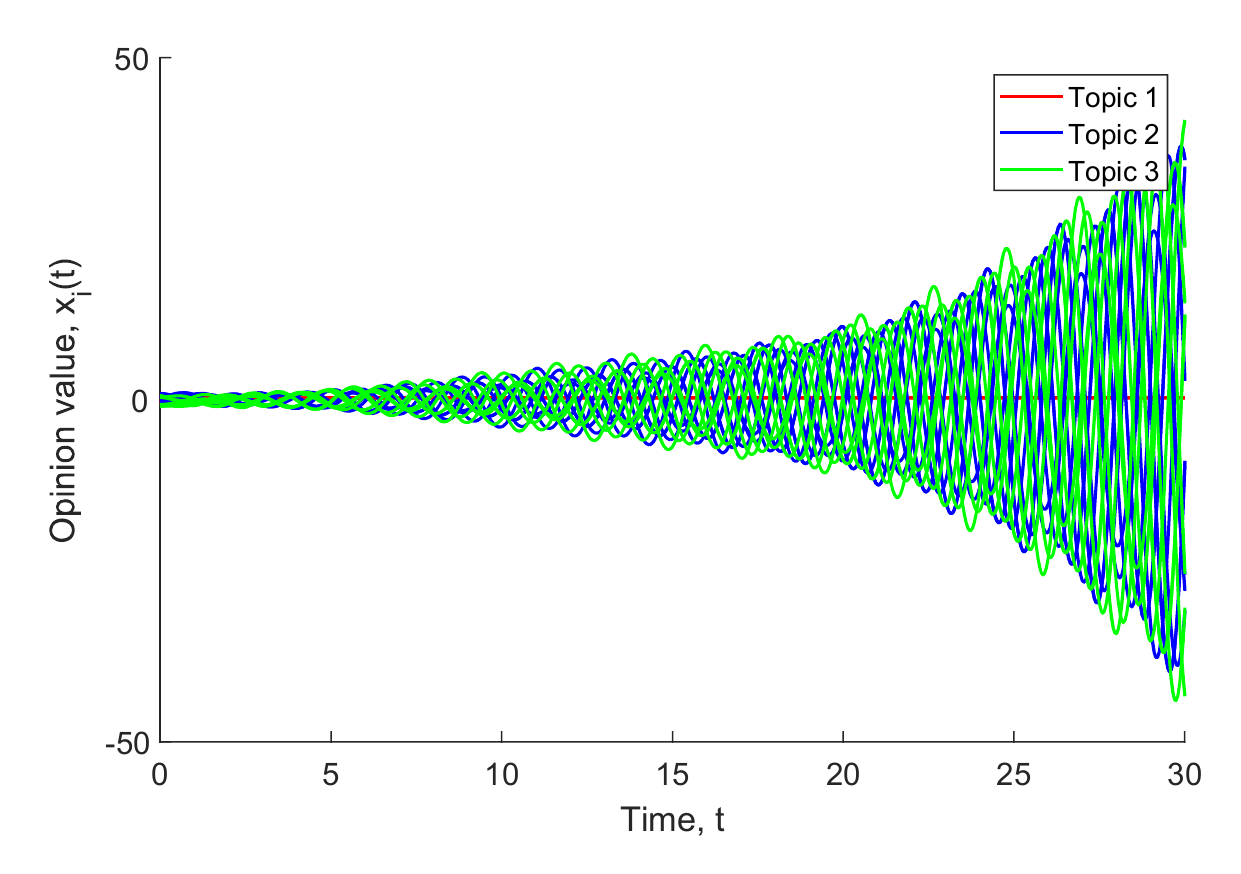}} \hfill
\subfloat{\includegraphics[height=0.425\linewidth,angle=-90]{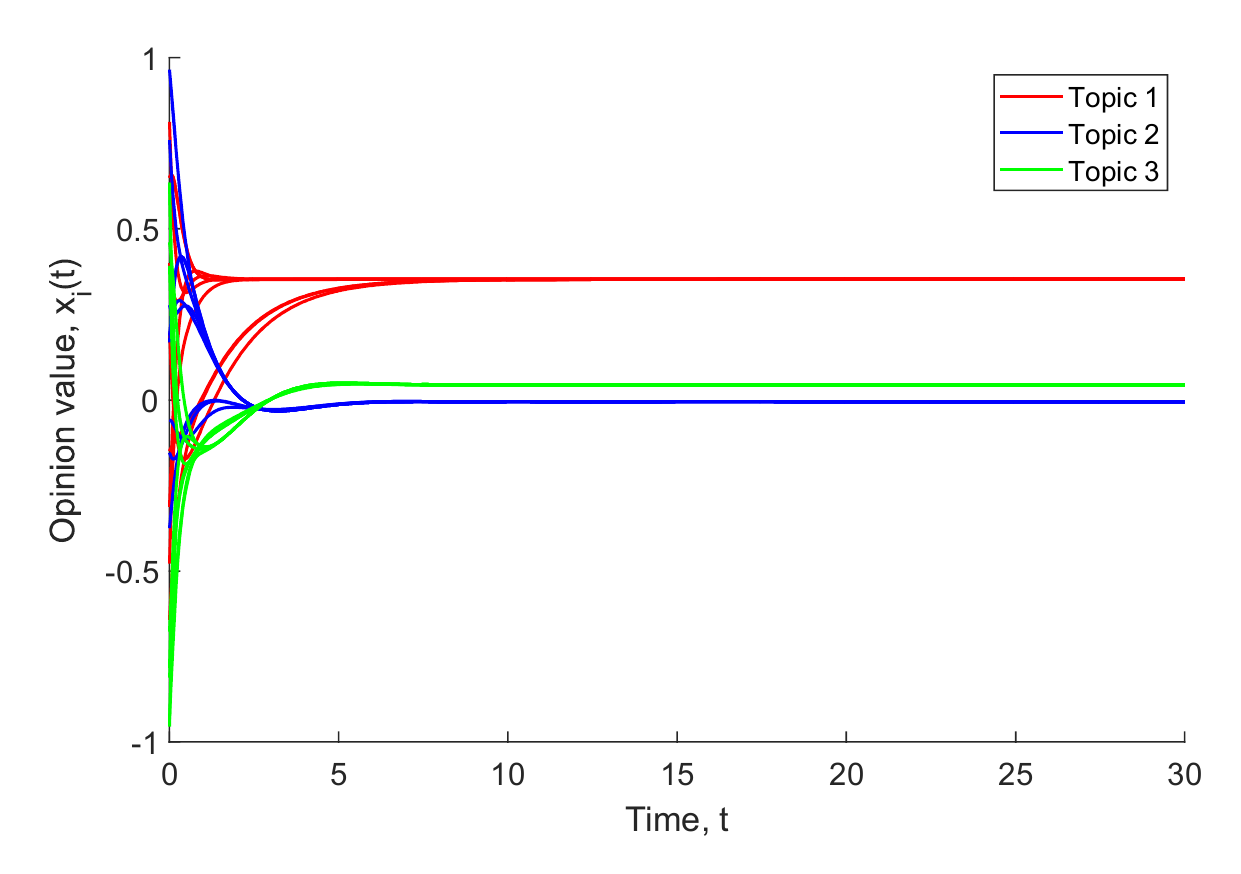}}
\caption{Tăng mức độ liên kết giữa các tác tử làm mô hình Ye 1 mất ổn định trong khi đó sẽ đẩy nhanh quá trình đồng thuận nhưng không làm thay đổi điểm đồng thuận của mô hình Ye 2.}
\label{fig:c8_modelYe_Fast}
\end{figure}

Mô hình đồng thuận đa chiều Ahn \cite{Ahn2020opinion} sử dụng các trọng số ma trận phụ thuộc vào biến trạng thái được cho dưới dạng:
\begin{align} \label{eq:c8_Ahn_model}
\dot{\m{x}}_i = \sum_{j\in N_i} \m{A}^{i,j}(\m{x}_j - \m{x}_i),~i=1,\ldots, n,
\end{align}
trong đó $\m{A}^{i,j} = [a_{p,q}^{i,j}(\m{x}_j - \m{x}_i)] \in \mb{R}^{d\times d}$ là ma trận trọng số phụ thuộc vào trạng thái tương đối giữa hai tác tử $i$ và $j$ trong hệ, tùy trường hợp mà được định nghĩa bởi:
\begin{itemize}
    \item Quan hệ phản hồi ngược (quan điểm càng khác nhau thì trọng số càng lớn):
    \begin{align}
    a_{p,q}^{i,j} = k_{p,q}^{i,j} \cdot {\rm sign}(x_{j,p} - x_{i,p}) \cdot {\rm sign}(x_{j,q} - x_{i,q}),
    \end{align}
    \item Quan hệ phản hồi thuận (quan điểm càng gần nhau thì trọng số càng lớn):
    \begin{align}
     a_{p,q}^{i,j} = \frac{k_{p,q}^{i,j}\cdot {\rm sign}(x_{j,p} - x_{i,p}) \cdot {\rm sign}(x_{j,q} - x_{i,q})}{(c_1 \|x_{j,p} - x_{i,p}\|+c_0)(c_1 \|x_{j,q} - x_{i,q}\|+c_0))},
    \end{align}
    trong đó $k_{p,q}^{i,j}=k_{q,p}^{i,j}>0$, $k_{p,q}^{i,j}=k_{p,q}^{j,i}>0$, và $c_1, c_0>0$ là các hằng số dương.
\end{itemize}
Trong trường hợp tổng quát, mô hình \eqref{eq:c8_Ahn_model} không tiệm cận tới giá trị đồng thuận. Phân tích chỉ ra hiện tượng ý kiến tiệm cận tới các cụm phụ thuộc vào đồ thị $G$ và cấu trúc của các ma trận $\m{A}^{i,j}$. \index{mô hình!Ahn}

\begin{example}[Mô hình Ahn] \label{VD:8.5}
\cite{Ahn2020opinion}. 
\begin{figure}[ht]
\centering
\begin{tikzpicture}[
roundnode/.style={circle, draw=black, thick, minimum size=2mm,inner sep= 0.25mm}
]
\node[roundnode] (node1) at (-2,0) [label=left:$1$] {};
\node[roundnode] (node2) at (-1,1) [label=above:$2$] {};
\node[roundnode] (node3) at (0,-1) [label=below:$3$] {};
\node[roundnode] (node4) at (1,0) [label=below:$4$] {};
\node[roundnode] (node5) at (1,1) [label=above:$5$] {};

\draw (node1) [line width=1pt] -- node [left] {} (node2);
\draw (node1) [line width=1pt] -- node [right] {} (node3);
\draw (node2) [line width=1pt] -- node [left] {} (node3);
\draw (node3) [line width=1pt] -- node [right] {} (node4);
\draw (node4) [line width=1pt] -- node [left] {} (node5);
\end{tikzpicture}
\caption{Đồ thị mô tả các tác tử trong ví dụ \ref{VD:8.5}.}
\label{c8:fig_example1_five_agents}
\end{figure}
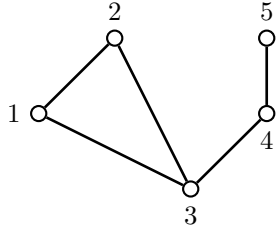
Xét hệ gồm 5 tác tử với đồ thị cho như bởi Hình~\ref{c8:fig_example1_five_agents}. Quan điểm ban đầu của các tác tử được cho bởi $\m{x}_1 =[1, 2, 3]^{\top}$, $\m{x}_2 =[2, 4, 4]^{\top}$, $\m{x}_3 =[3, 1, 5]^{\top}$, $\m{x}_4 =[4, 3, 2]^{\top}$, $\m{x}_5 =[5, 6, 1]^{\top}$. Có thể thấy rằng quan điểm ban đầu của các tác tử về ba chủ đề là khác nhau. Các trọng số ma trận giữa các tác tử được cho bởi 
\begin{align*}
\m{A}^{1,2} &=\left[\begin{array}{ccc}   
1 & 1   & 0  \\
1 & 1   & 0  \\
0 & 0   & 0  \\
\end{array} \right],~ \m{A}^{1,3} =\left[\begin{array}{ccc}   
1 & 0   & 0  \\
0 & 1   & 1  \\
0 & 1   & 1  \\
\end{array} \right]~, \m{A}^{2,3} =\left[\begin{array}{ccc}   
2 & 0   & 1  \\
0 & 2   & 1  \\
1 & 1   & 2  \\
\end{array} \right],\\
\m{A}^{3,4} &=\left[\begin{array}{ccc}   
1 & 1   & 1  \\
1 & 1   & 1  \\
1 & 1   & 1  \\
\end{array} \right],~
\m{A}^{4,5} =\left[\begin{array}{ccc}   
1 & 0   & 1  \\
0 & 1   & 0  \\
1 & 0   & 1  \\
\end{array} \right]
\end{align*}
đều là các ma trận đối xứng bán xác định dương. Với các ma trận trọng số ở trên, ta có thể dự đoán rằng các tác tử sẽ đạt được đồng thuận về tất cả các chủ đề. Hình~\ref{c8:fig_sim_psd_consensus} thể hiện mô phỏng về từng chủ đề.
\end{example}

\begin{figure}[t]
	\centering 
	\includegraphics[width=1\textwidth]{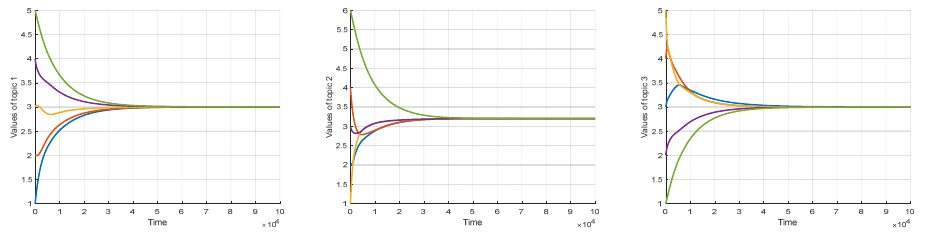}
	\caption{Đồng thuận với ma trận Laplace theo mô hình Ahn: Trái - Chủ đề $1$ ($x_{i,1},~i=1,\ldots, 5$). Giữa - Chủ đề $2$ ($x_{i,2},~i=1,\ldots, 5$). Phải - Chủ đề $3$  ($x_{i,3},~i=1,\ldots, 5$).}
	\label{c8:fig_sim_psd_consensus}
\end{figure}

\section{Mô hình Hegselmann-Krause}
\index{mô hình!Hegselmann-Krause}
Trong thực tế, một tác tử thường chỉ xem xét ý kiến của những tác tử có quan điểm gần với ý kiến của tác tử đó. Nhận xét này là cơ sở để đề xuất các mô hình với miền tự tin giới hạn, trong đó mỗi tác tử hoàn toàn không quan tâm tới các ý kiến nằm ngoài một miền cho trước. Hegselmann-Krause \cite{Hegselmann2002} định nghĩa miền tự tin của tác tử $i$ bởi $[x_i-d,x_i+d]\subset \mb{R}$, trong đó $d>0$ là giới hạn tin tưởng. Mỗi tác tử $i$ chỉ tương tác với một tập tác tử tin cậy $I_i(\m{x})=\{j|~|x_j-x_i |\le d \}$ theo mô hình Hegselmann-Krause (H-K):
\begin{align} 
    x_i[k+1] = C_i(\m{x}) = \frac{1}{|I_i(\m{x}[k])|}\sum_{j\in I_i(\m{x}[k])} x_j [k],~i=1, \ldots, n, \label{eq:c8_HK_Model1}
\end{align}
trong đó $|I_i(\m{x}[k])|$ kí hiệu số phần tử trong tập $I_i(\m{x}[k])$. Mô hình H-K là một mô hình phi tuyến và có thể coi là mô hình French-Degroot với ma trận trọng số thay đổi phụ thuộc vào biến trạng thái:
\begin{align}
    \m{A} = [a_{ij}(\m{x})],~a_{ij}=\frac{1}{|I_i(\m{x})|},~ \text{khi } j \in I_i(\m{x}), \text{ và } a_{ij}(\m{x})=0~\text{ngược lại.}
\end{align}
Với mô hình H-K, sự hội tụ của mô hình hoàn toàn phụ thuộc vào điều kiện đầu và khoảng cách $d$. Giả thuyết $2R$ phát biểu rằng với những điều kiện đầu lấy ngẫu nhiên theo phân bố đều trong khoảng $[0,1]$ và $d=R<1/2$ thì ý kiến của các tác tử hội tụ tới khoảng $1/2R$ cụm, phân cách bởi khoảng cách khoảng $2R$.

\begin{itemize}
    \item Với mọi điều kiện đầu $\m{x}[0]$, trạng thái của hệ \eqref{eq:c8_HK_Model1} ngừng thay đổi sau một số $k$ hữu hạn, $\m{x}[k]=\bar{\m{x}}, \forall k\ge k_*$, trong đó $\bar{\m{x}}$ và thời gian dừng phụ thuộc vào $\m{x}[0]$ và $d$. Tại trạng thái dừng, hai tác tử bất kỳ hoặc đạt đồng thuận $\bar{x}_i = \bar{x}_j$, hoặc là không tin tưởng nhau $|\bar{x}_i - \bar{x}_j|>d$. 
    \item Hệ \eqref{eq:c8_HK_Model1} bảo toàn thứ tự của các phần tử $x_1, \ldots, x_n$ theo thời gian, nghĩa là nếu theo thứ tự $j_1, \ldots, j_n$, ta có $x_{j_1}\le \ldots \le x_{j_n}$ thì $C_{j_1}(\m{x}) \le \ldots \le C_{j_n}(\m{x})$.
    \item Nếu ban đầu hai tác tử $i$ và $j$ nằm về hai thành phần liên thông khác nhau của đồ thị $G(\m{x}(k_0))$ thì chúng sẽ không bao giờ nằm cùng một thành phần liên thông trong đồ thị $G(\m{x}[k]),~k \ge k_0$. Khi $k$ tăng, các thành phần liên thông mạnh của $G(\m{x}[k])$ sẽ chia thành các thành phần nhỏ hơn nhưng sẽ không thể nhập lại với nhau.
\end{itemize}

\begin{example}[Mô hình Hegselmann-Krause] \label{eg:8.6} Xét hệ gồm 10 tác tử với các giá trị ý kiến ban đầu $\m{x}[0]$ được lấy ngẫu nhiên trong khoảng từ -2.5 tới 2.5. Với cùng một vector giá trị đầu và các giá trị khác nhau của giới hạn tin tưởng $d = 0.2,~ 0.4,~\ldots,~1.2$, ta thu được kết quả mô phỏng như trên Hình~\ref{fig:c8_H_K}. Mô phỏng cho thấy khi mở rộng miền tin tưởng, số lượng cụm\footnote{cluster} ở cuối mô phỏng giảm dần. Với $d=0.2$, kết quả cuối gồm 8 cụm. Số cụm giảm dần khi tăng dần $d=0.2$ tới $d=1$). Khi $d=1.2$, ta thấy rằng 10 tác tử đạt được đồng thuận sau 4 bước lặp.
\end{example}

\begin{figure}[h!]
\centering
\subfloat[$d=0.2$]{\includegraphics[height=5cm]{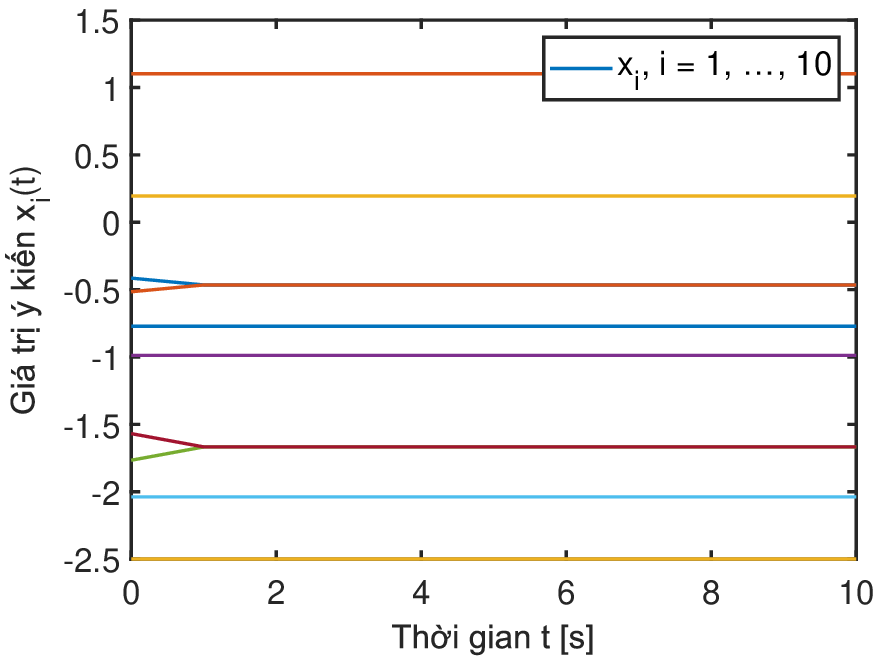}} \hfill
\subfloat[$d=0.4$]{\includegraphics[height=5cm]{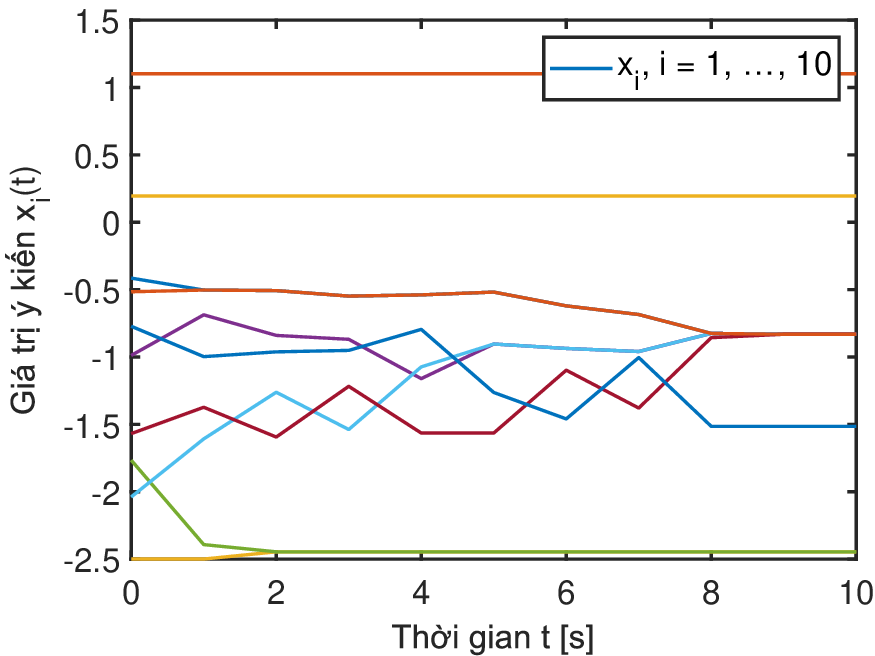}}\\
\subfloat[$d=0.6$]{\includegraphics[height=5cm]{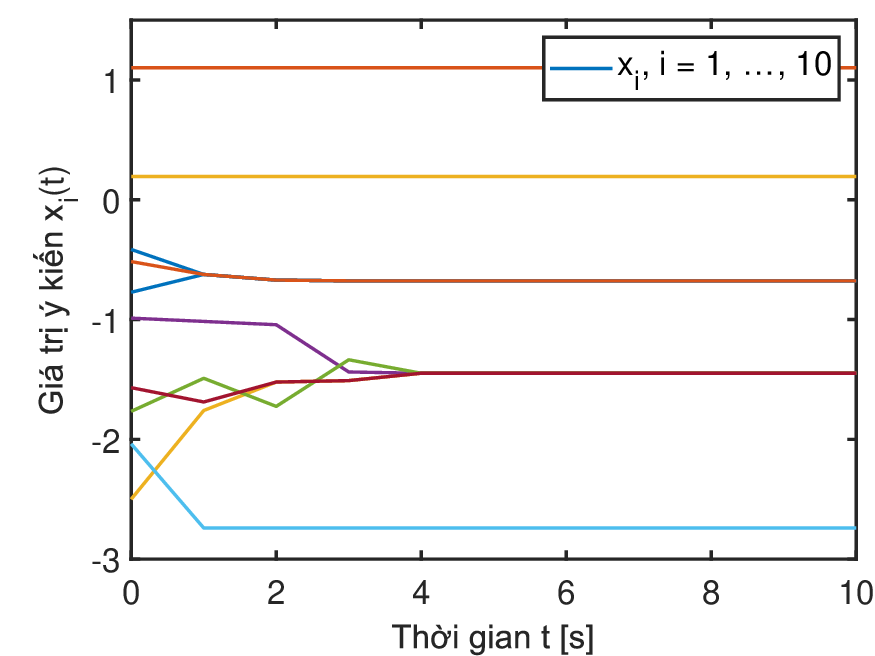}} \hfill
\subfloat[$d=0.8$]{\includegraphics[height=5cm]{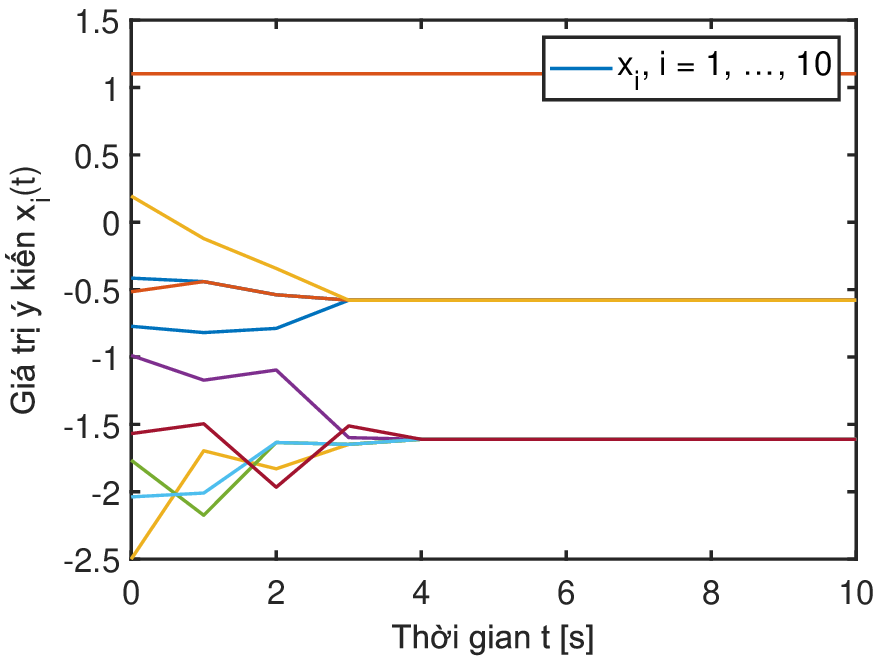}}\\
\subfloat[$d=1.0$]{\includegraphics[height=5cm]{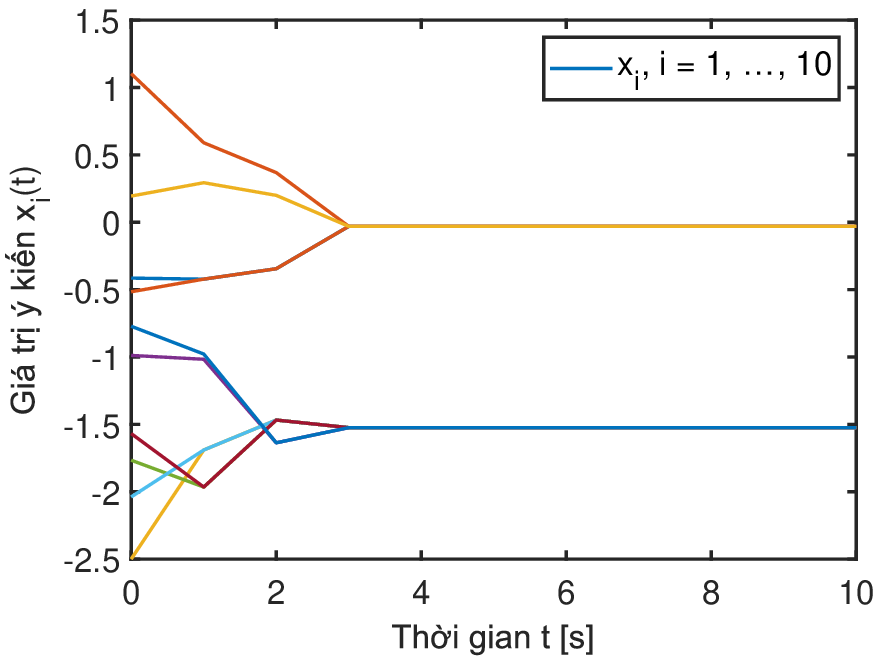}} \hfill
\subfloat[$d=1.2$]{\includegraphics[height=5cm]{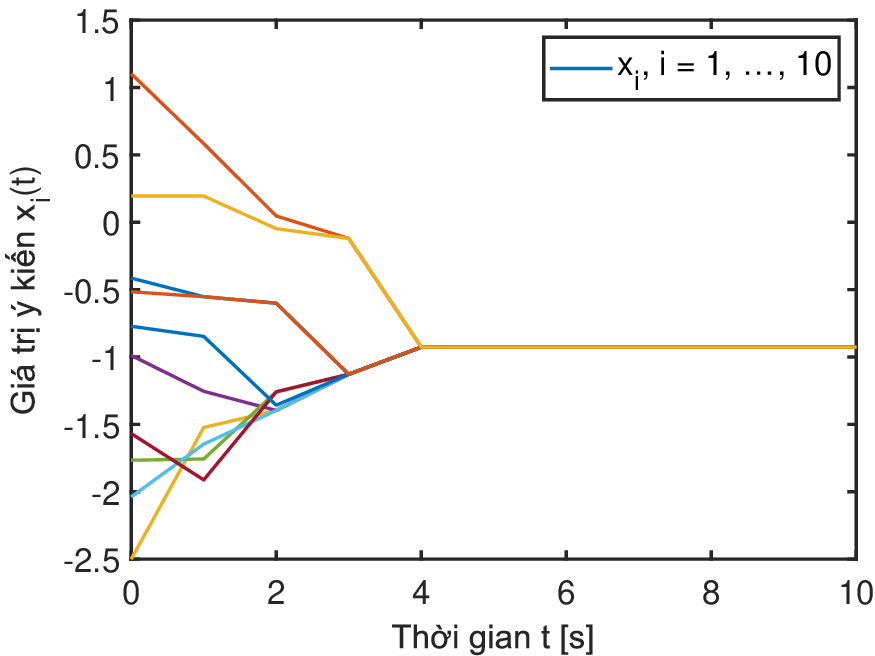}}
\caption{Mô phỏng mô hình Hegselmann - Krausse với $d = 0.2,~ 0.4,~ \ldots,~ 1.2$.}
\label{fig:c8_H_K}
\end{figure}

\section{Mô hình Altafini}
\index{mô hình!Altafini}
\begin{figure}[t]
\centering
\subfloat[]{
\begin{tikzpicture}[
roundnode/.style={circle, draw=black, thick, minimum size=2mm,inner sep= 0.25mm}
]
\node[roundnode] (node1) at (-1,0) {1};
\node[roundnode] (node2) at (-0,0) {2};
\node[roundnode] (node3) at (1,0) {3};
\node[roundnode] (node4) at (-1,-1) {4};
\node[roundnode] (node5) at (0,-1) {5};
\node[roundnode] (node6) at (1,-1) {6};

\draw[very thick] (node1) -- (node2) -- (node3) -- (node6);
\draw[very thick] (node4) -- (node5);
\draw[very thick, color = blue] (node1) -- (node4);
\draw[very thick, color = blue] (node2) -- (node4);
\draw[very thick, color = blue] (node2) -- (node5);
\draw[very thick, color = blue] (node3) -- (node5);
\end{tikzpicture}
} \qquad\qquad
\subfloat[]{
\begin{tikzpicture}[
roundnode/.style={circle, draw=black, thick, minimum size=2mm,inner sep= 0.25mm}
]
\node[roundnode] (node1) at (-1,0) {1};
\node[roundnode] (node2) at (0,0) {2};
\node[roundnode] (node3) at (1,0) {3};
\node[roundnode] (node4) at (-1,-1) {4};
\node[roundnode] (node5) at (0,-1) {5};
\node[roundnode] (node6) at (1,-1) {6};

\draw[very thick] (node1) -- (node2) -- (node3) -- (node6);
\draw[very thick] (node4) -- (node5);
\draw[very thick, color = blue] (node1) -- (node4);
\draw[very thick, color = blue] (node2) -- (node4);
\draw[very thick] (node2) -- (node5);
\draw[very thick, color = blue] (node3) -- (node5);
\end{tikzpicture}
}
\caption{(a) Đồ thị dấu cân bằng cấu trúc; (b) Đồ thị dấu không cân bằng cấu trúc.}
\label{fig:C8_signedGraph}
\end{figure}
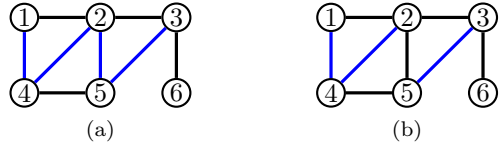
Khác với các mô hình đã đề cập ở trên, mô hình Altafini nghiên cứu mạng xã hội với hai khả năng tương tác: hợp tác và cạnh tranh, mô tả bởi một \emph{đồ thị dấu} (signed graph).\index{đồ thị!dấu} \index{Đồ thị!cân bằng cấu trúc dấu}

Một đồ thị có dấu được cho bởi $G=(V,E,W)$, trong đó $(V,E)$ mô tả một đồ thị trọng số, với  $\omega_{ji}\in W$ tương ứng với cạnh $(j,i)\in E$ có thể nhận giá trị dương (hợp tác), không (trung lập), và cạnh tranh (âm). Đồ thị được giả thiết là đối xứng dấu, $\omega_{ij}\omega_{ji}\ge 0$, tức là tương tác giữa hai tác tử (hợp tác, trung lập, và cạnh trang) là như nhau.

Định nghĩa ma trận kề $\m{A}=[a_{ij}]$ với các phần tử $a_{ij}=\omega_{ji}$, ma trận bậc $\m{D}=\text{diag}(d_{i})$, với $d_i=\sum_{j=1}^n|a_{ij}|$, và ma trận Laplace $\mcl{L}=\m{D}-\m{A}$. Một đồ thị dấu ${G}$ là cân bằng cấu trúc nếu có thể chia tập đỉnh của nó thành hai tập rời nhau $V=V_1 \cup V_2$ sao cho với mỗi cặp $i,j \in V,i\ne j$, ta có $a_{ij}\ge 0$ nếu $(j,i)\in (V_1\times V_1)\cup(V_2\times V_2)$; và $a_{ij} \le 0$, nếu $(j,i)\in (V_1\times V_2)\cup (V_2\times V_1)$.

Định nghĩa đồ thị dấu vô hướng $\hat{G}=(V,\hat{E},\hat{W})$ với qui ước nếu tồn tại $(i,j)\in E$ thì cũng tồn tại cạnh $(i,j)$, $(j,i) \in \hat{E}$ với cùng trọng số $\omega_{ij}$. Đồ thị dấu $G$ là \emph{cân bằng cấu trúc dấu} khi và chỉ khi đồ thị $G$ là đối xứng dấu và mọi chu trình trong $\hat{G}$ là dương (tức là tích trọng số của các cạnh trong mọi chu trình đều là số dương) \cite{Harary1953}. 

Giả sử $G$ là một đồ thị dấu liên thông mạnh thì $\lambda=0$ là một giá trị riêng của $\mcl{L}$ khi và chỉ khi $G$ là cân bằng cấu trúc dấu. Khi $G$ là liên thông mạnh nhưng không cân bằng cấu trúc dấu, ma trận $-\mcl{L}$ là một ma trận Hurwitz.

Giả sử mỗi đỉnh của đồ thị dấu mô tả một tác tử với ý kiến $x_i(t) \in \mb{R}$. Mô hình Altafini \cite{Altafini2013} đề xuất các tác tử cập nhật quan điểm theo phương trình:
\begin{align} \label{eq:c8_Altafini_Model1}
\dot{x}_i &= \sum_{j=1}^n|a_{ij}|(x_j {\rm sign}(a_{ij})-x_i) \nonumber\\
&= \sum_{j=1}^n(a_{ij}x_j-|a_{ij}| x_i), \; i=1,\ldots, n.
\end{align}
Mô hình \eqref{eq:c8_Altafini_Model1} luôn luôn hội tụ với mọi giá trị đầu $\m{x}(0)$. Giả sử đồ thị dấu $G$ cân bằng cấu trúc, với tập đỉnh được chia thành $V = V_1 \cup V_2$, ta có thể định nghĩa phép biến đổi Gauge:
\begin{align}
    x_i \mapsto y_i = \delta_i x_i,~ \delta_i = +1 \text{ nếu } i \in V_1, \text{ và } \delta_i=-1, \text{ nếu } i\in V_2,
\end{align}
và biến đổi mô hình \eqref{eq:c8_Altafini_Model1} về dạng mô hình Abelson với các trọng số không âm:
\begin{align}
    \dot{y}_i(t) = \sum_{j=1}^n|a_{ij}|(y_j(t)-y_i(t)), \forall i = 1, \ldots, n.
\end{align}
Như vậy, khi $G$ là cân bằng cấu trúc dấu:
\begin{itemize}
    \item Nếu $G$ là có gốc ra, ta có $y_i(t) \to \bm{\gamma}^\top \m{y}(0), \forall i = 1, \ldots, n$, với $\bm{\gamma}^\top$ là vector riêng bên trái được chuẩn hóa duy nhất ứng với giá trị riêng 0 của ma trận Laplace $\mcl{L}({G})$, trong đó ${G}=(V,E,{W})$ là đồ thị với các trọng số $w_{ij} = |\omega_{ij}|$. Từ đây suy ra mỗi giá trị $x_i$ hội tụ về một trong hai giá trị bằng nhau về độ lớn nhưng trái dấu.
    \item 	Nếu $G$ là đồ thị không có gốc ra, các tác tử hội tụ về các cụm khác nhau phụ thuộc vào tính liên thông trong $G$.
\end{itemize}
Ngược lại, khi $G$ không cân bằng cấu trúc dấu nhưng là liên thông mạnh, ma trận $-\mcl{L}$ là Hurwitz nên $x_i(t) \to 0, \forall i = 1, \ldots, n$.

\begin{example}[Mô hình Altafini] \label{eg:8.7}
Hình~\ref{fig:C8_signedGraph} mô tả hai đồ thị dấu, trong đó các cạnh màu đen thể hiện quan hệ hợp tác, cạnh màu xanh thể hiện quan hệ cạnh tranh. Đồ thị \ref{fig:C8_signedGraph}(a) là đồ thị dấu thỏa mãn tính cân bằng cấu trúc, với ma trận Laplace tương ứng cho bởi
\begin{align*}
    \mcl{L} = \begin{bmatrix}
     3 & -2  & 0 & 1   & 0   & 0\\
    -2 & 3.7 &-1 & 0.5 & 0.2 & 0 \\
     0 & -1  &1.7& 0   & 0.3 & -0.4\\
     1 & 0.5 & 0 & 2.5 & -1  & 0 \\
     0 & 0.2 & 0.3& -1 & 1.5 & 0\\
     0 & 0 & -0.4 & 0  & 0 & 0.4
    \end{bmatrix}.
\end{align*}

\begin{figure}[th!]
    \centering
    \subfloat[]{\includegraphics[height=5cm]{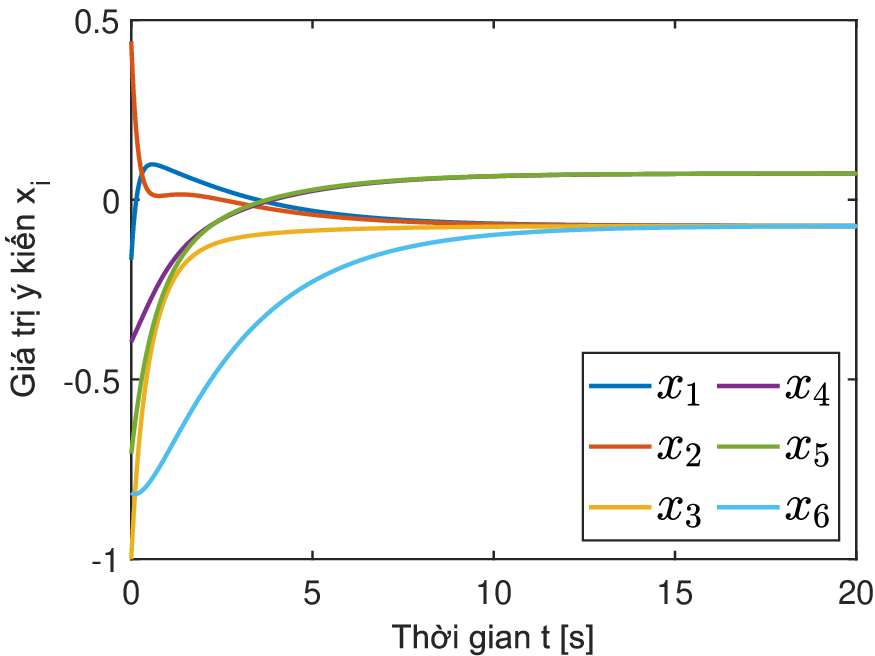}} \hfill
    \subfloat[]{\includegraphics[height=5cm]{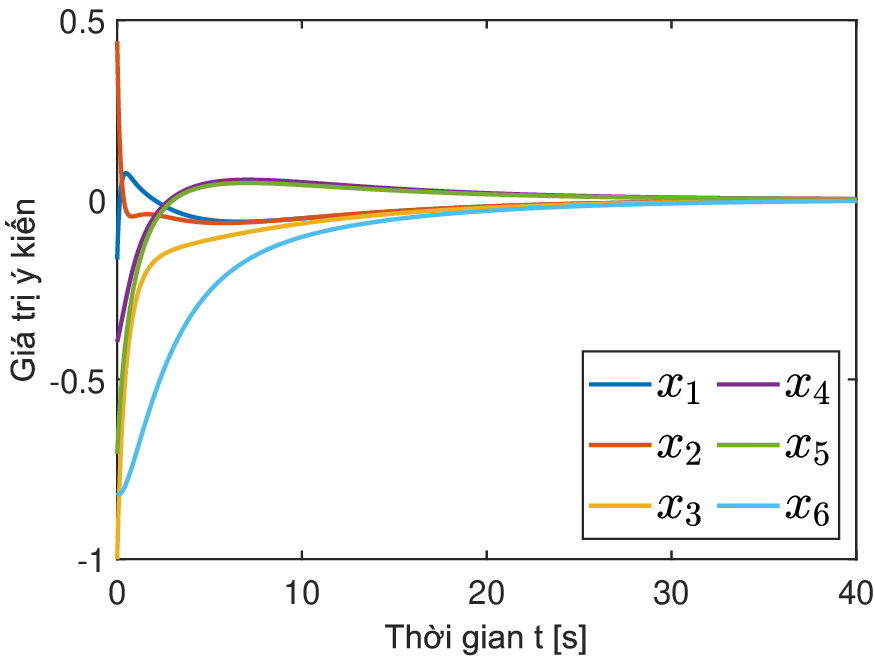}}
    \caption{Mô phỏng mô hình Altafini với các đồ thị trong Ví dụ \ref{eg:8.7}.}
    \label{fig:VD8.7}
\end{figure}

Mô phỏng thuật toán đồng thuận \eqref{eq:c8_Altafini_Model1} với đồ thị \ref{fig:C8_signedGraph}(a) cho ta kết quả như ở Hình~\ref{fig:VD8.7}(a). Dễ thấy các tác tử đạt được đồng thuận 2 phía với giá trị cuối đối lập nhau.

Tiếp theo, với đồ thị \ref{fig:C8_signedGraph}(b), ta có
\begin{align*}
    \mcl{L} = \begin{bmatrix}
     3 & -2  & 0 & 1   & 0   & 0\\
    -2 & 3.7 &-1 & 0.5 & -0.2 & 0 \\
     0 & -1  &1.7& 0   & 0.3 & -0.4\\
     1 & 0.5 & 0 & 2.5 & -1  & 0 \\
     0 & -0.2 & 0.3& -1 & 1.5 & 0\\
     0 & 0 & -0.4 & 0  & 0 & 0.4
    \end{bmatrix}.
\end{align*}
Đồ thị dấu lúc này không thỏa mãn tính cân bằng cấu trúc. Mô phỏng của hệ với đồ thị lúc này có dạng như trên Hình~\ref{fig:VD8.7}(b). Có thể thấy rằng $x_i(t) \to 0, \forall i=1, \ldots, 6$ giống như kết quả lý thuyết.
\end{example}


\section{Ghi chú và tài liệu tham khảo}
Trong chương này, chúng ta đã điểm qua một số mô hình động học ý kiến trong nghiên cứu mạng xã hội. Chú ý rằng phạm vi nghiên cứu của mạng xã hội rất rộng, bao gồm nhiều phương pháp nghiên cứu khác nhau (triết học, mô hình toán, phân tích dữ liệu). Nghiên cứu về mạng xã hội liên quan tới điều khiển đa tác tử liên quan tới các mô hình động học ý kiến \cite{Proskurnikov2017Tut,Proskurnikov2018Tut,Hegselmann2002,Altafini2013}, lan truyền dịch bệnh \cite{Mei2017dynamics,Pare2020modeling}, kinh tế \cite{Carvalho2019production,Trinh2024networked}, thuật toán PageRank \cite{Ishii2014pagerank}. 

\section{Bài tập}

\begin{exercise}[Mô hình DeGroot] \label{ex:8.1}
Xét mô hình DeGroot:
\begin{align} \label{ex:Degroot_model}
    \m{x}[k+1] = \m{A} \m{x}[k],~k=0,1,2,\ldots
\end{align}
trong đó $\m{x}[k] = [x_1[k],\ldots,x_n[k]]^\top$ và $\m{A}$ là một ma trận có các phần tử không âm (kí hiệu $\m{A}\geq 0$), ngẫu nhiên hàng ($\m{A}\m{1}_n=\m{1}_n$).
\begin{itemize}
\item[i.] Chứng minh rằng các giá trị riêng của ma trận $\m{A}$ đều nằm trên đĩa tròn bán kính 1 với tâm tại gốc tọa độ.
\item[ii.] Chứng minh rằng nếu ma trận $\m{A}$ là \emph{ma trận dương}\index{ma trận!dương}, nghĩa là $a_{ij}>0$, $\forall i,j=1,\ldots,n$,  thì vector riêng bên trái $\bm{\gamma}=[\gamma_1,\ldots,\gamma_n]^\top\in \mb{R}^n$ của $\m{A}$ tương ứng với giá trị dương 1 có tất cả các phần tử dương.
\item[iii.] Chứng minh rằng nếu ma trận $\m{A}$ là \emph{ma trận nguyên thủy}\index{ma trận!nguyên thủy}, nghĩa là tồn tại số nguyên  $k\ge 1$ sao cho ma trận $\m{A}^k$ là ma trận dương, thì vector riêng bên trái $\bm{\gamma}=[\gamma_1,\ldots,\gamma_n]^\top\in \mb{R}^n$ của $\m{A}$ tương ứng với giá trị dương 1 có tất cả các phần tử dương và thỏa mãn $\sum_{i=1}^n \gamma_i = 1$.
\item[iv.] Chứng minh rằng nếu có thêm giả thiết $\m{A}$ là ma trận tương ứng với đồ thị liên thông mạnh, không có chu kỳ $G$ thì tồn tại $\lim_{k \to +\infty}\m{A}^k = \m{1}_n\bm{\gamma}^\top$, với $\bm{\gamma} \in \mb{R}^n$ là vector riêng bên trái của $\m{A}$ tương ứng với giá trị dương 1 có tất cả các phần tử dương thỏa mãn $\bm{\gamma}^\top\m{1}_n = 1$.
\item[v.] Với các giả thiết ở ý (iv), hãy chứng minh rằng $\lim_{k \to +\infty}\m{x}[k] = x^*\m{1}_n$, với $x^*=\bm{\gamma}^\top\m{x}[0]$.
\end{itemize}
\end{exercise}

\begin{exercise} \label{ex:8.2} \cite{Bullo2019lectures}
Xét hệ mô tả bởi phương trình sai phân
\begin{align}
\m{x}[k+1] = \m{A}\m{x}[k]+\m{b},
\end{align}
trong đó ma trận $\m{A}\in \mb{R}^{n\times n}$ là \emph{hội tụ} (nghĩa là $\lim_{k \to +\infty}\m{A}^k = \m{0}_{n\times n}$). Khi đó, hãy chứng minh rằng
\begin{itemize}
\item[i.] Ma trận $\m{I}_n -\m{A}$ là khả nghịch;
\item[ii.] Hệ có điểm cân bằng duy nhất $\m{x}^*=(\m{I}_n -\m{A})^{-1}\m{b}$ và $\lim_{k \to +\infty}\m{x}[k] = \m{x}^*$.
\end{itemize}
\end{exercise}
\begin{exercise}[Trường hợp riêng của mô hình Friedkin-Johnsen] \label{ex:8.3}
Xét mô hình Friedkin-Johnsen (F-J) ở mục \eqref{c8:FJ_model}:
\begin{align}
\m{x}[k+1]= \bm{\Theta}\m{A}\m{x}[k] + (\m{I}_n - \bm{\Theta}) \m{x}[0],
\end{align}
trong đó $\bm{\Theta} = {\rm diag}(\theta_{11}, \ldots, \theta_{nn})$, với $\theta_{ii} \in [0,1]$ mô tả mức ảnh hưởng từ bên ngoài tới quan điểm của tác tử thứ $i$ và $\m{x}[0]\in \mb{R}^n$ là vector định kiến của các tác tử.
\begin{itemize}
\item[i.]  Hãy phân tích quá trình tiệm cận của các ý kiến trong hệ khi các tác tử không có định kiến và đồ thị $G$ là liên thông mạnh, không có chu kỳ.
\item[ii.] Hãy phân tích quá trình tiệm cận của các ý kiến trong hệ khi các tác tử đều có định kiến $0<\lambda_{ii}<1$ và đồ thị $G$ là liên thông mạnh, không có chu kỳ.
\item[iii.] Thực hiện mô phỏng cho hai trường hợp (i) và (ii).
\end{itemize}
\end{exercise}


\begin{exercise} \label{ex:8.4}
Xét ma trận \cite{Scardovi2010synchronization}
\begin{align}
\m{Q} = \begin{bmatrix}
-1 + (n-1)\nu & 1-\nu ^ -\nu & \ldots & -\nu \\
-1 + (n-1)\nu & 1-\nu ^ -\nu & \ddots & -\nu \\
 \vdots & \vdots & \ddots & \ddots & -\nu \\
-1 + (n-1)\nu & 1-\nu ^ -\nu & \ldots & -\nu \\
\end{bmatrix}
\end{align}
với $\nu = \frac{n-\sqrt{n}}{n(n-1)}$. Chứng minh rằng
$\m{Q}\m{1}_n = \m{0}_n$, $\m{Q}\m{Q}^\top = \m{I}_{n-1}$ và
\begin{align}
\m{Q}^\top\m{Q} = \m{I}_n - \frac{1}{n}\m{1}_n\m{1}_n^\top.
\end{align}
Giả thiết đồ thị $G$ là liên thông mạnh. Sử dụng phép biến đổi $\m{y} = \m{Q}\m{x}$, hãy phân tích quá trình tiệm cận của mô hình Abelson và mô hình Taylor ở \ref{c8:Abelson_Taylor} theo biến $\m{y}$, từ đó suy ra quá trình tiệm cận theo biến $\m{x}$.
\end{exercise}


\begin{exercise} \cite{Bullo2019lectures} \label{ex:8.5}
Xét $G$ là đồ thị vô hướng, liên thông có thể có khuyên với ma trận kề $\m{A}$. Xét ma trận $\bar{\m{A}} = \m{D}^{-1}\m{A}$, với $\m{D}$ là ma trận đường chéo với $d_i = \sum_{j=1}^n a_{ij}$, $i=1,\ldots,n$. 
\begin{itemize}
\item[i.] Chứng minh rằng ma trận $\bar{\m{A}}$ là ngẫu nhiên hàng và không thể phân rã.
\item[ii.] Tìm vector riêng trội bên trái của $\bar{\m{A}}$ tương ứng với giá trị riêng 1.
\item[iii.] Phân tích thuật toán đồng thuận $\m{x}[k+1]=\bar{\m{A}}\m{x}[k]$, với giả thiết thêm rằng tồn tại ít nhất một đỉnh $v_i\in V$ của đồ thị có khuyên ($a_{ii}>0$). 
\item[iv.] Tìm điều kiện của $G$ để hệ đồng thuận tại giá trị trung bình cộng với mọi điều kiện đầu $\m{x}[0] \in \mb{R}^n$.
\end{itemize}
\end{exercise}

\begin{exercise}[Mô hình mạng internet] \label{ex:8.6} \cite{Ishii2014pagerank} Giả sử mạng internet được hợp thành bởi các trang web và đường dẫn (hyperlink) liên kết với nhau, và có thể mô tả bởi đồ thị hữu hướng $G=(V,E)$. Tại mỗi đỉnh $v_i \in V$ (tương ứng với một trang web), giả sử có $d_i$ đường dẫn tới các trang web khác nhau (có thể là chính đỉnh $v_i$ hoặc một trong các đỉnh láng giềng), và người dùng có thể chọn chuyển từ trang web hiện tại tới một trong các trang web khác với khả năng như nhau ($\frac{1}{d_i}$).\index{thuật toán!PageRank}
\begin{itemize}
\item[i.] Gọi $\m{A}$ là ma trận kề của đồ thị $G$ và $\m{D}={\rm diag}(d_1,\ldots,d_n)$. Chứng minh rằng $\bar{\m{A}}=\m{D}^{-1}\m{A}$ là ma trận không âm và ngẫu nhiên cột, tức là $\bar{\m{A}}^\top\m{1}_n=\m{1}_n$ .

\item[ii.] Giả sử $G$ là đồ thị liên thông mạnh. Hãy chứng minh rằng $\bar{\m{A}}$ có vector riêng bên phải $\bm{\gamma} \in \mb{R}^n$ với các phần tử dương thỏa mãn $\bar{\m{A}}\bm{\gamma}=\bm{\gamma}$.

\item[iii.] Giả sử $G$ là đồ thị liên thông mạnh và không có chu kỳ, chứng minh rằng $\lim_{k \to +\infty} \bar{\m{A}}^k = \bm{\gamma}\m{1}_n^\top$, với $\bm{\gamma}$ là vector riêng bên phải tương ứng với giá trị riêng 1 (duy nhất) của $\bar{\m{A}}$, và đã được chuẩn hóa ($\bm{\gamma}^\top\m{1}_n = 1$). Các giá trị $\gamma_i>0$ thể hiện mức độ quan trọng của một trang web trong mạng internet. Do đó có thể sắp xếp thứ tự các trang web theo thứ tự giảm dần của ${\gamma}_i$, và $\bm{\gamma}$ gọi là vector riêng Pagerank.
\end{itemize}
\end{exercise}


\begin{exercise}[Thuật toán PageRank] \label{ex:8.7}\cite{Ishii2014pagerank} Thuật toán PageRank của Google là thuật toán lặp có dạng
\begin{align} \label{eq:c8_PageRank_alg}
\m{x}[k+1] = \bar{\m{A}}\m{x}[k],
\end{align}
trong đó $\m{x}[0] \in \mb{R}^n$ được chọn thỏa mãn $\m{1}_n^\top\m{x}[0]=1$. 
\begin{itemize}
\item[i.] Chứng minh rằng với các giả thiết ở ý (iii) Bài tập \ref{ex:8.6} thì vector riêng Pagerank có thể xác định theo thuật toán lặp \eqref{eq:c8_PageRank_alg} theo nghĩa $\lim_{k \to +\infty} \m{x}[k] = \bm{\gamma}$.
\item[ii.] Mạng internet thực tế thường có các nút treo (không có đường đi tới nút khác, ví dụ như một tệp tin được chia sẻ), hoặc bao gồm những thành phần liên thông mạnh khác nhau. Do đó, thuật toán gốc \eqref{eq:c8_PageRank_alg} được biến đổi dưới dạng
\begin{align} \label{eq:c8_PageRank_alg_modified}
\m{x}[k+1] = \left(\alpha \bar{\m{A}} + \frac{1-\alpha}{n}\m{1}_n\m{1}_n^\top \right)\m{x}[k] = \m{M}\m{x}[k],
\end{align}
với $\m{1}_n\m{1}_n^\top$ tạo các kết nối giả giữa các website để đảm bảo mạng liên thông mạnh, và $\alpha>0$ là một hằng số dương được chọn đủ nhỏ để không ảnh hưởng nhiều tới giá trị vector riêng PageRank. Chứng minh $\m{M}$ có một giá trị riêng 1 và các giá trị riêng khác của $\m{M}$ đều nhỏ hơn 1. 
\item[iii.] Phân tích quá trình tiệm cận của thuật toán \eqref{eq:c8_PageRank_alg_modified} và tìm $\lim_{k \to +\infty} \m{x}[k]$ khi $\m{x}[0]$ được chọn thỏa mãn $\m{1}_n^\top\m{x}[0]=1$.
\end{itemize}
\end{exercise}

\begin{exercise} \label{ex:8.8} Xét mô hình Friendkin-Johnsen đa chiều:
\begin{align} \label{ex:c8_MDFJ_Model}
\m{x}[k+1] = [(\bm{\Theta}\m{A})\otimes \m{C}]\m{x}[k] + [(\m{I}_n - \bm{\Theta}) \otimes \m{I}_d] \m{x}[0].
\end{align}
Chứng minh rằng với $\rho(\bm{\Theta}\m{A})\rho(\m{C}) < 1$ thì $\m{x}^\infty = \lim_{k \to +\infty}\m{x}[k] = (\m{I}_{dn}-\bm{\Theta}\m{A} \otimes \m{C})^{-1}[(\m{I}_n - \bm{\Theta})\otimes \m{I}_d]\m{x}[0]$.
\end{exercise}

\begin{exercise}[Mô hình Ahn] \label{ex:8.9} \cite{Ahn2020opinion}
Hãy phân tích quá trình tiệm cận của mô hình Ahn:
\begin{align}
\begin{bmatrix}
\dot{x}_{i,1}\\
\vdots \\
\dot{x}_{i,d}
\end{bmatrix} = \sum_{j\in N_i} \begin{bmatrix}
a_{1,1}^{i,j} & \ldots & a_{1,d}^{i,j} \\
a_{2,1}^{i,j} & \ldots & a_{2,d}^{i,j} \\
\vdots & \ldots & \vdots \\
a_{d,1}^{i,j} & \ldots & a_{d,d}^{i,j} \\
\end{bmatrix} \begin{bmatrix}
x_{j,1}-x_{i,1} \\
\vdots \\
x_{j,d}-x_{i,d} 
\end{bmatrix} = \sum_{j \in N_i} \m{A}_{ij}(\m{x}_j - \m{x}_i),
\end{align}
với $i=1,\ldots,n$ và các trọng số thỏa mãn
\begin{itemize}
\item các trọng số trên đường chéo: $a_{p,p}^{i,j}=k_{p,p}^{i,j}$, với $k_{p,p}^{i,j}=k_{p,p}^{j,i}>0$ là một hằng số, và
\item các trọng số chéo thỏa mãn: $a_{p,q}^{i,j}=k_{p,q}^{i,j} \cdot {\rm sgn}(x_{j,p}-x_{i,p}) \cdot {\rm sgn}(x_{j,q}-x_{i,q})$.
\end{itemize}
\end{exercise}

\begin{exercise}[Mô hình Ye] \label{ex:8.10} \cite{Ye2020Aut}
Xét mô hình động học ý kiến đa chiều thứ nhất của Ye:
\begin{align} \label{eq:Ye1}
\dot{\m{x}}_i = \sum_{j \in N_i} a_{ij}\m{C}(\m{x}_j - \m{x}_i) + (\m{C}-\m{I}_d)\m{x}_i + b_i(\m{x}_i(0)-\m{x}_i),\; i=1,\ldots,n,
\end{align}
trong đó các tác tử tương tác qua đồ thị $G$ có gốc ra.
\begin{itemize}
\item[i.] Với $\m{B}={\rm diag}(b_1,\ldots,b_n)$, hãy biểu diễn mô hình \eqref{eq:Ye1} dưới dạng ma trận.
\item[ii.] Giả sử ma trận $\m{C} \in \mb{R}^{d \times d}$ có giá trị riêng bán đơn giản 1 bội $p\geq 1$, tương ứng với các vector riêng bên phải và bên trái $\bm{\eta}_r$ và $\bm{\xi} \in \mb{R}^{d \times d}$ thỏa mãn điều kiện chuẩn hóa $\bm{\xi}^\top \bm{\eta} = 1$ với mọi $r=1,\ldots,p$. Các giá trị riêng khác của $\m{C}$ thỏa mãn ${\rm Re}(\mu_k(\m{C}))<1$, $\forall k>p$, và $c_{ii} \geq 0, \forall i=1,\ldots,p$. Nếu $b_i=0 \forall i=1,\ldots,n$, chứng minh rằng hệ tiệm cận tới đồng thuận khi và chỉ khi
\[{\rm Re}((1-\lambda_i)\mu_k)<1,~ \forall i=2,\ldots,n, ~\forall k = 1,\ldots,p, \]
trong đó $\lambda_i$ kí hiệu các giá trị riêng của ma trận Laplace $\mcl{L}$.

\item[iii.] Tìm giá trị tiệm cận của $\m{x}_i(t)$ khi $t\to\infty$ theo $\m{x}_i(0)$.
\end{itemize}
\end{exercise}

\begin{exercise}[Mô hình Ye] \label{ex:8.11}
Xét mô hình động học ý kiến đa chiều thứ hai của Ye \cite{Ye2020Aut}:
\begin{align} \label{eq:Ye2}
\dot{\m{x}}_i = \sum_{j \in N_i} a_{ij}(\m{x}_j - \m{x}_i) + (\m{C}-\m{I}_d)\m{x}_i + b_i(\m{x}_i(0)-\m{x}_i),\; i=1,\ldots,n,
\end{align}
trong đó các tác tử tương tác qua đồ thị $G$ có gốc ra.
\begin{itemize}
\item[i.] Với $\m{B}={\rm diag}(b_1,\ldots,b_n)$, hãy biểu diễn mô hình \eqref{eq:Ye2} dưới dạng ma trận.
\item[ii.] Giả sử ma trận $\m{C} \in \mb{R}^{d \times d}$ có giá trị riêng bán đơn giản 1 bội $p\geq 1$, tương ứng với các vector riêng bên phải và bên trái $\bm{\eta}_r$ và $\bm{\xi} \in \mb{R}^{d \times d}$ thỏa mãn điều kiện chuẩn hóa $\bm{\xi}^\top \bm{\eta} = 1$ với mọi $r=1,\ldots,p$. Các giá trị riêng khác của $\m{C}$ thỏa mãn ${\rm Re}(\mu_k(\m{C}))<1$, $\forall k>p$, và $c_{ii} \geq 0, \forall i=1,\ldots,p$. Nếu $b_i=0,~ \forall i=1,\ldots,n$, chứng minh rằng hệ tiệm cận tới đồng thuận.
\item[iii.] Tìm giá trị tiệm cận của $\m{x}_i(t)$ khi $t\to\infty$ theo $\m{x}_i(0)$.
\end{itemize}
\end{exercise}

\begin{exercise} \label{ex:8.12} Giả sử đồ thị dấu $(V,E)$ là đồ thị cân bằng cấu trúc dấu. Định nghĩa các ma trận $\m{A}=[a_{ij}]$, $\m{D}={\rm diag}(d_1,\ldots,d_n)$ với $d_i = \sum_{j=1}^n |a_{ij}|$, $\mcl{L}=\m{D}-\m{A}$. Chứng minh rằng $\Phi(\m{x}) = \m{x}^\top\mcl{L}\m{x} = \sum_{(i,j)\in E}|a_{ij}|(x_i-{\rm sgn}(a_{ij})x_j)^2$.
\end{exercise}

\begin{exercise}[Đồ thị cân bằng cấu trúc dấu] \label{ex:8.13} \cite{Harary1971graph} Chứng minh rằng đồ thị dấu vô hướng ,liên thông $G$ là cân bằng cấu trúc dấu nếu một trong các điều kiện tương đương sau thỏa mãn:
\begin{itemize}
\item[i.] Tất cả các chu trình $C=v_{i_1}\ldots v_{i_l}$ của $G$ thỏa mãn: 
\[{\rm sgn}(C)=\prod_{i=1}^l a_{i_ki_{k+1}} >0,\]
với qui ước $a_{i_li_{l+1}}=a_{i_li_{1}}$.
\item[ii.] Tồn tại ma trận đường chéo $\m{G}$ sao cho $\m{G}\m{A}\m{G}$ có tất cả các phần tử không âm.
\item[iii.] Ma trận Laplace của $G$ có một giá trị riêng bằng 0.
\end{itemize}
\end{exercise}

\begin{exercise}[Đồ thị không cân bằng cấu trúc dấu] \cite{Harary1971graph} \label{ex:8.14} Chứng minh rằng đồ thị dấu vô hướng, liên thông $G$ là không cân bằng cấu trúc dấu nếu một trong các điều kiện tương đương sau thỏa mãn:
\begin{itemize}
\item[i.] Tồn tại ít nhất một chu trình của $G$ có dấu âm.
\item[ii.] Không tồn tại ma trận đường chéo $\m{G}$ sao cho $\m{G}\m{A}\m{G}$ có tất cả các phần tử không âm.
\item[iii.] Ma trận Laplace của $G$ có tất cả giá trị riêng dương.
\end{itemize}
\end{exercise}

\begin{exercise}[Mô hình Altafini] \label{ex:8.15} \cite{Altafini2013} Xét đồ thị dấu vô hướng, liên thông $G$ mô tả một hệ gồm $n$ tác tử. Giả sử các tác tử cập nhật biến trạng thái $x_i$ theo mô hình Altafini:
\begin{align}
\dot{x}_i = - \sum_{j \in N_i} |a_{ij}|(x_i - {\rm sgn}(a_{ij})x_j),\; i=1,\ldots,n.
\end{align}
\begin{itemize}
\item[i.] Chứng minh rằng nếu đồ thị $G$ là cân bằng cấu trúc dấu thì ta có
\begin{align*}
\lim_{t \to +\infty} \m{x}(t) = \frac{1}{n} (\m{1}_n^\top \m{G}\m{x}(0))\m{G}\m{1}_n,
\end{align*}
với $\m{G}$ là ma trận đường chéo sao cho $\m{G}\m{A}\m{G}$ có tất cả các phần tử không âm (xem Bài tập \ref{ex:8.13}).
\item[ii.] Chứng minh rằng nếu đồ thị $G$ là không cân bằng cấu trúc dấu thì ta có
\begin{align*}
\lim_{t \to +\infty} \m{x}(t) = \m{0}_n,\; \forall \m{x}(0) \in \mb{R}^n.
\end{align*}
\end{itemize}
\end{exercise}

\chapter{Hệ đồng thuận trọng số ma trận}
\label{chap:MWC}
Trong chương này, một mở rộng tự nhiên của thuật toán đồng thuận thông thường sẽ được giới thiệu. Mô hình đồng thuận thông thường mô tả mối liên kết nhị nguyên giữa các tác tử trong hệ (hoặc có tương tác với nhau $a_{ij}>0$  hoặc không có tương tác với nhau $a_{ij}=0$). Trong thực tế, nếu chỉ xét hệ hợp tác, các tác tử có thể liên kết toàn bộ, liên kết một phần hoặc không có liên kết với nhau \cite{Trinh2026}. Để mô tả các tương tác khác nhau trong mạng, chúng ta có thể sử dụng các trọng số ma trận $\m{A}_{ij}$ thay thế cho các trọng số vô hướng trong thuật toán đồng thuận thông thường. Tương ứng, thuật toán này được gọi là thuật toán đồng thuận với trọng số ma trận\index{hệ đồng thuận!trọng số ma trận}.

Hệ đồng thuận trọng số ma trận \cite{Trinh2018matrix} được đề xuất dựa trên tổng quát hóa một sống nghiên cứu độc lập đã có: trọng số ma trận quay ở bài toán cyclic pursuit \cite{Ramirez2009distributed}, đồng thuận các hệ qui chiếu \cite{Thunberg2012,lee2016distributed}, trọng số ma trận đối xứng xác định dương ở bài toán định vị mạng \cite{Barooah2006graph}, trọng số ma trận đối xứng bán xác định dương trong đồng bộ hóa các mạng dao động ghép nối đa chiều \cite{Tuna2016aut,Tuna2009tac}, hay điều khiển đội hình và định vị mạng với ràng buộc vector hướng \cite{Zhao2015MSC,zhao2016aut}. 

Một số kết quả mở rộng dựa trên mô hình đồ thị trọng số ma trận \cite{Trinh2018matrix} bao gồm mô hình động học ý kiến nhiều chiều \cite{Ahn2020opinion,Ye2020Aut}, mô hình đồ thị trọng số ma trận với tương tác đối nghịch \cite{Pan2018bipartite}, mô hình kết hợp các ma trận cứng \cite{Fang2025simultaneous}, đồ thị trọng số ma trận ngẫu nhiên \cite{Le2024randomized}.

\section{Đồ thị với trọng số ma trận}
Xét đồ thị vô hướng $G$ với các trọng số ma trận, thể hiện bởi $(V,E,W)$ trong đó $V=\{v_1,\ldots,v_n\}$ thể hiện tập gồm $n = |V|$ đỉnh, $E = \{(i,j)|~v_i, v_j\in V, i\neq j\}$ là tập gồm $|E|=m$ cạnh, và $W = \{\m{A}_{ij}\in \mb{R}^{d\times d}|~(i,j) \in E, \m{A}_{ij} = \m{A}_{ij}^\top =\m{A}_{ji}\}$ là tập các trọng số ma trận kích thước $d\times d$, $d\ge 1$. Khi $d=1$, ta có lại đồ thị trọng số thực dương như trong hệ đồng thuận đã xét ở Chương~\ref{chap:consensus}.

Dựa trên trọng số ma trận, ta qui ước tương tác giữa hai tác tử $(i,j)$ như sau:
\begin{itemize}
    \item $\m{A}_{ij} = \m{0}_{d\times d}$: $i$ và $j$ không tương tác với nhau,
    \item $\m{A}_{ij}$ là xác định dương: $(i,j)$  là một cạnh xác định dương, tương ứng với một tương tác đầy đủ,
    \item $\m{A}_{ij}$ bán xác định dương với ${\rm rank}(\m{A}_{ij})<d$: $(i,j)$ là một cạnh bán xác định dương, tương ứng với một tương tác một phần.
\end{itemize}
\begin{example}[Tính đa lớp của đồ thị trọng số ma trận]
\label{eg:c9_scalar_and_matrix_weighted_graphs}
\begin{figure}[th!]
\centering
\subfloat[Đồ thị trọng số thực dương]{\resizebox{.45\linewidth}{!}{
\begin{tikzpicture}[
roundnode/.style={circle, draw=black, thick, minimum size=2mm,inner sep= 0.3mm},
squarednode/.style={rectangle, draw=black, thick, minimum size=3.5mm,inner sep= 0.25mm},
]
    \node[roundnode, fill = black]   (u1)   at  (0,0) {};     %
    \node[roundnode, fill = yellow]   (u2)   at  (1,1) {};
    \node[roundnode, fill = green]   (u3)   at  (2,0.5) {};
    \node[roundnode, fill = orange]   (u4)   at  (3,0.5) {};
    \draw [draw = black, very thick]
    (u1) edge [bend left=0] (u2)
    (u2) edge [bend left=0] (u3)
    (u3) edge [bend left=0] (u4)
    ;
    \node at (-0.75,-0.2) {1};     %
    \node at (1,1.3) {2};
    \node at (2,0.15) {3};
    \node at (3.8,0.5) {4};
    \node at (.25,0.65) {$a_{12}$};     %
    \node at (1.65,0.9) {$a_{23}$};     %
    \node at (2.5,0.7) {$a_{34}$};     %
\end{tikzpicture}
}}
\hfill 
\subfloat[Đồ thị trọng số ma trận]{\resizebox{.45\linewidth}{!}{
\begin{tikzpicture}[
roundnode/.style={circle, draw=black, thick, minimum size=2mm,inner sep= 0.3mm},
squarednode/.style={rectangle, draw=black, thick, minimum size=3.5mm,inner sep= 0.25mm},
]
    \node[roundnode, fill = black]   (u1)   at  (0,0) {};     %
    \node[roundnode, fill = yellow]   (u2)   at  (1,1) {};
    \node[roundnode, fill = green]   (u3)   at  (2,0.5) {};
    \node[roundnode, fill = orange]   (u4)   at  (3,0.5) {};
    \draw [draw = black, very thick]
    (u1) edge [bend left=0] (u2)
    (u2) edge [bend left=0] (u3)
    (u3) edge [bend left=0] (u4)
    ;
    \node at (-0.75,-0.2) {1};     %
    \node at (.9,1.3) {2};
    \node at (2,0.15) {3};
    \node at (3.8,0.5) {4};
    \node at (.22,0.65) {$\mathbf{A}_{12}$};     %
    \node at (1.65,1) {$\mathbf{A}_{23}$};     %
    \node at (2.5,0.7) {$\mathbf{A}_{34}$};     %
\end{tikzpicture}
}}\\
\subfloat[Tương tác trong đồ thị trọng số ma trận]{\resizebox{.65\linewidth}{!}{
\begin{tikzpicture}[
roundnode/.style={circle, draw=black, thick, minimum size=2mm,inner sep= 0.3mm},
squarednode/.style={rectangle, draw=black, thick, minimum size=3.5mm,inner sep= 0.25mm},
]
    \node[trapezium, trapezium left angle=40, trapezium right angle=140, fill=magenta!30, fill opacity=0.2, minimum width=6.5cm, minimum height=1cm]
         at (1.75,0.5) {}; 
    \node[trapezium, trapezium left angle=40, trapezium right angle=140, fill=cyan!30, fill opacity=0.2, minimum width=6.5cm, minimum height=1cm]
         at (1.75,2.5) {};
    \node[ellipse, minimum width = 0.45cm, minimum height = 2.6cm, align=center, draw = gray, thick, fill=gray!50, fill opacity=0.3] (e1) at (0,1) {}; 
    \node[ellipse, minimum width = 0.45cm, minimum height = 2.6cm, align=center, draw = yellow, thick, fill=yellow!50, fill opacity=0.3] (e1) at (1,2) {}; 
    \node[ellipse, minimum width = 0.45cm, minimum height = 2.6cm, align=center, draw = green, thick, fill=green!50, fill opacity=0.3] (e1) at (2,1.5) {}; 
    \node[ellipse, minimum width = 0.45cm, minimum height = 2.6cm, align=center, draw = orange, thick, fill=orange!50, fill opacity=0.3] (e1) at (3,1.5) {}; 
    \node[roundnode, fill = black]   (u1a)   at  (0,0) {};     %
    \node[roundnode, fill = yellow]   (u2a)   at  (1,1) {};
    \node[roundnode, fill = green]   (u3a)   at  (2,0.5) {};
    \node[roundnode, fill = orange]   (u4a)   at  (3,0.5) {};

    \node[roundnode, fill = black]   (u1b)   at  (0,2) {};     %
    \node[roundnode, fill = yellow]   (u2b)   at  (1,3) {};
    \node[roundnode, fill = green]   (u3b)   at  (2,2.5) {};
    \node[roundnode, fill = orange]   (u4b)   at  (3,2.5) {};

    \draw [draw = black, very thick]
    (u1a) edge [bend left=0] (u2a)
    (u2a) edge [bend left=0] (u3a)
    (u3a) edge [bend left=0] (u4a)
    ;
    \draw [draw = black, very thick]
    (u1b) edge [bend left=0] (u2b)
    (u2b) edge [bend left=0] (u3b)
    (u3b) edge [bend left=0] (u4b)
    ;
    \draw [draw = black, very thick]
    (u1a) edge [in=150, out=210,looseness=20] (u1a)
    (u1b) edge [in=150, out=210,looseness=20] (u1b)
    (u2a) edge [in=150, out=30,looseness=10] (u2a)
    (u2b) edge [in=150, out=30,looseness=10] (u2b)
    (u3a) edge [in=150, out=30,looseness=10] (u3a)
    (u3b) edge [in=150, out=30,looseness=10] (u3b)
    (u4a) edge [in=45, out=-45,looseness=10] (u4a)
    (u4b) edge [in=45, out=-45,looseness=10] (u4b)
    ;
    
    \draw [draw = red, very thick]
    (u1a) edge [bend left=0] (u2b)
    (u1b) edge [bend left=0] (u2a)
    (u2a) edge [bend left=0] (u3b)
    (u3a) edge [bend left=0] (u2b)
    (u4a) edge [bend left=0] (u3b)
    (u3a) edge [bend left=0] (u4b)
    ;

    \draw [draw = blue, very thick]
    (u1a) edge [bend left=0] (u1b)
    (u2b) edge [bend left=0] (u2a)
    (u3a) edge [bend left=0] (u3b)
    (u4a) edge [bend left=0] (u4b)
    ;

    \node at  (2.0,-0.2) {Lớp thứ 1};     %
    \node at  (3.5,3.1) {Lớp thứ 2};     %
    \node (it1) at  (-2,3.25) {Tương tác ngang};
    \node (it2) at  (0.5,2.4) { };
    \node[align=left] (it3) at  (6,2.5) {Tương tác dọc};
    \node (it4) at  (2.95,1.25) { };
    \node[align=left] (it5) at  (-2,.75) {Tương tác chéo};
    \node (it6) at  (0.5,1.2) { };
    \draw [draw = gray, very thick,-{Stealth[length=2mm]}]
    (it1) edge [bend left=10] (it2);
    \draw [draw = gray, very thick,-{Stealth[length=2mm]}]
    (it3) edge [bend right=20] (it4);
    \draw [draw = gray, very thick,-{Stealth[length=2mm]}]
    (it5) edge [bend left=30] (it6);
\end{tikzpicture}
}}
\caption{(a) Đồ thị thể hiện quan hệ liên thuộc của một đồ thị truyền thống (a) và của một đồ thị trọng số ma trận (b) là như nhau. Tuy nhiên, khi biểu diễn cụ thể từng tương tác tương ứng với các trọng số ma trận, đồ thị thể hiện quan hệ phân lớp với các tương tác ngang, dọc, và chéo (c). \label{fig:c9_explain_swg_mwg}}
\end{figure}
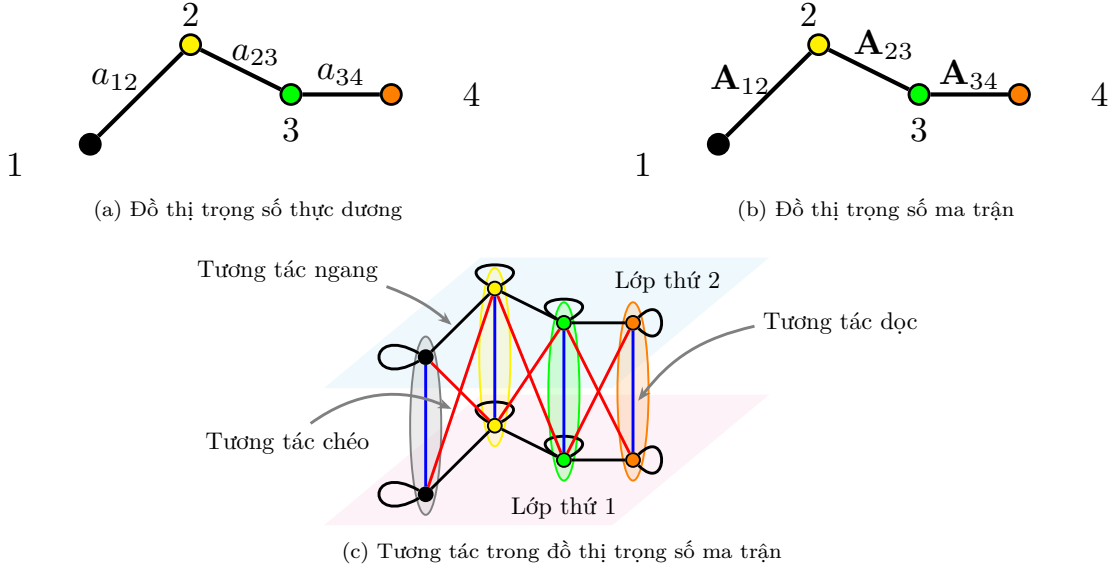
Xét đồ thị trọng số ma trận gồm 4 đỉnh. Mỗi đỉnh mô tả một tác tử $i$ với vector biến trạng thái hai chiều $\m{x}_i=\begin{bmatrix}
x_i^1 \\ x_i^2
\end{bmatrix}$, biểu diễn bởi hai nút tròn nhỏ. Các nút $\{x_i^1,i=1,\ldots,4\}$ và $\{x_i^2,i=1,\ldots,4\}$ thuộc về hai lớp khác nhau của đồ thị. Các ma trận tương tác $\m{A}_{ij}=\m{A}_{ji} = \begin{bmatrix}
a_{ij}^{11} & a_{ij}^{12} \\ a_{ij}^{21} & a_{ij}^{22}
\end{bmatrix}$ có kích thước $2 \times 2$, khi biểu diễn mỗi phần tử của ma trận $a_{ij}^{pq}$ bởi một cạnh nối $x_i^p$ và $x_j^q$ tương ứng trong thuật toán cập nhật (dạng đồng thuận trọng số ma trận) \cite{Trinh2024networked}:
\begin{align} \label{eq:c9_consensus}
 {x}_{ip}[k+1] &= \underbrace{\sum_{q=1}^d w_{ii}a_{i,i}^{p,q}x_{iq}[k]}_{\text{tương tác dọc}} + \underbrace{\sum_{\substack{j=1,\\ j\neq i}}^n w_{ij}a_{i,j}^{p,p}x_{jp}[k]}_{\text{tương tác ngang}} + \underbrace{\sum_{\substack{j=1,\\j\neq i}}^n\sum_{\substack{q=1,\\ q\neq p}}^d w_{ij}a_{i,j}^{p,q}x_{jq}[k]}_{\text{tương tác chéo}},
\end{align}
với $i=1,\ldots,n$, $p=1,\ldots,d$ và $k=0,1,2,\ldots$, chúng ta thu được mô tả ở Hình~\ref{fig:c9_explain_swg_mwg} (c). Các tương tác ngang là tương tác cùng lớp, tương tác dọc nối các nút ở cùng một tác tử, và tương tác chéo nối các nút khác lớp khác tác tử.
\end{example}

Một đường đi $P$ trong $G$ là một chuỗi các cạnh  $(i_1, i_2), \ldots, (i_{l-1},i_l)$ sao cho mỗi cạnh $(i_k,i_{k+1})$ là xác định dương hoặc bán xác định dương. Trong trường hợp mọi cạnh của $P$ đều là các cạnh xác định dương, ta gọi $P$ là một đường đi xác định dương của $G$.\index{đường đi!xác định dương}

Gọi $\bar{G}$ là đồ thị vô hướng tương ứng của $G$, định nghĩa bởi $G=(V,\bar{E})$ trong đó tồn tại cạnh $(i,j) \in \bar{E}$ khi và chỉ khi trọng số $\m{A}_{ij}$ của $(i,j)$ là bán xác định dương hoặc xác định dương. Với $\bar{{T}}$ là một cây trong $\bar{G}$ thì đồ thị ${T}$ với các đỉnh tương ứng trong ${G}$ cũng là một cây trong $G$. Ta gọi ${T}$ là một cây xác định dương nếu $T$ có mọi cạnh xác định dương. Cây  ${T}$ là một cây bao trùm xác định dương của $G$ nếu $T$ chứa mọi đỉnh của $G$ và có tất cả các cạnh xác định dương. \index{cây!xác định dương} \index{cây!bao trùm xác định dương}

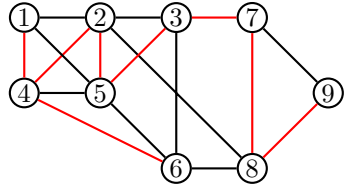
\begin{figure}[t]
    \centering
\begin{tikzpicture}[
roundnode/.style={circle, draw=black, thick, minimum size=2mm,inner sep= 0.3mm},
squarednode/.style={rectangle, draw=black, thick, minimum size=3.5mm,inner sep= 0.25mm},
]
    \node[roundnode]  (u1)   at  (0,1) {1};     %
    \node[roundnode]  (u2)   at  (1,1) {2};
    \node[roundnode]  (u3)   at  (2,1) {3};
    \node[roundnode]  (u4)   at  (0,0) {4};
    \node[roundnode]  (u5)   at  (1,0) {5};
    \node[roundnode]  (u6)   at  (2,-1) {6};
    \node[roundnode]  (u7)   at  (3,1) {7};
    \node[roundnode]  (u8)   at  (3,-1) {8};
    \node[roundnode]  (u9)   at  (4,0) {9};
    
    \draw [draw = black,  thick] (u1)--(u2)--(u3)--(u6)--(u8)--(u2);
    \draw [draw = black, thick] (u4)--(u5)--(u6);
    \draw [draw = black,  thick] (u7)--(u9);
    \draw [draw = red, thick] (u1)--(u4)--(u2)--(u5)--(u3)--(u7)--(u8)--(u9);
    \draw [draw = black, thick] (u5)--(u1);
    \draw [draw = red, thick] (u4)--(u6);
\end{tikzpicture}
    \caption{Ví dụ đồ thị trọng số ma trận trong đó cạnh màu đỏ thể hiện một cạnh xác định dương và cạnh màu đen thể hiện một cạnh xác định dương hoặc bán xác định dương. Đồ thị với các cạnh màu đỏ là một cây bao trùm xác định dương của $G$.}
    \label{fig:c9_positive_spanning_tree}
\end{figure}

Ma trận kề trọng số ma trận được định nghĩa bởi \index{ma trận!kề trọng số ma trận}
\begin{align}
\m{A} = 
\begin{bmatrix}
\m{0}_{d\times d}&{{\m{A}_{12}}}&{{\m{A}_{13}}}& \cdots &{{\m{A}_{1n}}}\\
{{\m{A}_{21}}}&\m{0}_{d \times d}&{{\m{A}_{23}}}& \cdots &{{\m{A}_{2n}}}\\
 \vdots & \ddots & \ddots & \ddots & \vdots \\
{{\m{A}_{n - 1,1}}}& \cdots &{{\m{A}_{n - 1,n - 2}}}&\m{0}_{d\times d}&{{\m{A}_{n - 1,n}}}\\
{{\m{A}_{n1}}}&{{\m{A}_{n2}}}& \cdots &{{\m{A}_{n,n - 1}}}& \m{0}_{d\times d}
\end{bmatrix}.
\end{align}
Tương ứng với mỗi đỉnh $i \in V$, ta định nghĩa bậc của đỉnh $i$ bởi $\m{D}_i = \sum_{j=1}^n \m{A}_{ij}$. Ma trận bậc của đồ thị được cho bởi $\m{D} = \text{blkdiag}(\m{D}_1,\ldots,\m{D}_n) = \text{blkdiag}(\m{A}(\m{1}_n \otimes \m{I}_d))$. Ma trận Laplace với trọng số ma trận được định nghĩa tương ứng bởi\index{ma trận!Laplace trọng số ma trận} \index{ma trận!bậc trọng số ma trận}
\begin{align*}
    \mcl{L} = \m{D} - \m{A}.
\end{align*}

Dễ thấy ma trận Laplace trọng số ma trận là đối xứng, bán xác định dương, đồng thời
\begin{align} \label{eq:kernel_of_L}
   \text{im}(\m{1}_n \otimes \m{I}_d) \subseteq \text{ker}(\mcl{L}).
\end{align}
Hơn nữa, ta có thể biểu diễn 
\begin{align}
    \mcl{L} = (\m{H}^\top \otimes \m{I}_d) \text{blkdiag}(\m{A}_k) (\m{H} \otimes \m{I}_d) = \bar{\m{H}}^\top \text{blkdiag}(\m{A}_k) \bar{\m{H}},
\end{align}
trong đó $\m{H}$ là ma trận liên thuộc của $G$, $\m{A}_k = \m{A}_{ij}$ nếu $e_k = (i,j)$.
\section{Thuật toán đồng thuận trọng số ma trận}
\subsection{Điều kiện đồng thuận}
Thuật toán đồng thuận trọng số ma trận có dạng:
\begin{align}
    \dot{\m{x}}_i = - \sum_{i=1}^n \m{A}_{ij}(\m{x}_i - \m{x}_j), ~\forall i=1, \ldots, n,
\end{align}
trong đó $\m{x}_i \in \mb{R}^d$ là biến trạng thái của tác tử $i$. Sử dụng ma trận $\mcl{L}$, thuật toán trên có thể được biểu diễn lại dưới dạng ma trận như sau:
\begin{align} \label{eq:MWC}
    \dot{\m{x}} = - \mcl{L} \m{x},~\m{x}(0) = \m{x}_0, 
\end{align}
trong đó $\m{x} = [\m{x}_1^\top, \ldots, \m{x}_n^\top]^\top \in \mb{R}^{dn}$. Dựa trên tính chất của ma trận $\mcl{L}$, ta có  
\begin{align}
    \m{x}(t) = \texttt{e}^{-\mcl{L}t} \m{x}(0).
\end{align}
Chú ý rằng trọng tâm hệ định nghĩa bởi $\bar{\m{x}}(t) = \frac{1}{n}\sum_{i=1}^n \m{x}_i(t)=\frac{1}{n} (\m{1}_n^\top \otimes \m{I}_d) \m{x}$ là bất biến theo thời gian do
\begin{align}
    \dot{\bar{\m{x}}}(t) = \frac{1}{n} (\m{1}_n^\top \otimes \m{I}_d) \dot{\m{x}} = -\frac{1}{n} (\m{1}_n^\top \otimes \m{I}_d) \mcl{L} \m{x} = \m{0}_d.
\end{align}

\begin{theorem}[Điều kiện đồng thuận trọng số ma trận]
Điều kiện cần và đủ để hệ đạt đồng thuận tại trọng tâm $\bar{\m{x}}$ với mọi giá trị đầu $\m{x}(0)$ là $\text{im}(\m{1}_n \otimes \m{I}_d) = \text{ker}(\mcl{L})$.
\end{theorem}
\begin{proof}
(Điều kiện cần) Giả sử $\text{im}(\m{1}_n \otimes \m{I}_d) = \text{ker}(\mcl{L})$. Ta có thể phân tích ma trận Laplace dưới dạng $\mcl{L} = \m{P} \m{\Lambda} \m{P}^\top$, trong đó
\begin{equation}
    \m{P} = \begin{bmatrix}
    \frac{1}{\sqrt{n}}(\m{1}_n \otimes \m{I}_d), \m{p}_{d+1}, \ldots, \m{p}_{dn}
    \end{bmatrix}
\end{equation}
là một ma trận trực chuẩn, tức là $\m{P}\m{P}^\top = \m{I}_{dn}$, và $\m{\Lambda} = \text{diag}(\lambda_1, \ldots, \lambda_{dn})$ trong đó $0= \lambda_1 = \ldots = \lambda_d < \lambda_{d+1} \leq \ldots \leq \lambda_{dn}$ là các giá trị riêng của ma trận $\mcl{L}$. Từ đây suy ra:
\begin{align}
    \m{x}(t) = \m{P} \texttt{e}^{-\m{\Lambda}t} \m{P}^{-1} \m{x}(0) \to \frac{1}{n}(\m{1}_n \otimes \m{I}_d)(\m{1}_n \otimes \m{I}_d)^\top \m{x}(0) = \m{1}_n \otimes \bar{\m{x}},
\end{align}
hay $\m{x}_i$ tiệm cận tới trọng tâm của hệ với $i=1,\ldots,n$.

(Điều kiện đủ) Ta sẽ chứng minh bằng phản chứng. Giả sử ${\rm im}(\m{1}_n \otimes \m{I}_d)\neq {\rm ker}(\mcl{L})$ thì theo \eqref{eq:kernel_of_L}, tồn tại vector $\m{0}_{dn} \neq \m{v} \notin {\rm im}(\m{1}_n \otimes \m{I}_d)$ sao cho $\mcl{L}\m{v} = \m{0}_{dn}$. Như vậy, chọn $\m{x}(0) = \m{v}$ thì $\dot{\m{x}} = \m{0}_{dn}, \forall t \geq 0$. Rõ ràng hệ không thể đạt được đồng thuận tại trọng tâm trong trường hợp này.
\end{proof}

Việc kiểm tra điều kiện về hạng không cho ta phương án để xây dựng đồ thị trọng số ma trận đảm bảo hệ tiệm cận tới đồng thuận. Một điều kiện đủ để hệ đạt đồng thuận tại trọng tâm được cho như sau:
\begin{corollary}
Hệ \eqref{eq:MWC} sẽ tiệm cận tới giá trị đồng thuận tại trọng tâm nếu $G$ chứa một cây bao trùm xác định dương.
\end{corollary}
\begin{proof}
Ta có thể sắp xếp các cạnh của đồ thị $G$ sao cho các cạnh $e_1, \ldots, e_{n-1}$ tương ứng là các cạnh của cây bao trùm xác định dương ${T}$ của $G$. Gọi $\m{H}$ là ma trận liên thuộc của $G$. Với cách phân chia này, ta viết lại $\m{H}$ dưới dạng
\begin{align}
    \m{H} = \begin{bmatrix}
    \m{H}_{E({T})}\\
    \m{H}_{{E}(G)\setminus E({T})}
    \end{bmatrix}.
\end{align}
Với $\m{v} = [\m{v}_1^\top, \ldots, \m{v}_n^\top]^\top \in \mb{R}^{dn}$ là một vector bất kỳ trong $\text{ker}(\mcl{L})$ thì cần có:
\begin{align}
\m{v}^\top \mcl{L} \m{v} = \m{v}^\top \bar{\m{H}}^\top \text{blkdiag}(\m{A}_k) \bar{\m{H}}\m{v} =\| \text{blkdiag}(\m{A}_k^{\frac{1}{2}}) \bar{\m{H}}\m{v} \|^2 = 0
\end{align}
Phương trình này tương đương với
\begin{align}
\text{blkdiag}(\m{A}_k^\frac{1}{2}) \bar{\m{H}} \m{v} = \text{blkdiag}(\m{A}_k^\frac{1}{2}) \begin{bmatrix}
    \bar{\m{H}}_{E({T})} \m{v}\\
    \bar{\m{H}}_{{E}(G)\setminus E({T})} \m{v}
    \end{bmatrix} = \m{0}_{dn}
\end{align}
Do các cạnh $e_i,~i=1, \ldots, n-1$ là xác định dương, các ma trận $\m{A}_{k},~k=1,\ldots, n-1$ là các ma trận xác định dương. Như vậy phương trình $\text{blkdiag}(\m{A}_1^\frac{1}{2},\ldots,\m{A}_{n-1}^\frac{1}{2})\bar{\m{H}}_{E({T})} \m{v} = \m{0}_{d(n-1)}$ tương đương với $\bar{\m{H}}_{E({T})} \m{v} = \m{0}_{d(n-1)}$. Mà $\text{ker}(\bar{\m{H}}_{E({T})} = {\rm im}(\m{1}_n \otimes \m{I}_d)$ do ${T}$ là một cây bao trùm của đồ thị, từ đây suy ra $\m{v} \in {\rm im}(\m{1}_n \otimes \m{I}_d) $. Do $\m{v}$ là một vector được chọn bất kỳ trong $\text{ker}(\mcl{L})$ nên suy ra  ${\rm ker}(\mcl{L}) \subseteq  \text{im}(\m{1}_n \otimes \m{I}_d)$. Kết hợp với \eqref{eq:kernel_of_L} suy ra  ${\rm ker}(\mcl{L}) = \text{im}(\m{1}_n \otimes \m{I}_d)$, từ đó suy ra điều phải chứng minh.
\end{proof}

\subsection{Hiện tượng phân cụm}
Mục này đưa ra một số điều kiện đủ để kiểm tra đồ thị trọng số ma trận tính đồng thuận hoặc phân cụm khi thực hiện thuật toán đồng thuận trọng số ma trận \eqref{eq:MWC}.

Một sự phân chia tập đỉnh ${V}$ thành các cụm $\mc{C}_1,\ldots, \mc{C}_l$ $(1\leq l \leq n)$ thỏa mãn: (i) $\mc{C}_i \bigcap \mc{C}_j = \emptyset$, for $i \neq j$, và (ii) $\bigcup_{k=1}^l \mc{C}_k = {V}$. Hệ động thuận \eqref{eq:MWC} sẽ đạt tới một trạng thái phân cụm nếu tồn tại phép chia $\mc{C}_1,\ldots, \mc{C}_l,$ sao cho mỗi tác tử thuộc về cùng một cụm thì đạt được đồng thuận, trong khi đó hai tác tử $i$ và $j$ ở hai cụm khác nhau thì có $\m{x}_i \neq \m{x}_j$. Mỗi $\mc{C}_i$, $i =1, \ldots, l$, gọi là một cụm của hệ.\index{hiện tượng phân cụm}

\begin{lemma}\label{lem:c9_convergenceMWC}
Với luật đồng thuận~\eqref{eq:MWC}, $\m{x}(t) \to \text{ker}(\mcl{L})$ khi $t\to +\infty$.
\end{lemma}
\begin{proof}
Xét hàm Lyapunov $V = \frac{1}{2} \m{x}^\top \mcl{L} \m{x} \geq 0$ thì 
\begin{equation}
    \dot{V} = \m{x}^\top \mcl{L} \dot{\m{x}} = - \m{x}^\top\mcl{L} \mcl{L}\m{x} = - \|\mcl{L} \m{x}\|^2 \leq 0.
\end{equation}
Do $\mcl{L}^{1/2}{\m{x}} \perp \text{ker}(\mcl{L})$, giả sử ${\rm dim}({\rm ker}(\mcl{L}))=r={\rm dim}({\rm ker}(\mcl{L}^{1/2}))\ge d$ thì ta có \[2\lambda_{dn}(\mcl{L})V=\lambda_{dn}(\mcl{L})\m{x}^\top \mcl{L} \m{x} \geq \m{x}^\top\mcl{L}^{1/2} \mcl{L} \mcl{L}^{1/2}\m{x} \ge \lambda_{r}(\mcl{L})\m{x}^\top \mcl{L} \m{x}=2\lambda_{r+1}(\mcl{L})V.\] 
Từ đây suy ra $\dot{V} \leq - 2\lambda_{r+1}(\mcl{L}) V$ , hay $V\to 0$ theo tốc độ hàm mũ \cite{Khalil2002}. Do $V(\m{x})=0$ khi và chỉ khi $\m{x}\in {\rm ker}(\mcl{L})$, ta có $\m{x}\to \text{ker}(\mcl{L})$. 
\end{proof}

\begin{lemma} \label{lem:pos_tree}
Giả sử ${T} \subset {G}$ là một cây xác định dương gồm $l$ đỉnh, với thuật toán đồng thuận \eqref{eq:MWC}, $\m{x}_i(t) \to \m{x}_j(t)$, $\forall i, j \in {T}$, khi $t \to +\infty$.
\end{lemma}
\begin{proof}
Biểu diễn $\m{x} = [\m{x}_{{T}}^\top, \m{x}_{{V}\setminus {V}({T})}^\top]^\top$ và $\m{H}= \begin{bmatrix}\m{H}_1 & \m{0}\\ \m{H}_2 & \m{0}\\ \m{H}_3 & \m{H}_4\end{bmatrix}$,
trong đó $[\m{H}_{1} ~ \m{0}] \in \mb{R}^{(l-1)\times n}$ ứng với các cạnh trong ${T}$, $[\m{H}_{2} ~  \m{0}]$ ứng với các cạnh nối các đỉnh thuộc ${V}({T})$ nhưng không chứa cây ${T}$, và $[\m{H}_3 ~ \m{H}_4]$ tương ứng với các cạnh còn lại trong ${E}(G)$. Chú ý rằng  $\m{H}_2$  phụ thuộc tuyến tính vào $\m{H}_1$ bởi $\m{H}_2 = \m{T} \m{H}_1$. Một điểm cân bằng bất kỳ của \eqref{eq:MWC} cần thỏa mãn $\bar{\m{H}}\m{x}^* = \m{0}$, hay
\begin{equation}
\text{blkdiag}(\m{A}_1^{{1}/{2}},\ldots,\m{A}^{{1}/{2}}_{l-1}) \bar{\m{H}}_{1} \m{x}_{{T}}^* = \m{0}.
\end{equation}
Vì ${T}$ là một cây xác định dương, $\bar{\m{H}}_1\m{x}_{{T}}^* = \m{0}$. Từ đây suy ra mỗi điểm cân bằng $\m{x}^*$ của \eqref{eq:MWC} phải thỏa mãn $\m{x}_{{T}}^* \in \text{ker}(\bar{\m{H}}_1) = \text{im}(\m{1}_{l} \otimes \m{I}_{d})$, hay mọi biến trạng thái của $l$ tác tử nằm trong cây xác định dương ${T}$ phải bằng nhau. Mặt khác, do $\m{x}(t) \to \m{x}^* \in \text{ker}(\mcl{L})$ nên các tác tử trong ${T}$ sẽ dần tiến tới đồng thuận.
\end{proof}

\begin{story}{Hyo-Sung Ahn và điều khiển phân tán}
Hyo-Sung Ahn nhận bằng Cử nhân (1998) và Thạc sỹ (2000) từ Đại học Yonsei, Hàn Quốc, chuyên ngành Thiên văn học, bằng Th.S. từ Đại học North Dakota, và bằng T.S. ngành Điện và Khoa học máy tính từ Đại học bang Utah, UT, (2006)). Hiện tại, ông là giáo sư tại Khoa Cơ khí và Robotics tại Viện Khoa học và Công nghệ Gwangju (GIST), Hàn Quốc.

Nghiên cứu bậc T.S. của Hyo-Sung Ahn về điều khiển học lặp đã được xuất bản bởi Springer. Từ 2007 đến nay, Ahn thành lập lab Điều khiển phân tán và các hệ tự trị tại GIST \url{https://dcas.gist.ac.kr/dcas/}, với hướng nghiên cứu chính là điều khiển hệ đa tác tử với ứng dụng trong điều khiển đội hình và tối ưu phân tán. Ahn là tác giả của sách chuyên khảo ``Formation Control: Approaches for Distributed Agents'' (2020), và ``Matrix-weighted graphs: Theory and Applications'' (2026) và giáo trình ``Control of Multi-agent Systems: Theory and Simulations with Python'' (2024)
\end{story}

Kết quả sau đây cung cấp một điều kiện đủ để xác định liệu hai đỉnh thuộc về cùng một cụm với luật đồng thuận \eqref{eq:MWC}.

\begin{theorem} \label{thm:cluster}
Với ${T}$ là một cây xác định dương, kí hiệu $\mc{C}({T})$ là cụm chứa:\\
(i) mọi đỉnh trong ${T}$,\\
(ii) các đỉnh $i \notin T$: $\mc{S}_i = \{ {P}_k=\{v^k_1\ldots v^k_{|{P}_k|}\} |~v^k_1 = i,~v^k_{|{P}_k|}  \in T, \text{ và } \forall j = 1,\ldots, {|{P}_k|-1},~v^k_j \notin T \}$, thỏa mãn
\begin{itemize}
\item[(a)] Với mỗi đường đi ${P}_k$, định nghĩa $\text{ker}({P}_k) = \bigcup_{j=1}^{{|{P}_k|-1}} \text{ker}(\m{A}_{v^k_jv^k_{j+1}})$,  ta có
\begin{equation} \label{eq:same_cluster}
\text{dim} \left(\bigcap\nolimits_{k=1}^{|\mc{S}_i|} \text{ker}({P}_k)\right) = 0.
\end{equation}
\item[(b)] Mỗi ${P}_{k} \in \mc{S}_i$ không chứa chu trình, tức là $v_l \neq v_m,$ $\forall v_l, v_m \in P_k$.
\end{itemize}
Khi đó, tập $\mc{S}_i$ là hữu hạn. Với luật đồng thuận \eqref{eq:MWC}, mọi tác tử trong $\mc{C}(T)$ sẽ tiệm cận tới cùng một giá trị.
\end{theorem}

\begin{proof}
Từ Bổ đề \ref{lem:pos_tree}, mọi $\m{x}_j$ ($j \in {V}(T)$) hội tụ tới một giá trị vector  chung $\m{x}^*_{T}$. Xét đỉnh $i \notin {V}(T)$ thỏa mãn điều kiện (ii) và  $\m{x}^*_i$ là giá trị tại điểm cân bằng của tác tử $i$. Khi đó, dựa trên định nghĩa của  ${\rm ker}({P}_k)$ thì
\begin{equation}\label{eq:same_cluster1}
\m{x}_i^* - \m{x}_{T}^* \in \text{ker}({P}_k),~ \forall P_k \in \mc{S}_i.
\end{equation}
Từ phương trình \eqref{eq:same_cluster}, nghiệm duy nhất của \eqref{eq:same_cluster1} là  $\m{x}_i^* - \m{x}_{T}^* = \m{0}$, hay $\m{x}_i^* = \m{x}_{T}^*$. Vì vậy, các tác tử thuộc cụm $\mc{C}(T)$ đạt được đồng thuận.

Giả sử ${P}_1$ và ${P}_2$ là hai đường đi khác nhau, và ${P}_2$ thu được nhờ thêm vào ${P}_1$ một chu trình. Khi đó, ${\rm ker}({P}_1) \subseteq \text{ker}({P}_2)$. Từ đây suy ra ${\rm ker}({P}_1) \cap {\rm ker}({P}_2) = \text{ker}({P}_1)$, tức là ta không cần phải xét tới đường đi chứa chu trình trong phương trình \eqref{eq:same_cluster}. Điều này có ý nghĩa là số phần tử của tập $\mc{S}_i$ cần xét là hữu hạn. Cuối cùng, xét một đỉnh $i$ không thỏa mãn cả hai điều kiện (i) and (ii). Gọi $\mc{S}_i$ là tập các đường đi từ $i$ tới  $T$ sao cho ${\rm dim}(\bigcap\nolimits_{k=1}^{|\mc{S}_i|} {\rm ker}({P}_k)) \geq 1$. Khi đó, tồn tại nghiệm  $\m{x}_i^* - \m{x}_{T}^* \neq 0$ thỏa mãn \eqref{eq:same_cluster1}.
\end{proof}

Từ Định lý \eqref{thm:cluster}, ta có hệ quả sau:
\begin{corollary} Xét cây xác định dương $T$ và một đỉnh $v_i \notin T$. Từ đỉnh  $v_i$ tới $v_j \in T$, nếu tồn tại ít nhất hai đường đi sao cho  phương trình~\eqref{eq:same_cluster} thỏa mãn, thì đỉnh $v_i$ thuộc cụm $\mc{C}(T)$.
\end{corollary}

Tương tự, xét hai cây xác định dương $T_1, T_2$ liên kết qua tập cạnh $\mc{S} = \{(i,j) \in E|~i \in T_1, j \in T_2\}$. Nếu $\sum_{(i,j) \in \mc{S}} \m{A}_{ij}$ là xác định dương thì $T_1$ và $T_2$ thuộc về cùng một cụm.

\begin{figure}[t]
\centering
\begin{tikzpicture}[
roundnode/.style={circle, draw=black, thick, minimum size=2mm,inner sep= 0.3mm},
squarednode/.style={rectangle, draw=black, thick, minimum size=3.5mm,inner sep= 0.25mm},
]
    \node[ellipse, minimum width = 2.25cm, minimum height = 1.2cm, align=center, draw = red, dashed, thick, fill=red!25, fill opacity=0.3, rotate=-20] (e1) at (-0.75,0.25) {}; 

    \node[roundnode]  (u1)   at  (0,0) {1};     %
    \node[roundnode]  (u2)   at  (-1,-1.6) {2};
    \node[roundnode]  (u3)   at  (1,-1.6) {3};
    \node[roundnode]  (u4)   at  (-1.5,0.5) {4};     

    \draw [draw = black, thick] (u1)--(u2)--(u3)--(u1);
	\draw [draw = red, thick] (u1)--(u4);
    \node at (-0.75,0) {\color{red} ${T}$};     %
\end{tikzpicture}
\qquad\qquad
\begin{tikzpicture}[
roundnode/.style={circle, draw=black, thick, minimum size=2mm,inner sep= 0.3mm},
squarednode/.style={rectangle, draw=black, thick, minimum size=3.5mm,inner sep= 0.25mm},
]
    \node[ellipse, minimum width = 2.75cm, minimum height = 3.5cm, align=center, draw = red, dashed, thick, fill=red!25, fill opacity=0.3, rotate=-25] (e1) at (-1.1,-0.5) {}; 

    \node[roundnode]  (u1)   at  (0,0) {1};     %
    \node[roundnode]  (u2)   at  (-1,-1.6) {2};
    \node[roundnode]  (u3)   at  (1,-1.6) {3};
    \node[roundnode]  (u4)   at  (-1.5,0.5) {4};     

    \draw [draw = black, thick] (u1)--(u2)--(u3)--(u1);
	\draw [draw = red, thick] (u1)--(u4);
    \node at (-1.1,-0.5) {\color{red} ${C}(T)$};     %
\end{tikzpicture}
\caption{Đồ thị minh họa hệ bốn tác tử trong Ví dụ~\ref{ex:MWC1}. }
\label{fig:MWC_ex1}
\end{figure}
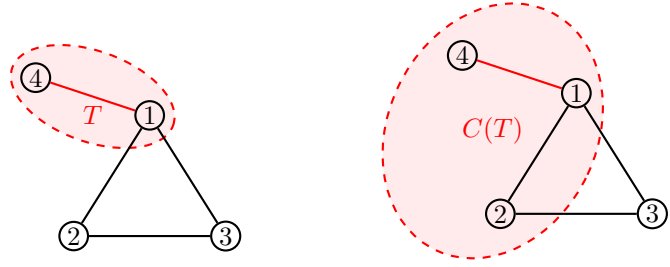
\begin{figure}[t]
    \subfloat[Thay đổi theo trục $x$.]{\includegraphics[height=4cm]{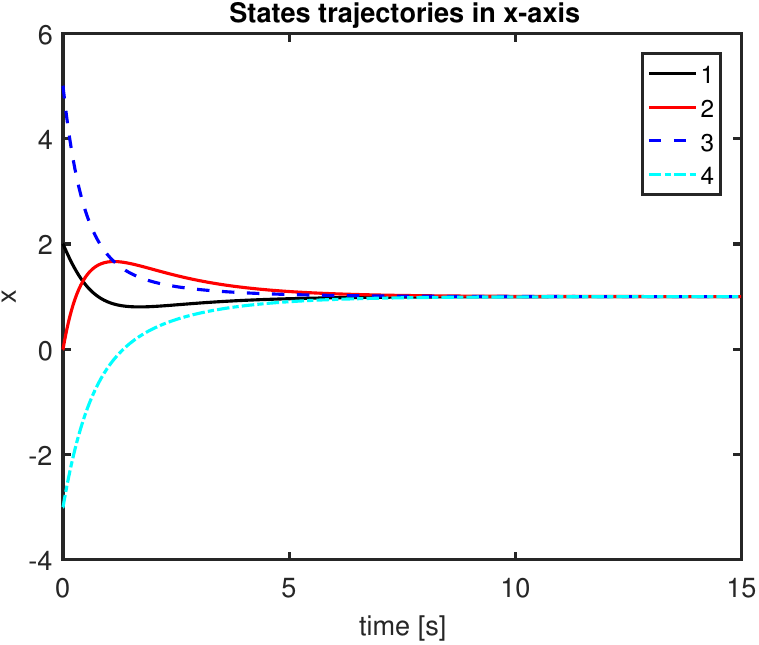}} \hfill
    \subfloat[Thay đổi theo trục $y$.]{\includegraphics[height=4cm]{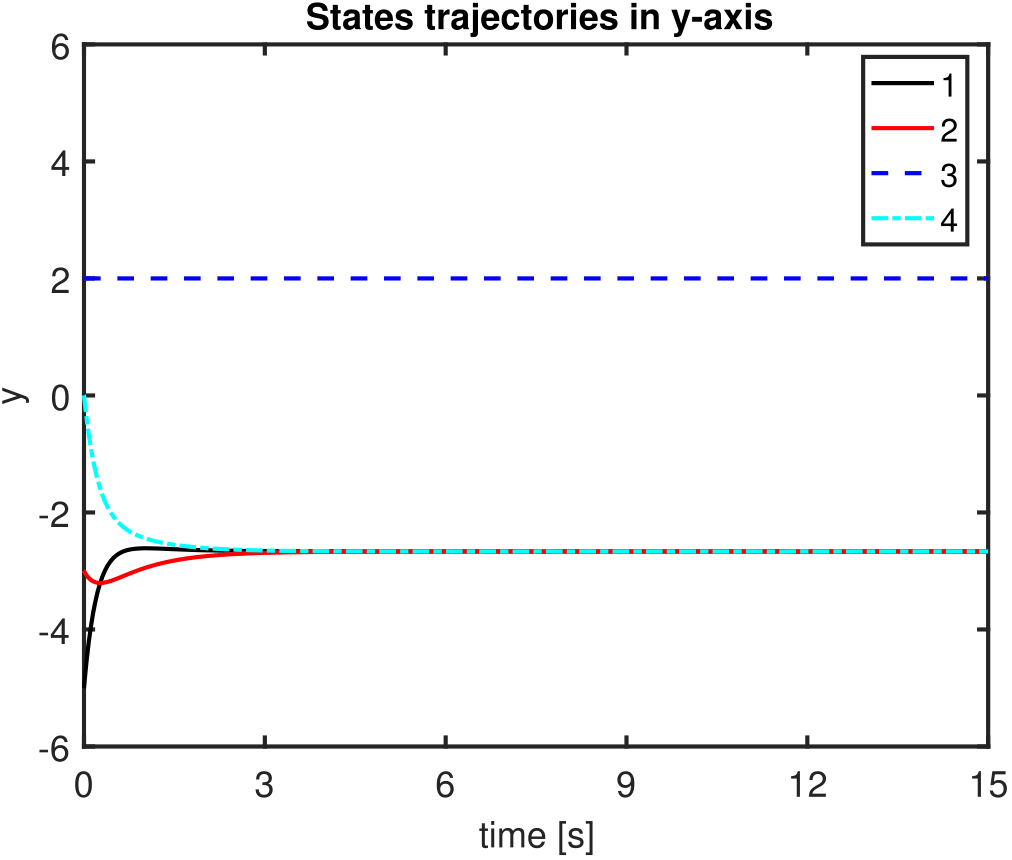}} \hfill
    \subfloat[Thay đổi theo trục $z$.]{\includegraphics[height=4cm]{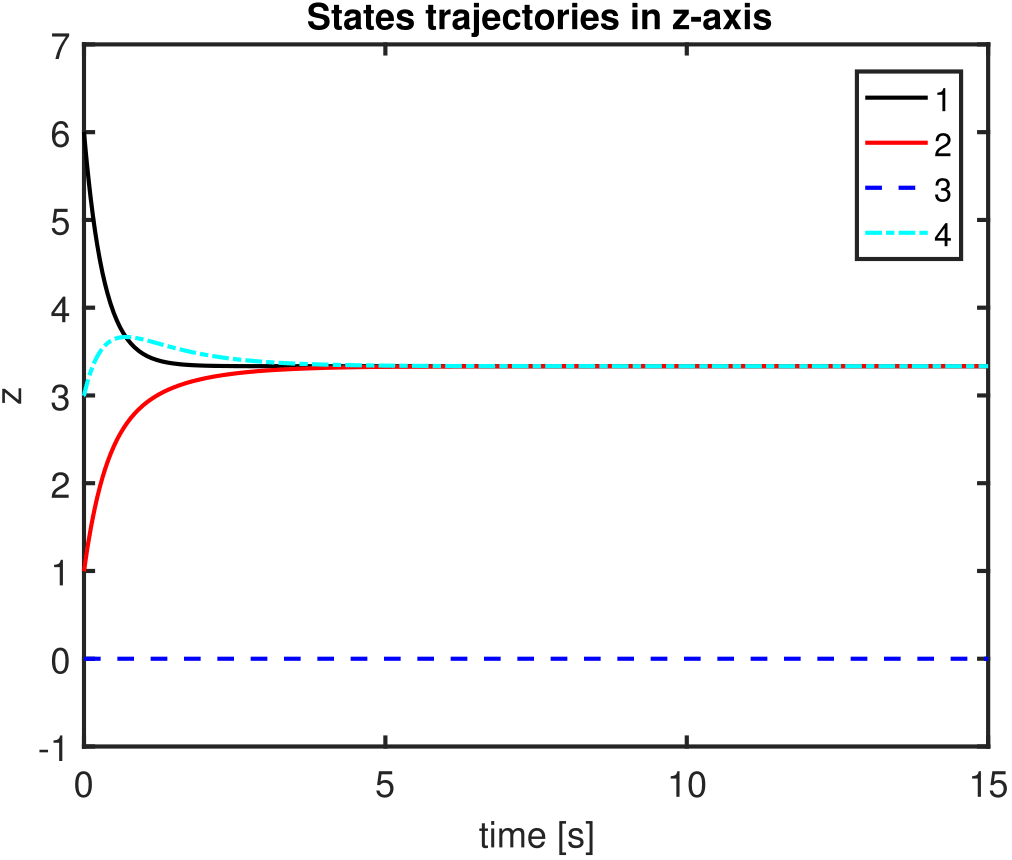}}
    \caption{Mô phỏng sự thay đổi của các biến trạng thái của hệ theo thời gian với luật đồng thuận \eqref{eq:MWC} ở Ví dụ \ref{ex:MWC1}.}
    \label{fig:sim_MWC_ex}
\end{figure}

Hai ví dụ dưới đây minh họa cho các kết quả được trình bày ở mục này. 
\begin{example}\label{ex:MWC1} Để minh họa Định lý \ref{thm:cluster}, xét hệ gồm 4 tác tử trong $\mb{R}^{3}$ như trên Hình~\ref{fig:sim_MWC_ex}. Các trọng số ma trận của đồ thị được cho bởi 
\[\m{A}_{12}=\begin{bmatrix} 0&0&0\\0&1&0\\0&0&1 \end{bmatrix},~ 
\m{A}_{13}=\begin{bmatrix} 1&0&0\\0&0&0\\0&0&0 \end{bmatrix},~\m{A}_{23}=\begin{bmatrix} 1&0&0\\0&0&0\\0&0&1 \end{bmatrix}, \text{ và } \m{A}_{14} = \begin{bmatrix} 1&0&0\\0&2&0\\0&0&1 \end{bmatrix}.\]

Dễ thấy $\m{A}_{14}$ là ma trận xác định dương còn các ma trận khác là bán xác định dương. Từ đây, ta xác định được một cây xác định dương $T$ chứa đỉnh $v_1$ và $v_4$ của đồ thị. Hơn nữa, $\text{ker}(\m{A}_{12}) = \text{span} \{ [1,0,0]^\top \}$, $\text{ker}(\m{A}_{13}) = \text{span}\{ [0,1,0]^\top, [0,0,1]^\top\}$, và $\text{ker}(\m{A}_{23}) = \text{span}\{ [0,1,0]^\top\}$.

Tồn tại hai đường đi từ $v_2$ tới $v_1$ (hay cũng là tới cây $T$) là $P_1 = v_2v_1$  và  ${P}_2 = v_2v_3v_1$. Từ định nghĩa ta có:
\begin{align*}
\text{ker}({P}_1) &= \text{ker}(\m{A}_{12}) = \text{span}\{ [1,0,0]^\top \},\\
\text{ker}({P}_2) &=  \text{ker}(\m{A}_{13}) \bigcup \text{ker}(\m{A}_{23}) = \text{span} \{ [0,1,0]^\top, [0,0,1]^\top \}.
\end{align*}
Do đó,  $\text{ker}(P_1) \bigcap \text{ker}(P_2) = \{ \m{0} \}$, tức là hai tác tử thuộc về cùng một cụm $\mc{C}(T)$ theo Định lý~\ref{thm:cluster} (ii).

Đối với đỉnh $v_3$, tồn tại 2 cạnh (đường đi) từ đỉnh 3 tới $\mc{C} = \{v_1, v_2, v_4\}$ là  $P_3 = v_3v_1$ và $P_4 = v_3v_2$. Do $\text{ker}(P_3) \bigcap \text{ker}(P_4) = \text{ker}(\m{A}_{23}) = \text{span} \{ [0,1,0]^\top \}$, đỉnh $v_3$ không thuộc $\mc{C}$. 

Mô phỏng luật đồng thuận \eqref{eq:MWC} được cho ở Hình~\ref{fig:sim_MWC_ex}. Kết quả cho thấy  $\m{x}_1^* = \m{x}_2^* = \m{x}_4^* \neq \m{x}_3^*$, giống như đã phân tích ở trên.
\end{example}

\begin{example}\label{ex:MWC2} Ta minh họa một trường hợp chỉ xuất hiện ở thuật toán đồng thuận trọng số ma trận. Xét hệ gồm năm tác tử trong $\mb{R}^2$ với đồ thị trọng số ma trận $G$ cho trên Hình Fig.~\ref{fig:MWC_ex2}. Các trọng số ma trận được cho bởi $$\m{A}_{12} = \m{A}_{24}= \begin{bmatrix}
1 & 0\\0 &0
\end{bmatrix}, \m{A}_{13} = \m{A}_{34}= \m{A}_{35} = \begin{bmatrix}
0 & 0\\0 &1
\end{bmatrix},~\text{và }\m{A}_{25} = \begin{bmatrix}
1 & -1\\-1 &1
\end{bmatrix}.$$
Do các trọng số đều là các ma trận bán xác định dương, không tồn tại cây bao trùm xác định dương nào trong $G$. Sử dụng Định lý~\ref{thm:cluster}, ta có thể tính được $\text{ker}(\m{A}_{12}) = \text{ker}(\m{A}_{24}) = \text{span}\{[1,0]^\top\}$, $\text{ker}(\m{A}_{13}) = \text{ker}(\m{A}_{34}) = \text{ker}(\m{A}_{35})= \text{span}\{[0,1]^\top\}$, và $\text{ker}(\m{A}_{25})= \text{span}\{[1,1]^\top\}$. Tồn tại 4 đường đi từ $v_4$ tới $v_1$ như sau:
\begin{itemize}
\item $P_1 = v_4v_2v_1$: $\text{ker}(P_1) = \text{ker}(\m{A}_{24}) \cup \text{ker}(\m{A}_{12}) = \text{span}\{[1,0]^\top\}$.
\item $P_2 = v_4v_3v_1$: $\text{ker}(P_2) = \text{ker}(\m{A}_{34}) \cup \text{ker}(\m{A}_{13}) = \text{span}\{[0,1]^\top\}$.
\item $P_3 = v_4v_2v_5v_3v_1$: $\text{ker}(P_3) = \text{ker}(\m{A}_{24})\cup \text{ker}(\m{A}_{25}) \cup \text{ker}(\m{A}_{12})\cup \text{ker}(\m{A}_{35}) = \mb{R}^2$.
\item $P_4 = v_4v_3v_5v_3v_1$: $\text{ker}(P_4) = \text{ker}(\m{A}_{34})\cup \text{ker}(\m{A}_{35}) \cup \text{ker}(\m{A}_{25})\cup \text{ker}(\m{A}_{12}) = \mb{R}^2$.
\end{itemize}
Do đó $\cap_{k=1}^4P_k = \text{span}\{[1,0]^\top\} \cap \text{span}\{[0,1]^\top\} = \emptyset$, hay $\text{dim}(\cap_{k=1}^4P_k) = 0$. Từ định lý \ref{thm:cluster}, ta thấy rằng $v_1$ và $v_4$ ở cùng một cụm  $\mc{C}_1$. Kiểm tra có thể thấy được $v_5$ không thuộc cụm này. Từ đây ta kết luận rằng đồ thị gồm bốn cụm: $\mc{C}_1 = \{v_1,v_4\}, \mc{C}_2 = \{v_2\}, \mc{C}_3 = \{v_3\}$, và $\mc{C}_4 = \{v_5\}$.

Trong ví dụ này, mặc dù hai đỉnh 2 và 3 đều là láng giềng của đỉnh 1, chúng không cùng thuộc một cụm với đỉnh 1. Trong khi đó, đỉnh  4 ở cùng một cụm với đỉnh 1 mặc dù chúng không có một liên kết trực tiếp nào. Kết quả mô phỏng của hệ năm tác tử với thuật toán đồng thuận \eqref{eq:MWC} được cho trong Hình~\ref{fig:sim_MWC_ex2}.
\end{example}

\begin{figure}[t]
\centering
\begin{tikzpicture}[
roundnode/.style={circle, draw=black, thick, minimum size=2mm,inner sep= 0.3mm},
squarednode/.style={rectangle, draw=black, thick, minimum size=3.5mm,inner sep= 0.25mm},
]
    \node[ellipse, minimum width = 1.65cm, minimum height = 1.4cm, align=center, draw = red, dashed, thick, fill=red!25, fill opacity=0.3] (e1) at (0.45,0) {}; 

    \node[roundnode]  (u1)   at  (0,0) {1};     %
    \node[roundnode]  (u2)   at  (1.5,1) {2};
    \node[roundnode]  (u3)   at  (1.5,-1) {3};
    \node[roundnode]  (u5)   at  (3,0) {5};
    \node[roundnode]  (u4)   at  (0.75,0) {4};     

    \draw [draw = black, thick] (u1)--(u2)--(u3)--(u4)--(u2)--(u5)--(u3)--(u1);

    \node at (0.16,0.37) {\color{red} $\mc{C}$};     %
\end{tikzpicture}
\caption{Đồ thị gồm 5 đỉnh ở Ví dụ~\ref{ex:MWC2}. $v_1,v_4$ ở cùng một cụm mặc dù bị ngăn cách bởi các đỉnh không thuộc cùng cụm $v_2,v_3$.}
\label{fig:MWC_ex2}
\end{figure}
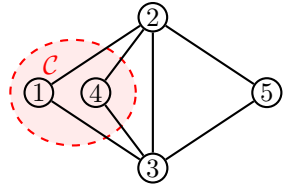
\begin{figure}[t]
\centering
    \subfloat[Thay đổi theo trục $x$]{\includegraphics[width=0.345\textwidth]{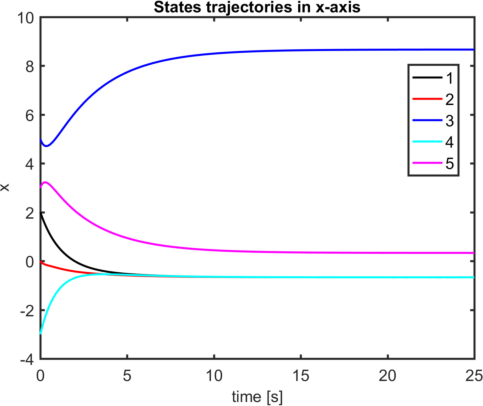}} \hfill
    \subfloat[Thay đổi theo trục $y$]{\includegraphics[width=0.345\textwidth]{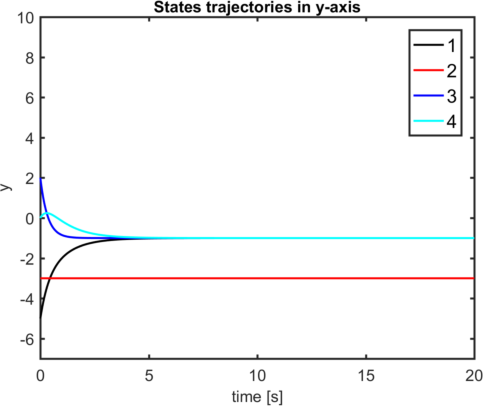}} \hfill
    \subfloat[Quỹ đạo của các tác tử.]{\includegraphics[width=0.28\textwidth]{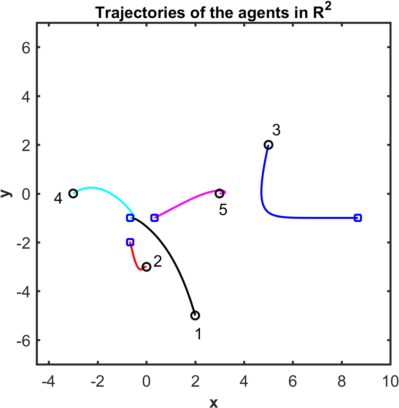}}
    \caption{Ví dụ \ref{ex:MWC2}: Thay đổi của biến trạng thái của hệ với luật đồng thuận  \eqref{eq:MWC}.}
    \label{fig:sim_MWC_ex2}
\end{figure}

\section{Đồng thuận trọng số ma trận với hệ leader-follower}
\subsection{Trường hợp leader không cập nhật trạng thái}
Xét trường hợp trong hệ có một số tác tử đặc biệt không cập nhật thông tin nhờ luật đồng thuận \eqref{eq:MWC} mà giữ nguyên biến trạng thái của mình. Ta gọi các tác tử này là các leader của hệ, và không mất tính tổng quát, kí hiệu các tác tử này bởi $i=1,\ldots,l$ với $l\geq 1$. Các tác tử khác cập nhật biến trạng thái theo \eqref{eq:MWC} và gọi là các follower.

Giả sử đồ thị $G$ mô tả tương tác giữa các tác tử trong hệ là một đồ thị trọng số ma trận thỏa mãn $ \text{ker} (\mcl{L} ) = \text{im} (\m{1}_n \otimes \m{I}_d)$. Với sự phân chia đồ thị thành các leader và các follower, ta có thể viết lại ma trận Laplace trọng số ma trận dưới dạng
\begin{align}
    \mcl{L} = \begin{bmatrix}
    \mcl{L}_{ll} & \mcl{L}_{lf}\\
    \mcl{L}_{fl} & \mcl{L}_{ff}
    \end{bmatrix},
\end{align}
trong đó $\mcl{L}_{ll} \in \mb{R}^{dl\times dl}$, $\mcl{L}_{ff} \in \mb{R}^{df\times df}$, $\mcl{L}_{lf} = \mcl{L}_{lf}^\top \in \mb{R}^{dl\times df}$, và $f = n-d$. 

Giả sử $G'$ là đồ thị trọng số ma trận mô tả tương tác giữa các tác tử follower và kí hiệu ma trận Laplace của $G'$ là $\mcl{L}'$ thì ta có:
\begin{equation}
    \mcl{L}_{ff} = \mcl{L}' + \text{blkdiag}(\mcl{L}_{fl}(\m{1}_l \otimes \m{I}_d))
\end{equation}
Giả sử $\text{ker}(\mcl{L}')=\text{im}(\m{1}_f\otimes \m{I}_d)$. Ta chứng minh rằng $\mcl{L}_{ff}$ là đối xứng, xác định dương khi và chỉ khi ma trận $(\m{1}_f \otimes \m{I}_d)^\top \mcl{L}_{fl}(\m{1}_l \otimes \m{I}_d)$ là đối xứng, xác định dương.

(Điều kiện đủ) Giả sử  $(\m{1}_f \otimes \m{I}_d)^\top \mcl{L}_{fl}(\m{1}_l \otimes \m{I}_d)$ là đối xứng, xác định dương thì do $\mcl{L}_{ff} = \mcl{L}' + \text{blkdiag}(\mcl{L}_{fl}(\m{1}_l \otimes \m{I}_d))$, với một vector $\m{w} \in \mb{R}^{df}$ bất kỳ, ta có
\begin{align} \label{eq:MWC_LF_eigen_vector}
    \m{w}^\top \mcl{L}_{ff} \m{w} = \m{w}^\top\mcl{L}'\m{w} + \m{w}^\top \mcl{L}_{fl}\text{blkdiag}(\m{1}_l \otimes \m{I}_d) \m{w}
\end{align}
Giả sử $\m{w} \in \text{ker}(\mcl{L}_{ff})$ thì do $\mcl{L}'$ và $\text{blkdiag}(\mcl{L}_{fl}(\m{1}_l \otimes \m{I}_d))$ đều là các ma trận đối xứng, bán xác định dương nên \[\m{w} \in \text{ker}(\mcl{L}') \bigcap \text{ker}(\text{blkdiag}(\mcl{L}_{fl}(\m{1}_l \otimes \m{I}_d))).\] Từ đây suy ra $\m{w} \in \text{im}(\m{1}_f\otimes \m{I}_d)$, tức là $\m{w} = \m{1}_f \otimes \bmm{\omega} = (\m{1}_f \otimes \m{I}_d)\bmm{\omega}$ với $\bmm{\omega} \neq \m{0}_d$ là một vector trong $\mb{R}^d$. Từ đây suy ra 
\begin{align*}
    0 &= \m{w}^\top \mcl{L}_{fl}\text{blkdiag}(\m{1}_l \otimes \m{I}_d) \m{w} \\
      &= \bmm{\omega}^\top (\m{1}_f \otimes \m{I}_d)^\top \text{blkdiag}(\mcl{L}_{fl}(\m{1}_l \otimes \m{I}_d)) (\m{1}_f \otimes \m{I}_d) \bmm{\omega}\\
      &= \bmm{\omega}^\top (\m{1}_f \otimes \m{I}_d)^\top \mcl{L}_{fl}(\m{1}_l \otimes \m{I}_d) \bmm{\omega}\\
      &= \bmm{\omega}^\top \Big(\sum_{j=1}^l\sum_{i=l+1}^{n} \m{A}_{ij} \Big) \bmm{\omega}.
\end{align*}
Do $(\m{1}_f \otimes \m{I}_d)^\top \mcl{L}_{fl}(\m{1}_l \otimes \m{I}_d) = \sum_{j=1}^l\sum_{i=l+1}^{n} \m{A}_{ij} $ là đối xứng và xác định dương, nghiệm duy nhất thỏa mãn là $\bmm{\omega} = \m{0}_d$. Điều này tương đương với việc $\mcl{L}_{ff}$ là đối xứng, xác định dương.

(Điều kiện cần) Giả sử  $(\m{1}_f \otimes \m{I}_d)^\top \mcl{L}_{fl}(\m{1}_l \otimes \m{I}_d)$ không là đối xứng, xác định dương. Trong phương trình \eqref{eq:MWC_LF_eigen_vector}, chọn $\m{w} = \m{1}_f \otimes \bmm{\omega}$ với $\bmm{\omega}$ thỏa mãn $(\m{1}_f \otimes \m{I}_d)^\top \mcl{L}_{fl}(\m{1}_l \otimes \m{I}_d)\bmm{\omega} = \m{0}_d$ thì $\m{w} \neq \m{0}_d$ là một vector riêng ứng với giá trị riêng bằng 0 của $\mcl{L}_{ff}$. Điều này chứng tỏ  $\mcl{L}_{ff}$ không xác định dương.

Hệ đồng thuận dạng leader-follower có thể viết dưới dạng:
\begin{align}
    \dot{\m{x}}_i &= \m{0}_d,~i=1, \ldots, l,\\
    \dot{\m{x}}_i &= - \sum_{j=1}^n \m{A}_{ij}(\m{x}_i - \m{x}_j),~i=l+1, \ldots, n.
\end{align}
Đặt $\m{x} = [m{x}_1^\top,\ldots,\m{x}_n^\top]^\top$, $\m{x}_L = [\m{x}_1^\top,\ldots,\m{x}_{\rm F}^\top]^\top$, $\m{x}_{\rm F} = [\m{x}_{l+1}^\top,\ldots,\m{x}_{n}^\top]^\top$, ta  biểu diễn hệ dưới dạng 
\begin{align} \label{eq:MWC_LF}
    \begin{bmatrix}
        \dot{\m{x}}_L\\
        \dot{\m{x}}_F
    \end{bmatrix} = - \begin{bmatrix}
    \m{0}_{dl\times dl} & \m{0}_{dl\times df}\\ \mcl{L}_{fl} & \mcl{L}_{ff}
    \end{bmatrix}
    \begin{bmatrix}
    \m{x}_L\\ \m{x}_F
    \end{bmatrix}.
\end{align}
Dễ thấy điểm cân bằng của hệ cần thỏa mãn
\begin{align}
\m{x}_L^* &= \m{x}_L(0), \nonumber \\
    \mcl{L}_{fl}\m{x}_L^* + \mcl{L}_{ff} \m{x}_F^* &= \m{0}_{df}.
\end{align}
Với điều kiện $\mcl{L}_{ff}$ là đối xứng, xác định dương thì 
\begin{align}
    \m{x}_F^* = -\mcl{L}_{ff}^{-1}\mcl{L}_{fl}\m{x}_L(0).
\end{align}
Chúng ta sẽ chứng minh rằng các tác tử sẽ dần tiệm cận tới điểm cân bằng duy nhất $\m{x}^* = [\m{x}_L^\top(0), -\mcl{L}_{ff}^{-1}\mcl{L}_{fl}\m{x}_L^\top(0)]^\top$ này. Thật vậy, với phép đổi biến $\tilde{\m{x}}_F=\m{x}_F - \m{x}_F^*$ thì từ phương trình \eqref{eq:MWC_LF}, ta thu được:
\begin{align}
    \dot{\tilde{\m{x}}}_F &= - \mcl{L}_{fl}\m{x}_L - \mcl{L}_{ff} \m{x}_F \nonumber\\
    &=  \mcl{L}_{ff} \m{x}_F^*- \mcl{L}_{ff} \m{x}_F \nonumber \\
    &= - \mcl{L}_{ff} {\tilde{\m{x}}}_F.
\end{align}
Do ma trận $-\mcl{L}_{ff}$ là đối xứng, xác định âm nên ta suy ra ${\tilde{\m{x}}}_F = \m{0}_{df}$ là ổn định theo hàm mũ. Như vậy, $\m{x}_F(t) \to \m{x}_F^*$ khi $t \to +\infty$. 

Xét trường hợp đặc biệt, các leader có cùng điều kiện đầu $\m{x}_1(0) = \ldots = \m{x}_l(0) = \m{x}^*$. Lúc này, ta viết lại $\m{x}_L = \m{1}_l \otimes \m{x}_1(0)$. Mặt khác, do $\mcl{L}_{ff} (\m{1}_f\otimes \m{x}_l)= \mcl{L}' (\m{1}_f\otimes \m{x}_l) + \text{blkdiag}(\mcl{L}_{fl}(\m{1}_l \otimes \m{I}_d)) (\m{1}_f\otimes \m{x}_l) = \mcl{L}_{fl} (\m{1}_l\otimes \m{x}_l)$ nên $\m{x}_F^* = -\mcl{L}_{ff}^{-1}\mcl{L}_{fl}(\m{1}_l\otimes \m{x}_l)= \m{1}_f\otimes \m{x}_l$. Do đó, trong trường hợp này, các follower sẽ dần đạt tới đồng thuận với leader. 

Để ý rằng $\m{1}_f\otimes \m{I}_d = -\mcl{L}_{ff}^{-1} \mcl{L}_{fl} (\m{1}_l\otimes \m{I}_d)$ nên tổng các $d\times d$ hàng của ma trận $\m{R}=-\mcl{L}_{ff}^{-1} \mcl{L}_{fl} \in \mb{R}^{df \times dl}$ đều bằng ma trận đơn vị $\m{I}_d$. Do đó, trong trường hợp tổng quát thì
\begin{align}
    \m{x}_i^* =  \sum_{j=1}^l \m{R}_{ij} \m{x}_{j}(0),
\end{align}
với $\sum_{j=1}^l\m{R}_{ij} = \m{I}_d$. Nếu như đồ thị là vô hướng với các trọng số hằng, ma trận $-\mcl{L}^{-1}_{ff}\mcl{L}_{fl}$ là một ma trận với mọi phần tử không âm. Các tác tử sẽ hội tụ tới bao lồi của các điểm $\m{x}_1(0), \ldots, \m{x}_l(0)$. Trong trường hợp trọng số ma trận, điểm cân bằng của các tác tử follower có thể nằm trong hoặc ngoài bao lồi của vị trí các tác tử leader.

\subsection{Trường hợp leader di chuyển}
Ta xét trường hợp các leader chuyển động với vận tốc hằng, $\dot{\m{x}}_i = \m{v}_L, ~i=1,\ldots, l$ và $\dot{\m{v}}_L = \m{0}_d$. Luật đồng thuận \eqref{eq:MWC_LF} lúc này không đưa hệ về trạng thái đồng thuận. Thay cho \eqref{eq:MWC_LF}, các follower sử dụng luật đồng thuận dạng PI cho bởi:
\begin{align}
    \dot{\m{x}}_i      &= - \sum_{j=1}^n \m{A}_{ij}(\m{x}_i - \m{x}_j) + \bmm{\eta}_i, \\
    \dot{\bmm{\eta}}_i &= - \sum_{j=1}^n \m{A}_{ij}(\m{x}_i - \m{x}_j),~i=l+1, \ldots, n.
\end{align}
Viết lại hệ với biến trạng thái $\m{r}_i=[\tilde{\m{x}}_i^\top,\bm{\eta}_i^\top]^\top \in \mb{R}^{2d}$ và $\m{r} = [\m{r}_{l+1}^\top, \ldots, \m{r}_n^\top]^\top \in \mb{R}^{df}$, trong đó $\tilde{\m{x}}_i = \m{x}_i - \m{x}_i^*$, và
\begin{align}
    \m{x}_F^*(t) &= -\mcl{L}_{ff}^{-1}\mcl{L}_{fl}\m{x}_L(t), \\
    \dot{\m{x}}_F^*(t) &= -\mcl{L}_{ff}^{-1}\mcl{L}_{fl} \dot{\m{x}}_L(t) = -\mcl{L}_{ff}^{-1}\mcl{L}_{fl} (\m{1}_l \otimes \m{v}_L) = \m{1}_f \otimes \m{v}_L,
\end{align}
thì
\begin{align}
    \dot{\m{r}} &= \begin{bmatrix}
    -\mcl{L}_{ff} & \m{I}_{df}\\
    -\mcl{L}_{ff} & \m{0}_{df \times df}
    \end{bmatrix} \m{r} = \m{M} \m{r}.
\end{align}
Các giá trị riêng của ma trận $\m{M}$ được cho bởi
\begin{align}
    \text{det}\big( s \m{I}_{2df} - \m{M} \big) &= \text{det}\big( s(s\m{I}_{df} + \mcl{L}_{ff}) + \mcl{L}_{ff} \big) \nonumber\\
    &= \prod_{k=1}^{df} (s^2 + \mu_k s + \mu_k) = 0, \label{eq:MWC_Lff_PI}
\end{align}
trong đó $\mu_k > 0,~\forall k = l+1, \ldots, df$ là các giá trị riêng của ma trận $\mcl{L}_{ff}$. Dễ thấy mọi nghiệm của \eqref{eq:MWC_Lff_PI} đều phải có phần thực âm, điều này chứng tỏ ma trận $\m{M}$ là Hurwitz. Từ đây suy ra, $\m{x}_i(t) \to \m{x}_i^*(t)$ và $\m{v}_i(t) \to \m{v}_L(t)$ khi $t \to +\infty$.

Trong trường hợp vận tốc của leader thay đổi theo thời gian, luật điều khiển PI không đưa hệ về đồng thuận. Một trong các phương pháp có thể sử dụng để thiết kế luật đồng thuận dựa trên điều khiển trượt \cite{Nguyen2022leaderless} và sẽ là nội dung của Bài tập \ref{ex:9.11}.

\section{Ghi chú và tài liệu tham khảo}
Những kết quả trình bày ở chương này phần lớn được trình bày từ \cite{Trinh2017ASCC,Trinh2018matrix}. Một số phát triển lý thuyết gần đây tập trung vào khía cạnh điều khiển của hệ trọng số ma trận: tính điều khiển được và quan sát được \cite{Pan2020controllability}, đồng thuận trọng số ma trận bậc hai \cite{Miao2021second,Nguyen2021VCCA}, đồng thuận trọng số ma trận không liên tục \cite{Tran2021discrete}, đồ thị thay đổi theo thời gian \cite{Pan2021consensus}, hay có trễ \cite{Pham2021delay}. Vấn đề tổng hợp đồ thị trọng số ma trận xác định dương với chất lượng cho trước được nghiên cứu trong \cite{De2020H2,foight2020performance}. Tuy nhiên, hiện chưa có nhiều kết quả về lý thuyết đồ thị và thuật toán trên đồ thị trọng số ma trận. 

\section{Bài tập}
\label{sec:c9_exercise}

\begin{exercise}\label{ex:9.1} Hãy viết các ma trận kề trọng số ma trận, ma trận bậc trọng số ma trận, và ma trận Laplace trọng số ma trận cho đồ thị trên Hình \ref{fig:c9_ex9.1}, với 
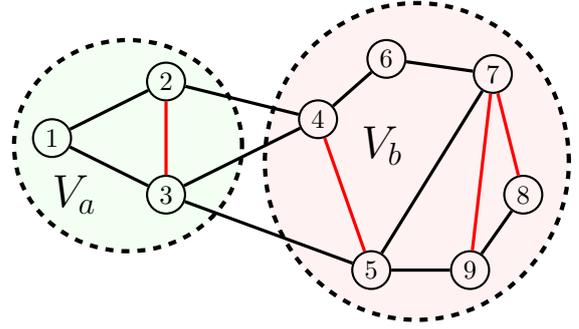
\begin{SCfigure}[][t!]
\caption{Đồ thị trọng số ma trận trong Bài tập \ref{ex:9.1} và \ref{ex:9.2}}
\label{fig:c9_ex9.1}
\hspace{4cm}
\begin{tikzpicture}[
roundnode/.style={circle, draw=black, thick, minimum size=5 mm,inner sep= 0.35mm},
squarednode/.style={rectangle, draw=black, thick, minimum size=5 mm,inner sep= 0.35 mm},
]
    \node[draw = black, dashed, ultra thick, ellipse, minimum width = 3cm, minimum height = 2.8 cm, align=center, fill=green!5] (e1) at (2,0.65) {};
    \node[draw = black, dashed, ultra thick, ellipse, minimum width = 4cm, minimum height = 4.2 cm, align=center, fill=red!5] (e1) at (5.8,0.44) {};
    \node[roundnode]   (u1)   at  (1,0.75) {$1$};     %
    \node[roundnode]   (u2)   at  (2.5,1.5) {$2$};     %
    \node[roundnode]   (u4)   at  (4.5,1) {$4$};     %
    \node[roundnode]   (u3)   at  (2.5,0) {$3$};    %
    \node[roundnode]   (u5)   at  (5.2,-1) {$5$}; %
    \node[roundnode]   (u6)   at  (5.4,1.8) {$6$};%
    \node[roundnode]   (u7)   at  (6.8,1.6) {$7$};%
    \node[roundnode]   (u8)   at  (7.2,0) {$8$};%
    \node[roundnode]   (u9)   at  (6.5,-1){$9$};%
    
	\node (ux) at (1.2,0) {\begin{LARGE} $V_a$ \end{LARGE}};
    \node (ux1) at (5.25,0.65) {\begin{LARGE} $V_b$ \end{LARGE}};
    \draw [very thick]
    (u1) edge [bend left=0] (u2)
    (u1) edge [bend left=0] (u3)
    (u2) edge [bend left=0, draw = red] (u3)
    (u3) edge [bend right=0] (u5)
    (u4) edge [bend right=0] (u6)
    (u5) edge [bend right=0] (u7)
    (u6) edge [bend right=0] (u7)
    (u7) edge [bend right=0, draw = red] (u8)
    (u7) edge [bend right=0, draw = red] (u9)
    ;
    \draw [draw = black, very thick]
    (u3) edge [bend left=0] (u4)
    (u2) edge [bend left=0] (u4)
    (u4) edge [bend left=0, draw = red] (u5)
    (u8) edge [bend right=0] (u9)
    (u9) edge [bend right=0] (u5)
    ; 
\end{tikzpicture}
\end{SCfigure}
\begin{align*}
\m{A}_{12} &= \m{A}_{67} = \begin{bmatrix}
1 & \frac{\sqrt{3}}{2} & 0 \\ \frac{\sqrt{3}}{2} & 1 & 0 \\ 0 & 0 & 0
\end{bmatrix},~\m{A}_{13} = \m{A}_{24} = \m{A}_{34} = \m{A}_{59} = \begin{bmatrix}
0 & 0 & 0 \\ 0 & 1 & 0 \\ 0 & 0 & 1
\end{bmatrix}, \\
\m{A}_{23} &= \begin{bmatrix}
2 & 1 & 0 \\ 1 & 1 & 0 \\ 0 & 0 & 1
\end{bmatrix},~\m{A}_{35} = \m{A}_{57} = \begin{bmatrix}
1 & 0 & 0 \\ 0 & 0 & 0 \\ 0 & 0 & 0
\end{bmatrix},~\m{A}_{45} = \begin{bmatrix}
1 & 0 & 0 \\ 0 & 2 & 0 \\ 0 & 0 & 3
\end{bmatrix}, \\
\m{A}_{78} &= \m{A}_{79} = \begin{bmatrix}
1 & 0 & 0 \\ 0 & 1 & 0 \\ 0 & 0 & 1
\end{bmatrix},~\m{A}_{46} = \m{A}_{89} = \begin{bmatrix}
1 & 1 & 1 \\ 1 & 1 & 1 \\ 1 & 1 & 1
\end{bmatrix},
\end{align*}
Đồ thị có cây bao trùm hay không? Hãy xác định hạng của ma trận Laplace trọng số ma trận tương ứng. 
\end{exercise}


\begin{exercise}\label{ex:9.2} Giả sử các đỉnh của đồ thị trọng số ma trận ở Bài tập \ref{ex:9.1} được chia làm hai tập $V_1$ (leader), $V_2$ (follower) như trên Hình \ref{fig:c9_ex9.1}. Hãy xác định các ma trận $\mcl{L}_{ll},\mcl{L}_{fl},\mcl{L}_{ff}$. Ma trận $\mcl{L}_{ff}$ có xác định dương hay không?
\end{exercise}


\begin{exercise}\label{ex:9.3} Chứng minh công thức
\begin{align}
\mcl{L} = (\m{H}^\top \otimes \m{I}_d) \text{blkdiag}(\m{A}_k) (\m{H} \otimes \m{I}_d) = \bar{\m{H}}^\top \text{blkdiag}(\m{A}_k) \bar{\m{H}}
\end{align}
cho đồ thị trọng số ma trận vô hướng, trong đó $\m{H}$ kí hiệu ma trận liên thuộc.
\end{exercise}

\begin{exercise}\label{ex:9.4} Giả sử ma trận Laplace trọng số ma trận có hạng ${\rm rank}(\mcl{L})=dn-r$, với $r>d$. Hãy xác định điểm $\m{x}^* \in {\rm ker}(\mcl{L})$ mà $\m{x}(t)$ tiệm cận tới với thuật toán đồng thuận trọng số ma trận~\eqref{eq:MWC}.
\end{exercise}

\begin{exercise}\label{ex:9.5} Hãy lấy ví dụ chứng minh rằng điều kiện trong Định lý \ref{thm:cluster} chỉ là một điều kiện đủ.
\end{exercise}

\begin{exercise}\label{ex:9.6} Giả sử ma trận Laplace trọng số ma trận có hạng ${\rm rank}(\mcl{L})=dn-d$.
\begin{itemize}
\item[i.] Hãy chứng minh rằng ma trận bậc trọng số ma trận $\m{D}$ là đối xứng xác định dương.
\item[ii.] Hãy chứng minh ma trận $\m{D}^{-1/2}\mcl{L}\m{D}^{-1/2}$ có các giá trị riêng không âm.
\item[iii.] \cite{Tran2021discrete} Hãy xác định khoảng giá trị của $\alpha$ để $\m{I}_{dn}-\alpha\m{D}^{-1/2}\mcl{L}\m{D}^{1/2}$ là ma trận có các giá trị riêng nằm trong đường tròn đơn vị.
\end{itemize}
\end{exercise}

\begin{exercise} \label{ex:9.7}
Giả sử ma trận $\m{X}=\m{X}^\top \in \mb{R}^{d \times d}$ thuộc tập $\mc{S}$ gồm các ma trận đối xứng xác định dương/âm, ma trận đối xứng bán xác định dương/âm và ma trận không. Định nghĩa hàm dấu của ma trận như sau:
\begin{align*}
{\rm sign}(\m{X}) = \left\lbrace \begin{array}{ll}
1, & \m{X} \text{ là xác định dương hoặc bán xác định dương}, \\
-1, & \m{X} \text{ là xác định âm hoặc bán xác định âm}, \\
0, & \m{X}=\m{0}.
\end{array} \right.
\end{align*}
Kí hiệu $|\m{X}|={\rm sign}(\m{X})\m{X}$. Nếu $\m{X}_1,\ldots,\m{X}_n \in \mc{S}$ thì 
\begin{itemize}
\item[i.] Tồn tại ma trận $\m{G}$ để $\m{G}{\rm blkdiag}(\m{X}_1,\ldots,\m{X}_n)$ là ma trận đối xứng, bán xác định dương.
\item[ii.] \cite{Trinh2024msc} Nếu $\mcl{L} \in \mb{R}^{dn \times dn}$ là ma trận Laplace của một đồ thị trọng số ma trận thì ma trận 
\[\m{X}\mcl{L}({\rm diag}({\rm sign}(\m{X}_1),\ldots,{\rm sign}(\m{X}_n))\otimes \m{I}_d)\] có các giá trị riêng không âm.
\end{itemize}
\end{exercise}

\begin{exercise}\cite{Pan2018bipartite} \label{ex:9.8} Xét đồ thị dấu trọng số ma trận (signed matrix-weighted Laplacian) trong đó các ma trận $\m{A}_{ij}=\m{A}_{ij}^\top \in \mb{R}^{d \times d}$ có thể là ma trận đối xứng xác định dương/âm,  ma trận đối xứng bán xác định dương/âm và ma trận không. Giả sử đồ thị không trọng số $(V,E)$ là đồ thị cân bằng cấu trúc dấu (xem định nghĩa ở Chương 8 - mô hình Altafini).
\begin{itemize}
\item[i.] Định nghĩa các ma trận $\m{A}=[\m{A}_{ij}]$, $\m{D}={\rm blkdiag}(\m{D}_1,\ldots,\m{D}_n)$ với $\m{D}_i = \sum_{j=1}^n {\rm sign}(\m{A}_{ij})\m{A}_{ij}= \sum_{i=1}^n |\m{A}_{ij}|$, $\mcl{L}=\m{D}-\m{A}$, trong đó hàm dấu của ma trận được định nghĩa ở Bài tập \ref{ex:9.6}. Chứng minh tồn tại ma trận $\m{G}$ để $\mcl{L}'=\m{G}(\m{D}-\m{A})\m{G}^{-1}$ là một ma trận đối xứng bán xác định dương.
\item[ii.] Giả sử đồ thị mô tả hệ gồm $n$ tác tử. Giả sử mỗi tác tử có vector biến trạng thái $\m{x}_i \in \mb{R}^d$, và cập nhật biến trạng thái theo công thức:
\begin{align}
\dot{\m{x}}_i = - \sum_{i=1}^n |\m{A}_{ij}|(\m{x}_i - {\rm sign}(\m{A}_{ij}) \m{x}_j),~i=1,\ldots,n.
\end{align}
Hãy biểu diễn hệ dưới dạng ma trận, với biến trạng thái $\m{x}=[\m{x}_1^\top,\ldots,\m{x}_n^\top]^\top\in\mb{R}^{dn}$. 
\item[iii.] Thực hiện phép đổi biến $\m{y}=\m{G}\m{x}$, hãy viết phương trình biểu diễn hệ theo $\m{y}$. Chứng minh rằng nếu ${\rm rank}(\mcl{L})=dn-d$ thì $\m{y}$ tiệm cận tới một điểm trong không gian đồng thuận.
\item[iv.] Từ hành vi tiệm cận của $\m{y}$, hãy kết luận về hành vi tiệm cận của $\m{x}$.
\end{itemize}
\end{exercise}

\begin{exercise}[Định lý ma trận - cây]\cite{Trinh2026} \label{ex:9.9}
Giả sử $G=(V,E,W)$ là đồ thị trọng số ma trận vô hướng gồm $n$ đỉnh với ma trận Laplace trọng số ma trận $\mcl{L}$. Gọi $0 = \lambda_1 = \ldots = \lambda_d \leq \lambda_{d+1} \leq \ldots \leq \lambda_{dn}$ là các giá trị riêng của $\mcl{L}$ and $\mcl{L}_{[v]}$ kí hiệu ma trận thu được từ ma trận Laplace sau khi xóa đi $d$ hàng $d$ cột tương ứng với đỉnh $v \in V$ bất kỳ. Chứng minh rằng: \index{định lý!ma trân-cây đồ thị trọng số ma trận}
\begin{itemize}
\item[i.] ${\rm trace}(\m{L}) = \sum_{i=d+1}^{dn}\lambda_i$,
\item[ii.] ${\rm det}(\m{L}_{[v]}) = \frac{1}{n}\prod_{i=d+1}^{dn}\lambda_i$.
\end{itemize}
\end{exercise}

\begin{exercise}[Đồ thị trọng số ma trận hữu hướng, cân bằng] \label{ex:9.10} Giả sử $G=(V,E,W)$ là đồ thị trọng số ma trận gồm $n$ đỉnh với ma trận Laplace trọng số ma trận $\mcl{L}$. Đồ thị $G$ được gọi là cân bằng tổng quát nếu ồn tại các số thực dương $p_1, \ldots, p_n$ sao cho
\begin{equation}
p_i \sum_{j=1}^n \m{A}_{ij} = \sum_{j=1}^n p_j \m{A}_{ji},  \label{eq:c9_MWC_balanced_condition}
\end{equation}
hay tồn tại vector $\m{p} = [p_1, \ldots, p_n]^\top$ với $p_i$ là các số thực dương thỏa mãn
\begin{equation}
(\m{p}^\top \otimes \m{I}_d) \mcl{L} = \m{0}_{d \times nd}.
\end{equation}
\begin{itemize}
\item[i.] Đặt $\m{P}= \text{diag}(\m{p}) \in \mb{R}^{n \times n}$ và $\bar{\m{P}} = \m{P} \otimes \m{I}_d$, chứng minh rằng ma trận
\begin{equation} \label{eq:MWCQ}
\m{Q} = \bar{\m{P}}\m{L} + \m{L}^\top \bar{\m{P}},
\end{equation}
là đối xứng, bán xác định dương.

\item[ii.] Nếu ${\rm rank}(\mcl{L})=dn-d$ thì ${\rm ker}(\mcl{L})={\rm im}(\m{1}_n\otimes\m{I}_d)$ và hệ sẽ tiệm cận tới tập đồng thuận với thuật đồng thuận:
\begin{align}
\dot{\m{x}}_i = -\sum_{j \in N_i}\m{A}_{ij}(\m{x}_i - \m{x}_j),\; i=1,\ldots,n.
\end{align}
\end{itemize}
Hãy tìm điểm đồng thuận theo giá trị đầu $\m{x}(0)$.
\item[iii.] Hãy chứng minh ý (ii) sử dụng hàm Lyapunov $V=\m{x}^\top\bar{\m{P}}\m{x}$.
\end{exercise}

\begin{exercise}\cite{Miao2021second} \label{ex:9.11}
Giả sử $G=(V,E,W)$ là đồ thị trọng số ma trận gồm $n$ đỉnh với ma trận Laplace trọng số ma trận $\mcl{L}$. Giả sử hệ gồm $n$ tác tử tích phân bậc hai với biến vị trí và vận tốc $\m{p}_i,\m{v}_i\in\mb{R}^{d}$, $i=1,\ldots,n$. Giả sử các tác tử tương tác qua đồ thị $G$, với thuật toán đồng thuận:
\begin{subequations}
\begin{align}
\dot{\m{p}}_i &= \m{v}_i,\\
\dot{\m{v}}_i &= \m{u}_i = -k_p\mcl{L}\m{p}_i - k_v\m{v}_i,\; i=1,\ldots,n.
\end{align}
\end{subequations}
\begin{itemize}
\item[i.] Hãy tìm điều kiện để vị trí của các tác tử tiệm cận tới tập đồng thuận và vận tốc của các tác tử hội tụ.
\item[ii.] Hãy phân tích hệ với các giả thiết ở ý (i) dựa trên lý thuyết Lyapunov.
\end{itemize}
\end{exercise}

\begin{exercise}\cite{Nguyen2022leaderless} \label{ex:9.12}
Giả sử $G=(V,E,W)$ là đồ thị trọng số ma trận gồm $n$ đỉnh với ma trận Laplace trọng số ma trận $\mcl{L}$. Giả sử hệ gồm $n$ tác tử tích phân bậc nhất với biến trạng thái $\m{x}_i\in\mb{R}^{d}$, $i=1,\ldots,n$. Giả sử các tác tử tương tác qua đồ thị $G$.
\begin{itemize}
\item[i.] Hãy chứng minh hệ tiệm cận tới tập đồng thuận với thuật toán:
\begin{align*}
\dot{\m{x}}_i = -\sum_{j=1}^n \m{A}_{ij} {\rm sign}\left( \m{A}_{ij}(\m{x}_i - \m{x}_j) \right),\; i=1,\ldots,n,
\end{align*}
\item[ii.] Hãy chứng minh hệ tiệm cận tới tập đồng thuận với thuật toán:
\begin{align*}
\dot{\m{x}}_i = -{\rm sign}\left(\sum_{j=1}^n \m{A}_{ij}(\m{x}_i - \m{x}_j)\right),\; i=1,\ldots,n. 
\end{align*}
\end{itemize}
\end{exercise}

\begin{exercise}\cite{Nguyen2022leaderless} \label{ex:9.13}
Giả sử $G=(V,E,W)$ là đồ thị trọng số ma trận gồm $n$ đỉnh với ma trận Laplace trọng số ma trận $\mcl{L}$. Giả sử hệ gồm $n$ tác tử tích phân bậc hai với biến vị trí và vận tốc $\m{p}_i,\m{v}_i\in\mb{R}^{d}$, $i=1,\ldots,n$. Giả sử các tác tử tương tác qua đồ thị $G$, với thuật toán đồng thuận:
\begin{subequations}
\begin{align}
\dot{\m{p}}_i &= \m{v}_i,\\
\dot{\m{v}}_i &= \m{u}_i = -k{\rm sgn}(\m{s}_i)-\sum_{j=1}^n\m{A}_{ij}(\m{p}_i - \m{p}_j) - \lambda {\rm sig}^{\alpha}(\m{e}_i),\\
\m{e}_i &= \m{v}_i + \sum_{j \in N_i}\m{A}_{ij}(\m{x}_i - \m{x}_j), \\
\m{s}_i &= \m{e}_i + \lambda \int_0^t{\rm sig}^{\alpha}(\m{e}_i(\tau))d\tau,
\; i=1,\ldots,n,
\end{align}
\end{subequations}
với ${\rm sig}^{\alpha}(x)={\rm sgn}(x)|x|^{\alpha}$. 
\begin{itemize}
\item[i.] Hãy biểu diễn hệ ở dạng ma trận.
\item[ii.] Chứng minh $\m{s}_i \to \m{0}$ và $\m{e}\to \m{0}$ trong thời gian hữu hạn, từ đó suy ra hệ tiệm cận tới tập đồng thuận.
\end{itemize}
\end{exercise}

\begin{exercise}[Bộ quan sát nhiễu] \label{ex:9.14}
Giả sử $G=(V,E,W)$ là đồ thị trọng số ma trận gồm $n$ đỉnh với ma trận Laplace trọng số ma trận $\mcl{L}$. Giả sử hệ gồm $n$ tác tử tích phân bậc hai với biến vị trí và vận tốc $\m{p}_i,\m{v}_i\in\mb{R}^{d}$, $i=1,\ldots,n$:
\begin{align}
\dot{\m{p}}_i &= \m{v}_i + \m{Q}_i(t)\m{d}_i,\\
\dot{\m{v}}_i &= \m{w}_i, \; i=1,\ldots,n,
\end{align}
trong đó $\m{Q}_i\m{d}_i$ là thành phần nhiễu không bù trực tiếp được bởi tín hiệu điều khiển $\m{w}_i$ của mỗi tác tử. Thông tin về nhiễu gồm hai phần: $\m{Q}_i(t)\in \mb{R}^{d \times r}$ là ma trận thay đổi theo thời gian và giả sử là đã biết (ma trận hồi quy), còn $\m{d}_i \in \mb{R}^r$ là vector tham số chưa biết bởi mỗi tác tử.
\begin{itemize}
\item[i.] Giả sử mỗi tác tử thực hiện bộ quan sát nhiễu:
\begin{align}
\dot{\m{z}}_i &= -\m{Q}_i^\top\left(\m{Q}_i(\m{z}_i + \m{Q}_i^\top\m{p}_i) + \m{v}_i \right) - \dot{\m{Q}}_i^\top\m{p}_i, \\
\hat{\m{d}}_i &= \m{z}_i + \m{Q}_i^\top\m{p}_i,\; i=1,\ldots,n.
\end{align}
Hãy chứng minh rằng bộ quan sát có khả năng ước lượng nhiễu. Nói cách khác, chứng minh $\m{e}_i=\hat{\m{d}}_i - \m{d}$ hội tụ.
\item[ii.] Đặt $\m{s}_i=\m{v}_i + \gamma \sum_{j=1}^n \m{A}_{ij}(\m{p}_i - \m{p}_j) + \m{Q}_i \hat{\m{d}}_i$ với $\gamma>0$ là một hằng số dương. Giả sử các tác tử cập nhật biến trạng thái theo luật đồng thuận kết hợp bộ quan sát nhiễu:
\begin{subequations}\label{eq:DO_MWC}
\begin{align}
\m{w}_i &= -\m{s}_i - \gamma \sum_{j=1}^n \m{A}_{ij}(\m{p}_i - \m{p}_j) - \beta_i(t) {\rm sgn}(\m{s}_i),\\
\dot{\beta}_i(t) &= \gamma_i \|\m{s}_i\|_1,\; i=1,\ldots,n.
\end{align}
\end{subequations}
với $\|\m{s}_i\|_1 = \sum_{k=1}^d |s_{ik}|$, $\gamma_i>0$ là một hằng số dương. Chứng minh $\m{s}_i \to \m{0}$ và $\lim_{t\to \infty} \beta_i{t}$ tồn tại và bị chặn.
\item[iii.] Chứng minh $\m{x}(t) \to {\rm ker}(\m{L})$ khi $t\to +\infty$ với thuật toán \eqref{eq:DO_MWC}. 
\end{itemize}
\end{exercise}

\appendix 

\part{Phụ lục}
\chapter{Đại số tuyến tính}
\label{append:matrix_theory}
Các kết quả trong phụ lục về đại số tuyến tính này có thể tham khảo tại các tài liệu tiếng Anh \cite{Strang1988,NicholsonLAA} hoặc tiếng Việt . 

\section{Một số định nghĩa và phép toán cơ bản}
Ma trận $\m{A} \in \mb{C}^{m\times n}$ bao gồm  $mn$ phần tử $a_{ij} ~(i=1,\ldots,m; j=1,\ldots,n)$ được  viết thành $m$ hàng và $n$ cột dưới dạng
\begin{align*}
\m{A} = \begin{bmatrix}
a_{11} & a_{12} & \ldots & a_{1n}\\
a_{21} & a_{22} & \ldots & a_{2n}\\
\vdots & \vdots & & \vdots\\
a_{m1} & a_{m2} & \ldots & a_{mn}
\end{bmatrix}.
\end{align*}
Ta cũng sử dụng kí hiệu $\m{A} = (a_{ij})_{m\times n}$. Một vector cột gồm $n$ phần tử là một ma trận với $n$ hàng với \[\m{x} = \begin{bmatrix}
x_{1}\\x_2\\ \vdots \\ x_n
\end{bmatrix}.\]

Với $\m{A}$ và $\m{B}$ là hai ma trận có cùng kích thước, ta có thể định nghĩa các phép toán cộng, trừ ma trận và nhân với một đại lượng vô hướng. Ma trận
\[\m{C} = k_1\m{A} + k_2\m{B},\]
có các phần tử $c_{ij} = k_1 a_{ij} + k_2 b_{ij}$, còn phép trừ ma trận $\m{A} -\m{B} = \m{A} + (-\m{B})$. Các phép cộng ma trận thỏa mãn tính kết hợp và giao hoán.
\begin{align*}
    \m{A} + (\m{B} + \m{C}) &= (\m{A} + \m{B}) + \m{C}\\
    \m{A} + \m{B} &= \m{B} + \m{A}
\end{align*}

Phép nhân hai ma trận $\m{A}$ có kích thước $m\times n$ và $\m{B}$ có kích thước $n\times p$ được cho bởi
\[\m{C} = \m{A}\m{B},\]
trong đó $c_{ij} = \sum_{k=1}^n a_{ik} b_{kj}$. Phép nhân ma trận thỏa mãn tính kết hợp nhưng thông thường không thỏa mãn tính giao hoán. 

Ma trận chuyển vị $\m{B} = (b_{ij})$ của ma trận $\m{A}$ kích thước $m\times n$ là một ma trận kích thước $n\times m$ với các phần tử $b_{ij} = a_{ji}$. Ta kí hiệu $\m{B} = \m{A}^\top$. Ma trận $\m{A}$ là đối xứng nếu $\m{A} = \m{A}^\top$. Ma trận $\m{A}$ là phản xứng nếu $\m{A} = - \m{A}^\top$. 

Định nghĩa ma trận đơn vị kích thức $n\times n$ bởi
\[\m{I}_n =\begin{bmatrix}
1 & 0 & \ldots & 0\\
0 & 1 & \ddots & \ddots\\
\vdots & \ddots & \ddots & 0\\
0 & \ldots & 0 & 1
\end{bmatrix} \in \mb{R}^{n\times n}.\]
Với mọi ma trận $\m{A}$ kích thước $m\times n$ và $\m{B}$ kích thước $n\times m$ thì $\m{A}\m{I}_n = \m{A}$ và $\m{I}_n\m{B} = \m{B}$.

Một ma trận với toàn bộ các phần tử bằng 0 gọi là ma trận không. Ma trận không kích thước $n\times n$ được kí hiệu bởi $\m{0}_n$.

Kí hiệu các cột của ma trận $\m{A}$ bởi $\m{a}_i = [a_{11},\ldots, a_{m1}]^\top$ thì $\m{A} = [\m{a}_1, \ldots, \m{a}_n]$. Toán tử vec được định nghĩa bởi:
\begin{align}
    \text{vec}(\m{A}) = [\m{a}_1^\top,\ldots, \m{a}_n^\top]^\top \in \mb{C}^{mn}.
\end{align}

\section{Định thức và ma trận nghịch đảo}
Định thức của ma trận vuông $\m{A}$ kích thước $n\times n$, kí hiệu $\text{det}(\m{A}) = |\m{A}|$ có thể  được tính bởi công thức Laplace
\begin{align*}
\left| A \right| =& {a_{11}}\left| {\begin{array}{*{20}{c}}
{{a_{22}}}& \cdots &{{a_{2n}}}\\
 \vdots &{}& \vdots \\
{{a_{n2}}}& \cdots &{{a_{nn}}}
\end{array}} \right| - {a_{12}}\left| {\begin{array}{*{20}{c}}
{{a_{21}}}&{{a_{23}}}& \cdots &{{a_{2n}}}\\
{{a_{31}}}&{{a_{33}}}& \cdots &{{a_{3n}}}\\
 \vdots & \vdots &{}& \vdots \\
{{a_{n1}}}&{{a_{n3}}}& \cdots &{{a_{nn}}}
\end{array}} \right| \\&+ {a_{13}}\left| {\begin{array}{*{20}{c}}
{{a_{21}}}&{{a_{22}}}&{{a_{24}}}& \cdots &{{a_{2n}}}\\
{{a_{31}}}&{{a_{32}}}&{{a_{34}}}& \cdots &{{a_{3n}}}\\
 \vdots & \vdots & \vdots &{}& \vdots \\
{{a_{n1}}}&{{a_{n2}}}&{{a_{n4}}}& \cdots &{{a_{nn}}}
\end{array}} \right| -  \cdots \\
=& \sum_{j=1}^n (-1)^{i+j} a_{ij} |\m{A}_{ij}|,
\end{align*}
với $\m{A}_{ij}$ là ma trận $\m{A}$ sau khi bỏ đi hàng $i$ và cột $j$.

Ma trận vuông  $\m{A}$ kích thước $n\times n$ là không suy biến (suy biến) khi và chỉ khi $|\m{A}| \ne 0$ (tương ứng $|\m{A}| = 0$). Khi $\m{A}$ là không suy biến thì tồn tại duy nhất ma trận $\m{B}$ (kích thước $n\times n$ ) thỏa mãn $\m{A}\m{B} = \m{B}\m{A} = \m{I}_n$ gọi là ma trận nghịch đảo của $\m{A}$, kí hiệu $\m{B} = \m{A}^{-1} = \frac{\m{A}_{adj}}{\text{det}(\m{A})}$, trong đó $\m{A}_{adj} = [|\m{A}_{ij}|]^\top$ với $|\m{A}_{ij}|$ là định thức của $\m{A}_{ij}$. 

\section{Giá trị riêng, vector riêng, định lý Cayley - Hamilton}
Đa thức $|s\m{I}_n - \m{A}|$ được gọi là đa thức đặc tính của ma trận $\m{A}$. Mỗi  nghiệm $\lambda_i$ của đa thức này được gọi là một giá trị riêng của ma trận $\m{A}$. Với mỗi giá trị riêng của $\m{A}$, luôn tồn tại ít nhất một vector riêng $\m{x}$ thỏa mãn phương trình
\begin{align*}
    \m{A} \m{x} = \lambda_i \m{x}.
\end{align*}
Nếu $\lambda_k$ là một nghiệm đơn của đa thức đặc tính thì tồn tại duy nhất một vector riêng  $\m{x}$ với $|\m{x}| = 1$. Ngược lại, nếu $\lambda_k$ là một nghiệm bội của đa thức thì có thể tồn tại nhiều hơn một vector riêng được chuẩn hóa ứng với giá trị riêng này. Với mỗi ma trận $\m{A}$ thì
\[|\m{A}| = \prod_{i=1}^n \lambda_i. \] 
Ma trận $\m{A}$ thỏa mãn phương trình đặc tính của chính nó (định lý Cayley - Hamilton), tức là nếu 
\begin{align}
    |s\m{I}_n - \m{A}| = s^n + \alpha_{n-1} s^{n-1} + \ldots + \alpha_0,
\end{align}
thì
\begin{align}
    \m{A}^n + \alpha_{n-1} \m{A}^{n-1} + \ldots + \alpha_0 \m{I}_n = \m{0}_n.
\end{align}

\section{Định lý Gerschgorin}
Với $\m{A} = [a_{ij}] \in \mb{C}^{n\times n}$, vị trí của các giá trị riêng của $\m{A}$ có thể được ước lượng dựa trên Định lý Gerschgorin \index{định lý!Gerschgorin}: Với mỗi giá trị riêng $s_k$ của $\m{A}$, luôn tồn tại một chỉ số $i=1, 2, \ldots, n$, sao cho $s_k$ nằm trong đường tròn tâm $a_{ii}$ bán kính $R_i = \sum_{j=1,j\ne i}^n |a_{ij}|$. Như vậy, mọi giá trị riêng của $\m{A}$ đều nằm trong phần mặt phẳng hợp bởi các đĩa tròn Gerschhorin $C_1 \cup C_2 \cup \ldots C_n$, trong đó
\begin{align}
    C_i = \{ s \in \mb{C}|~|s - a_{ii}| \leq R_i \}, ~i=1,\ldots, n.
\end{align}

\section{Chéo hóa ma trận và dạng Jordan}
Không gian ảnh của ma trận $\m{A} \in \mb{R}^{m\times n}$, kí hiệu bởi im$(\m{A})$ được định nghĩa bởi  $\text{im}(\m{A}) = \{\m{A}\m{x} \in \mb{R}^m|~\m{x} \in \mb{R}^n\}.$
Hạt nhân của $\m{A}$ được định nghĩa bởi 
$\text{ker}(\m{A}) = \{\m{x} \in \mb{R}^n |~\m{A} \m{x} = \m{0}_m\}.$
Dễ thấy $\text{im}(\m{A}^\top)$ trực giao với $\text{ker}(\m{A})$ do với mọi $\m{y} \in \text{im}(\m{A}^\top), \m{x} \in \text{ker}(\m{A})$ thì $\m{y} = \m{A}^\top \m{x}_1$ và $\m{y}^\top\m{x} = \m{x}_1^\top\m{A} \m{x} = 0$. 

Hai ma trận $\m{A}, \m{B}$ có kích thước $n\times n$ là đồng dạng khi và chỉ khi tồn tại một ma trận không suy biến $\m{P}$ có kích thước $n\times n$ sao cho $\m{B} = \m{P}\m{A}\m{P}^{-1}$. Hai ma trận đồng dạng có cùng các giá trị riêng. Ma trận $\m{A}$ là chéo hóa được nếu tồn tại ma trận không suy biến $\m{P}$ sao cho
\begin{align*}
    \m{A} = \m{P}\m{\Lambda} \m{P}^{-1}
\end{align*}
với $\m{\Lambda}$ là một ma trận đường chéo với các phần tử trên đường chéo chính là các giá trị riêng của $\m{A}$. 

Với mỗi ma trận vuông $\m{A}$ luôn tồn tại ma trận $\m{P}$ sao cho
\[
\m{P}\m{A} \m{P} ^{-1}= \m{J}= \text{blkdiag}(\m{J}_k),
\]
trong đó $\m{J}$ là dạng Jordan của $\m{A}$ và blkdiag kí hiệu ma trận chéo với các ma trận $\m{J}_k$ nằm trên đường chéo chính. Mỗi khối $\m{J}_k$ của ma trận $\m{J}$ tương ứng với một giá trị riêng $\lambda_k$ của $\m{A}$. Nếu $\lambda_k$ là một giá trị riêng đơn, hoặc là một giá trị riêng bội $m$ với đầy đủ $m$ vector riêng độc lập tuyến tính thì 
\[\m{J}_k = \text{diag}(\lambda_k,\ldots, \lambda_k) \in \m{C}^{m\times m}.\]
Nếu $\lambda_k$ là một giá trị riêng bội $m$ với duy nhất một vector riêng độc lập tuyến tính thì
\[\m{J}_k = \left[ {\begin{array}{*{20}{c}}
{{\lambda _k}}&1&0& \cdots &0\\
0&{{\lambda _k}}& \ddots & \ddots & \vdots \\
 \vdots & \ddots & \ddots &1 & 0\\
\vdots & \cdots & 0 & \ddots & 1 \\ 
0& \cdots & \cdots & 0& {{\lambda _k}}
\end{array}} \right] \in \mb{C}^{m\times m}\]
Các vector riêng và vector riêng suy rộng tương ứng với giá trị riêng $\lambda_k$ được tìm từ việc giải lần lượt các phương trình:
\begin{align}
    (\m{A} - \lambda_k\m{I}) \m{v}_1 &= \m{0}, \\
    (\m{A} - \lambda_k\m{I}) \m{v}_2 &= \m{v}_1, \\
    \vdots \nonumber \\
    (\m{A} - \lambda_k\m{I}) \m{v}_m &= \m{v}_{m-1}. 
\end{align}
\section{Ma trận hàm mũ}
Ma trận mũ của ma trận vuông $\m{A}$ được định nghĩa bởi
\begin{align*}
    \mathtt{e}^{\m{A}} = \m{I} + \m{A} + \frac{1}{2! } \m{A}^2 + \frac{1}{3! } \m{A}^3 + \ldots,
\end{align*}
trong đó ta hiểu chuổi hội tụ theo nghĩa mỗi phần tử $ij$ trong ma trận hội tụ. Quy ước $\mathtt{e}^{\m{0}} = \m{I}$. Từ định lý Cayley - Hamilton, ta cũng suy ra $\mathtt{e}^{\m{A}}$ có thể biểu diễu thành tổ hợp tuyến tính của các ma trận $\m{I}_n, \m{A}, \ldots, \m{A}^{n-1}$. Khi $\m{A} = \text{diag}(x_1,\ldots,x_n)$ thì $\mathtt{e}^{\m{A}} = \text{diag}(\mathtt{e}^{x_1},\ldots, \mathtt{e}^{x_n})$.

Nếu $\m{A}$ và $\m{B}$ là hai ma trận vuông giao hoán được thì 
\begin{align}
    \mathtt{e}^{\m{A}} e^{\m{B}} = \mathtt{e}^{\m{B}} \mathtt{e}^{\m{A}} = \mathtt{e}^{\m{A}+\m{B}}.
\end{align}
Từ đây suy ra $\mathtt{e}^{\m{A}} \mathtt{e}^{-\m{A}} = \m{I}$ và với mỗi đa thức $p(\m{A})$ thì $p(\m{A})\mathtt{e}^{\m{A}t} = \m{A} p(\m{A})$. Hơn nữa, với $t$ là biến thời gian thì:
\begin{equation}
\begin{aligned}
    \frac{d}{dt} \mathtt{e}^{\m{A}t} &= \m{A} \mathtt{e}^{\m{A}t} = \mathtt{e}^{\m{A}t} \m{A} \\
    \mathtt{e}^{\m{A}t_1} \mathtt{e}^{\m{A}t_2} & =  \mathtt{e}^{\m{A}(t_1+t_2)}\\
    \mathtt{e}^{\m{A}t} \mathtt{e}^{-\m{A}t} & = \mathtt{e}^{\m{0}t} = \m{I}.
\end{aligned}
\end{equation}

\section{Tích Kronecker}
Tích Kronecker của hai ma trận $\m{A} \in \mb{C}^{m\times n}$ và $\m{B} \in \mb{C}^{p\times q}$, kí hiệu $\m{A} \otimes \m{B}$ là một ma trận kích thước $mp\times nq$ thỏa mãn:
\begin{align}
    \m{A}\otimes \m{B} = \begin{bmatrix}
    a_{11} \m{B} & a_{12}\m{B}&\ldots &a_{1n}\m{B}\\
    a_{21} \m{B} & a_{22}\m{B}&\ldots &a_{2n}\m{B}\\
    \vdots & \vdots & & \vdots\\
    a_{m1}\m{B} & a_{m2} \m{B} & \ldots & a_{mn}\m{B}
    \end{bmatrix}.
\end{align}
Với $\m{A}, \m{B}, \m{C}, \m{D}$ là các ma trận với kích thước phù hợp, một số tính chất của tích Kronecker được cho dưới đây:
\begin{equation}
\begin{aligned}
        \m{A} \otimes(\m{B}+\m{C}) &= \m{A}\otimes \m{B} + \m{A}\otimes \m{C}\\
         (\m{B}+\m{C})\otimes \m{A} &= \m{B}\otimes \m{A} + \m{C}\otimes \m{A}\\
        ( \m{A}\otimes \m{B}) \otimes \m{C} &= \m{A} \otimes (\m{B}\otimes \m{C})\\
        k (\m{A}\otimes \m{B}) &= k\m{A} \otimes \m{B}\\
         ( \m{A}\otimes \m{B}) (\m{C} \otimes \m{D}) &= ( \m{A} \m{C})\otimes (\m{B} \m{D})\\
         (\m{A}\otimes \m{B})^\top &= \m{A}^\top \otimes \m{B}^\top\\
         \text{rank} (\m{A}\otimes \m{B}) &= \text{rank}(\m{A}) \text{rank}(\m{B})
       \end{aligned}
\end{equation}
Giả sử các ma trận vuông $\m{A} \in \mb{C}^{m\times m}$ có các cặp giá trị riêng - vector riêng tương ứng là $(\lambda_i, \m{u}_i),~i = 1,\ldots, n$, và $\m{B} \in \mb{C}^{n \times n}$ có các cặp giá trị riêng - vector riêng tương ứng là $(\mu_j , \m{v}_j)$, $j=1, \ldots, n$ thì $\m{A} \otimes \m{B}$ có các cặp giá trị riêng - vector riêng là  $( {\lambda_i} {\mu_j}, \m{u}_i \otimes \m{v}_j)$, $i = 1,\ldots, n; j=1,\ldots, m$. Ta cũng có các tính chất:
\begin{equation}
\begin{aligned}
         (\m{A}\otimes \m{B})^{-1} &= \m{A}^{-1} \otimes \m{B}^{-1}\\
         |\m{A}\otimes \m{B}| &= |\m{A}|^m|\m{B}|^n\\
         \text{trace}(\m{A}\otimes \m{B}) &= \text{trace}(\m{A}) \text{trace}(\m{B})
\end{aligned}.
\end{equation}

\section{Ma trận xác định dương và ma trận bán xác định dương}
Ma trận $\m{A} \in \mb{R}^{n\times n}$ đối xứng được gọi là xác định dương khi và chỉ khi với mỗi vector $\m{x} \in \mb{R}^n$ sao cho $\m{x}\ne \m{0}_n$ thì dạng toàn phương 
\begin{align}
    \m{x}^\top \m{A} \m{x} = \sum_{i,j=1}^n a_{ij} x_i x_j >0.
\end{align}
Ma trận $\m{A} \in \mb{R}^{n\times n}$ đối xứng được gọi là bán xác định dương khi và chỉ khi với mỗi vector $\m{x} \in \mb{R}^n$ sao cho $\m{x}\ne \m{0}_n$ thì $ \m{x}^\top \m{A} \m{x} $ là không âm. 

\section{Chuẩn của vector và ma trận}
Chuẩn của vector $\m{x}$, kí hiệu bởi $\|\m{x}\|$ thỏa mãn (i) $\|\m{x}\| \ge 0$, (ii) $\|a\m{x}\| = |a| \|\m{x}\|$ với mọi $a\in \mb{R}$, và (iii) $\|\m{x} + \m{y}\| \leq \|\m{x}\| + \|\m{y}\|$, $\forall \m{x}, \m{y}$. Với $\m{x} = [x_1, \ldots, x_n]^\top$ thì các chuẩn thường dùng là
\begin{align*}
    \|\m{x}\|_2 = \Big(\sum_{i=1}^n x_i^2\Big)^{1/2}, ~ \|\m{x}\|_\infty = \max_{i}|x_i|, ~\text{ và} ~\|\m{x}\|_1 = \sum_{i=1}^n |x_i|.
\end{align*}
Nếu không có kí hiệu chỉ số dưới cụ thể, ta hiểu $\|\m{x}\|$ là chuẩn 2 của vector $\m{x}$.

Bất đẳng thức Schwarz: $|\m{x}^\top \m{y}| \le \|\m{x}\|\|\m{y}\|$, trong đó đẳng thức xảy ra khi và chỉ khi $\m{x} = \lambda \m{y}$ với $\lambda\in \mb{C}$.

Chuẩn của ma trận $\m{A}\in \mb{R}^{m\times n}$ được định nghĩa thông qua chuẩn của vector như sau: 
\[ \|\m{A}\| = \max_{\|\m{x}\|=1} \|\m{A}\m{x}\|.\]
Tương ứng với các chuẩn vector ở trên, ta có các chuẩn ma trận tương ứng là $\lambda_{\max}(\m{A}^\top \m{A})^{1/2}$, $\max_{i} (\sum_{j=1}^n |a_{ij}|)$ và $\max_{j}(\sum_{i=1}^n|a_{ij}|)$. Đồng thời, một số tính chất của chuẩn ma trận được cho bởi
\begin{align*}
    \|\m{A}\m{x}\| \le \|\m{A}\| \|\m{x}\|,~ \|\m{A}+\m{B}\| \leq \|\m{A}\| +\m{B},~\text{ và }~ \|\m{A}\m{B}\| \leq \|\m{A}\| \|\m{B}\|.
\end{align*}
\section{Lý thuyết Perron-Frobenius}
Ma trận $\m{A} \in \mb{R}^{n\times n}, n\ge 2$ là ma trận không âm\index{ma trận!không âm} (ma trận dương) nếu $a_{ij}\ge 0~(\text{tương ứng} ~a_{ij}>0), \forall i, j = 1, \ldots, n$. Bán kính quang phổ\index{bán kính quang phổ} của $\m{A}$ là bán kính của đĩa tròn nhỏ nhất trong mặt phẳng phức có tâm ở gốc tọa độ và chứa mọi giá trị riêng của $\m{A}$,
\begin{align}
    \rho (\m{A}) = {\max}\{ |\lambda|~|~\lambda ~\text{ là giá trị riêng của } ~\m{A} \}.
\end{align}

Ma trận $\m{A}$ là không rút gọn được (irreducible)\index{ma trận!không rút gọn được} nếu $\sum_{k=0}^{n-1}\m{A}^k$ là ma trận dương\index{ma trận!dương}, ngược lại $\m{A}$ là thu gọn được. Nếu tồn tại $k\in \mb{N}$ sao cho $\m{A}^k$ là ma trận dương thì $\m{A}$ gọi là ma trận nguyên thủy (primitive)\index{ma trận!nguyên thủy}.

Với $\m{A}$ là không âm thì:
\begin{itemize}
    \item tồn tại một giá trị riêng thực trội (hay giá trị riêng Perron\index{giá trị riêng!Perron}) $\lambda$ sao cho với mọi giá trị riêng $\mu$ khác của $\m{A}$ thì $\lambda \ge |\mu| \ge 0$.
    \item các vector riêng bên phải và bên trái tương ứng với $\lambda$ có thể được chọn là các vector không âm.
    \item nếu có thêm giả thuyết $\m{A}$ là không thu gọn được thì giá trị riêng trội là đơn và $\lambda>0$. Đồng thời, các vector riêng bên phải và bên trái tương ứng với giá trị riêng $\lambda$ là duy nhất (không tính tới sai khác về tỉ lệ).
    \item nếu có thêm giả thuyết $\m{A}$ là nguyên thủy thì giá trị riêng trội thỏa mãn $\lambda > |\mu|$ với mọi giá trị riêng $\mu \ne \lambda$.
\end{itemize}

\chapter{Tập lồi và hàm lồi}
\label{append:convex_set_functions}
Các nội dung về tập lồi, hàm lồi trong chương này được trình bày từ \cite{Nguyen2023}. 
\section{Tập lồi}
Với $\m{x}_1,\m{x}_2$ là hai điểm trong $\mb{R}^d$, đoạn thẳng nối $\m{x}_1$ và $\m{x}_2$ là tập các điểm
\begin{align}
\m{x} = \lambda_1 \m{x}_1 + (1-\lambda)\m{x}_2,\quad \forall \lambda \in [0,1].
\end{align}

Tập $M \subseteq \mb{R}^d$ là tập lồi nếu $M$ chứa tất cả các đoạn thẳng nối hai điểm bất kỳ trong $M$. Nói cách khác, $M$ là tập lồi nếu với mọi $\m{x}_1,\m{x}_2 \in \mb{R}^d$ thì \index{tập lồi}
\begin{align}
\m{x} = \lambda_1 \m{x}_1 + (1-\lambda)\m{x}_2 \in M,\quad \forall \lambda \in [0,1].
\end{align}

Với $\m{x}_i\in \mb{R}^d$, $i=1,\ldots,n$, bao lồi hay tổ hợp lồi của các điểm $\m{x}_1,\ldots,\m{x}_n$ được định nghĩa là
\begin{align} \label{eq:app_convex_hull}
\mc{H}(\m{x}_1,\ldots,\m{x}_n)=\{\m{x}\in \mb{R}^d|~\m{x}=\sum_{i=1}^n \lambda_i \m{x}_i,~\forall \lambda_i\in [0,1], \sum_{i=1}^n\lambda_i = 1\}.
\end{align}
Nếu trong định nghĩa \eqref{eq:app_convex_hull}, $\lambda_i \in (0,1)$ thì $\mc{H}(\m{x}_1,\ldots,\m{x}_n)$ là tập lồi chặt. \index{tập lồi chặt}

\section{Hàm lồi}
Hàm ${f}:M\to \mb{R}$ xác định trên tập lồi $M \subseteq \mb{R}^d$ gọi là hàm lồi nếu \index{hàm!lồi}
\begin{align}
{f}(\lambda\m{x}_1+(1-\lambda)\m{x}_2) \leq \lambda{f}(\m{x}_1) + (1-\lambda){f}(\m{x}_2), \; \forall \m{x}_1,\m{x}_2 \in \mb{R}^d,\forall \lambda \in [0,1].
\end{align}

Nếu 
\begin{align}
{f}(\lambda\m{x}_1+(1-\lambda)\m{x}_2) < \lambda {f}(\m{x}_1) + (1-\lambda) {f}(\m{x}_2), \; \forall \m{x}_1,\m{x}_2 \in M,\forall \lambda \in (0,1),
\end{align}
thì $f$ là hàm lồi chặt.\index{hàm!lồi chặt}

Hàm $f$ là hàm lõm (hàm lõm chặt) nếu $-f$ là hàm lồi (tương ứng, hàm lồi chặt). \index{hàm!lõm} \index{hàm!lõm chặt}

Với $f_1,f_2$ là hai hàm lồi xác định trên tập lồi $M_1,M_2 \subseteq \mb{R}^n$ và số thực $\lambda \geq 0, \lambda_1, \lambda_2 \geq 0$, thì các hàm sau là các hàm lồi:
\begin{subequations}
\begin{align}
(\lambda f_1)(x) \triangleq \lambda f_1(x),\; \text{trên tập lồi } X_1, \\
(\lambda_1 f_1+\lambda_2 f_2)(x) \triangleq f_1(x) + f_2(x),\; \text{trên tập lồi } x\in X_1 \cap X_2, \\
\max\{f_1,f_2\}(x) \triangleq \max\{f_1(x),f_2(x)\},\; \text{trên tập lồi } X_1 \cap X_2.
\end{align}
\end{subequations}

\begin{theorem}[Hàm lồi khả vi]
Với $f$ là hàm khả vi xác định trên tập lồi mở $M\subseteq \mb{R}^d$ thì
\begin{itemize}
\item[i.]  $f$ là hàm lồi trên $M$ khi và chỉ khi
\[f(y)-f(x) \geq \nabla f(x) (y-x)\; \forall x,y\in M.\]
\item[ii.] $f$ là hàm lõm trên $M$ khi và chỉ khi
\[f(y)-f(x) \geq \nabla f(x) (y-x)\; \forall x,y\in M.\]
\end{itemize}
\end{theorem}

\begin{theorem}[Hàm lồi khả vi hai lần]
Với $f$ là hàm khả vi hai lần trên tập lồi mở $M\subseteq \mb{R}^d$ thì
\begin{itemize}
\item[i.]  $f$ là hàm lồi trên $M$ khi và chỉ khi ma trận Hess $\bigtriangledown^2 f(x)$ là bán xác định dương trên $M$.
\item[ii.] $f$ là hàm lồi chặt trên $M$ nếu $\bigtriangledown^2 f(x)$ là xác định dương trên $M$.
\end{itemize}
\end{theorem}

\begin{theorem}[Cực trị của hàm lồi]
Với $f$ là hàm lồi trên tập lồi $\emptyset \neq M \subseteq \mb{R}^d$ và $x^*$ là một cực trị địa phương của $f$ thì:
\begin{itemize}
\item[i.] $x^*$ cũng là điểm tối ưu toàn cục.
\item[ii.] Nếu $x^*$ là nghiệm tối ưu địa phương chặt hoặc $f$ là hàm lồi chặt thì $x^*$ là nghiệm tối ưu toàn cục duy nhất của $f$.
\end{itemize}
\end{theorem}
\chapter{Lý thuyết điều khiển}
\label{append:control_theory}
Một số tài liệu tham khảo về lý thuyết điều khiển tuyến tính bao gồm  \cite{Ogata2009modern,antsaklis2006linear} và \cite{Nguyen2009}. Một số kết quả về lý thuyết điều khiển phi tuyến ở mục này được trình bày từ \cite{Khalil2002}.

\section{Hệ tuyến tính}
Xét hệ liên tục, tuyến tính, bất biến và nhân quả cho bởi phương trình:
\begin{align}
    \dot{\m{x}}(t) &= \m{A} \m{x} (t)+ \m{B} \m{u}(t), \label{eq:Appendix1}\\
    \m{y} (t)&= \m{C} \m{x}(t),~ \m{x}(0) = \m{x}_0, \label{eq:Appendix2}
\end{align}
trong đó $\m{x} \in \mb{R}^d$ là biến trạng thái, $\m{u} \in \mb{R}^p$ là biến đầu vào, $\m{y} \in \mb{R}^q$ là biến đầu ra, và  $\m{A} \in \mb{R}^{d \times d}$, $\m{B} \in \mb{R}^{d \times p}$ và $\m{C} \in \mb{R}^{q \times d}$ là các ma trận của hệ.

Nghiệm của hệ \eqref{eq:Appendix1}--\eqref{eq:Appendix2}, với $t\ge 0$, được cho bởi
\begin{align}
    \m{x}(t) &= \mathtt{e}^{\m{A}t} \m{x}_0 + \int_0^t \mathtt{e}^{\m{A}(t-\tau)}\m{B} \m{u}(\tau) d\tau,\\
    \m{y}(t) &= \m{C} \mathtt{e}^{\m{A}t} \m{x}_0 + \m{C} \int_0^t  \mathtt{e}^{\m{A}(t-\tau)}\m{B} \m{u}(\tau) d\tau,
\end{align}
trong đó $\mathtt{e}^{\m{A}}$ là ma trận hàm mũ của ma trận vuông $\m{A}$.

Xét hệ \eqref{eq:Appendix1}--\eqref{eq:Appendix2} khi $\m{u} = \m{0}$, tức là $\dot{\m{x}}(t) = \m{A} \m{x} (t)$. Gọi $\lambda_k,k=1, \ldots, d$ là các giá trị riêng của ma trận $\m{A}$, thỏa mãn phương trình đặc tính det$(s\m{I}_d - \m{A})=0$. Hệ là ổn định Hurwitz nếu Re$\{\lambda_k\} < 0 , \forall k$. Nếu Re$\{\lambda_k\} \leq 0, \forall k$ và các giá trị riêng có phần thực bằng 0 đều có bội đại số bằng bội hình học. Giả sử ma trận $\m{A}$ có dạng Jordan $\m{J} = \m{U}^{-1} \m{A} \m{U}$, với $\m{U} \in \mb{C}^{d \times d}$ thì 
nghiệm tự do của hệ được cho bởi $\m{x}(t) = \mathtt{e}^{\m{A}t} \m{x}_0 = \m{U}\mathtt{e}^{\m{Jt}}\m{U}^{-1}\m{x}_0$.

Định nghĩa ma trận điều khiển \[\mcl{C} = [\m{B}, \m{A}\m{B}, \ldots, \m{A}^{n-1}\m{B}],\] và ma trận quan sát \[\mcl{O} = \begin{bmatrix} \m{C}\\ \m{C}\m{A} \\ \vdots \\ \m{C} \m{A}^{n-1}
\end{bmatrix}.\]

Hệ \eqref{eq:Appendix1}--\eqref{eq:Appendix2} được gọi là ổn định được (stabilizable) nếu tồn tại ma trận $\m{B} \in\mb{R}^{p\times d}$ sao cho ma trận $(\m{A} - \m{B}\m{K})$ là ma trận Hurwitz (có mọi giá trị riêng nằm về bên trái trục ảo). Hệ \eqref{eq:Appendix1}--\eqref{eq:Appendix2} là điều khiển được hoàn toàn khi và chỉ
\begin{align*}
    \text{rank}(\mcl{C}) = n.
\end{align*}
Nếu hệ điều khiển được hoàn toàn, ta luôn  tìm được ma trận $\m{K}$ để ma trận $(\m{A} - \m{B}\m{K})$ có  $d$ giá trị riêng chọn trước bất kỳ.

Hệ \eqref{eq:Appendix1}--\eqref{eq:Appendix2} được gọi là có thể dò được (detectable) được nếu tồn tại ma trận $\m{L} \in\mb{R}^{q\times d}$ sao cho ma trận $(\m{A} - \m{L}\m{C})$ là ma trận Hurwitz. Hệ \eqref{eq:Appendix1}--\eqref{eq:Appendix2} là quan sát được hoàn toàn khi và chỉ 
\begin{align*}
    \text{rank}(\mcl{O}) = n.
\end{align*}
Nếu hệ quan sát được hoàn toàn, ta luôn tìm được ma trận $\m{L}$ để ma trận $(\m{A} - \m{L}\m{C})$ có $d$ giá trị riêng chọn trước bất kỳ.

\section{Lý thuyết ổn định Lyapunov}
Trong mục này, ta điểm lại một số kết quả cơ bản của phương pháp Lyapunov trong phân tích ổn định các hệ thống tự trị (autonomous systems). Chứng minh và các ví dụ minh họa của các kết quả này, độc giả quan tâm có thể xem tại các sách tham khảo kinh điển về lý thuyết điều khiển phi tuyến như \cite{Khalil2002,Slotine1991applied}.

Xét hệ tự trị cho bởi phương trình:
\begin{align}
    \dot{\m{x}} &= \m{f}(\m{x}), ~ \label{eqAppendix:NonLinearSystems}
\end{align}
trong đó $\m{f}: D \to \mb{R}^d$ là một hàm liên tục từ tập $D \subseteq \mb{R}^d$ tới $\mb{R}^d$. Không mất tính tổng quát, giả sử rằng $\m{x}=\m{0}_d \in D$ là một điểm cân bằng của hệ, tức là $\m{f}(\m{0}_d)=\m{0}_d$. 

\begin{Definition}
Điểm cân bằng $\m{x} = \m{0}_d$ của \eqref{eqAppendix:NonLinearSystems} gọi là
\begin{itemize}
    \item Ổn định nếu, với mỗi $\epsilon>0$, tồn tại $\delta=\delta(\epsilon)>0$ sao cho
    \begin{equation*}
        \|\m{x}(0)\|<\delta \Rightarrow \|\m{x}(t)\|<\epsilon, \; \forall t\geq 0.
    \end{equation*}
    Ngược lại, nếu điều kiện trên không thỏa mãn, $\m{x}=\m{0}_d$ là một điểm cân bằng không ổn định.
    \item Ổn định tiệm cận nếu nó là ổn định, đồng thời $\delta$ có thể được chọn sao cho
    \begin{equation*}
\|\m{x}(0)\|<\delta \Rightarrow \lim_{t\to\infty} \m{x}(t) = \m{0}_d.
    \end{equation*}
\end{itemize}
\end{Definition}

Ta có một số định lý sau để kiểm tra tính ổn định của hệ tại $\m{x} = \m{0}_d$
\begin{theorem}[Phương pháp trực tiếp của Lyapunov]
Giả sử tồn tại hàm $V: D \to \mb{R}$ khả vi liên tục thỏa mãn:
\begin{itemize}
    \item $V(\m{x})$ là xác định dương đối với $\m{x}$: $V(\m{0}_d) = 0$ và $V(\m{x})>0, \forall \m{x} \in D\setminus \{\m{0}_d\}$,
    \item $\dot{V}(\m{x})$ là bán xác định âm đối với $\m{x}$: $\dot{V}(\m{x}) \leq 0, \forall \m{x} \in D$.
\end{itemize}
Khi đó, $\m{x} = \m{0}_d$ là ổn định. 

Nếu có thêm điều kiện $\dot{V}(\m{x})$ là  xác định âm đối với $\m{x}$ (tức là $\dot{V}(\m{0}_d)=0$ và $\dot{V}(\m{x}) < 0, \forall \m{x} \in D \setminus \{\m{0}\}$) thì gốc tọa độ là ổn định tiệm cận.

Cuối cùng, nếu các điều kiện trên thỏa mãn với $D = \mb{R}^d$, đồng thời hàm $V$ thỏa mãn $\|\m{x}\| \to \infty \Rightarrow V(\m{x}) \to \infty$ thì $\m{x} = \m{0}_d$ tương ứng là ổn định/ổn định tiệm cận toàn cục.
\end{theorem}

Để chứng minh điểm cân bằng $\m{x} = \m{0}_d$ của hệ là không ổn định, ta có thể sử dụng định lý Chetaev, phát biểu như sau:
\begin{theorem}[Định lý Chetaev]
Giả sử tồn tại phiếm hàm $V: D \to \mb{R}$ khả vi liên tục thỏa mãn:
\begin{itemize}
    \item $V(\m{0}_d)=0$ và $V(\m{x})>0$ với $\m{x}$ thuộc một lân cận nhỏ bất kỳ của gốc tọa độ,
    \item Trong tập $B_r = \{ \m{x} \in \mb{R}^d| ~ \|\m{x}\| \leq r\}$ với $r>0$ đủ nhỏ sao cho $B_r \subset D$, tồn tại tập $U = \{\m{x} \in B_r| V(\m{x})>0\}$ sao cho $\dot{V}(\m{x}) >0, \forall \m{x} \in U$.
\end{itemize}
Khi đó, $\m{x} = \m{0}_d$ là không ổn định.
\end{theorem}

Trong định lý Lyapunov, việc tìm hàm $V$ thỏa mãn $\dot{V}$ xác định âm thường gặp nhiều khó khăn. Định lý bất biến LaSalle cung cấp một công cụ để chứng minh tính ổn định tiệm cận khi $\dot{V}$ là bán xác định âm. 

Một tập là bất biến đối với \eqref{eqAppendix:NonLinearSystems} nếu
\begin{align*}
    \m{x}(0) \in M \Rightarrow \m{x}(t) \in M,~\forall t \geq 0.
\end{align*}
Định nghĩa khoảng cách từ $\m{x}(t)$ tới tập bất biến $M$ là khoảng cách nhỏ nhất từ $\m{x}(t)$ tới một điểm bất kỳ trong $M$:
\begin{align*}
\text{dist}(\m{x}(t),M) = \inf_{\m{y} \in M} \| \m{y} - \m{x}(t) \|.    
\end{align*}
Khi đó, $\m{x}(t)$ là tiến tới tập $M$ khi $t$ tiến tới vô cùng nếu với mỗi $\epsilon>0$ cho trước, tồn tại $T>0$ sao cho
\begin{align*}
    \text{dist}(\m{x}(t),M) < \epsilon, ~\forall t > T.
\end{align*}

\begin{theorem}[Định lý bất biến LaSalle] \label{Thm:LaSalle}
Gọi $\Omega \subset D$ là một tập compact và bất biến đối với \eqref{eqAppendix:NonLinearSystems}.
Giả sử tồn tại hàm $V: D \to \mb{R}$ khả vi liên tục thỏa mãn $\dot{V}(\m{x})$ là bán xác định âm đối với mọi $\m{x} \in \Omega$. Định nghĩa $E = \{\m{x} \in D|~ \dot{V} = 0\}$ và $M$ là tập bất biến lớn nhất trong $E$. Khi đó mọi nghiệm của  \eqref{eqAppendix:NonLinearSystems} xuất phát trong $\Omega$ sẽ tiến tới $M$ khi $t \to \infty$. 

Giả sử không có nghiệm nào là dừng trong $E$, ngoại trừ nghiệm tầm thường $\m{x}(t) \equiv \m{0}_d$, thì gốc tọa độ là ổn định tiệm cận. 
\end{theorem}

Trong định lý \ref{Thm:LaSalle}, nếu $D=\mathbb{R}^d$ và hàm $V$ thỏa mãn thêm điều kiện $\| \m{x}\| \to \infty \Rightarrow V(\m{x}) \to \infty$, đồng thời không có nghiệm nào là dừng trong $E = \{\m{x} \in \mb{R}^d| ~\dot{V} = 0\}$, ngoại trừ nghiệm tầm thường $\m{x}(t) \equiv \m{0}_d$, thì gốc tọa độ là ổn định tiệm cận. 

\section{Bổ đề Barbalat}
Bổ đề Barbalat là một công cụ cho phân tích tính hội tụ của một hệ động học, thường được sử dụng trong trường hợp hệ là phụ thuộc thời gian.

Một hàm $\m{f}: \mb{R}^d \to \mb{R}^n$ là liên tục đều nếu với mỗi $\epsilon>0$, tồn tại $\delta>0$ sao cho:
\begin{align}
    \|\m{x}-\m{y}\|<\delta \Rightarrow \|\m{f}(\m{x}) - \m{f}(\m{y})\|<\epsilon,~\forall \m{x}, \m{y} \in \mb{R}^d.
\end{align}

\begin{theorem}[Bổ đề Barbalat]
Giả sử hàm số $\m{f}(t)$ có giới hạn hữu hạn khi $t\to\infty$ và nếu $\dot{\m{f}}$ là liên tục đều hoặc $\ddot{\m{f}}$ là bị chặn thì $\lim_{t\to \infty} \dot{\m{f}}(t) = \m{0}_n$.
\end{theorem}

\chapter{Mô phỏng MATLAB\texttrademark}
\label{append:MATLAB}
Các tài liệu hướng dẫn sử dụng MATLAB/SIMULINK với định hướng điều khiển có thể tham khảo tại \url{https://ctms.engin.umich.edu/CTMS/index.php?aux=Home}. 
\section{Hàm biểu diễn các đội hình 2D và 3D} \label{append:visualize_formation}
\begin{lstlisting}
function y = PlotFormation(n,d,t,p,H,o)
% n: number of agents
% d: space's dimension (or number of state variables of each agent)
% t: simulation time interval
% p: configuration in time
% H: incident matrix
% o: option
set(0,'defaultTextInterpreter','tex');

if ((d*n==size(p,1)) && (n == length(H(1,:))))
if o==1  % Draw trajectory, initial and end formation
  figure;
  hold on
  m = length(H(:,1));
  r = 1:n;
  t_end = length(t);
  if d==2
      plot(p(1,:), p(2,:), 'Color', 'k', 'LineWidth', 1);
      plot(p(1,1), p(2,1), 'o', 'MarkerSize', 12, 'MarkerEdgeColor', 'b', 'MarkerFaceColor', 'w');
      plot(p(1,t_end), p(2,t_end), 'o', 'MarkerSize', 12, 'MarkerEdgeColor', 'r', 'MarkerFaceColor', 'w');
    for k=1:m
      index1 = r(H(k,:)<0); index2 = r(H(k,:)>0);
      % drop % from next line if you would like to draw initial formation
      % line([p(index1*2-1,1), p(index2*2-1,1)], [p(index1*2,1), p(index2*2,1)], 'Color', color1, 'LineWidth', 1, 'LineStyle','--');
      line([p(index1*2-1,t_end), p(index2*2-1,t_end)], [p(index1*2,t_end), p(index2*2,t_end)], 'Color', 'r', 'LineWidth', 1, 'LineStyle','-');
    end
    for i=1:n
      plot(p(2*i-1,:), p(2*i,:), 'Color', 'k', 'LineWidth', 1);
      plot(p(2*i-1,1), p(2*i,1), 'o', 'MarkerSize', 12, 'MarkerEdgeColor', 'b', 'MarkerFaceColor', 'w');
      plot(p(2*i-1,t_end), p(2*i,t_end), 'o', 'MarkerSize', 12, 'MarkerEdgeColor', 'r', 'MarkerFaceColor', 'w');
      text(p(2*i-1,1), p(2*i,1), sprintf('%d',i), 'FontSize',9, 'Color', 'k', 'HorizontalAlignment','center');
      text(p(2*i-1,t_end), p(2*i,t_end), sprintf('%d',i), 'FontSize', 9, 'Color', 'k', 'HorizontalAlignment', 'center');
    end
  else
    if d==3
      plot3(p(1,1:t_end), p(2,1:t_end), p(3,1:t_end), 'Color', 'k', 'LineWidth', 1); 
      plot3(p(1,1), p(2,1), p(3,1), 'o', 'MarkerSize', 12, 'MarkerEdgeColor', 'b', 'MarkerFaceColor', 'w');
      plot3(p(1,t_end), p(2,t_end), p(3,t_end), 'o', 'MarkerSize', 12, 'MarkerEdgeColor', 'r', 'MarkerFaceColor', 'w');
     for k=1:m
      index1 = r(H(k,:)<0); index2 = r(H(k,:)>0);
      % drop % in next line if you would like to draw initial formation
      % line([p(index1*3-2,1), p(index2*3-2,1)], [p(index1*3-1,1), p(index2*3-1,1)], [p(index1*3,1),p(index2*3,1)], 'Color', color1, 'LineWidth', 1, 'LineStyle','--');
      line([p(index1*3-2,t_end), p(index2*3-2,t_end)], [p(index1*3-1,t_end), p(index2*3-1,t_end)], [p(index1*3,t_end), p(index2*3,t_end)], 'Color', 'r', 'LineWidth', 1, 'LineStyle','-');
     end
     for i=1:n
      plot3(p(3*i-2,1:t_end), p(3*i-1,1:t_end), p(3*i,1:t_end), 'Color', 'k', 'LineWidth', 1); 
      plot3(p(3*i-2,1), p(3*i-1,1), p(3*i,1), 'o', 'MarkerSize', 12, 'MarkerEdgeColor', 'b', 'MarkerFaceColor', 'w');
      plot3(p(3*i-2,t_end), p(3*i-1,t_end), p(3*i,t_end), 'o', 'MarkerSize', 12, 'MarkerEdgeColor', 'r', 'MarkerFaceColor', 'w');
      text(p(3*i-2,1), p(3*i-1,1), p(3*i,1), sprintf('%d',i), 'FontSize', 9, 'Color', 'k', 'HorizontalAlignment', 'center');
      text(p(3*i-2,t_end), p(3*i-1,t_end), p(3*i,t_end), sprintf('%d',i), 'FontSize', 9, 'Color', 'k', 'HorizontalAlignment', 'center');
      view(3);
      zlabel z;
     end
    end
  end
  xlabel x;
  ylabel y;
  axis equal;
  box on;
  leg{1} = 'p_i(t)';leg{2} = 'p_i(0)'; leg{3} = sprintf('p_i(%d)', t(t_end));
  legend(leg,'NumColumns',1,'Location','southeast');
  
elseif o==2 % Draw initial formation
  figure;
  hold on
  m = length(H(:,1));
  r = 1:n;
  if d==2
      plot(p(1,1), p(2,1), 'o', 'MarkerSize', 12, 'MarkerEdgeColor', 'b', 'MarkerFaceColor', 'w');
    for k=1:m
      index1 = r(H(k,:)<0); index2 = r(H(k,:)>0);
      line([p(index1*2-1,1), p(index2*2-1,1)], [p(index1*2,1), p(index2*2,1)], 'Color', 'b', 'LineWidth', 1, 'LineStyle','-');
    end
    for i=1:n
      plot(p(2*i-1,1), p(2*i,1), 'o', 'MarkerSize', 12, 'MarkerEdgeColor', 'b', 'MarkerFaceColor', 'w');
      text(p(2*i-1,1), p(2*i,1), sprintf('%d',i), 'FontSize',9, 'Color','k', 'HorizontalAlignment','center');
    end
  else
    if d==3
      plot3(p(1,1), p(2,1), p(3,1), 'o', 'MarkerSize', 12, 'MarkerEdgeColor', 'b', 'MarkerFaceColor', 'w');
     for k=1:m
      index1 = r(H(k,:)<0); index2 = r(H(k,:)>0);
      line([p(index1*3-2,1), p(index2*3-2,1)], [p(index1*3-1,1), p(index2*3-1,1)], [p(index1*3,1),p(index2*3,1)], 'Color', 'b', 'LineWidth', 1, 'LineStyle','-');
     end
     for i=1:n
      plot3(p(3*i-2,1), p(3*i-1,1), p(3*i,1), 'o', 'MarkerSize', 12, 'MarkerEdgeColor', 'b', 'MarkerFaceColor', 'w');
      text(p(3*i-2,1), p(3*i-1,1), p(3*i,1), sprintf('%d',i), 'FontSize', 9, 'Color', 'k', 'HorizontalAlignment', 'center');
     end
     zlabel z;   view(3)
    end
  end
  xlabel x
  ylabel y
  axis equal
  box on
  leg{1} = 'p_i(0)';
  legend(leg,'NumColumns',1,'Location','southeast');
else
    y = 1;
end
    y = 1;
else
    % Print error
    fprintf('Error! Please check input data \n')
    y = 0;
end
end
\end{lstlisting}

\section{Biểu diễn sự thay đổi của đội hình theo thời gian}
\begin{lstlisting}
%% Mo phong dieu khien doi hinh theo sai lech vi tri
global p_d L
% Khoi tao vi tri ban dau
n = 9;
p0 = 20*(rand(2*n,1)-0.5);
% Ma tran Laplace
L = n*eye(n) - ones(n,n);

% Doi hinh dat
P_d = [0 0; 1 1; 1 -1; 3 0; 0 3; 0 -3; -1 1; -1 -1;-3 0]';

p_d = P_d(:);
[t,p] = ode45(@control_law,[0:0.01:1],p0);
p = p';
t_end = length(t);

figure(1)
hold on
str = '#DEDAE0';
color = sscanf(str(2:end),'%2x%2x%2x',[1 3])/255;
for i=1:n
    plot3(t(1,1),p(2*i-1,1),p(2*i,1),'Color', 'r',"Marker","x");
    plot3(t,p(2*i-1,1:t_end),p(2*i,1:t_end), 'Color', color, 'LineWidth', 1,"LineStyle","-");
    plot3(t(t_end,1),p(2*i-1,t_end),p(2*i,t_end),'ob');
end

H = [-1 1 0 0 0 0 0 0 0;
     -1 0 1 0 0 0 0 0 0;
     -1 0 0 0 0 0 0 1 0;
     -1 0 0 0 0 0 1 0 0;
     0 -1 0 1 0 0 0 0 0;
     0 -1 0 0 1 0 0 0 0;
     0 0 -1 1 0 0 0 0 0;
     0 0 -1 0 0 1 0 0 0;
     0 0 0 -1 1 0 0 0 0;
     0 0 0 -1 0 1 0 0 0;
     0 0 0 0 -1 0 1 0 0;
     0 0 0 0 -1 0 0 0 1;
     0 0 0 0 0 -1 0 1 0;
     0 0 0 0 0 -1 0 0 1;
     0 0 0 0 0 0 -1 0 1;
     0 0 0 0 0 0 0 -1 1];

for i=1:length(H)
    Px=[]; Py=[];Px1=[]; Py1=[];
    for j=1:9
        if H(i,j)~=0
            Px = [Px,p(2*j-1,t_end)];
            Py = [Py,p(2*j,t_end)];
            Px1 = [Px1,p(2*j-1,1)];
            Py1 = [Py1,p(2*j,1)];
        end
    end
    line([t(t_end,1),t(t_end,1)],Px,Py,'Color','b');
    line([t(1,1),t(1,1)],Px1,Py1,'Color','k');
end
box on
xlabel t; ylabel x; zlabel y
legend({'Vi tri dau','Qui dao','Vi tri cuoi'});
view(3)

%%
function dpdt = control_law(t,xi)
    global L p_d
    dpdt = -kron(L,eye(2))*(xi-p_d);
end
\end{lstlisting}

\begin{figure}[h!]
    \centering
    \includegraphics[height=6.5cm]{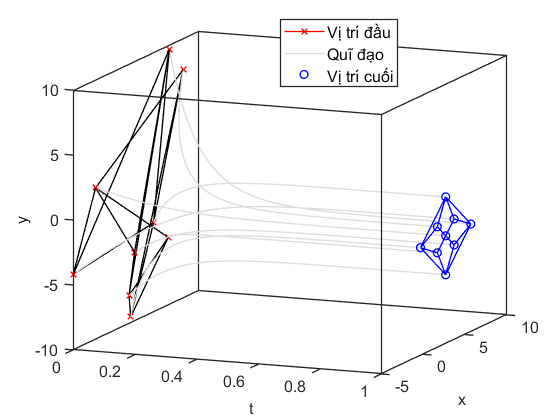}
    \caption{Thay đổi đội hình theo thời gian}
    \label{fig:C2_FCTime}
\end{figure}

\addcontentsline{toc}{chapter}{Chỉ mục}
\printindex
\bibliographystyle{abbrv}
\bibliography{references}

\newpage
\thispagestyle{empty}
\mbox{}
\newpage

\begingroup
\thispagestyle{empty} 
\begin{tikzpicture}[remember picture,overlay]
\node[inner sep=0pt] (background) at (current page.center) {\includegraphics[width=\paperwidth]{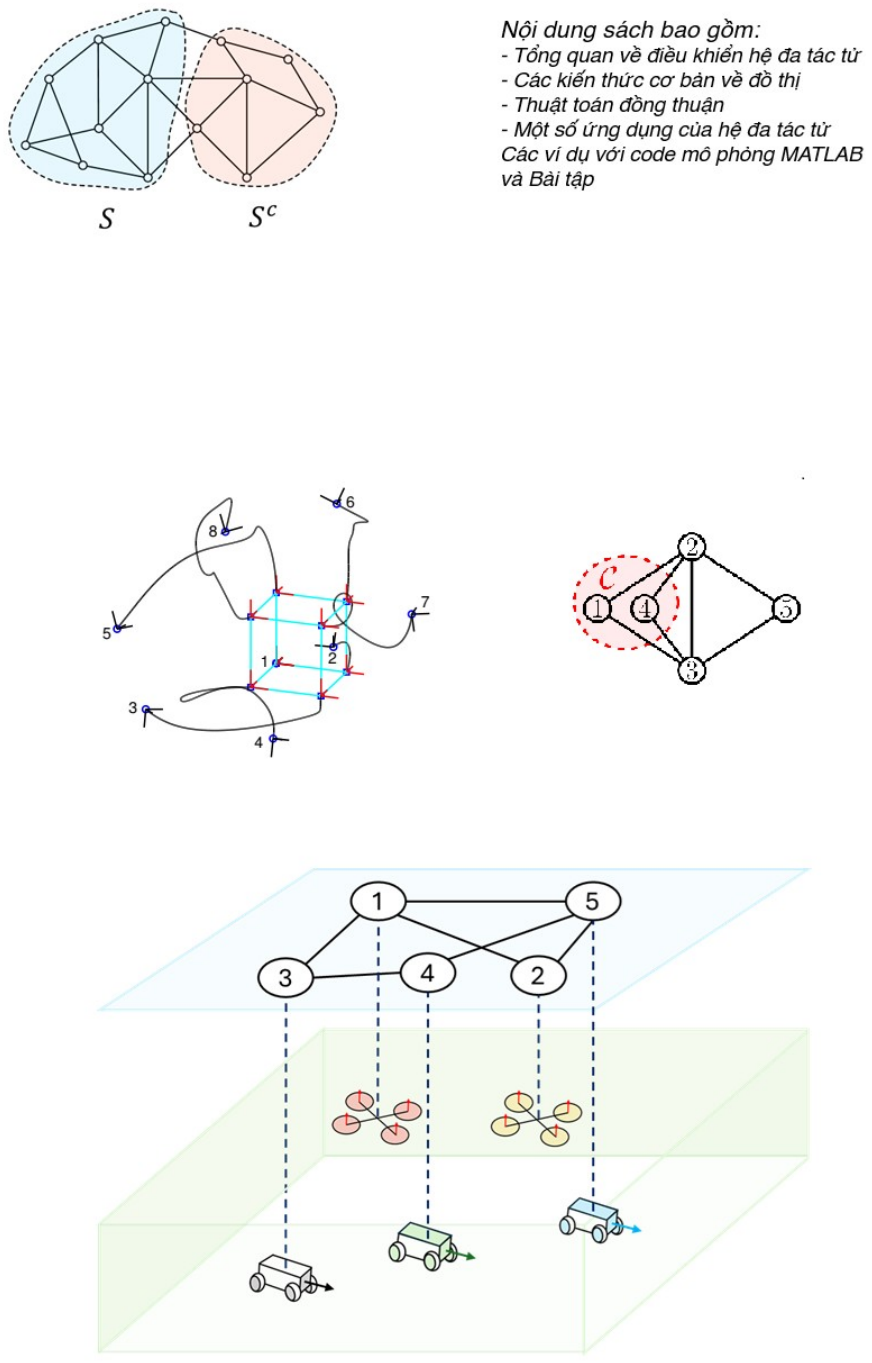}};
\node[inner sep=0pt] (background) at (current page.center) {};
\draw (current page.center) node [fill=blue!10,fill opacity=0.8,text opacity=.8,inner sep=2cm]{\Huge\centering\bfseries\sffamily\parbox[c][][t]{\paperwidth}{\centering \textcolor{blue!10}{Control of multiagent systems}\\[16pt]
{\Large \textcolor{blue!10}{Minh Hoang Trinh, Hieu Minh Nguyen}}}};
\end{tikzpicture}
\vfill
\endgroup

\end{document}